\documentclass[twocolumn,amsmath,amssymb,longbibliography,aps,prb]{revtex4-2}

\usepackage{graphicx}
\usepackage{hyperref}
\hypersetup{
  colorlinks=true,
  linkcolor=blue,   
  citecolor=blue,  
  urlcolor=black     
}
\usepackage{bm}
\usepackage{color}
\usepackage{soul}
\usepackage{braket}
\usepackage{lipsum}
\usepackage{mathdots}
\usepackage[version=3]{mhchem}
\usepackage{appendix}
\usepackage{ulem}
\usepackage{xspace}
\usepackage{url}
\usepackage{txfonts}
\usepackage{subfiles}

\newcommand  {\eqn}[1]{(\ref{eqn:#1})} 
\renewcommand{\(}     {\left(}
\renewcommand{\)}     {\right)}
\renewcommand{\[}     {\left[}
\renewcommand{\]}     {\right]}
\renewcommand{\_}[1]  {_\textrm{#1}}

\newcommand{\CSS}  {\ce{Co3Sn2S2}\xspace}
\newcommand{\TMA}  {\ce{Ti2MnAl}\xspace}
\newcommand{\SRO}  {\ce{SrRuO3}\xspace}

\begin{document}

\title{
Magnetic Weyl semimetals: 
Interplay of band topology and magnetism
}

\newcommand{\affiA}{$^1$Institute for Solid State Physics, The University of Tokyo, Kashiwa 277-8581, Japan}
\newcommand{\affiB}{$^2$Advanced Science Research Center, Japan Atomic Energy Agency, Tokai, Ibaraki 319-1195, Japan}
\newcommand{\affiC}{$^3$Physics Division, Sophia University, Tokyo 102-8554, Japan}

\newcommand{\affiD}{$^4$Department of Physics, Kyushu University, Fukuoka 819-0395, Japan}

\newcommand{\affiE}{$^5$Quantum and Spacetime Research Institute, Kyushu University, Fukuoka 819-0395, Japan}


\author{Akihiro Ozawa$^1$}
\email{akihiroozawa@issp.u-tokyo.ac.jp} 
\altaffiliation{\\Present address:
Institut f\"ur Theoretische Physik, Universit\"at Leipzig, Br\"uderstra\ss e 16, 04103 Leipzig, Germany}
\author{Yasufumi Araki$^2$}
\email{araki.yasufumi@jaea.go.jp}
\author{Koji Kobayashi$^3$}
\email{k-koji@sophia.ac.jp}
\author{Kentaro Nomura$^{4,5}$}
\email{nomura.kentaro@phys.kyushu-u.ac.jp}

\affiliation{
    \affiA\\
    \affiB\\
    \affiC\\
    \affiD\\
    \affiE\\
}

\newcommand{\abstbody}{
We review recent theoretical and experimental developments in magnetic Weyl semimetals,
focusing on the electromagnetic responses emerging from the interplay of their electronic band topology and magnetism.
We begin by introducing the fundamental topological properties of electrons in Weyl semimetals,
and provide an overview of the characteristic phenomena arising from their band topology, such as the anomalous Hall effect and chiral magnetic effect.
The materials exhibiting the magnetic Weyl semimetal state,
including ferromagnetic, antiferromagnetic, and noncentrosymmetric systems, are reviewed.
The possible mechanisms for their magnetism
are discussed in connection with the Weyl electrons.
Non-uniform magnetic textures and magnetization dynamics are expected to exhibit a topological interplay with the Weyl electrons,
manifesting as spinmotive force and spin torques.
We also review magnetotransport phenomena
such as domain wall magnetoresistance, studied by mesoscopic scale calculations.
Finally, we discuss the spin transport properties studied in magnetic Weyl semimetals.
The topological nature of Weyl electrons reviewed here is important not only for fundamental physics,
but also for potential application in low-dissipative electronics and spintronics devices.
}

 \begin{abstract}
  \abstbody
 \end{abstract}

\maketitle

\begingroup
\hypersetup{linkcolor=black}
\tableofcontents
\endgroup


\section{Overview}\indent
\label{sec:overiew}
Topological phases in electronic systems
have been extensively investigated as a new class of quantum phases.
Weyl semimetals are significant case of such topological phases,
characterized by the pointlike band touching in momentum space \cite{Wan2011, Burkov2011, Armitage2018,burkov2018weyl}.
Due to the peculiar band topology in momentum space,
Weyl semimetals exhibit various magnetoelectric properties, yielding cross correlation between the electric 
(e.g., electric field, current)
and the magnetic 
(e.g., magnetic field, magnetization, spin textures) degrees of freedom.
To achieve such cross-correlated magnetoelectric effects,
Weyl semimetals showing magnetism,
which are called magnetic Weyl semimetals,
are attracting a great attention.
This paper is devoted to the review of the series of studies on magnetic Weyl semimetals.

The whole history of the research on Weyl semimetals can be traced back to Weyl's pioneering theory in 1929 \cite{weyl1929gravitation}.
After the theoretical description of the relativistic massive fermion by Dirac \cite{dirac1928quantum},
Weyl introduced the relativistic description of the massless chiral fermion in the context of high-energy physics,
which is now called the Weyl fermion.
The Weyl fermion is labeled by the quantum number called the chirality,
which characterizes its spin being either parallel (right-handed, $+1$) or antiparallel (left-handed, $-1$) to its momentum.
Whereas the Dirac fermion allows a mass term mixing right- and left-handed chiralities,
the Weyl fermion is massless 
so that each chirality is kept isolated.
Such a single-chirality description was anticipated by Pauli to explain the beta decay by incorporating neutrino, 
which is used to be considered as
the elementary particle without mass~\cite{bethe1934neutrino}.
While the descriptions of the Dirac and Weyl fermions were originally defined in the continuum,
they were also successfully formulated and studied on hypothetical lattices,
for the computational simulations of quantum field theories such as quantum chromodynamics (QCD) \cite{kogut1979introduction,gattringer2009quantum,ratti2018lattice}.

As an analog of elementary particles in condensed matter,
the quest for relativistic fermions, including Weyl fermions, also has a long history.
It was proposed by Wallace in 1947~\cite{wallence1947} that a single layer of honeycomb lattice extracted from graphite hosts two-dimensional~(2D) relativistic Dirac fermions,
which is the origin of the vast series of recent studies on graphene~\cite{Neto2009,Sarma2011-pc,Bistritzer2011}. 
The idea of 2D Dirac fermions was later applied also to 
organic conductors~\cite{Katayama2006,Kobayashi2007},
transition metal dichalcogenides \cite{Manzeli2017-fi},
etc.

In three dimensions (3D),
Volovik pointed out in 1987 the existence of the Weyl-type quasiparticle excitation in the A-phase of the superfluid $^3$He \cite{volovik1987zeros,volovik2003universe}.
For crystals without inversion symmetry,
Murakami proposed in 2007 the emergence of gapless linear dispersion in the band structure,
at the critical point of topological phase transition
between topological insulator and normal insulator~\cite{Murakami2007}.
Both in 2D and 3D,
exploring peculiar properties originating from these relativistic fermions has become one of the central interests in condensed matter physics in recent years.
%

The studies on Weyl-type excitations in magnetic materials started in the early 2000s.
In 2001, Shindou and Nagaosa theoretically proposed the Weyl fermion excitations on a lattice model with a triple-$Q$ antiferromagnetic ordering~\cite{shindou2001orbital}.
In 2003, experimental observations by Fang {\it et al.} on the ferromagnetic \ce{SrRuO_3} confirmed the effect of Weyl fermions on the anomalous Hall effect \cite{Fang2003}.
Later in 2011,
the terminology ``Weyl semimetal'' appeared in two theoretical studies,
one on the antiferromagnetic pyrochlore lattice by Wan {\it et al.} \cite{Wan2011},
and the other on a magnetic superlattice of topological insulator by Burkov and Balents \cite{Burkov2011}.
Following these pioneering works, 
theoretical and experimental studies in search of
Weyl semimetal phases in various materials have been intensely conducted until now. 
At the early stage,
the studies on Weyl semimetals were mostly focused on nonmagnetic materials with broken inversion symmetry,
such as \ce{TaAs} \cite{Huang15a, xu2015discovery, lv2015experimental, Yang15Weyl, lv2015observation, huang2015observation}.
To realize the magnetic Weyl semimetal phase 
with time-reversal symmetry broken,
a great deal of effort was devoted. 
The first experimental identification of Weyl fermions in magnetic material was in the chiral antiferromagnet \ce{Mn3Sn} using the angle-resolved photoemission spectroscopy (ARPES)\cite{Kuroda2017}
in 2017.
Nowadays,
both theoretical and experimental studies have achieved a wide variety of magnetic Weyl semimetal materials with magnetic orderings,
including ferromagnetic, antiferromagnetic, ferrimagnetic, and helimagnetic orderings.

Owing to the great success in the material realization for magnetic Weyl semimetals, 
experimental verifications of the characteristic properties suggested in the early-stage theories,
originating from the band topology of Weyl fermions,
are intensely going on.
The most typically observed phenomenon in magnetic Weyl semimetals is the anomalous Hall effect~\cite{Burkov2011},
which arises as the intrinsic effect from the band topology known as the Berry curvature~\cite{Nagaosa2010,xiao2010berry}.
An extremely strong anomalous Hall effect has been reported in some species of magnetic Weyl semimetal materials, in comparison with that in 
conventional
magnetic materials.

Besides the anomalous Hall effect,
various magnetoelectric effects
have been expected in magnetic Weyl semimetals,
motivated by the idea developed for the long time in relativistic quantum field theory.
Such magnetoelectric effects manifest the cross-correlated phenomena between the electric and magnetic degrees of freedom.
For example, one may achieve a control of spin by electric current or voltage~(see section 3).
From the practical point of view,
these effects are expected to be applied to the development of future electronics and spintronics devices with high functionalities and low power consumption.
To make use of such characteristic properties,
numerous theoretical and experimental efforts are now in progress in magnetic Weyl semimetals.

In this paper,
we review the recent development in the theoretical and experimental studies on the magnetic Weyl semimetals, where the interplay between band topology and magnetism becomes significant. 
In Sec.~\ref{sec:fundamentals}, 
we review the fundamental properties of the Weyl fermions in crystalline systems from the theoretical aspect.
We first introduce the theoretical formalism of the Weyl fermions,
and list up the possible magnetoelectric effects arising from the topological nature of Weyl fermions.
In Sec.~\ref{sec:materials},
we explain the species of magnetic materials exhibiting the Weyl fermions.
We list up various ferromagnets, antiferromagnets, and several other types of magnets,
and explain the current status of experiments therein.
In Sec.~\ref{sec:magnetism},
we discuss the mechanism for forming the magnetic orderings in magnetic Weyl semimetal materials.
We especially focus on
the cases where the topological nature of electrons also contributes to the magnetism.
In Sec.~\ref{sec:texture},
we move on to the magnetoelectric effects related to nonuniform magnetic textures,
such as magnetic domain walls, spirals, and skyrmions,
which commonly appear in magnetic materials.
The magnetoelectric effects on such textures play the indispensable roles in manipulating the magnetic orderings, leading to future spintronics functionalities.
In Sec.~\ref{sec:transport},
we review the transport properties in magnetic Weyl semimetals under a disorder effect,
referring to some numerical calculation studies.
We focus on the magnetotransport properties, i.e., the electric transport related to the magnetism,
and discuss their robustness against disorder.
In Sec.~\ref{sec:spin-transport},
we see the studies on spin transport properties in magnetic Weyl semimetals,
which will be important in the electric manipulation of spins for spintronics applications.
Finally, in Sec.~\ref{sec:conclusion},
we conclude our review with giving prospects on the future studies on magnetic Weyl semimetals.

Throughout this review,
we take the International System of Units (SI),
where the speed of light $c$, Planck constant $h = 2\pi \hbar$,
and the Boltzmann constant $k_{\rm B}$ are written explicitly.
We mainly explain the studies characteristic to \textit{magnetic} Weyl semimetals.
For further review on the current status applying also to nonmagnetic Weyl semimetals,
the preceding review paper may also be referred \cite{Armitage2018,burkov2018weyl}.

\section{Topological properties of magnetic Weyl semimetals}~
\label{sec:fundamentals}

In this section, 
we review the fundamentals of the theory of magnetic Weyl semimetals.
First, we review relativistic quantum mechanics, focusing on the properties of Dirac and Weyl fermions.
Next, we recall Bloch’s theorem for crystals and explain how Dirac and Weyl electrons arise in solid states.
We then introduce a specific model of Weyl fermions to demonstrate the effects of magnetic ordering, topological properties, symmetries, transport phenomena, and surface states.
Finally, we explain the electronic responses to electromagnetic fields,
in connection with the chiral gauge fields and the chiral anomaly characteristic to Weyl fermions.

\subsection{Relativistic quantum mechanics: Dirac and Weyl Hamiltonian} \label{sec:relativistic} \indent

Before visiting the theory of Weyl fermion in solid states, 
we begin with the fundamental theory of relativistic quantum mechanics.
In contrast to the Klein--Gordon equation for the relativistic bosons with spin 0 \cite{klein1926quantentheorie,gordon1926comptoneffekt},
Dirac proposed a linearized 
wave equation for the fermions with spin $1/2$ satisfying the Lorentz symmetry,
by introducing the multi-component spinor representation of the fermion \cite{dirac1928quantum}.
This formalism consists of the Dirac equation,
\begin{align}
    i\hbar\frac{\partial}{\partial t} \Psi(\boldsymbol{r},t) = H_D(\boldsymbol{p}) \Psi(\boldsymbol{r},t),
    \label{eq:Dirac-eq}
\end{align}
with the four-component wave function (spinor) $\Psi(\boldsymbol{r},t)$
and the Dirac Hamiltonian

\begin{eqnarray}
H_D({\bm p})= cp_x \alpha_1+ cp_y \alpha_2+ cp_z \alpha_3+mc^2 \alpha_4,
  \label{eq:MassiveDirac}
\end{eqnarray}
where $p_{i=x,y,z} = -i\hbar \partial_i$ is the momentum operator,
$c$ is the speed of light,
and $m$ is the mass.
The matrices $\alpha_{\mu=1,2,3,4}$ in Eq.~(\ref{eq:MassiveDirac}) are defined to satisfy the Clifford algebra, i.e., the anticommutation relation $\{\alpha_\mu,\alpha_\nu\} = 2\delta_{\mu\nu}$,
so that the relativistic relationship between energy and momentum, $E^2 = c^2 |\bm{p}|^2 + (mc)^4$, is satisfied.
The energy eigenvalues of the Dirac Hamiltonian 
are thus
$E(\boldsymbol{p}) = \pm c \sqrt{|\boldsymbol{p}|^2 + (mc)^2}$,
which acquires a gap around zero energy for $m \neq 0$.
In $(3+1)$-dimensions, the matrices $\alpha_{\mu=1,2,3,4}$ 
can be represented in the $4\times 4$ form,
\begin{align}
    \alpha_{i=1,2,3} = \begin{pmatrix}
        0 & \sigma_i \\
        \sigma_i & 0
    \end{pmatrix}, \quad 
    \alpha_4 = \begin{pmatrix}
        1 & 0 \\
        0 & -1
    \end{pmatrix}
    , \label{eq:alpha-Dirac}
\end{align}
with $\sigma_{i=1,2,3}$ the Pauli matrices corresponding to the spin.
This representation is called the Dirac(--Pauli) representation.
In this Dirac representation,
the upper two components of the spinor $\Psi$ correspond to the electron with a positive energy,
and the lower two components to negative energy.
The $\alpha$-matrices can be written as the Kronecker products of two Pauli matrices, $\sigma$ and $\tau$,
$(\alpha_1,\alpha_2,\alpha_3,\alpha_4) = (\tau_x \sigma_x, \tau_x \sigma_y, \tau_x \sigma_z, \tau_z \sigma_0)$.
Here, the Pauli matrices ${\bm \tau}$ act on the upper and lower energy components of the spinor $\Psi$.

The matrix representation of the Clifford algebra is not uniquely the Dirac representation.
Another typical form related by a unitary transformation is the Weyl representation,
\begin{align}
    \alpha_{i=1,2,3} = \begin{pmatrix}
        \sigma_i &  0 \\
         0 & -\sigma_i
    \end{pmatrix}, \quad
    \alpha_4 = \begin{pmatrix}
        0 & 1 \\
        1 & 0
    \end{pmatrix}. \label{eq:alpha-Weyl}
\end{align}
With the Weyl representation, the matrix form of the Dirac Hamiltonian becomes,
\begin{align}
    H_D(\boldsymbol{p}) 
    = \begin{pmatrix}
        c \bm{\sigma}\cdot \bm{p} & mc^2 \\
        mc^2 & -c \bm{\sigma}\cdot \bm{p}
    \end{pmatrix}. \label{eq:weyl-repr}
\end{align}
In this Weyl representation,
the Dirac equation [Eq.~(\ref{eq:Dirac-eq})] leads,
\begin{align}
    i\hbar\frac{\partial}{\partial t}
        \Psi({\bm r},t )
    =
    \begin{pmatrix}
        c \bm{\sigma}\cdot \bm{p} & mc^2 \\
        mc^2 & -c \bm{\sigma}\cdot \bm{p}
    \end{pmatrix}
     \Psi({\bm r},t )
    \label{eq:Dirac-Weyl-eq}.
\end{align}
In case $m=0$,
Eq.~(\ref{eq:Dirac-Weyl-eq}) becomes block diagonal with
\begin{eqnarray}
  H_{\eta =\pm}({\bm p}) = \eta c {\bm \sigma}\cdot {\bm p},
  \label{eq:low-energy}
\end{eqnarray}
with the label $\eta=\pm$.
This leads to the two-component equation,
\begin{align}
    i\hbar\frac{\partial}{\partial t} \psi_\eta = H_{\eta}({\bm p}) \psi_\eta,
\end{align}
where $\psi_{\eta=\pm}$ are the two-component spinors with $\Psi = \begin{pmatrix} \psi_+ \\ \psi_- \end{pmatrix} $,
corresponding to the wave function for Weyl fermions.
We emphasize $\eta$ distinguishes the species of the Weyl fermions.
Here, $\psi_+$ corresponds to the \textit{right-handed} fermion state
with its momentum parallel to spin $({\bm p} \parallel {\bm \sigma})$,
whereas $\psi_-$ corresponds to the \textit{left-handed} state (${\bm p} \parallel {-\bm \sigma}$).
Such right- or left-handedness labeled by $\eta$ is called the chirality.
This is the case Weyl considered in his pioneering study~\cite{weyl1929gravitation}.
Each $2\times 2$ Hamiltonian
\begin{align}
H_{\eta}({\bm p})=\eta c
\begin{pmatrix}
    p_z& p_x-ip_y\\
    p_x+ip_y & -p_z
  \end{pmatrix}
\end{align}
is now called the Weyl Hamiltonian,
which governs the massless relativistic fermion in terms of the 2-component spinor $\psi_\pm$.
In the four-component spinor representation,
the right-handed state $\Psi_+ = \begin{pmatrix} \psi_+ \\ 0 \end{pmatrix}$
and the left-handed state state $\Psi_- = \begin{pmatrix} 0 \\ \psi_- \end{pmatrix}$
are given as the eigenstates of the matrix $\Gamma_5 = - i \alpha_1 \alpha_2 \alpha_3 = \mathrm{diag} (+1,+1,-1,-1)$,
called the chirality operator, with the eigenvalues $\pm 1$.
The energy eigenvalues of the Weyl fermions become $E_\pm(\boldsymbol{p}) = \pm c|\boldsymbol{p}|$, which is 
linear and gapless. 

\subsection{Electronic bands in solid states: insulator, metal, and  semimetal}\indent

\begin{figure}[htb]
    \centering
    \includegraphics[width=0.7\linewidth]{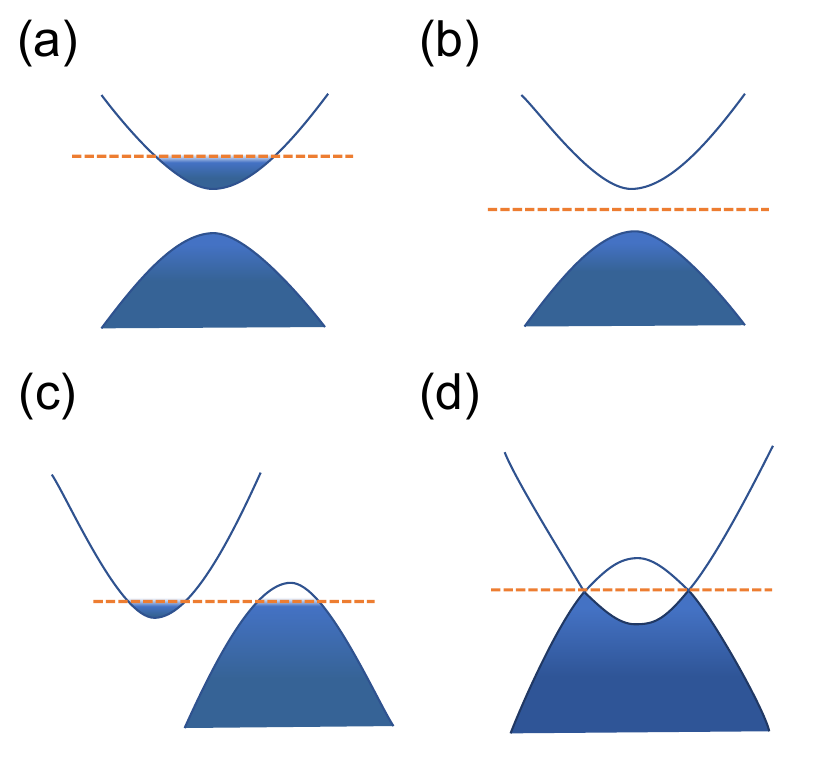}
    \caption{
(Color-online) 
Schematic illustrations of the electronic band structures of 
(a)~metal,
(b)~insulator,
(c)~semimetal, and 
(d)~(ideal)~Weyl semimetals.
The orange dashed lines represent the Fermi level (chemical potential).
}
    \label{fig:semimetal}
\end{figure}

The theoretical formulations of Dirac and Weyl fermions explained above were originally introduced to describe the elementary particles in the relativistic regime in high-energy physics.
Nevertheless,
such relativistic descriptions are applicable to quasiparticle excitations in solid states in some cases,
even though the dynamics of electrons in solid states themselves are originally described nonrelativistically.
The materials called Dirac or Weyl semimetals apply to such cases.

To understand the behavior of electrons in solid states,
we first briefly review the electronic band theory in crystalline systems based on Bloch's theorem.
Since the crystals are spatially periodic, the electron wave functions
are governed by the Hamiltonian which is
invariant under the translational operation with primitive lattice vector(s) ${\bm T}$,
$\hat{H}({\bm r}) = \hat{H}({\bm r} + {\bm T})$.
The Bloch's theorem demands that the eigenfunction of this Hamiltonian should satisfy
$\phi({\bm r}) =u_{\bm k}({\bm r}) e^{+i {\bm k}\cdot {\bm r}}$,
where 
$u_{\bm k}({\bm r})$ is a periodic function satisfying $u_{\bm k}({\bm r}+{\bm T})=u_{\bm k}({\bm r})$.
Here, ${\bm k }$ is the crystal wave vector,
which is defined in the reciprocal space (Brillouin zone) determined by ${\bm T}$.
The wave vector is usually identified as the momentum ${\bm p} = \hbar {\bm k}$.
The eigenequation for the Bloch wave function $u_{n{\bm k}}({\bm r})$ becomes $\hat{H}({\bm k},{\bm r}) u_{n{\bm k}}({\bm r}) = E_{n{\bm k}} u_{n{\bm k}}({\bm r})$, with $\hat{H}({\bm k},{\bm r}) \equiv e^{-i{\bm k}\cdot {\bm r}} \hat{H}({\bm r}) e^{+i{\bm k}\cdot {\bm r}}$.
By considering the microscopic ${\bm r}$-dependence in $u_{n{\bm k}}({\bm r})$ as the internal degrees of freedom for the vector $\ket{u_{n {\bm k}}}$,
this eigen equation is written with the bracket notation as
$H(\bm k) \ket{u_{n {\bm k}}}=E_{n{\bm k}}\ket{u_{n {\bm k}}}$.
Here $n$ is the band index labeling the eigenstate of $H(\bm k)$.
The structures of the bands $E_{n{\bm k}}$ and the eigenstates $\ket{u_{n {\bm k}}}$ are determined by the internal degrees of freedom of the electrons, such as spin, orbital, and sublattices,
which govern the electronic properties characteristic to each material.

The band structure around the Fermi level $E_{\rm F}$, up to which electrons are filled,
determines the metallicity of materials.
If $E_{\rm F}$ is located inside the band gap, which is the energy region without any energy eigenstates, the system is an insulator.
In Fig.~\ref{fig:semimetal}, electronic band structures for 
(a)~metal,
(b)~insulator,
(c)~semimetal, and
(d)~Weyl semimetal
are schematically shown.
The band structure of (a)~metal shows a nonzero density of states~(DOS)
$\rho(E_{\rm F}) = \frac{1}{V} \sum_{n,{\bm k}} \delta(E_{n{\bm k}} - E_{\rm F})$
near $E_{\rm F}$,
and hence the electrons can be easily excited beyond $E_{\rm F}$ to contribute to electric conduction.
Typically, the longitudinal conductivity $\sigma_{xx}$ depends on the DOS at $E_{\rm F}$.
Meanwhile,
a large number of electrons near the Fermi level $E_{\rm F}$ are scattered by the impurities,
causing the Joule heating.
In contrast, (b)~insulator shows a wide band gap between the valence (lower) and conduction (upper) bands, and needs a relatively large excitation energy for the electric conduction.
If the band gap is narrow,
some carriers can be excited and contribute to electric conduction,
which is identified as a semiconductor.
As the intermediate situation between metal and insulator,
(c) semimetal is characterized with the valence and conduction bands having a small overlap in energy, to exhibit relatively small Fermi surfaces and low carrier density.
One typical example for semimetal is elemental bismuth,
which shows various exotic electronic properties from its low carrier density ~\cite{fuseya2015}.

\subsection{Electronic bands of Weyl semimetal}

Weyl semimetal is identified as the extreme case of semimetals.
By shrinking the energy overlap between the valence and conduction bands in semimetals,
one ultimately reaches the situation where these bands touch at some points in the momentum space.
In particular, we consider the case where the energy-momentum dispersion around the band touching point is linear,
as shown in Fig.~\ref{fig:semimetal}(d).
In this case, the low-momentum excitations around this point are necessarily described by the Weyl Hamiltonian in Eq.~(\ref{eq:low-energy}),
except for the speed of light $c$ replaced with a material-dependent Fermi velocity $v\_F$,
\begin{align}
    H_\eta(\boldsymbol{k}) &= \eta \hbar v_{\rm F} \boldsymbol{k} \cdot \boldsymbol{\sigma}.
    \label{eqn:Weyl_sml}
\end{align}
Here, the Pauli matrix ${\bm \sigma}$ acts on the two degrees of freedom forming the band touching.
Note that ${\bm \sigma}$ does not necessarily correspond to the spin degrees of freedom appearing in the relativistic Weyl fermions,
but may involve other degrees of freedom,
such as atomic orbital, sublattice, etc.
Depending on the crystal structure of each material,
the Fermi velocity $v_{\rm F}$ may become anisotropic (direction dependent).
This linear band touching is the definition of the Weyl semimetal,
and the band touching point is called the Weyl point.
We here note that this Hamiltonian uses all three Pauli matrices,
and hence we cannot open a gap in the spectrum.
Even if we add a mass term, such as $\Delta \sigma_z$,
this induces a shift of the Weyl point in the $k_z$-direction and cannot open a gap.
This is why the Weyl points are robust against perturbation.
From the discussion of band topology, the Weyl points in crystals should always appear in pair(s), as we shall explain in Sec.~\ref{sec:band-topology}.
To realize the Weyl point structure in the electronic bands,
the time-reversal or spatial inversion symmetry needs to be broken.
The breaking of time-reversal symmetry is necessarily satisfied in magnetic materials.

Ideally, the terminology ``Weyl semimetal'' applies to the case where $E_{\rm F}$ is located close enough to the energy of the Weyl point,
so that
the DOS gets suppressed and the electric conduction becomes almost negligible.
In reality, in most materials showing the Weyl points,
$E_{\rm F}$ is not necessarily located close to the Weyl point.
In such cases, the system acquires a finite Fermi surface and a sizable DOS,
and becomes metallic.
Such a system is sometimes called ``Weyl metal'' instead of semimetal,
whereas ``Weyl semimetal'' is also conventionally used in a vast literature.

\begin{figure}[htb]
    \centering
    \includegraphics[width=0.8\linewidth]{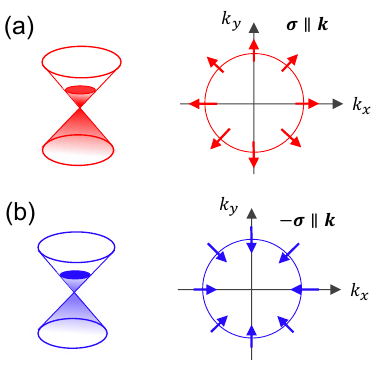}
    \caption{
(Color-online)~
The spin configurations in momentum space for the Weyl fermions with spin-momentum locking,
described by the Hamiltonian in Eq.~(\ref{eqn:Weyl_sml}).
Panels (a) and (b) correspond to the Weyl fermions with the positive $(\eta = +)$ and negative $(\eta = -)$, respectively.
}
    \label{fig:sml}
\end{figure}

The Weyl Hamiltonian of Eq.~(\ref{eqn:Weyl_sml}) shows the correlation between momentum ${\bm k}$ and the internal degrees of freedom ${\bm \sigma}$.
The correlation becomes significant if
the Fermi energy is located near the energy of the Weyl points.
If $\boldsymbol{\sigma}$ corresponds to the spin degrees of freedom, the correlation between $\boldsymbol{k}$ and $\boldsymbol{\sigma}$ is
called the ``spin-momentum locking'' in Weyl semimetal.
Figure~\ref{fig:sml} schematically shows the spin configurations of the electrons, described as the eigenstates of $H_\eta({\bm k})$ in momentum space.
Here, $E_{\rm F}$ is set slightly beyond the Weyl points, which yields the Fermi surfaces spherical around the Weyl points.
For $\eta = +$,
the directions of spin and momentum are aligned parallel (${\bm \sigma} \parallel {\bm k}$), as shown in Fig.~\ref{fig:sml}(a). 
In contrast, for $\eta = -$,
the spin and momentum are aligned antiparallel~($ - {\bm \sigma} \parallel {\bm k}$), as shown in Fig.~\ref{fig:sml}(b).
In reality, the structure and presence or absence of spin-momentum locking depend on the spin-orbit coupling for each material.
Several characteristic spintronic functionalities have been proposed based on the spin-momentum locking.
We note that this spin-momentum locking structure is not necessarily present in realistic magnetic Weyl materials where ${\bm \sigma}$ is not real spin, which will be explained in Sec.~\ref{sec:materials}.

\subsection{Early-stage proposals of magnetic Weyl semimetals}\indent
\label{subsec:early}

Before going into the details of the electronic properties of magnetic Weyl semimetals,
we review the early-stage proposals of Weyl semimetals.
The earliest proposals of Weyl semimetals were based on the breaking of time-reversal symmetry by the spontaneous magnetic ordering.

After the preceding studies on Weyl-type excitations in magnetic materials \cite{shindou2001orbital,fang2012multi},
it was in 2011 when the terminology ``Weyl semimetal'' first appeared in literature.
One of the earliest theoretical studies was done by Wan \textit{et al.},
focusing on the iridates such as \ce{Y2Ir2O7} having the pyrochlore lattice structure \cite{Wan2011}.
In this work, they studied the spontaneous magnetic ordering and the corresponding band structure by first-principles calculations.
Among the several configurations of spins on the pyrochlore lattice,
they found that the so-called ``all-in-all-out'' ordering, where the spins are aligned noncollinear in each unit cell, gives rise to the Weyl point structure.
By the downfolding to the tight-binding model,
they also found the existence of the surface states characteristic to Weyl semimetals,
called Fermi arcs,
which we shall explain in Sec.~\ref{sec:fermi-arc}.
this paper is recognized as a milestone paper
as the earliest proposal of Weyl semimetal phase and the corresponding Fermi arc surface states.
Recent theoretical studies have further considered the relation between the possible spin configurations on such a pyrochlore lattice and the nodal structures in detail,
treating the electron correlation by the mean-field and renormalization group analyses~\cite{goswami2017competing,szabo2021interacting,ladovrechis2021competing}.

In the same year 2011,
another notable theoretical work on magnetic Weyl semimetal was conveyed by Burkov and Balents \cite{Burkov2011}.
They considered the multilayer superlattice of magnetic topological insulators,
where the 2D Dirac fermions on the surface of each layer can tunnel to its adjacent layers and become 3D.
Depending on the magnitudes of the interlayer hoppings,
they found the emergence of the Weyl fermions in the 3D band structure.
The important finding in this work is the semi-quantized anomalous Hall conductivity, which is one of the important characteristics in magnetic Weyl semimetals and shall be explained in Sec.~\ref{sec:anomalous-hall}.
Motivated by these two pioneering works,
a lot of theoretical and experimental studies on Weyl semimetals have been performed until now.

\begin{figure}[htb]
    \centering
    \includegraphics[width=0.8\linewidth]{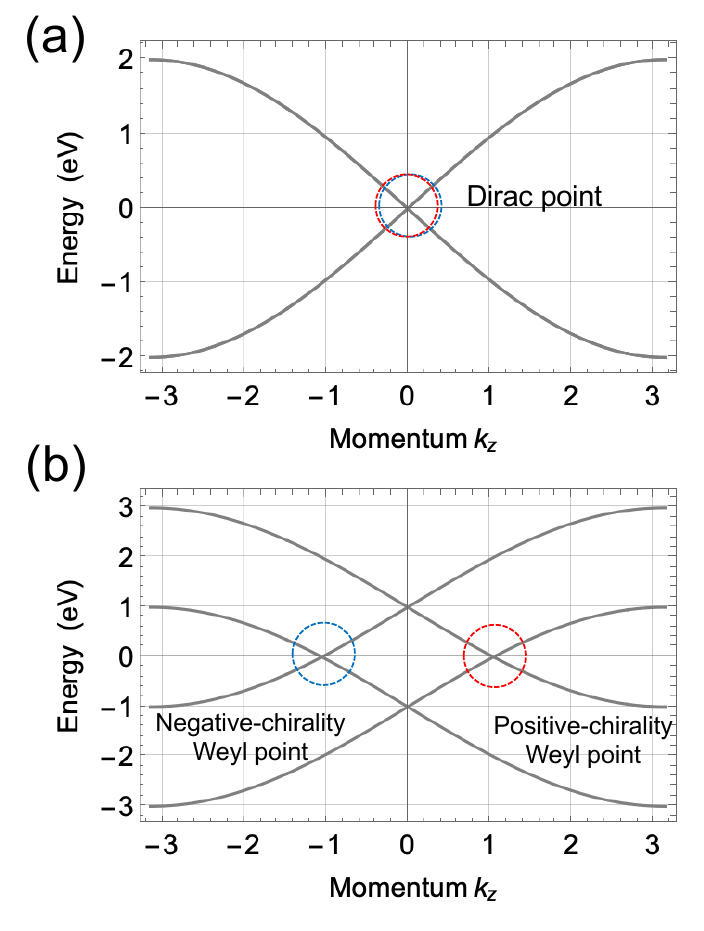}
    \caption{(Color-online)~
    The band structures of (a)~the Dirac semimetal state and (b)~magnetic Weyl semimetal state,
    calculated from the Wilson--Dirac model [Eq.~(\ref{eqn:HWD_k})].
    }
    \label{fig:wd_band}
\end{figure}

\subsection{Lattice model of Weyl fermions}
\label{sec:lattice}

To understand the features of Weyl fermions in crystals,
one needs a Hamiltonian defined on a lattice respecting the spetial periodicity.
While the lattice structure is unique to each material,
the Hamiltonian formulated on a hypothetical cubic lattice can also reproduce the Weyl (or Dirac) fermions.
Here, we review one of the tight-binding formulations of Weyl (and Dirac) fermions, called the Wilson--Dirac model.
Some other examples of tight-binding models are explained in detail in Appendix.
Such lattice fermion formulations were invented originally for the numerical simulations,
especially for the interactions of quarks and gluons in the context of lattice QCD \cite{creutz1983quarks,rothe2012lattice}.

\subsubsection{Dirac semimetal}

The Dirac and Weyl fermions
introduced in Sec.~\ref{sec:relativistic} can be reproduced as low-energy excitations in a Hamiltonian defined on a lattice.
Among several possible formulations of lattice Hamiltonians for Dirac fermions,
one of the commonly used models is the Wilson--Dirac model \cite{Wilson1977},
\begin{align}
 H\_{WD}(\bm{k}) &= t\sum_{i = x,y,z}  \sin{ (k_i a)}\alpha_i  + m(\bm{k}) \alpha_4,   \label{eqn:HWD_k}\\
 m(\bm{k}) &= m_0 + m_2 \sum_{i=x,y,z} \left[ 1 -\cos {(k_i a)} \right], \label{eq:WD-mass}
\end{align}
where $a$ is the lattice spacing of the cubic lattice.
Here, $t$ and $m_{2}(>0)$ are the hopping parameters between the neighboring lattice sites.
With the Weyl representation [Eq.~(\ref{eq:alpha-Weyl})], the matrix form of $H_{\rm WD}({\bm k})$ becomes,
 \begin{align}
    H_{\rm WD}({\bm k}) =
    \begin{pmatrix}
        t \sum_i \sin{(k_i a)} \sigma_i & m({\bm k})\\
           m({\bm k}) & -t \sum_i \sin{(k_i a)} \sigma_i \\
    \end{pmatrix}, \label{eq:weyl-repr-lattice}
\end{align}

By expanding this Hamiltonian up to linear in $\boldsymbol{k}$ around $\boldsymbol{k} = 0$,
it reduces as,
\begin{align}
    H_{\rm WD}(\boldsymbol{k}) &= ta \sum_i k_i \alpha_i + m_0 \alpha_4 + O(k^2), \label{eq:WD-linearized}
\end{align}
which corresponds to the Dirac Hamiltonian [Eq.~(\ref{eq:MassiveDirac})] with the Fermi velocity $c = ta/\hbar$ and the mass $m = m_0/c^2$.
Thus, we can say that the Wilson--Dirac model describes the Dirac fermions in the low-energy region $(|E| \ll t)$.

Let us consider the energy spectrum of the Wilson--Dirac model in more detail.
Regardless of the representation of the $\alpha$-matrices,
owing to their anticommutation relations,
the eigenenergies of $H_{\rm WD}(\bm{k})$ in Eq.~(\ref{eqn:HWD_k}) become,
\begin{align}
 E\_{WD}(\bm{k}) &= \pm \sqrt{ \sum_{i} t^2 \sin^2{(k_i a)}+ m^2(\bm{k})},
 \label{eqn:E_WD}
\end{align}

where both the positive- and negative-energy states are doubly degenerate.
In the massless limit $m_0 = 0$,
the energy spectrum becomes gapless and linear at $\boldsymbol{k} = 0$.
This case is shown in Fig.~\ref{fig:wd_band}(a),
as a function of as a function of $k_z$ with $k_x=k_y=0$ fixed.
If such a gapless and doubly degenerate dispersion is realized in solid states,
this state is called the ``Dirac semimetal'',
and the gapless point is called the ``Dirac point'' or ``Dirac node''.
In the Dirac semimetal phase,
the Bloch state for each ${\bm k}$-point is degenerate,
which is guaranteed by the time-reversal and spatial inversion symmetry. 
We will discuss the role of these symmetries for the Dirac and Weyl semimetal states later in Sec.~\ref{sec:symmetry}.

The role of the coefficient $m_2$,
which does not appear in the linearized Dirac Hamiltonian [Eq.~(\ref{eq:WD-linearized})], is crucial to get a single Dirac point.
If we take $m_2 = m_0 = 0$,
the momentum-dependent mass term $m(\boldsymbol{k})$ becomes zero not only at $\boldsymbol{k} = (0,0,0)$ but also at $8(= 2^3)$ time-reversal invariant momenta,
\begin{align}
    a\boldsymbol{k} &= (0,0,0),(0,0,\pi),(0,\pi,0),(0,\pi,\pi), \\
    & \quad \quad (\pi,0,0),(\pi,0,\pi),(\pi,\pi,0),(\pi,\pi,\pi). \nonumber
\end{align}
Thus, the energy spectrum shows eight gapless nodes at these $\boldsymbol{k}$-points,
which means that this model exhibits eight species (or ``flavors'') of Dirac fermions at low energy.
To get a single species of Dirac fermion in the lattice model,
we need to gap out the seven unwanted $\boldsymbol{k}$-points,
which is achieved by setting $m_2 > 0$.
The term with this $m_2$ is called the ``Wilson term''.
In solid states,
the Wilson term corresponds to the effect of band inversion, e.g., by spin-orbit coupling.
which is essential in describing the Dirac semimetals and topological insulators as well~\cite{qi2011topological}.

Multiple Dirac points can also emerge in so-called topological Dirac semimetals, where they are protected by the crystalline symmetries of the material~\cite{Young2012}. 
Such Dirac points have been experimentally observed in materials like \ce{Na3Bi}~\cite{wang2012dirac,Liu2014} and \ce{Cd3As2}~\cite{wang2013three,neupane2014observation,uchida2017quantum,uchida2019ferromagnetic}.
The details of the modeling for the topological Dirac semimetal are explained in Appendix.

\subsubsection{Magnetic Weyl semimetal on lattice}

By introducing the effect of magnetic ordering that breaks the time-reversal symmetry,
the magnetic Weyl semimetal state can also be straightforwardly obtained from the Wilson--Dirac model.
We introduce the exchange coupling term between the magnetization ${\bm M}$ and the spins of conduction electrons ${\bf \Sigma}$,
\begin{align}
 H\_{exc} &= J\bm{M} \cdot {\bm \Sigma}.\label{eqn:H-exc} 
\end{align}
Here, the spin operator $\boldsymbol{\Sigma}$ in this model is defined as,
$\Sigma_{x} = -i \alpha_y \alpha_z$,
$\Sigma_{y} = -i \alpha_z \alpha_x$, and 
$\Sigma_{z} = -i \alpha_x \alpha_y$,
which reduces to $\boldsymbol{\Sigma} = {\rm diag} (\boldsymbol{\sigma}, \boldsymbol{\sigma})$ in the Weyl representation.

Due to the breaking of time-reversal symmetry,
the exchange coupling term splits the double degeneracy of the Dirac point,
leading to a pair of Weyl points to reach the magnetic Weyl semimetal state.
Figure \ref{fig:wd_band}(b) shows the band structure with the exchange coupling term,
with the magnetization $\boldsymbol{M}$ pointing in the $z$-direction.
The Dirac point is split into two Weyl points with positive and negative chiralities, $\boldsymbol{K}_\pm = (0,0, \pm K)$, shown as the red and blue points in Fig.~\ref{fig:wd_band}(b).
The distance $2K$ between the Weyl points depends on the magnitude of the magnetization $|{\bm M}|$ (see Appendix~\ref{sec:app:lattice:WD} for detail).
In particular, for $J|\boldsymbol{M}| \ll t$,
we can approximate,
\begin{align}
    K \approx  \frac{J|{\bm M}|}{\hbar v_{\rm F}},
\end{align}
where we use the relation $v\_F = at/\hbar$.
This relation clearly exhibits that the Dirac point at $\boldsymbol{k}=0$ is split into two Weyl points $\boldsymbol{K}_\pm (\neq 0)$ once the magnetization $\boldsymbol{M}$ is switched on.

Around each
Weyl point $\boldsymbol{K}_\pm$,
the Hamiltonian can be linearized as $H({\bm k}) = {\rm diag}\{ H_+({\bm k}), H_-({\bm k})\} + O(|\boldsymbol{k} - \boldsymbol{K}_\pm|^2)$, with the $2\times 2$ blocks
\begin{align}
    H_{\eta = \pm}(\boldsymbol{k}) = \eta \hbar v\_F (\boldsymbol{k} - \boldsymbol{K}_\eta) \cdot \boldsymbol{\sigma}, \label{eq:weyl-shifted}
\end{align}
which corresponds to the Weyl Hamiltonian given in Eq.~(\ref{eqn:Weyl_sml}).
This block diagonal form of $H({\bm k})$ is identical to that of massless relativistic Dirac Hamiltonian [$m=0$ in Eq.~(\ref{eq:MassiveDirac})],
except for the momentum shift ${\bm K}_\pm$.

We have so far treated the lattice Hamiltonian in momentum space,
which is useful to describe translationally symmetric systems.
On the other hand, to consider the systems without translational symmetry,
such as interfaces, non-uniform magnetic textures (magnetic domain walls, spirals, etc.), or disorder,
we need to switch from momentum to the real-space representation.
Noting that the Wilson--Dirac model is defined on the hypothetical cubic lattice with the lattice spacing $a$ in real space,
the 4-component creation operators $c^\dag_{\bm k}$ appearing in the Hamiltonian operator $\mathcal{H} =\sum_{\bm k} c^{\dag}_{\bm k} H({\bm k}) c_{{\bm k}}$ can be converted to the real-space representation as,
\begin{align}
    c_{\bm k} &= N^{-3/2} \sum_{\bm r} e^{-i {\bm r}\cdot{\bm k}} c_{\bm r},
\end{align}
where $\boldsymbol{r}$ runs over the lattice sites on the cubic lattice, and $N$ denotes the number of sites in each direction $(x,y,z)$.
With this form, the real-space Hamiltonian reads,
\begin{align}
 \mathcal{H} &=  -\frac{1}{2} \sum_{\bm r} \sum_{i=x,y,z} 
        \left[ c^{\dag}_{{\bm r}} 
           \left( it \alpha_{i}
              +m_2 \alpha_4 
           \right)
           c_{{\bm r}+{\bm e}_i}  + \mathrm{H.c.}
        \right]  \nonumber \\
   & \quad\quad + \sum_{\bf r} c^{\dag}_{\bm r} 
        \left[\left(m_0 +  3m_{2}\right) \alpha_4
           +J{\bm M} \cdot {\bm \Sigma}
        \right] c_{\bm r},
 \label{eqn:H_real}
\end{align}
where ${\bm e}_i = a\hat{i}$
is the lattice vector in the direction of $i(=x,y,z)$.
For example, in the systems with nonuniform magnetic textures,
the magnetization $\boldsymbol{M} = \boldsymbol{M}(\boldsymbol{r})$ in the exchange coupling term becomes position dependent.
To demonstrate the characteristics of Weyl semimetals,
we will use these real-space and momentum-space representations of the Wilson--Dirac model in the following sections.

\subsection{Topology of Weyl semimetal}
\label{sec:band-topology}
So far, we have seen the fundamentals of Weyl semimetals based on the low-energy effective models in continuum and on the lattice,
and reviewed their early-stage studies.
Here, we proceed to explain the topological nature of Weyl semimetals,
characterized by the Berry curvature and topological charge defined in momentum space,
which are now understood as the origin of various magnetoelectric effects in Weyl semimetals.

\subsubsection{Berry curvature}

As mentioned above, the low-energy excitations around the Weyl point with chirality $\eta$ are denoted by the Weyl Hamiltonian,
\begin{align}
H_{\eta}({\bm k})=\eta \hbar v_{\rm F}
\begin{pmatrix}
    k_z& k_x-ik_y\\
    k_x+ik_y & -k_z
  \end{pmatrix}.
\end{align}
By diagonalizing this Hamiltonian, we obtain two energy eigenvalues, $E_{\pm,\boldsymbol{k}} = \pm v_{\rm F}k$,
with $k = |\boldsymbol{k}|$.
Corresponding to these eigenvalues, the eigenstates are given in the form of 2-component spinors,
\begin{align}
    \ket{u_{+,\boldsymbol{k}}} &= \frac{1}{\sqrt{2k(k + \eta k_z)}}
    \begin{pmatrix}
        k_z + \eta k \\
        k_x+ i k_y
    \end{pmatrix}
    , \nonumber
    \\
    \ket{u_{-,\boldsymbol{k}}} &= \frac{1}{\sqrt{2k(k+\eta k_z)}}
    \begin{pmatrix}
        k_x- i k_y \\
        -k_z - \eta k
    \end{pmatrix}
    . \label{eq:weyl-eigenstate}
\end{align}
Note that Eq.~(\ref{eq:weyl-eigenstate}) is not a unique form for the eigenstates,
because the gauge degree of freedom by multiplying a phase factor $e^{i\varphi(\boldsymbol{k})}$ can be arbitrarily chosen.
We can see that Eq.~(\ref{eq:weyl-eigenstate}) becomes singular along half of the $k_z$-axis with $k_z \leq 0$ (for $\eta = +$) or $k_z \geq 0$ (for $\eta = -$),
along which $k + \eta k_z =0$ is satisfied.
While the position of such a singularity line can be arbitrarily shifted by a gauge transformation,
its existence starting from the Weyl point $k = 0$ is inevitable.
Thus, this singularity line can be seen as a topological object arising from the Weyl point structure,
which is called the Dirac string.

To quantify the band topology around the Weyl points,
we introduce the two geometric quantities in momentum space,
the Berry connection and the Berry curvature.
The Berry connection $\boldsymbol{a}_\pm(\boldsymbol{k})$ for each band $(\pm)$ is defined with the momentum-spce gradient as,
\begin{align}
    \bm{a}_\pm (\bm{k})= i \bra{u_{\pm,\boldsymbol{k}}}  \nabla_{\bm{k}} \ket{u_{\pm,\boldsymbol{k}}}.
\end{align}
As explained above, the Berry connection becomes singular along the Dirac string,
and is gauge dependent.
On the other hand, the Berry curvature $\boldsymbol{b}_\pm(\boldsymbol{k})$, defined as the curl of the Berry connection,
\begin{align}
    \boldsymbol{b}_\pm(\boldsymbol{k}) &= \nabla_{\bm{k}} \times \bm{a}_\pm (\bm{k}) \\
    &= i\bra{\nabla_{\bm{k}} u_{\pm,\boldsymbol{k}}} \times \ket{\nabla_{\bm{k}} u_{\pm,\boldsymbol{k}}}
    = -{\rm Im} \bra{\nabla_{\bm{k}} u_{\pm,\boldsymbol{k}}} \times \ket{\nabla_{\bm{k}} u_{\pm,\boldsymbol{k}}}, \nonumber
\end{align}
is a gauge-independent quantity.
Such a correspondence between $\boldsymbol{a}_\pm(\boldsymbol{k})$ and $\boldsymbol{b}_\pm(\boldsymbol{k})$ in momentum space is analogous to the vector potential $\boldsymbol{A}(\boldsymbol{r})$ and the magnetic field $\boldsymbol{B}(\boldsymbol{r}) = \nabla_{\bm r} \times \boldsymbol{A}(\bm r)$ in real space.

With the eigenstates of Weyl Hamiltonian given in Eq.~(\ref{eq:weyl-eigenstate}),
we obtain an explicit form of the Berry curvature,
\begin{align}
{\bm b}_{\pm}({\bm k})=\mp \eta \frac{\bm k}{2k^3}.
\end{align}
This configuration implies that the Weyl point $\boldsymbol{k} =0$ behaves as the source or sink of the flux of the vector field ${\bm b}_{\pm}({\bm k})$.

\subsubsection{Topological charge of Weyl points}

From the analogy between the Berry curvature and the magnetic field,
the Weyl point is sometimes referred to as a ``magnetic monopole'' in momentum space.
While real magnetic monopoles do not exist,
the monopoles of the Berry curvature are allowed,
which come from the band crossing in momentum space.
By integrating the Berry curvature over a closed surface $\partial\Gamma$ of the 3D region $\Gamma$ enclosing the Weyl point,
the monopole charge $\nu_\pm$ is quantized,
\begin{align}
\nu_\pm = \frac{1}{2\pi}\int_{\partial\Gamma} {\bm b}_{\pm} \cdot d {\bm S} = \frac{1}{2\pi}\int_{\Gamma} (\boldsymbol{\nabla}_k \cdot { \bm b_{\pm}} ) dV = \mp \eta,
\end{align}
which is known as the \textit{topological charge} of the Weyl point.
For the occupied band below $E=0$ (taking the band index $-$ out of $\pm$),
the topological charge $\nu_-$ corresponds to the chirality $\eta$ of each Weyl point.

Other than the topological charge $\pm 1$,
one can also consider a band-touching point with a higher topological charge $|\nu| \geq 2$.
Such a band-touching point,
which is no longer described by the Weyl Hamiltonian with the linear dispersion,
is called a multi-Weyl point.
For instance, a band-touching point with a topological charge $|\nu| =2$ is called a double Weyl point.
The existence of such multi-Weyl points has also been proposed in several materials from their symmetries \cite{xu2011chern,fang2012multi,huang2016new},
while we shall not go into their details in this review.

Although the Berry curvature seems to be just a mathematical object here,
it contributes to physical phenomena intrinsic to magnetic Weyl semimetals,
such as the anomalous Hall effect,
as we shall see below.

\subsubsection{Number of Weyl points}
The Wilson--Dirac model introduced above shows a pair of Weyl points,
one with the positive chirality and the other with the negative chirality.
In the Weyl semimetal materials theoretically proposed and experimentally synthesized,
there are often many Weyl points residing in the Brillouin zone (see Sec.~\ref{sec:materials} for detailed explanations).
There is an important topological condition on the number of Weyl points in lattice (crystalline) systems,
which is known as the Nielsen--Ninomiya's theorem \cite{nielsen1981absence,nielsen1983adler}.
It demands that the number of Weyl points with positive and negative chiralities should be equal in the Brillouin zone.
In other words, the topological charge of the Weyl points should cancel in total in the Brillouin zone.
Thus,
the number of the Weyl points should be always an even number,
which is also known as the doubling of fermions in the context of lattice quantum field theory \cite{kogut1979introduction,creutz1983quarks}.

The Nielsen--Ninomiya's theorem is understood from the periodicity of the Brillouin zone.
If there are odd number of Weyl points,
the topological charge in the Brillouin zone is not cancelled in total,
and
there should be a finite flux of Berry curvature piercing the boundary of the first Brillouin zone.
As a result, this Berry flux enters the second Brillouin zone,
which makes the Berry flux configurations in the first and second Brillouin zones different,
violating the identity of the Brillouin zones.
Therefore, the total topological charge in the Brillouin zone should be exactly zero in lattice systems, as seen in Fig.~\ref{fig:chern_number}(a).
The cancellation of topological charge is satisfied
even with the multi-Weyl points with higher topological charges \cite{fang2012multi}.

\subsubsection{Breaking of time-reversal and inversion symmetry}
\label{sec:symmetry}

The configurations of the Berry curvature and the Weyl points of positive/negative topological charges are governed by the symmetry of the system \cite{berry1984,volovik2003universe,Murakami2007,
Burkov2011}.
Here we focus on time-reversal $(\mathcal{T})$ and spatial inversion $(\mathcal{P})$ symmetries.
The time-reversal symmetry is broken by magnetic interaction, such as the exchange coupling between the magnetic ordering and conduction electrons' spins,
the Zeeman coupling to an external magnetic field, etc.
On the other hand, the inversion symmetry breaking originates from the noncentrosymmetric crystalline structure in the bulk, the surface or interface structure, an external electric field, etc.
From the analysis of the Berry curvature and topological charge, we explain that at least one of these two symmetries should be broken to keep the Weyl points isolated in momentum space.

\begin{figure}[t]
\centering
\includegraphics[width=1\hsize]{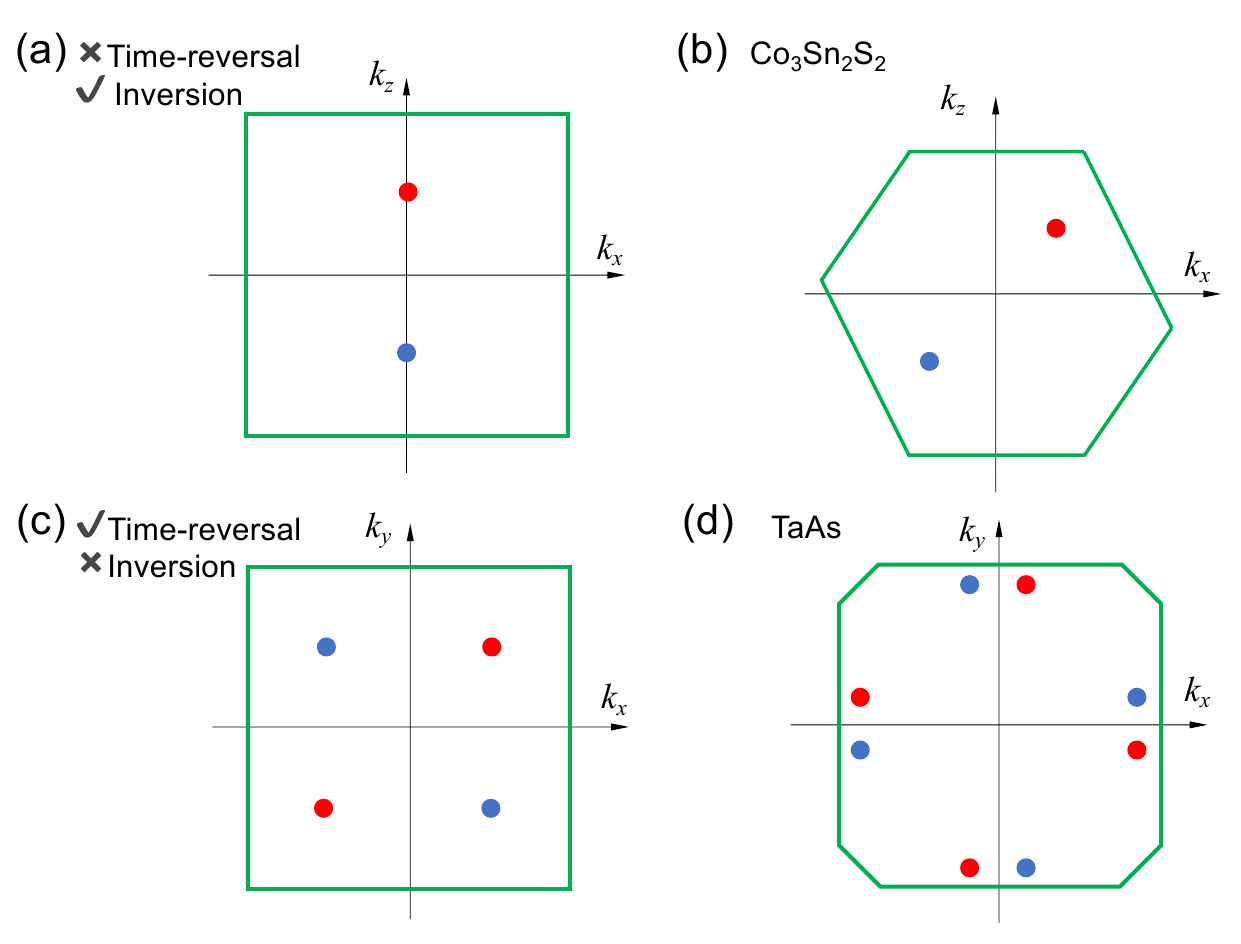}
\caption{(Color online)
Schematic illustrations of the momentum-space configuration of Weyl points in Weyl semimetal states.
Red and blue points correspond to the Weyl points with positive and negative chiralities, respectively.
(a)~The hypothetical Weyl semimetal state satisfying inversion symmetry but not time-reversal symmetry,
which minimally supports a single pair of Weyl points.
(b)~The ferromagnetic Weyl semimetal \CSS,
whose 2 Weyl points out of 6 reside on the $k_x k_z$-plane.
(c)~The hypothetical Weyl semimetal state satisfying time-reversal symmetry but not inversion symmetry,
which minimally supports two pairs of Weyl points.
(d)~The noncentrosymmetic Weyl semimetal \ce{TaAs},
whose 8 Weyl points out of 24 reside on the $k_x k_y$-plane.
}
\label{fig:symmetry}
\end{figure}

We first start with the case where the inversion symmetry is present but the time-reversal symmetry is broken,
which applies to most of the magnetic Weyl semimetal materials proposed and synthesized so far.
In this case,
the inversion symmetry imposes the relation
$\ket{u_{n\boldsymbol{k}}} = \mathcal{P} \ket{u_{n,\boldsymbol{-k}}}$
for the Bloch wave function $\ket{u_{n\boldsymbol{k}}}$
with the band index $n$.
Here, the unitary operator $\mathcal{P}$ denotes the mapping of the internal degrees of freedom (orbitals, sublattices, etc.) under the spatial inversion.
Thus, the Berry curvature $\boldsymbol{b}_n(\boldsymbol{k}) = -{\rm Im}\bra{\nabla_{\bm k} u_{n{\bm k}}} \times \ket{\nabla_{\bm k} u_{n{\bm k}}}$ and the topological charge density $\rho_n(\boldsymbol{k}) = \nabla_{\bm k} \cdot \boldsymbol{b}_n(\boldsymbol{k})$ satisfy the relations,
\begin{align}
{\bm b}_n({\bm k})={\bm b}_n(-{\bm k}), \quad
\rho_n({\bm k}) = -\rho_n(-{\bm k}).
\label{eq:mc_tr}
\end{align}
Therefore, if there is a Weyl point with a positive topological charge at momentum $\boldsymbol{K}_0$,
it accompanies another Weyl point with a negative topological charge at $-\boldsymbol{K}_0$,
as we have demonstrated with the Wilson--Dirac model [Fig.~\ref{fig:symmetry}(a)].
In reality,
magnetic Weyl semimetal materials with inversion symmetry exhibit multiple Weyl points.
Even in such cases, the positive- and negative-charge Weyl points always reside at the locations opposite to each other in momentum space.
For example, the ferromagnetic \CSS exhibits six Weyl points in the whole Brillouin zone as shown in Fig.~\ref{fig:symmetry}(b),
where all the Weyl points reside on the 
plane satisfying the mirror reflection symmetry~\cite{Liu2018}.

We next consider the case of nonmagnetic Weyl semimetals,
where the time-reversal symmetry is present but the inversion symmetry is broken.
In this case, the Bloch wave function satisfies the relation
$\ket{u_{n\boldsymbol{k}}} = \mathcal{T} \ket{u_{n,\boldsymbol{-k}}}$,
where $\mathcal{T}$ is the antiunitary operator for the time-reversal operation of the internal degrees of freedom (spin, orbital, etc.).
Keeping in mind that $\mathcal{T}$ contains the operation of complex conjugation,
we reach the relations for 
$\boldsymbol{b}_n(\boldsymbol{k})$ and $\rho_n(\boldsymbol{k})$,
\begin{align}
{\bm b}_n({\bm k})=-{\bm b}_n(-{\bm k}), \quad
\rho_n({\bm k})=\rho_n(-{\bm k}).
\label{eq:mc_inv}
\end{align}
These relations suggest that a Weyl point at momentum $\boldsymbol{K}_0$ accompanies another Weyl point with the same topological charge at $-\boldsymbol{K}_0$.
Thus, to satisfy the cancellation of net topological charge in the Brillouin zone,
we need at least four Weyl points,
two with topological charge $+1$ and the other two with $-1$,
as shown in Fig.~\ref{fig:symmetry}(c).
For instance, the noncentrosymmetric Weyl semimetal \ce{TaAs}~\cite{Huang15a},
which is the first Weyl semimetal material that was successfully synthesized, exhibits 24 Weyl points [Fig.~\ref{fig:symmetry}(d)].

Recent theoretical and experimental studies have found several Weyl semimetal materials where both the time-reversal and inversion symmetries are broken,
in the family of $R {\rm Al} X \ (R={\rm Nd},{\rm Pr},{\rm Ce}; \ X = {\rm Si},{\rm Ge})$ \cite{chang2018magnetic},
the inverse Heusler magnet \ce{Ti2MnAl}~\cite{shi2018prediction}, etc.
Such noncentrosymmetric magnetic Weyl semimetal states do not satisfy either of the relations Eqs.~(\ref{eq:mc_tr}) or (\ref{eq:mc_inv}).

To realize Weyl fermions in solids,
the time-reversal or inversion symmetries should be broken.
If both time-reversal and inversion symmetries are preserved,
the relations of Eqs.~(\ref{eq:mc_tr}) and (\ref{eq:mc_inv}) demand $\rho_n(\boldsymbol{k}) =0$,
which implies that the Weyl points cannot exist isolated with its topological charge $+1$ or $-1$.
A band-crossing point in such cases should consist of four internal degrees of freedom,
which can be regarded as the overlap of two Weyl nodes with topological charges $+1$ and $-1$ at the same momentum.
Such a band crossing point is 
nothing but
the Dirac point, as we showed in Fig.~\ref{fig:wd_band}(a).

\subsection{Characteristic properties related to band topology}\indent

In the previous subsection, 
we have seen the topological natures of the Weyl fermions at low energy around the Weyl points.
We now proceed to the characteristic properties of magnetic Weyl semimetals by considering the overall band topology in the Brillouin zone.

\begin{figure}[t]
\centering
\includegraphics[width=1\hsize]{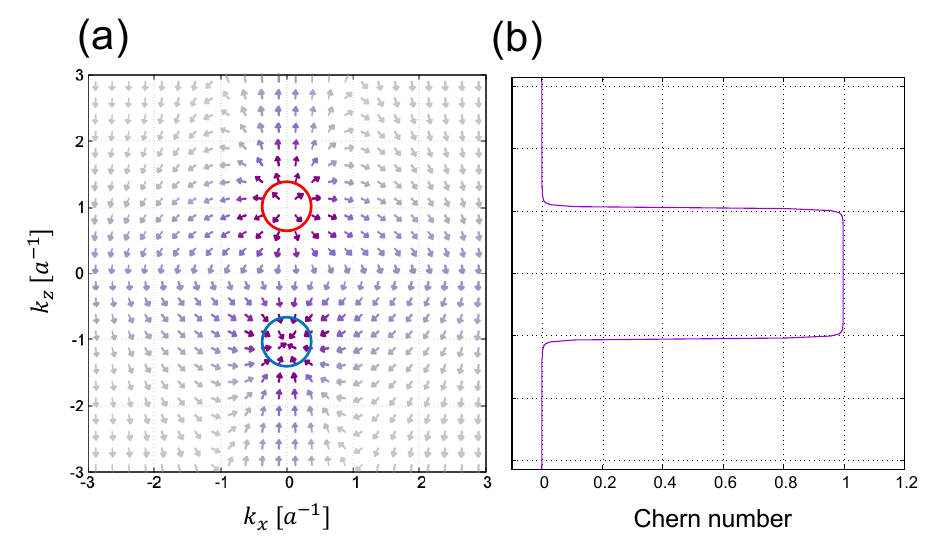}
\caption{(Color online)
(a)~Berry curvature distribution ${\bm b}_-(\boldsymbol{k})$ and (b)~Chern number $\nu(k_z)$, calculated from the Wilson--Dirac model on lattice.
The Weyl point with chirality $+(-)$ behaves as a sink~(source) of the Berry curvature vector field in momentum space.
}
\label{fig:chern_number}
\end{figure}

\subsubsection{Anomalous Hall effect}
\label{sec:anomalous-hall}

One of the most important consequences of the topological nature of magnetic Weyl semimetals is the anomalous Hall effect.
The anomalous Hall effect,
which gives the transverse electric current $j_x = \sigma_{xy} E_y$ not proportional to external magnetic field,
is permitted only if time-reversal symmetry is broken,
typically in 
metals
with spontaneous magnetic orderings.
The origin of the anomalous Hall effect is classified into the extrinsic and intrinsic mechanisms~\cite{karplus1954,Nagaosa2010}.
The extrinsic mechanism originates from the skew or side-jump scatterings by impurities,
which are contributed by the carriers on the Fermi surface.
On the other hand, the intrinsic one comes from the structure of the electron wave function itself, namely, the topological nature,
from all the occupied states in the Fermi sea.
We here explain the relation between the band topology and the intrinsic anomalous Hall effect in magnetic Weyl semimetals,
based on the lattice model introduced above.

Generally, the intrinsic contribution to the anomalous Hall conductivity $\sigma_{xy}$ in 3D system is given by the sum of 
the $z$-component of the Berry curvature $b_n^z(\boldsymbol{k})$ of all the occupied states, 
\begin{align}
\sigma_{yx}= -\frac{e^2}{\hbar} \sum_n \int_{\mathrm{BZ}} \frac{d^3\bm{k}}{(2 \pi)^3} f(E_{n{\bm{k}}}-E_{\rm F}) b_n^z(\bm{k}) ,
\label{eq:tknn}
\end{align}
which is known as the Thouless--Kohmoto--Nightingale--den Nijs (TKNN) formula \cite{thouless1982quantized,kohmoto1985topological}.
Here $f(E_{n{\bm{k}}}-E_{\rm F}) = [e^{(E_{n{\bm k}}-E_{\rm F})/k_{\rm B} T} +1]^{-1}$ is the Fermi distribution function,
with the temperature $T$.
For a better understanding of this relation,
let us consider the integrations in the $k_x k_y$-plane and along the $k_z$-axis separately.
Then, the anomalous Hall conductivity is given as 
\begin{align}
\sigma_{yx}= -\frac{e^2}{(2 \pi)^2 \hbar}  \int_{-\pi/a}^{\pi/a} dk_z \; \nu(k_z),
\label{eq:kb}
\end{align}
with
\begin{align}
\nu(k_z)= \frac{1}{2\pi} \sum_n \int dk_x dk_y \; f(E_{n{\bm k}}-E_{\rm F}) b_n^z(k_x, k_y, k_z). 
\end{align}
If the bands are gapped on this plane at $E_{\rm F}$,
this $\nu(k_z)$ at zero-temperature corresponds to the topological index known as the Chern number.
This $\nu(k_z)$ characterizes the overall band topology in the 2D system,
here corresponding to the
2D Brillouin zone for $(k_x,k_y)$
with the fixed $k_z$.
In other words, by treating $k_z$ as a fixed parameter in the 3D Hamiltonian $H(k_x,k_y,k_z)$, one can locally define the 2D Brilluoin zone.
The Chern number
takes an integer value $(\in \mathbb{Z})$,
which is demanded by the topological nature of the 2D class-A systems (in the Altland--Zirnbauer's classification \cite{altland1997nonstandard})
where time-reversal, particle-hole, and chiral symmetries are all broken \cite{schnyder2008classification}.
The Chern number is topologically robust against small perturbations such as impurities, and
change in the Chern number should always accompany the closing of the band gap.
2D insulator with a nonzero Chern number $\nu \neq 0$ is often called the Chern insulator, or the quantum anomalous Hall insulator,
showing the quantized anomalous Hall conductivity $\sigma_{yx} = -(e^2/h) \nu$ in 2D \cite{haldane1988model}.

\begin{figure*}[t]
    \centering
    \includegraphics[width=1.0\linewidth]{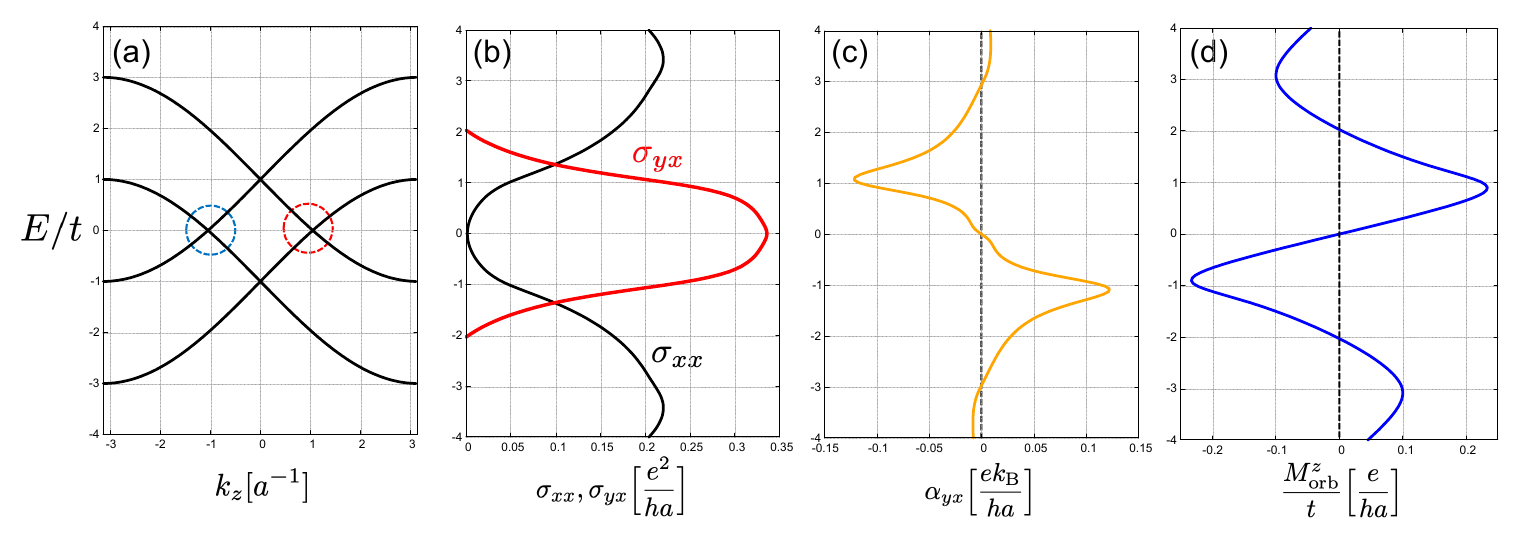}
    \caption{(Color-online)~
    (a) Band structure and (b-d) transport coefficients of a magnetic Weyl semimetal, numerically calculated from the Wilson--Dirac model.
    (b)~Longitudinal conductivity $\sigma_{xx}$ and 
    anomalous Hall conductivity $\sigma_{yx}$,
    (c)~anomalous Nernst coefficient $\alpha_{yx}$, and 
    (d)~orbital magnetization $M_{\rm orb}^z$.
        }
    \label{fig:transport}
\end{figure*}

\begin{figure}[tb]
\centering
\includegraphics[width=1.0\hsize]{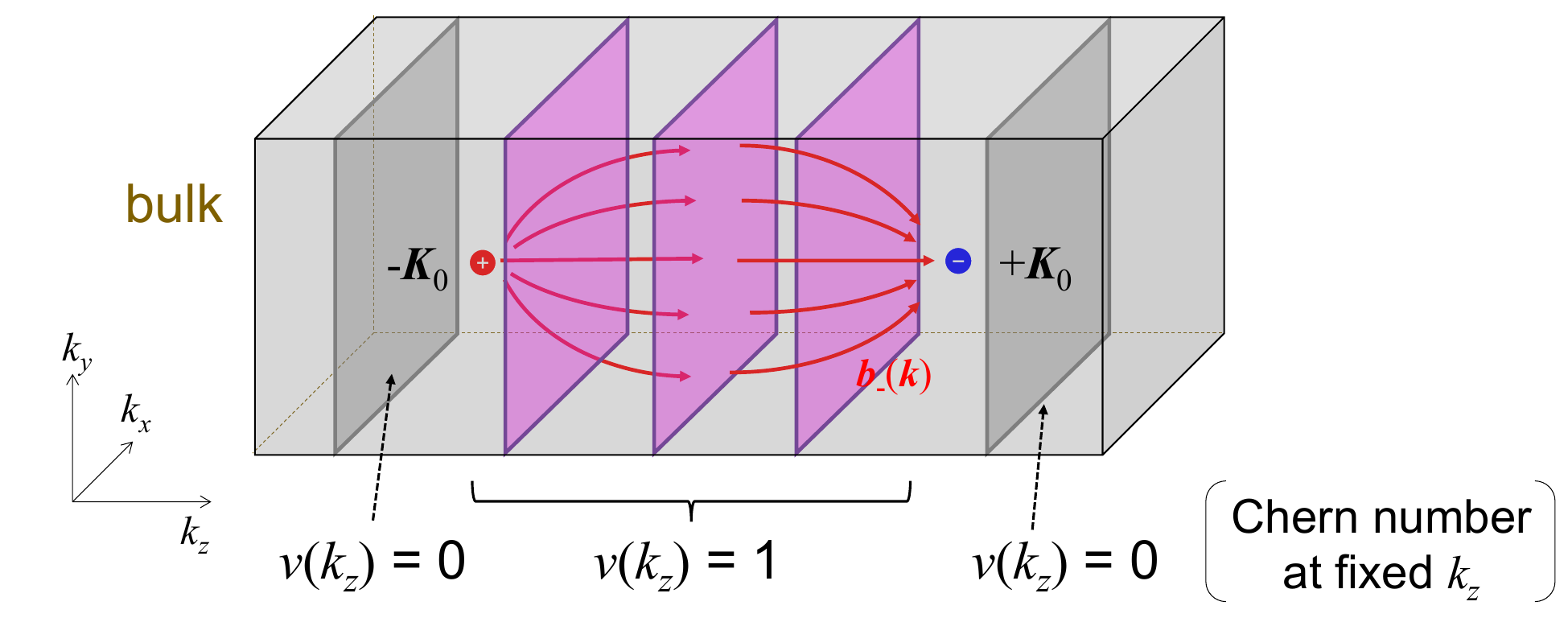}
\caption{
Schematic picture for the configurations of the Berry curvature and the Chern number in a magnetic Weyl semimetal.
The Chern number becomes finite $[\nu(k_z) = 1]$ on a 2D $k_x k_y$-plane at each fixed $k_z$ between the Weyl points,
which contributes to the anomalous Hall conductivity $\sigma_{yx}$.
}
\label{fig:bulk_bz}
\end{figure} 

Based on the above general idea with the Berry curvature and the Chern number,
we now proceed to consider the intrinsic anomalous Hall effect in magnetic Weyl semimetals.
Since the Weyl points serve as source or sink of the Berry curvature,
we can naively expect that the Weyl points contribute to a large anomalous Hall conductivity.
For the concrete demonstration,
let us consider the Wilson--Dirac model introduced above,
with the pair of Weyl points at $(0,0,\pm K_0)$.
We fix $E_{\rm F}=0$,
and take the zero-temperature limit.
First, we evaluate the Chern number on the $k_x k_y$-plane at the fixed $k_z$.
%
The distribution of the Berry curvature for the occupied state $\boldsymbol{b}_{-}(\boldsymbol{k})$ is shown in Fig.~\ref{fig:chern_number}(a), 
which is calculated by the Fukui-Hatsugai-Suzuki method~\cite{fukui2005chern}.
The 2D slice becomes a Chern insulator with $\nu_z(k_z) = 1$ for $-K_0 < k_z < K_0$,
and a trivial insulator with $\nu_z(k_z) = 0$ for $|k_z| > K_0$ ,
as shown in Fig.~\ref{fig:chern_number}(b).
The gap closing occurs at the Weyl points, $k_z = \pm K_0$,
to change the Chern number [see Fig.~\ref{fig:bulk_bz} for its schematics].
By integrating the Chern number along the $k_z$-axis,
the intrinsic anomalous Hall conductivity is given as,
\begin{align}
\sigma_{yx}= \frac{e^2}{(2 \pi)^2\hbar} 2K_{\rm{0}}. \label{eq:AHE-K0}
\end{align}
This relation indicates that the anomalous Hall conductivity in magnetic Weyl semimetal is proportional to the distance $2K_{\rm{0}}$ between the Weyl points with the different chirality \cite{Burkov2011,burkov2011topological,Zyuzin2012}. 
Note that, in the present model, the splitting of the Weyl points $2K_0$ originates from the magnetic ordering which breaks time-reversal symmetry.

The above relation between the band structure and $\sigma_{yx}$ can be numerically examined with the Wilson--Dirac model,
as shown in Fig.~\ref{fig:transport}(b).
 $\sigma_{yx}$ maximizes at $E_{\rm F}=0$, corresponding to the energy of the Weyl points.
 This value from the numerical calculation is identical to the value estimated theoretically from Eq.~(\ref{eq:AHE-K0}).
In the Weyl semimetal materials in reality,
there are often multiple pairs of Weyl points residing in the Brillouin zone, as shown in Fig.~\ref{fig:symmetry}.
In such cases, the Weyl points near $E_{\rm F}$ mainly contribute to the anomalous Hall effect.
If the Weyl points are located at the energy away from $E_{\rm F}$,
their Berry curvature contribution to the anomalous Hall effect is suppressed \cite{burkov2014anomalous}.

The anomalous Hall transport is often characterized by the anomalous Hall angle,
\begin{align}
\theta_{yx}=\frac{\sigma_{yx}}{\sigma_{xx}},
\end{align}
which is the ratio between $\sigma_{yx}$ and the longitudinal conductivity $\sigma_{xx}$.
In ideal magnetic Weyl semimetals,
the longitudinal conductivity $\sigma_{xx}$ gets suppressed due to the small desity of states near the energy of the Weyl points.
This feature is reproduced by the calculation with the Kubo formula based on the Wilson--Dirac model,
as shown in Fig.~\ref{fig:transport}(b).
On the other hand,
the anomalous Hall conductivity $\sigma_{xy}$ becomes nonzero,
because the intrinsic contribution is given not by the electrons at $E_{\rm F}$ but by all the occupied electronic states.
Thus, the anomalous Hall angle $\theta_{yx}$ can take an extremely large value in ideal magnetic Weyl semimetals,
which was reported in several experiments as we shall see in Sec.~\ref{sec:materials}.
The intrinsic anomalous Hall current does not require any longitudinal current,
and hence the Joule heating from the resistivity ideally gets suppressed.
Besides such an intrinsic contribution,
the extrinsic contributions shall be reviewed later in Sec.~\ref{sec:transport} based on the numerical studies.

As a consequence of the Berry curvature emerging from the Weyl points,
the anomalous Nernst effect is also often seen in magnetic Weyl semimetals.
The anomalous Nernst effect is the thermoelectric effect that induces the transverse electric current $j_y$ in response to the temperature gradient $\partial_x T$, as $j_y = \alpha_{yx} (-\partial_x T)$.
The intrinsic contribution to the transverse thermoelectric coefficient $\alpha_{yx}$ is attributed to the Berry curvature \cite{xiao2006berry},
\begin{align}
    \alpha_{yx} &= \frac{e k_{\rm B}}{\hbar} \int_{\mathrm{BZ}} \frac{d^3\bm{k}}{(2 \pi)^3} \sum_n s(E_{n{\bm{k}}}-\mu) b_n^z(\bm{k}), \label{eq:anomalous-nernst}
\end{align}
with the entropy density
\begin{align}
    s(\epsilon) &= -f(\epsilon) \ln f(\epsilon) - [1-f(\epsilon)] \ln [1-f(\epsilon)],
\end{align}
with the chemical potential $\mu$ at finite temperature.
At low temperature,
$\alpha_{yx}$ is related to the anomalous Hall conductivity $\sigma_{yx}$ by the Mott's relation,
\begin{align}
    \alpha_{yx} &= -\frac{\pi^2}{3} \frac{k^2_{\rm B}T}{e} \frac{\partial \sigma_{yx}}{\partial{\mu}} .
\end{align}
Due to the large Berry curvature from the Weyl points,
$\alpha_{yx}$ becomes also large in magnetic Weyl semimetals \cite{sharma2016nernst,matsushita2025intrinsic},
although the energy for its maxima is deviated from the energy of the Weyl points.
The enhancement of the anomalous effect was reported in several magnetic Weyl semimetal materials,
as we shall see in detail in Sec.~\ref{sec:materials}.
Such an effect is expected to be useful for the application to the thermoelectric conversion,
i.e., the conversion from waste heat to electricity.

\subsubsection{Fermi arc surface state}
\label{sec:fermi-arc}

Similarly to the Dirac surface state of topological insulators,
Weyl semimetals also show the gapless surface state related to the band topology in the bulk.
Unlike conventional 2D electronic states forming a closed Fermi contour,
the surface states of Weyl semimetals,
residing on the 2D projected Brillouin zone,
form an open line at $E_{\rm F}$,
which is called the Fermi arc~\cite{Wan2011}.
The presence of the Fermi arc surface state is understood from the Chern number in the bulk.
Let us consider the case where the two Weyl points are separated along the $k_z$-axis, as we have seen above for the demonstration of the anomalous Hall effect
A 2D slice at fixed $k_z$ between the Weyl points acquires a finite Chern number, $\nu(k_z) = 1$,
as shown in Fig.~\ref{fig:bulk_bz}.
This leads to a 1D chiral edge state,
which is unidirectionally dispersed along the edge ($k_x$-direction) and becomes zero energy at a certain $k_x$,
for each $k_z$ in $-K_0 <k_z < K_0$.
By collecting these zero-energy points,
we reach the Fermi arc connecting the the projected Weyl points at $k_z = K_0$ and $k_z = -K_0$.
While the presence of the Fermi arc is topologically robust,
we should note that the detailed shape of the Fermi arc between the projected Weyl points depends on the boundary condition,
which is governed by the structure of surface termination.

To elucidate the presence of the surface Fermi arcs,
here we numerically demonstrate their structure with the
Wilson--Dirac model on the lattice.
As we have introduced above,
we consider the case where the magnetization points in the $z$-direction
with the two Weyl points split along the $k_z$-axis in momentum space.
To obtain the Fermi arc states,
we here treat the system in the slab structure,
where the 2D layers on the $yz$-plane are stacked by $L$ times along $x$-direction,
imposing the open boundary condition for the $x$-direction and the periodic boundary condition for the $y$- and $z$-directions.
With the Fourier transformation
\begin{align}
    c_{\boldsymbol{k}_\perp,x} = \frac{1}{L^2} \sum_{y,z} e^{i(k_y y + k_z z)} c_{\boldsymbol{r} =(x,y,z)},
    \quad [\boldsymbol{k}_\perp = (k_y,k_z)]
\end{align}
the lattice Hamiltonian in Eq.~(\ref{eqn:H_real}) reduces as
\begin{align}
    \mathcal{H} &= \sum_{x,x'} \sum_{\boldsymbol{k}_\perp} c_{\boldsymbol{k}_\perp,x}^\dag H_{x,x'}(\boldsymbol{k}_\perp) c_{\boldsymbol{k}_\perp,x'} .
\end{align}
Here, $H(\boldsymbol{k}_\perp) = [H_{x,x'}(\boldsymbol{k}_\perp)]_{x,x' = 1, \cdots, L}$ is given as a $4L \times 4L$ matrix,
\begin{align}
H(\boldsymbol{k}_\perp) =
\begin{pmatrix}
     H^\perp(\boldsymbol{k}_\perp) & \Gamma & 0  & ...\\
     \Gamma^{\dag} & H^\perp(\boldsymbol{k}_\perp) & \Gamma & ...\\
      0 & \Gamma^{\dag} & H^\perp(\boldsymbol{k}_\perp) & ... \\
      \vdots & \vdots & \vdots & \ddots &
  \end{pmatrix}.
  \label{eq:slabham}
\end{align}
with the $4\times 4$ intralayer block
\begin{align}
H^\perp(\boldsymbol{k}_\perp) &=
\begin{pmatrix}
    A(\boldsymbol{k}_\perp) \sigma_0 + J\boldsymbol{M}\cdot\boldsymbol{\sigma}& \sum_{i=y,z} t \sin(k_i a) \sigma_i   \\
    \sum_{i=y,z} t \sin(k_i a) \sigma_i   & -A(\boldsymbol{k}_\perp) \sigma_0 + J\boldsymbol{M}\cdot\boldsymbol{\sigma}
  \end{pmatrix} , \\
  & [ A(\boldsymbol{k}_\perp)=M_0+m_2(3-\cos{k_y a }-\cos{k_z a }) ] \nonumber
\end{align}
and the interlayer block

\begin{align}
    \Gamma =\frac{1}{2}
\begin{pmatrix}
     -m_2 \sigma_0& -i t \sigma_z    \\
     +i t \sigma_z   & r \sigma_0 
  \end{pmatrix}.
\end{align}

\begin{figure}[t]
\centering
\includegraphics[width=1.0\hsize]{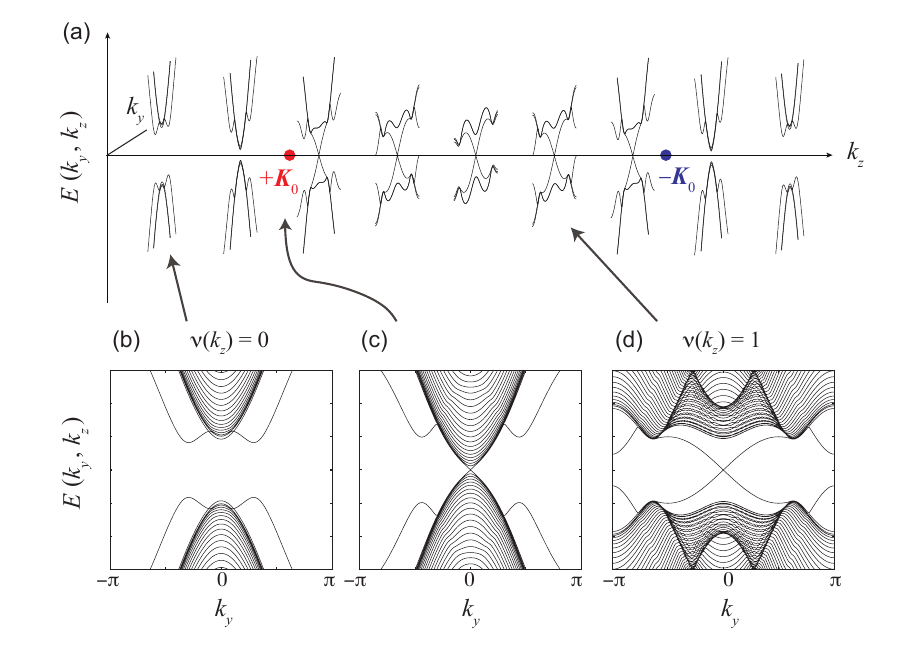}
\caption{
Schematic images for the surface Fermi arcs,
with the Weyl points located at $k_z = \pm K_0$.
(a) Slices $E(k_y,k_z)$ of the band structure at fixed $k_z$'s, with the open boundary condition in the $x$-direction.
(b) A slice for $|k_z| > K_0$, where the band structure becomes a trivial insulator $[\nu(k_z) = 0]$.
(c) A slice for $|k_z| = K_0$, where the band gap closes to form a Weyl point.
(d) A slice for $|k_z| < K_0$, where the band structure becomes a Chern insulator $[\nu(k_z) = 0]$ forming the chiral edge states.
Collection of these edge states form the Fermi arc connecting between $k_z = + K_0$ and $k_z = -K_0$.
}
\label{fig:weyl_band}
\end{figure}

By diagonalizing the matrix Hamiltonian $H^{\rm WD}(\boldsymbol{k}_\perp)$, we obtain the energy eigenvalues and eigenstates including both bulk and surface states.
In Fig.~\ref{fig:weyl_band}, we show the band structure obtained from this slab system with fixed $k_z$.
There are $4L$ bands, i.e., $4$ from spinor components times $L$ from layer degrees of freedom.
For $|k_z| > K_0$,
the bands are gapped, corresponding to the trivial insulator with the bulk Chern number $\nu(k_z) =0$ [Fig.~\ref{fig:weyl_band}(a)].
In contrast, we find that the gapless states appear at $k_z$ between the Weyl points $(-K_0 < k_z < K_0)$,
which correspond to the chiral edge states of Chern insulator with $\nu(k_z) = 1$ as we mentioned above [Fig.~\ref{fig:weyl_band}(d)].
We find two gapless branches for each $k_z$,
corresponding to the edge states localized at the boundaries $x=1$ and $L$.
At the Weyl points $k_z = \pm K_0$,
we find that the gap closes and the edge states vanish,
which occurs at the topological phase transition between $\nu(k_z) =0$ and $1$ [Fig.~\ref{fig:weyl_band}(c)].
As a result, the edge states found for $-K_0 < k_z < K_0$ form the Fermi arc connecting the projected Weyl points.

The structure of the surface Fermi arcs has also been theoretically studied by evaluating the spectral function, with a fully analytical calculation of the Green's function at the boundary~\cite{Zhang2024-dn}.
Experimentally, the presence of surface Fermi arcs
has been confirmed by the photoemission spectroscopy measurements in several Weyl semimetals.
Their contribution to the electron transport is also measured in some thin-film samples,
as we shall review in Sec.~\ref{sec:materials}.
In particular, under a magnetic field,
it has been found that the Fermi arcs contribute to the quantum oscillation in the magnetotransport properties.
Semiclassically,
it has been considered as a consequence of the interference on a closed trajectory, known as the Weyl orbit,
which is formed by the Fermi arcs on the two surfaces interconnected via the bulk Landau levels~\cite{potter2014quantum,moll2016transport, Galletti2019absence}.

The surface Fermi arc has been found to be important also for the 3D quantum Hall effect \cite{Nguyen2021quantum, Zhang2022, Lu2026}.

\subsubsection{Orbital magnetization}
\label{sec:orbital-magnetization}

The nontrivial band topology in magnetic Weyl semimetals results also in the bulk physical quantity, the orbital magnetization.
The orbital magnetization comes from the orbital magnetic moment of electrons,
$\boldsymbol{m}_{\rm orb} = -(e/2) \boldsymbol{r} \times \boldsymbol{v}$,
where $\boldsymbol{r}$ and $\boldsymbol{v}$ denote the position and velocity of an electron, respectively.
However, in the bulk of periodic systems like crystals,
the position operator $\boldsymbol{r} \approx i \nabla_{\bm k}$ is gauge dependent,
and hence the orbital magnetization cannot be uniquely defined in this way.
This problem was solved by the so-called ``modern theory'' of orbital magnetization,
which is gauge invariant by including the $\boldsymbol{k}$-space derivatives of the Bloch wavefunctions,
i.e., the geometric quantities \cite{xiao2005berry,thonhauser2005orbital,ceresoli2006orbital,resta2010electrical}.

With the eigenfunctions $|u_{n\boldsymbol{k}}\rangle$ and eigenvalues $E_{n\boldsymbol{k}}$ of the Hamiltonian $H(\boldsymbol{k})$,
the orbital magnetization in the modern theory at finite temperature is given as,
\begin{align}
    \boldsymbol{M}_{\rm orb} &= \sum_n \int_{\rm BZ} \frac{d^d\boldsymbol{k}}{(2\pi)^d} f(E_{n\boldsymbol{k}}-\mu) \boldsymbol{m}^{\rm orb}_{n}({\bm k}) \\
    & \hspace{10pt} + \frac{e k_{\rm B} T}{\hbar} \sum_n \int_{\rm BZ} \frac{d^d\boldsymbol{k}}{(2\pi)^d} \ln \left[1+e^{-(E_{n\boldsymbol{k}}-\mu)/ k_{\rm B}T}\right] \boldsymbol{b}_{n}({\bm k}), \nonumber
\end{align}
where $d$ denotes the dimensionality of the system.
The first term is attributed to the orbital magnetic moment of a wave packet,
\begin{align}
    \boldsymbol{m}^{\rm orb}_{n}({\bm k}) &= \frac{e}{2\hbar} {\rm Im} \bra{\nabla_{\bm k} u_{n{\bm k}}} \times [H(\boldsymbol{k}) - E_{n\boldsymbol{k}}] \ket{\nabla_{\bm k} u_{n{\bm k}}},
\end{align}
whereas the second term is considered as the contribution from the Berry curvature $\boldsymbol{b}_{n}({\bm k})$.
In the zero-temperature limit $(\mu \rightarrow E_{\rm F})$,
these two terms can be unified as,
\begin{align}
    \boldsymbol{M}_{\rm orb} &= \frac{e}{2\hbar} \sum_n \int_{\rm BZ} \frac{d^d\boldsymbol{k}}{(2\pi)^d} f(E_{n\boldsymbol{k}}-E_{\rm F}) \\
    & \hspace{30pt} \times {\rm Im} \bra{\nabla_{\bm k} u_{n{\bm k}}} \times [H(\boldsymbol{k}) + E_{n\boldsymbol{k}} -2E_{\rm F}] \ket{\nabla_{\bm k} u_{n{\bm k}}}. \nonumber
\end{align}

Let us here consider the orbital magnetization in a magnetic Weyl semimetal,
based on the Wilson--Dirac model on lattice defined by Eq.~(\ref{eqn:HWD_k}) with the exchange coupling in Eq.~(\ref{eqn:H-exc}).
We fix the magnetization $\boldsymbol{M}$ to the $z$-direction,
so that the Weyl points are located along the $z$-axis as $(0,0,\pm K_0)$.
As we have seen in the discussion of the anomalous Hall effect,
we first consider the 2D slice on the $k_x k_y$-plane with the fixed $k_z$.
If $E_{\rm F}$ is close enough to the Weyl points $(E_{\rm F} \approx 0)$ so that $E_{\rm F}$ captures only the Weyl cones,
the orbital magnetization integrated over this plane is given with the Chern number $\nu(k_z)$,
\begin{align}
    M_{\rm orb}^z(k_z) = -\frac{e E_{\rm F}}{2\pi\hbar} \nu(k_z) = \frac{E_{\rm F}}{e} \sigma_{yx}(k_z),
\end{align}
in the zero-temperature limit,
where $\sigma_{yx}(k_z) = -(e^2/h) \nu_z(k_z) $ is the anomalous Hall conductivity on the 2D plane at $k_z$.
Thus, the orbital magnetization in 3D is also related to the anomalous Hall conductivity in 3D,
\begin{align}
    M_{\rm orb}^z = \int \frac{dk_z}{2\pi} M_{\rm orb}^z(k_z) = \frac{E_{\rm F}}{e} \sigma_{yx} = -\frac{eE_{\rm F}}{(2\pi^2)\hbar} 2K_0 . \label{eq:M_orb-3D}
\end{align}
$\boldsymbol{M}_{\rm orb}$ and $\sigma_{yx}$ satisfy the thermodynamic relation, 
so-called Streda formula~\cite{streda1982theory,Nomura2015,xiao2010berry},
\begin{align}
    \sigma_{yx} = e \left( \frac{\partial M_{\rm orb}^z}{\partial \mu}\right)_{B,T} ,
    \label{eq:AHE-orbital}
\end{align}
where the derivative is taken with the fixed magnetic field $\boldsymbol{B}$ and temperature $T$.
This formula is satisfied at finite $T$ as well.

\begin{figure}[t]
\centering
\includegraphics[width=1.0\hsize]{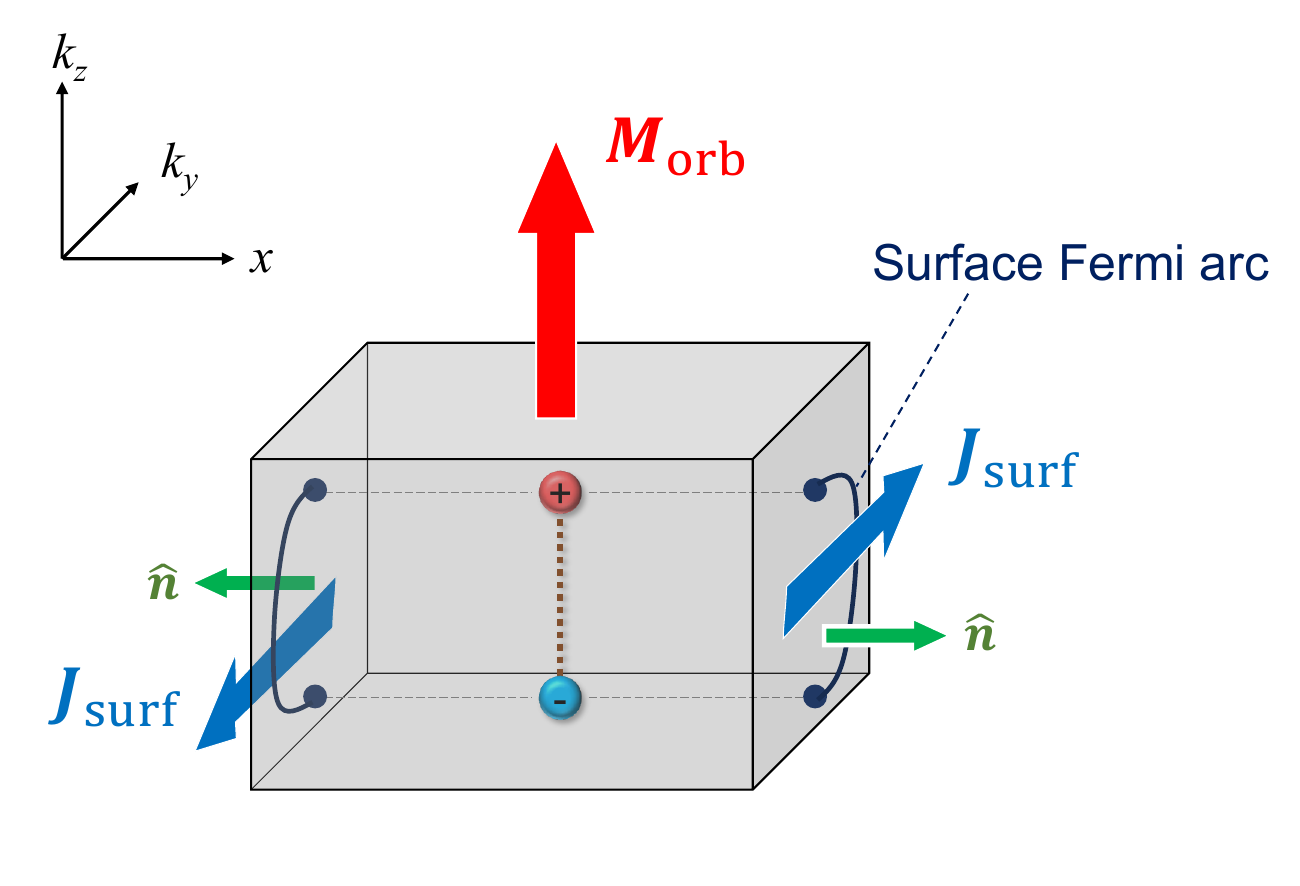}
\caption{
Schematic picture of the orbital magnetization $\boldsymbol{M}_{\rm orb}$, the surface Fermi arcs,
and the surface current $\boldsymbol{J}_{\rm surf}$ carried by them.}	
\label{fig:orbital-mag}
\end{figure}

While the orbital magnetization is defined as a bulk quantity,
it also has a correspondence to the edge states,
particularly the Fermi-arc surface states in the magnetic Weyl semimetals discussed here.
The orbital magnetization corresponds to the circulating current, with its density
\begin{align}
    \boldsymbol{j} &= \boldsymbol{\nabla} \times \boldsymbol{M}_{\rm orb}(\boldsymbol{r}).
\end{align}
If we take a surface of the 3D system characterized by the unit vector $\hat{\boldsymbol{n}}$ normal to the surface,
the surface serves as the boundary between the regions with $\boldsymbol{M}_{\rm orb}(\boldsymbol{r}) \neq 0$ and $\boldsymbol{M}_{\rm orb}(\boldsymbol{r}) = 0$,
and hence it hosts the 2D current density,
\begin{align}
    \boldsymbol{J}_{\rm surf} &= \boldsymbol{M}_{\rm orb} \times \hat{\boldsymbol{n}}. \label{eq:J_surf}
\end{align}
In magnetic Weyl semimetals,
the surface current shown above is attributed to the Fermi-arc surface states [see Fig.~\ref{fig:orbital-mag} ].
For example,
let us take again the Wilson--Dirac model with the Weyl points at $(0,0,\pm K_0)$,
which shows the orbital magnetization $\boldsymbol{M}_{\rm orb} \parallel z$ in the bulk.
If we take a surface with $\hat{\boldsymbol{n}} = \hat{x}$,
it hosts a Fermi arc connecting the projected Weyl points $(k_y,k_z) = (0, \pm K_0)$,
with the unidirectional dispersion along the $y$-direction.
Thus, even though there is no external force applied to this system,
the Fermi-arc states carry a current along the $y$-direction in equilibrium,
which corresponds to the magnetization current given by Eq.~(\ref{eq:J_surf}).
This surface current is an equilibrium current,
which cannot be extracted by any transport measurements.

\subsection{Gauge fields for Weyl fermions} \label{sec:chiral-gauge-field}

To further discuss the magnetoelectric phenomena arising from the Weyl fermions,
we introduce the gauge field describing the external electromagnetic perturbations.
In addition to the conventional U(1) gauge field for the electromagnetic fields, the concept of a chiral~(or axial) gauge field, which describes the coupling between Weyl fermions and other degrees of freedom, plays an important role in understanding the magnetoelectric phenomena. 
Before going to the specific phenomena in Weyl semimetals,
we introduce the concept of chiral gauge field to describe these phenomena.

\subsubsection{Conventional and chiral gauge fields}

The Weyl fermion with chirality $\eta$ interacting with the electromagnetic field is described by the minimally-coupled Hamiltonian,
\begin{align}
    H_\eta(\boldsymbol{k}) = \eta \hbar v_{\rm F} \bm{\sigma}\cdot\left(\bm{k}+\frac{e}{\hbar}{\bm A} \right),
\end{align}
where the gauge field (vector potential) ${\bm A}({\bm r},t)$ accounts for the electromagnetic field.
On the other hand, the chiral gauge field does not correspond to the real electromagnetic fields, but describes the effect of a perturbation that shifts the positions of Weyl points. 
To understand the idea of the chiral gauge field in analogy with the conventional gauge field,
here we start with a simple model with exchange coupling to magnetization ${\bm M}$,
$H_\eta(\boldsymbol{k}) = \eta \hbar v_{\rm F} \bm{\sigma}\cdot\bm{k}+J \bm{\sigma}\cdot {\bm M}$.
This model can be rewritten as,
\begin{eqnarray}
    H_\eta(\boldsymbol{k}) = \eta \hbar v_{\rm F} \bm{\sigma}\cdot 
    \Bigl( \bm{k}+\eta\frac{e}{\hbar}{\bm A}_5 \Bigr),
    \label{eq:Hamiltonian-A5}
\end{eqnarray}
with ${\bm A}_5 = (J/ev\_F){\bm M}$.
This form implies that the Weyl point for each chirality $\eta$ is located at ${\bm K}_\eta = -\eta (e/\hbar){\bm A}_5 = -\eta(J/\hbar v\_F){\bm M}$,
as shown by Eq.~(\ref{eq:weyl-shifted}).
Meanwhile, we can see that the appearance of $\boldsymbol{A}_5$ in Eq.~(\ref{eq:Hamiltonian-A5}) is quite similar to that of $\boldsymbol{A}$ in Eq.~(\ref{eq:hamiltonian-A}), except for the $\eta$-dependence in the coupling coefficient for $\boldsymbol{A}_5$.
In this sense, the coupling to ${\bm M}$ is regarded as the chirality-dependent gauge field, i.e., the \textit{chiral} gauge field ${\bm A}_5$
~(Since ${\bm A}_5$ is an axial vector, it is also called the axial gauge field).
By combining the conventional and chiral gauge fields,
the total Hamiltonian reads,
\begin{align}
    H_\eta(\boldsymbol{k}) &= \eta \hbar v_{\rm F} \boldsymbol{\sigma}\cdot \left(\boldsymbol{k} + \frac{e}{\hbar}\boldsymbol{A}_\eta \right),
    \label{eq:hamiltonian-A}
\end{align}
with ${\bm A}_{\eta} = {\bm A} + \eta {\bm A}_{5} \ (\eta = \pm)$.

The idea of chiral gauge field was originally introduced in the context of realtivistic quantum field theory \cite{Bardeen1969PhysRev,Landsteiner2016notes}.
While the chiral gauge field does not physically exist in the relativistic quantum mechanics,
recent theoretical studies have proposed that the chiral gauge field
can be emulated in Weyl semimetals~\cite{Liu2013PRB,Pikulin2016PRX,ilan2020pseudo}.
As seen above with the simple model,
the effect of external perturbations inducing chirality-dependent shift of Weyl points $(\boldsymbol{K}_\eta)$,
such as the perturbation of magnetization,
can be treated as the chiral gauge field $\boldsymbol{A}_5$ for Weyl fermions.
Several other perturbations,
such as a lattice strain \cite{Cortijo2015,Pikulin2016PRX}, gradient of chemical composition \cite{Kariyado2019jpsj}, circularly polarized light \cite{Ebihara2016,Bucciantini2017},
have also been studied in terms of the chiral gauge field.

Once the external perturbation on a Weyl semimetal is written in terms of the chiral gauge field,
its effect on the electrons can be treated just like the conventional gauge field for the electromagnetism.
Similarly to the electromagnetic fields $\boldsymbol{E}=-\partial_t\boldsymbol{A}$
and $\boldsymbol{B} = \boldsymbol{\nabla}\times\boldsymbol{A}$,
we can treat the effect of the chiral gauge field as the \textit{chiral} electromagntic fields,
$\boldsymbol{E}_5=-\partial_t\boldsymbol{A}_5$
and $\boldsymbol{B}_5 = \boldsymbol{\nabla}\times\boldsymbol{A}_5$,
which act on the Weyl fermions of different chiralities $(\eta = \pm)$ with the opposite signs.
The effects of the conventional and chiral electromagnetic fields on the Weyl fermions are peculiar, 
which shall be reviewed in detail in Sec.~\ref{sec:magnetoelectric-effects}.
In particular,
the magnetoelectric effects arising from perturbations in magnetic ordering,
such as magnetization dynamics and nonuniform magnetic textures,
shall be discussed in Sec.~\ref{sec:texture}.

\subsubsection{Gauge field and symmetry}
For the detailed understanding of the theoretical fundamentals of the chiral gauge field,
we shall briefly review the gauge symmetry of Weyl fermions.
(The readers interested in the phenomenology of Weyl semimetals may skip to Sec.~\ref{sec:magnetoelectric-effects}.)
Similarly to the conventional gauge field corresponding to the U(1) gauge symmetry,
the chiral gauge field also corresponds to another U(1) gauge symmetry, known as the \textit{chiral} gauge symmetry, for the Weyl (or Dirac) fermions.

We start with the Lagrangian formalism of the massless Dirac fermion (Weyl fermion) \cite{dirac1928quantum,weyl1929gravitation,peskin1995introduction}
with the Lagrangian density,
\begin{align}
    \mathcal{L}_{\mathrm{Weyl}} &= \sum_{\eta =\pm} \psi_\eta^\dag \left[i\hbar\partial_t -\eta v_{\rm F} \boldsymbol{\sigma}\cdot\boldsymbol{\hat{p}}\right] \psi_\eta 
    = i\hbar \bar{\Psi} \left[\gamma_0\partial_t - v_{\rm F} \boldsymbol{\gamma}\cdot\boldsymbol{\nabla} \right] \Psi, \label{eq:weyl-lagrangian}
\end{align}
where we take the momentum operator $\boldsymbol{\hat{p}} = -i\hbar\boldsymbol{\nabla}$.
In relativistic quantum mechanics,
the Fermi velocity $v_{\rm F}$ is replaced with the speed of light $c$.
Here, $\psi_{\eta = \pm}(\boldsymbol{r},t)$ are the Weyl spinor fields
with $\eta$ labeling the fermion chirality.
In the Weyl representation,
the 4-component Dirac spinor is constructed as $\Psi = \begin{pmatrix} \psi_+ \\ \psi_- \end{pmatrix}$,
with $\bar{\Psi} = \Psi^\dag \gamma_0$,
and the $\gamma$-matrices read
\begin{align}
    \gamma_0 = \begin{pmatrix}
        0 & 1 \\
        1 & 0
    \end{pmatrix}, \ 
    \gamma_{i=x,y,z} = \begin{pmatrix}
        0 & -\sigma_i \\
        \sigma_i & 0
    \end{pmatrix}
    .
\end{align}
The chirality operator,
\begin{align}
    \gamma_5 = i \gamma_0 \gamma_x \gamma_y \gamma_z = \begin{pmatrix}
        1 & 0 \\
        0 & -1
    \end{pmatrix},
\end{align}
acts on each component $\psi_\pm$ in $\Psi$
as a factor $\pm 1$, respectively.

As long as the Dirac mass is zero,
$\psi_+$ and $\psi_-$ 
are independent of each other.
Therefore, apparently at the classical level, they satisfy two distinct U(1) gauge symmetries,
\begin{align}
    \psi_\pm \rightarrow e^{i\theta_\pm} \psi_\pm,
\end{align}
which rotate the phases of $\psi_\pm$ independently.
These transformations can be written for the Dirac spinor $\Psi$ as
\begin{align}
    \Psi \rightarrow 
    \begin{pmatrix}
        e^{i\theta_+} & 0 \\
        0 & e^{i\theta_-}
    \end{pmatrix}
    \Psi
    = e^{i(\theta + \theta_5 \gamma_5)} \Psi,
\end{align}
with $\theta = \frac{1}{2}(\theta_+ + \theta_-)$ and $\theta_5 = \frac{1}{2}(\theta_+ - \theta_-)$.
Here, $\theta$ gives the rotation of the overall phase factor,
which is the conventional U(1) gauge transformation [$\mathrm{U(1)_V}$],
corresponding to the conservation of particle number (or electric charge).
On the other hand, the chirality-dependent rotation by $\theta_5$ is termed as the chiral gauge transformation [$\mathrm{U(1)_A}$].
While the Weyl fermions apparently satisfy this U(1) chiral gauge symmetry,
conservation of the ``chiral charge'' is essentially violated by the quantum effect,
which is known as the chiral anomaly (see Sec.~\ref{sec:anomaly}).

The coupling to the U(1) gauge field (4-vector) $A = (A_0, \boldsymbol{A})$
is introduced with the local $\mathrm{U(1)_V}$ gauge symmetry.
Similarly, by considering the local $\mathrm{U(1)_A}$ chiral gauge transformation,
we can introduce the coupling to the chiral gauge field (4-vector)  $A_5 = (A_{5,0}, \boldsymbol{A}_5)$.
The Lagrangian coupled with $A$ and $A_5$ becomes
\begin{align}
    \mathcal{L}_{\mathrm{Weyl}} = i\hbar \bar{\Psi} \left[ \gamma_0 \left(\partial_t-i\frac{e}{\hbar}\mathcal{A}_0 \right) - v_{\rm F} \boldsymbol{\gamma}\cdot \left(\boldsymbol{\nabla} -i\frac{e}{\hbar}\boldsymbol{\mathcal{A}} \right) \right] \Psi,
\end{align}
with $\mathcal{A}= (\mathcal{A}_0,\boldsymbol{\mathcal{A}})= A + A_5 \gamma_5$.
The chiral gauge field $A_5 = (A_{5,0}, \boldsymbol{A}_5)$ is coupled to the chiral charge $\rho_5$ and chiral current $\boldsymbol{j}_5$,
defined as
\begin{align}
    \rho_5 &= -e\Psi^\dag \gamma_5 \Psi = \rho_+ -\rho_-, \\
    \boldsymbol{j}_5 &= -e v_{\rm F}\Psi^\dag \gamma_0\boldsymbol{\gamma}\gamma_5 \Psi = \boldsymbol{j}_+ - \boldsymbol{j}_-,
\end{align}
with $\rho_\pm$ and $\boldsymbol{j}_\pm$ the charge and current of left- or right-handed Weyl fermions.
In terms of the Weyl spinors, this Lagrangian can be written as
\begin{align}
    \mathcal{L}_{\mathrm{Weyl}} &= \sum_{\eta =\pm} \psi_\eta^\dag \left[i\hbar\partial_t - eA_{\eta,0} +\eta v_{\rm F} \boldsymbol{\sigma}\cdot(\boldsymbol{p}+e\boldsymbol{A}_\eta)\right] \psi_\eta,
\end{align}
where
$A_{\eta,\mu} = A_{\mu} +\eta A_{5,\mu} \ (\mu = 0,x,y,z)$
serves as the gauge field for each chirality $\eta = \pm$.
This Lagrangian corresponds to the Hamiltonian form for each chirality sector $\eta$, given by Eq.~(\ref{eq:hamiltonian-A}).

\subsection{Magnetoelectric effects of Weyl fermions} \label{sec:magnetoelectric-effects}
Once the ordinary electromagnetic fields $(\boldsymbol{E},\boldsymbol{B})$ are coupled to the Weyl fermions,
they yield various effects attributed to the topological nature of the Weyl fermions~\cite{gorbar2018anomalous,gorbar2021electronic, ahmad2025geometry}.
Theoretically, the effects from chiral electromagnetic fields $(\boldsymbol{E}_5,\boldsymbol{B}_5)$ can be treated in a manner analogous to $(\boldsymbol{E},\boldsymbol{B})$.
Here, we compare the effects of $(\boldsymbol{E},\boldsymbol{B})$ and those  from $(\boldsymbol{E}_5,\boldsymbol{B}_5)$
in the three typical phenomena:
the quantum Hall effect (QHE), the chiral magnetic effect (CME), and the chiral anomaly.
In this part, we identify the Fermi level $E_{\rm F}$ as the chemical potential $\mu$,
in accordance with the conventionally used notation in the relativistic quantum field theory.

\subsubsection{Quantum Hall effect (QHE)} \label{sec:quantum-hall}

\begin{figure}[htb]
\centering
\includegraphics[width=1.0\hsize]{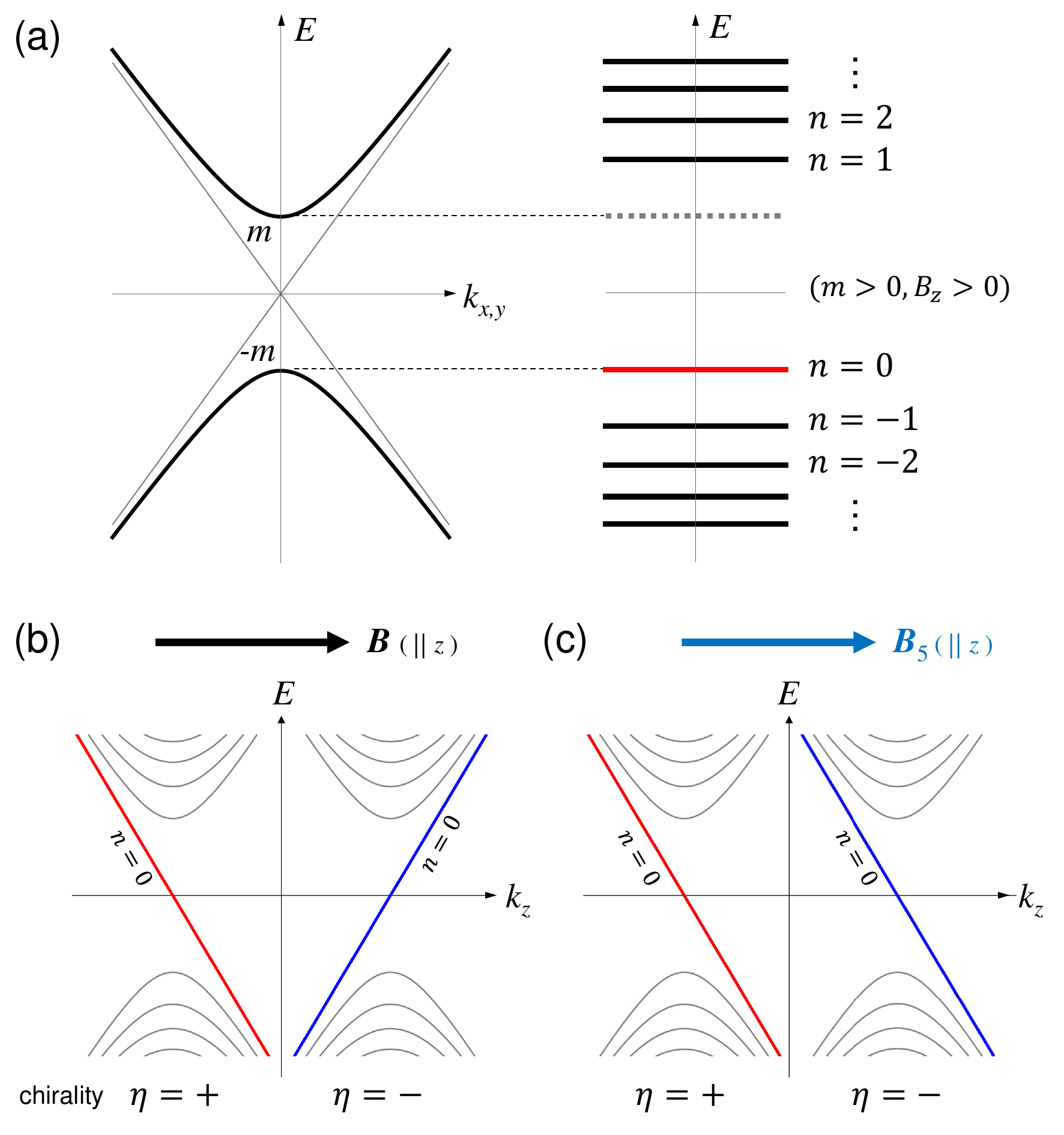}
\caption{
The structures of the Landau levels (LLs):
(a) The LLs for the 2D massive Dirac fermions.
(b) The LLs for the Weyl fermions under the conventional magnetic field $\boldsymbol{B}$ in 3D, and (c) those under the chiral magnetic field $\boldsymbol{B}_5$.
}	
\label{fig:landau-levels}
\end{figure} 

For 2D fermion systems,
an out-of-plane magnetic field $\boldsymbol{B} = B \hat{z}$ leads to the discretization of the level structure,
known as the Landau quantization \cite{laughlin1981quantized}.
As a result, the Hall conductivity is quantized by the unit of $e^2/h$,
which is the well-known quantum Hall effect \cite{klitzing1980new,tsui1982two}.
For the 2D massive Dirac fermions,
which is described by the $2\times 2$ Hamiltonian $H^{\rm 2D}(k_x,k_y) = \hbar v_{\rm F}(k_x \sigma_x + k_y \sigma_y) + m\sigma_z$,
the Landau levels (LLs) $E^{\rm 2D}_n \ (n \in \mathbb{Z})$ form a characteristic structure \cite{jackiw1984fractional,haldane1988model},
\begin{align}
    E^{\rm 2D}_{n \neq 0} =\mathrm{sgn}(n)\sqrt{2\hbar v_{\rm F}^2 e |nB| +m^2} , \ E^{\rm 2D}_0 = -\mathrm{sgn}(B) m,
\end{align}
as shown in Fig.~\ref{fig:landau-levels}(a).
Each LL has the degeneracy of the factor $B/\phi_0$ per unit area,
where $\phi_0 = h/e$ is the magnetic flux quantum.
While the $\pm n$-th LLs $(|n| \geq 1)$ satsify particle-hole symmetry,
the zeroth LL does not have its opposite-energy counterpart,
whose sign of the energy depends on the signs of $m$ and $B$.
Due to this unpaired zeroth LL, the Hall codncutivity becomes a half-odd integer,
$\sigma_{\rm H}^{\mathrm{2D}} = (e^2/h)(n+1/2)$ for a single Dirac cone.
In realistic 2D Dirac electron systems like graphene,
this value is multiplied by the degeneracies of spin and valley degrees of freedom,
which is well observed in experiments \cite{novoselov2005two,zhang2005experimental}.

Let us now proceed to the case for 3D Weyl fermions.
For the sake of concreteness,
here we assume the isotropic Weyl Hamiltonian $H_\eta(\boldsymbol{k}) = \eta \hbar v_{\rm F} \boldsymbol{k}\cdot\boldsymbol{\sigma}$
with the chirality $\eta = \pm$,
corresponding to the Lagrangian of Eq.~(\ref{eq:weyl-lagrangian}).
If the magnetic field is applied along the $z$-axis,
i.e., $\boldsymbol{B} = B\hat{z}$,
the translational symmetry only in the $z$-direction is preserved,
and hence $k_z$ serves as a good quantum number in addition to the LL index $n$.
The LL structure is given by substituting $m$ in the 2D Dirac fermions with $v_{\rm F}k_z$,
yielding \cite{nielsen1983adler,Yang2011prb}
\begin{align}
    E_{\eta,n\neq 0}(k_z) &= \mathrm{sgn}(n)\hbar v_{\rm F} \sqrt{2 \frac{e}{\hbar}|nB| +k_z^2}, \nonumber \\
    E_{\eta,0}(k_z) &= -\eta \ \mathrm{sgn}(B) \hbar v_{\rm F} k_z.
\end{align}
Each LL is dispersed along the direction of $\boldsymbol{B}$.
While the nonzero LLs are in the quadratic dispersions,
the zeroth LL shows the linear dispersion, whose direction depends on the chirality $\eta$ of the Weyl fermion.
Therefore, for the pair of left- and right-handed Weyl fermions,
the dispersion directions of their zeroth LLs are opposite to each other, as shown in Fig.~\ref{fig:landau-levels}(b).
The Hall conductivity from these LLs is given by integrating $\sigma\_H^{\mathrm{2D}}(k_z)$,
which is the quantized Hall conductivity at the 2D plane with fixed $k_z$.
If the LLs are well separated and the Fermi level $\mu$ is crossing only the zeroth LL, i.e., $|\mu| < v_{\rm F} \sqrt{2e|B|/\hbar}$,
the Hall conductivity
is governed by the zeroth LL.
The Hall conductivity $\sigma\_H^{\mathrm{2D}}(k_z)$ becomes $+e^2/2h$ if the zeroth LL is occupied $[E_{\eta,0}(k_z) < \mu]$,
and $-e^2/2h$ if unoccupied $[E_{\eta,0}(k_z) > \mu]$.
Thus, the 3D Hall conductivity for each Weyl node becomes linear in $\mu$ and independent of $\eta$,
\begin{align}
    \sigma_{\rm H}^{\mathrm{3D}} &= \int_{-\infty}^{\infty} \frac{dk_z}{2\pi} \frac{e^2}{2h} \mathrm{sgn}[\mu - E_{\eta,0}(k_z)] = \frac{e^2}{h^2} \frac{\mu}{v_{\rm F}}.
\end{align}
When an electric field $\boldsymbol{E}$ is applied to the system,
each chirality contributes to the Hall current,
$\boldsymbol{j}_\eta^{\mathrm{(H)}} = \sigma_{\rm H}^{\mathrm{3D}} \hat{\boldsymbol{B}} \times \boldsymbol{E}$,
yielding the net charge current,
\begin{align}
    \boldsymbol{j}^{\mathrm{(H)}} &=
    \boldsymbol{j}_+^{\mathrm{(H)}} + \boldsymbol{j}_-^{\mathrm{(H)}}
     = \frac{2 e^2}{h^2} \frac{\mu}{v_{\rm F}} \hat{\boldsymbol{B}} \times \boldsymbol{E}.
\end{align}

Let us now consider the effect of the chiral magnetic field $\boldsymbol{B}_5 (=B_5 \hat{z})$ instead of the ordinary magnetic field.
In this case, the Weyl fermions are subject to the chirality-dependent magnetic field $\boldsymbol{B}_\eta = \eta \boldsymbol{B}_5$,
and hence the structure of the LLs becomes \cite{Liu2013PRB,Grushin2016PRX,Pikulin2016PRX},
\begin{align}
    E_{\eta,n\neq 0}(k_z) &= \mathrm{sgn}(n) \hbar v_{\rm F} \sqrt{2\frac{e}{\hbar}|nB_5| +k_z^2}, \nonumber \\
    E_{\eta,0}(k_z) &= -\mathrm{sgn}(B_5) \hbar v_{\rm F} k_z.
\end{align}
The non-zero LL structure is identical to that from $\boldsymbol{B}$,
whereas the difference appears in the zeroth LLs.
Both the left- and right-handed Weyl fermions exhibit the zeroth LLs dispersed antiparallel to $\boldsymbol{B}_5$,
which are known as the chiral Landau levels,
as shown in Fig.~\ref{fig:landau-levels}(c).
They yield the chirality-dependent Hall current,
\begin{align}
    j_\eta^{({\rm H})} &= \sigma_{\rm H}^{\mathrm{3D}} \eta \hat{\boldsymbol{B}}_5 \times \boldsymbol{E}.
\end{align}
The net contribution to the Hall current vanishes, i.e., $\boldsymbol{j}^{\mathrm{(H)}} = \boldsymbol{j}_+^{\mathrm{(H)}} + \boldsymbol{j}_-^{\mathrm{(H)}} =0$,
whereas it gives rise to the chiral Hall current,
\begin{align}
    \boldsymbol{j}_5^{\mathrm{(H)}} &=
    \boldsymbol{j}_+^{\mathrm{(H)}} - \boldsymbol{j}_-^{\mathrm{(H)}}
    = \frac{2e^2}{h^2} \frac{\mu}{v_{\rm F}} \hat{\boldsymbol{B}}_5 \times \boldsymbol{E}
\end{align}
in the quantum Hall regime.
This chiral current is capable of generating the electron spin polarization,
which can exert a spin torque on magnetic textures \cite{kurebayashi2019theory,kurebayashi2021jpsj},
as we shall see in Sec.~\ref{sec:texture}.

In the case under ${\bm E}_5$ and ${\bm B}_5$,
the Hall current for each chirality becomes,
\begin{align}
    j_\eta^{\mathrm{(H)}} &= \sigma_{\rm H}^{\mathrm{3D}} (\eta\hat{\boldsymbol{B}}_5) \times (\eta \boldsymbol{E}_5) = \sigma_{\rm H}^{\mathrm{3D}} \hat{\boldsymbol{B}}_5 \times \boldsymbol{E}_5,
\end{align}
and hence gives a net charge current,
\begin{align}
    \boldsymbol{j}^{\mathrm{(H)}} &=
    \boldsymbol{j}_+^{\mathrm{(H)}} + \boldsymbol{j}_-^{\mathrm{(H)}}
    = \frac{2e^2}{h^2} \frac{\mu}{v_{\rm F}} \hat{\boldsymbol{B}}_5 \times \boldsymbol{E}_5.
\end{align}
This Hall current is expected to be driven by the dynamics of magnetic textures,
which shall be seen in Sec.~\ref{sec:spin-motive-force}.

While we have so far considered the Landau quantization of linearly dispersed bands,
the band dispersions in realistic materials have nonlinearity.
It was theoretically suggested that
the LLs under such nonlinearity get deformed and energetically shifted depending on $|{\bm B}|$,
which may leads to the nonmonotonic $|{\bm B}|$-dependence in the quantum Hall conductance~\cite{Zhang2024}.

\begin{figure*}[tb]
\centering
\includegraphics[width=1.0\hsize]{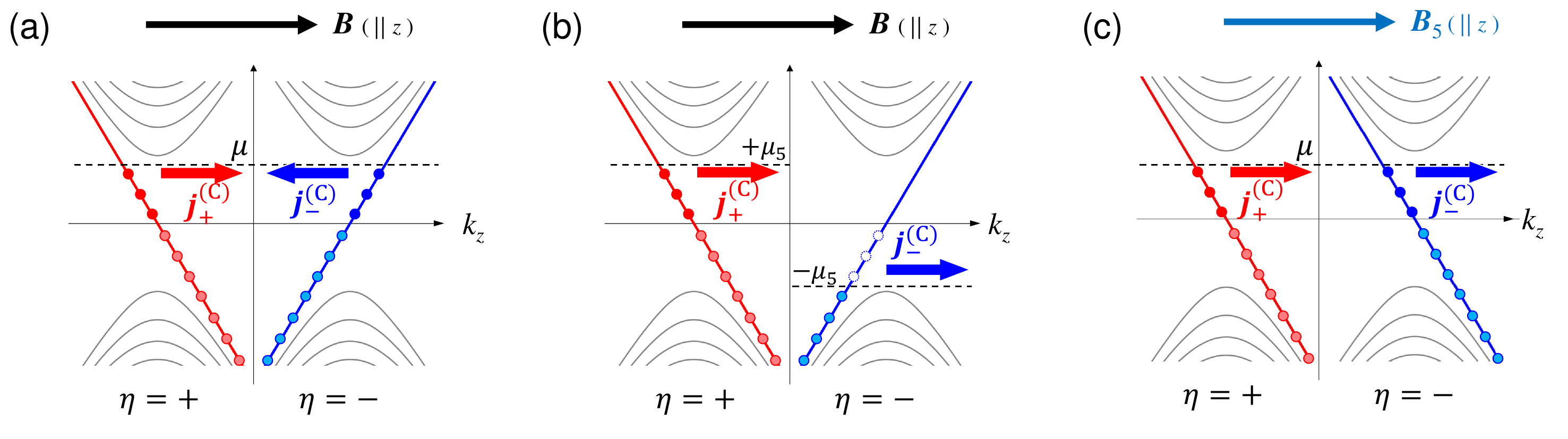}
\caption{
Schematic pictures of the chiral magnetic effect focusing on the zeroth Landau level.
(Note that each electron in the occupied state carries a negative charge $-e$.)
(a) Under the magnetic field $\boldsymbol{B}$ with the finite chemical potential $\mu$,
the Weyl fermions give the chiral current $\boldsymbol{j}^{\rm (C)}_5 = \boldsymbol{j}^{\rm (C)}_+ - \boldsymbol{j}^{\rm (C)}_-$.
(b) Under the magnetic field $\boldsymbol{B}$ with the chiral chemical potential $\mu_5$,
the Weyl fermions give the charge current $\boldsymbol{j}^{\rm (C)} = \boldsymbol{j}^{\rm (C)}_+ + \boldsymbol{j}^{\rm (C)}_-$.
(c) Under the chiral magnetic field $\boldsymbol{B}_5$ with the finite chemical potential $\mu$,
the Weyl fermions give the charge current $\boldsymbol{j}^{\rm (C)} = \boldsymbol{j}^{\rm (C)}_+ + \boldsymbol{j}^{\rm (C)}_-$.
}	
\label{fig:CME}
\end{figure*} 

\subsubsection{Chiral magnetic effect~(CME)} \label{sec:chiral-magnetic-effect}

Since the zeroth LL is unidirectionally dispersed along the magnetic field direction,
it contributes to the current along the field direction for each chirality,
\begin{align}
    \boldsymbol{j}_\eta^{\mathrm{(C)}} &= \eta \frac{e^2 \mu}{h^2} \boldsymbol{B},
    \label{eq:CME-each}
\end{align}
which is known as the CME \cite{nielsen1983adler,metlitski2005anomalous,Fukushima2008}.
For the emergence of the CME,
the Landau quantization is not necessary,
and it is derived also from the linear response theory \cite{kharzeev2009chiral}, the semiclassical approach known as the chiral kinetic theory \cite{son2013kinetic},
the axion electrodynamics \cite{goswami2013axionic,chen2013axion}, etc.

The net charge and chiral currents can be constructed from Eq.~(\ref{eq:CME-each}).
Under the nonzero chemical potential $\mu (\neq 0)$, the CME contributes to the chiral current along the magnetic field,
\begin{align}
    \boldsymbol{j}^{\rm (C)}_5 &=  \boldsymbol{j}^{\rm (C)}_{+}-\boldsymbol{j}^{\rm (C)}_{-} = \frac{2e^2 \mu}{h^2} \boldsymbol{B},
\end{align}
leading to the local imbalance of fermion chirality [see Fig.~\ref{fig:CME}(a)].
This effect is sometimes called the ``chiral separation effect''.
On the other hand,
the net charge current from the CME appears to be present
once we assume the chirality-dependent chemical potential $\mu_\eta$ [see Fig.~\ref{fig:CME}(b)],
\begin{align}
    \boldsymbol{j}^{\rm (C)} &= \boldsymbol{j}^{\rm (C)}_{+}+\boldsymbol{j}^{\rm (C)}_{-} = \frac{2 e^2 \mu_5}{h^2} \boldsymbol{B}. \label{eq:CME-B-tot}
\end{align}
Here, the chemical potential difference $\mu_5 = (\mu_+ - \mu_-)/2$
is known as the chiral chemical potential in the relativistic quantum field theory.
In solid states, such a CME current should vanish in equilibrium,
because it is cancelled by the Fermi sea contribution from the whole Brillouin zone,
including the region away from the Weyl points \cite{Vazifeh2013,yamamoto2015generalized}.
The CME for the charge current is permitted in nonequilibrium systems \cite{sekine2016chiral}.
There are some proposals to obtain the non-equilibrium charge current from the CME.
One case is the so-called gyrotropic magnetic effect,
where a charge current responds to the time-dependent magnetic field oscillating faster than the momentum relaxation time \cite{zhong2016gyrotropic,ma2015chiral}.
Here, the response coefficient is different from $e^2/2\pi^2$ expected in equilibrium.
Another case is with the chiral chemical potential $\mu_5 \propto \boldsymbol{E}\cdot\boldsymbol{B}$
induced dynamically by the chiral anomaly, which leads to negative longitudinal magnetoresistivity,
as we shall revisit below.

For the case of ${\bm B}_5$, it
also induces the formation of the LLs.
Since the dispersions of the zeroth LLs for $\eta=+$ and $\eta=-$ are in the same direction,
the CME current becomes independent of $\eta$,
\begin{align}
    \boldsymbol{j}_\eta^{\mathrm{(C)}} &= \frac{e^2 \mu}{h^2} \boldsymbol{B}_5.
\end{align}
Therefore, the chiral current vanishes, ${\bm j}^{\rm (C)}_5={\bm j}^{\rm (C)}_{+}-{\bm j}^{\rm (C)}_{-}=0$.
On the other hand, 
it yields the net charge current along $\boldsymbol{B}_5$ [see Fig.~\ref{fig:CME}(c)],
\begin{align}
    \boldsymbol{j}^{\rm (C)} &= \boldsymbol{j}^{\rm (C)}_{+}+\boldsymbol{j}^{\rm (C)}_{-} = \frac{2 e^2 \mu}{h^2} \boldsymbol{B}_5. \label{eq:CME-B5-tot}
\end{align}
This effect is sometimes called the chiral pseudomagnetic effect,
or the axial magnetic effect \cite{zhou2013topological,huang2017topological}.
In contrast to $\boldsymbol{j}^{\mathrm{(C)}}$ from $\boldsymbol{B}$ in Eq.~(\ref{eq:CME-B-tot}),
which is cancelled in equilibrium of solid states,
$\boldsymbol{j}^{\mathrm{(C)}}$ from $\boldsymbol{B}_5$ in Eq.~(\ref{eq:CME-B5-tot}) is a physical current that is present even in equilibrium.
However, since the equilibrium current does not exert a work,
this current cannot be extracted and measured as a transport current.
By the numerical calculations with lattice models of Weyl fermions~\cite{Pikulin2016PRX,Shitade2021prb},
it is confirmed that such an equilibrium current is locally present in the region with a finite $\boldsymbol{B}_5$ coming from the lattice strain,
whereas the charge transport is absent in total.
The cancellation of charge transport is due to the counterflow current on the boundary,
conveyed by the Fermi-arc states.

\subsubsection{Chiral anomaly} \label{sec:anomaly}

\begin{figure}[tb]
\centering
\includegraphics[width=1.0\hsize]{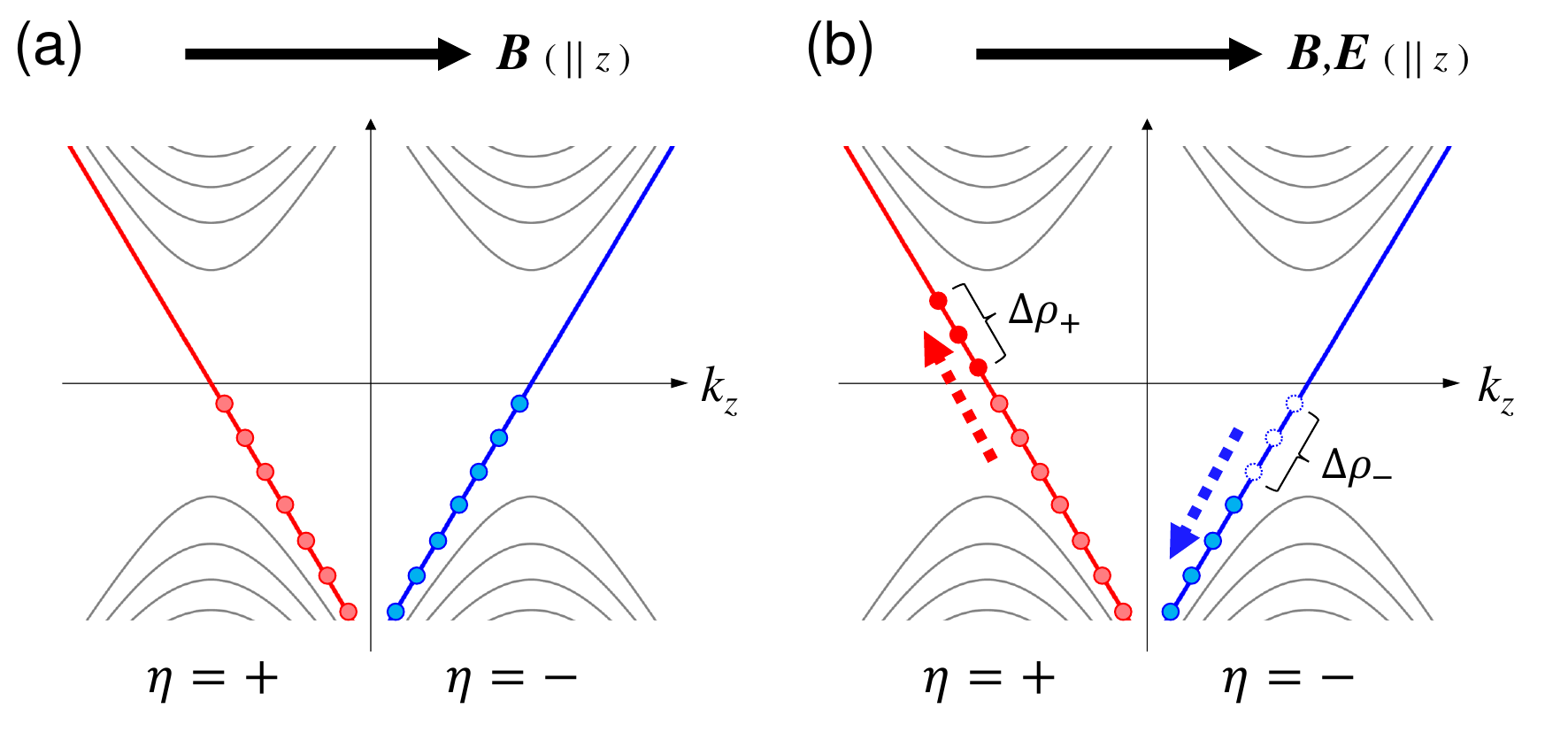}
\caption{
Schematic pictures showing the chiral anomaly under the electromagnetic fields.
The Fermi level is taken as $\mu =0$ for simplicity of illustration.
(a) The magnetic field $\boldsymbol{B}$ induces the Landau quantization,
giving rise to the linearly dispersed zeroth Landau levels.
(b) Once the electric field $\boldsymbol{E}$ is applied in parallel to $\boldsymbol{B}$,
the force $-e\boldsymbol{E}$ on each electron leads to the shift of the occupied states,
yielding the imbalance of chiral charge $\Delta \rho_+ - \Delta \rho_-$.
}	
\label{fig:anomaly}
\end{figure} 

The chiral anomaly is the violation of the $\mathrm{U(1)_A}$ chiral gauge symmetry
from the quantum effect of Weyl fermions.
The ordinary electromagnetic fields $(\boldsymbol{E},\boldsymbol{B})$ leads to the nonconservation of chiral charge,
\begin{align}
    \partial_t \rho_5 + \boldsymbol{\nabla} \cdot\boldsymbol{j}_5 = -\frac{2e^3}{h^2} \boldsymbol{E}\cdot\boldsymbol{B},
\end{align}
leading to the imbalance of the fermion chirality \cite{Adler1969PhysRev,Bell1969pcac}.
This effect can again be understood in terms of the LLs \cite{nielsen1983adler}, as schematically shown in Fig.~\ref{fig:anomaly}:
if $\boldsymbol{E} = E\hat{z}$ is applied in parallel to $\boldsymbol{B}=B\hat{z}$,
the momenta $k_z$ for the occupied states get shifted by the force $-eE$,
leading to increase of the fermions with $\eta = +$ and decrease of the fermions with $\eta = -$.

Such a charge transfer between the Weyl nodes of different chiralities is usually relaxed by the scattering between these nodes.
As a result, the system eventually reaches the steady state with the charge imbalance between the Weyl nodes,
which is regarded as the dynamically induced chiral chemical potential
$\mu_5 = (\mu_+ - \mu_-)/2 \propto \tau_\chi \boldsymbol{E}\cdot\boldsymbol{B}$~\cite{nielsen1983adler}.
Here, $\tau_\chi$ is the relaxation time for the inter-node scattering processes.
This $\mu_5$ leads to a finite charge current from the chiral magnetic effect [Eq.~(\ref{eq:CME-B-tot})],
\begin{align}
    \boldsymbol{j}^{\mathrm{(C)}} \propto \tau_\chi(\boldsymbol{E}\cdot\boldsymbol{B})\boldsymbol{B}.
\end{align}
It serves as a $\boldsymbol{B}$-dependent correction to the (Drude's) conduction current $\boldsymbol{j}^{\mathrm{(D)}} = \sigma \boldsymbol{E}$,
which can be captured as the the increase of the conductivity $\sigma$ (i.e., the decrease of the resistivity) with respect to the longitudinal magnetic field $\boldsymbol{B}$ \cite{nielsen1983adler,son2013chiral}.
Such an effect, termed as the negative longitudinal magnetoresistivity,
has been measured as a typical evidence of the presence of Weyl fermions in materials,
while it is still controversial whether it is truly from the chiral anomaly~\cite{Armitage2018,Liang2018}.

In a manner similar to $(\boldsymbol{E},\boldsymbol{B})$,
the chiral electromagnetic fields $(\boldsymbol{E}_5,\boldsymbol{B}_5)$ also contribute to the chiral anomaly.
Taking into account the contribution from $(\boldsymbol{E}_5,\boldsymbol{B}_5)$,
the anomaly relations apparently become,
\begin{align}
    \partial_t \rho_5 + \boldsymbol{\nabla} \cdot\boldsymbol{j}_5 &= -\frac{2e^3}{h^2}\left(\boldsymbol{E}\cdot\boldsymbol{B} + \boldsymbol{E}_5\cdot\boldsymbol{B}_5\right), \label{eq:covariant-anomaly-chiral}\\
    \partial_t \rho + \boldsymbol{\nabla} \cdot\boldsymbol{j} &= -\frac{2e^3}{h^2}\left(\boldsymbol{E}\cdot\boldsymbol{B}_5 + \boldsymbol{E}_5\cdot\boldsymbol{B}\right), \label{eq:covariant-anomaly-normal}
\end{align}
which is called the \textit{covariant anomaly} \cite{bardeen1984consistent,Landsteiner2016notes}.
The second relation yields the nonconservation of the total charge $Q = \int d^3 \boldsymbol{r} \ \rho$.
Here, $Q$ measures the charge of electrons belonging to the Weyl fermion states, and this unconserved $Q$ is compensated by the states away from the Weyl points.

To certify the charge conservation in total,
we need to modify the definitions of $(\rho_5,\boldsymbol{j}_5)$ and $(\rho,\boldsymbol{j})$,
so that they include the contributions from the whole momentum space.
For this purpose,
we may add to the Lagrangian a counterterm accounting for the contribution at the boundaries of momentum space,
which is 
known as the Bardeen--Zumino term
 \cite{bardeen1984consistent}.
In particular, if the Lorentz invariance is required,
$(\rho_5,\boldsymbol{j}_5)$ and $(\rho,\boldsymbol{j})$ get modified as,
\begin{align}
    \rho_5^{\rm con} &= \rho_5 + \frac{2e^3}{3 h^2} \boldsymbol{A}_5 \cdot \boldsymbol{B}_5 \\
    \boldsymbol{j}_5^{\rm con} &= \boldsymbol{j}_5 + \frac{2e^3}{3 h^2} (\boldsymbol{A}_5 \times \boldsymbol{E}_5 - A_{5,0} \boldsymbol{B}_5)\\
    \rho^{\rm con} &= \rho + \frac{2e^3}{h^2} \boldsymbol{A}_5 \cdot \boldsymbol{B} \label{eq:consistent-rho} \\
    \boldsymbol{j}^{\rm con} &= \boldsymbol{j} + \frac{2e^3}{h^2} (\boldsymbol{A}_5 \times \boldsymbol{E} - A_{5,0} \boldsymbol{B}). \label{eq:consistent-j}
\end{align}
For instance, the correction term for $\rho$ in Eq.~(\ref{eq:consistent-rho}) comes from the charge density of the zeroth Landau level,
which is occupied for $\boldsymbol{k}$ between the Weyl points.
The correction term for $\boldsymbol{j}$ in Eq.~(\ref{eq:consistent-j}) stands for the anomalous Hall current and the chiral magnetic current
carried by the Fermi-sea electrons.
The divergence of those currents compensates the change in the local charge $\partial_t \rho$ appearing in the covariant anomaly. 
By using these modified quantities,
the relations in Eqs.~(\ref{eq:covariant-anomaly-chiral}) and (\ref{eq:covariant-anomaly-normal}) get corrected,
\begin{align}
    \partial_t \rho^{\rm con}_5 + \boldsymbol{\nabla} \cdot\boldsymbol{j}^{\rm con}_5 &= -\frac{2e^3}{h^2}\left(\boldsymbol{E}\cdot\boldsymbol{B} + \frac{1}{3}\boldsymbol{E}_5\cdot\boldsymbol{B}_5\right), \\
    \partial_t \rho^{\rm con} + \boldsymbol{\nabla} \cdot\boldsymbol{j}^{\rm con} &= 0,
\end{align}
which is termed as the \textit{consistent anomaly}.
The factor $1/3$ in the consistent anomaly comes from the cubic symmetry in the system.
In solid states, this factor $1/3$ is modified,
depending on how to separate the chirality degrees of freedom on lattice.
The Bardeen--Zumino correction in the consistent anomaly was confirmed on a hypothetical lattice model of Weyl semimetal,
by both the analytical derivation \cite{gorbar2017origin} and the numerical calculation \cite{Behrends2019}.
Its effect on the collective electron dynamics in Weyl semimetals has been theoretically studied from various aspects,
such as the plasma oscillations \cite{gorbar2017consistent,gorbar2017chiral,gorbar2017second} and the hydrodynamic flow \cite{gorbar2018consistent,gorbar2018hydrodynamic}.


Besides the chiral anomaly associated with the electromagnetic fields,
the relativistic fermions also host the \textit{gravitational} anomaly,
associated with the Riemann curvature tensors for the gravitational field \cite{chernodub2022thermal}.
While the chiral anomaly appears in the conservation laws for the charge and current,
the gravitational anomaly enters the conservation laws for the energy-momentum tensor.
Thus, the gravitational anomaly gives a contribution to the thermal transport in Weyl (and Dirac) semimetals,
while we do not go into its detail here.

\subsubsection{
Theoretical implications of chiral anomaly
}


Theoretically, both the anomalous Hall effect and the chiral magnetic effect for the Weyl fermions are described in terms of the chiral anomaly in a unified manner \cite{Zyuzin2012,Sekine2020-km}.
By integrating out the fermionic degrees of freedom from the Weyl Lagrangian Eq.~(\ref{eq:weyl-lagrangian}),
the magnetoelectric effect from the Weyl fermions is described by the action,
\begin{align}
    S_\theta &= \frac{e^2}{32\pi^2} \int dt d^3{\bm r} \ \theta(\boldsymbol{r},t) \epsilon^{\alpha\beta\gamma\delta} F_{\alpha\beta} F_{\gamma\delta}, \ (\alpha,\beta,\gamma,\delta = 0,x,y,z)
\end{align}
known as the $\theta$-term,
with the field strength $F_{\alpha\beta} = \partial_\alpha A_\beta - \partial_\beta A_\alpha$.
Here, $\epsilon^{\alpha\beta\gamma\delta}$ is the Levi--Civita symbol for complete antisymmetric summation.
In this formalism, we take the natural unit $\hbar = v_{\rm F} = 1$.
Here, $\theta(r,t)=2e{\bm A_5}\cdot {\bm r}-2{\mu_5}t$ is attributed to the \textit{axion} field in the relativistic quantum field theory.
By applying $j^\alpha = -e^{-1} \delta S_\theta / \delta A_\alpha$,
it yields the anomalous Hall effect,
\begin{align}
    \boldsymbol{j} &= \frac{e^2}{4\pi^2} \boldsymbol{\nabla}\theta \times \boldsymbol{E} = \frac{e^2}{2 \pi^2} \boldsymbol{A}_5 \times \boldsymbol{E},
\end{align}
and the chiral magnetic effect,
\begin{align}
    \boldsymbol{j} &= -\frac{e^2}{4\pi^2}\dot{\theta}\boldsymbol{B} = \frac{e^2 \mu_5}{2\pi^2} \boldsymbol{B} .
\end{align}

Some of the magnetoelectric effects shown in this section have been confirmed experimentally as the evidences of the presence of Weyl fermions in materials,
as we shall review in the following sections.


\section{Magnetic Weyl semimetal materials}
~\label{sec:materials}

\begin{table*}[t]
    \centering
    \caption{List of typical materials exhibiting the magnetic Weyl semimetal state.
    FM and AFM in the column ``Magnetic orderings'' stand for ferromagnet and antiferromagnet, respectively.
    $T_{\rm C,N}$ corresponds to the Curie temperature for FMs and the N\'{e}el temperature for AFMs.}
    \label{tab:materials}
    \begin{tabular}{cccccc}
             \hline
         Year observed & Material    & Magnetic ordering & Crystal structure (space group) & $T_{\rm C,N} \approx$    \\ \hline
         2003~\cite{Fang2003} &\ce{SrRuO3}   & FM & Perovskite $(Pnma)$ & $160\ {\rm K}$ \\ 
         2016~\cite{hirschberger2016} &\ce{GdPtBi} & Forced FM & Half Heusler $(F\bar{4}3m)$ & $(8.8{\rm K})$\footnote{$T_{\rm N}$ for the spontaneous AFM state.\label{fn:AFM}} \\
         2017~\cite{Kuroda2017} &\ce{Mn3Sn} & Chiral AFM & Stacked kagome $(P6_3/m mc)$ & $420\ {\rm K}$ \\
         2018~\cite{Sakai2018} & \ce{Co2MnGa} & FM & Heusler $(Fm\bar{3}m)$  & $700\ {\rm K}$ \\
         2018~\cite{Liu2018} &\ce{Co3Sn2S2} & FM  & Stacked kagome $(R\bar{3}m)$ & $177 \ {\rm K}$   \\
         2018~\cite{soh2018magnetic} & \ce{EuCd2Sb2} & Forced FM & Trigonal $(P\bar{3}m1)$ & $(7.4 \ {\rm K})$\footref{fn:AFM} \\
         2021~\cite{Gaudet2021} & \ce{NdAlSi} & Helical (nearly ferrimagnetic) & Noncentrosymmetric tetragonal $(I4_1 md)$ & $7 \ {\rm K}$ \\
         2024~\cite{belopolski2024took} & $({\rm Cr},{\rm Bi})_2 {\rm Te}_3$ & FM & Trigonal layered $(R\bar{3}m)$ & $150 \ {\rm K}$  \\
         \hline
    \end{tabular}
\end{table*}

In this section, we review the realization of the magnetic Weyl semimetals with various magnetic orderings.
After the early-stage theoretical predictions explained in Sec.~\ref{sec:fundamentals}, 
theoretical and experimental studies in search of Weyl semimetal states have been conducted with various materials.
Prior to magnetic Weyl semimetals,
non-magnetic Weyl semimetals with broken inversion symmetry were experimentally confirmed in several materials,
such as the family of \ce{TaAs}~\cite{Xu2015_1,Xu2015_2}.

Experimental synthesis and verification of magnetic Weyl semimetal materials eventually started from the late 2010s.
Nowadays,
various magnetic materials, not only ferromagnetic but also antiferromagnetic or ferrimagnetic, have been confirmed to exhibit Weyl semimetal states, as summarized in Table \ref{tab:materials}.
In the following, we review the electronic structure, magnetic ordering, and transport properties in those magnetic Weyl semimetal materials.

\subsection{Ferromagnetic systems}

\begin{figure}[t]
\centering
\includegraphics[width=1\hsize]{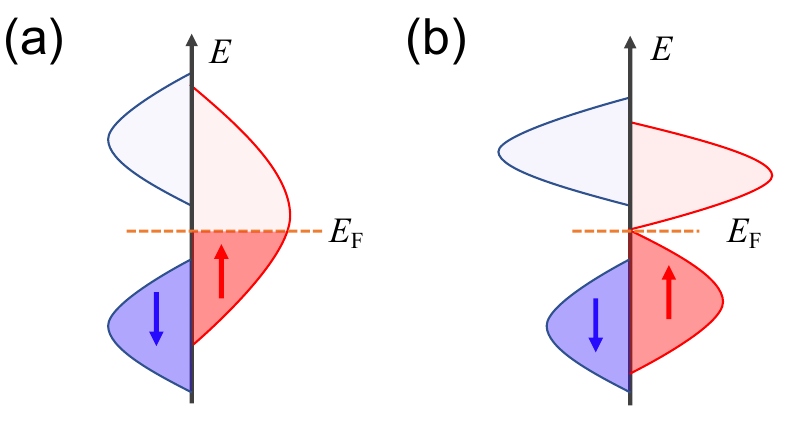}
\caption{
Schematics for the spin-resolved density of states of (a) half metal and (b) spin gapless semiconductor.
}	
\label{fig:half-metal}
\end{figure}

We first review the materials hosting the Weyl fermions under the ferromagnetic orderings.
Since the ferromagnetic orderings can be easily controlled by external magnetic fields,
there have been versatile studies on these materials regarding both the electronic structures and magnetic properties.
Prior to the interest in Weyl semimetals,
some of these materials, such as the Heusler alloys, were studied as candidates for half metals or spin gapless semiconductors.
These states, having (a)~metallic or (b)~semiconductor-like band structure for one spin state and fully gapped for the other [see Fig.~\ref{fig:half-metal}],
have been attracting great attention as a generator of fully spin-polarized electric current.
In such materials,
the majority spin state mainly contributes to the formation of the Weyl nodes.
We mainly review the Weyl semimetal materials with spontaneous magnetization without applying magnetic field.
We also briefly mention some materials showing the Weyl fermions in the forced ferromagnetic states under a magnetic field.

\subsubsection{Ferromagnetic Perovskite \SRO}

The ferromagnetic oxide \SRO is the material whose Weyl point structure was theoretically and experimentally considered at the earliest stage.
\SRO is the $4d$ metallic oxide in a perovskite structure,
which becomes ferromagnetic below the Curie temperature $T_C \sim 160 \ \mathrm{K}$.
Prior to the current interest in its topological band structure,
\SRO had been intensely studied both theoretically and experimentally as the system where the electron correlation plays important roles in the ferromagnetism and metal-to-insulator transition \cite{Singh1996,Mazin2000,Rondinelli2008,Toyota2005,Kim2004}. 
In 2003,
a large anomalous Hall conductivity showing the nonmonotonic temperature dependence was experimentally reported \cite{Fang2003}.
By the first-principles calculations of the band structure including the temperature dependence of the spontaneous magnetization,
they identified the hot spot of Berry curvature in momentum space as
the origin of the anomalous Hall conductivity.
This hot spot, serving as the magnetic monopole in momentum space,
is nothing but the band crossing point that is now called the Weyl point.
This finding was reported earlier than the theoretical predictions of Weyl semimetal in 2011 \cite{Wan2011,Burkov2011},
although the terminology ``Weyl (semi)metal'' or ``Weyl point'' was not introduced in the original paper.
From the tight-binding model calculations later on \cite{Chen2013},
it was predicted that \SRO possesses many pairs of Weyl points coexisting with large Fermi surfaces.

After the finding of the Berry curvature effect probed by the AHE,
the effects from the Weyl fermions were reported by various experiments.
The spin wave modes therein were measured by the inelastic neutron scattering,
where the spin wave gap 
showed a temperature dependence inconsistent with the magnetization curve \cite{Itoh2016}.
This deviation was theoretically attributed to the modification of the spin-spin correlation function from the Berry curvature,
which is proportional to the current-current correlation function for the AHE \cite{Onoda2008}.
In the measurements of the current-induced magnetization switching processes,
the switching efficiency was reported to be largely enhanced
in comparison to those in conventional ferromagnetic metals \cite{Feigenson2007,Yamanouchi2019,yamanouchi2022sciadv}.
From the theoretical analysis, its origin was identified as the intrinsic spin torque component arising from the Berry curvature,
which was termed as the ``topological Hall torque'' (see Sec.~\ref{sec:texture} for its detail) \cite{Araki2021prl}.

In recent experiments,
the crystal quality of \SRO was highly improved,
raising its residual resistivity ratio (RRR) up to $\sim 80$.
Synthesis of such high-quality crystals enabled further measurements of the transport signals from the Weyl fermions,
such as the quantum oscillation and the negative longitudinal magnetoresistivity \cite{Takiguchi2020}, and also the 2D transport from the surface Fermi arc states \cite{Kaneta-Takada2022}.


\subsubsection{Co-based Heusler alloys}

\begin{figure}[htb]
\centering
\caption{
Weyl fermions in the full-Heusler  \ce{Co2MnGa}.
(a) Crystal structure.
(b) Nodal-line structure (red, blue, and yellow curves) in the Brillouin zone obtained by the first-principles calculations without spin-orbit coupling.
(c) Image of the Weyl point structure obtained by the first-principles calculations with spin-orbit coupling.
(d) The anomalous Nernst $(S_{yx})$ and Hall $(\rho_{yx})$ coefficients measured experimentally.
(a) and (b) are cited from Ref.~\cite{chang2017topological}.
(c) and (d) are cited from Ref.~\cite{Sakai2018}.
}	
\label{fig:Co2MnGa}
\end{figure}

The Heusler compounds $X_2 YZ$ generally show the cubic lattice structure,
with $X$, $Y$, and $Z$ atoms forming the fcc-type sublattices, respectively.
Among them,
the family of \ce{Co}-based alloys ${\rm Co}_2 YZ \ (Y= {\rm V, Zr, Nb, Ti,
Mn, Hf}; \ Z={\rm Si, Ge, Sn, Ga, Al})$
have long been explored as the ferromagnetic materials showing half-metallic electronic states and high Curie temperatures \cite{felser2007spintronics,kubler2007understanding,umetsu2008magnetic}.
Such properties are important for spintronics applications,
to make use of the material as an efficient generator of 
spin polarized current.
By the first-principles calculations,
the Weyl point structure and the giant anomalous Hall effect from that are predicted in several species of such Heusler compounds \cite{wang2016time,manna2018colossal}.

\ce{Co2MnGa} is one of such materials studied from the early stage.
Figure.~\ref{fig:Co2MnGa}(a) shows its crystal structure.
If spin-orbit coupling is neglected,
the calclated band structure shows the multiple nodal rings connected in momentum space~\cite{chang2017topological},
in the ferromagnetic state below the Curie temperature $T_c \approx 700 {\rm K}$ [see Fig.~\ref{fig:Co2MnGa}(b)].
Once spin-orbit coupling is included,
these nodal rings are 
partially
gapped out,
reducing to the Weyl point structure [see Fig.~\ref{fig:Co2MnGa}(c)].
As shown in Fig.~\ref{fig:Co2MnGa}(d)], the contribution of the Weyl points was confirmed by the transport measurements,
revealing the giant anomalous Hall~[bottom panel]
and Nernst effects~[top panel] from the Berry curvature \cite{Sakai2018,sumida2020spin,xu2020anomalous}.
The negative longitudinal magnetoresistance from the chiral anomaly \cite{Sakai2018}
and the spin Hall effect \cite{isshiki2022determination} were also reported.
The ARPES measurements also confirmed the the Weyl point and nodal line structure,
accompanied by the drumhead surface states corresponding to the bulk nodal lines \cite{belopolski2019discovery}.

\ce{Co2MnAl} is another material that is intensely studied with the interest for the Weyl fermions.
It becomes ferromagnetic under the Curie temperature $T_c \approx 700 {\rm K}$ \cite{umetsu2008magnetic}.
Its giant anomalous Hall effect,
present at room temperature, was experimentally reported before the finding of the Weyl points \cite{chen2004anomalous,vilanova2011exploring}.
The contributions of the Berry curvature~\cite{kubler2012berry} and Weyl points~\cite{kubler2016weyl} to the giant anomalous Hall effect were suggested theoretically by the later calculations.
It is also reported that the anomalous Hall angle can reach a giant value $\theta_{\rm AHE}=\sigma_{\rm AHE}/\sigma_{xx} \approx 20 \% $ 
even at room temperature~\cite{li2020giant}.
This material shows the topological phase transition between the Weyl point and nodal ring structures,
following the direction of the magnetization.
Such a transition was experimentally pointed out by the transport measurement with the single-crystal samples free from structural disorder \cite{li2020giant}.

\subsubsection{Stacked Kagome lattice shandite \CSS}
\label{sec:co3sn2s2}

\begin{figure}[t]
\centering
\caption{
Weyl fermions in ferromagnetic \ce{Co3Sn2S2}.
(a) Crystal structure.
(b)~Band structure obtained from the first-principles calculations.
(c)~side and (d)~top view of nodal structure in the Brillouin zone. 
Green rings represent the nodal rings in the absence of spin-orbit coupling,
and red and blue points represent the Weyl points in the presence of spin-orbit coupling.
(a)-(d) are reprinted from Ref.~\cite{Muechler2020}.
}	
\label{fig:Co3Sn2S2}
\end{figure}

\begin{figure}[htb]
\centering
\caption{
The directions of the hoppings in the effective tight-binding model~\cite{Ozawa2019}:
(a) the intralayer first-nearest neighbor hopping $t_1$ between Co sites~(top panel), and the intralayer second-nearest neighbor hopping $t_2$ between Co sites~(bottom panel).
(b) the interlayer hopping $t_z$ between Co sites and hopping $t_{dp}$ between Co and Sn1.
(c)~The Weyl points configurations and (b)~band structure.
(e)~The energy dependence of the density of states~(left panel) and the anomalous Hall conductivity~(right panel).
(a)-(d) are reprinted from Ref.~\cite{Ozawa2024}, and (e) is reprinted from Ref.~\cite{ozawa2024effective}.
}	
\label{fig:ozanomu}
\end{figure}

The ferromagnetic kagome-lattice material \CSS is reported as a candidate for magnetic Weyl semimetal~\cite{Liu2018, huang2025co3sn2s2}.
It is of the structure same as shandite (\ce{Ni3Pb2S2})
which belongs
to the rhombohedral space group $R\bar{3}m$ [Fig.~\ref{fig:Co3Sn2S2}(a)].
In the unit cell,
there are three sublattices of Co atoms forming layers of kagome lattice, two sublattices of Sn atoms, and two sublattices of S atoms.
The ferromagnetic ordering is formed by the $3d$-electrons of Co,
which arises below the Curie temperature $T_{C}$ $\approx$ 173K.
It shows an extremely strong easy-axis magnetic anisotropy along the $c$-axis,
with the anisotropy field around $20 {\rm T}$ \cite{schnelle2013ferromagnetic,kassem2016quasi}.
Prior to the current interest in its Weyl-semimetal characters, 
\CSS was studied from the viewpoint of 
half-metallic electronic structure~\cite{weihrich2004magnetischer,weihrich2005half,weihrich2006half,weihrich2007structure} and the 
itinerant magnetism~\cite{kassem2015single,kassem2016quasi,kassem2017low}.
Further discussion on the relation between the itinerant magnetism and Weyl points shall be given in Sec.~\ref{sec:low-spin}.

In 2018, it was experimentally reported that this material shows an outstandingly large anomalous Hall effect \cite{Liu2018}.
The first-principles calculations predicted that the three pairs of Weyl points are located about $60{\rm meV}$ above the Fermi level \cite{Liu2018,Xu2018,Muechler2020} [Fig.~\ref{fig:Co3Sn2S2}(b)].
The Weyl point structure was measured by the ARPES \cite{Wang2018}.
The Berry curvature from these Weyl points gives
a significant contribution to the anomalous Hall effect.
Moreover,
it shows the anomalous Hall angle much larger than those in other Weyl magnets,
$\theta_{\rm{AHE}}={\it{\sigma}}_{\rm{AHE}}/{\it{\sigma}}_{xx} \approx \rm{20} \%$.
This is because of the semimetallic property with a small longitudinal conductivity $\sigma_{xx}$ due to the small DOS around the Weyl points.
Thus, \CSS serves as an ideal magnetic Weyl semimetal, which is not much contaminated by metallic conduction.
As an ideal Weyl semimetal,
the presence of the Fermi arc surface states was also confirmed by the ARPES \cite{Liu2019}.

Let us review the electronic structure of \CSS in more detail.
Near the L-point on the high symmetry line,
the valence and conduction bands are inverted.
In the absence of spin-orbit coupling, the gapless points form closed loops (nodal rings) of one-dimensional gapless nodes in Brillouin zone, as shown by the green lines in Fig.~\ref{fig:Co3Sn2S2} (c) and (d).
Once the spin-orbit coupling is included, each nodal ring gets gapped out except at two points, which results in the Weyl point structure,
as shown by the red and blue points in Fig.~\ref{fig:Co3Sn2S2} (c) and (d).

The low-energy band structure including the Weyl points was successfully reproduced by the tight-binding model composed of only the Co-$d$ electrons on the kagome lattice and Sn-$p$ electrons between the kagome layers~\cite{Ozawa2019}.
The total Hamiltonian of this model is composed of three terms,
$H_{0}=H_{\text{dp}}+H_{\rm{so}}+H_{\rm exc}$.
Here, 
$H_{\text{dp}}$ is the spin-independent hopping term,
$H_{\rm so}$ is the spin-orbit coupling term,
and 
$H_{\rm exc}$ is the exchange coupling term.
As shown in Fig.~\ref{fig:ozanomu}, $H_{\rm dp}$ incorporates 
(a)~the intralayer nearest neighbor hopping $t_1$, ~second nearest neighbor hopping $t_2$, 
(b) interlayer hopping between the adjacent kagome layers $t_z$, 
and d-p hybridization between Co and Sn $t^{\rm dp}$.
$\epsilon_{\rm p}$ is the on-site potential of Sn-$p$ electron.
(See Appendix \ref{sec:app:lattice:CSS} for the details of the construction of this model.)
Figure.~\ref{fig:ozanomu}(c) shows the Weyl point configurations computed with this model, similarly reproducing those by the first principles calculations~\cite{Liu2018}.
The Weyl point structure around $E_{\rm F}$ is formed only by the spin-up state [Fig.~\ref{fig:ozanomu}(d)].
As shown in Fig.~\ref{fig:ozanomu}(e), the up-spin DOS shows the minimum around $E_{\rm F}$,
whereas the down-spin DOS is almost negligible.
Thus, the system can be seen as a half metal or a spin gapless semiconductor.
Around $E_{\rm F}$, as shown in Fig.~\ref{fig:ozanomu}(f),
the calculated $\sigma_{xy}$ is maximized around the Weyl points due to the Berry curvature contribution, while the DOS and thus $\sigma_{xx}$ minimizes.
Its value $\sigma_{xy}\approx 1000~{\rm S/cm}$ around $E_{\rm F}$ is very close to that observed in the experiment.

In addition to the anomalous Hall effect, 
recent studies have found various functional properties of \CSS originating from its band topology.
One of them is the large anomalous Nernst coefficient,
which is due to the Berry curvature from the Weyl points around the Fermi level \cite{guin2019,Yang2020}.
Since the sign of the Nernst coefficient is sensitive to the Fermi level,
one can control its sign by the chemical substitution \cite{yanagi2021prb}.
Based on this idea,
a thermoelectric device using a thermopile structure of \CSS was tested \cite{noguchi2024bipolarity}.
This device is composed of multiple samples of non-doped and In-doped \CSS,
which show positive and negative Nernst coefficients, respectively.
By connecting them alternately in a single plane,
they contribute to the Nernst voltage additively in response to the temperature gradient,
which leads to a large thermoelectric conversion efficiency in a single device. 
The spin Hall effect arising from the spin Berry curvature has also been studied theoretically and experimentally in \CSS \cite{seki2023enhancement,ozawa2024effective},
which shall be reviewed in detail in Sec.~\ref{sec:spin-transport}.

Due to its large easy-axis magnetic anisotropy,
the magnetic domain structure is also an important topic in \CSS.
The domain structure in \CSS was
observed by the Lorentz transmission electron microscopy (TEM)~\cite{Sugawara2019} and the magneto-optical Kerr effect (MOKE)~\cite{Lee2022}. 
The magnetotransport measurement during the magnetization switching process also gives an implication on the nucleation of magnetic domains~\cite{Shiogai2022}.
For manipulation of the domain structure,
it was reported that a circularly polarized light switches the magnetic domains in \CSS films~\cite{yoshikawa2022non,yoshikawa2025all}.
Besides, the current-induced reduction of the coercivity was also experimentally reported, 
suggesting the current-induced torque exerting on the domain walls~\cite{wang2023magnetism}  
(see Sec.~\ref{sec:texture} for further discussions).

The physical properties reviewed above, including the anomalous Hall effect, have been seen
in both bulk and thin film samples.
In particular,
thin films of \CSS have been intensely studied to implement diverse functionalities in integrated devices.
Theoretically,
it was predicted that a single kagome layer of \CSS shows the quantized anomalous Hall conductivity \cite{Muechler2020}.
An experimental study showed that the anomalous Hall conductivity arises above the thickness of $\sim 10\ {\rm nm}$,
reaching the quantized value per a kagome layer \cite{Ikeda2021}.
Another experimental study found that the \CSS thin films show the metallic conduction independent of the film thickness, which may be attributed to the Fermi arc surface states \cite{ikeda2021two}.
In addition, in nano-flake samples with the thickness of $\approx 250 \ $nm, the high electron mobility and enhanced anomalous Hall angle reaching $32\%$ were experimentally reported~\cite{tanaka2020topological}.
A very recent experiment demonstrated the fabrication of an ultrathin ($\approx2.3 \ $nm) \CSS single-crystal sample~\cite{Zhang2025-xk}.
With Fe doping, the anomalous Hall angle was greatly improved to $48\%$.

Theoretical calculations based on DFT reported that the surface termination~(Sn, or S) of \CSS thin films affects the stable magnetic orderings and the transport properties~\cite{Nakazawa2024magnetic,Nakazawa2024topo}.
With these findings,
it will be important and challenging to understand the behaviors of electronic structure including Weyl points, magnetism, and transport properties,
in \CSS thin films.

\subsubsection{Topological insulator-based ferromagnet }
\label{sec:TI-based}

\begin{figure}
\centering
\caption{
(a) Schematic picture of the  topological insulator (TI) multilayers, and its phase diagram parametrized by the intra- and inter-layer tunneling amplitudes $\Delta_{S,D}$.
Reprinted from Ref.~\onlinecite{burkov2011topological}.
(b) Phase diagram of magnetically doped topological insulator calculated theoretically.
It shows the TI, magnetic TI (MTI), and Weyl semimetal (WSM) states depending on the temperature and the magnetic doping concentration.
Reprinted from Ref.~\onlinecite{kurebayashi2014weyl}.
}	
\label{fig:CrBiSbTe}
\end{figure}

Topological insulator materials have long been expected as candidates for realizing the topological semimetal states.
It was theoretically suggested at the earliest stage that the topological semimetal state arises at the phase transition between the topological and trivial insulator states,
because such a phase transition requires the closing of the topological band gap \cite{Murakami2007}.
In particular, realization of magnetic Weyl semimetal based on topological insulators has been an important challenge for a long time.
While introducing magnetism for topological insulators has been well studied and successful over a decade~\cite{tokura2019magnetic,Bernevig2022},
by using the proximity effect, magnetic doping, or designing van der Waals superstructures such as \ce{MnBi2Te4},
Weyl semimetal state therein was not confirmed until a very recent experiment \cite{belopolski2024took}.
In the following,
we review its theoretical proposals and the recent status of experimental studies.

The quest for topological insulator-based magnetic Weyl semimetal is traced back to one of the earliest theoretical proposal by Burkov and Balents \cite{Burkov2011}.
They considered a multilayer superlattice of magnetic topological insulator and normal insulator thin films.
Due to the breaking of time-reversal symmetry by magnetism,
the 2D Dirac fermions at each interface acquire a mass gap,
showing the quantized anomalous Hall effect.
Depending on the intra- and inter-layer tunnelings among those interfacial Dirac fermions,
the band structure becomes three-dimensional,
exhibiting a pair of Weyl points with broken time-reversal symmetry [see Fig.~\ref{fig:CrBiSbTe}(a)].
There was also a theoretical proposal of Weyl semimetal state considering the magnetic doping in topological insulator \cite{kurebayashi2014weyl}.
In this work,
the effect of magnetism on the band structure was studied by the mean-field theory.
The dependence on the concentration of magnetic dopants was obtained as a phase diagram,
which exhibited a magnetic Weyl semimetal state in a certain range of doping concentration around $10 \%$, 
as shown in Fig.~\ref{fig:CrBiSbTe}(b).
The mechanism for this magnetism shall be discussed in detail in Sec.~\ref{sec:high-spin}.

More than 10 years after the theoretical proposals explained above,
a recent experimental study reported the emergence of the Weyl semimetallic state in the magnetically doped topological insulator material $\mathrm{(Cr,Bi,Sb)_2 Te_3}$ \cite{belopolski2024took}.
The existence of Weyl fermions was validated by the measurement of the negative longitudinal magnetoresistivity indicating the effect of chiral anomaly.
Notably, the bulk anomalous Hall angle $\theta_{yx} = \sigma_{yx}/\sigma_{xx}$ of this material reached greater than 50\%.
With the first-principles calculations,
the origin of such a large $\theta_{yx}$ is supposed to be the large $\sigma_{yx}$ from the large separation of Weyl points,
and the small $\sigma_{xx}$ from the Weyl points located very close to $E_{\rm F}$.
This experiment also found the phase transitions among the different phases,
including the Chern insulator, Weyl semimetal, and magnetic semiconductor phases,
which depend on the doping concentrations of Cr, Sb, and In atoms.
Since these dopants affect magnetism, carrier concentration, and spin-orbit coupling in the system,
we may understand these phase transitions in terms of the rich parameter dependences considered in the earlier theoretical studies shown in the previous paragraphs.



\subsubsection{Weyl fermions in forced ferromagnetic states}

The materials reviewed above exhibit Weyl fermions in their ferromagnetic ground states,
without any external magnetic field.
There are some other materials that show Weyl fermions in the forced ferromagnetic states,
which are induced by applying an external magnetic field.
We here review some notable features in such materials.

GdPtBi is one of such materials studied from the early stages in the quest for magnetic Weyl semimetals.
It is in the half-Heusler structure,
where the $4f$-electrons of Gd show magnetism.
The first-principles calculations predicted that its band structure in the antiferromagntic ground state is similar to the topological insulator \ce{HgTe} \cite{li2015electronic}.
Once the spins align ferromagnetically by the external magnetic field,
it is predictd to exhibit the Weyl points in the band structure.
By the transport measurements under a magnteic field,
the negative longitudinal magnetoresistivity \cite{hirschberger2016,liang2018experimental} and the planar Hall effect \cite{kumar2018planar} were observed,
which imply the contributions of Weyl fermions in the forced ferromagnetic state.
The anomalous Hall effect was also observed,
which persists even in the antiferromagntic ground state where the Weyl points are absent \cite{suzuki2016large}.
In this material, pressure-induced tuning of the Weyl point positions was suggested through transport measurements of the anomalous Hall effect and the anomaly-induced magnetoresistivity \cite{sun2021pressure}.
Studies on other rare-earth-based half-Heusler compounds \cite{mun2022r},
such as TbPtBi, HoPtBi \cite{pavlosiuk2020anomalous}, and DyPdBi \cite{bhattacharya2023first},
are also expecting the presence of Weyl fermions in their forced ferromagnetic states.


\ce{EuCd2As2} and \ce{EuCd2Sb2} are also notable compounds that show Weyl fermions in the forced ferromagnetic states.
The Eu atoms form layers of hexagonal lattices,
whose $4f$-electrons are responsible for magnetic orderings.
Below the Neel temperatures $T_N = 9{\rm K}$ for \ce{EuCd2As2} and $7.4{\rm K}$ for \ce{EuCd2Sb2},
they show the antiferromagnetic ordering of alternating in-plane spin polarizations.
In this antiferromagnetic ground state,
these materials show the massive Dirac fermions around $E_{\rm F}$ \cite{wang2016anisotropic,rahn2018coupling,soh2018magnetic,li2021engineering}.
Once a magnetic field of a few tesla is applied to the out-of-plane $(z)$ direction,
the spins become ferromagnetically aligned,
and the electronic bands show the Weyl points along the $k_z$-axis in the vicinity of $E_{\rm F}$.
The first-principles calculations found that \ce{EuCd2As2} shows just a single pair of Weyl nodes \cite{wang2019single},
whereas \ce{EuCd2Sb2} shows multiple pairs of type-II Weyl nodes \cite{su2020magnetic}.
The signature of the Weyl fermions was seen in the Shubnikov--de Haas oscillation behavior \cite{Soh2019,su2020magnetic},
the anomalous Hall effect \cite{Soh2019,roychowdhury2023anomalous,ohno2022maximizing},
and the Berry curvature effect on the magnetoresistivity \cite{nakamura2024berry}.
Due to the anisotropic Weyl point configuration reflecting the threefold rotational symmetry of crystal,
\ce{EuCd2Sb2} shows the in-plane anomalous Hall effect even though the magnetic field is applied in the same plane \cite{nakamura2024plane}.
Moreover, the ARPES measurements on \ce{EuCd2As2} found the Weyl fermions even in the paramagnetic phase,
which are suggested to be the consequence of ferromagnetic spin fluctuations \cite{Ma2019}.





For the above-mentioned materials,
stabilizing the ferromagnetic ordering without magnetic field is also an important challenge,
which should be tackled by chemical substitutions, hydrostatic pressures, etc.

\subsection{Antiferromagnetic systems}

Besides the various species of ferromagnetic Weyl semimetal materials mentioned above,
there are also several species of materials that exhibit Weyl fermions under the antiferromagnetic ordering.
It is surprising that those antiferromagnetic Weyl semimetal materials show the topological properties almost similar to those in ferromagnetic Weyl semimetals,
even though the net magnetization therein is nearly zero.
Following the recent trend of antiferromagnetic spintronics,
which take advantage of the absence of stray magnetic fields and the ultrafast spin dynamics in antiferromagents,
application of the antiferromagnetic Weyl semimetal materials to spintronics is also intensely studied in recent experiments.
In the following,
we review the recent status of the studies of such antiferromagnetic Weyl semimetal materials.

\subsubsection{Chiral antiferromagnet: \ce{Mn3Sn}}

\begin{figure}
\centering
\caption{
Weyl fermions in noncollinear antiferromagnetic \ce{Mn3Sn}.
(a) Configurations of Mn and Sn atoms on the AB-stacked kagome lattice.
Red arrows represent the chiral antiferromagnetic ordering of spins.
(b) Momentum-space configuration of Weyl points (WP1/WP2),
(c) Berry curvatire distribution,
(d) the electronic band structure,
(e) the anomalous Hall conductivity,
and (f) the orbital magnetization,
obtained from the tight-binding model calculations.
The results in (b-f),
reprinted from Ref.~\onlinecite{ito2017anomalous},
are obtained under the inverse-triangular spin configuration shown in (a).
(g) The directions of magnetic octupole moment and the orbital magnetization $(\boldsymbol{M}_{\rm orb})$ under each spin configuration.
}	
\label{fig:Mn3Sn}
\end{figure}

Among an enormous number of antiferromagnetic materials,
the Mn-based antiferromagnet \ce{Mn3Sn} is the most well-studied material exhibiting the Weyl point structure and the related magnetoelectric properties.
In \ce{Mn3Sn} and its sibling material \ce{Mn3Ge}, the \ce{Mn} atoms form the
AB-stacking kagome lattice structure,
in the hexagonal space group $P 6_3/m mc$ [see Fig.~\ref{fig:Mn3Sn}(a)].
Prior to the interst in Weyl fermions,
these materials have long been studied with interest in their frustrated magnetism \cite{kren1975study,nagamiya1982triangular,tomiyoshi1982magnetic,brown1990determination}.
It was discovered that the magnetic moments of \ce{Mn} sites in \ce{Mn3Sn}
form the non-collinear antiferromagnetic ordering,
where the magnetic moments are aligned in-plane with the inverse triangular structure.
This antiferromagnetic ordering is present below the Neel temperature $T_N = 430 \ {\rm K}$ and is stable down to $T \approx 250 \ {\rm K}$.
While the magnetic moment on each \ce{Mn} site is around $3 \mu_B$ (with $\mu_B$ the Bohr magneton),
its sum over the three sublattices of the kagome lattice is almost calcelled,
showing a weak net magnetization of $\sim 0.002 \mu_B$ per site.



Traditionally, it was believed that the anomalous Hall conductivity is proportinal to the magnetization,
yielding no anomalous Hall effect in antiferromagnets due to the vanishing net magnetization.
A theory against this traditional intuition was proposed in 2014,
demonstrating a large anomalous Hall conductivity under a noncollinear antiferromagnetic ordering on the kagome lattice \cite{Chen2014}.
This model includes the broken mirror symmetry and spin-orbit coupling,
which was intended for the realistic non-collinear antiferromagnetic material \ce{Mn3Ir}.
The model calculation gave a nontrivial distribution of Berry curvature,
which was identified as the origin of the nonzero anomalous Hall conductivity by the intrinsic mechanism.
After this proposal,
the anomalous Hall effects in \ce{Mn3Ge} and \ce{Mn3Sn}, exhibiting the non-collinear antiferromagnetic ordering, were also demonstrated by first-principles calculations~\cite{kubler2014non}.
Such an anomalous Hall effect with (almost) vanishing net magnetization was first reported experimentally for \ce{Mn3Sn} in 2015 \cite{nakatsuji2015large}.

As the origin of the Berry curvature and the anomalous Hall effect in \ce{Mn3Sn},
the Weyl fermions in its electronic bands were predicted by the first-principles calculations \cite{Yang2017}.
The Weyl fermions were confirmed experimentally in 2017 \cite{Kuroda2017},
which were identified from the consistency between the ARPES measurements and the first-principles calculations,
and also from the magnetoresistivity measurements implying the effect of chiral anomaly.
The presence of Weyl fermions was later supported by the tight-binding model calculations on kagome lattice \cite{ito2017anomalous,Liu2017prl} [see Fig.~\ref{fig:Mn3Sn}(b-d)].
Since the large Fermi surfaces coexist with the Weyl points around the Fermi level,
\ce{Mn3Sn} is not strictly a Weyl semimetal but a metal,
while the Berry curvature from the Weyl points still give a large contribution to the anomalous Hall effect.

Instead of the magnetization in ferromagnets,
the Weyl fermions in \ce{Mn3Sn} are governed by the magnetic octupole moment in its spin texture.
The octupole moment is 
derived from the inverse-triangular spin texture by the cluster multipole theory,
which transforms in a same way as the magnetic dipole moment under the hexagonal space group $D_{6h}$ of \ce{Mn3Sn} crystal \cite{Suzuki2017}.
It means that the octupole moment plays a role same as the dipole moment, corresponding to the magnetization in ferromagnets.
It was confirmed from the tight-binding model calculation that the positions of the Weyl points depend on the direction of the octupole moment \cite{ito2017anomalous}.
As a result,
electron system shows the orbital magnetization (explained in Sec.~\ref{sec:orbital-magnetization}) along the direction of the octupole moment [see Fig.~\ref{fig:Mn3Sn}(g)],
which is directly related to the anomalous Hall effect as shown by Eq.~(\ref{eq:AHE-orbital}) [see Fig.~\ref{fig:Mn3Sn}(e,f)].
The direction of the octupole moment was successfully observed by the imaging of the MOKE \cite{Higo2018,uchimura2022observation}.

Motivated by the finding of Weyl fermions,
various phenomena related to the nontrivial band geometry have been studied in \ce{Mn3Sn}.
Besides the intrinsic anomalous Hall effect,
there have been several experimental findings of the anomalous transport phenomena arising from the band geometry.
One is the large anomalous Nernst effect due to the Berry curvature from the Weyl points,
which showed a good consistency with the first-principles calculations \cite{ikhlas2017large}.
This effect is now being used for imaging the distribution of the octupole domain structure,
by measureing a response to the application of a local thermal gradient \cite{isshiki2024observation,reichlova2019imaging}.
Another case is the interconversion between charge current and spin, measured with the heterostructure of \ce{Mn3Sn} and ferromagnetic \ce{NiFe} \cite{kimata2019magnetic}.
In contrast to the well-known spin Hall effect in nonmagnetic metals such as \ce{Pt},
the spin polarization measured here changes sign under the reversal of the noncollinear magnetic ordering,
which is named as the magnetic spin Hall effect.
While the general description of the magnetic spin Hall effect is still controversial,
it is implied that the magnetic spin Hall effect measured in \ce{Mn3Sn} is related to the intrinsic effect corresponding to the band geometry \cite{kimata2019magnetic}.

\ce{Mn3Sn}, supporting a large Berry curvature around the Weyl points,
is expected to exhibit a significant effect also from quantum metric,
which is the rank-2 tensor characterizing the momentum-space geometry \cite{resta2011insulating}.
This is because the magnitude of quantum metric is necessarily larger than that of Berry curvature \cite{mera2021kahler,ozawa2021relations}.
One of the theoretically proposed effects of this quantum metric structure is the nonlinear anomalous Hall effect,
of the second order of injected current \cite{gao2014field}.
With the heterostructure of \ce{Mn3Sn} and \ce{Pt} films,
such a second-order Hall effect was experimentally reported,
whose origin was identified to be the quantum metric structure by the symmetry-based analysis and the tight-binding model calculations
 \cite{han2024room}.








\subsection{Noncentrosymmetric magnetic Weyl semimetals}

The ferromagnetic and antiferromagnetic Weyl semimetal materials reviewed above have inversion centers in their crystal structures,
keeping the spatial inversion symmetry unbroken.
On the other hand,
recent theoretical and experimental studies have found several magnetic materials that show Weyl electrons without inversion symmetry in their crystal structures.
Such noncentrosymmetric magnetic Weyl semimetals are anticipated to provide the functionalities even richer than the centrosymmetric ones,
such as the induction of pure spin current by a magnetic field \cite{wang2019generation},
the spin-orbit torque in the bulk \cite{kurebayashi2021jpsj, meguro2025topological},
etc.
In this subsection, we review the current status of the studies on such noncentrosymmetric magnetic Weyl semimetals.

\subsubsection{Rare-earth based compounds}

\begin{figure}[htb]
\centering
\caption{
Weyl fermions in the compounds $R {\rm Al} X$, with $R$ the rare-earth element and $X = {\rm Si}, {\rm Ge}$.
(a) Schematic of the crystal structure. $RE$ stands for the rare-earth element, and Ge sites can be substituted with Si.
(b) Momentum-space distribution of the Weyl points in CeAlGe predicted from the first-principles calculation.
(c) Schematic of the mechanism for generating the Weyl fermions with both time-reversal $(\mathcal{T})$ and inversion $(\mathcal{I})$ symmetries broken.
Reprinted from Ref.~\onlinecite{chang2018magnetic}.
}	
\label{fig:NdAlSi}
\end{figure}

The family of the compounds $R {\rm Al} X$, with $R=$ rare-earth element and $X = {\rm Si, Ge}$,
have been studied as the major candidate materials for Weyl semimetals that show both time-reversal and inversion symmetries broken.
These materials show the crystal structure of the space group $I 4_1 m d$ [see Fig.~\ref{fig:NdAlSi}(a)],
which does not have an inversion center.
It was reported in 2017 that the nonmagnetic phase of \ce{LaAlGe} shows the Weyl fermions without inversion symmetry,
by the ARPES measurements \cite{xu2017discovery}.
This finding implies that,
once the $f$-electrons in $R$ form any magnetic ordering,
it may shift the energies of the Weyl points as the effective Zeeman coupling while keeping the Weyl node structure [see Fig.~\ref{fig:NdAlSi}(c)].
Motivated by this finding,
a prediction of the noncentrosymmetric magnetic Weyl state, with $R = {\rm Ce, Pr}$, was given by the first-principles calculations in 2018 \cite{chang2018magnetic} [see Fig.~\ref{fig:NdAlSi}(b)].
This study predicted the coexistence of the type-I and type-II Weyl fermions around $E_{\rm F}$.
After the theoretical prediction of Weyl fermions in these compounds,
series of experimental studies are currently in progress to reveal their rich characteristics,
in both magnetic and transport properties, as summarized in the following paragraphs.

In \ce{NdAlSi},
a rich magnetic phase structure was found from the magnetoresistivity measurements, including the paramagnetic, antiferromagnetic, and ferrimagnetic phases \cite{wang2022ndalsi}.
In this material,
the neutron diffraction measurement revealed the helimagnetic ordering below $7 \;{\rm K}$ \cite{Gaudet2021},
which is theoretically expected as a consequence of the nesting of Fermi pockets of multiple Weyl nodes \cite{nikolic2021dynamics}.
The presence of the multiple Fermi pockets was suggested also by the measurements of Shubnikov-de Haas oscillations \cite{Gaudet2021,wang2022ndalsi}.
Furthermore, the ARPES measurements directly observed bulk Weyl dispersions and surface Fermi arcs,
especially the formation of new Weyl points in the helimagnetic phase \cite{li2023emergence}.
Its paramagnetic phase also shows the intriguing transport properties arising from the spin fluctuations with nested Fermi surfaces,
such as the enhancement of the Nernst effect and the suppression of electron conduction \cite{yamada2024nernst,zhang2024abnormally}.
Nevertheless,
the anomalous Hall effect does not clearly occur in this compound.
A recent experiment with this material
observed an emergence of voltage, known as the spinmotive force~\cite{Barnes2007prl},
arising from the current-induced sliding dynamics of magnetic domain walls~\cite{yamada2026emergent}.
It remains an open question whether or not this spinmotive force is attributed to the nature of Weyl fermions,
like the scenario proposed theoretically so far (see Sec.~\ref{sec:spin-motive-force}).

NdAlGe shares a similar magnetic structure with NdAlSi,
displaying a rich phase diagram including the ferrimagnetic and spin glass phases \cite{wang2020correlation,zhao2022field}.
A stripe domain structure of the ferrimagnetic ordering was observed by the neutron scattering experiment  below $7\;{\rm K}$ \cite{yang2023stripe}.
Unlike \ce{NdAlSi},
it is reported that \ce{NdAlGe} sharply exhibits the anomalous Hall effect,
which is suggested to be arising from both the intrinsic contribution from the Berry curvature and the extrinsic contribution from the magnetic fluctuations  \cite{yang2023stripe,dhital2023multi,kikugawa2024anomalous}.

A large anomalous Hall effect was measured also in \ce{PrAlGe} \cite{meng2019large,yang2020transition,destraz2020magnetism}.
This material becomes ferromagnetic below the Curie temperature $T_c = 16 \;{\rm K}$,
where the ARPES measurements directly observed bulk Weyl cones and Fermi arcs \cite{sanchez2020observation}.
Importantly, the Weyl points in PrAlGe are located close to $E_{\rm F}$, suggesting potential for finding the transport properties characteristic to Weyl fermions.
In addition to the anomalous Hall effect, 
the second harmonic generation by the linearly polarized laser was also reported, which is the direct consequence of the breaking of time-reversal and inversion symmetries~\cite{shoriki2024large}.

\ce{CeAlSi} displays noncollinear in-plane ferromagnetic order confirmed by the inelastic neutron diffraction,
under which the electrons show a large anomalous Hall effect \cite{yang2021noncollinear}.
The ARPES measurements provided evidence for the Fermi arcs and Weyl cones even in the paramagnetic phase \cite{sakhya2023observation}.
Furthermore, tuning of the positions of Weyl points was demonstrated by applying pressure, which can lead to topological phase transitions \cite{cheng2024tunable}.
The presence of chiral domain walls was also investigated in \ce{CeAlSi} \cite{Sun2021},
which is suggested to suppress the anomalous Hall effect due to the scattering of electrons \cite{Piva2023}.

Other species of materials in this family,
such as \ce{SmAlSi} \cite{yao2023large} and \ce{CeAlGe} \cite{Puphal2020,drucker2023topology,piva2023topological} are also intensely studied,
while clear experimental signatures of Weyl fermions have not been conclusively established in these materials.

\subsubsection{Inverse Heusler alloy: \TMA}

\begin{figure}[htb]
\centering
\caption{
Weyl fermions in \ce{Ti2MnAl}.
(a) Inverse Heusler crystal structure, and the compensated ferrimagnetic ordering of spins on Ti and Mn sites.
(b) Configurations of Weyl points obtained from the first-principles calculations.
(c) Band structure and (d) anomalous Hall conductivity obtained from the tight-binding model calculations.
(a) and (b) are cited from Ref.~\cite{shi2018prediction}.
(c) and (d) are cited from Ref.~\cite{meguro2024effective}.
}	
\label{fig:Ti2MnAl}
\end{figure}

The inverse Heusler compounds of the chemical formula $X_2 YZ$ with the space group $F \bar{4} 3m$ have been intensely studied as candidates for spin gapless semiconductor,
in which one of the two spin channels becomes a gapless semiconductor while the other becomes fully gapped \cite{skaftouros2013search}.
Their magnetism has also been studied.
In particular, for the \ce{Ti}-based alloys ${\rm Ti}_2 YZ$,
it was shown from the first-principles calclations that their magnetism follows the generalized Slater--Pauling's rule, $M_t = Z_t - 18$,
where $M_t$ is the number of spin magnetic moments and $Z_t$ is the number of valence electrons in each unit cell \cite{skaftouros2013generalized,fang2014magnetic}.
From this relation,
it was predicted that the magnetic moment for the series ${\rm Ti}_2 {\rm Mn}Z \; (Z={\rm Al, Ga, In})$ becomes fully compensated $(M_t = 0)$.
After these theoretical predictions,
an experimental study on \ce{Ti2MnAl} confirmed that it indeed becomes a spin gapless semiconductor with the compensated ferrimagnetic properties \cite{feng2015z},
which are expected to be useful for spintronics devices with low operation current.

Further first-principles calculations discovered that the compensated ferrimagnetic state of \TMA becomes a Weyl semimetal \cite{shi2018prediction}.
In \TMA, 
magnetic moments at two Ti sites are anti-parallel to that at the Mn site, showing a collinear ferrimagnetic order with zero net magnetization [see Fig.~\ref{fig:Ti2MnAl}(a)]. 
The transition temperature reaches $\sim 650 {\rm K}$~\cite{feng2015z}, which is comparable to those of other Heusler Weyl systems, such as \ce{Co2MnGa}.
The Weyl points in this material are distributed close to $E_{\rm F}$,
with both the time reversal and spatial inversion symmetries broken~[see Fig.~\ref{fig:Ti2MnAl}(b)].
These Weyl points contribute to the large anomalous Hall effect, despite its vanishing total spin magnetization.
With the negligibly small DOS around the Weyl points,
its longidutinal conductivity becomes small,
which is expected to give a large anomalous Hall angle compared to other MWSMs.
Corresponding to the anomalous Hall effect,
it is predicted to show a finite orbital magnetization parallel to the direction of the spins,
which was calculated from a tight-binding model \cite{meguro2024effective} [see Fig.~\ref{fig:Ti2MnAl}(c)(d)].
This tight-binding model is composed of the electron hopping, the spin-orbit coupling, and the coupling to the ferrimagnetically ordered spins on each lattice site.
The calculation result implies that,
once a magnetic field is applied externally,
it couples to the orbital magnetization and enables the manipulation of the ferrimagnetic order in \TMA,
even though its net spin magnetization is zero.

Moreover, from the band topology including the Weyl points,
it is proposed that \TMA can host a giant current-induced torque on its ferrimagnetic ordering \cite{meguro2025topological}.
This effect will be useful for reducing power consumption in future spintronics devices,
which shall be revisited in detail in Sec.~\ref{sec:texture}.


\section{Magnetism and magnetic properties of Weyl semimetals}
\label{sec:magnetism}


So far we have seen the electronic properties in a wide variety of magnetic Weyl semimetal materials.
In these materials,
spontaneous magnetic ordering breaks time-reversal symmetry and lifts the spin degeneracy of electronic band structure,
giving rise to the Weyl point structure.
Therefore, formation of magnetic orderings is essential in realizing various electromagnetic responses, such as the anomalous Hall effect, emerging from the Weyl points.

It is thus important to understand
how and what kind of magnetic orderings appear in magnetic Weyl semimetal materials. 
The detailed mechanism of the magnetic orderings strongly depends on the elements, structures of crystals, orbitals, etc., in each compound~\cite{yosida1996theory}.
As we have mentioned in the previous section,
there have been many theoretical attempts to determine the ground-state magnetic orderings in magnetic Weyl semimetal materials, with the first-principles calculations and the model calculations.

For an overall understanding on the magnetism in Weyl semimetal materials,
in this section, we review fundamental mechanisms of magnetism,
classified into the itinerant magnetism in low-spin state and the localized magnetism in high-spin states.
We also explain the mechanisms responsible for non-uniform magnetic textures in magnetic Weyl semimetal materials,
i.e.,
the magnetic anisotropy and the Rudermann-Kittel-Kasuya-Yosida~(RKKY) interaction.
After those discussions on the ground-state magnetic textures,
we also review the magnetic excitations beyond the ground states, known as spin waves or magnons,
which also show characteristic features in magnetic Weyl semimetals.


\subsection{Mechanisms of magnetism in magnetic Weyl semimetals}

Interactions between magnetic moments in magnetic metals can be understood in terms of underlying electronic interactions, such as Coulomb interactions and spin-orbit coupling.
We review the two general mechanisms for magnetism,
(i) the Stoner mechanism for the itinerant magnetism in the low-spin state,
and (ii) the carrier-mediated mechanism for the localized magnetism in the high-spin state.
We discuss how these mechanisms can lead to an ideal Weyl semimetal state, characterized by the low carrier density near $E_{\rm F}$.

\subsubsection{Itinerant magnetism in low-spin state 
}
\label{sec:low-spin}

Magnetic Weyl semimetals reported so far are primarily based on $3d$ transition metals, where the Coulomb interaction plays a crucial role in their physical properties.
The typical examples are the stacked kagome-lattice materials such as \CSS, \ce{Mn3Sn}, etc.,
and Heusler-type alloys such as \ce{Co2MnGa}, \ce{Ti2MnAl}, etc.
The electronic structures including the Weyl points in these materials are generally complicated, 
because multiple orbitals and sublattices participate in the electronic structures.
Depending on the competition between the Coulomb interaction and crystal field splitting, 
$3d$ transition metal magnets are classified into two species: low-spin and high-spin states~\cite{yosida1996theory}. 
Here,
we explain the mechanism of magnetism in the low-spin state,
in connection with the Weyl point structure with small Fermi surfaces.
Magnetic Weyl semimetals of high-spin states
shall be explained later.

If the crystal field splitting energies are large enough compared to the Hund coupling energies,
the low-spin states are favored,
where the low-energy orbitals of both spin-up and down states are occupied by the valence electrons.
In this case, the itinerant electrons around $E_{\rm F}$ only contribute to the net magnetization.
The Stoner's theory demands that ferromagnetism of itinerant electrons is favored under the condition \cite{stoner1938collective},
\begin{equation}
    U D(E_{\rm F})> 1.
    \label{eq:stoner}
\end{equation}
Here, $D(E_{\rm F})$ is the density of states at $E_{\rm F}$ in the nonmagnetic state,
and $U$ is the energy of the on-site Coulomb (Hubbard) interaction.
Phenomenologically,
this condition is understood as the tendency to reduce the interaction energy at $E_{\rm F}$ by forming a spin polarization.
It is also derived from the Ginzburg-Landau theory with the on-site Hubbard model~\cite{altland2010condensed}.

Since the $d$-electrons show relatively localized wave functions compared to those of $s$- or $p$-electrons,
the bandwidth tend to be small and it results in a large density of states.
The Stoner ferromagnetism reduces the DOS relative to the nonmagnetic state and can consequently shift $E_{\rm F}$ toward the energy of the Weyl points, where the DOS is minimal, in certain magnetic Weyl semimetals.

For example,
the ferromagnetism in \CSS is attributed to the Stoner's criterion \cite{Ozawa2019, Ozawa2022}.
While the pristine \CSS is ferromagnetic,
it was experimentally seen that this ferromagnetic ordering gets suppressed by the chemical dopings, substituting Co with Fe or Ni atoms, and Sn with In atoms \cite{weihrich2006half,kubodera2006ni,kassem2015single,kassem2016quasi,kassem2017low}.
For all cases, the out-of-plane components are suppressed once the carriers are doped.
This tendency is understood by the DOS at the non-magnetic phase appearing the Stoner's criterion.
For the pristine \CSS,
the DOS in the nonmagnetic state shows a peak near $E_{\rm F}$,
and hence the ferromagnetic state is preferred by the Stoner's criterion.
Once we dope \ce{In} with holes, or \ce{Ni} with electrons, $E_{\rm F}$ deviates from this peak of the DOS,
which reduces $D(E_{\rm F})$.
As a result, the Stoner's criterion Eq.~(\ref{eq:stoner}) is not satisfied,
which leads to the suppression of ferromagnetic ordering.

\begin{figure}
    \centering
    \caption{The density of states computed by an effective model of \CSS.
    (a)~Non-magnetic state and
    (b)~ferromagnetic state.
    Reprinted from Ref.~\cite{Ozawa2019}.
    }
    \label{fig:MT}
\end{figure}

\subsubsection{Localized magnetism in high-spin state}

\label{sec:high-spin}

In contrast to the low-spin states,
if the spin splitting due to Hund’s coupling exceeds the crystal-field splitting, electrons well below $E_{\rm F}$ become spin-polarized within each orbital.
Such a state is called the high-spin state,
where the spin polarization on each site can be seen as the localized magnetic moment. 
Here, the coupling between the localized magnetic moment ${\bm S}_i$ and the spin of conduction electron ${\bm s}_i$ 
at site $i$
is described as the exchange interaction $-J{\bm S}_i\cdot{\bm s}_i$.
Then,
the localized moments interact with each other
mediated by the itinerant electrons,
and become ferromagnetically ordered under the condition \cite{yu2010quantized},
\begin{align}
    J^2 \chi_{\rm s} \chi_{\rm e} (E_{\rm F}) > 1,
    \label{eq:pauli-magnetism}
\end{align}
with the spin susceptibility of itinerant electrons $\chi_{\rm e}$, localized spin susceptibility $\chi_{\rm s}$.
A finite DOS of the conduction electrons yields a finite value of $\chi_{\rm e}$,
known as the Pauli susceptibility,
which enhances the left-hand side of Eq.~(\ref{eq:pauli-magnetism}) and can lead to a ferromagnetic order.
Such a carrier-mediated
magnetism is seen
in diluted magnetic semiconductors, such as $({\rm Ga}_{1-x} {\rm Mn}_x) {\rm As}$ \cite{munekata1989diluted, ohno1991new,ohno1992magnetotransport,ohno1996}.
Several magnetic Weyl semimetal materials such as \ce{Fe3Sn2} and \ce{Mn3Sn} belong to the high-spin state with a large itinerant carrier density,
while the mechanisms for their magnetism are not specified and remain an open question.

On the other hand, even in the absence of conduction carriers,
the susceptibility $\chi_{\rm e}$ of itinerant electrons can become finite by another mechanism, known as the Van Vleck paramagnetism~\cite{vanvleck1932}, 
giving rise to the spontaneous magnetic ordering.
The Van Vleck paramagnetism arises from virtual transitions between different angular momentum states of electrons,
and does not require the existence of Fermi surfaces.
$\chi_{\rm e}$ can be enhanced if the electron orbitals are largely perturbed,
by crystal field, spin-orbit coupling, etc.
The ferromagnetism from the Van Vleck mechanism is expected on the surface of magnetically doped topological insulators,
such as Cr-doped \ce{Bi2Se3}~\cite{yu2010quantized}.
Even though the topological insulators have a low carrier density due to the bulk gap and the surface Dirac dispersion,
the spin susceptibility is enhanced by the strong spin-orbit coupling, which mixes the different angular momentum states.
This leads to the ferromagnetic orderings coupled to the surface Dirac fermions, seen via the quantized anomalous Hall effect in experiments~\cite{chen2010massive,chang2013experimental,zhang2013topology,chang2015high,Li2015-mj}.

As well as on the surface,
the Van Vleck paramagnetism is expected to be important in realizing the magnetic Weyl semimetal state in the bulk.
One example is that from magnetically doped topological insulator $({\rm Cr},{\rm Bi})_2 {\rm Te}_3$,
as previously introduced in Sec.~\ref{sec:TI-based}.
Prior to its experimental observation \cite{belopolski2024took},
a mean-field calculation for the doped magnetic moments predicted the formation of the Weyl semimetal state under the ferromagnetic ordering \cite{kurebayashi2014weyl}.
This ferromagnetic ordering is attributed to the Van Vleck paramagnetism inside the topological gap,
because the spin-orbit coupling for the band inversion largely contributes to the interband virtual transition.
Indeed,
the calculation found a positive correlation between the Curie temperature and the spin-orbit coupling strength.

\subsection{Magnetic anisotropy}

Magnetic anisotropy 
is a property that characterizes
how much the magnetic moments are energetically likely or unlikely to align to the crystalline axis or plane.
It is generally determined by the combination of the crystal field splitting and the spin-orbit coupling in the bulk \cite{brooks1940ferromagnetic}.
In micro-structured samples, the shape of the sample also contributes to the magnetic anisotropy,
due to the magnetostatic energy from the stray magnetic field \cite{holstein1940field}.
Magnetic anisotropy is essential in keeping the magnetic orderings stable,
which is important for manipulating the magnetic orderings in spintronics devices.
It also determines the size of magnetic textures such as domain walls,
by the competition with the spin-spin exchange coupling energy.

\begin{figure}[t]
\centering
\caption{
Magnetic anisotropy of the ferromagnetic \CSS calculated with a tight-binding model.
(a) Variation of the total energy  $\Delta E$ of the electron system depending on the magnetization direction (tilted from the $c$-axis by the angle $\theta_1$).
(b) Change in the nodal structure between the nodal ring and Weyl points depending on $\theta_1$.
Reprinted from Ref.~\cite{Ozawa2019}.
}	
\label{fig:ozawa-css-anisotropy}
\end{figure}

In topological materials including Weyl semimetals,
the band inversion for their topological band structure arises mostly from spin-orbit coupling,
and hence
the topological band structure including the gap or nodal structure
is related to the magnetic orderings.
Thus, the direction of magnetization tends to govern the total energy of the electron system,
which leads to the magnetic anisotropy energy.
As a typical example,
the tight-binding model calculation of a Chern insulator on a kagome lattice demonstrated a relation between the magnetic anisotropy and the topological band structure  \cite{watanabe2022magnetic}.

For some magnetic Weyl semimetal materials,
the idea on the relation between the nodal structure and magnetic anisotropy is suggested.
For example,
the ferromagnetic \CSS thin film is reported to show quite a strong easy-axis anisotropy,
reaching up to $20\;{\rm T}$,
perpendicular to the kagome plane \cite{schnelle2013ferromagnetic,kassem2016quasi}.
This anisotropy was reproduced by the tight-binding model calculation,
where the magnetic anisotropy was discussed in connection with the Weyl point structure near $E_{\rm F}$~\cite{Ozawa2019} (see Fig.~\ref{fig:ozawa-css-anisotropy}):
while the electron system shows the nodal rings under the in-plane magnetization direction $(\theta_1 = \pi/2)$,
these nodal rings are gapped out to show the Weyl points under the out-of-plane magnetization $(\theta_1 = 0,\pi)$.
Such a gap opening lowers the energy of electronic states below $E_{\rm F}$,
which may make the out-of-plane magnetization energetically more stable.

On the other hand,
another Weyl semimetal material  \ce{Ti2MnAl},
which exhibits the compensated ferrimagnetic ordering with magnetic moments aligned collinearly,
is expected to show vanishingly small magnetic anisotropy,
which was proposed by the tight-binding model calculation \cite{meguro2024effective}.
This behavior is also related to the Weyl point structure,
where the Weyl points arise irrelevant of spin-orbit coupling
and are almost independent of the directions of magnetic moments.

If the Weyl nodes coexist with other metallic Fermi surfaces,
the Weyl fermion contributions to the magnetic anisotropy becomes less dominant.
In such cases, the magnetic anisotropy is rather governed by the crystalline structure,
like in most of the magnetic materials.
For instance,
\ce{SrRuO3} \cite{kanbayasi1976magnetocrystalline} show almost cubic anisotropy,
and hence the magnetic anisotropy becomes sensitive to the sample shape and lattice strain.
The inverse-triangular spin structure in the antiferromagnetic \ce{Mn3Sn},
characterized by the magnetic octupole moment,
shows the sixfold in-plane anisotropy due to its kagome-lattice structure.
By breaking crystalline symmetry by the uniaxial strain,
it was observed that this sixfold anisotropy reduces down to the twofold anisotropy,
which enables the deterministic switching of the spin structure and the anomalous Hall effect \cite{ikhlas2022piezomagnetic,yoon2023handedness}.

\subsection{Spin-spin interaction and magnetic textures}
In metallic magnets,
the interaction between localized spins is mediated by the spins of itinerant electrons,
which is known as the RKKY interaction \cite{ruderman1954indirect,kasuya1956theory,yosida1957magnetic}.
In the absence of SOC for the itinerant electrons,
the RKKY interaction takes the isotropic Heisenberg-like form ($\propto \boldsymbol{S}_1 \cdot \boldsymbol{S}_2$ for two spins $\boldsymbol{S}_{1,2}$),
and it becomes either ferromagnetic or antiferromagnetic depending on the distance between the spins.
The structure of the RKKY interaction is influenced by the SOC,
which can give rise to anisotropic or asymmetric spin-spin interaction.

Since the SOC is essential in most of the Weyl semimetal materials,
theoretical studies have found that the Weyl semimetals can exhibit 
RKKY intearctions that are anisotropic in spin space  \cite{Chang2015rkky,hosseini2015ruderman,araki2016spin,duan2019signature,nikolic2021dynamics}.
For the Weyl fermions with the simple spin-momentum locking structure of $H_\eta(\boldsymbol{k}) = \eta \hbar v\_F \boldsymbol{k}\cdot\boldsymbol{\sigma}$,
the RKKY interaction contains not only the isotropic Heisenberg-like structure,
but also the Ising-like term of $\propto (\boldsymbol{S}_1 \cdot \boldsymbol{r}_{12})(\boldsymbol{S}_2 \cdot \boldsymbol{r}_{12})$,
where $\boldsymbol{r}_{12}$ is the relative position between two magnetic moments \cite{Chang2015rkky,hosseini2015ruderman,araki2016spin} [see Fig.~\ref{fig:araki-rkky}(a)].
This Ising-like interaction can give rise to not only the uniform ferromagnetic ordering,
but also some topological spin textures, such as vortex and hedgehog structures.
In addition, if the inversion symmetry is broken,
a term like the Dzyaloshinskii--Moriya (DM) interaction of $\propto \boldsymbol{D}_{12}\cdot (\boldsymbol{S}_1 \times \boldsymbol{S}_2)$ also appears in the RKKY interaction,
where the DM vector $\boldsymbol{D}_{12}$ is governed by the pattern of inversion symmetry breaking,
including the relative positions of the two spins \cite{Chang2015rkky,hosseini2015ruderman}.
Such a DM-like interaction tends to favor chiral spin textures,
including spirals and skyrmions.
The three types of RKKY interaction terms,
i.e., the Heisenberg, Ising, and DM-like terms,
show different oscillation and damping behaviors with respect to the distance $|\boldsymbol{r}_{12}|$,
which is proposed from a theoretical calculation with a low-energy effective model  \cite{Chang2015rkky} [see Fig.~\ref{fig:araki-rkky}(c)].

If $E_{\rm F}$ is close to the Weyl points,
the strengths of the RKKY interactions mentioned above become significantly enhanced
due to the infrared divergence of the interband processes around the Weyl points \cite{araki2016spin}.
The divergence of the interaction strengths is predicted also at the phase boundary between type-I and II Weyl semimetal phases \cite{duan2019signature}.
Moreover, if the system possesses multiple numbers of Weyl points in the Brillouin zone,
the nesting between different Weyl nodes gives rise to the RKKY interactions of a finite wave vector $\boldsymbol{q}$,
which corresponds to the displacement between the Weyl points in momentum space \cite{nikolic2021dynamics}.
It was suggested that the finite-$\boldsymbol{q}$ spin textures observed in noncentrosymmetric \ce{NdAlSi} \cite{Gaudet2021} and \ce{SmAlSi} \cite{yao2023large} are the consequences of such inter-node nesting processes.

\begin{figure}[t]
\centering
\caption{
Effect of the Weyl electrons on the Rudermann-Kittel-Kasuya-Yosida~(RKKY) interaction.
(a) Schematic picture of the Ising-type RKKY interaction.
Under the spin-momentum locking structure of Weyl electrons,
electron propagation with the momentum $\boldsymbol{k}$ parallel to the spin direction $\boldsymbol{\sigma}$ is favored,
while that with $\boldsymbol{k}\perp\boldsymbol{\sigma}$ is prohibited.
(b) The spin-wave dispersion calculated in a hypothetical ferromagnetic Weyl semimetal,
with the magnetization along the $z$-direction.
The spin wave becomes highly anisotropic due to the spin-momentum locking of Weyl electrons \cite{araki2016spin}.
(c) The RKKY interaction calculated with the low-energy effective model of Weyl semimetal,
classified into the Dzyaloshinskii--Moriya (DM), Ising, and Heisenberg-like terms \cite{Chang2015rkky}.
The three terms show different oscillation and damping behavior with respect to the normalized distance $\zeta$.
}
\label{fig:araki-rkky}
\end{figure}

\subsection{Spin waves in magnetic Weyl semimetals}

The 
fluctuations of spins in magnetic materials 
cause collective
spin-wave excitations,
which are regarded as \textit{magnon} quasiparticles \cite{bloch1930theorie}.
Since spin waves carry spin and heat currents in magnetic materials,
the role of spin waves has been intensely studied also in spintronics toward thermoelectric and information processing device applications \cite{chumak2015magnon}.

In metallic magnets,
the spin-wave dispersion is generally determined by the spin-spin correlation function of itinerant electrons.
Thus, several theoretical studies have shown that the band topology of electrons have certain effects on the spin-wave dispersion~\cite{Onoda2008,araki2016spin,nikolic2021universal}.
In particular,
under the spin-momentum locking structure,
the momentum-space Berry curvature contributes to the spin-wave dispersion,
just like the intrinsic anomalous Hall conductivity \cite{Onoda2008}.
This is because the spin-spin correlation function for the spin waves becomes proportional to the current-current correlation function for the conductivity,
due to the correspondence between the current operator $\boldsymbol{j} = -e\boldsymbol{v}$ and the spin operator $\boldsymbol{\sigma}$ by the spin-momentum locking.
This contribution was experimentally confirmed in the Weyl ferromagnet \ce{SrRuO3},
from the temperature dependence of the spin-wave gap measured by the inelastic neutron scattering \cite{Itoh2016,jenni2019interplay}.
Such a contribution in the spin-wave dispersion was suggested also in the half-metallic Weyl semimetal \CSS,
observed by the inelastic neutron scattering and the Brillouin light scattering
\cite{liu2021spin,neubauer2022spin,toyoda2022weyl}.

The anisotropic spin-spin interactions discussed in the previous subsection can also affect the spin-wave dispersions.
In particular, the Ising-type interaction is supposed to give a strong anisotropy in the spin-wave dispersion in the ferromagnetic state;
the longitudinal propagation along the magnetization direction gets strongly dispersed, while the transverse dispersion is rather weak \cite{araki2016spin} [see Fig.~\ref{fig:araki-rkky}(b)].
The DM-type interaction under the broken inversion symmetry
leads to the asymmetric dispersion of spin waves,
which may contribute to the nonreciprocity of spin-wave propagation.
It was theoretically proposed that the damping of the spin waves also gets influeunced by the Weyl electrons \cite{nikolic2021universal}.
Experimental observation of these features is left for future studies.


\section{Magnetic textures and dynamics} 
\label{sec:texture}
Due to the interplay of various magnetic interactions mentioned so far,
a rich variety of magnetic textures can arise in magnetic materials.
The most typical example is a magnetic domain wall that separates two different magnetic domains.
Magnetic spirals, with spatially periodic spin textures,
and magnetic skyrmions, exhibiting the swirling texture of spins,
are also typically seen.
These magnetic textures are intensely studied not only in the context of magnetism,
but also of spintronics,
to control the electron transport with these textures
and to utilize
these textures as carriers of information~\cite{Parkin2008,nagaosa2013topological,Maekawa2017}.
Interplay of electron transport with these magnetic textures has been theoretically and experimentally studied:
the topological Hall effect (THE) \cite{lee2009unusual,neubauer2009topological,kanazawa2011large} from skyrmion textures,
the current-induced spin transfer torque (STT) \cite{Slonczewski1989prb,Berger1996prb,Tserkovnyak2005rmp} 
acting on magnetic moments,
the spinmotive force (SMF; also known as the emergent electric field) \cite{Volovik1987jopc,Stern1992prl,Barnes2007prl} acting on electrons driven by magnetic texture dynamics,
etc.
These effects have been studied mainly in conventional metallic magnets,
where the momentum-space topology of electrons is not taken into account.
For their theoretical treatments,
the idea of emergent electromagnetic fields or spin electromagnetic fields,
which is introduced by the gauge transformation in spin SU(2) space,
has been frequently used \cite{Volovik1987jopc,nagaosa2012physscr,Tatara2008}.

In this section, we focus on the effects of magnetic textures in magnetic Weyl semimetals.
With the picture of chiral gauge field introduced in Sec.~\ref{sec:chiral-gauge-field},
the interplay between the Weyl fermions and magnetic textures has been theoretically proposed from the early-stage studies of magnetic Weyl semimetals.
As the presence of magnetic textures in magnetic Weyl semimetal materials has been eventually confirmed in recent experiments,
the theoretically proposed effects are now expected to be observed in such materials.
Here, we first review the experimental observations of magnetic textures in Weyl semimetal materials.
We then proceed to the magnetoelectric effects arising at magnetic textures in Weyl semimetals,
including both the theoretical studies and experimental observations.

\subsection{Observation of magnetic textures}

Magnetic domains and domain walls, which commonly arise in magnetic materials,
were observed in magnetic Weyl semimetals with various experimental techniques.
As long as the material shows a net magnetization,
one can observe the domain structure by detecting the spatial distribution of the stray magnetic field.
For this purpose,
magnetic force microscopy (MFM) is a powerful technique by picking up the field distribution mechanically.
For the Weyl magnets,
it was applied to the ferromagnetic \ce{Co3Sn2S2} \cite{Howlader2020} and \ce{EuB6} \cite{Li2022} to observe the domain structures.
Superconducting quantum interference device (SQUID), which detects the magnetic field by the Josephson effect in a superconducting ring,
is also frequently used.
In the \ce{CeAlSi},
not only the stable ferromagnetic domains but also the metastable nonmagnetic domains were successfully observed by SQUID \cite{Xu2021}.
The domain imaging by the magnetic field sensing was conducted also with the nitrogen-vacancy (NV) center in \ce{Mn3Sn} \cite{tsukamoto2025}.

For the imaging of magnetic domain structure,
another powerful tool is the microscopy using the magneto-optical Kerr effect (MOKE).
Since the MOKE is related to the anomalous Hall conductivity,
the magentic Weyl semimetals hosting a large Berry curvature are fundamentally suitable for the MOKE imaging~\cite{Kargarian2015-uq,Cheskis2020-oi}.
The domain structure in magentic Weyl semimetals was first observed in the antiferromagnetic \ce{Mn3Sn} \cite{Higo2018}.
Even though its net magnetization is almost vanishing,
its sixfold spin configuration gives rise to a large MOKE via the Berry curvature.
The MOKE was also used in the ferromagnetic \ce{Co3Sn2S2} \cite{Lee2022,yoshikawa2022commphys} and \ce{CeAlSi} \cite{Sun2021} to observe the domain structures.


Since the transport properties in Weyl semimetals, 
including anomalous Hall and Nernst effects,
are significantly affected by the magnetic orderings,
measurements of the transport signals are also used to understand the domain structures.
By measuring the spatial profile of the anomalous Nernst effect in response to the local temperature gradient applied by the atomic force microscopy,
the domain structure was successfully observed up to sub-hundred nanometers in the ferromagnetic \ce{Co2MnGa} \cite{Budai2023}.
The domain wall magnetoresistance (MR),
which shall be explained theoretically in Sec.~\ref{sec:transport},
also helps understanding of the domain structure.
By the measurement of the MR in the ferromagnetic \ce{Co3Sn2S2},
the nucleation process of magnetic domains during the magnetization switching process was inferred \cite{Shiogai2022,Fujiwara2024-my}.
The onset of magnetic domain walls in the MR signals was pointed out also in the ferromagnetic \ce{CeAlSi} \cite{Piva2023} and \ce{Co2MnAl} \cite{Wang2023jmmm}.


Besides the domain wall structure,
various types of magnetic textures were suggested experimentally in Weyl magnets.
From the measurement of the magnetic susceptibility in \ce{Co3Sn2S2},
the phase diagram with the magnetic field $(H)$ and the temperature $(T)$ was obtained,
which showed the skyrmion-like $A$-phase under a moderate magnetic field around 150 K \cite{Wu2020}.
The skyrmion-like multiple-$q$ state was also suggested in \ce{CeAlGe} by the measurements of neutron scattering and the THE \cite{Puphal2020}.
The single-$q$ helix was observed in the noncentrosymmetric \ce{NdAlSi} by the neutron scattering,
which was attributed to the nesting
of Weyl nodes in momentum space \cite{Gaudet2021}.

\subsection{Magnetic textures and chiral electromagnetic fields}

From here, we discuss how the magnetic textures observed above may influence the behavior of the Weyl fermions.
In Sec.~\ref{sec:chiral-gauge-field},
we have seen that perturbations inducing the chirality-dependent shift of Weyl points in momentum space can be regarded as the chiral gauge field.
Since the configurations of Weyl points in magnetic Weyl semimetals are governed by their magnetic orderings,
we may regard that the perturbations in the magnetic orderings effectively serve as the chiral gauge field for the Weyl fermions $\boldsymbol{A}_5(\boldsymbol{r},t)$.
With this correspondence,
spatially non-uniform magnetic textures  can yield the chiral magnetic field component $\boldsymbol{B}_5(\boldsymbol{r}) = \boldsymbol{\nabla} \times \boldsymbol{A}_5(\boldsymbol{r})$,
whereas magnetization dynamics leads to the chiral electric field component $\boldsymbol{E}_5(t) = -\partial_t \boldsymbol{A}_5(t)$.
With the chiral electromagnetic fields from magnetic textures and dynamics,
several phenomena are expected as consequences of the magnetoelectric effects explained in Sec.~\ref{sec:magnetoelectric-effects} \cite{Liu2013PRB,Araki2018prapplied,Araki2020adp}.

\begin{figure}
\centering
\includegraphics[width=0.9\hsize]{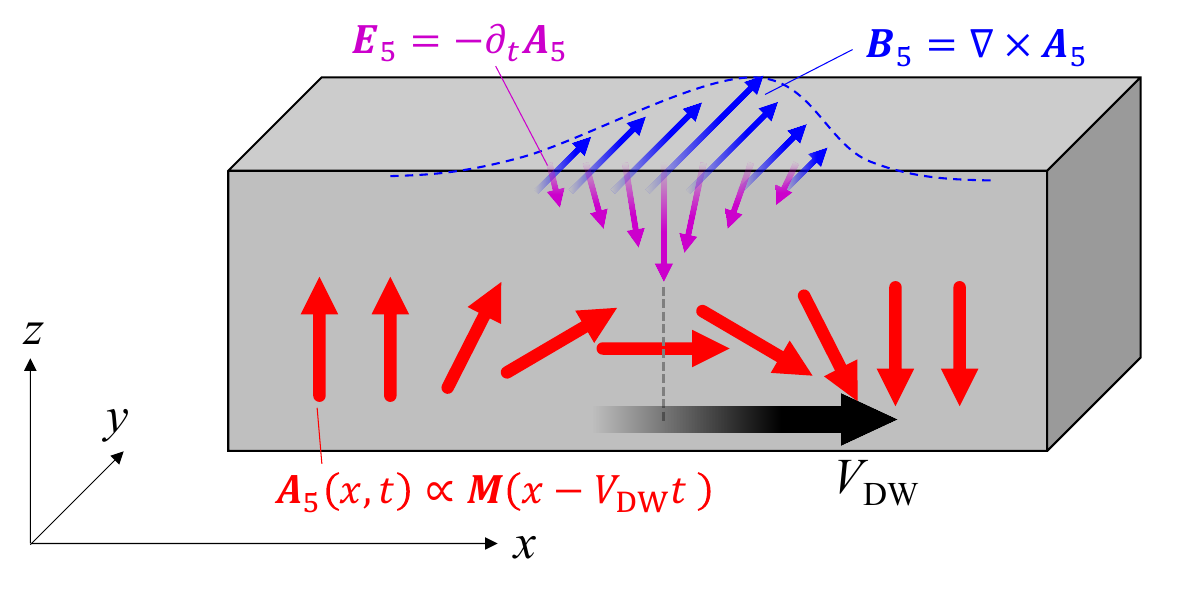}
\caption{
Schematic image of the chiral electromagnetic fields $\boldsymbol{B}_5$ and $\boldsymbol{E}_5$ emerging from the magnetic domain wall,
under the isotropic spin-momentum locking structure.
The N\'{e}el-type domain wall $\boldsymbol{M}(x)$ [Eq.~(\ref{eq:Neel-DW})] yields the chiral magnetic field $\boldsymbol{B}_5 = \boldsymbol{\nabla} \times\boldsymbol{A}_5$.
When the domain wall is sliding with the velocity $V_{\rm DW}$,
it gives the chiral electric field $\boldsymbol{E}_5 = -\partial_t \boldsymbol{A}_5$.
Both of them are localized around the center of the domain wall.
}	
\label{fig:domainwall}
\end{figure}

\begin{figure}
\centering
\caption{
Chiral gauge field structure investigated with the tight-binding model of ferromagnetic \CSS, adapted from \cite{Ozawa2024}.
(a)~Trajectory of Weyl points $(WP_1^\pm)$ in momentum space under the gradual change of magnetization direction $\boldsymbol{m}$ from the $c$-axis.
Red and blue points indicate the Weyl points with positive and negative chiralities, respectively. 
(c)~Spatial profile of the chiral magnetic field $\bm{B}_5(x)$ arising from the N\'{e}el type magnetic domain wall $\bm{m}(x)$.
}	
\label{fig:css_B5}
\end{figure}

For the concreteness of discussion,
let us first take the Weyl fermion with the fully isotropic spin-momentum-locking  structure,
$H_\eta({\bm k}) = \hbar v_{\rm F} {\bm k}\cdot{\bm \sigma}$.
Then, the exchange coupling $J \boldsymbol{M}\cdot\boldsymbol{\sigma}$ can be rewritten as the coupling to the chiral gauge field $\boldsymbol{A}_5 = (J/e v_{\rm F}) \boldsymbol{M}$,
as we have seen in Eq.~(\ref{eq:Hamiltonian-A5}).
As an example, we consider a N\'{e}el-type magnetic domain wall structure localized at $x = 0$,
\begin{align}
    \boldsymbol{M}(x) &= M_s \left( \mathrm{sech}\frac{x}{w}, 0, -\tanh\frac{x}{w} \right), \label{eq:Neel-DW}
\end{align}
as schematically shown in Fig.~\ref{fig:domainwall}.
Here, $M_s$ is the magnitude of the saturation magnetization,
and $w$ is the width of the domain wall.
Since this domain wall structure yields the spatial texture of $\boldsymbol{A}_5(x)$,
it gives
the chiral magnetic field $\boldsymbol{B}_5$ localized at the domain wall,
\begin{align}
    \boldsymbol{B}_5 = \frac{J}{e v\_F} \boldsymbol{\nabla}\times\boldsymbol{M} = \frac{JM_s}{ev\_F w} \left( 0, \mathrm{sech}^2\frac{x}{w}, 0 \right).
\end{align}
Whereas, the chiral electric field $\boldsymbol{E}_5$ arises from the dynamics of magnetization.
Let us consider the case where the domain wall texture Eq.~(\ref{eq:Neel-DW}) is in a sliding motion along the $x$-axis with the velocity $V_{\mathrm{DW}}$,
which can be induced by the magnetization switching due to the external magnetic field, current-induced spin torque, etc.
In this case, $\boldsymbol{M}$ becomes time-dependent, replacing $x$ with $\tilde{x} \equiv x - V_{\mathrm{DW}} t$.
As a consequence, there arises a chiral electric field $\boldsymbol{E}_5$ localized at the domain wall,
\begin{align}
    \boldsymbol{E}_5 = -\frac{J}{ev\_F} \partial_t \boldsymbol{M} = \frac{JM_s V_{\mathrm{DW}}}{ev\_F w} \left( \tanh\frac{\tilde{x}}{w} \mathrm{sech}\frac{\tilde{x}}{w}, 0, -\mathrm{sech}^2\frac{\tilde{x}}{w} \right).
\end{align}
Note that the magnitudes of both $\boldsymbol{B}_5$ and $\boldsymbol{E}_5$ inversely scale with $w$.

We have so far used the isotropic spin-momentum locking structure to demonstrate the structures of the chiral electromagnetic fields.
Note that the idea of the chiral electromagnetic fields for magnetic textures is applicable regardless of the spin-momentum locking structure,
as long as the Weyl point positions get smoothly shifted as functions of the magnetic ordering.

For instance,
the idea of chiral gauge field
applies to the ferromagnetic \ce{Co3Sn2S2},
even though it is almost fully spin polarized and the spin-momentum locking is absent near $E_{\rm F}$.
It was demonstrated by the first-principles and the tight-binding model calculations that the tilt of its magnetization 
eventually shifts the positions of the Weyl points \cite{Ghimire2019,Ozawa2024}.
Once the magnetization is tilted from $\boldsymbol{m} \parallel \hat{z} \ (\theta = 0^\circ)$ to $\boldsymbol{m} \parallel -\hat{z} \ (\theta = 180^\circ)$,
the two Weyl points among the six in the Brillouin zone [see Fig.~\ref{fig:css_B5}(a)] follow the momentum-space trajectories shown in Fig.~\ref{fig:css_B5}(b),
obtained from the tight-binding model calculation \cite{Ozawa2024}.
By attributing such shifts of Weyl points to the chiral gauge field,
we may apply the idea of the chiral electromagnetic fields to the magnetic textures in this material.
For instance, a N\'{e}el-type magnetic domain wall leads to the chiral magnetic field $\bm{B}_5$ localized around the domain wall, as shown in Fig.~\ref{fig:css_B5}(c).
It is estimated that a single domain wall with the width $10\ \mathrm{nm}$ induces 
the strength of the $|\boldsymbol{B}_5| \approx 260 \ \mathrm{T}$,
and that its sliding motion with the velocity $V_{\mathrm{DW}} = 100 \ \mathrm{m/s}$ gives $|\boldsymbol{E}_5| \approx 26 \ \mathrm{kV/m}$~\cite{Ozawa2024}.
Both of them scale fairly higher than the ordinary electromagnetic fields $\bm{E}$ and $\bm{B}$.
We shall discuss a consequence of these chiral electromagnetic fields in Sec~\ref{sec:spin-motive-force}.
It is also noted that the chiral electromagnetic fields reachable by other types of perturbations in Weyl semimetals such as lattice strain \cite{ilan2020pseudo}.

\subsection{Localized states at magnetic textures}
\label{sec:localized}

\begin{figure}
\centering
\caption{
Localized states of Weyl fermions under the magentic domain wall structure $\boldsymbol{M}(x)$,
obtained by the analytical calculation \cite{Araki2016prb}.
(a),(b) Energy-momentum dispersions of localized states $E_N(k_y, k_z)$.
The splitting of Weyl points is parametrized by $2k_\Delta$.
(c) Probability distributions of localized states around the domain wall $(x=0)$ along the $x$-axis.
(d) Distribution of the equilibrium current $j_z(x)$ from the chiral pseudomagnetic effect.
(e) Schematic image for the correspondence between the localized states and the surface Fermi arcs.
The magnetization current $\boldsymbol{j} = \boldsymbol{\nabla} \times \boldsymbol{M}_{\rm orb}$,
corresponding to the localized current given in panel (d),
is also shown.
Panels (a)-(d) are reprinted from Ref.~\onlinecite{Araki2016prb}.}	
\label{fig:fermiarc-localized}
\end{figure}

As we have seen in Sec.~\ref{sec:quantum-hall},
the chiral magnetic field $\boldsymbol{B}_5$ for the Weyl fermions leads to the Landau quantization of the energy levels.
The zeroth LLs become the chiral LLs,
whose energy-momentum dispersion is unidirectional along $\boldsymbol{B}_5$ for both positive- and negative-chirality fermions.
The wave functions of the Landau-quantized electrons are localized within the order of the magnetic length $l_\phi = \sqrt{\hbar/e|\boldsymbol{B}_5|}$.
Therefore, 
magnetic textures showing $\boldsymbol{B}_5$ host the localized electronic states with the Landau-quantized spectrum.

Under some magnetic domain wall structures $\boldsymbol{M}(x)$,
the wave functions of localized states were demonstrated both analytically and numerically \cite{Araki2016prb,Araki2018prb}.
The dispersions of the obtained localized states,
in the 2D momentum space projected onto the plane of domain wall (here $k_y k_z$-plane),
are shown in Figs.~\ref{fig:fermiarc-localized}(a) and (b),
with their real-space distributions shown in Fig.~\ref{fig:fermiarc-localized}(c).
In the direction of $\boldsymbol{B}_5$ (here $k_z$-direction),
the dispersions of the localized states are in the structure compatible to that of the Landau levels.
Among the several localized states labeled by the discrete integers $N$,
the state $N=0$ appears as the chiral Landau level dispersed along $\boldsymbol{B}_5$ [see Fig.~\ref{fig:fermiarc-localized}(b)],
while it is present within the path connecting the two Weyl points [see Fig.~\ref{fig:fermiarc-localized}(a)].
Such a structure is similar to the Fermi-arc states on the surface.
This implies that the zero-energy localized states at the domain wall,
can also be understood as a mixture of the two branches of surface Fermi-arc states from the different magnetic domains,
as schematically shown in Fig.~\ref{fig:fermiarc-localized}(e).
The localized states at magnetic domain walls were obtained also by the first-principle calculations of the ferromagnetic \ce{EuB6} \cite{Li2022}.
Even in the models of nonmagnetic Weyl and Dirac fermions,
there were several reports on the emergence of chiral LLs localized around $\boldsymbol{B}_5$ by the numerical calculations \cite{Grushin2016PRX,Liu2013PRB,ilan2020pseudo,Pikulin2016PRX,Kariyado2019jpsj}.

Due to the localized states induced by $\boldsymbol{B}_5$ from the magnetic texture,
we 
expect an electric charge localized at the magnetic texture \cite{Araki2016prb,Araki2018prb}.
If the spacing of the LLs are large enough so that $E_{\rm F}$ crosses only the zeroth LLs,
the localized charge becomes linear in $E_{\rm F}$,
because the zeroth LLs (chiral LLs) have the 1D linear dispersion and exhibit the constant density of states in the low-energy regime.
Note that the Coulomb screening by the bulk electrons may hinder the measurement of this localized charge,
if the bulk electron system has large Fermi surfaces in addition to the Weyl points.

Since the loclized states are the chiral LLs from $\boldsymbol{B}_5$,
they also give rise to the equilibrium charge current $\boldsymbol{j}^{\mathrm{(C)}} \propto \boldsymbol{B}_5$ due to the chiral pseudomagnetic effect, mentioned in Eq.~(\ref{eq:CME-B5-tot})
[see Fig.~\ref{fig:fermiarc-localized}(d)].
The superscript (C) denotes the contribution from the chiral (pseudo-)magnetic effect introduced in Sec.~\ref{sec:chiral-magnetic-effect}.
It corresponds to the magnetization current $\boldsymbol{j}^{\mathrm{(C)}}({\rm r}) = \boldsymbol{\nabla} \times \boldsymbol{M}_{\mathrm{orb}}(\boldsymbol{r})$,
with the orbital magnetization $\boldsymbol{M}_{\mathrm{orb}}$ of the Weyl fermions.
In the present case with a magnetic domain wall,
the surfaces of both the two magnetic domains facing at the vacuum exhibit the surface Fermi arc states,
which carry the current flowing oppositely to $\boldsymbol{j}^{\mathrm{(C)}}$ in the domain wall [see Fig.~\ref{fig:fermiarc-localized}(e)].
Thus, the currents in the domain wall and on the surfaces cancel each other,
leading to the vanishing net current in equilibrium.
This discussion applies also to the energy current,
exhibiting the energy current locally finite but zero in total \cite{Shitade2021prb}.

We have so far focused on the magnetic Weyl semimetal states,
and considered the localized states at domain walls exhibiting the structure similar to the topological surface states, here the Fermi arcs.
The similar discussion applies to the topological electronic states other than Weyl semimetals.
In particular, there was a theoretical discussion considering the ferromagnetic nodal-line semimetal state \cite{araki2021jpsj}.
Nodal-line semimetals have almost dispersionless 2D states localized at the surface, which are called the drumhead surface states \cite{Kim2015,Chan2016prb,Yamakage2016jpsj}.
Once the magnetic domain wall structure is introduced,
it was demonstrated numerically that the domain wall supports the localized states,
whose energy spectrum and topological origin were found to be same as those of the drumhead surface states.

\subsection{Electric responses to magnetization dynamics} \label{sec:spin-motive-force}
In addition to the spatial texture of the magnetization seen above,
once we consider the dynamics in the magnetization,
the electrons therein are driven out of equilibrium and are capable of generating nonequilibrium physical quantities in response.
In ferromagnetic metals where the momentum-space topology is trivial,
the effect of the magnetization dynamics is well studied.
The magnetic texture dynamics yields the SMF,
also known as the emergent electric field,
on the conduction electrons \cite{Volovik1987jopc,Stern1992prl,Barnes2007prl}.
Under the dynamics of magnetic texture $\bm{M}(\bm{r},t) = M_s \bm{m}(\bm{r},t)$,
with $M_s$ the saturation magnetizaion,
the emergent electric field becomes,
\begin{align}
    \boldsymbol{e}^\pm(\bm{r},t) &= \pm P \frac{\hbar}{2e}\left( 
\boldsymbol{m}\times\dot{\boldsymbol{m}} + \beta\dot{\boldsymbol{m}} \right) \cdot \boldsymbol{\nabla}\boldsymbol{m}, \label{eq:spinmotive-force}
\end{align}
Here, $\pm$ distinguishes the majority and minority spin states of electrons, and $P \ (\leq 1)$ is the ratio of spin polarization.
The first term originates from the adiabatic dynamics of electron fixed to the majority (or minority) spin state, and the second term is from the nonadiabatic dynamics from the interband transition of spin \cite{Saslow2007prb,Duine2008prb,Duine2009prb,Tserkovnyak2008prb}.
The phenomenological parameter $\beta$ characterizing the nonadiabaticity is usually much smaller than unity,
of the order close to the Gilbert damping parameter $\alpha \sim 0.01$.
While this form requires a nonuniform magnetic texture $(\nabla{\bm m})$,
the dynamics of uniform magnetization can also yield a SMF by the spin-orbit coupling,
in the systems with broken inversion symmetry, such as in noncentrosymmetric crystals or at the interfaces with the Rashba spin-orbit coupling \cite{Ryu1996prl,Kim2012prl,Tatara2013prb,Yamane2013prb,ciccarelli2015magnonic}.
Unconventional transport properties arising from magnetization dynamics, such as the emergent inductance \cite{Nagaosa2019jjap,kurebayashi2021electromagnetic,Ieda2021prb,Yamane2022prl,yokouchi2020emergent,anan2025emergent,oh2024emergent},
have also been studied based on the SMF.

It should be noted that the chiral electromagnetic fields $(\boldsymbol{E}_5, \boldsymbol{B}_5)$ for the Weyl fermions are 
conceptually distinct from the emergent electromagnetic fields explained above.
The idea of emergent electromagnetic fields applies to the cases where
the majority and minority spin states are largely split in energy, so that the electrons remain in the majority (or minority) spin state
almost adiabatically under the magnetization dynamics.
On the other hand, in Weyl semimetals, 
we assume that the Weyl dispersion is robust against perturbations from magnetization dynamics, 
where the electrons are supposed to remain (approximately) adiabatic within each chirality $(\pm)$ state instead of the majority or minority spin state.
Under this condition, the idea of chiral electromagnetic fields is ensured.

Our interest herein is
whether the magentic Weyl semimetals show the effect compatible to the SMF.
For the Weyl electrons,
the description with the chiral electric field $\boldsymbol{E}_5 = -\partial_t \boldsymbol{A}_5$ is useful to understand the effect of magnetization dynamics.
Since $\boldsymbol{E}_5$ acts on the chirality $\pm$ states oppositely,
it can contribute to the net electric current only if the inversion symmetry is broken and the chirality is imbalanced.
In noncentrosymmetric Weyl semimetals,
it is theoretically proposed that the momentum-space dynamics of Weyl points leads to the adiabatic charge pumping analogous to the Thouless pumping in 1D \cite{Ishizuka2016prl,Ishizuka2017prb}.
This proposal applies to the dynamics of uniform magnetization in noncentrosymmetric magnetic Weyl semimetals \cite{Harada2023prb},
which results in the net charge current in response to the precession of magnetization.

In centrosymmetric magnetic Weyl semimetals,
$\boldsymbol{E}_5$ can contribute to the net current once there is a magnetic texture that breaks the inversion symmetry.
This can be understood as the combination effect from $\boldsymbol{E}_5 = -\partial_t {\bm A}_5$ and $\boldsymbol{B}_5 = \boldsymbol{\nabla} \times \boldsymbol{A}_5$,
both of which are present under the dynamics of magnetic texture.
They result in the ordinary Hall effect, 
yielding a finite net charge current $ \boldsymbol{j}^{\rm (H)} \propto \hat{\boldsymbol{B}}_5 \times \boldsymbol{E}_5 $
as seen in Sec.~\ref{sec:quantum-hall}
\cite{Araki2018prapplied,Ozawa2024}.
In particular,
suppose the domain wall is in a sliding motion,
e.g., in the $x$-direction with the velocity $V_{\rm DW}$.
In such case, $\bm{A}_5$ satisfies $\bm{A}_5(\boldsymbol{r},t) = \bm{A}_5(\tilde{x} = x - V_{\rm DW} t)$,
and hence the chiral electromagnetic fields
\begin{align}
    \boldsymbol{E}_5 = -\partial_t \boldsymbol{A}_5 = V_{\rm DW} \partial_{\tilde{x}}\bm{A}_5, \\
    \boldsymbol{B}_5 = \boldsymbol{\nabla}\times\boldsymbol{A}_5 = \hat{\bm x} \times \partial_{\tilde{x}}\bm{A}_5
\end{align}
become perpendicular to each other,
contributing to this Hall current along the sliding direction (here $x$-direction; see also Fig.~\ref{fig:domainwall}).
The induced current or voltage appears to be
compatible to the nonadiabatic SMF [$\beta$-term in Eq.(\ref{eq:spinmotive-force})],
while its strength is not limited by the spin polarization ratio $P$ and can largely exceed the SMF in conventional metals.
For example,
by the calculations with the tight-binding model of the ferromagnetic \ce{Co3Sn2S2},
it was estimated that
the Hall voltage arising from the domain wall motion scales about 800 times larger than that from the conventional SMF \cite{Ozawa2024}.

Besides $\bm{j}^{(\rm H)}$ explained above, the contribution from the chiral magnetic current $\bm{j}^{(\rm C)}$ is also suggested,
once the external magnetic field $\bm{B}$ is taken into account \cite{Hannukainen2020}.
The combination of $\boldsymbol{E}_5$ and $\boldsymbol{B}_5$ leads to the chirality imbalance $\mu_5 \propto \boldsymbol{E}_5\cdot\boldsymbol{B}_5$ 
due to the chiral anomaly, in a manner similar to the conventional $\bm{E} \cdot \bm{B}$
[see Eq.~(82) in Sec.~\ref{sec:anomaly}].
This $\mu_5$ leads to the nonequilibrium charge current $\boldsymbol{j}^{\rm (C)} \propto \mu_5 \boldsymbol{B}$ due to the chiral magnetic effect.
In contrast to the above $\boldsymbol{j}^{\rm (H)}$ predicted from the sliding motion of a domain wall,
this anomaly-induced $\boldsymbol{j}^{\rm (C)}$ is predicted from the rotational dynamics of a domain wall,
whose consequence is similar to the adiabatic SMF [the first term in Eq.~(\ref{eq:spinmotive-force})].
This $\mu_5$ can contribute also to the anomalous drift current nonlinear in $\boldsymbol{E}_5$,
which is predicted under the structural transition of a domain wall between the Bloch and N\'{e}el types \cite{Heidari2023}.

\subsection{Electrically induced spin torques}
In the above section, we have seen the process where the magnetization dynamics is converted into electric current and voltage.
As its inverse process,
the conversion from the electric current or voltage into the magnetization dynamics
is also widely discussed in the context of spintronics.
Such a process is regarded as the current (or voltage)-induced torques on the magnetic moments.
The dynamics of a magnetic moment, with its direction $\boldsymbol{m}(\bm{r},t)$, is governed by the Landau--Lifshitz--Gilbert equation,
\begin{align}
\dot{\boldsymbol{m}}(\bm{r},t) &= -\gamma \boldsymbol{m} \times \boldsymbol{B}_{\rm eff} + \alpha \boldsymbol{m} \times \dot{\boldsymbol{m}} + \boldsymbol{t}, 
\end{align}
where $\gamma$ is the gyromagnetic ratio and $\alpha$ is the Gilbert damping constant.
$\boldsymbol{B}_{\rm eff}$ is the effective magnetic field for the magnetic moment,
due to an external magnetic field, magnetic anisotropy, etc.
Inducing the torque $\boldsymbol{t}(\bm{r},t)$ from the external inputs,
such as electric current or voltage,
is essential
to manipulate magnetization dynamics and to switch magnetic orderings in devices.
Owing to the characteristics of Weyl electons,
such as the spin-orbit coupling, band topology, and the suppression of dissipation,
the spin torques in magnetic Weyl semimetals are affected by the topological magnetoelectric response and are fundamentally distinct from those developed in conventional magnetic metals.
In the following,
we review the studies on such electrically induced spin torques in magnetic Weyl semimetals,
by comparing them with the well-established theories of conventional spin torques.


\begin{figure}
\centering
\includegraphics[width=\hsize]{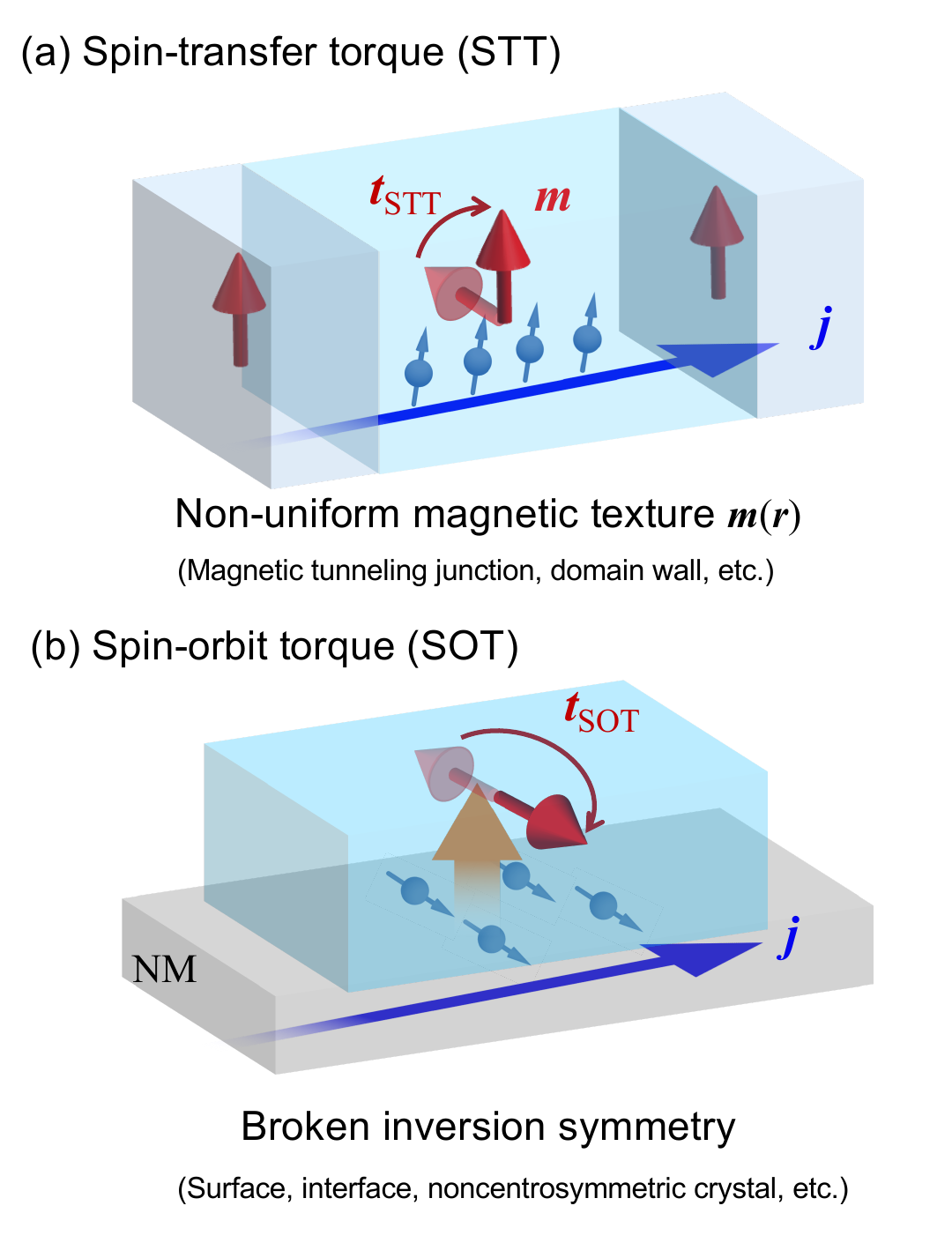}
\caption{
Schematic images of the conventional current-induced torques.
(a) The STT occurs where the magnetic texture $\boldsymbol{m}(\boldsymbol{r})$ breaks the inversion symmetry,
whereas (b) the SOT arises where the inversion symmetry is intrinsically broken regardless of the magnetization.
NM denotes the nonmagnetic material with strong SOC, such as heavy metals, topological insulators, etc.
}	
\label{fig:STT-SOT}
\end{figure}

\subsubsection{Spin torques on magnetic textures}
\label{sec:STT}

We first discuss the case where inversion symmetry is (locally) broken by the configuration of magnetic moments,
e.g., in magnetic tunnel junctions, magnetic textures, etc.
In metallic magnets, the transfer of spin angular momentum occurs between the spin-polarized conduction electrons and the local moments \cite{Slonczewski1989prb,Berger1996prb,Tserkovnyak2005rmp,Zhang2005prl,Barnes2005prl} [see Fig.~\ref{fig:STT-SOT}(a)].
This torque is known as the STT,
which has been widely studied to manipulate magnetizations in spintronics devices.
For instance,
when a spin-polarized electric current $\boldsymbol{j}$ is injected to a magnetic texture $\boldsymbol{m}(\boldsymbol{r}) = \bm{M}(\bm{r}) / M_s$ in a ferromagnetic metal,
the STT on $\boldsymbol{m}(\boldsymbol{r})$ is given as
\begin{align}
    \boldsymbol{t}_{\mathrm{STT}} (\bm r) &= -(\boldsymbol{u}\cdot\boldsymbol{\nabla})\boldsymbol{m} + \beta\boldsymbol{m}\times(\boldsymbol{u}\cdot\boldsymbol{\nabla})\boldsymbol{m} 
    \label{eq:STT}
\end{align}
at $O(\bm{\nabla}\bm{m})$.
Here, $\boldsymbol{u} = (\gamma\hbar P/2eM_s)\boldsymbol{j}$ is the vector having the dimension of velocity.
The first and second terms come from adiabatic and nonadiabatic spin dynamics of electrons, respectively,
when the electrons travel through the magnetic texture \cite{thiaville2005micromagnetic,Tserkovnyak2006prb,Kohno2006jpsj}.
The STT can be regarded as an inverse process of SMF.
Similarly to the theory of SMF [Eq.~(\ref{eq:spinmotive-force})],
the adiabatic part comes from the dynamics of electrons fixed to either the majority or minority spin states under a large spin spitting,
whereas the nonadiabatic part comes from the interband dephasing.

In magnetic Weyl semimetals,
the theories of the STT cannot be applied as they are,
because of the nontrivial SOC structure and band topology around the Weyl points.
To theoretically understand the spin torques,
one needs to evaluate the electrically-induced spin polarization of the Weyl electrons,
which serves as the effective magnetic field coupling to the magnetic moments.

\begin{figure}
\centering
\includegraphics[width=\hsize]{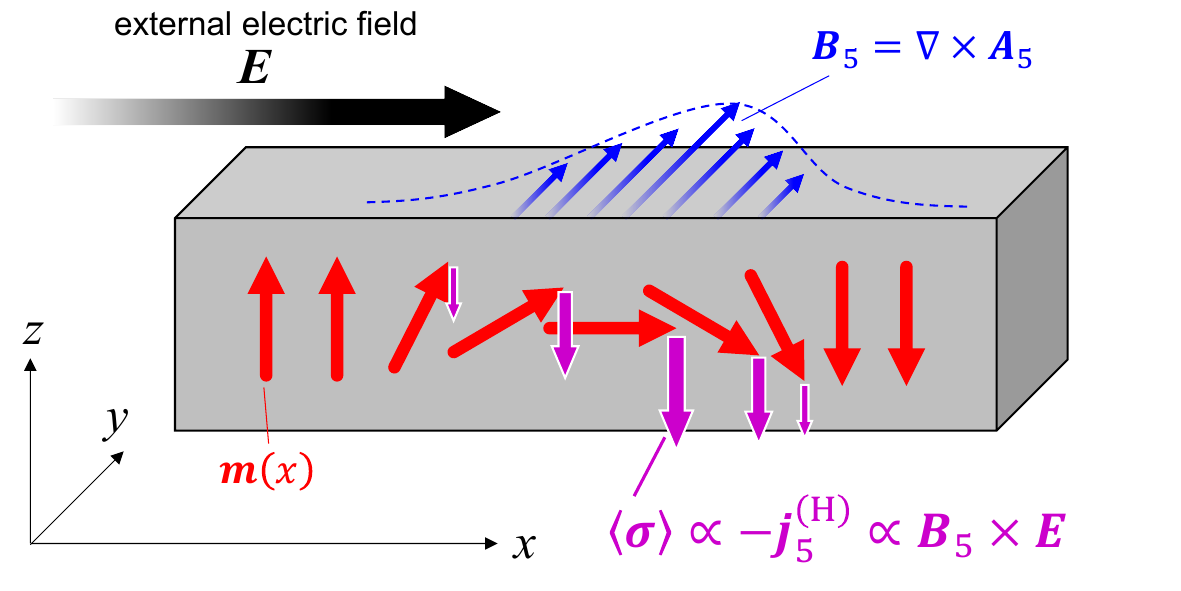}
\caption{
Schematic image of the electrically induced spin torque at a domain wall,
proposed in a Weyl semimetal with the isotropic spin-momentum locking structure \cite{kurebayashi2019theory,kurebayashi2021jpsj}.
Once the electric field $\boldsymbol{E}$ is applied to the domain wall,
it induces the chiral Hall current $\boldsymbol{j}^{\rm (H)} \propto \boldsymbol{B}_5 \times \boldsymbol{E}$
corresponding the spin polarization $\langle \boldsymbol{\sigma} \rangle \propto -\boldsymbol{j}_5^{\rm (H)}$,
which exerts torques on the magnetic moments.
}	
\label{fig:stt-schematic}
\end{figure}

The early theoretical works were based on the isotropic spin-momentum locking structure \cite{kurebayashi2019theory,kurebayashi2021jpsj}.
With the simple relation $\boldsymbol{j}_5 = -e v\_F \boldsymbol{\sigma}$
between the electron spin polarization $\boldsymbol{\sigma}$ and the chiral current $\boldsymbol{j}_5$,
the spin torques were evaluated by the linear response theory with respect to the electric field $\boldsymbol{E}$.
In the presence of a magnetic texture,
the chiral magnetic field $\boldsymbol{B}_5$ from the magnetic texture gives rise to a chiral current from the ordinary Hall effect, $\boldsymbol{j}_5^{\rm (H)} = \sigma\_H \boldsymbol{B}_5 \times\boldsymbol{E} / |\boldsymbol{B}_5|$.
This chiral current is equivalent to the spin polaization $\langle \boldsymbol{\sigma} \rangle = -\boldsymbol{j}_5^{\rm (H)} / ev\_F$, 
as schematically illustrated in Fig.~\ref{fig:stt-schematic},
exerting a torque $\boldsymbol{t}_{\rm Weyl} = \boldsymbol{m} \times J\langle \boldsymbol{\sigma} \rangle$ on the magnetic texture.
In particular, under the isotropic spin-momentum locking structure,
the magnetic texture gives $\bm{B}_5 \propto \bm{\nabla}\times\bm{m}$,
and hence this torque takes the structure,
\begin{align}
    \bm{t}^{\rm Weyl}_{\rm STT} &\propto \bm{m} \times ( \bm{B}_5 \times \bm{E} ) \nonumber \\ 
    &\propto \bm{m} 
\times (\bm{E}\cdot \bm{\nabla})\bm{m} - \bm{m}\times \bm{\nabla}(\bm{E}\cdot\bm{m}).\label{eq:Weyl-STT}     
\end{align}
Compared with the conventional STT in Eq.~(\ref{eq:STT}),
the first term is compatible to the nonadiabatic STT led by the factor $\beta$.
On the other hand, the term compatible to the adiabatic STT,
of the form $\bm{t}\propto (\bm{E}\cdot \bm{\nabla})\bm{m}$,
is prohibited in Eq.~(\ref{eq:Weyl-STT}),
because of the U(1) chiral gauge invariance of the Weyl fermions under the spin-momentum locking structure.
From the quantitative aspect,
unlike the nonadiabatic STT limited to $\beta < 1$ in conventional magnetic metals,
the strength of the torque of Eq.~(\ref{eq:Weyl-STT}) in Weyl semimetals can reach $\beta > 1$.
This is due to the dominance of SOC and the robustness of Weyl fermions against disorder around the Weyl points \cite{kurebayashi2021jpsj}.
We note that, in realistic lattice system, where U(1) chiral gauge symmetry is broken, 
the different form of the spin torque beyond Eq.(\ref{eq:Weyl-STT}) is expected.

The momentum-space Berry curvature also intrinsically contributes to the torque on magnetic textures,
which was proposed from the semiclassical theory \cite{Araki2021prl}.
In reseponse to the electric field $\boldsymbol{E}$,
we have seen that each electron acquires the anomalous velocity $\boldsymbol{v}^{\rm A}(\bm{k}) = (e/\hbar)\boldsymbol{E}\times\boldsymbol{b}(\bm{k})$
from the Berry curvature $\boldsymbol{b}$,
contributing to the intrinsic anomalous Hall effect.
Under the spin-momentum locking structure,
the anomalous velocity is converted to the spin polarization,
$\langle \bm{\sigma}^{\rm A}(\bm{k}) \rangle \propto \boldsymbol{v}^{\rm A}(\bm{k})$.
The local spin polarization $\langle \bm{\sigma}_{\rm tot}^{\rm A}(\bm{r}) \rangle = \frac{1}{V} \sum_{\bm{k}} f(\bm{r},\bm{k}) \langle \bm{\sigma}^{\rm A}(\bm{k}) \rangle $,
summed over the local electron distribution $f(\bm{r},\bm{k})$,
becomes nonzero if the inversion symmetry is broken.
Even if the crystal structure itself is centrosymmetric,
a gradient of magnetic texture $\nabla {\bm m}({\bm r})$ breaks the inversion symmetry of $f(\bm{r},\bm{k})$,
and hence it gives a torque $\boldsymbol{t}_{\rm THT}(\bm{r}) = \boldsymbol{m}(\bm{r}) \times J\langle \bm{\sigma}_{\rm tot}^{\rm A}(\bm{r}) \rangle$ of the order $O(\nabla {\bm m})$
exerting on the magnetic texture.
This torque is
explicitly named as the \textit{topological Hall torque},
whose structure is again compatible to the $\beta$-term in the STT.
Due to the intrinsic origin of the anomalous spin polarization $\langle \bm{\sigma}^A \rangle$,
the topological Hall torque does not require dissipative longitudinal current,
and it is (ideally) free from power loss by the Joule heating.
From the model calculation,
the efficiency of the topological Hall torque, in terms of the $\beta$-parameter, is also estimated to exceed the upper limit $\beta > 1$.
Such an effect requires a large Berry curvature and the spin-momentum locking structure,
which are likely to be satisfied in Weyl materials.
It was experimentally confirmed in the ferromagnetic Weyl metal \ce{SrRuO3} \cite{yamanouchi2022sciadv},
which shall be revisited in the end of this section.


It was also confirmed numerically
that the $\beta$-term-like torque is present regardless of the structure of spin-momentum locking:
in Ref.~\onlinecite{kurebayashi2019theory},
the spin polarizations $\langle \boldsymbol{\sigma} \rangle$ induced by the electric field $(E_x)$ around the domain walls were calculated, by using the hypothetical lattice models with the Weyl-type (isotropic) spin-momentum locking,
the Rashba-type spin-momentum locking, and the $s_z$-conserved (Ising-type) spin-momentum locking,
all of which showed similar structures of current-induced effective fields around the domain wall.

\subsubsection{Spin torques in noncentrosymmetric systems}

We next consider the case where the electron system does not have inversion symmetry regardless of the presence of absence of magnetic textures.
This statement applies to the electron systems at surfaces or interfaces,
and also to the bulk electrons in noncentrosymmetric crystals [see Fig.~\ref{fig:STT-SOT}(b)].
In such cases, an electric field or current can induce a spin torque
even if the magnetization is uniform.
Such a torque arises from the charge-to-spin conversion by spin-orbit coupling, and hence is classified as the SOT \cite{obata2008prb,manchon2008prb,miron2010current,miron2011perpendicular,liu2012prb,liu2012spin}.
The major origins of the SOT are
the spin current from the spin Hall effect \cite{hirsch1999spin},
the current-induced spin polarization like the Rashba--Edelstein effect \cite{edelstein1990spin}, etc.
Generally, the form of SOT is governed by the symmetries of crystal structure \cite{vzelezny2017spin}.



In noncentrosymmetric Weyl semimetals,
the SOT-like contribution acting on uniform magnetization is also studied theoretically.
One study is based on
the model of isotropic spin-momentum locking structure,
$H_\eta(\boldsymbol{k}) = \eta \hbar v_{\rm F} \boldsymbol{k}\cdot\boldsymbol{\sigma} -\eta \Delta$,
where $\Delta$ corresponding to the chiral chemital potential parametrizes the breaking of inversion symmetry \cite{kurebayashi2021jpsj}.
In response to the electric field $\boldsymbol{E}$,
it was found that the spin density $\langle \boldsymbol{\sigma} \rangle \propto \tau E_{\rm F} \Delta \boldsymbol{E}$ is induced,
with $\tau$ the transport relaxation time,
which results in the SOT $\boldsymbol{t}^{\rm Weyl}_{\rm SOT} = J\boldsymbol{m}\times \langle \boldsymbol{\sigma} \rangle$.
This effect is due to the shift of electron distribution on the Fermi surfaces
analogous to 
the Rashba--Edelstein effect:
while the spin-momentum locking structures on the Fermi surfaces of the chiralities $\eta = \pm$ are opposite to each other,
the inversion symmetry breaking with $\Delta$ gives the imbalance in the sizes of these Fermi surfaces,
which yields $\langle \boldsymbol{\sigma} \rangle \neq 0$ not fully cancelled.
This SOT depends on the relaxation time $\tau$ and is classified as a dissipative process,
which requires a large longitudinal current density to obtain a large torque.

\begin{figure}
\centering
\caption{
Model calculation results of topological spin-orbit torque (SOT) in the ferrimagnetic \ce{Ti2MnAl}.
Momentum-space distributions of (a) the Berry curvature and (b) the mixed Berry curvature,
both showing singular behaviors at the Weyl points (WPs).
Fermi level dependences of (c) the density of states (DOS) and (d) the efficiencies of topological SOT $(\theta^{\rm top}_{yx})$ and dissipative SOT $(\theta^{\rm dis}_{xx})$ per injected current density.
Reprinted and edited from Ref.~\cite{meguro2025topological}.
}	
\label{fig:topological-SOT}
\end{figure}

On the other hand,
a non-dissipative SOT almost independent of $\tau$ was predicted in a model calculation for the noncentrosymmetric ferrimagnetic Weyl semimetal candidate \ce{Ti2MnAl} \cite{meguro2025topological}.
This non-dissipative SOT comes from the band topology of electrons,
and hence is called the topological SOT.
Within the whole effective field $\boldsymbol{h}_{\rm SOT} = \langle \partial \mathcal{H} / \partial{\boldsymbol{n}} \rangle$ exerting an SOT on the magnetic ordering $\boldsymbol{n}$ (including the ferrimagnetic ordering),
the topological SOT part is given as \cite{freimuth2014spin,hanke2017mixed,li2015intraband,kurebayashi2014antidamping},
\begin{align}
    \boldsymbol{h}_{\rm SOT}^{\rm top} &= e\hbar \int \frac{d^3 \boldsymbol{k}}{(2\pi)^3}  \sum_m f_m(\boldsymbol{k}) \hat{\Omega}_m^{\boldsymbol{n} \boldsymbol{k}}(\boldsymbol{k}) \boldsymbol{E} .
\end{align}
Here, the tensor quantity $\hat{\Omega}_m^{\boldsymbol{n} \boldsymbol{k}}$ is defined as,
\begin{align}
    \Omega_m^{n_i k_j}(\boldsymbol{k}) &= 2\, {\rm Im} \sum_{l\, (\neq m)} \frac{\langle u_m | \partial_{n_i}\mathcal{H} | u_l \rangle \langle u_l | \partial_{k_j}\mathcal{H} | u_m \rangle}{(E_m - E_l)^2} ,
\end{align}
with the energy $E_m$ and Bloch wave function $|u_m\rangle$ for each band $m$.
It characterizes the band topology in the mixed phase space of spin and momentum,
which is known as the \textit{mixed} Berry curvature.
Similarly to the momentum-space Berry curvature $\boldsymbol{b}(\boldsymbol{k})$ [Fig.~\ref{fig:topological-SOT}(a)],
the mixed Berry curvature also becomes singular at the Weyl points  [Fig.~\ref{fig:topological-SOT}(b)],
thus yielding a sizable contribution to the topological SOT in \ce{Ti2MnAl}.
The topological SOT is advantageous over the dissipative SOT in that it does not require a large longitudinal current $(j_x)$,
and hence is irrelevant to the power loss by Joule heating. 
Indeed, the calculations show that the torque efficiency per injected current density for the topological SOT $(\theta^{\rm top}_{yx} = h^{\rm top}_{{\rm SOT},y} / j_x)$ is maximized near $E_{\rm F}$ in the vicinity of the Weyl points,
even though the density of states is minimized,
and is about 5 times larger than that for the dissipative SOT $(\theta^{\rm dis}_{xx})$ [see Figs.~\ref{fig:topological-SOT}(c)(d)].
This $\theta^{\rm top}_{yx}$ can reach 100-1000 times the SOT efficiencies in metallic antiferromagnets, such as \ce{Mn2Au}~\cite{vzelezny2014relativistic,vzelezny2017spin} and \ce{CuMnAs} \cite{wadley2016electrical},
which mainly originate from the dissipative SOT.


In addition to the STT- and SOT-like contributions listed above,
another type of electrically-induced spin torque,
in response to the local chemical potential,
is predicted in magnetic Weyl semimetals \cite{Nomura2015,kurebayashi2021jpsj}.
Under the application of a magnetic field $\boldsymbol{B}$,
the magnetization can be switched locally by the modulation of chemical potential $\delta \mu$,
corresponding to the local gate voltage,
which is named the \textit{charge-induced torque}.
This effect was derived thermodynamically from the anomalous magnetoelectric coupling characteristic to the Weyl fermions \cite{Nomura2015}.
In particular, under the isotropic spin-momentum locking structure,
the charge-induced torque can be understood in terms of the chiral magnetic effect,
where the magnetic field induces the chiral current proportional to $\delta\mu$, $\boldsymbol{j}_5^{\rm (C)} \propto \delta\mu \boldsymbol{B}$,
yielding a finite spin polarization $\langle \boldsymbol{\sigma} \rangle \propto \boldsymbol{j}_5^{\rm (C)}$ \cite{kurebayashi2021jpsj}.
In contrast to the STT and SOT,
the charge-induced torque is a thermodynamic effect without any electric current,
and hence it is essentially dissipationless.

\subsubsection{Experimentally observed spin torques in Weyl semimetals}

\begin{figure}
\centering
\caption{
Measurement of the Berry-curvature induced torque (topological Hall torque) in a ferromagnetic Weyl metal \ce{SrRuO3}.
(a) Schematic of the experimental setup to measure the current-induced domain wall motion.
(b) Temperature dependence of the current-induced torque efficiency $|\eta_J|$ measured experimentally.
The black curve shows the upper bound of the nonadiabatic spin-transfer torque $(\beta=1)$,
which cannot explain the observed value (red points).
(c) Theoretically calculated values of the torque efficiency $|\eta_E|$ based on the theory of topological Hall torque (blue curve).
Comparison with the experimentally measured values (red points) shows a good agreement.
Reprinted and edited from Ref.~\cite{yamanouchi2022sciadv}.
}	
\label{fig:THT-SRO}
\end{figure}

Experimentally,
electrically induced
dynamics of magnetic textures, especially magntic domains,
have been observed in several magnetic Weyl semimetal materials.
In the chiral antiferromagnet \ce{Mn3Sn},
the current-induced reconfiguration of magnetic domains was observed by the measurement of the anomalous Hall effect \cite{Sugimoto2020}.
The dynamics of the $60^\circ$-domain walls characteristic to the chiral antiferromagnet was studied earlier with the phenomenological equation of motion \cite{Liu2017prl},
while the contribution of the Weyl fermions therein has not yet been fully understood.

In the ferromagnetic \ce{SrRuO3},
the early-stage measurements of the current-induced magnetization switching reported that the torque on a domain wall becomes unusually large compared to the conventional STT \cite{Feigenson2007,Yamanouchi2019}.
However, its origin remained unsolved for almost 15 years.
With the detailed measurements of its temperature dependence,
such a large torque was finally identified as the topological Hall torque from the Berry curvature around the Weyl points \cite{yamanouchi2022sciadv}
theoretically explained in Sec.~\ref{sec:STT}.
Indeed, the model calculation based on the theory of topological Hall torque well reproduced the temperature dependence of torque efficiency observed in the experiment
(see Fig.~\ref{fig:THT-SRO}).

A similar observation of magnetization switching was also conducted in the ferromagnetic \ce{Co3Sn2S2},
which revealed the torque efficiency even larger than that in \ce{SrRuO3} \cite{wang2023magnetism}.
The theories of spin torques in Weyl semimetals so far are not applicable to such a largely spin-polarized (i.e., half-metallic) Weyl semimetal,
and hence further theoretical 
studies are needed.

\section{Magneto-transport and Hall transport in Weyl semimetals} 
\label{sec:transport}

We have so far reviewed the interplay between electronic and magnetic properties in magnetic Weyl semimetals~(MWSM) originating from bulk band topology.
In addition to these intriguing properties,
transport in the mesoscopic regime forms a distinct and crucial frontier~\cite{datta1997electronic}.
This is because transport properties at short length scales can be significantly modified by phase coherence of wave functions, quantum confinement, surface states, and interfacial coupling with electrodes.


Before proceeding to studies specific to Weyl semimetals,
we briefly review the history of transport properties in systems hosting Dirac fermions.
In early works in the 1980s and 1990s,
generic features of Dirac electron systems were theoretically explored.
In particular, the robustness of the DOS around the Dirac points against disorder was established~\cite{fradkin1986critical1, fradkin1986critical2},
and notable transport phenomena such as minimal conductivity \cite{Ludwig1994-af, Shon1998-ab} and the half-integer quantum Hall effect were predicted~\cite{zheng2002hall, Gusynin2005}.

Subsequently, in 2005, experimental measurements of transport properties in graphene~\cite{novoselov2005two,zhang2005experimental} further highlighted the significance of Dirac electrons.
These developments stimulated further theoretical studies not only in 2D Dirac electron systems, including topological insulator surfaces~\cite{Hasan10topological},
but also in 3D Dirac electron systems such as Dirac semimetals~(DSM)~\cite{Murakami2007}
and organic conductors~\cite{Katayama2006,Kajita14molecular}.
The findings in these theoretical and experimental studies 
mostly originate from the vanishing DOS and the robustness against disorder,
which are direct consequences of the Dirac point structure.



Motivated by the preceding studies explained above,
mesoscopic transport properties of MWSMs are also intensely studied,
to establish an understanding on interplay among topological nature, magnetism, and quantum coherence.
This section is devoted to the review on such studies,
mainly about theoretical studies concerning transport in nanoscale devices.
We first discuss the robustness of Weyl semimetals against disorder, and then 
review two major topics:
(i) the magnetoresistance effect, i.e., the modulation of conductance depending on orientations of magnetization and 
(ii) the anomalous Hall effect involving both intrinsic and extrinsic mechanisms, with the surface state contributions.

\subsection{Robustness of Dirac/Weyl semimetals against disorder} \label{sec:transport:robustness}
 Robustness against disorder is one of the hallmark features of topological systems.
 Nevertheless, a definitive consensus regarding the robustness of Dirac/Weyl semimetals against disorder has yet to be established.
 This ambiguity stems from the difficulty of properly defining the semimetal phase through an observable quantity in the presence of disorder.
 This situation contrasts with 2D topological systems such as quantum Hall states or $\mathbb{Z}_2$ topological insulators, where quantized transport phenomena \cite{Hasan10topological} serve as unambiguous indicators.

 One of the defining characteristics of ideal Dirac/Weyl semimetals is the vanishing DOS at the energy of the Dirac/Weyl nodes. 
 If this definition is adopted strictly, 
an ideal semimetal cannot exist in realistic systems;
even infinitesimal disorder---particularly long-range correlated disorder---induces a finite DOS, thereby destabilizing the semimetal phase \cite{Nandkishore14rare, Skinner14coulomb, Ominato15quantum, Syzranov15unconventional, Pixley16rare, Louvet17new}.

 On the other hand, numerous theoretical and numerical studies have demonstrated the semimetal--metal transition at finite disorder strength \cite{Goswami11quantum, Kobayashi14density, Ominato14quantum, Sbierski14z2, Sbierski14quantum, Syzranov15criticalTransport, Roy14diffusive, Roy16erratum, Pixley15anderson, Sbierski15quantum, Chen15disorder, Liu16effect, Bera16dirty, Pixley16disorder, Shapourian16phase, Syzranov16criticalExponents, Louvet16on, Pixley16uncovering, Syzranov16multifractality, Syzranov18high, Fu17accurate, Roy18global}.
 The transition can be characterized by, for example, using the scaling law \cite{Kobayashi14density} of the functional form of DOS $\rho(E)$ instead of the DOS just at the Dirac/Weyl point $E=0$, $\rho(0)$.
 This transition might be a sharp crossover from the standpoint of the rigid definition based on $\rho(0)$,
but it is practically indistinguishable from the transition because a small finite DOS does not destroy the features of Dirac/Weyl semimetals other than $\rho(0)=0$.
 Indeed, numerical investigation of the spectral function under disorder (Fig.~\ref{fig:transport:spectral}) shows that Weyl nodes remain sharply discernible up to the critical disorder strength, and ballistic transport feature alives as well \cite{Kobayashi20ballistic}.

 From a practical perspective, the effect of disorder on measurable properties is more important than a mathematically rigorous definition of semimetals.
 In particular, robustness of transport phenomena characteristic to Dirac/Weyl semimetals under disorder is of central interest.
 The longitudinal conductivity near Dirac/Weyl points cannot be a defining feature of a semimetal, as its value depends strongly on the model, the type of disorder, and the computational methods or approximations employed \cite{Hosur12charge, Nandkishore14rare, Ominato14quantum, Sbierski14quantum, Syzranov15criticalTransport, Ominato15quantum, Ziegler16quantum, Klier19from}.
 Furthermore, real samples are finite in size, and contributions from surface states (Fermi arcs) cannot be neglected, making it difficult to isolate the bulk contribution from Dirac/Weyl points in experiments. 
 Consequently, identifying experimentally accessible transport signatures that are characteristic of semimetals remains an important challenge.

\begin{figure}[tbp]
 \centering
 \vspace{-3mm}
\caption{(Color online)
  Density plot of the spectral function $\rho_{(k_x,0,0)}(E)$ of disordered WSMs with different disorder strengths: 
  (a) clean, (b) weak, (c) slightly below critical, and (d) just at critical.
  The horizontal axis is $k_x$ and the vertical axis is $E$.
}
\label{fig:transport:spectral}
\end{figure}


\subsection{Magnetoresistance in MWSMs} \label{sec:transport:MR}
 In this section, we review magnetoresistance effects in MWSMs.
 Magnetoresistance refers to a change in resistivity induced by magnetization or by an applied magnetic field.
 In conventional metals, three representative types of magnetoresistance are well known~\cite{Zutic2004spintronics}:
anisotropic magnetoresistance (AMR), giant magnetoresistance (GMR), and tunneling magnetoresistance (TMR).
 AMR arises from the dependence of resistivity on the relative angle between the current and magnetization, although the effect is typically small.
 GMR appears in magnetic multilayers, where resistance depends on the relative magnetization orientation of adjacent ferromagnetic layers.
 TMR occurs in magnetic tunnel junctions, where spin-dependent tunneling across an insulating barrier produces a large resistance change.
 In MWSMs, the coupling between Weyl cones and magnetization can induce two mechanisms that enhance magnetoresistance: less-overlapping \cite{Ominato17anisotropic} and helicity mismatch \cite{Kobayashi2018helicity}.
 As a result, anisotropic magnetoresistance and magnetoresistance across thick domain walls, which are generally small in conventional metals, can become very large in MWSMs.

\subsubsection{Anisotropic magnetoresistance} \label{sec:transport:LOL}
 The less-overlapping mechanism induces an anisotropic magnetoresistance effect~\cite{Ominato17anisotropic}.
 This effect typically arises in junctions between a MWSM and a DSM, or between two MWSMs whose Weyl points are located at different positions.
 The transmission probability through the junction strongly depends on the splitting of the Weyl points along the direction perpendicular to the current.

\begin{figure}[tbp]
 \centering
  \includegraphics[width=0.98\linewidth]{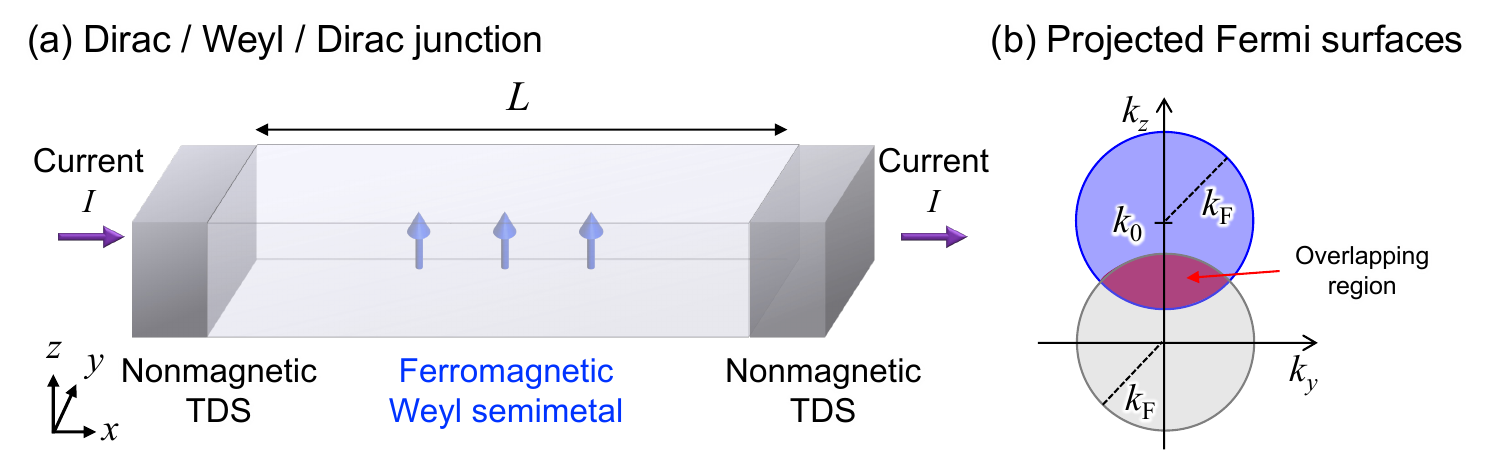}
 \vspace{-3mm}
\caption{
  (a) Schematic of the DSM/MWSM/DSM junction.
  A MWSM of length $L$ in the $x$ direction is sandwiched between semi-infinite DSM electrodes.
  The system sizes in the $y$ and $z$ directions are infinite.
  (b) Projected Fermi surfaces of the MWSM (blue) and DSMs (gray).
  The radius of the Fermi surfaces is $k\_F$.
}
 \label{fig:transport:LOL:DWD}
\end{figure}

 We consider transport through a junction of DSM/MWSM/DSM,
as illustrated in Fig.~\ref{fig:transport:LOL:DWD}(a).
 We take the transport direction to be the $x$ axis, and the boundaries between the MWSM and the DSMs are assumed to be normal to the $x$ axis.
 In MWSM, we consider the ferromagnetic moment pointing in the $z$ direction.
 Thus, the separation of the Weyl points is along the $z$ axis in momentum space.
 The length of the MWSM region is $L$, and
we neglect finite-size effects in the $y$ and $z$ directions.

 For analytical calculations, we adopt a two-band low-energy effective model for MWSMs
and focus on the positive chirality part,
\begin{align}
 H(\bm{k}) = + \bm{k} \cdot \bm{\sigma} - k_0 \sigma_z = 
 \begin{pmatrix}
  k_z - k_0 & k_x - i k_y \\
  k_x + i k_y & -k_z + k_0
 \end{pmatrix}
 \,,
\end{align}
where the Weyl point is located at $(k_x,k_y,k_z)=(0,0,k_0)$.
 In DSMs, $k_0=0$.
 In this expression, $\bm{\sigma}$ does not necessarily correspond to the spin degree of freedom, and hence the following discussion applies to any representation of Weyl Hamiltonians.

 Transport in this system can be understood from the Fermi surfaces projected onto the $k_y$--$k_z$ plane
[see Fig.~\ref{fig:transport:LOL:DWD}(b)].
 Current-carrying states are labeled with transverse wavenumbers $(k_y,k_z)$ within the projected Fermi surface of the DSM,
$k_y^2 + k_z^2 < k\_F^2$,
where $k\_F$ is the Fermi wavenumber.
 The longitudinal wavenumber is then given by 
$k_x^2 = k\_F^2 - k_y^2 - k_z^2$ in the DSM,
and that in the MWSM is given by 
$k_x'^2 = k\_F^2 - k_y^2 - (k_z-k_0)^2$.

 The projected Fermi surface of the DSM is divided into \textit{overlapping} and \textit{non-overlapping} regions, depending on its intersection with the projected Fermi surface of the MWSM (see Fig.~\ref{fig:transport:LOL:DWD}).
 In the overlapping region, $k_x'^2 > 0$,
propagating modes contribute to the current.
 In contrast, in the non-overlapping region, $k_x'^2 < 0$,
the current is carried by evanescent modes with purely imaginary longitudinal wavenumber.

 By matching the wavefunctions---the eigenspinors on the Fermi surfaces of the DSM and MWSM---across the interfaces,
the transmission probability $T(k_y,k_z)$ through the junction is given by
\begin{align}
  &T(k_y,k_z)= \!
%
    \frac{1}
    { 1
     +\[\( k_x^2 - k_z k_0 \)^2 / k_x^2 k_x'^2 - 1\] \! \sin^2(k_x' L)}.
%
 \label{eqn:transport:LOL:T}
\end{align}
 In the overlapping region,
transmission probability oscillates with the length $L$ of the MWSM.
 In the non-overlapping region,
transmission probability eventually decays with increasing $L$.

\begin{figure}[htb]
 \begin{center}
  \includegraphics[width=0.95\linewidth]{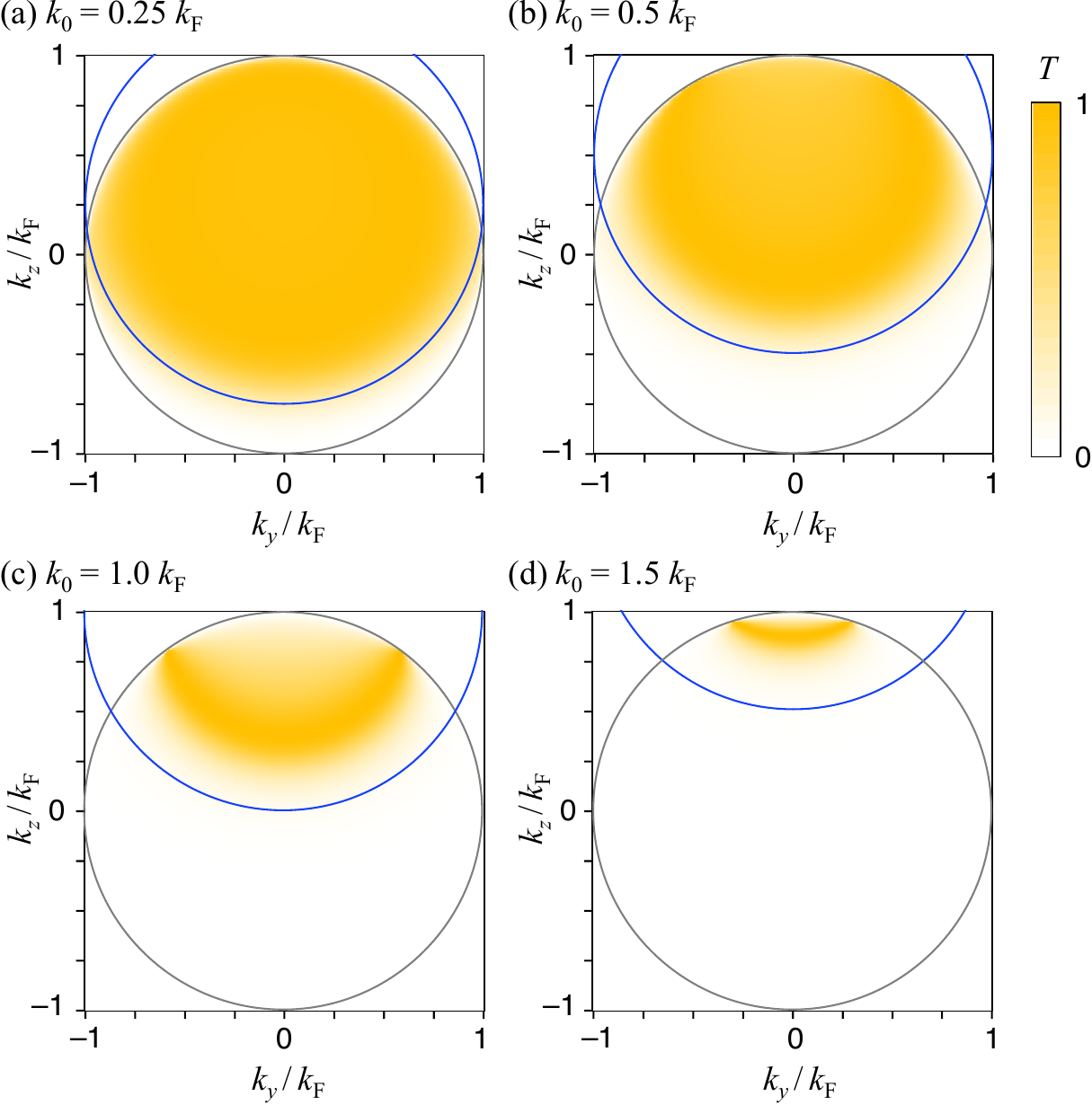}
 \end{center}
\caption{Transmission probability $T$ in the DSM/MWSM/DSM junction for
 (a) $k_0 / k\_F =0.25$,
 (b) $0.5$,
 (c) $1.0$, and
 (d) $1.5$, with $L = 4/k\_F$.
 The gray and blue circles are the projected Fermi surfaces of DSM and WSM, respectively.
}
\label{fig:transport:LOL:results}
\end{figure}

 Figure~\ref{fig:transport:LOL:results} shows the transmission probabilities as functions of $(k_y,k_z)$, with varying the separation of Weyl points $k_0/k\_F$.
 The region where transmission occurs decreases as the overlapping region becomes smaller.
 This behavior characterizes the \textit{less-overlapping} mechanism.
 Due to this mechanism, transport in MWSMs can exhibit a pronounced anisotropy with respect to the separation direction of Weyl points.
 In MWSMs where the positions of the Weyl points change depending on the magnitude and direction of magnetization, this mechanism leads to a magnetoresistance effect.

 Furthermore, for a large separation $k_0/k\_F \gtrsim 1$, the transmission probability $T$ is significantly suppressed even within the overlapping region.
 The states on the Fermi surface of a MWSM are eigenstates of the helicity operator, 
$ \bm{\sigma}\cdot(\bm{k}-\bm{k}_0) \over |\bm{k}-\bm{k}_0| $,
with eigenvalue $+1$ ($-1$) for the Weyl node of the positive (negative) chirality.
 Therefore, large displacements of Fermi surfaces lead to \textit{helicity mismatch}, resulting in the reflection of electron waves.

\subsubsection{Magnetoresistance through domain walls}\label{sec:transport:DWresistance}
 In conventional ferromagnetic metals, domain wall magnetoresistance---the resistance for an electric current passing through a magnetic domain wall---is observed~\cite{Kent2001}.
 However, its magnitude is typically several orders of magnitude smaller than the magnetoresistance observed in the absence of domain walls, such as TMR.
 In contrast,
the domain wall magnetoresistance can be huge in MWSMs in which the position of the Weyl point changes simultaneously with the change of the magnetization direction \cite{Kobayashi2018helicity}.
 This huge magnetoresistance arises from the mechanisms characteristic to the MWSMs: less-overlapping and helicity mismatch.
%
 This domain wall magnetoresistance effect is insensitive to the detail, i.e., the type of the domain wall or its thickness, of the junction and not suppressed even in the presence of disorder,
in contrast to conventional magnetoresistance effects such as magnetic tunnel junctions.

 Specifically, we consider the transport through a domain wall between two MWSMs with relative magnetization angle $\theta\_{DW}$.
 For numerical simulations,
we employ a Wilson--Dirac type model 
with N{\'e}el type magnetic domain wall.
 We assume that the direction of splitting of the Weyl points is parallel to the direction of the magnetization $\bm{M}$.
 (In other words, the $J_0$ term is dominant in the exchange coupling.
 See Appendix~\ref{sec:app:lattice:WD:exchange} for detail).
 Figure~\ref{fig:transport:helicitymismatch}(a) shows a schematic relation between the real space magnetization texture and the Weyl points configuration in momentum space.
%
%
 The two-terminal conductance along the $x$ direction is calculated via the transfer matrix method \cite{Pendry1992universality, Slevin2001numerical},
assuming the ideal metallic leads
\cite{Kobayashi2013disordered, Kobayashi2025backsolution}.

 The calculated conductance is plotted in Fig.~\ref{fig:transport:helicitymismatch}(b) 
with varying $\theta\_{DW}$ 
and the Fermi energy $E\_F$.
%
 When the relative angle of the magnetizations increases from $\theta\_{DW}=0$ to $\theta\_{DW}= \pi/2$,
the conductance in the MWSM region ($E\_F \lesssim 1$) rapidly decreases due to the less overlapping mechanism.
 When the angle $\theta\_{DW}$ further increases 
to $\theta\_{DW}=\pi$, the area of the overlapping region again increases.
 However, the transport remain supressed even 
when the Fermi surfaces are completely overlapped.
 This suppression originates from the \textit{helicity mismatch} mechanism, which is robust even for a thick domain wall.


\begin{figure}[htbp]
 \centering
 \vspace{-7mm}
\caption{
  (a) N\'{e}el type domain wall structure.
  The relative magnetization angle between domains is $\theta\_{DW}$.
  (b) The conductance $G$ \cite{Kobayashi2018helicity} as a function of $\theta\_{DW}$ and the Fermi energy $E\_F$.
  A huge magnetoresistance arises in the blue region, which is in good agreement with the energy range where Weyl semimetal arises.
  For the detailed setting, see Ref.~\onlinecite{Kobayashi2018helicity}.
}
\label{fig:transport:helicitymismatch}
\end{figure}

\subsubsection{Domain wall resistance in finite sized MWSMs} \label{sec:transport:DW_CSS}
 So far, we have focused on magnetoresistance arising from bulk Weyl cones in MWSMs.
 However, in finite-sized systems, surface states originating from Fermi arcs can significantly affect transport and may even dominate it, especially when bulk conduction is suppressed.
 Such surface contributions become particularly important in the presence of magnetic inhomogeneities, such as domain walls.
 For instance, in certain systems such as single-layer kagome materials, edge transport plays an important role in the domain wall resistance~\cite{Kobayashi19robust}.
 Surface (edge) transport in topological systems has been extensively studied, particularly in magnetic topological insulators, where it plays a central role in transport phenomena~\cite{tokura2019magnetic}.
 In contrast, surface transport in MWSMs remains relatively unexplored, despite its potential importance due to the coexistence of bulk Weyl states and Fermi arc surface states.

\begin{figure}[tbp]
 \centering
  \includegraphics[width=\linewidth]{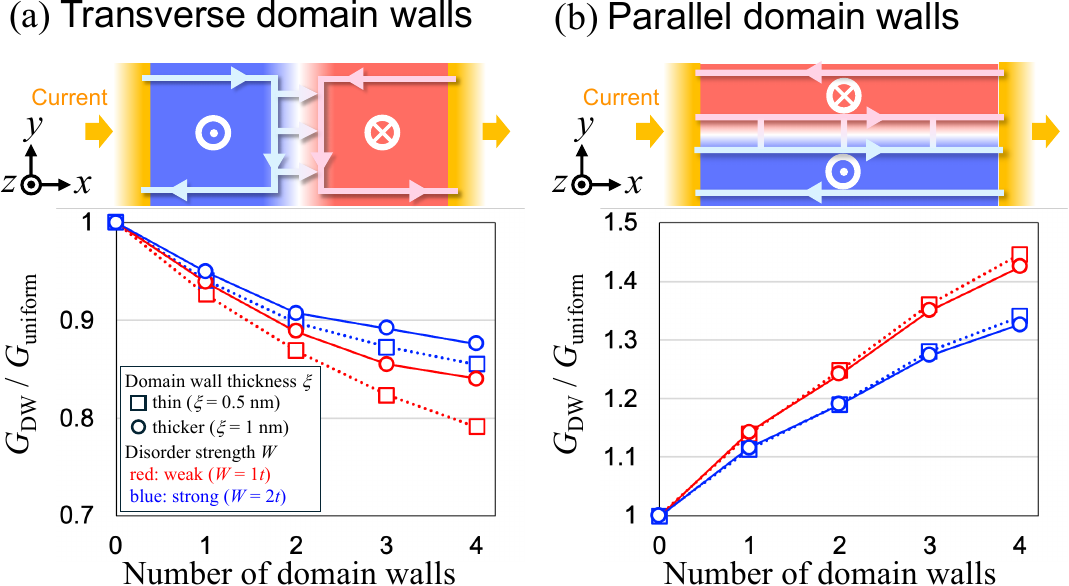}
 \vspace{-7mm}
\caption{
   $G\_{DW}/G\_{uniform}$ in an effective tight-binding model of \CSS as a function of the number of domain walls, for
  domain walls that are
  (a) transverse and
  (b) paralel to the current.
  Squares and circles represent thinner and thicker domain walls, respectively.
  Symbol colors correspond to weak (red) and strong (blue) disorder.
  The system size is $40.3$ nm $\times$ $23.2$ nm $\times$ $5.27$ nm.
  The statistical error bars are smaller than symbol sizes.
}
\label{fig:transport:DW_CSS}
\end{figure}

 Motivated by this situation, we investigate the transport properties of MWSMs with both surface states and magnetic domain walls.
 As a concrete example, we employ an effective model~\cite{ozawa2024effective} for the kagome layered MWSM \ce{Co3Sn2S2} (see also Appendix \ref{sec:app:lattice:CSS}).
 Figure~\ref{fig:transport:DW_CSS} shows the numerically simulated change of the two-terminal conductance for varying the number of domain walls.

 When the domain walls are transverse to the current direction, bulk transport is strongly suppressed, as discussed in the previous section. 
 Consequently, surface transport becomes dominant. 
 In this configuration, chiral surface states emerge along the domain walls,
since domains with opposite magnetization have opposite topological (Chern) numbers.
 If the magnetization changes sharply between two domains (e.g., due to a thin non-magnetic layer), a large magnetoresistance is expected.
 This is because the surface states in adjacent domains have opposite spin polarizations, leading to strong reflection at the domain wall. 
 In contrast, for thick domain walls, spin mixing allows transmission across the wall more easily.
 Therefore, as shown in Fig.~\ref{fig:transport:DW_CSS}(a), the conductance decreases as the number of domain walls increases.

 On the other hand, when the domain walls are parallel to the current direction, the transport behavior is qualitatively different:
the conductance increases with increasing number of domain walls [Fig.~\ref{fig:transport:DW_CSS}(b)].
 In this configuration, each domain wall introduces additional conducting channels running along it, thereby enhancing the total conductance.
 In contrast to the transverse case, the conductance is insensitive to the domain wall thickness, because mixing between parallel channels plays only a minor role.
 These features are robust against disorder, as evidenced by the weak dependence on disorder strength in both configurations.
 This robustness highlights the topological nature of the underlying transport mechanisms.

\subsection{Suppression of extrinsic Hall effect} \label{sec:transport:extrinsic}

 The anomalous Hall effect \cite{Nagaosa2010, xiao2010berry} is one of the iconic transport properties of MWSMs
as we have reviewed in Secs.~\ref{sec:fundamentals} and \ref{sec:materials}.
 The anomalous Hall effect is divided into two contributions: intrinsic and extrinsic.
 The extrinsic anomalous Hall effect is attributed to impurity scattering,
whereas the intrinsic anomalous Hall effect originates from the Berry curvature and can occur even in the clean limit.
 The extrinsic anomalous Hall effect can be distinguished by the relation between the Hall and longitudinal conductances~\cite{Nagaosa2010};
the impurity-induced (skew scattering) Hall conductance is proportional to the longitudinal conductance, $G\_H^{\rm ext} \propto G_{xx}$.
 In contrast, the intrinsic anomalous Hall effect in MWSMs exhibits a non-monotonic yet characteristic relationship \cite{Kobayashi2021intrinsic}.
 Similar unique relationships between Hall and longitudinal conductances are also found in other topological or Dirac systems, such as quantum Hall insulators \cite{Aoki87quantized,Huckestein95scaling} and the valley Hall conductivity in graphene \cite{Ando15theory}.



%
 In WSMs, 
the intrinsic and extrinsic contributions have been predicted to show peculiar behaviors distinct from normal metals.
 The intrinsic contribution $G\_H^{\mathrm{int}}$ shows a non-monotonic but characteristic behavior with respect to $G_{xx}$ \cite{Kobayashi2021intrinsic}.
 Such a behavior is also known in quantum Hall states \cite{Aoki87quantized,Huckestein95scaling} and in graphene~\cite{Ando15theory}.
 On the other hand, the extrinsic contribution is found to be much smaller than the intrinsic one
based on an analytical calculation in a low-energy effective model \cite{burkov2014anomalous} and numerical transport calculations in a finite-size system~\cite{Kobayashi2021intrinsic}.
 This is a sharp contrast to ordinary metals where the extrinsic contribution is as large as the intrinsic one.

\begin{figure}[htbp]
 \centering
  \includegraphics[width=\linewidth]{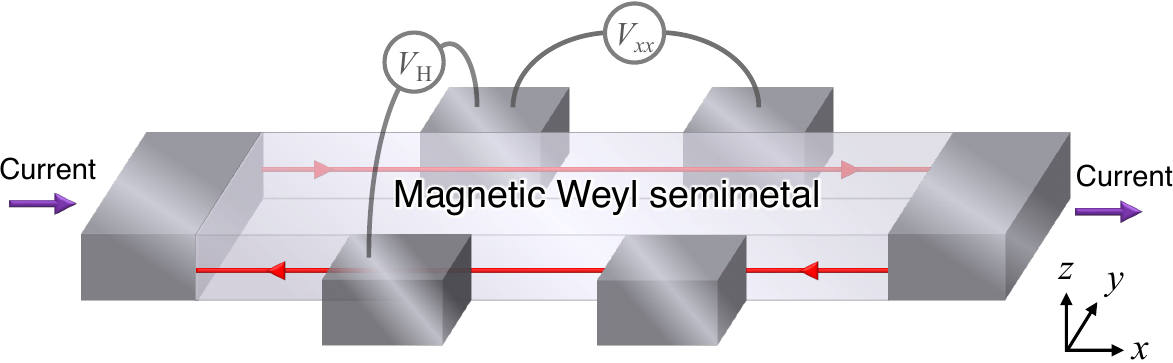}
\caption{
  The 6-terminal Hall bar geometry.
  Voltage probes are attached on the side surfaces, where chiral surface states emerge.
  The periodic boundary conditions are imposed in the $z$ direction, for simplicity.
}
\label{fig:transport:Hallbar}
\end{figure}

 Here, we review the numerical calculation showing the suppression of the extrinsic contribution.
 We consider a half-metallic MWSM realized by introducing a strong exchange interaction into a topological Dirac semimetal (see Appendix~\ref{sec:app:lattice:TDSM}).
 We assume a short-range impurity potential $U$,
randomly distributed on the lattice sites with a certain density and a fixed strength.
 We also consider spin-orbit-coupled scattering from the impurity potential, taking the form $\({\bf S} \times {\bf p}\) \cdot \nabla U$ \cite{Nikolic07extrinsically}.
%
 In a 6-terminal Hall bar geometry (Fig.~\ref{fig:transport:Hallbar}),
the Hall and longitudinal conductances are defined as,
\begin{align}
 G_{xx} =  \frac{V_{xx}}{V_{xx}^2 + V\_{H}^2},\quad
 G\_{H} =  \frac{V\_{H}}{V_{xx}^2 + V\_{H}^2},
\end{align}
where the voltages $V_{xx}$ and $V\_{H}$ are numerically calculated using the recursive Green's function method \cite{datta1997electronic,Datta2005quantum,Takane16disorder} via the Landauer-B\"{u}ttiker formula.

 In the calculated Hall conductance,
the intrinsic contribution is insensitive to the spin-orbit-coupled scattering by impurities.
 Thus, the extrinsic contribution is identified as the difference between the Hall conductances calculated with and without the spin-orbit-coupled scattering.
 The estimated extrinsic contribution to the Hall conductance $G\_H^{\mathrm{ext}}(E)$ is shown against the Fermi energy $E$ in Fig.~\ref{fig:transport:extrinsic}(a),
together with the longitudinal conductance averaged with and without spin-orbit-coupled scattering $G_{xx}^{\mathrm{ave}}(E)$.
 The suppression of the extrinsic effect in MWSMs becomes clear by plotting $G\_H^{\mathrm{ext}}(E)$ against $G_{xx}^{\mathrm{ave}}(E)$ for different energy ranges,
as shown in Fig.~\ref{fig:transport:extrinsic}(b).
 Although the extrinsic Hall conductance $G\_H^{\mathrm{ext}}$ in the WSM region $(E/t \approx 1)$ is finite and slightly increasing with increasing longitudinal conductance $G_{xx}^{\mathrm{ave}}$, 
its Hall angle 
$G\_H^{\mathrm{ext}}/G_{xx}^{\mathrm{ave}}$
is significantly smaller than that in the metallic region $(E/t \lesssim -4)$.
 This shows that the extrinsic contribution is negligible in the MWSM region and hence the intrinsic contribution is dominant even in finite size systems.

\begin{figure}[tbp]
 \centering
\caption{
  (a) Longitudinal conductance $G_{xx}^{\mathrm{ave}}(E)$ (blue line) and extrinsic contribution of Hall conductance $G\_H^{\mathrm{ext}}(E)$ (red line).
  (b) Relations between conductances for the Weyl semimetal region (green circle)
  and for the half-metallic region (orange square).
}
\label{fig:transport:extrinsic}
\end{figure}



\section{Spin transport in Dirac/Weyl electron systems}
\label{sec:spin-transport}


Besides the electric charge transport properties,
the transport of electron spins in materials is also attracting a significant attention \cite{Maekawa2017}.
The electrons in ferromagnetic metal are spin polarized,
and hence an electric current therein carries spin angular momentum polarized along the magnetization direction.
On the other hand, although the electrons in nonmagnetic materials have no spin polarization,
they can still carry spin angular momentum, termed a spin current.
A spin current consists of two counterpropagating currents with opposite spin polarizations,
which does not carry a net charge current.
Spin current is responsible for the transfer of spin angular momenta among magnets and for exerting spin torques,
and hence the spin transport properties are intensely studied in various materials to make use of them in future spintronics devices.

This section is devoted to the review on recent theoretical and experimental studies on spin transport properties in magnetic Weyl semimetal materials.
We mainly focus on the spin Hall effect (SHE),
i.e., the transverse response of a spin current to an electric field,
in which the contribution of band topology is essential.
We first give a general overview on the SHE,
and then proceed to the studies on the SHE and some other spin transport properties in magnetic Weyl semimetals.

\subsection{Overview of spin Hall effect}

The SHE is intensely studied in spintronics for the electrical manipulation of magnetization by spin current \cite{sinova2015spin}.
The inverse SHE, i.e., the conversion from a spin current into a transverse electric current or voltage,
is equally essential for the electrical readout of spin current injected from other materials~\cite{Saitoh2006}.
The SHE does not demand the breaking of time-reversal or inversion symmetries,
and can occur even in nonmagnetic materials.
Heavy metals with strong SOC, such as Pt~\cite{Guo2008}, Ta~\cite{Liu2012-pw}, and W~\cite{pai2012spin}
are often used in devices to implement the SHE.

The SHE is characterized by the spin Hall conductivity tensor $\sigma_{\mu\lambda}^{s_\alpha}$,
which relates the applied electric field $E_\lambda$ to the resulting spin current $j^{s_\alpha}_\mu$ as $j^{s_\alpha}_\mu = \sigma_{\mu\lambda}^{s_\alpha} E_\lambda$.
Here $\mu$ and $\lambda$ label the spatial directions, and $\alpha$ denotes the direction of spin polarization for the spin current.
The spin Hall angle $\theta_{\mu\lambda}^{s_\alpha} = (2e/\hbar)(j^{s_\alpha}_\mu / j_\lambda)$,
defined as the dimensionless ratio between the input electric current $j_\lambda$ and the output spin current $j^{s_\alpha}_\mu$,
is also a commonly used quantity.
In spatially isotropic systems such as 
Pt, Ta, and W,
$\sigma_{\mu\lambda}^{s_\alpha}$ and $\theta_{\mu\lambda}^{s_\alpha}$ become totally antisymmetric tensors,
i.e.,
$\sigma_{\mu\lambda}^{s_\alpha} = \epsilon_{\alpha\mu\lambda} \sigma_{\rm SH}$ and $\theta_{\mu\lambda}^{s_\alpha} = \epsilon_{\alpha\mu\lambda} \theta_{\rm SH}$.
In such heavy metals,
the spin Hall angle $\theta_{\rm SH}$ is of the order of 0.01--0.1.
Its value is rather small, because it requires a large input current due to the metallic conduction.

Similar to the AHE,
the mechanisms for the SHE are also classified into the two contributions.
One is the extrinsic contribution from the scattering by spin-orbit coupled disorders \cite{dyakonov1971possibility}.
The other is the intrinsic contribution from the band topology~\cite{murakami2003dissipationless,murakami2004spin}.
With the conventional definition of the spin current operator $j_\mu^{s_\alpha} = \tfrac{1}{2} \{ s_\alpha, v_\mu \}$,
with the spin operator $s_\alpha$ and the velocity operator $v_\mu = \partial_{k_\mu} \mathcal{H} / \hbar$ $(\alpha,\mu = x,y,z)$,
the intrinsic contribution to the spin Hall conductivity is given from the Kubo formula as,
\begin{align}
    \sigma_{\mu\lambda}^{s_\alpha{\rm (int)}} &= e\hbar \int \frac{d^d{\bm k}}{(2\pi)^d} \sum_{n} f_n(\boldsymbol{k}) b_{n,\mu\lambda}^{s_\alpha}(\boldsymbol{k}).
\end{align}
Here,
\begin{align}
    b_{n,\mu\lambda}^{s_\alpha}(\boldsymbol{k}) &= -2 {\rm Im} \sum_{m (\neq n)} \frac{ \langle u_n | j_\mu^{s_\alpha} | u_m \rangle \langle u_m | v_\lambda | u_n \rangle }{(E_n - E_m)^2}
\end{align}
is called the ``spin Berry curvature'',
which has a structure similar to the Berry curvature and reflects the band topology in connection with SOC.
(Note that the above definition of $j_\mu^{s_\alpha}$ is exact only if $s_\alpha$ is a conserved quantity.
Otherwise, it cannot be defined as a Noether's current,
and the above definition may be considered as a phenomenological form subject to spin relaxation effect~\cite{Shi2006}.)

The intrinsic SHE was well considered for the 2D topological insulators in their early-stage studies,
where the 2D bulk band topology is associated with the quantized spin Hall conductivity and the helical edge states
\cite{murakami2004spin,Kane2005,bernevig2006quantum,wu2006helical}.
Among the 3D electron systems accompanying the intrinsic SHE,
one typical case is the topological Dirac semimetal introduced in Sec.~\ref{sec:lattice},
which is characterized by the pair of Dirac points with spin degeneracy, protected by crystalline symmetries~\cite{wang2012dirac,wang2013three}.
Similarly to the case of the AHE and Berry curvature in magnetic Weyl semimetals,
the spin-degenerate Dirac points serve as source and sink of the spin Berry curvature,
and hence a contribution to the intrinsic SHE is predicted in the topological Dirac semimetal \cite{burkov2016z,taguchi2020spin}.
Like the anomalous Hall conductivity following Eq.~(\ref{eq:AHE-K0}) in Weyl semimetals,
the spin Hall conductivity in topological Dirac semimetals is estimated to be propotional to the momentum-space spacing of Dirac points.
The idea of spin Berry curvature is applied also to magnetic Weyl semimetals to study the SHE,
which we shall review in the following parts.

\begin{figure}[htb]
 \centering
\caption{
Energy dependence of the (a)~anomalous Hall conductivity in the ferromagnetic state and 
(b)~spin Hall conductivity in the paramagnetic state.
Adapted from Ref.~\cite{Lau2023}
}
\label{fig:she_css}
\end{figure}

\subsection{Spin transport in magnetic Dirac/Weyl semimetals}
\label{sec:spintransport}


The enhanced spin Hall conductivity from the spin Berry curvature is experimentally observed in the Dirac semimetal state.
In \CSS, for instance,
it was reported that the spin Hall effect in the paramagnetic Dirac semimetal state is related to the anomalous Hall effect in the ferromagnetic Weyl semimetal state \cite{Lau2023}.
As explained in Sec.~\ref{sec:co3sn2s2},
\CSS behaves as a Weyl semimetal and shows a giant anomalous Hall effect in the ferromagnetic phase.
Figure~\ref{fig:she_css}(a) shows the energy dependence of the anomalous Hall conductivity calculated by the effective tight-binding model,
which shows the maximum near the energy of the Weyl points.
On the other hand,
beyond the Curie temperature where the magnetic order vanishes, the Weyl points merge into doubly degenerate Dirac points to form a non-magnetic Dirac semimetal phase.
In this non-magnetic state, a strong SHE emerges from the spin Berry curvature associated with the Dirac points, as shown in Fig.~\ref{fig:she_css}(b).
Indeed, the experimentally observed spin Hall conductivity at room temperature showed the maximum under a certain ratio of Ni doping in the \CSS film,
with the measurement of spin-torque ferromagnetic resonance.
This is attributed to the doping-induced shift of Fermi level to the Dirac point,
which enhances the contribution from the spin Berry curvature to the spin Hall conductivity.

The studies on the SHE is now being extended to the magnetic Weyl semimetal states,
to make use of them as an efficient source of spin current.
In a \CSS-based device,
it was experimentally observed that the spin-charge conversion efficiency gets enhanced with the development of ferromagnetic ordering at low temperatures,
which is attributed to the spin current from the SHE and the highly spin-polarized current from the AHE~\cite{seki2023enhancement}.
Moreover,
it was also found by the tight-binding model calculation that the spin Hall conductivity in \CSS develops the components that do not belong to the totally antisymmetric tensor, $\sigma_{\mu\lambda}^{s_\alpha} = \epsilon_{\alpha\mu\lambda} \sigma_{\rm SH}$.
Such components emerge from the violation of spatial isotropy and shows a notable dependence on the magnetic field direction \cite{ozawa2024effective}.

In the ferromagnetic Weyl metal~\ce{SrRuO3},
it was reported by the magnetization switching experiment that the lattice strain from the \ce{SrTiO3} substrate leads to the enhancement of SHE \cite{horiuchi2025single}.
With the theoretical model calculation,
the origin of such an enhancement of SHE was found to be the hot spots of spin Berry curvature emerging from the band inversion due to the lattice strain.

In the antiferromagnetic \ce{Mn3Sn},
the direct and inverse SHEs were experimentally measured,
where the SHE showed the component that is odd in the magnetic field direction~\cite{kimata2019magnetic}.
Such a behavior is in a sharp contrast to the conventional SHE, which is even under time-reversal,
and is explicitly called the magnetic SHE.
The theoretical understanding on the origin of such magnetic SHE,
in connection with the band topology of Weyl electrons,
is not well established and is left for future studies.

Besides the SHE reviewed so far,
the spin Nernst effect,
i.e., the transverse response of spin current $\boldsymbol{j}^s$ to a temperature gradient ${\bm \nabla}T$,
is also attracting an interest in topological materials.
Similar to the Berry curvature contribution in the intrinsic anomalous Nernst effect [Eq.~(\ref{eq:anomalous-nernst})],
the intrinsic contribution to the spin Nernst effect is described in terms of the spin Berry curvature \cite{fujimoto2024microscopic}.
Theoretical calculations with the effective models of topological Dirac semimetal and magnetic Weyl semimetal demonstrated the enhancement of the spin Nernst effect due to the contribution of spin Berry curvature,
with a sharp dependence on the energy of Dirac or Weyl points relative to the Fermi level
\cite{matsushita2025intrinsic}.
These calculations suggest that the tuning of the energy levels of Dirac or Weyl points,
by chemical substitution or modulation of magnetization,
is essential in achieving a large spin Nernst effect.
With such an enhancement of spin Nernst effect,
the Dirac and Weyl semimetals may provide an efficient control of spins by heat,
which is an important objective in the rapidly growing research field of \textit{spin caloritronics} \cite{bauer2012spin}.



\section{Summary and perspective}
\label{sec:conclusion}

In this review, we have surveyed the recent progress achieved in the theoretical and experimental investigation of magnetic Weyl semimetals.
The central defining feature of magnetic Weyl semimetals lies in the intimate connection between their non-trivial band topology and magnetic ordering.
This profound connection serves as the origin for diverse emergent phenomena detailed throughout this article.
The most prominent manifestation of this link is the magnetization-dependent electron transport driven by the Berry curvature,
emerging notably as the intrinsic anomalous Hall and Nernst effects.
Conversely, the magnetic properties are subject to the Weyl electrons via spin-orbit coupling,
leading to the emergence of various uniform and non-uniform magnetic orderings, magnetic anisotropy, the Dzyaloshinskii–Moriya interaction,
and anomalous spin-wave excitations.
Furthermore, the interaction between Weyl electrons and non-uniform magnetic textures enables their mutual manipulation,
resulting in phenomena such as spinmotive force and current-induced spin torques,
which are conceptually described by the chiral gauge fields for Weyl fermions.
At mesoscopic scales, magneto-transport properties are also significantly affected by the magnetization-dependent bulk Weyl-point structure and the corresponding surface Fermi arc states.
It has been highly active to realize magnetic Weyl semimetal materials exhibiting these multifunctional properties,
with significant experimental progress demonstrated in both ferromagnetic and antiferromagnetic systems.

From a fundamental perspective,
deeper material-specific analysis is crucial for fully comprehending the electronic influence of Weyl fermions.
Regarding static magnetic properties, novel magnetic properties originating from Weyl fermions, such as magnetic anisotropy and formation of magnetic textures, remain to be explored. 
Non-equilibrium dynamical processes also form a frontier of research in magnetic Weyl semimetals,
such as 
current-induced switching processes,
collective spin-wave excitations,
and magneto-optical responses.
In particular, spin dynamics in antiferromagnets and compensated ferrimagnets coupled with Weyl fermions would be an important target.
This is because they are supposed to be attributed to the axion electrodynamics peculiar to the relativistic quantum field theory~\cite{Sekine2020-km},
which will provide a testbed for linking condensed matter phenomena to high-energy physics concepts.
To explore the above-mentioned properties and phenomena,
theoretical methodologies such as 
Ginzburg--Landau theory based on the path-integral formalism~\cite{altland2010condensed},
non-equilibrium Green's functions~\cite{rammer1986quantum}, 
and Floquet theory~\cite{Oka2019},
will be indispensable.

Furthermore, the strong combination of topology and magnetism in magnetic Weyl semimetals poses them as a highly promising platform for realizing next-generation spintronics devices.
Their topologically enhanced magnetoelectric properties will offer pathways toward ultra-low-power consumption and high-speed operation,
which can help overcome the major obstacles in conventional spintronics~\cite{Maekawa2017}.
For instance, we have seen the large magnetoresistance arising from the Weyl point structure,
which will potentially advance the integration of magnetic random access memories~(MRAMs).
The current-induced spin torques leveraged by the band topology of Weyl fermions
would be suited for improving the operation speed and power efficiency of MRAMs and spin torque oscillators.
To translate these fundamental properties into functional devices,
material design and verification of topological magnetoelectric properties are becoming increasingly essential from both theoretical and experimental perspectives.

\acknowledgments

The authors would like to appreciate 
G.~Bauer,  
K.~Fujiwara,  
S.~Fukami,
M.~Hayashi,
J.~Han,
H.~Horiuchi,
H.~Ishizuka,  
J.~Ieda,  
T.~Koretsune,
D.~Kurebayashi,
Y.C.~Lau,
T.~Meguro,  
T.~Misawa,  
Y.~Motome,  
R.~Nakai,  
K.~Nakazawa, 
S.~Ohya,
S.~Okumura,
Y.~Ominato,  
T.~Morimoto,  
T.~Ohtsuki,  
T.~Oka,  
T.~Sato,
T.~Seki,
R.~Shimano, 
A.~Tsukazaki,
Y.~Yamane,
and M.~Yamanouchi
for fruitful collaborations and valuable discussions.
The authors are supported by JSPS KAKENHI, 
Grants 
No. JP20H01830 (K.N.), 
No. JP22K03446 (K.K.), 
No. JP22K03538 (Y.A.),
No. JP22H05114 (K.K.), 
No. JP23K19194 (A.O.), and 
No. JP25H01250 (K.N.), 
and by JST CREST, Grant No. JPMJCR19T3 (A.O.).

\appendix
\appendix
\section{Lattice models for magnetic Weyl semimetal}
\label{sec:app:lattice}
 In this appendix, we explain the mathematical structure of the lattice models for MWSMs.
 Effective models describing realistic MWSM materials can be complex,
as with models derived from first-principles calculations.
 Nevertheless, the essential physics near Weyl points can be captured within a four-band model based on the Wilson--Dirac model introduced in Sec.~\ref{sec:lattice}.
 In addition, in some simple cases, the model can be reduced to a minimal two-band model.
 Accordingly, we focus on two- and four-band models describing MWSMs.
 We introduce both momentum-space and real-space representations.
 While the momentum-space representation is convenient for describing bulk properties and topological invariants, it assumes translational symmetry and cannot capture spatial inhomogeneities.
 In contrast, the real-space formulation allows us to incorporate non-uniform magnetic textures (e.g., domain walls), local disorder potentials, and boundaries, 
which are essential for studying realistic systems and numerical simulations.

 Throughout this appendix, unless otherwise specified, we set the lattice constant as the length unit, $a=1$, so that the wave vectors $\bm{k}$ are dimensionless.

\subsection{Two-band model} \label{sec:two-band}
\subsubsection{Momentum-space representation}
 To describe a Weyl semimetal state with broken time-reversal symmetry,
one minimally requires two bands.
 On a cubic lattice,
the two-band model is described in momentum space as
\begin{align}
 H\_{tb}(\bm{k}) &= \sum_{\mu=x,y} (t_\mu \sin{k_\mu}) \tau_\mu  + m(\bm{k}) \tau_z  + c(\bm{k}) \tau_0 , \label{eqn:app:lattice:H2x2_k}\\
 m(\bm{k}) &= m_0 +\!\! \sum_{\mu=x,y,z} m_{2\mu} \( 1 - \cos{k_\mu} \) , \\ 
 c(\bm{k}) &= c_0 + \sum_{\mu=x,y,z} c_\mu \cos{k_\mu} .
\end{align}
 Here, $\tau_{\mu=x,y,z}$ are the Pauli matrices,
which do not necessarily correspond to the real spin degrees of freedom but could represent other degrees of freedom, such as orbitals or sublattices.
 The parameters $t_{\mu}$ and $m_{2\mu}$ account for the hoppings between the nearest-neighbor sites.
 $m_0$ controls the energy gap or inversion of two bands at $\bm{k}=(0,0,0)$, and is thus referred to as the \textit{mass} parameter.
 The band inversion within the Brillouin zone is controlled by $m(\bm{k})$, and the term involving $m_{2\mu}$ is called the \textit{Wilson (mass)} term.
 The $c(\bm{k})$ term causes energy modulation but does not affect the topological properties.

\begin{figure}[tbp]
 \centering
  \includegraphics[width=0.98\linewidth]{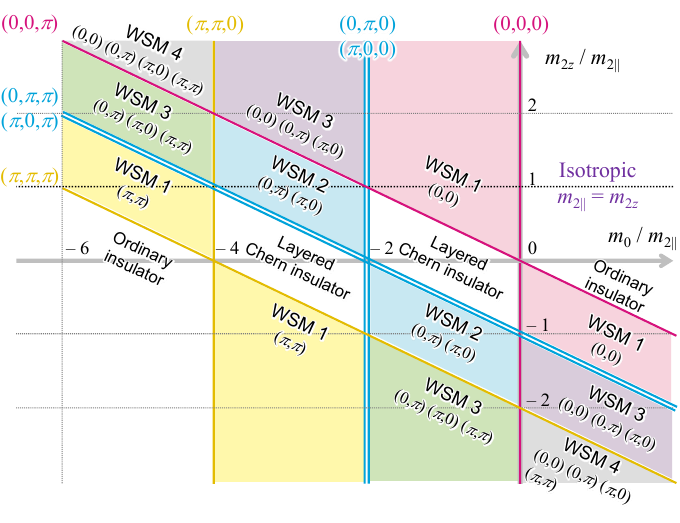}
  \vspace{-3mm}
\caption{
   The phase diagram of $2\times 2$ model
  in the $m_0$--$m_{2z}$ parameter space.
   The parameters are scaled by $m_{2||}=m_{2x}=m_{2y}$.
   The WSM phase with $n$ pairs of Weyl points are denoted as ``WSM $n$''
  with the positions of Weyl points projected on the $(k_x,k_y)$ plane.
   The colored solid lines are the topological phase boundary with gap closing points $(k_x, k_y, k_z)$.
}
\label{fig:app:lattice:PD2x2}
\end{figure}

 This model can exhibit zero to four pairs of Weyl points~\cite{Yoshimura2016comparative} split along the $z$ axis, depending on the mass parameters $m_0$ and $m_{2\mu}$.
 The phase without Weyl points can be either a trivial or a layered Chern insulator phase.
 The position of Weyl points can be identified by the band-touching points.
 The energy eigenvalues of the model are given by
\begin{align}
 E\_{tb}^{\pm}(\bm{k}) &= c^2(\bm{k}) \pm \sqrt{
  \sum_{\mu=x,y} t_\mu^2 \sin^2{k_\mu}
  + m^2(\bm{k})
 }.
\end{align}
 For the two bands to be degenerate, $(k_x,k_y)$ should be $(0,0), (0,\pi), (\pi,0)$, or $(\pi,\pi)$.
 For each $(k_x,k_y)$, a pair of Weyl points arises at $k_z = \pm k_0^{(k_x,k_y)}$ if $k_0^{(k_x,k_y)}$ given by
\begin{align}
 k_0^{(k_x,k_y)} &= \arccos \( \frac{m^{(k_x,k_y)}}{m_{2z}} \)
 \label{eqn:app:lattice:k0kxky} \\
 m^{(k_x,k_y)} &= m_0 + 2 \delta_{k_x,\pi} m_{2x} + 2 \delta_{k_y,\pi} m_{2y} + m_{2z} ,
 \label{eqn:app:lattice:mkxky}
\end{align}
has a real solution.
 This condition is satisfied for $|m^{(k_x,k_y)}/m_{2z}| \leq 1$.
 Note that at the boundary $|m^{(k_x,k_y)}/m_{2z}| = 1$, the band-touching point is not a Weyl point, since the dispersion becomes quadratic in the $z$ direction rather than linear.
 As a result, the number of Weyl point pairs varies according to $m_0$ and the anisotropy of $m_{2\mu}$,
as illustrated by the phase diagram in Fig.~\ref{fig:app:lattice:PD2x2}.

 For instance, if $m_{2\mu}$ are isotropic, $m_{2\mu} = m_2$, 
and $m_0$ has the value in the range $m_0 \in (-2 m_2, 0)$,
a single pair of Weyl points emerges at $(0,0,\pm k_0^{(0,0)})$, where
\begin{align}
 k_0^{(0,0)} &= \arccos \(\frac{m_0}{m_2} + 1\).
\end{align}
 The energy in the vicinity of the Weyl point can be linearized as
\begin{align}
 E\_{tb}^{\pm}(\bm{k}) &\simeq c^2(\bm{k})
  \pm \!\! \sqrt{
   \sum_{\mu=x,y} t_\mu^2 k_\mu^2
   +  \[ - m_0 \(m_0 + 2 m_2\) \] {k'_z}^2
  },
\end{align}
where $k_z = k_0 + k'_z$.
 In this case, if $k_z$ is fixed to a value between the Weyl points,
i.e., $-k_0^{(0,0)} < k_z < k_0^{(0,0)}$,
the model reduces to a 2D Chern insulator with a Chern number $\nu = 1$.

\subsubsection{Real-space representation}
 The real-space representation of the two-band model, Eq.~\eqn{app:lattice:H2x2_k}, is given by
\begin{align}
 \mathcal{H}\_{tb} = & 
  \sum_{\bm{r}} c^\dag_{\bm{r}} \[ c_0 \tau_0 + \(m_0 + \sum_{\mu=x,y,z} m_{2\mu} \) \tau_z \]c_{\bm{r}} \nonumber\\
  & + \sum_{\bm{r}}
   \[ \sum_{\mu=x,y,z} 
     c^\dag_{\bm{r}+\hat{\mu}} \( \frac{c_\mu}{2} \tau_0 - \frac{m_{2\mu}}{2} \tau_z \) c_{\bm{r}}
   \right. \nonumber\\
  & \hspace{10mm} + 
   \left.
     \sum_{\mu=x,y} c^\dag_{\bm{r}+\hat{\mu}} \frac{i t_\mu}{2} \tau_\mu c_{\bm{r}} 
    + \mathrm{H.c.} 
   \]  ,
\end{align}
where $c^{(\dag)}_{\bm{r}}$ is the two-component annihilation (creation) operator on the lattice site $\bm{r}$,
with the lattice vector $\hat{\bm{\mu}}$.
 The inhomogeneities (on-site disorder) of the system can be introduced in the form of potential disorder ($V_0 \tau_0$) or mass disorder ($V_m \tau_z$) as
\begin{align}
 & \mathcal{H}\_{V} 
  = \sum_{\bm{r}} c^\dag_{\bm{r}} \[ V_0(\bm{r}) \tau_0 + V_m(\bm{r}) \tau_z \] c_{\bm{r}}.
\end{align}
 In general, it is difficult to correctly incorporate the effects of magnetization into this two-band model. 
 Therefore, for MWSMs with magnetic textures such as domain walls, 
it is necessary to use the four-band model discussed in the following subsection.

\subsection{Wilson--Dirac type (four-band) model} \label{sec:app:lattice:WD}
\subsubsection{Momentum-space representation}
 The Wilson--Dirac type model consisting of four degrees of freedom is the most fundamental model for the Dirac and Weyl fermion systems on lattice.
 The original model was proposed in the context of lattice quantum field theory~\cite{Wilson1977} as explained in Sec.~\ref{sec:lattice}.
 In condensed matter physics, it was first used to systematically describe effective models of topological insulators~\cite{Qi08topological, liu2010model}.
 It is also used as an abstract model of topological insulators and Dirac semimetals~\cite{Ryu12disorder}.
 Although the model preserves time-reversal symmetry, one can construct the model for MWSMs by introducing time-reversal breaking (exchange coupling) terms.

 On a cubic lattice,
the Wilson--Dirac model is expressed in momentum space as
\begin{align}
 H\_{WD}(\bm{k}) &= \sum_{\mu=x,y,z} (t_\mu \sin{k_\mu}) \alpha_\mu  + m(\bm{k}) \alpha_4,
 \label{eqn:app:lattice:WD:HWDk}\\
 m(\bm{k}) &= m_0 +\!\! \sum_{\mu=x,y,z} m_{2\mu} \(1 - \cos{k_\mu} \).
\end{align}
 Here, $\alpha_{\mu=x,y,z}$ and $\alpha_4$ are $4\times 4$ Dirac matrices corresponding to those in the Dirac Hamiltonian Eq.~\eqref{eq:MassiveDirac}.
 The meanings of the parameters $t_{\mu}$, $m_{2\mu}$, and $m_0$ are the same as those in the two-band model.
 As discussed in the two-band model, a term of the form~\cite{wang2012dirac,wang2013three}
\begin{align}
 H\_{c}(\bm{k}) = \( c_0 + \sum_{\mu=x,y,z} c_\mu \cos{k_\mu} \) \bm{1}_4,
\label{eqn:app:lattice:WDHc}
\end{align}
with $\bm{1}_4$ the $4\times 4$ identity matrix,
can be added without affecting the topological properties.
 In the following, we omit this term for simplicity.

\begin{figure}[tbp]
 \centering
  \includegraphics[width=0.98\linewidth]{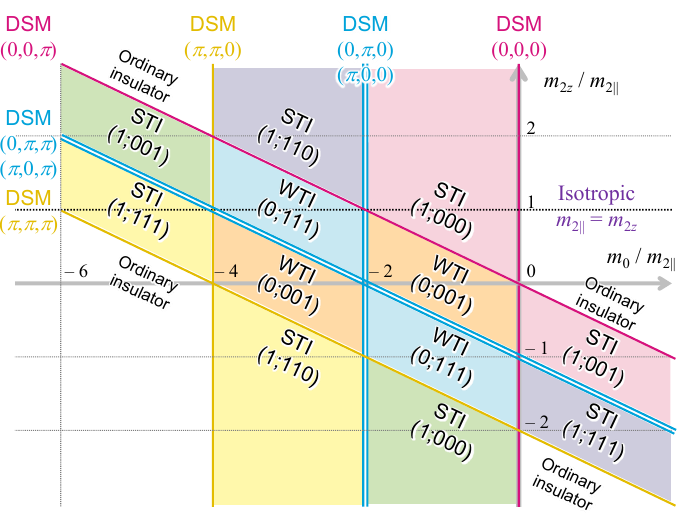}
\caption{
   The phase diagram of Wilson--Dirac model
  in the $m_0$--$m_{2z}$ parameter space.
   The parameters are scaled by $m_{2||}=m_{2x}=m_{2y}$.
   The strong and weak topological insulator phases are denoted as STI and WTI, respectively, with its topological indices $(\nu_0; \nu_x, \nu_y, \nu_z)$.
   The colored solid lines are DSM states on the topological phase boundary.
}
\label{fig:app:lattice:PD_WD}
\end{figure}

 The energy eigenvalues of this model are given by
\begin{align}
 E\_{WD}^{\pm}(\bm{k}) &= \pm \sqrt{ \sum_{\mu=x,y,z} t_\mu^2 \sin^2{k_\mu} + m^2(\bm{k}) }.
\end{align}
 Both the positive and negative energy states are twofold degenerate.
 If the mass terms are absent, $m_0 = m_{2\mu} = 0$,
this model exhibits eight Dirac points at the high-symmetry points in the Brillouin zone: $\bm{k} = (0,0,0)$, $(\pi,0,0)$, $(0,\pi,0)$, $(\pi,\pi,0)$, $(0,0,\pi)$, $(\pi,0,\pi)$, $(0,\pi,\pi)$, $(\pi,\pi,\pi)$.
 To eliminate the Dirac points except $(0,0,0)$
so that the model reproduces the Dirac Hamiltonian Eq.~(\ref{eq:MassiveDirac}) at low energy around $\bm{k} = (0,0,0)$,
the term with $m_{2\mu}$ $(\neq 0)$ is necessary.
 This term is known as the Wilson term in lattice quantum field theory, and $m_0$ corresponds to the mass in the original Dirac Hamiltonian.
%
%
%
 Depending on these mass parameters, the model describes strong and weak topological insulator phases~\cite{Yoshimura2016comparative},
as shown in the phase diagram in Fig.~\ref{fig:app:lattice:PD_WD}.
 The phase boundaries between topologically distinct states correspond to Dirac semimetals where the band gap closes, $E\_{WD}^{\pm}(\bm{k})=0$.

 The explicit representation of Dirac matrices $\alpha_{\mu=x,y,z,4}$ is not unique,
provided that the Clifford algebra with Euclidean metric
$\{ \alpha_\mu, \alpha_\nu \} = 2\delta_{\mu\nu}$ holds.
 Some commonly used representations are summarized in Table~\ref{tab:app:lattice:WD:rep}.
 In this table, the Dirac matrices are expressed as Kronecker products of two Pauli matrices, $\bm{\sigma}$ and $\bm{\tau}$, 
which typically correspond to the spin and orbital degrees of freedom, respectively.
 For studies on the general features of topological insulators and Dirac/Weyl semimetals, the spatially isotropic Dirac(--Pauli) or Weyl representations are often employed
because the topological features depend on the algebraic structure of the model and are irrelevant to the choice of the representation.
 On the other hand, for material-specific studies,
the representation should be properly chosen so that the model efficiently describe characteristic properties of each material,
such as spin and orbital structures of the bands, spin-momentum locking, and coupling to magnetization.
 Since the spin and orbital structures are usually anisotropic due to the crystal structure,
the low-energy effective models of realistic materials typically have anisotropic representations.
%
%
 In addition, we comment on the relationship between the Dirac matrices discussed here and the Dirac gamma matrices $\gamma^{\nu=0,1,2,3}$ in quantum field theory.
 The set of $\gamma$ matrices satisfy the Clifford algebra in the Minkowski metric and can be related by $\gamma^{0,1,2,3} = \alpha_{4,1,2,3} \alpha_4$.
 This means that the $\gamma^{\nu=1,2,3}$ are non-Hermitian, while the $\alpha$ matrices are Hermitian.

\begin{table}[htbp]
   \centering
   \caption{
    Representations of the Dirac matrices $\alpha_\mu$ and the spin operators $(\bm{s}_0, \bm{s}_1)$
    in terms of Kronecker products of Pauli matrices $\bm{\tau}$ and $\bm{\sigma}$.
    The rightmost three columns are the representations in the low-energy effective tight-binding models for specific materials.
    %
}
  \begin{tabular}{c||c|c|c|c|c}
  \hline
    & Dirac rep. & Weyl rep. & \ce{Na3Bi}~\cite{wang2012dirac} & \ce{Bi2Se3} & \ce{TlBiSe2} \\
    &            &           & \ce{Cd3As2}~\cite{wang2013three} & family~\cite{liu2010model} & family~\cite{yan2010theoretical} \\
   \hline
   \hline
   $\alpha_x$ & $\tau_x \sigma_x$ & $\tau_z \sigma_x$ & $\tau_x \sigma_z$  & $\tau_x \sigma_x$  & $-\tau_x \sigma_y$ \\
   $\alpha_y$ & $\tau_x \sigma_y$ & $\tau_z \sigma_y$ & $\tau_y \sigma_0$  & $\tau_x \sigma_y$  & $\tau_x \sigma_x$  \\
   $\alpha_z$ & $\tau_x \sigma_z$ & $\tau_z \sigma_z$ & $-\tau_x \sigma_x$ & $\tau_y \sigma_0$  & $\tau_y \sigma_0$  \\
   $\alpha_4$ & $\tau_z \sigma_0$ & $\tau_x \sigma_0$ & $\tau_z \sigma_0$  & $\tau_z \sigma_0$  & $\tau_z \sigma_0$  \\
   \hline
   $s_{0x}$   & $\tau_0 \sigma_x$ & $\tau_0 \sigma_x$ & $\tau_z \sigma_x$  & $\tau_z \sigma_y$  & $\tau_z \sigma_x$  \\
   $s_{0y}$   & $\tau_0 \sigma_y$ & $\tau_0 \sigma_y$ & $\tau_0 \sigma_y$  & $-\tau_z \sigma_x$ & $\tau_z \sigma_y$  \\
   $s_{0z}$   & $\tau_0 \sigma_z$ & $\tau_0 \sigma_z$ & $\tau_z \sigma_z$  & $\tau_0 \sigma_z$  & $\tau_0 \sigma_z$  \\
   \hline
   $s_{1x}$   & $\tau_z \sigma_x$ & $\tau_x \sigma_x$ & $\tau_0 \sigma_x$  & $\tau_0 \sigma_y$  & $\tau_0 \sigma_x$  \\
   $s_{1y}$   & $\tau_z \sigma_y$ & $\tau_x \sigma_y$ & $\tau_z \sigma_y$  & $-\tau_0 \sigma_x$ & $\tau_0 \sigma_y$  \\
   $s_{1z}$   & $\tau_z \sigma_z$ & $\tau_x \sigma_z$ & $\tau_0 \sigma_z$  & $\tau_z \sigma_z$  & $\tau_z \sigma_z$  \\
   \hline
  \end{tabular}
 \label{tab:app:lattice:WD:rep}
\end{table}

\subsubsection{Exchange coupling} \label{sec:app:lattice:WD:exchange}
 To derive the MWSM state from the Wilson--Dirac model,
it is necessary to introduce an exchange coupling term that breaks time-reversal symmetry.
 To construct the exchange coupling term,
the forms of the spin operators should be specified.
 Under spatial rotation, $\alpha_\mu$ ($\mu=1,2,3$) transforms as a vector, while $\alpha_4$ is a scalar.
 Since the Hamiltonian preserves the time-reversal (TR) symetry and $\sin{k_\mu}$ are odd under TR,
$\alpha_\mu$ ($\mu=1,2,3$) are TR odd, while $\alpha_4$ is TR even.
 Using these matrices, one can construct two sets of axial-vector operators $\bm{s}_0$ and $\bm{s}_1$, which are odd under TR, as
\begin{align}
    \bm{s}_0 = -i
    \begin{pmatrix}
        \alpha_y \alpha_z \\
        \alpha_z \alpha_x \\
        \alpha_x \alpha_y
    \end{pmatrix},
    \quad
    \bm{s}_1 = \bm{s}_0 \alpha_4 = -i
    \begin{pmatrix}
        \alpha_y \alpha_z \alpha_4 \\
        \alpha_z \alpha_x \alpha_4 \\
        \alpha_x \alpha_y \alpha_4
    \end{pmatrix}
    .
\end{align}
 Each of these two sets of matrices, $\{s_{0\mu}\}_{\mu=x,y,z}$ and $\{s_{1\mu}\}_{\mu=x,y,z}$, satisfies the Clifford algebra.
 Since the exchange coupling strength can be anisotropic reflecting the anisotropy of the material,
the exchange coupling term with the magnetization $\bm{M}$ is expressed as
\begin{align}
 H\_{exc} &= \sum_{\mu=x,y,z} M_\mu (J_{0\mu} s_{0\mu} + J_{1\mu} s_{1\mu}) .
  \label{eqn:app:lattice:WD:Hexc}
\end{align}

 If the system is isotropic,
these two sets of matrices satisfy the spin algebra,
$[s_a^i, s_a^j] = 2i \epsilon^{ijk} s_a^k$ with $i,j,k = x,y,z$ and $a = 0,1$.
 The exchange coupling term for an isotropic sytem is written as
\begin{align}
 H\_{exc}^{\mathrm{isotropic}} &= J_0 \bm{M} \cdot \bm{s}_0 + J_1 \bm{M} \cdot \bm{s}_1.
\end{align}
 In realistic materials,
$J_0$ and $J_1$ can be attributed to the orbital or sublattice dependence of the exchange coupling.
 For example,
if the exchange couplings for the $s$-orbital $(J_s)$ and for the $p$-orbital $(J_p)$ are different,
$J_0$ and $J_1$ are given by 
$J_0 = \tfrac{1}{2}(J_s + J_p)$ and $J_1 = \tfrac{1}{2}(J_s - J_p)$.



 The roles of $J_0$ and $J_1$ terms in the band structure are different~\cite{burkov2011topological};
typically, the $J_0$ term splits a Dirac point into a pair of Weyl points, while
the $J_1$ term transforms a Dirac point into a nodal ring.
 To understand the role of $J_0$ term,
we first consider the case $J_0 \neq 0$ and $J_1 = 0$.
 For simplicity, we set $\bm{M} = M \hat{\bm{z}}$.
 Then, the four energy eigenvalues $E_{\eta_1 \eta_2}(\bm{k}) \ (\eta_1,\eta_2 = \pm)$ are given by
\begin{align}
 E_{\mathrm{WD}J_0}^{\eta_1 \eta_2}(\bm{k}) &= \eta_1 \sqrt{
  \sum_{\mu=x,y} t_\mu^2 \sin^2{k_\mu} 
  + \[ \epsilon_z(\bm{k}) + \eta_2 |J_0 M| \]^2
 },
\end{align}
with $\epsilon_z(\bm{k}) = \sqrt{ t_z^2 \sin^2{k_z} + m^2(\bm{k}) }$.
 The two bands with $\eta_2 = - $ touch at the momenta satisfying $\sin{k_x} = \sin{k_y} = 0$ and $\epsilon_z(\bm{k}) = |J_0 M|$.
 Thus, the Weyl points can arise at eight points at most:
$\bm{k} = (0,0,\pm k_0^{(0,0)})$, $(0,\pi,\pm k_0^{(0,\pi)})$, $(\pi,0,\pm k_0^{(\pi,0)})$, and $(\pi,\pi,\pm k_0^{(\pi,\pi)})$,
if $\epsilon_z(\bm{k}) = |J_0 M|$ is satisfied with a real $k_0^{(k_x,k_y)}$.
 Using dimensionless notations
$\bar{m}^{(k_x,k_y)} = m^{(k_x,k_y)}/t_z$,
$\bar{m}_{2z} = m_{2z}/t_z$, and
$\bar{J}_0 = J_0/t_z$,
the positions of the Weyl points $k_0^{(k_x,k_y)}$ are given as
\begin{align}
 k_0^{(k_x,k_y)} &= 
  \arccos\[
   \frac{
    \bar{m}^{(k_x,k_y)} \bar{m}_{2z}
    - \sqrt{D^{(k_x,k_y)}}
   }{\bar{m}_{2z}^2 - 1}
  \] , \\
 D^{(k_x,k_y)} &=  
  \(\bar{m}^{(k_x,k_y)}\)^2
  - \(\bar{m}_{2z}^2 \!-\! 1\) \(1 \!-\! \bar{J}_0^2 M^2\) ,
\end{align}
for $m_{2z}^2 \ne t_{z}^2$, and
\begin{align}
 k_0^{(k_x,k_y)} &= 
  \arccos\[
   \frac{
    1 + (\bar{m}^{(k_x,k_y)})^2 - \bar{J}_0^2 M_0^2
    }{2 \bar{m}^{(k_x,k_y)} \bar{m}_{2z}}
  \],
\end{align}
for $m_{2z}^2 = t_{z}^2$.
 Here $m^{(k_x,k_y)}$ is the same as that for the two-band model, Eq.~\eqn{app:lattice:mkxky}.

 For instance, by setting $m_0=0$ and $m_{2z}^2\ne t_z^2$,
the system exhibits a pair of Weyl points at $\bm{k} = (0,0,\pm k_0)$,
with
\begin{align}
 k_0 &= 
  \arccos\[
   \frac{
    \bar{m}_{2z}^2  - \sqrt{1 + \bar{J}_0^2 M^2 \( \bar{m}_{2z}^2 - 1 \)}
   }{\bar{m}_{2z}^2 - 1} 
  \].
\end{align}
 If the exchange coupling is perturbatively small $(|J_0 M| \ll |t_z|)$,
the expression simplifies to $k_0 \simeq |J_0 M / t_z|$.
%
%
%

 On the other hand, $J_1$ term does not give rise to the MWSM state 
but induces a nodal line semimetal state \cite{burkov2011topological}.
 For the case $J_0 = 0$ and $J_1 \neq 0$ with $\bm{M} = M \hat{\bm{z}}$,
the energy eigenvalues are given by
\begin{align}
 E_{\mathrm{WD}J_1}^{\eta_1 \eta_2}(\bm{k}) &= 
  \eta_1 \sqrt{
    t_z^2 \sin^2{k_z}
    +\[ \epsilon_{xy}(\bm{k}) + \eta_2 |J_1 M| \]^2 
  },
\end{align}
with $\epsilon_{xy}(\bm{k}) = \sqrt{ \sum_{\mu=x,y} t_\mu^2 \sin^2{k_\mu} + m^2(\bm{k}) }$.
 The two bands with $\eta_2 = - $ touch at the momenta satisfying $\sin{k_z} = 0$ and $\epsilon_{xy}(\bm{k}) = |J_1 M|$.
 The band touching can occur along closed contours on the $k_z = 0$ and $\pi$ planes,
when $(k_x,k_y)$ for each contour satisfies
\begin{align}
 \sum_{\mu=x,y} t_\mu^2 \sin^2{k_\mu}
 + \( m^{(k_z)} - \!\! \sum_{\mu=x,y} m_{2\mu} \cos{k_\mu} \)^2 
 &= (J_1 M)^2,
\end{align}
with
\begin{align}
 m^{(k_z)} = m_0 + m_{2x} + m_{2y} + 2 \delta_{k_z,\pi} m_{2z}.
\end{align}

 For instance, by setting $m_0=0$ 
and assuming a perturbatively small exchange coupling, $|J_1 M| \ll |t_x t_y|$,
the system exhibits an elliptic nodal ring structure on the $k_z = 0$ plane around $(k_x,k_y)=(0,0)$,
described by
\begin{align}
 \(\frac{k_x}{t_y}\)^2 + \(\frac{k_y}{t_x}\)^2 = \(\frac{J_1 M}{t_x t_y}\)^2.
\end{align}
 If both $J_0$ and $J_1$ terms are nonzero,
the larger component dominates the nodal structure \cite{ominato2019phase}.

\subsubsection{Real-space representation}
 The real-space Hamiltonian of Wilson--Dirac model is described as
\begin{align}
 &\mathcal{H}\_{WD} 
   = \sum_{\bm{r}} c^\dag_{\bm{r}}  
      \(m_0 + \!\sum_{\mu=x,y,z}\! m_{2\mu}\)
                   \alpha_4
                   c_{\bm{r}} \\
 & \quad + \sum_{\bm{r}} \sum_{\mu=x,y,z} 
        \[ c^\dag_{\bm{r}+\bm{\mu}} 
           \(  {i t_{\mu} \over 2}  \alpha_{\mu}
              -{m_{2\mu} \over 2} \alpha_4 
           \)
           c_{\bm{r}}  + \textrm{h.c.}
        \]  \nonumber
 \label{eqn:app:lattice:WD:H_real}
\end{align}
where $c^{(\dag)}_{\bm{r}}$ is the four-component annihilation (creation) operator on the lattice site $\bm{r}$,
with the lattice vector $\hat{\bm{\mu}}$.
 The on-site disorder that preserves the symmetry are introduced in the form of potential disorder ($V_0 \bm{1}_4$) or mass disorder ($V_m \alpha_4$) as
\begin{align}
 & \mathcal{H}\_{V} 
  = \sum_{\bm{r}} c^\dag_{\bm{r}} \[ V_0(\bm{r}) \bm{1}_4 + V_m(\bm{r}) \alpha_4 \] c_{\bm{r}}.
\end{align}
%
 The exchange coupling term for the spatially nonuniform magnetization $\bm{M}(\bm{r})$ is written as
\begin{align}
 \mathcal{H}\_{exc} &
  = \sum_{\bm{r}} c^\dag_{\bm{r}} 
     \[ 
      \sum_{\mu=x,y,z} M_\mu(\bm{r}) (J_{0\mu} s_{0\mu} + J_{1\mu} s_{1\mu})
     \] c_{\bm{r}}.
\end{align}

\subsection{Description of topological Dirac semimetals} \label{sec:app:lattice:TDSM}
 So far, the mathematical structure of models describing Dirac/Weyl electron systems has been thoroughly discussed in the previous two sections.
 We here proceed to review a description of a specific system: topological Dirac semimetal (TDSM).
 The TDSM is characterized by pairs of Dirac points protected by crystalline symmetries.
 From the theoretical calculations and experimental observations,
\ce{Na3Bi} \cite{wang2012dirac,Liu2014} and 
\ce{Cd3As2} \cite{wang2013three,neupane2014observation,uchida2017quantum,uchida2019ferromagnetic} 
are known to exhibit the TDSM state,
both of which have a pair of Weyl points on a certain crystal axis protected by rotational symmetries.

 The low-energy effective model for \ce{Na3Bi} and \ce{Cd3As2} describing the TDSM state around the $\Gamma$ point is derived from the $\bm{k} \cdot \bm{p}$ expansion \cite{wang2012dirac,wang2013three},
\begin{align}
 H\_{TDSM}^{\Gamma}(\bm{k}) &= \epsilon_0^{\Gamma}(\bm{k}) \bm{1}_4 
  + A (k_x \alpha_x + k_y \alpha_y) 
  + M^{\Gamma}(\bm{k}) \alpha_4,
 \label{eqn:app:lattice:TDSM0}
\end{align}
with $\epsilon_0^{\Gamma}(\bm{k}) = C_0 + C_1 k_z^2 + C_2 (k_x^2 + k_y^2)$ and $M^{\Gamma}(\bm{k}) = M_0 - M_1 k_z^2 - M_2(k_x^2 + k_y^2)$.
 Here, the rotational symmetry is imposed around the $z$-axis.
 This model consists of four degrees of freedom spanned by the spin-orbital eigenstates.
 The specific representation of the Dirac matrices is shown in Table~\ref{tab:app:lattice:WD:rep},
where the Pauli matrices $\bm{\tau}$ and $\bm{\sigma}$ act on the orbital and spin degrees of freedom, respectively.
 With this representation,
the Hamiltonian 
is decomposed into two separate blocks corresponding to the spin-up and spin-down states $(\sigma_z = \pm)$,
\begin{align}
 H\_{TDSM}^{\Gamma,\sigma_z=+}(\bm{k}) &= \epsilon_0^{\Gamma}(\bm{k}) \tau_0 + A (k_x \tau_x + k_y \tau_y) + M^{\Gamma}(\bm{k}) \tau_z , \\
 H\_{TDSM}^{\Gamma,\sigma_z=-}(\bm{k}) &= \epsilon_0^{\Gamma}(\bm{k}) \tau_0 + A (-k_x \tau_x + k_y \tau_y) + M^{\Gamma}(\bm{k}) \tau_z .
\end{align}
This model shows a pair of topologically protected Dirac points at $\bm{k} = (0,0,\pm k_0)$ with $k_0 = \sqrt{M_0 / M_1}$.

%
 The lattice model reproducing the low-energy band structure of TDSM given by Eq.~\eqn{app:lattice:TDSM0} has been used for numerical calculations of various electronic properties of TDSMs \cite{wang2013three, Kobayashi2021ferromagnetic, araki2021jpsj, Kobayashi2021intrinsic, matsushita2025intrinsic}.
 The TDSM model on a hypothetical tetragonal lattice with lattice constants $(a, a, c)$ is given as
\begin{align}
 H\_{TDSM}(\bm{k}) 
  &=
      \epsilon_0(\bm{k}) \bm{1}_4
     + \frac{A}{a} \sum_{\mu=x,y} \sin(k_\mu a) \alpha_\mu 
     + M(\bm{k}) \alpha_4 ,
\end{align}
with
$\epsilon_0(\bm{k}) 
= C_0 
+ 2 (C_1/c^2) [1 - \cos(k_z c)] 
+ 2 (C_2/a^2) [2 - \cos(k_x a) - \cos(k_y a)]$
and
$M(\bm{k})
= M_0
- 2 (M_1/c^2) [1 - \cos(k_z c)]
- 2 (M_2/a^2) [2 - \cos(k_x a) - \cos(k_y a)]$.
 This model is equivalent to the Wilson--Dirac type model, Eq.~\eqn{app:lattice:WD:HWDk}, with substitutions
$t_x , t_y \to A/a$, 
$t_z \to 0$,
$m_0 \to M_0$, 
$m_{2x} , m_{2y} \to - 2 M_2/a^2$, 
$m_{2z} \to -2 M_1/c^2$, 
$c_0 \to C_0$, 
$c_x , c_y \to 2 C_2/a^2$, and
$c_z \to 2 C_1/c^2$.
 As with its low-energy form,
this lattice Hamiltonian is block diagonal with the spin-up and down blocks $H\_{TDSM}^{\sigma_z =\pm}(\bm{k})$,
each of which is identical with the two-band WSM model, Eq.~\eqn{app:lattice:H2x2_k}.
 Therefore, the properties of a TDSM without magnetization can be understood in terms of the two copies of two-band WSMs, with spin polarization $\sigma_z = +$ or $\sigma_z = -$ for each copy,
by superposing those of the pair of spin-polarized two-band WSMs.
 For instance, each Dirac point is the superposition of two Weyl points with $\sigma_z = +$ and $\sigma_z = -$, and the Fermi arcs are doubled correspondingly.


 As discussed in Sec.~\ref{sec:app:lattice:WD:exchange}, 
the exchange coupling term for this TDSM model consists of two types of terms, Eq.~\eqn{app:lattice:WD:Hexc}.
 Doe to the lack of $\alpha_z$ term,
the effects of exchange couplings on the nodal structure are slightly modified from those in the Wilson--Dirac model with three $\alpha_\mu$ terms~\cite{araki2021jpsj}.
 The coupling $J_0$ mostly leads to the Weyl point structure,
except for the case $\bm{M} \perp z$ yielding the nodal ring structure.
 On the other hand, $J_1$ primaliry leads to the nodal ring structure,
except for the case $\bm{M} \parallel z$,
where opposite signs of energy shifts are induced to $\sigma_z = +$ and $-$ WSMs.

\subsection{Effective tight-binding model for stacked kagome lattice \CSS}
\label{sec:app:lattice:CSS}

\begin{figure}
    \centering
    \caption{Figure 1. (Color online) (a) Original unit cell of \CSS and (b) unit cell of our model. 
    Co is responsible for the ferromagnetic order. 
    (c) Each layer of \CSS. 
    Kagome layer contains Co atoms and Sn2 atoms. 
    Those Kagome layers sandwich two types of triangle layers formed by Sn1 atoms and S atoms. 
    (d) The energy relation and occupied electrons anticipated in our model.
    Reprinted from \cite{Ozawa2019}.
    }
    \label{fig:co3sn2s2-model}
\end{figure}

A minimal tight-binding model for the ferromagnetic Weyl semimetal \CSS~\cite{Ozawa2019} is briefly introduced in Sec.~\ref{sec:co3sn2s2} in this review.
This model was employed to analyze the spin Hall effect~\cite{Lau2023} (see Sec.~\ref{sec:spin-transport}), the light-induced AHE~\cite{yoshikawa2025}, and also the second-order conductivity~\cite{jia2026nonlinear} in \CSS.
In this part, we explain its detail.

The crystal structure of \CSS is shown in Fig.~\ref{fig:co3sn2s2-model}(a).
\CSS consists of a primitive rhombohedral unit cell including three Co atoms, two Sn atoms, and two S atoms.
Co atoms form kagome layers responsible for ferromagnetic ordering.
One of the two Sn atoms (Sn1) forms a layer of triangular lattice,
whereas the other Sn atom (Sn2) resides at the center of hexagon of the kagome lattice formed by Co sites.
There are also two layers of triangular lattices of S atoms sandwiched between the kagome layers of Co [see Fig.~\ref{fig:co3sn2s2-model}(c)].

To describe the low-energy electronic structure including Weyl points,
the minimal model is constructed using the partially occupied orbitals near $E_{\rm F}$.
Specifically, the model includes one $d$ orbital from each Co atom on the kagome layer and one 
$p$ orbital from each Sn1 atom.
They are selected by the discussion as follows.

First, we note that the atomic orbitals of each site are subject to the crystal field splitting, as shown in Fig.~\ref{fig:co3sn2s2-model}(d).
Splitting energies in the kagome layer are supposed to be larger than those in the triangular lattices of Sn and S atoms,
because the interatomic distance in the kagome lattice is shorter than those in the triangular lattices. 
As a result, the five-fold degeneracy of Co's $d$-orbitals is lifted.
%
As long as the crystal field splitting energies are large enough compared to the Hund coupling energies,
the low-spin state is favored,
with the low-energy orbitals occupied by the valence electrons
(see also Sec.~\ref{sec:low-spin}).
Under this condition,
the electron configuration is assumed as shown in Fig.~\ref{fig:co3sn2s2-model}(d).
The fourth lowest $d$-orbital of Co among the five is partially occupied,
and there remains one unpaired electron in $3\times 2=6$ states on three Co atoms per unit cell. 
If spins are ferromagnetically polarized,
this configuration gives the magnetic moment $\mu_{\rm B}/3$ per one Co site,
which is consistent with the value $\approx 0.3\mu_{\rm B}$ suggested by first-principles calculations and experiments~\cite{Liu2018}.

Then, we extract the orbitals close to $E_{\rm F}$ for the effective model.
Here, $E_{\rm F}$ is determined by the $3/8$-filling condition.
Close to $E_{\rm F}$,
there are the partially occupied $d_{3z^2 -r^2}$ orbitals of Co and the occupied $p_z$ orbital of Sn1. 
All other orbitals are far from $E_{\rm F}$ and neglected. 
Thus, this model is formed by the three Co sites with $d$-orbitals and one Sn1 site with a $p$-orbital,
as shown in Fig.~\ref{fig:co3sn2s2-model}(b).
The primitive lattice vectors are defined as
${\bm a}_1=(\frac{a}{\sqrt{3}},0,\frac{c}{3})$,
${\bm a}_2=(-\frac{a}{2\sqrt{3}},\frac{a}{2},\frac{c}{3})$,
and
${\bm a}_3=(-\frac{a}{2\sqrt{3}},-\frac{a}{2},\frac{c}{3})$,
with lattice parameters $a=5.37$~\AA\ and $c=13.18$~\AA.

With the above selection of orbitals,
the total Hamiltonian of this model is given as,
\begin{align}
    H_{0}=H_{\text{d-p}}+H_{\rm so}+H_{\rm exc}.
\end{align}
Here $H_{\text{d-p}}$ denotes the spin-independent hopping term,
$H_{\rm so}$ the spin-orbit coupling,
and $H_{\rm exc}$ the exchange coupling to the mean field of the magnetic order parameter.

The spin-independent hopping term is
\begin{align}
 H_{\text{d-p}}=&- \sum_{ijs}t_{ij} d^{\dagger}_{is}d_{js}
 + \sum_{\langle ij \rangle s} t^{\rm dp}_{ij}(d^{\dagger}_{is}p_{js}+p^{\dagger}_{is}d_{js})
 +\epsilon_{\rm p}\sum_{is} p^{\dagger}_{is}p_{is}.
 \label{eq:d-p}
\end{align}
Here, $d_{is}$ annihilates a Co-$d$ electron at site $i$ with spin $s$,
and $p_{is}$ for Sn1-$p$.
The first term represents the Co--Co hoppings within and between kagome layers,
the second term the Co--Sn hopping,
and the third term the on-site energy of the Sn1 orbital.
For the Co--Co hoppings $t_{ij}$,
we include the nearest-neighbor hopping $t_1$,
the second-nearest-neighbor hopping $t_2$ within each kagome layer,
and the interlayer hopping $t_z$ between adjacent kagome layers
[See Fig.~\ref{fig:ozanomu}(a) and (b) in main text].
Each Co site has four neighbors for $t_1$, $t_2$, and $t_z$.
For the Co--Sn hopping $t^{\rm dp}_{ij}$,
we consider Sn1 sites located above and below the kagome layers.
Due to the odd parity of the Sn1 $p_z$ orbital,
$t^{\rm dp}_{ij}=\pm t^{\rm dp}$ depends on the relative positions of Co and Sn1 sites.
Specifically, 
the lattice vectors for $t_1$ are
${\bm b}_{AB} =
({\bm a}_2 - {\bm a}_1)/2
$,
${\bm b}_{BC} =
({\bm a}_3 -{\bm a}_2)/2
$,
${\bm b}_{CA} =
({\bm a}_1 - {\bm a}_3)/2
$.
Here, we used the sublattice indices A,B, and C.
With these vectors,
the lattice vectors for $t_2$ are 
${\bm d}_{AB} =
{\bm b}_2 - {\bm b}_1
$,
${\bm d}_{BC} =
{\bm b}_3 -{\bm b}_2
$,
${\bm d}_{CA} =
{\bm b}_1 - {\bm b}_3
$.
Then, the lattice vectors for $t_z$ are 
${\bm c}_{AB} =
({\bm a}_1 + {\bm a}_2)/2
$,
${\bm c}_{BC} =
({\bm a}_2 +{\bm a}_3)/2
$,
${\bm c}_{CA} =
({\bm a}_3 + {\bm a}_1)/2
$.
The on-site energy of the Co-$d$ orbital is set to zero, and $\epsilon_{\rm p}$ is measured from it.

The spin-orbit coupling term consists of intra-layer and inter-layer contributions,
$H_{\rm so}=H^{\rm intra}_{\rm so}+H^{\rm inter}_{\rm so}$, given by
\begin{align}
    H^{\rm intra}_{\rm so} &=-i t^{\rm intra}_{\rm so}
    \sum_{\langle\!\langle ij \rangle\!\rangle ss'}
    \nu_{ij} d^{\dagger}_{is} \sigma^z_{ss'} d_{js'}, 
    \label{eqn:KM} \\
    H^{\rm inter}_{\rm so} &=-i t^{\rm inter}_{\rm so}
    \sum_{\langle ij \rangle ss'}
    {\bm \lambda}_{ij} \cdot
    d^{\dagger}_{is} {\bm \sigma}_{ss'} d_{js'}.
    \label{eqn:stgRashba}
\end{align}
Here ${\bm \sigma}$ denotes the Pauli matrices for the electron spin.
The intra-layer term arises from the in-plane electric field generated by Sn2 atoms
located at the centers of kagome hexagons.
The factor $\nu_{ij}=\pm1$ depends on the clockwise or counterclockwise hopping direction
around the adjacent Sn2 atom.
This term breaks spin SU(2) symmetry while preserving U(1) symmetry with respect to $\sigma^z$,
and is analogous to that in the Kane--Mele model~\cite{Kane2005,Guo2009}.
The inter-layer term describes spin-dependent hopping between adjacent kagome layers.
The vector ${\bm \lambda}_{ij}$ represents the direction of the effective magnetic field
and is defined by the geometry of the primitive lattice vectors,
${\bm \lambda}_{\rm AB} = -{\bm \lambda}_{\rm BA} = \tfrac{{\bm a}_2}{2} \times \tfrac{{\bm a}_1}{2} / |\tfrac{{\bm a}_2}{2} \times \tfrac{{\bm a}_1}{2}|$,
${\bm \lambda}_{\rm BC} = -{\bm \lambda}_{\rm CB} = \tfrac{{\bm a}_3}{2} \times \tfrac{{\bm a}_2}{2} / |\tfrac{{\bm a}_3}{2} \times \tfrac{{\bm a}_2}{2}|$,
and
${\bm \lambda}_{\rm CA} = -{\bm \lambda}_{\rm AC} = \tfrac{{\bm a}_1}{2} \times \tfrac{{\bm a}_3}{2} / |\tfrac{{\bm a}_1}{2} \times \tfrac{{\bm a}_3}{2}|$.
This term breaks the conservation of $\sigma^z$ due to the lack of inversion symmetry between layers.

Finally, the exchange coupling is given by
\begin{align}
    H_{\rm exc}=-J \sum_{iss'} {\bm m}_{i} \cdot
    (d^{\dagger}_{is} {\bm \sigma}_{ss'} d_{is'}
    + p^{\dagger}_{is} {\bm \sigma}_{ss'} p_{is'}).
    \label{eq:exc}
\end{align}
Here $J$ is the exchange coupling constant and ${\bm m}_i$ is the mean field for the magnetic moment at site $i$, originating from the on-site Hubbard interaction.
For simplicity, we assume the same coupling strength for Co and Sn1 sites.
Previous self-consistent calculations showed that the ground state exhibits
ferromagnetic ordering in the low-carrier-density regime~\cite{Ozawa2022}.
In such a uniform ferromagnetic state, we take 
${\bm m}_i={\bm m}$ with $|{\bm m}|=1$.

By the Fourier transformation from the real-space to the momentum-space representation,
we can calculate the band structure from this tight-binding model.
In total, there are 8 bands in total, consisting of $4$ site $\times$ $2$ spin degrees of freedom.
We here consider the case with the out-of-plane magnetization, ${\bm m} = \hat{\bm z}$.
In the absence of the spin-orbit coupling,
each band is fully polarized to spin-up or down, as shown by the red and blue dotted lines in Fig.~\ref{fig:ozanomu}(d).
The high symmetry points in momentum space are listed in table~\ref{table:css}.
The band crossing close to $E_{\rm F}$ forms nodal rings around the three L-points in the Brillouin zone.
Once we introduce the spin-orbit coupling,
these nodal rings are gapped out as shown by the green lines in Fig.~\ref{fig:ozanomu}(d),
except for a pair of Weyl points on each ring.
Thus, this model exhibits three pairs of Weyl points as shown in Fig.~\ref{fig:ozanomu}(c),
which is almost consistent with the first-principles calculation results \cite{Liu2018,Xu2018,Muechler2020}.
Once the exchange coupling $J$ is set to zero,
all the bands become spin degenerate and the Weyl points reduce to the Dirac points.
Such a case corresponds to the paramagnetic state,
which is achieved by the carrier doping from chemical substitution [see Sec.~\ref{sec:low-spin}].

\begin{table}[htbp]
\caption{
High-symmetry points of the rhombohedral lattice class RHL$_1$ in the basis of reciprocal lattice vectors
$\mathbf{b}_1$, $\mathbf{b}_2$, and $\mathbf{b}_3$.
The parameters are
$\eta=(1+4\cos\alpha)/(2+4\cos\alpha)$ and
$\nu=3/4-\eta/2$
(see Table~14 of Ref.~\cite{setyawan2010high}).
In Co$_3$Sn$_2$S$_2$, $\alpha$ is parameterized $\alpha=59.916^\circ$~\cite{Liu2018}.
}
\centering
\begin{tabular}{lccc}
\hline
 &
$\mathbf{b}_1$ &
$\mathbf{b}_2$ &
$\mathbf{b}_3$ \\
\hline
\hline
${\rm \Gamma}$ & 0   & 0   & 0 \\
${\rm L}$      & 1/2 & 0   & 0 \\
${\rm W}$      & $\eta$ & 1/2 & 1-$\eta$ \\
${\rm U}$      & $\eta$ & $\nu$ & $\nu$ \\
${\rm T}$      & 1/2 & 1/2 & 1/2 \\
\hline
\end{tabular}
\label{table:css}
\end{table}

\bibliography{ref_jpsj}

\begin{thebibliography}{481}%
\makeatletter
\providecommand \@ifxundefined [1]{%
 \@ifx{#1\undefined}
}%
\providecommand \@ifnum [1]{%
 \ifnum #1\expandafter \@firstoftwo
 \else \expandafter \@secondoftwo
 \fi
}%
\providecommand \@ifx [1]{%
 \ifx #1\expandafter \@firstoftwo
 \else \expandafter \@secondoftwo
 \fi
}%
\providecommand \natexlab [1]{#1}%
\providecommand \enquote  [1]{``#1''}%
\providecommand \bibnamefont  [1]{#1}%
\providecommand \bibfnamefont [1]{#1}%
\providecommand \citenamefont [1]{#1}%
\providecommand \href@noop [0]{\@secondoftwo}%
\providecommand \href [0]{\begingroup \@sanitize@url \@href}%
\providecommand \@href[1]{\@@startlink{#1}\@@href}%
\providecommand \@@href[1]{\endgroup#1\@@endlink}%
\providecommand \@sanitize@url [0]{\catcode `\\12\catcode `\$12\catcode `\&12\catcode `\#12\catcode `\^12\catcode `\_12\catcode `\%12\relax}%
\providecommand \@@startlink[1]{}%
\providecommand \@@endlink[0]{}%
\providecommand \url  [0]{\begingroup\@sanitize@url \@url }%
\providecommand \@url [1]{\endgroup\@href {#1}{\urlprefix }}%
\providecommand \urlprefix  [0]{URL }%
\providecommand \Eprint [0]{\href }%
\providecommand \doibase [0]{https://doi.org/}%
\providecommand \selectlanguage [0]{\@gobble}%
\providecommand \bibinfo  [0]{\@secondoftwo}%
\providecommand \bibfield  [0]{\@secondoftwo}%
\providecommand \translation [1]{[#1]}%
\providecommand \BibitemOpen [0]{}%
\providecommand \bibitemStop [0]{}%
\providecommand \bibitemNoStop [0]{.\EOS\space}%
\providecommand \EOS [0]{\spacefactor3000\relax}%
\providecommand \BibitemShut  [1]{\csname bibitem#1\endcsname}%
\let\auto@bib@innerbib\@empty
\bibitem [{\citenamefont {Wan}\ \emph {et~al.}(2011)\citenamefont {Wan}, \citenamefont {Turner}, \citenamefont {Vishwanath},\ and\ \citenamefont {Savrasov}}]{Wan2011}%
  \BibitemOpen
  \bibfield  {author} {\bibinfo {author} {\bibfnamefont {X.}~\bibnamefont {Wan}}, \bibinfo {author} {\bibfnamefont {A.~M.}\ \bibnamefont {Turner}}, \bibinfo {author} {\bibfnamefont {A.}~\bibnamefont {Vishwanath}},\ and\ \bibinfo {author} {\bibfnamefont {S.~Y.}\ \bibnamefont {Savrasov}},\ }\bibfield  {title} {\bibinfo {title} {Topological semimetal and {Fermi}-arc surface states in the electronic structure of pyrochlore iridates},\ }\href@noop {} {\bibfield  {journal} {\bibinfo  {journal} {Phys. Rev. B}\ }\textbf {\bibinfo {volume} {83}},\ \bibinfo {pages} {205101} (\bibinfo {year} {2011})}\BibitemShut {NoStop}%
\bibitem [{\citenamefont {Burkov}\ and\ \citenamefont {Balents}(2011)}]{Burkov2011}%
  \BibitemOpen
  \bibfield  {author} {\bibinfo {author} {\bibfnamefont {A.~A.}\ \bibnamefont {Burkov}}\ and\ \bibinfo {author} {\bibfnamefont {L.}~\bibnamefont {Balents}},\ }\bibfield  {title} {\bibinfo {title} {{Weyl} semimetal in a topological insulator multilayer},\ }\href@noop {} {\bibfield  {journal} {\bibinfo  {journal} {Phys. Rev. Lett.}\ }\textbf {\bibinfo {volume} {107}},\ \bibinfo {pages} {127205} (\bibinfo {year} {2011})}\BibitemShut {NoStop}%
\bibitem [{\citenamefont {Armitage}\ \emph {et~al.}(2018)\citenamefont {Armitage}, \citenamefont {Mele},\ and\ \citenamefont {Vishwanath}}]{Armitage2018}%
  \BibitemOpen
  \bibfield  {author} {\bibinfo {author} {\bibfnamefont {N.~P.}\ \bibnamefont {Armitage}}, \bibinfo {author} {\bibfnamefont {E.~J.}\ \bibnamefont {Mele}},\ and\ \bibinfo {author} {\bibfnamefont {A.}~\bibnamefont {Vishwanath}},\ }\bibfield  {title} {\bibinfo {title} {Weyl and dirac semimetals in three-dimensional solids},\ }\href@noop {} {\bibfield  {journal} {\bibinfo  {journal} {Rev. Mod. Phys.}\ }\textbf {\bibinfo {volume} {90}},\ \bibinfo {pages} {015001} (\bibinfo {year} {2018})}\BibitemShut {NoStop}%
\bibitem [{\citenamefont {Burkov}(2018)}]{burkov2018weyl}%
  \BibitemOpen
  \bibfield  {author} {\bibinfo {author} {\bibfnamefont {A.}~\bibnamefont {Burkov}},\ }\bibfield  {title} {\bibinfo {title} {Weyl metals},\ }\href@noop {} {\bibfield  {journal} {\bibinfo  {journal} {Annual Review of Condensed Matter Physics}\ }\textbf {\bibinfo {volume} {9}},\ \bibinfo {pages} {359} (\bibinfo {year} {2018})}\BibitemShut {NoStop}%
\bibitem [{\citenamefont {Weyl}(1929)}]{weyl1929gravitation}%
  \BibitemOpen
  \bibfield  {author} {\bibinfo {author} {\bibfnamefont {H.}~\bibnamefont {Weyl}},\ }\bibfield  {title} {\bibinfo {title} {Gravitation and the electron},\ }\href@noop {} {\bibfield  {journal} {\bibinfo  {journal} {Proceedings of the National Academy of Sciences}\ }\textbf {\bibinfo {volume} {15}},\ \bibinfo {pages} {323} (\bibinfo {year} {1929})}\BibitemShut {NoStop}%
\bibitem [{\citenamefont {Dirac}(1928)}]{dirac1928quantum}%
  \BibitemOpen
  \bibfield  {author} {\bibinfo {author} {\bibfnamefont {P.~A.~M.}\ \bibnamefont {Dirac}},\ }\bibfield  {title} {\bibinfo {title} {The quantum theory of the electron},\ }\href@noop {} {\bibfield  {journal} {\bibinfo  {journal} {Proceedings of the Royal Society of London. Series A, Containing Papers of a Mathematical and Physical Character}\ }\textbf {\bibinfo {volume} {117}},\ \bibinfo {pages} {610} (\bibinfo {year} {1928})}\BibitemShut {NoStop}%
\bibitem [{\citenamefont {Bethe}\ and\ \citenamefont {Peierls}(1934)}]{bethe1934neutrino}%
  \BibitemOpen
  \bibfield  {author} {\bibinfo {author} {\bibfnamefont {H.}~\bibnamefont {Bethe}}\ and\ \bibinfo {author} {\bibfnamefont {R.}~\bibnamefont {Peierls}},\ }\bibfield  {title} {\bibinfo {title} {The “neutrino”},\ }\href@noop {} {\bibfield  {journal} {\bibinfo  {journal} {Nature}\ }\textbf {\bibinfo {volume} {133}},\ \bibinfo {pages} {532} (\bibinfo {year} {1934})}\BibitemShut {NoStop}%
\bibitem [{\citenamefont {Kogut}(1979)}]{kogut1979introduction}%
  \BibitemOpen
  \bibfield  {author} {\bibinfo {author} {\bibfnamefont {J.~B.}\ \bibnamefont {Kogut}},\ }\bibfield  {title} {\bibinfo {title} {An introduction to lattice gauge theory and spin systems},\ }\href@noop {} {\bibfield  {journal} {\bibinfo  {journal} {Reviews of Modern Physics}\ }\textbf {\bibinfo {volume} {51}},\ \bibinfo {pages} {659} (\bibinfo {year} {1979})}\BibitemShut {NoStop}%
\bibitem [{\citenamefont {Gattringer}\ and\ \citenamefont {Lang}(2009)}]{gattringer2009quantum}%
  \BibitemOpen
  \bibfield  {author} {\bibinfo {author} {\bibfnamefont {C.}~\bibnamefont {Gattringer}}\ and\ \bibinfo {author} {\bibfnamefont {C.}~\bibnamefont {Lang}},\ }\href@noop {} {\emph {\bibinfo {title} {Quantum chromodynamics on the lattice: an introductory presentation}}},\ Vol.\ \bibinfo {volume} {788}\ (\bibinfo  {publisher} {Springer Science \& Business Media},\ \bibinfo {year} {2009})\BibitemShut {NoStop}%
\bibitem [{\citenamefont {Ratti}(2018)}]{ratti2018lattice}%
  \BibitemOpen
  \bibfield  {author} {\bibinfo {author} {\bibfnamefont {C.}~\bibnamefont {Ratti}},\ }\bibfield  {title} {\bibinfo {title} {Lattice qcd and heavy ion collisions: a review of recent progress},\ }\href@noop {} {\bibfield  {journal} {\bibinfo  {journal} {Reports on Progress in Physics}\ }\textbf {\bibinfo {volume} {81}},\ \bibinfo {pages} {084301} (\bibinfo {year} {2018})}\BibitemShut {NoStop}%
\bibitem [{\citenamefont {Wallace}(1947)}]{wallence1947}%
  \BibitemOpen
  \bibfield  {author} {\bibinfo {author} {\bibfnamefont {P.~R.}\ \bibnamefont {Wallace}},\ }\bibfield  {title} {\bibinfo {title} {The band theory of graphite},\ }\href {https://doi.org/10.1103/PhysRev.71.622} {\bibfield  {journal} {\bibinfo  {journal} {Phys. Rev.}\ }\textbf {\bibinfo {volume} {71}},\ \bibinfo {pages} {622} (\bibinfo {year} {1947})}\BibitemShut {NoStop}%
\bibitem [{\citenamefont {Castro~Neto}\ \emph {et~al.}(2009)\citenamefont {Castro~Neto}, \citenamefont {Guinea}, \citenamefont {Peres}, \citenamefont {Novoselov},\ and\ \citenamefont {Geim}}]{Neto2009}%
  \BibitemOpen
  \bibfield  {author} {\bibinfo {author} {\bibfnamefont {A.~H.}\ \bibnamefont {Castro~Neto}}, \bibinfo {author} {\bibfnamefont {F.}~\bibnamefont {Guinea}}, \bibinfo {author} {\bibfnamefont {N.~M.~R.}\ \bibnamefont {Peres}}, \bibinfo {author} {\bibfnamefont {K.~S.}\ \bibnamefont {Novoselov}},\ and\ \bibinfo {author} {\bibfnamefont {A.~K.}\ \bibnamefont {Geim}},\ }\bibfield  {title} {\bibinfo {title} {The electronic properties of graphene},\ }\href {https://doi.org/10.1103/RevModPhys.81.109} {\bibfield  {journal} {\bibinfo  {journal} {Rev. Mod. Phys.}\ }\textbf {\bibinfo {volume} {81}},\ \bibinfo {pages} {109} (\bibinfo {year} {2009})}\BibitemShut {NoStop}%
\bibitem [{\citenamefont {Sarma}\ \emph {et~al.}(2011)\citenamefont {Sarma}, \citenamefont {Adam}, \citenamefont {Hwang},\ and\ \citenamefont {Rossi}}]{Sarma2011-pc}%
  \BibitemOpen
  \bibfield  {author} {\bibinfo {author} {\bibfnamefont {S.~D.}\ \bibnamefont {Sarma}}, \bibinfo {author} {\bibfnamefont {S.}~\bibnamefont {Adam}}, \bibinfo {author} {\bibfnamefont {E.~H.}\ \bibnamefont {Hwang}},\ and\ \bibinfo {author} {\bibfnamefont {E.}~\bibnamefont {Rossi}},\ }\bibfield  {title} {\bibinfo {title} {Electronic transport in two-dimensional graphene},\ }\href@noop {} {\bibfield  {journal} {\bibinfo  {journal} {Rev. Mod. Phys.}\ }\textbf {\bibinfo {volume} {83}},\ \bibinfo {pages} {407} (\bibinfo {year} {2011})}\BibitemShut {NoStop}%
\bibitem [{\citenamefont {Bistritzer}\ and\ \citenamefont {MacDonald}(2011)}]{Bistritzer2011}%
  \BibitemOpen
  \bibfield  {author} {\bibinfo {author} {\bibfnamefont {R.}~\bibnamefont {Bistritzer}}\ and\ \bibinfo {author} {\bibfnamefont {A.~H.}\ \bibnamefont {MacDonald}},\ }\bibfield  {title} {\bibinfo {title} {Moir{\'e} bands in twisted double-layer graphene},\ }\href@noop {} {\bibfield  {journal} {\bibinfo  {journal} {Proceedings of the National Academy of Sciences}\ }\textbf {\bibinfo {volume} {108}},\ \bibinfo {pages} {12233} (\bibinfo {year} {2011})}\BibitemShut {NoStop}%
\bibitem [{\citenamefont {Katayama}\ \emph {et~al.}(2006)\citenamefont {Katayama}, \citenamefont {Kobayashi},\ and\ \citenamefont {Suzumura}}]{Katayama2006}%
  \BibitemOpen
  \bibfield  {author} {\bibinfo {author} {\bibfnamefont {S.}~\bibnamefont {Katayama}}, \bibinfo {author} {\bibfnamefont {A.}~\bibnamefont {Kobayashi}},\ and\ \bibinfo {author} {\bibfnamefont {Y.}~\bibnamefont {Suzumura}},\ }\bibfield  {title} {\bibinfo {title} {Pressure-induced zero-gap semiconducting state in organic conductor {$\alpha$}-({BEDT}-{TTF})$_{2}${I}$_{3}${salt}},\ }\href {https://doi.org/10.1143/JPSJ.75.054705} {\bibfield  {journal} {\bibinfo  {journal} {J. Phys. Soc. Jpn.}\ }\textbf {\bibinfo {volume} {75}},\ \bibinfo {pages} {054705} (\bibinfo {year} {2006})}\BibitemShut {NoStop}%
\bibitem [{\citenamefont {Kobayashi}\ \emph {et~al.}(2007)\citenamefont {Kobayashi}, \citenamefont {Katayama}, \citenamefont {Suzumura},\ and\ \citenamefont {Fukuyama}}]{Kobayashi2007}%
  \BibitemOpen
  \bibfield  {author} {\bibinfo {author} {\bibfnamefont {A.}~\bibnamefont {Kobayashi}}, \bibinfo {author} {\bibfnamefont {S.}~\bibnamefont {Katayama}}, \bibinfo {author} {\bibfnamefont {Y.}~\bibnamefont {Suzumura}},\ and\ \bibinfo {author} {\bibfnamefont {H.}~\bibnamefont {Fukuyama}},\ }\bibfield  {title} {\bibinfo {title} {Massless fermions in organic conductor},\ }\href@noop {} {\bibfield  {journal} {\bibinfo  {journal} {J. Phys. Soc. Jpn.}\ }\textbf {\bibinfo {volume} {76}},\ \bibinfo {pages} {034711} (\bibinfo {year} {2007})}\BibitemShut {NoStop}%
\bibitem [{\citenamefont {Manzeli}\ \emph {et~al.}(2017)\citenamefont {Manzeli}, \citenamefont {Ovchinnikov}, \citenamefont {Pasquier}, \citenamefont {Yazyev},\ and\ \citenamefont {Kis}}]{Manzeli2017-fi}%
  \BibitemOpen
  \bibfield  {author} {\bibinfo {author} {\bibfnamefont {S.}~\bibnamefont {Manzeli}}, \bibinfo {author} {\bibfnamefont {D.}~\bibnamefont {Ovchinnikov}}, \bibinfo {author} {\bibfnamefont {D.}~\bibnamefont {Pasquier}}, \bibinfo {author} {\bibfnamefont {O.~V.}\ \bibnamefont {Yazyev}},\ and\ \bibinfo {author} {\bibfnamefont {A.}~\bibnamefont {Kis}},\ }\bibfield  {title} {\bibinfo {title} {{2D} transition metal dichalcogenides},\ }\href@noop {} {\bibfield  {journal} {\bibinfo  {journal} {Nat. Rev. Mater.}\ }\textbf {\bibinfo {volume} {2}},\ \bibinfo {pages} {17033} (\bibinfo {year} {2017})}\BibitemShut {NoStop}%
\bibitem [{\citenamefont {Volovik}(1987)}]{volovik1987zeros}%
  \BibitemOpen
  \bibfield  {author} {\bibinfo {author} {\bibfnamefont {G.}~\bibnamefont {Volovik}},\ }\bibfield  {title} {\bibinfo {title} {Zeros in the fermion spectrum in superfluid systems as diabolical points},\ }\href@noop {} {\bibfield  {journal} {\bibinfo  {journal} {JETP Lett}\ }\textbf {\bibinfo {volume} {46}},\ \bibinfo {pages} {98} (\bibinfo {year} {1987})}\BibitemShut {NoStop}%
\bibitem [{\citenamefont {Volovik}(2003)}]{volovik2003universe}%
  \BibitemOpen
  \bibfield  {author} {\bibinfo {author} {\bibfnamefont {G.~E.}\ \bibnamefont {Volovik}},\ }\href@noop {} {\emph {\bibinfo {title} {The universe in a helium droplet}}},\ Vol.\ \bibinfo {volume} {117}\ (\bibinfo  {publisher} {OUP Oxford},\ \bibinfo {year} {2003})\BibitemShut {NoStop}%
\bibitem [{\citenamefont {Murakami}(2007)}]{Murakami2007}%
  \BibitemOpen
  \bibfield  {author} {\bibinfo {author} {\bibfnamefont {S.}~\bibnamefont {Murakami}},\ }\bibfield  {title} {\bibinfo {title} {Phase transition between the quantum spin hall and insulator phases in 3d: emergence of a topological gapless phase},\ }\href@noop {} {\bibfield  {journal} {\bibinfo  {journal} {New Journal of Physics}\ }\textbf {\bibinfo {volume} {9}},\ \bibinfo {pages} {356} (\bibinfo {year} {2007})}\BibitemShut {NoStop}%
\bibitem [{\citenamefont {Shindou}\ and\ \citenamefont {Nagaosa}(2001)}]{shindou2001orbital}%
  \BibitemOpen
  \bibfield  {author} {\bibinfo {author} {\bibfnamefont {R.}~\bibnamefont {Shindou}}\ and\ \bibinfo {author} {\bibfnamefont {N.}~\bibnamefont {Nagaosa}},\ }\bibfield  {title} {\bibinfo {title} {Orbital ferromagnetism and anomalous hall effect in antiferromagnets on the distorted fcc lattice},\ }\href@noop {} {\bibfield  {journal} {\bibinfo  {journal} {Physical review letters}\ }\textbf {\bibinfo {volume} {87}},\ \bibinfo {pages} {116801} (\bibinfo {year} {2001})}\BibitemShut {NoStop}%
\bibitem [{\citenamefont {Fang}\ \emph {et~al.}(2003)\citenamefont {Fang}, \citenamefont {Nagaosa}, \citenamefont {Takahashi}, \citenamefont {Asamitsu}, \citenamefont {Mathieu}, \citenamefont {Ogasawara}, \citenamefont {Yamada}, \citenamefont {Kawasaki}, \citenamefont {Tokura},\ and\ \citenamefont {Terakura}}]{Fang2003}%
  \BibitemOpen
  \bibfield  {author} {\bibinfo {author} {\bibfnamefont {Z.}~\bibnamefont {Fang}}, \bibinfo {author} {\bibfnamefont {N.}~\bibnamefont {Nagaosa}}, \bibinfo {author} {\bibfnamefont {K.~S.}\ \bibnamefont {Takahashi}}, \bibinfo {author} {\bibfnamefont {A.}~\bibnamefont {Asamitsu}}, \bibinfo {author} {\bibfnamefont {R.}~\bibnamefont {Mathieu}}, \bibinfo {author} {\bibfnamefont {T.}~\bibnamefont {Ogasawara}}, \bibinfo {author} {\bibfnamefont {H.}~\bibnamefont {Yamada}}, \bibinfo {author} {\bibfnamefont {M.}~\bibnamefont {Kawasaki}}, \bibinfo {author} {\bibfnamefont {Y.}~\bibnamefont {Tokura}},\ and\ \bibinfo {author} {\bibfnamefont {K.}~\bibnamefont {Terakura}},\ }\bibfield  {title} {\bibinfo {title} {The anomalous hall effect and magnetic monopoles in momentum space},\ }\href {https://doi.org/10.1126/science.1089408} {\bibfield  {journal} {\bibinfo  {journal} {Science}\ }\textbf {\bibinfo {volume} {302}},\ \bibinfo {pages} {92} (\bibinfo {year} {2003})}\BibitemShut {NoStop}%
\bibitem [{\citenamefont {Huang}\ \emph {et~al.}(2015{\natexlab{a}})\citenamefont {Huang}, \citenamefont {Xu}, \citenamefont {Belopolski}, \citenamefont {Lee}, \citenamefont {Chang}, \citenamefont {Wang}, \citenamefont {Alidoust}, \citenamefont {Bian}, \citenamefont {Neupane}, \citenamefont {Zhang}, \citenamefont {Jia}, \citenamefont {Bansil}, \citenamefont {Lin},\ and\ \citenamefont {Hasan}}]{Huang15a}%
  \BibitemOpen
  \bibfield  {author} {\bibinfo {author} {\bibfnamefont {S.-M.}\ \bibnamefont {Huang}}, \bibinfo {author} {\bibfnamefont {S.-Y.}\ \bibnamefont {Xu}}, \bibinfo {author} {\bibfnamefont {I.}~\bibnamefont {Belopolski}}, \bibinfo {author} {\bibfnamefont {C.-C.}\ \bibnamefont {Lee}}, \bibinfo {author} {\bibfnamefont {G.}~\bibnamefont {Chang}}, \bibinfo {author} {\bibfnamefont {B.}~\bibnamefont {Wang}}, \bibinfo {author} {\bibfnamefont {N.}~\bibnamefont {Alidoust}}, \bibinfo {author} {\bibfnamefont {G.}~\bibnamefont {Bian}}, \bibinfo {author} {\bibfnamefont {M.}~\bibnamefont {Neupane}}, \bibinfo {author} {\bibfnamefont {C.}~\bibnamefont {Zhang}}, \bibinfo {author} {\bibfnamefont {S.}~\bibnamefont {Jia}}, \bibinfo {author} {\bibfnamefont {A.}~\bibnamefont {Bansil}}, \bibinfo {author} {\bibfnamefont {H.}~\bibnamefont {Lin}},\ and\ \bibinfo {author} {\bibfnamefont {M.~Z.}\ \bibnamefont {Hasan}},\ }\bibfield  {title} {\bibinfo {title} {{A {W}eyl {F}ermion semimetal with surface {F}ermi arcs in the transition metal
  monopnictide {T}a{A}s class}},\ }\href {https://doi.org/10.1038/ncomms8373} {\bibfield  {journal} {\bibinfo  {journal} {Nat. Commun.}\ }\textbf {\bibinfo {volume} {6}},\ \bibinfo {pages} {7373} (\bibinfo {year} {2015}{\natexlab{a}})}\BibitemShut {NoStop}%
\bibitem [{\citenamefont {Xu}\ \emph {et~al.}(2015{\natexlab{a}})\citenamefont {Xu}, \citenamefont {Belopolski}, \citenamefont {Alidoust}, \citenamefont {Neupane}, \citenamefont {Bian}, \citenamefont {Zhang}, \citenamefont {Sankar}, \citenamefont {Chang}, \citenamefont {Yuan}, \citenamefont {Lee} \emph {et~al.}}]{xu2015discovery}%
  \BibitemOpen
  \bibfield  {author} {\bibinfo {author} {\bibfnamefont {S.-Y.}\ \bibnamefont {Xu}}, \bibinfo {author} {\bibfnamefont {I.}~\bibnamefont {Belopolski}}, \bibinfo {author} {\bibfnamefont {N.}~\bibnamefont {Alidoust}}, \bibinfo {author} {\bibfnamefont {M.}~\bibnamefont {Neupane}}, \bibinfo {author} {\bibfnamefont {G.}~\bibnamefont {Bian}}, \bibinfo {author} {\bibfnamefont {C.}~\bibnamefont {Zhang}}, \bibinfo {author} {\bibfnamefont {R.}~\bibnamefont {Sankar}}, \bibinfo {author} {\bibfnamefont {G.}~\bibnamefont {Chang}}, \bibinfo {author} {\bibfnamefont {Z.}~\bibnamefont {Yuan}}, \bibinfo {author} {\bibfnamefont {C.-C.}\ \bibnamefont {Lee}}, \emph {et~al.},\ }\bibfield  {title} {\bibinfo {title} {Discovery of a weyl fermion semimetal and topological fermi arcs},\ }\href@noop {} {\bibfield  {journal} {\bibinfo  {journal} {Science}\ }\textbf {\bibinfo {volume} {349}},\ \bibinfo {pages} {613} (\bibinfo {year} {2015}{\natexlab{a}})}\BibitemShut {NoStop}%
\bibitem [{\citenamefont {Lv}\ \emph {et~al.}(2015{\natexlab{a}})\citenamefont {Lv}, \citenamefont {Weng}, \citenamefont {Fu}, \citenamefont {Wang}, \citenamefont {Miao}, \citenamefont {Ma}, \citenamefont {Richard}, \citenamefont {Huang}, \citenamefont {Zhao}, \citenamefont {Chen} \emph {et~al.}}]{lv2015experimental}%
  \BibitemOpen
  \bibfield  {author} {\bibinfo {author} {\bibfnamefont {B.}~\bibnamefont {Lv}}, \bibinfo {author} {\bibfnamefont {H.}~\bibnamefont {Weng}}, \bibinfo {author} {\bibfnamefont {B.}~\bibnamefont {Fu}}, \bibinfo {author} {\bibfnamefont {X.~P.}\ \bibnamefont {Wang}}, \bibinfo {author} {\bibfnamefont {H.}~\bibnamefont {Miao}}, \bibinfo {author} {\bibfnamefont {J.}~\bibnamefont {Ma}}, \bibinfo {author} {\bibfnamefont {P.}~\bibnamefont {Richard}}, \bibinfo {author} {\bibfnamefont {X.}~\bibnamefont {Huang}}, \bibinfo {author} {\bibfnamefont {L.}~\bibnamefont {Zhao}}, \bibinfo {author} {\bibfnamefont {G.}~\bibnamefont {Chen}}, \emph {et~al.},\ }\bibfield  {title} {\bibinfo {title} {Experimental discovery of weyl semimetal taas},\ }\href@noop {} {\bibfield  {journal} {\bibinfo  {journal} {Physical Review X}\ }\textbf {\bibinfo {volume} {5}},\ \bibinfo {pages} {031013} (\bibinfo {year} {2015}{\natexlab{a}})}\BibitemShut {NoStop}%
\bibitem [{\citenamefont {Yang}\ \emph {et~al.}(2015)\citenamefont {Yang}, \citenamefont {Liu}, \citenamefont {Sun}, \citenamefont {Peng}, \citenamefont {Yang}, \citenamefont {Zhang}, \citenamefont {Zhou}, \citenamefont {Zhang}, \citenamefont {Guo}, \citenamefont {Rahn}, \citenamefont {Prabhakaran}, \citenamefont {Hussain}, \citenamefont {Mo}, \citenamefont {Felser}, \citenamefont {Yan},\ and\ \citenamefont {Chen}}]{Yang15Weyl}%
  \BibitemOpen
  \bibfield  {author} {\bibinfo {author} {\bibfnamefont {L.~X.}\ \bibnamefont {Yang}}, \bibinfo {author} {\bibfnamefont {Z.~K.}\ \bibnamefont {Liu}}, \bibinfo {author} {\bibfnamefont {Y.}~\bibnamefont {Sun}}, \bibinfo {author} {\bibfnamefont {H.}~\bibnamefont {Peng}}, \bibinfo {author} {\bibfnamefont {H.~F.}\ \bibnamefont {Yang}}, \bibinfo {author} {\bibfnamefont {T.}~\bibnamefont {Zhang}}, \bibinfo {author} {\bibfnamefont {B.}~\bibnamefont {Zhou}}, \bibinfo {author} {\bibfnamefont {Y.}~\bibnamefont {Zhang}}, \bibinfo {author} {\bibfnamefont {Y.~F.}\ \bibnamefont {Guo}}, \bibinfo {author} {\bibfnamefont {M.}~\bibnamefont {Rahn}}, \bibinfo {author} {\bibfnamefont {D.}~\bibnamefont {Prabhakaran}}, \bibinfo {author} {\bibfnamefont {Z.}~\bibnamefont {Hussain}}, \bibinfo {author} {\bibfnamefont {S.-K.}\ \bibnamefont {Mo}}, \bibinfo {author} {\bibfnamefont {C.}~\bibnamefont {Felser}}, \bibinfo {author} {\bibfnamefont {B.}~\bibnamefont {Yan}},\ and\ \bibinfo {author} {\bibfnamefont {Y.~L.}\ \bibnamefont {Chen}},\
  }\bibfield  {title} {\bibinfo {title} {{Weyl semimetal phase in the non-centrosymmetric compound TaAs}},\ }\href {https://doi.org/10.1038/nphys3425} {\bibfield  {journal} {\bibinfo  {journal} {Nat. Phys.}\ }\textbf {\bibinfo {volume} {11}},\ \bibinfo {pages} {728} (\bibinfo {year} {2015})}\BibitemShut {NoStop}%
\bibitem [{\citenamefont {Lv}\ \emph {et~al.}(2015{\natexlab{b}})\citenamefont {Lv}, \citenamefont {Xu}, \citenamefont {Weng}, \citenamefont {Ma}, \citenamefont {Richard}, \citenamefont {Huang}, \citenamefont {Zhao}, \citenamefont {Chen}, \citenamefont {Matt}, \citenamefont {Bisti} \emph {et~al.}}]{lv2015observation}%
  \BibitemOpen
  \bibfield  {author} {\bibinfo {author} {\bibfnamefont {B.}~\bibnamefont {Lv}}, \bibinfo {author} {\bibfnamefont {N.}~\bibnamefont {Xu}}, \bibinfo {author} {\bibfnamefont {H.}~\bibnamefont {Weng}}, \bibinfo {author} {\bibfnamefont {J.}~\bibnamefont {Ma}}, \bibinfo {author} {\bibfnamefont {P.}~\bibnamefont {Richard}}, \bibinfo {author} {\bibfnamefont {X.}~\bibnamefont {Huang}}, \bibinfo {author} {\bibfnamefont {L.}~\bibnamefont {Zhao}}, \bibinfo {author} {\bibfnamefont {G.}~\bibnamefont {Chen}}, \bibinfo {author} {\bibfnamefont {C.}~\bibnamefont {Matt}}, \bibinfo {author} {\bibfnamefont {F.}~\bibnamefont {Bisti}}, \emph {et~al.},\ }\bibfield  {title} {\bibinfo {title} {Observation of weyl nodes in taas},\ }\href@noop {} {\bibfield  {journal} {\bibinfo  {journal} {Nature Physics}\ }\textbf {\bibinfo {volume} {11}},\ \bibinfo {pages} {724} (\bibinfo {year} {2015}{\natexlab{b}})}\BibitemShut {NoStop}%
\bibitem [{\citenamefont {Huang}\ \emph {et~al.}(2015{\natexlab{b}})\citenamefont {Huang}, \citenamefont {Zhao}, \citenamefont {Long}, \citenamefont {Wang}, \citenamefont {Chen}, \citenamefont {Yang}, \citenamefont {Liang}, \citenamefont {Xue}, \citenamefont {Weng}, \citenamefont {Fang} \emph {et~al.}}]{huang2015observation}%
  \BibitemOpen
  \bibfield  {author} {\bibinfo {author} {\bibfnamefont {X.}~\bibnamefont {Huang}}, \bibinfo {author} {\bibfnamefont {L.}~\bibnamefont {Zhao}}, \bibinfo {author} {\bibfnamefont {Y.}~\bibnamefont {Long}}, \bibinfo {author} {\bibfnamefont {P.}~\bibnamefont {Wang}}, \bibinfo {author} {\bibfnamefont {D.}~\bibnamefont {Chen}}, \bibinfo {author} {\bibfnamefont {Z.}~\bibnamefont {Yang}}, \bibinfo {author} {\bibfnamefont {H.}~\bibnamefont {Liang}}, \bibinfo {author} {\bibfnamefont {M.}~\bibnamefont {Xue}}, \bibinfo {author} {\bibfnamefont {H.}~\bibnamefont {Weng}}, \bibinfo {author} {\bibfnamefont {Z.}~\bibnamefont {Fang}}, \emph {et~al.},\ }\bibfield  {title} {\bibinfo {title} {Observation of the chiral-anomaly-induced negative magnetoresistance in 3d weyl semimetal taas},\ }\href@noop {} {\bibfield  {journal} {\bibinfo  {journal} {Physical Review X}\ }\textbf {\bibinfo {volume} {5}},\ \bibinfo {pages} {031023} (\bibinfo {year} {2015}{\natexlab{b}})}\BibitemShut {NoStop}%
\bibitem [{\citenamefont {Kuroda}\ \emph {et~al.}(2017)\citenamefont {Kuroda}, \citenamefont {Tomita}, \citenamefont {Suzuki}, \citenamefont {Bareille}, \citenamefont {Nugroho}, \citenamefont {Goswami}, \citenamefont {Ochi}, \citenamefont {Ikhlas}, \citenamefont {Nakayama}, \citenamefont {Akebi} \emph {et~al.}}]{Kuroda2017}%
  \BibitemOpen
  \bibfield  {author} {\bibinfo {author} {\bibfnamefont {K.}~\bibnamefont {Kuroda}}, \bibinfo {author} {\bibfnamefont {T.}~\bibnamefont {Tomita}}, \bibinfo {author} {\bibfnamefont {M.-T.}\ \bibnamefont {Suzuki}}, \bibinfo {author} {\bibfnamefont {C.}~\bibnamefont {Bareille}}, \bibinfo {author} {\bibfnamefont {A.}~\bibnamefont {Nugroho}}, \bibinfo {author} {\bibfnamefont {P.}~\bibnamefont {Goswami}}, \bibinfo {author} {\bibfnamefont {M.}~\bibnamefont {Ochi}}, \bibinfo {author} {\bibfnamefont {M.}~\bibnamefont {Ikhlas}}, \bibinfo {author} {\bibfnamefont {M.}~\bibnamefont {Nakayama}}, \bibinfo {author} {\bibfnamefont {S.}~\bibnamefont {Akebi}}, \emph {et~al.},\ }\bibfield  {title} {\bibinfo {title} {Evidence for magnetic weyl fermions in a correlated metal},\ }\href@noop {} {\bibfield  {journal} {\bibinfo  {journal} {Nature materials}\ }\textbf {\bibinfo {volume} {16}},\ \bibinfo {pages} {1090} (\bibinfo {year} {2017})}\BibitemShut {NoStop}%
\bibitem [{\citenamefont {Nagaosa}\ \emph {et~al.}(2010)\citenamefont {Nagaosa}, \citenamefont {Sinova}, \citenamefont {Onoda}, \citenamefont {MacDonald},\ and\ \citenamefont {Ong}}]{Nagaosa2010}%
  \BibitemOpen
  \bibfield  {author} {\bibinfo {author} {\bibfnamefont {N.}~\bibnamefont {Nagaosa}}, \bibinfo {author} {\bibfnamefont {J.}~\bibnamefont {Sinova}}, \bibinfo {author} {\bibfnamefont {S.}~\bibnamefont {Onoda}}, \bibinfo {author} {\bibfnamefont {A.~H.}\ \bibnamefont {MacDonald}},\ and\ \bibinfo {author} {\bibfnamefont {N.~P.}\ \bibnamefont {Ong}},\ }\bibfield  {title} {\bibinfo {title} {Anomalous hall effect},\ }\href@noop {} {\bibfield  {journal} {\bibinfo  {journal} {Rev. Mod. Phys.}\ }\textbf {\bibinfo {volume} {82}},\ \bibinfo {pages} {1539} (\bibinfo {year} {2010})}\BibitemShut {NoStop}%
\bibitem [{\citenamefont {Xiao}\ \emph {et~al.}(2010)\citenamefont {Xiao}, \citenamefont {Chang},\ and\ \citenamefont {Niu}}]{xiao2010berry}%
  \BibitemOpen
  \bibfield  {author} {\bibinfo {author} {\bibfnamefont {D.}~\bibnamefont {Xiao}}, \bibinfo {author} {\bibfnamefont {M.-C.}\ \bibnamefont {Chang}},\ and\ \bibinfo {author} {\bibfnamefont {Q.}~\bibnamefont {Niu}},\ }\bibfield  {title} {\bibinfo {title} {Berry phase effects on electronic properties},\ }\href {https://doi.org/10.1103/RevModPhys.82.1959} {\bibfield  {journal} {\bibinfo  {journal} {Rev. Mod. Phys.}\ }\textbf {\bibinfo {volume} {82}},\ \bibinfo {pages} {1959} (\bibinfo {year} {2010})}\BibitemShut {NoStop}%
\bibitem [{\citenamefont {Klein}(1926)}]{klein1926quantentheorie}%
  \BibitemOpen
  \bibfield  {author} {\bibinfo {author} {\bibfnamefont {O.}~\bibnamefont {Klein}},\ }\bibfield  {title} {\bibinfo {title} {Quantentheorie und f{\"u}nfdimensionale relativit{\"a}tstheorie},\ }\href@noop {} {\bibfield  {journal} {\bibinfo  {journal} {Zeitschrift f{\"u}r Physik}\ }\textbf {\bibinfo {volume} {37}},\ \bibinfo {pages} {895} (\bibinfo {year} {1926})}\BibitemShut {NoStop}%
\bibitem [{\citenamefont {Gordon}(1926)}]{gordon1926comptoneffekt}%
  \BibitemOpen
  \bibfield  {author} {\bibinfo {author} {\bibfnamefont {W.}~\bibnamefont {Gordon}},\ }\bibfield  {title} {\bibinfo {title} {Der comptoneffekt nach der schr{\"o}dingerschen theorie},\ }\href@noop {} {\bibfield  {journal} {\bibinfo  {journal} {Zeitschrift f{\"u}r Physik}\ }\textbf {\bibinfo {volume} {40}},\ \bibinfo {pages} {117} (\bibinfo {year} {1926})}\BibitemShut {NoStop}%
\bibitem [{\citenamefont {Fuseya}\ \emph {et~al.}(2015)\citenamefont {Fuseya}, \citenamefont {Ogata},\ and\ \citenamefont {Fukuyama}}]{fuseya2015}%
  \BibitemOpen
  \bibfield  {author} {\bibinfo {author} {\bibfnamefont {Y.}~\bibnamefont {Fuseya}}, \bibinfo {author} {\bibfnamefont {M.}~\bibnamefont {Ogata}},\ and\ \bibinfo {author} {\bibfnamefont {H.}~\bibnamefont {Fukuyama}},\ }\bibfield  {title} {\bibinfo {title} {Transport properties and diamagnetism of dirac electrons in bismuth},\ }\href@noop {} {\bibfield  {journal} {\bibinfo  {journal} {Journal of the Physical Society of Japan}\ }\textbf {\bibinfo {volume} {84}},\ \bibinfo {pages} {012001} (\bibinfo {year} {2015})}\BibitemShut {NoStop}%
\bibitem [{\citenamefont {Fang}\ \emph {et~al.}(2012)\citenamefont {Fang}, \citenamefont {Gilbert}, \citenamefont {Dai},\ and\ \citenamefont {Bernevig}}]{fang2012multi}%
  \BibitemOpen
  \bibfield  {author} {\bibinfo {author} {\bibfnamefont {C.}~\bibnamefont {Fang}}, \bibinfo {author} {\bibfnamefont {M.~J.}\ \bibnamefont {Gilbert}}, \bibinfo {author} {\bibfnamefont {X.}~\bibnamefont {Dai}},\ and\ \bibinfo {author} {\bibfnamefont {B.~A.}\ \bibnamefont {Bernevig}},\ }\bibfield  {title} {\bibinfo {title} {Multi-weyl topological semimetals stabilized by point group symmetry},\ }\href@noop {} {\bibfield  {journal} {\bibinfo  {journal} {Physical review letters}\ }\textbf {\bibinfo {volume} {108}},\ \bibinfo {pages} {266802} (\bibinfo {year} {2012})}\BibitemShut {NoStop}%
\bibitem [{\citenamefont {Goswami}\ \emph {et~al.}(2017)\citenamefont {Goswami}, \citenamefont {Roy},\ and\ \citenamefont {Das~Sarma}}]{goswami2017competing}%
  \BibitemOpen
  \bibfield  {author} {\bibinfo {author} {\bibfnamefont {P.}~\bibnamefont {Goswami}}, \bibinfo {author} {\bibfnamefont {B.}~\bibnamefont {Roy}},\ and\ \bibinfo {author} {\bibfnamefont {S.}~\bibnamefont {Das~Sarma}},\ }\bibfield  {title} {\bibinfo {title} {Competing orders and topology in the global phase diagram of pyrochlore iridates},\ }\href@noop {} {\bibfield  {journal} {\bibinfo  {journal} {Physical Review B}\ }\textbf {\bibinfo {volume} {95}},\ \bibinfo {pages} {085120} (\bibinfo {year} {2017})}\BibitemShut {NoStop}%
\bibitem [{\citenamefont {Szab{\'o}}\ \emph {et~al.}(2021)\citenamefont {Szab{\'o}}, \citenamefont {Moessner},\ and\ \citenamefont {Roy}}]{szabo2021interacting}%
  \BibitemOpen
  \bibfield  {author} {\bibinfo {author} {\bibfnamefont {A.~L.}\ \bibnamefont {Szab{\'o}}}, \bibinfo {author} {\bibfnamefont {R.}~\bibnamefont {Moessner}},\ and\ \bibinfo {author} {\bibfnamefont {B.}~\bibnamefont {Roy}},\ }\bibfield  {title} {\bibinfo {title} {Interacting spin-3 2 fermions in a luttinger semimetal: Competing phases and their selection in the global phase diagram},\ }\href@noop {} {\bibfield  {journal} {\bibinfo  {journal} {Physical Review B}\ }\textbf {\bibinfo {volume} {103}},\ \bibinfo {pages} {165139} (\bibinfo {year} {2021})}\BibitemShut {NoStop}%
\bibitem [{\citenamefont {Ladovrechis}\ \emph {et~al.}(2021)\citenamefont {Ladovrechis}, \citenamefont {Meng},\ and\ \citenamefont {Roy}}]{ladovrechis2021competing}%
  \BibitemOpen
  \bibfield  {author} {\bibinfo {author} {\bibfnamefont {K.}~\bibnamefont {Ladovrechis}}, \bibinfo {author} {\bibfnamefont {T.}~\bibnamefont {Meng}},\ and\ \bibinfo {author} {\bibfnamefont {B.}~\bibnamefont {Roy}},\ }\bibfield  {title} {\bibinfo {title} {Competing magnetic orders and multipolar weyl fermions in 227 pyrochlore iridates},\ }\href@noop {} {\bibfield  {journal} {\bibinfo  {journal} {Physical Review B}\ }\textbf {\bibinfo {volume} {103}},\ \bibinfo {pages} {L241116} (\bibinfo {year} {2021})}\BibitemShut {NoStop}%
\bibitem [{\citenamefont {Creutz}(1983)}]{creutz1983quarks}%
  \BibitemOpen
  \bibfield  {author} {\bibinfo {author} {\bibfnamefont {M.}~\bibnamefont {Creutz}},\ }\href@noop {} {\emph {\bibinfo {title} {Quarks, gluons and lattices}}}\ (\bibinfo  {publisher} {Cambridge University Press},\ \bibinfo {year} {1983})\BibitemShut {NoStop}%
\bibitem [{\citenamefont {Rothe}(2012)}]{rothe2012lattice}%
  \BibitemOpen
  \bibfield  {author} {\bibinfo {author} {\bibfnamefont {H.~J.}\ \bibnamefont {Rothe}},\ }\href@noop {} {\emph {\bibinfo {title} {Lattice gauge theories: an introduction}}}\ (\bibinfo  {publisher} {World Scientific Publishing Company},\ \bibinfo {year} {2012})\BibitemShut {NoStop}%
\bibitem [{\citenamefont {Wilson}(1977)}]{Wilson1977}%
  \BibitemOpen
  \bibfield  {author} {\bibinfo {author} {\bibfnamefont {K.~G.}\ \bibnamefont {Wilson}},\ }\bibinfo {title} {Quarks and strings on a lattice},\ in\ \href {https://doi.org/10.1007/978-1-4613-4208-3_6} {\emph {\bibinfo {booktitle} {New Phenomena in Subnuclear Physics: Part A}}},\ \bibinfo {editor} {edited by\ \bibinfo {editor} {\bibfnamefont {A.}~\bibnamefont {Zichichi}}}\ (\bibinfo  {publisher} {Springer US},\ \bibinfo {address} {Boston, MA},\ \bibinfo {year} {1977})\ pp.\ \bibinfo {pages} {69--142}\BibitemShut {NoStop}%
\bibitem [{\citenamefont {Qi}\ and\ \citenamefont {Zhang}(2011)}]{qi2011topological}%
  \BibitemOpen
  \bibfield  {author} {\bibinfo {author} {\bibfnamefont {X.-L.}\ \bibnamefont {Qi}}\ and\ \bibinfo {author} {\bibfnamefont {S.-C.}\ \bibnamefont {Zhang}},\ }\bibfield  {title} {\bibinfo {title} {Topological insulators and superconductors},\ }\href@noop {} {\bibfield  {journal} {\bibinfo  {journal} {Reviews of modern physics}\ }\textbf {\bibinfo {volume} {83}},\ \bibinfo {pages} {1057} (\bibinfo {year} {2011})}\BibitemShut {NoStop}%
\bibitem [{\citenamefont {Young}\ \emph {et~al.}(2012)\citenamefont {Young}, \citenamefont {Zaheer}, \citenamefont {Teo}, \citenamefont {Kane}, \citenamefont {Mele},\ and\ \citenamefont {Rappe}}]{Young2012}%
  \BibitemOpen
  \bibfield  {author} {\bibinfo {author} {\bibfnamefont {S.~M.}\ \bibnamefont {Young}}, \bibinfo {author} {\bibfnamefont {S.}~\bibnamefont {Zaheer}}, \bibinfo {author} {\bibfnamefont {J.~C.}\ \bibnamefont {Teo}}, \bibinfo {author} {\bibfnamefont {C.~L.}\ \bibnamefont {Kane}}, \bibinfo {author} {\bibfnamefont {E.~J.}\ \bibnamefont {Mele}},\ and\ \bibinfo {author} {\bibfnamefont {A.~M.}\ \bibnamefont {Rappe}},\ }\bibfield  {title} {\bibinfo {title} {Dirac semimetal in three dimensions},\ }\href@noop {} {\bibfield  {journal} {\bibinfo  {journal} {Physical review letters}\ }\textbf {\bibinfo {volume} {108}},\ \bibinfo {pages} {140405} (\bibinfo {year} {2012})}\BibitemShut {NoStop}%
\bibitem [{\citenamefont {Wang}\ \emph {et~al.}(2012)\citenamefont {Wang}, \citenamefont {Sun}, \citenamefont {Chen}, \citenamefont {Franchini}, \citenamefont {Xu}, \citenamefont {Weng}, \citenamefont {Dai},\ and\ \citenamefont {Fang}}]{wang2012dirac}%
  \BibitemOpen
  \bibfield  {author} {\bibinfo {author} {\bibfnamefont {Z.}~\bibnamefont {Wang}}, \bibinfo {author} {\bibfnamefont {Y.}~\bibnamefont {Sun}}, \bibinfo {author} {\bibfnamefont {X.-Q.}\ \bibnamefont {Chen}}, \bibinfo {author} {\bibfnamefont {C.}~\bibnamefont {Franchini}}, \bibinfo {author} {\bibfnamefont {G.}~\bibnamefont {Xu}}, \bibinfo {author} {\bibfnamefont {H.}~\bibnamefont {Weng}}, \bibinfo {author} {\bibfnamefont {X.}~\bibnamefont {Dai}},\ and\ \bibinfo {author} {\bibfnamefont {Z.}~\bibnamefont {Fang}},\ }\bibfield  {title} {\bibinfo {title} {Dirac semimetal and topological phase transitions in a 3 bi (a= na, k, rb)},\ }\href@noop {} {\bibfield  {journal} {\bibinfo  {journal} {Physical Review B—Condensed Matter and Materials Physics}\ }\textbf {\bibinfo {volume} {85}},\ \bibinfo {pages} {195320} (\bibinfo {year} {2012})}\BibitemShut {NoStop}%
\bibitem [{\citenamefont {Liu}\ \emph {et~al.}(2014)\citenamefont {Liu}, \citenamefont {Zhou}, \citenamefont {Zhang}, \citenamefont {Wang}, \citenamefont {Weng}, \citenamefont {Prabhakaran}, \citenamefont {Mo}, \citenamefont {Shen}, \citenamefont {Fang}, \citenamefont {Dai} \emph {et~al.}}]{Liu2014}%
  \BibitemOpen
  \bibfield  {author} {\bibinfo {author} {\bibfnamefont {Z.}~\bibnamefont {Liu}}, \bibinfo {author} {\bibfnamefont {B.}~\bibnamefont {Zhou}}, \bibinfo {author} {\bibfnamefont {Y.}~\bibnamefont {Zhang}}, \bibinfo {author} {\bibfnamefont {Z.}~\bibnamefont {Wang}}, \bibinfo {author} {\bibfnamefont {H.}~\bibnamefont {Weng}}, \bibinfo {author} {\bibfnamefont {D.}~\bibnamefont {Prabhakaran}}, \bibinfo {author} {\bibfnamefont {S.-K.}\ \bibnamefont {Mo}}, \bibinfo {author} {\bibfnamefont {Z.}~\bibnamefont {Shen}}, \bibinfo {author} {\bibfnamefont {Z.}~\bibnamefont {Fang}}, \bibinfo {author} {\bibfnamefont {X.}~\bibnamefont {Dai}}, \emph {et~al.},\ }\bibfield  {title} {\bibinfo {title} {Discovery of a three-dimensional topological dirac semimetal, na3bi},\ }\href@noop {} {\bibfield  {journal} {\bibinfo  {journal} {Science}\ }\textbf {\bibinfo {volume} {343}},\ \bibinfo {pages} {864} (\bibinfo {year} {2014})}\BibitemShut {NoStop}%
\bibitem [{\citenamefont {Wang}\ \emph {et~al.}(2013)\citenamefont {Wang}, \citenamefont {Weng}, \citenamefont {Wu}, \citenamefont {Dai},\ and\ \citenamefont {Fang}}]{wang2013three}%
  \BibitemOpen
  \bibfield  {author} {\bibinfo {author} {\bibfnamefont {Z.}~\bibnamefont {Wang}}, \bibinfo {author} {\bibfnamefont {H.}~\bibnamefont {Weng}}, \bibinfo {author} {\bibfnamefont {Q.}~\bibnamefont {Wu}}, \bibinfo {author} {\bibfnamefont {X.}~\bibnamefont {Dai}},\ and\ \bibinfo {author} {\bibfnamefont {Z.}~\bibnamefont {Fang}},\ }\bibfield  {title} {\bibinfo {title} {Three-dimensional dirac semimetal and quantum transport in cd 3 as 2},\ }\href@noop {} {\bibfield  {journal} {\bibinfo  {journal} {Physical Review B—Condensed Matter and Materials Physics}\ }\textbf {\bibinfo {volume} {88}},\ \bibinfo {pages} {125427} (\bibinfo {year} {2013})}\BibitemShut {NoStop}%
\bibitem [{\citenamefont {Neupane}\ \emph {et~al.}(2014)\citenamefont {Neupane}, \citenamefont {Xu}, \citenamefont {Sankar}, \citenamefont {Alidoust}, \citenamefont {Bian}, \citenamefont {Liu}, \citenamefont {Belopolski}, \citenamefont {Chang}, \citenamefont {Jeng}, \citenamefont {Lin} \emph {et~al.}}]{neupane2014observation}%
  \BibitemOpen
  \bibfield  {author} {\bibinfo {author} {\bibfnamefont {M.}~\bibnamefont {Neupane}}, \bibinfo {author} {\bibfnamefont {S.-Y.}\ \bibnamefont {Xu}}, \bibinfo {author} {\bibfnamefont {R.}~\bibnamefont {Sankar}}, \bibinfo {author} {\bibfnamefont {N.}~\bibnamefont {Alidoust}}, \bibinfo {author} {\bibfnamefont {G.}~\bibnamefont {Bian}}, \bibinfo {author} {\bibfnamefont {C.}~\bibnamefont {Liu}}, \bibinfo {author} {\bibfnamefont {I.}~\bibnamefont {Belopolski}}, \bibinfo {author} {\bibfnamefont {T.-R.}\ \bibnamefont {Chang}}, \bibinfo {author} {\bibfnamefont {H.-T.}\ \bibnamefont {Jeng}}, \bibinfo {author} {\bibfnamefont {H.}~\bibnamefont {Lin}}, \emph {et~al.},\ }\bibfield  {title} {\bibinfo {title} {Observation of a three-dimensional topological dirac semimetal phase in high-mobility cd3as2},\ }\href@noop {} {\bibfield  {journal} {\bibinfo  {journal} {Nature communications}\ }\textbf {\bibinfo {volume} {5}},\ \bibinfo {pages} {3786} (\bibinfo {year} {2014})}\BibitemShut {NoStop}%
\bibitem [{\citenamefont {Uchida}\ \emph {et~al.}(2017)\citenamefont {Uchida}, \citenamefont {Nakazawa}, \citenamefont {Nishihaya}, \citenamefont {Akiba}, \citenamefont {Kriener}, \citenamefont {Kozuka}, \citenamefont {Miyake}, \citenamefont {Taguchi}, \citenamefont {Tokunaga}, \citenamefont {Nagaosa} \emph {et~al.}}]{uchida2017quantum}%
  \BibitemOpen
  \bibfield  {author} {\bibinfo {author} {\bibfnamefont {M.}~\bibnamefont {Uchida}}, \bibinfo {author} {\bibfnamefont {Y.}~\bibnamefont {Nakazawa}}, \bibinfo {author} {\bibfnamefont {S.}~\bibnamefont {Nishihaya}}, \bibinfo {author} {\bibfnamefont {K.}~\bibnamefont {Akiba}}, \bibinfo {author} {\bibfnamefont {M.}~\bibnamefont {Kriener}}, \bibinfo {author} {\bibfnamefont {Y.}~\bibnamefont {Kozuka}}, \bibinfo {author} {\bibfnamefont {A.}~\bibnamefont {Miyake}}, \bibinfo {author} {\bibfnamefont {Y.}~\bibnamefont {Taguchi}}, \bibinfo {author} {\bibfnamefont {M.}~\bibnamefont {Tokunaga}}, \bibinfo {author} {\bibfnamefont {N.}~\bibnamefont {Nagaosa}}, \emph {et~al.},\ }\bibfield  {title} {\bibinfo {title} {Quantum hall states observed in thin films of dirac semimetal cd3as2},\ }\href@noop {} {\bibfield  {journal} {\bibinfo  {journal} {Nature communications}\ }\textbf {\bibinfo {volume} {8}},\ \bibinfo {pages} {2274} (\bibinfo {year} {2017})}\BibitemShut {NoStop}%
\bibitem [{\citenamefont {Uchida}\ \emph {et~al.}(2019)\citenamefont {Uchida}, \citenamefont {Koretsune}, \citenamefont {Sato}, \citenamefont {Kriener}, \citenamefont {Nakazawa}, \citenamefont {Nishihaya}, \citenamefont {Taguchi}, \citenamefont {Arita},\ and\ \citenamefont {Kawasaki}}]{uchida2019ferromagnetic}%
  \BibitemOpen
  \bibfield  {author} {\bibinfo {author} {\bibfnamefont {M.}~\bibnamefont {Uchida}}, \bibinfo {author} {\bibfnamefont {T.}~\bibnamefont {Koretsune}}, \bibinfo {author} {\bibfnamefont {S.}~\bibnamefont {Sato}}, \bibinfo {author} {\bibfnamefont {M.}~\bibnamefont {Kriener}}, \bibinfo {author} {\bibfnamefont {Y.}~\bibnamefont {Nakazawa}}, \bibinfo {author} {\bibfnamefont {S.}~\bibnamefont {Nishihaya}}, \bibinfo {author} {\bibfnamefont {Y.}~\bibnamefont {Taguchi}}, \bibinfo {author} {\bibfnamefont {R.}~\bibnamefont {Arita}},\ and\ \bibinfo {author} {\bibfnamefont {M.}~\bibnamefont {Kawasaki}},\ }\bibfield  {title} {\bibinfo {title} {Ferromagnetic state above room temperature in a proximitized topological dirac semimetal},\ }\href@noop {} {\bibfield  {journal} {\bibinfo  {journal} {Physical Review B}\ }\textbf {\bibinfo {volume} {100}},\ \bibinfo {pages} {245148} (\bibinfo {year} {2019})}\BibitemShut {NoStop}%
\bibitem [{\citenamefont {Xu}\ \emph {et~al.}(2011)\citenamefont {Xu}, \citenamefont {Weng}, \citenamefont {Wang}, \citenamefont {Dai},\ and\ \citenamefont {Fang}}]{xu2011chern}%
  \BibitemOpen
  \bibfield  {author} {\bibinfo {author} {\bibfnamefont {G.}~\bibnamefont {Xu}}, \bibinfo {author} {\bibfnamefont {H.}~\bibnamefont {Weng}}, \bibinfo {author} {\bibfnamefont {Z.}~\bibnamefont {Wang}}, \bibinfo {author} {\bibfnamefont {X.}~\bibnamefont {Dai}},\ and\ \bibinfo {author} {\bibfnamefont {Z.}~\bibnamefont {Fang}},\ }\bibfield  {title} {\bibinfo {title} {Chern semimetal and the quantized anomalous hall effect in hgcr 2 se 4},\ }\href@noop {} {\bibfield  {journal} {\bibinfo  {journal} {Physical review letters}\ }\textbf {\bibinfo {volume} {107}},\ \bibinfo {pages} {186806} (\bibinfo {year} {2011})}\BibitemShut {NoStop}%
\bibitem [{\citenamefont {Huang}\ \emph {et~al.}(2016)\citenamefont {Huang}, \citenamefont {Xu}, \citenamefont {Belopolski}, \citenamefont {Lee}, \citenamefont {Chang}, \citenamefont {Chang}, \citenamefont {Wang}, \citenamefont {Alidoust}, \citenamefont {Bian}, \citenamefont {Neupane} \emph {et~al.}}]{huang2016new}%
  \BibitemOpen
  \bibfield  {author} {\bibinfo {author} {\bibfnamefont {S.-M.}\ \bibnamefont {Huang}}, \bibinfo {author} {\bibfnamefont {S.-Y.}\ \bibnamefont {Xu}}, \bibinfo {author} {\bibfnamefont {I.}~\bibnamefont {Belopolski}}, \bibinfo {author} {\bibfnamefont {C.-C.}\ \bibnamefont {Lee}}, \bibinfo {author} {\bibfnamefont {G.}~\bibnamefont {Chang}}, \bibinfo {author} {\bibfnamefont {T.-R.}\ \bibnamefont {Chang}}, \bibinfo {author} {\bibfnamefont {B.}~\bibnamefont {Wang}}, \bibinfo {author} {\bibfnamefont {N.}~\bibnamefont {Alidoust}}, \bibinfo {author} {\bibfnamefont {G.}~\bibnamefont {Bian}}, \bibinfo {author} {\bibfnamefont {M.}~\bibnamefont {Neupane}}, \emph {et~al.},\ }\bibfield  {title} {\bibinfo {title} {New type of weyl semimetal with quadratic double weyl fermions},\ }\href@noop {} {\bibfield  {journal} {\bibinfo  {journal} {Proceedings of the National Academy of Sciences}\ }\textbf {\bibinfo {volume} {113}},\ \bibinfo {pages} {1180} (\bibinfo {year} {2016})}\BibitemShut {NoStop}%
\bibitem [{\citenamefont {Nielsen}\ and\ \citenamefont {Ninomiya}(1981)}]{nielsen1981absence}%
  \BibitemOpen
  \bibfield  {author} {\bibinfo {author} {\bibfnamefont {H.~B.}\ \bibnamefont {Nielsen}}\ and\ \bibinfo {author} {\bibfnamefont {M.}~\bibnamefont {Ninomiya}},\ }\bibfield  {title} {\bibinfo {title} {Absence of neutrinos on a lattice:(i). proof by homotopy theory},\ }\href@noop {} {\bibfield  {journal} {\bibinfo  {journal} {Nuclear Physics B}\ }\textbf {\bibinfo {volume} {185}},\ \bibinfo {pages} {20} (\bibinfo {year} {1981})}\BibitemShut {NoStop}%
\bibitem [{\citenamefont {Nielsen}\ and\ \citenamefont {Ninomiya}(1983)}]{nielsen1983adler}%
  \BibitemOpen
  \bibfield  {author} {\bibinfo {author} {\bibfnamefont {H.~B.}\ \bibnamefont {Nielsen}}\ and\ \bibinfo {author} {\bibfnamefont {M.}~\bibnamefont {Ninomiya}},\ }\bibfield  {title} {\bibinfo {title} {The adler-bell-jackiw anomaly and weyl fermions in a crystal},\ }\href@noop {} {\bibfield  {journal} {\bibinfo  {journal} {Physics Letters B}\ }\textbf {\bibinfo {volume} {130}},\ \bibinfo {pages} {389} (\bibinfo {year} {1983})}\BibitemShut {NoStop}%
\bibitem [{\citenamefont {Berry}(1984)}]{berry1984}%
  \BibitemOpen
  \bibfield  {author} {\bibinfo {author} {\bibfnamefont {M.~V.}\ \bibnamefont {Berry}},\ }\bibfield  {title} {\bibinfo {title} {Quantal phase factors accompanying adiabatic changes},\ }\href@noop {} {\bibfield  {journal} {\bibinfo  {journal} {Proceedings of the Royal Society of London. A. Mathematical and Physical Sciences}\ }\textbf {\bibinfo {volume} {392}},\ \bibinfo {pages} {45} (\bibinfo {year} {1984})}\BibitemShut {NoStop}%
\bibitem [{\citenamefont {Liu}\ \emph {et~al.}(2018)\citenamefont {Liu}, \citenamefont {Sun}, \citenamefont {Kumar}, \citenamefont {Muechler}, \citenamefont {Sun}, \citenamefont {Jiao}, \citenamefont {Yang}, \citenamefont {Liu}, \citenamefont {Liang}, \citenamefont {Xu}, \citenamefont {Kroder}, \citenamefont {S\"u\ss{}}, \citenamefont {Borrmann}, \citenamefont {Shekhar}, \citenamefont {Wang}, \citenamefont {Xi}, \citenamefont {Wang}, \citenamefont {Schnelle}, \citenamefont {Wirth}, \citenamefont {Chen}, \citenamefont {Goennenwein},\ and\ \citenamefont {Felser}}]{Liu2018}%
  \BibitemOpen
  \bibfield  {author} {\bibinfo {author} {\bibfnamefont {E.}~\bibnamefont {Liu}}, \bibinfo {author} {\bibfnamefont {Y.}~\bibnamefont {Sun}}, \bibinfo {author} {\bibfnamefont {N.}~\bibnamefont {Kumar}}, \bibinfo {author} {\bibfnamefont {L.}~\bibnamefont {Muechler}}, \bibinfo {author} {\bibfnamefont {A.}~\bibnamefont {Sun}}, \bibinfo {author} {\bibfnamefont {L.}~\bibnamefont {Jiao}}, \bibinfo {author} {\bibfnamefont {S.-Y.}\ \bibnamefont {Yang}}, \bibinfo {author} {\bibfnamefont {D.}~\bibnamefont {Liu}}, \bibinfo {author} {\bibfnamefont {A.}~\bibnamefont {Liang}}, \bibinfo {author} {\bibfnamefont {Q.}~\bibnamefont {Xu}}, \bibinfo {author} {\bibfnamefont {J.}~\bibnamefont {Kroder}}, \bibinfo {author} {\bibfnamefont {V.}~\bibnamefont {S\"u\ss{}}}, \bibinfo {author} {\bibfnamefont {H.}~\bibnamefont {Borrmann}}, \bibinfo {author} {\bibfnamefont {C.}~\bibnamefont {Shekhar}}, \bibinfo {author} {\bibfnamefont {Z.}~\bibnamefont {Wang}}, \bibinfo {author} {\bibfnamefont {C.}~\bibnamefont {Xi}}, \bibinfo {author}
  {\bibfnamefont {W.}~\bibnamefont {Wang}}, \bibinfo {author} {\bibfnamefont {W.}~\bibnamefont {Schnelle}}, \bibinfo {author} {\bibfnamefont {S.}~\bibnamefont {Wirth}}, \bibinfo {author} {\bibfnamefont {Y.}~\bibnamefont {Chen}}, \bibinfo {author} {\bibfnamefont {S.~T.~B.}\ \bibnamefont {Goennenwein}},\ and\ \bibinfo {author} {\bibfnamefont {C.}~\bibnamefont {Felser}},\ }\bibfield  {title} {\bibinfo {title} {Giant anomalous {Hall} effect in a ferromagnetic kagome-lattice semimetal},\ }\href@noop {} {\bibfield  {journal} {\bibinfo  {journal} {Nat. Phys.}\ }\textbf {\bibinfo {volume} {14}},\ \bibinfo {pages} {1125} (\bibinfo {year} {2018})}\BibitemShut {NoStop}%
\bibitem [{\citenamefont {Chang}\ \emph {et~al.}(2018)\citenamefont {Chang}, \citenamefont {Singh}, \citenamefont {Xu}, \citenamefont {Bian}, \citenamefont {Huang}, \citenamefont {Hsu}, \citenamefont {Belopolski}, \citenamefont {Alidoust}, \citenamefont {Sanchez}, \citenamefont {Zheng} \emph {et~al.}}]{chang2018magnetic}%
  \BibitemOpen
  \bibfield  {author} {\bibinfo {author} {\bibfnamefont {G.}~\bibnamefont {Chang}}, \bibinfo {author} {\bibfnamefont {B.}~\bibnamefont {Singh}}, \bibinfo {author} {\bibfnamefont {S.-Y.}\ \bibnamefont {Xu}}, \bibinfo {author} {\bibfnamefont {G.}~\bibnamefont {Bian}}, \bibinfo {author} {\bibfnamefont {S.-M.}\ \bibnamefont {Huang}}, \bibinfo {author} {\bibfnamefont {C.-H.}\ \bibnamefont {Hsu}}, \bibinfo {author} {\bibfnamefont {I.}~\bibnamefont {Belopolski}}, \bibinfo {author} {\bibfnamefont {N.}~\bibnamefont {Alidoust}}, \bibinfo {author} {\bibfnamefont {D.~S.}\ \bibnamefont {Sanchez}}, \bibinfo {author} {\bibfnamefont {H.}~\bibnamefont {Zheng}}, \emph {et~al.},\ }\bibfield  {title} {\bibinfo {title} {Magnetic and noncentrosymmetric weyl fermion semimetals in the r alge family of compounds (r= rare earth)},\ }\href@noop {} {\bibfield  {journal} {\bibinfo  {journal} {Physical Review B}\ }\textbf {\bibinfo {volume} {97}},\ \bibinfo {pages} {041104} (\bibinfo {year} {2018})}\BibitemShut {NoStop}%
\bibitem [{\citenamefont {Shi}\ \emph {et~al.}(2018)\citenamefont {Shi}, \citenamefont {Muechler}, \citenamefont {Manna}, \citenamefont {Zhang}, \citenamefont {Koepernik}, \citenamefont {Car}, \citenamefont {Van Den~Brink}, \citenamefont {Felser},\ and\ \citenamefont {Sun}}]{shi2018prediction}%
  \BibitemOpen
  \bibfield  {author} {\bibinfo {author} {\bibfnamefont {W.}~\bibnamefont {Shi}}, \bibinfo {author} {\bibfnamefont {L.}~\bibnamefont {Muechler}}, \bibinfo {author} {\bibfnamefont {K.}~\bibnamefont {Manna}}, \bibinfo {author} {\bibfnamefont {Y.}~\bibnamefont {Zhang}}, \bibinfo {author} {\bibfnamefont {K.}~\bibnamefont {Koepernik}}, \bibinfo {author} {\bibfnamefont {R.}~\bibnamefont {Car}}, \bibinfo {author} {\bibfnamefont {J.}~\bibnamefont {Van Den~Brink}}, \bibinfo {author} {\bibfnamefont {C.}~\bibnamefont {Felser}},\ and\ \bibinfo {author} {\bibfnamefont {Y.}~\bibnamefont {Sun}},\ }\bibfield  {title} {\bibinfo {title} {Prediction of a magnetic weyl semimetal without spin-orbit coupling and strong anomalous hall effect in the heusler compensated ferrimagnet ti 2 mnal},\ }\href@noop {} {\bibfield  {journal} {\bibinfo  {journal} {Physical Review B}\ }\textbf {\bibinfo {volume} {97}},\ \bibinfo {pages} {060406} (\bibinfo {year} {2018})}\BibitemShut {NoStop}%
\bibitem [{\citenamefont {Karplus}\ and\ \citenamefont {Luttinger}(1954)}]{karplus1954}%
  \BibitemOpen
  \bibfield  {author} {\bibinfo {author} {\bibfnamefont {R.}~\bibnamefont {Karplus}}\ and\ \bibinfo {author} {\bibfnamefont {J.}~\bibnamefont {Luttinger}},\ }\bibfield  {title} {\bibinfo {title} {Hall effect in ferromagnetics},\ }\href@noop {} {\bibfield  {journal} {\bibinfo  {journal} {Physical Review}\ }\textbf {\bibinfo {volume} {95}},\ \bibinfo {pages} {1154} (\bibinfo {year} {1954})}\BibitemShut {NoStop}%
\bibitem [{\citenamefont {Thouless}\ \emph {et~al.}(1982)\citenamefont {Thouless}, \citenamefont {Kohmoto}, \citenamefont {Nightingale},\ and\ \citenamefont {den Nijs}}]{thouless1982quantized}%
  \BibitemOpen
  \bibfield  {author} {\bibinfo {author} {\bibfnamefont {D.~J.}\ \bibnamefont {Thouless}}, \bibinfo {author} {\bibfnamefont {M.}~\bibnamefont {Kohmoto}}, \bibinfo {author} {\bibfnamefont {M.~P.}\ \bibnamefont {Nightingale}},\ and\ \bibinfo {author} {\bibfnamefont {M.}~\bibnamefont {den Nijs}},\ }\bibfield  {title} {\bibinfo {title} {Quantized hall conductance in a two-dimensional periodic potential},\ }\href@noop {} {\bibfield  {journal} {\bibinfo  {journal} {Physical review letters}\ }\textbf {\bibinfo {volume} {49}},\ \bibinfo {pages} {405} (\bibinfo {year} {1982})}\BibitemShut {NoStop}%
\bibitem [{\citenamefont {Kohmoto}(1985)}]{kohmoto1985topological}%
  \BibitemOpen
  \bibfield  {author} {\bibinfo {author} {\bibfnamefont {M.}~\bibnamefont {Kohmoto}},\ }\bibfield  {title} {\bibinfo {title} {Topological invariant and the quantization of the hall conductance},\ }\href@noop {} {\bibfield  {journal} {\bibinfo  {journal} {Annals of Physics}\ }\textbf {\bibinfo {volume} {160}},\ \bibinfo {pages} {343} (\bibinfo {year} {1985})}\BibitemShut {NoStop}%
\bibitem [{\citenamefont {Altland}\ and\ \citenamefont {Zirnbauer}(1997)}]{altland1997nonstandard}%
  \BibitemOpen
  \bibfield  {author} {\bibinfo {author} {\bibfnamefont {A.}~\bibnamefont {Altland}}\ and\ \bibinfo {author} {\bibfnamefont {M.~R.}\ \bibnamefont {Zirnbauer}},\ }\bibfield  {title} {\bibinfo {title} {Nonstandard symmetry classes in mesoscopic normal-superconducting hybrid structures},\ }\href@noop {} {\bibfield  {journal} {\bibinfo  {journal} {Physical Review B}\ }\textbf {\bibinfo {volume} {55}},\ \bibinfo {pages} {1142} (\bibinfo {year} {1997})}\BibitemShut {NoStop}%
\bibitem [{\citenamefont {Schnyder}\ \emph {et~al.}(2008)\citenamefont {Schnyder}, \citenamefont {Ryu}, \citenamefont {Furusaki},\ and\ \citenamefont {Ludwig}}]{schnyder2008classification}%
  \BibitemOpen
  \bibfield  {author} {\bibinfo {author} {\bibfnamefont {A.~P.}\ \bibnamefont {Schnyder}}, \bibinfo {author} {\bibfnamefont {S.}~\bibnamefont {Ryu}}, \bibinfo {author} {\bibfnamefont {A.}~\bibnamefont {Furusaki}},\ and\ \bibinfo {author} {\bibfnamefont {A.~W.}\ \bibnamefont {Ludwig}},\ }\bibfield  {title} {\bibinfo {title} {Classification of topological insulators and superconductors in three spatial dimensions},\ }\href@noop {} {\bibfield  {journal} {\bibinfo  {journal} {Physical Review B—Condensed Matter and Materials Physics}\ }\textbf {\bibinfo {volume} {78}},\ \bibinfo {pages} {195125} (\bibinfo {year} {2008})}\BibitemShut {NoStop}%
\bibitem [{\citenamefont {Haldane}(1988)}]{haldane1988model}%
  \BibitemOpen
  \bibfield  {author} {\bibinfo {author} {\bibfnamefont {F.~D.~M.}\ \bibnamefont {Haldane}},\ }\bibfield  {title} {\bibinfo {title} {Model for a quantum hall effect without landau levels: Condensed-matter realization of the" parity anomaly"},\ }\href@noop {} {\bibfield  {journal} {\bibinfo  {journal} {Physical review letters}\ }\textbf {\bibinfo {volume} {61}},\ \bibinfo {pages} {2015} (\bibinfo {year} {1988})}\BibitemShut {NoStop}%
\bibitem [{\citenamefont {Fukui}\ \emph {et~al.}(2005)\citenamefont {Fukui}, \citenamefont {Hatsugai},\ and\ \citenamefont {Suzuki}}]{fukui2005chern}%
  \BibitemOpen
  \bibfield  {author} {\bibinfo {author} {\bibfnamefont {T.}~\bibnamefont {Fukui}}, \bibinfo {author} {\bibfnamefont {Y.}~\bibnamefont {Hatsugai}},\ and\ \bibinfo {author} {\bibfnamefont {H.}~\bibnamefont {Suzuki}},\ }\bibfield  {title} {\bibinfo {title} {Chern numbers in discretized brillouin zone: Efficient method of computing (spin) hall conductances},\ }\href@noop {} {\bibfield  {journal} {\bibinfo  {journal} {Journal of the Physical Society of Japan}\ }\textbf {\bibinfo {volume} {74}},\ \bibinfo {pages} {1674} (\bibinfo {year} {2005})}\BibitemShut {NoStop}%
\bibitem [{\citenamefont {Burkov}\ \emph {et~al.}(2011)\citenamefont {Burkov}, \citenamefont {Hook},\ and\ \citenamefont {Balents}}]{burkov2011topological}%
  \BibitemOpen
  \bibfield  {author} {\bibinfo {author} {\bibfnamefont {A.}~\bibnamefont {Burkov}}, \bibinfo {author} {\bibfnamefont {M.}~\bibnamefont {Hook}},\ and\ \bibinfo {author} {\bibfnamefont {L.}~\bibnamefont {Balents}},\ }\bibfield  {title} {\bibinfo {title} {Topological nodal semimetals},\ }\href@noop {} {\bibfield  {journal} {\bibinfo  {journal} {Physical Review B—Condensed Matter and Materials Physics}\ }\textbf {\bibinfo {volume} {84}},\ \bibinfo {pages} {235126} (\bibinfo {year} {2011})}\BibitemShut {NoStop}%
\bibitem [{\citenamefont {Zyuzin}\ and\ \citenamefont {Burkov}(2012)}]{Zyuzin2012}%
  \BibitemOpen
  \bibfield  {author} {\bibinfo {author} {\bibfnamefont {A.~A.}\ \bibnamefont {Zyuzin}}\ and\ \bibinfo {author} {\bibfnamefont {A.~A.}\ \bibnamefont {Burkov}},\ }\bibfield  {title} {\bibinfo {title} {Topological response in weyl semimetals and the chiral anomaly},\ }\href {https://doi.org/10.1103/PhysRevB.86.115133} {\bibfield  {journal} {\bibinfo  {journal} {Phys. Rev. B}\ }\textbf {\bibinfo {volume} {86}},\ \bibinfo {pages} {115133} (\bibinfo {year} {2012})}\BibitemShut {NoStop}%
\bibitem [{\citenamefont {Burkov}(2014)}]{burkov2014anomalous}%
  \BibitemOpen
  \bibfield  {author} {\bibinfo {author} {\bibfnamefont {A.}~\bibnamefont {Burkov}},\ }\bibfield  {title} {\bibinfo {title} {Anomalous hall effect in weyl metals},\ }\href@noop {} {\bibfield  {journal} {\bibinfo  {journal} {Physical review letters}\ }\textbf {\bibinfo {volume} {113}},\ \bibinfo {pages} {187202} (\bibinfo {year} {2014})}\BibitemShut {NoStop}%
\bibitem [{\citenamefont {Xiao}\ \emph {et~al.}(2006)\citenamefont {Xiao}, \citenamefont {Yao}, \citenamefont {Fang},\ and\ \citenamefont {Niu}}]{xiao2006berry}%
  \BibitemOpen
  \bibfield  {author} {\bibinfo {author} {\bibfnamefont {D.}~\bibnamefont {Xiao}}, \bibinfo {author} {\bibfnamefont {Y.}~\bibnamefont {Yao}}, \bibinfo {author} {\bibfnamefont {Z.}~\bibnamefont {Fang}},\ and\ \bibinfo {author} {\bibfnamefont {Q.}~\bibnamefont {Niu}},\ }\bibfield  {title} {\bibinfo {title} {Berry-phase effect in anomalous thermoelectric transport},\ }\href@noop {} {\bibfield  {journal} {\bibinfo  {journal} {Physical review letters}\ }\textbf {\bibinfo {volume} {97}},\ \bibinfo {pages} {026603} (\bibinfo {year} {2006})}\BibitemShut {NoStop}%
\bibitem [{\citenamefont {Sharma}\ \emph {et~al.}(2016)\citenamefont {Sharma}, \citenamefont {Goswami},\ and\ \citenamefont {Tewari}}]{sharma2016nernst}%
  \BibitemOpen
  \bibfield  {author} {\bibinfo {author} {\bibfnamefont {G.}~\bibnamefont {Sharma}}, \bibinfo {author} {\bibfnamefont {P.}~\bibnamefont {Goswami}},\ and\ \bibinfo {author} {\bibfnamefont {S.}~\bibnamefont {Tewari}},\ }\bibfield  {title} {\bibinfo {title} {Nernst and magnetothermal conductivity in a lattice model of weyl fermions},\ }\href@noop {} {\bibfield  {journal} {\bibinfo  {journal} {Physical Review B}\ }\textbf {\bibinfo {volume} {93}},\ \bibinfo {pages} {035116} (\bibinfo {year} {2016})}\BibitemShut {NoStop}%
\bibitem [{\citenamefont {Matsushita}\ \emph {et~al.}(2025)\citenamefont {Matsushita}, \citenamefont {Ozawa}, \citenamefont {Araki}, \citenamefont {Fujimoto},\ and\ \citenamefont {Sato}}]{matsushita2025intrinsic}%
  \BibitemOpen
  \bibfield  {author} {\bibinfo {author} {\bibfnamefont {T.}~\bibnamefont {Matsushita}}, \bibinfo {author} {\bibfnamefont {A.}~\bibnamefont {Ozawa}}, \bibinfo {author} {\bibfnamefont {Y.}~\bibnamefont {Araki}}, \bibinfo {author} {\bibfnamefont {J.}~\bibnamefont {Fujimoto}},\ and\ \bibinfo {author} {\bibfnamefont {M.}~\bibnamefont {Sato}},\ }\bibfield  {title} {\bibinfo {title} {Intrinsic spin nernst effect in topological dirac and magnetic weyl semimetals},\ }\href@noop {} {\bibfield  {journal} {\bibinfo  {journal} {Physical Review B}\ }\textbf {\bibinfo {volume} {111}},\ \bibinfo {pages} {245131} (\bibinfo {year} {2025})}\BibitemShut {NoStop}%
\bibitem [{\citenamefont {Zhang}\ and\ \citenamefont {Nagaosa}(2024{\natexlab{a}})}]{Zhang2024-dn}%
  \BibitemOpen
  \bibfield  {author} {\bibinfo {author} {\bibfnamefont {X.-X.}\ \bibnamefont {Zhang}}\ and\ \bibinfo {author} {\bibfnamefont {N.}~\bibnamefont {Nagaosa}},\ }\bibfield  {title} {\bibinfo {title} {Surface spectroscopy and surface-bulk hybridization of weyl semimetals},\ }\href@noop {} {\bibfield  {journal} {\bibinfo  {journal} {Proc. Natl. Acad. Sci. U. S. A.}\ }\textbf {\bibinfo {volume} {121}},\ \bibinfo {pages} {e2313488121} (\bibinfo {year} {2024}{\natexlab{a}})}\BibitemShut {NoStop}%
\bibitem [{\citenamefont {Potter}\ \emph {et~al.}(2014)\citenamefont {Potter}, \citenamefont {Kimchi},\ and\ \citenamefont {Vishwanath}}]{potter2014quantum}%
  \BibitemOpen
  \bibfield  {author} {\bibinfo {author} {\bibfnamefont {A.~C.}\ \bibnamefont {Potter}}, \bibinfo {author} {\bibfnamefont {I.}~\bibnamefont {Kimchi}},\ and\ \bibinfo {author} {\bibfnamefont {A.}~\bibnamefont {Vishwanath}},\ }\bibfield  {title} {\bibinfo {title} {Quantum oscillations from surface fermi arcs in weyl and dirac semimetals},\ }\href@noop {} {\bibfield  {journal} {\bibinfo  {journal} {Nature communications}\ }\textbf {\bibinfo {volume} {5}},\ \bibinfo {pages} {5161} (\bibinfo {year} {2014})}\BibitemShut {NoStop}%
\bibitem [{\citenamefont {Moll}\ \emph {et~al.}(2016)\citenamefont {Moll}, \citenamefont {Nair}, \citenamefont {Helm}, \citenamefont {Potter}, \citenamefont {Kimchi}, \citenamefont {Vishwanath},\ and\ \citenamefont {Analytis}}]{moll2016transport}%
  \BibitemOpen
  \bibfield  {author} {\bibinfo {author} {\bibfnamefont {P.~J.}\ \bibnamefont {Moll}}, \bibinfo {author} {\bibfnamefont {N.~L.}\ \bibnamefont {Nair}}, \bibinfo {author} {\bibfnamefont {T.}~\bibnamefont {Helm}}, \bibinfo {author} {\bibfnamefont {A.~C.}\ \bibnamefont {Potter}}, \bibinfo {author} {\bibfnamefont {I.}~\bibnamefont {Kimchi}}, \bibinfo {author} {\bibfnamefont {A.}~\bibnamefont {Vishwanath}},\ and\ \bibinfo {author} {\bibfnamefont {J.~G.}\ \bibnamefont {Analytis}},\ }\bibfield  {title} {\bibinfo {title} {Transport evidence for fermi-arc-mediated chirality transfer in the dirac semimetal cd3as2},\ }\href@noop {} {\bibfield  {journal} {\bibinfo  {journal} {Nature}\ }\textbf {\bibinfo {volume} {535}},\ \bibinfo {pages} {266} (\bibinfo {year} {2016})}\BibitemShut {NoStop}%
\bibitem [{\citenamefont {Galletti}\ \emph {et~al.}(2019)\citenamefont {Galletti}, \citenamefont {Schumann}, \citenamefont {Kealhofer}, \citenamefont {Goyal},\ and\ \citenamefont {Stemmer}}]{Galletti2019absence}%
  \BibitemOpen
  \bibfield  {author} {\bibinfo {author} {\bibfnamefont {L.}~\bibnamefont {Galletti}}, \bibinfo {author} {\bibfnamefont {T.}~\bibnamefont {Schumann}}, \bibinfo {author} {\bibfnamefont {D.~A.}\ \bibnamefont {Kealhofer}}, \bibinfo {author} {\bibfnamefont {M.}~\bibnamefont {Goyal}},\ and\ \bibinfo {author} {\bibfnamefont {S.}~\bibnamefont {Stemmer}},\ }\bibfield  {title} {\bibinfo {title} {{Absence of signatures of Weyl orbits in the thickness dependence of quantum transport in cadmium arsenide}},\ }\href {https://doi.org/10.1103/PhysRevB.99.201401} {\bibfield  {journal} {\bibinfo  {journal} {Phys. Rev. B}\ }\textbf {\bibinfo {volume} {99}},\ \bibinfo {pages} {201401} (\bibinfo {year} {2019})}\BibitemShut {NoStop}%
\bibitem [{\citenamefont {Nguyen}\ \emph {et~al.}(2021)\citenamefont {Nguyen}, \citenamefont {Kobayashi}, \citenamefont {Wichmann},\ and\ \citenamefont {Nomura}}]{Nguyen2021quantum}%
  \BibitemOpen
  \bibfield  {author} {\bibinfo {author} {\bibfnamefont {D.-H.-M.}\ \bibnamefont {Nguyen}}, \bibinfo {author} {\bibfnamefont {K.}~\bibnamefont {Kobayashi}}, \bibinfo {author} {\bibfnamefont {J.-E.~R.}\ \bibnamefont {Wichmann}},\ and\ \bibinfo {author} {\bibfnamefont {K.}~\bibnamefont {Nomura}},\ }\bibfield  {title} {\bibinfo {title} {{Quantum Hall effect induced by chiral Landau levels in topological semimetal films}},\ }\href {https://doi.org/10.1103/PhysRevB.104.045302} {\bibfield  {journal} {\bibinfo  {journal} {Phys. Rev. B}\ }\textbf {\bibinfo {volume} {104}},\ \bibinfo {pages} {045302} (\bibinfo {year} {2021})}\BibitemShut {NoStop}%
\bibitem [{\citenamefont {Zhang}\ and\ \citenamefont {Nagaosa}(2022)}]{Zhang2022}%
  \BibitemOpen
  \bibfield  {author} {\bibinfo {author} {\bibfnamefont {X.-X.}\ \bibnamefont {Zhang}}\ and\ \bibinfo {author} {\bibfnamefont {N.}~\bibnamefont {Nagaosa}},\ }\bibfield  {title} {\bibinfo {title} {Anisotropic three-dimensional quantum hall effect and magnetotransport in mesoscopic weyl semimetals},\ }\href@noop {} {\bibfield  {journal} {\bibinfo  {journal} {Nano Lett.}\ }\textbf {\bibinfo {volume} {22}},\ \bibinfo {pages} {3033} (\bibinfo {year} {2022})}\BibitemShut {NoStop}%
\bibitem [{\citenamefont {Lu}\ and\ \citenamefont {Zhang}(2026)}]{Lu2026}%
  \BibitemOpen
  \bibfield  {author} {\bibinfo {author} {\bibfnamefont {M.}~\bibnamefont {Lu}}\ and\ \bibinfo {author} {\bibfnamefont {X.-X.}\ \bibnamefont {Zhang}},\ }\bibfield  {title} {\bibinfo {title} {In-plane anomalous features in the {3D} quantum hall regime},\ }\href@noop {} {\bibfield  {journal} {\bibinfo  {journal} {Phys. Rev. Lett.}\ }\textbf {\bibinfo {volume} {136}},\ \bibinfo {pages} {026602} (\bibinfo {year} {2026})}\BibitemShut {NoStop}%
\bibitem [{\citenamefont {Xiao}\ \emph {et~al.}(2005)\citenamefont {Xiao}, \citenamefont {Shi},\ and\ \citenamefont {Niu}}]{xiao2005berry}%
  \BibitemOpen
  \bibfield  {author} {\bibinfo {author} {\bibfnamefont {D.}~\bibnamefont {Xiao}}, \bibinfo {author} {\bibfnamefont {J.}~\bibnamefont {Shi}},\ and\ \bibinfo {author} {\bibfnamefont {Q.}~\bibnamefont {Niu}},\ }\bibfield  {title} {\bibinfo {title} {Berry phase correction to electron density of states in solids},\ }\href@noop {} {\bibfield  {journal} {\bibinfo  {journal} {Physical review letters}\ }\textbf {\bibinfo {volume} {95}},\ \bibinfo {pages} {137204} (\bibinfo {year} {2005})}\BibitemShut {NoStop}%
\bibitem [{\citenamefont {Thonhauser}\ \emph {et~al.}(2005)\citenamefont {Thonhauser}, \citenamefont {Ceresoli}, \citenamefont {Vanderbilt},\ and\ \citenamefont {Resta}}]{thonhauser2005orbital}%
  \BibitemOpen
  \bibfield  {author} {\bibinfo {author} {\bibfnamefont {T.}~\bibnamefont {Thonhauser}}, \bibinfo {author} {\bibfnamefont {D.}~\bibnamefont {Ceresoli}}, \bibinfo {author} {\bibfnamefont {D.}~\bibnamefont {Vanderbilt}},\ and\ \bibinfo {author} {\bibfnamefont {R.}~\bibnamefont {Resta}},\ }\bibfield  {title} {\bibinfo {title} {Orbital magnetization in periodic insulators},\ }\href@noop {} {\bibfield  {journal} {\bibinfo  {journal} {Physical review letters}\ }\textbf {\bibinfo {volume} {95}},\ \bibinfo {pages} {137205} (\bibinfo {year} {2005})}\BibitemShut {NoStop}%
\bibitem [{\citenamefont {Ceresoli}\ \emph {et~al.}(2006)\citenamefont {Ceresoli}, \citenamefont {Thonhauser}, \citenamefont {Vanderbilt},\ and\ \citenamefont {Resta}}]{ceresoli2006orbital}%
  \BibitemOpen
  \bibfield  {author} {\bibinfo {author} {\bibfnamefont {D.}~\bibnamefont {Ceresoli}}, \bibinfo {author} {\bibfnamefont {T.}~\bibnamefont {Thonhauser}}, \bibinfo {author} {\bibfnamefont {D.}~\bibnamefont {Vanderbilt}},\ and\ \bibinfo {author} {\bibfnamefont {R.}~\bibnamefont {Resta}},\ }\bibfield  {title} {\bibinfo {title} {Orbital magnetization in crystalline solids: Multi-band insulators, chern insulators, and metals},\ }\href@noop {} {\bibfield  {journal} {\bibinfo  {journal} {Physical Review B—Condensed Matter and Materials Physics}\ }\textbf {\bibinfo {volume} {74}},\ \bibinfo {pages} {024408} (\bibinfo {year} {2006})}\BibitemShut {NoStop}%
\bibitem [{\citenamefont {Resta}(2010)}]{resta2010electrical}%
  \BibitemOpen
  \bibfield  {author} {\bibinfo {author} {\bibfnamefont {R.}~\bibnamefont {Resta}},\ }\bibfield  {title} {\bibinfo {title} {Electrical polarization and orbital magnetization: the modern theories},\ }\href@noop {} {\bibfield  {journal} {\bibinfo  {journal} {Journal of Physics: Condensed Matter}\ }\textbf {\bibinfo {volume} {22}},\ \bibinfo {pages} {123201} (\bibinfo {year} {2010})}\BibitemShut {NoStop}%
\bibitem [{\citenamefont {Streda}(1982)}]{streda1982theory}%
  \BibitemOpen
  \bibfield  {author} {\bibinfo {author} {\bibfnamefont {P.}~\bibnamefont {Streda}},\ }\bibfield  {title} {\bibinfo {title} {Theory of quantised hall conductivity in two dimensions},\ }\href@noop {} {\bibfield  {journal} {\bibinfo  {journal} {Journal of Physics C: Solid State Physics}\ }\textbf {\bibinfo {volume} {15}},\ \bibinfo {pages} {L717} (\bibinfo {year} {1982})}\BibitemShut {NoStop}%
\bibitem [{\citenamefont {Nomura}\ and\ \citenamefont {Kurebayashi}(2015)}]{Nomura2015}%
  \BibitemOpen
  \bibfield  {author} {\bibinfo {author} {\bibfnamefont {K.}~\bibnamefont {Nomura}}\ and\ \bibinfo {author} {\bibfnamefont {D.}~\bibnamefont {Kurebayashi}},\ }\bibfield  {title} {\bibinfo {title} {Charge-induced spin torque in anomalous hall ferromagnets},\ }\href {https://doi.org/10.1103/PhysRevLett.115.127201} {\bibfield  {journal} {\bibinfo  {journal} {Phys. Rev. Lett.}\ }\textbf {\bibinfo {volume} {115}},\ \bibinfo {pages} {127201} (\bibinfo {year} {2015})}\BibitemShut {NoStop}%
\bibitem [{\citenamefont {Bardeen}(1969)}]{Bardeen1969PhysRev}%
  \BibitemOpen
  \bibfield  {author} {\bibinfo {author} {\bibfnamefont {W.~A.}\ \bibnamefont {Bardeen}},\ }\bibfield  {title} {\bibinfo {title} {Anomalous ward identities in spinor field theories},\ }\href {https://doi.org/10.1103/PhysRev.184.1848} {\bibfield  {journal} {\bibinfo  {journal} {Phys. Rev.}\ }\textbf {\bibinfo {volume} {184}},\ \bibinfo {pages} {1848} (\bibinfo {year} {1969})}\BibitemShut {NoStop}%
\bibitem [{\citenamefont {Landsteiner}(2016)}]{Landsteiner2016notes}%
  \BibitemOpen
  \bibfield  {author} {\bibinfo {author} {\bibfnamefont {K.}~\bibnamefont {Landsteiner}},\ }\bibfield  {title} {\bibinfo {title} {Notes on anomaly induced transport},\ }\href {https://doi.org/10.5506/APhysPolB.47.2617} {\bibfield  {journal} {\bibinfo  {journal} {Acta Physica Polonica B}\ }\textbf {\bibinfo {volume} {47}},\ \bibinfo {pages} {2617} (\bibinfo {year} {2016})}\BibitemShut {NoStop}%
\bibitem [{\citenamefont {Liu}\ \emph {et~al.}(2013)\citenamefont {Liu}, \citenamefont {Ye},\ and\ \citenamefont {Qi}}]{Liu2013PRB}%
  \BibitemOpen
  \bibfield  {author} {\bibinfo {author} {\bibfnamefont {C.-X.}\ \bibnamefont {Liu}}, \bibinfo {author} {\bibfnamefont {P.}~\bibnamefont {Ye}},\ and\ \bibinfo {author} {\bibfnamefont {X.-L.}\ \bibnamefont {Qi}},\ }\bibfield  {title} {\bibinfo {title} {Chiral gauge field and axial anomaly in a weyl semimetal},\ }\href {https://doi.org/10.1103/PhysRevB.87.235306} {\bibfield  {journal} {\bibinfo  {journal} {Phys. Rev. B}\ }\textbf {\bibinfo {volume} {87}},\ \bibinfo {pages} {235306} (\bibinfo {year} {2013})}\BibitemShut {NoStop}%
\bibitem [{\citenamefont {Pikulin}\ \emph {et~al.}(2016)\citenamefont {Pikulin}, \citenamefont {Chen},\ and\ \citenamefont {Franz}}]{Pikulin2016PRX}%
  \BibitemOpen
  \bibfield  {author} {\bibinfo {author} {\bibfnamefont {D.~I.}\ \bibnamefont {Pikulin}}, \bibinfo {author} {\bibfnamefont {A.}~\bibnamefont {Chen}},\ and\ \bibinfo {author} {\bibfnamefont {M.}~\bibnamefont {Franz}},\ }\bibfield  {title} {\bibinfo {title} {Chiral anomaly from strain-induced gauge fields in dirac and weyl semimetals},\ }\href {https://doi.org/10.1103/PhysRevX.6.041021} {\bibfield  {journal} {\bibinfo  {journal} {Phys. Rev. X}\ }\textbf {\bibinfo {volume} {6}},\ \bibinfo {pages} {041021} (\bibinfo {year} {2016})}\BibitemShut {NoStop}%
\bibitem [{\citenamefont {Ilan}\ \emph {et~al.}(2020)\citenamefont {Ilan}, \citenamefont {Grushin},\ and\ \citenamefont {Pikulin}}]{ilan2020pseudo}%
  \BibitemOpen
  \bibfield  {author} {\bibinfo {author} {\bibfnamefont {R.}~\bibnamefont {Ilan}}, \bibinfo {author} {\bibfnamefont {A.~G.}\ \bibnamefont {Grushin}},\ and\ \bibinfo {author} {\bibfnamefont {D.~I.}\ \bibnamefont {Pikulin}},\ }\bibfield  {title} {\bibinfo {title} {Pseudo-electromagnetic fields in 3d topological semimetals},\ }\href {https://doi.org/10.1038/s42254-019-0121-8} {\bibfield  {journal} {\bibinfo  {journal} {Nature Reviews Physics}\ }\textbf {\bibinfo {volume} {2}},\ \bibinfo {pages} {29} (\bibinfo {year} {2020})}\BibitemShut {NoStop}%
\bibitem [{\citenamefont {Cortijo}\ \emph {et~al.}(2015)\citenamefont {Cortijo}, \citenamefont {Ferreir\'os}, \citenamefont {Landsteiner},\ and\ \citenamefont {Vozmediano}}]{Cortijo2015}%
  \BibitemOpen
  \bibfield  {author} {\bibinfo {author} {\bibfnamefont {A.}~\bibnamefont {Cortijo}}, \bibinfo {author} {\bibfnamefont {Y.}~\bibnamefont {Ferreir\'os}}, \bibinfo {author} {\bibfnamefont {K.}~\bibnamefont {Landsteiner}},\ and\ \bibinfo {author} {\bibfnamefont {M.~A.~H.}\ \bibnamefont {Vozmediano}},\ }\bibfield  {title} {\bibinfo {title} {Elastic gauge fields in weyl semimetals},\ }\href {https://doi.org/10.1103/PhysRevLett.115.177202} {\bibfield  {journal} {\bibinfo  {journal} {Phys. Rev. Lett.}\ }\textbf {\bibinfo {volume} {115}},\ \bibinfo {pages} {177202} (\bibinfo {year} {2015})}\BibitemShut {NoStop}%
\bibitem [{\citenamefont {Kariyado}(2019)}]{Kariyado2019jpsj}%
  \BibitemOpen
  \bibfield  {author} {\bibinfo {author} {\bibfnamefont {T.}~\bibnamefont {Kariyado}},\ }\bibfield  {title} {\bibinfo {title} {Counting pseudo landau levels in spatially modulated dirac systems},\ }\href {https://doi.org/10.7566/JPSJ.88.083701} {\bibfield  {journal} {\bibinfo  {journal} {Journal of the Physical Society of Japan}\ }\textbf {\bibinfo {volume} {88}},\ \bibinfo {pages} {083701} (\bibinfo {year} {2019})}\BibitemShut {NoStop}%
\bibitem [{\citenamefont {Ebihara}\ \emph {et~al.}(2016)\citenamefont {Ebihara}, \citenamefont {Fukushima},\ and\ \citenamefont {Oka}}]{Ebihara2016}%
  \BibitemOpen
  \bibfield  {author} {\bibinfo {author} {\bibfnamefont {S.}~\bibnamefont {Ebihara}}, \bibinfo {author} {\bibfnamefont {K.}~\bibnamefont {Fukushima}},\ and\ \bibinfo {author} {\bibfnamefont {T.}~\bibnamefont {Oka}},\ }\bibfield  {title} {\bibinfo {title} {Chiral pumping effect induced by rotating electric fields},\ }\href {https://doi.org/10.1103/PhysRevB.93.155107} {\bibfield  {journal} {\bibinfo  {journal} {Phys. Rev. B}\ }\textbf {\bibinfo {volume} {93}},\ \bibinfo {pages} {155107} (\bibinfo {year} {2016})}\BibitemShut {NoStop}%
\bibitem [{\citenamefont {Bucciantini}\ \emph {et~al.}(2017)\citenamefont {Bucciantini}, \citenamefont {Roy}, \citenamefont {Kitamura},\ and\ \citenamefont {Oka}}]{Bucciantini2017}%
  \BibitemOpen
  \bibfield  {author} {\bibinfo {author} {\bibfnamefont {L.}~\bibnamefont {Bucciantini}}, \bibinfo {author} {\bibfnamefont {S.}~\bibnamefont {Roy}}, \bibinfo {author} {\bibfnamefont {S.}~\bibnamefont {Kitamura}},\ and\ \bibinfo {author} {\bibfnamefont {T.}~\bibnamefont {Oka}},\ }\bibfield  {title} {\bibinfo {title} {Emergent weyl nodes and fermi arcs in a floquet weyl semimetal},\ }\href {https://doi.org/10.1103/PhysRevB.96.041126} {\bibfield  {journal} {\bibinfo  {journal} {Phys. Rev. B}\ }\textbf {\bibinfo {volume} {96}},\ \bibinfo {pages} {041126} (\bibinfo {year} {2017})}\BibitemShut {NoStop}%
\bibitem [{\citenamefont {Peskin}\ and\ \citenamefont {Halzen}(1995)}]{peskin1995introduction}%
  \BibitemOpen
  \bibfield  {author} {\bibinfo {author} {\bibfnamefont {M.~E.}\ \bibnamefont {Peskin}}\ and\ \bibinfo {author} {\bibfnamefont {F.}~\bibnamefont {Halzen}},\ }\href@noop {} {\emph {\bibinfo {title} {An Introduction to Quantum Field Theory}}}\ (\bibinfo  {publisher} {Addison-Wesley},\ \bibinfo {year} {1995})\BibitemShut {NoStop}%
\bibitem [{\citenamefont {Gorbar}\ \emph {et~al.}(2018{\natexlab{a}})\citenamefont {Gorbar}, \citenamefont {Miransky}, \citenamefont {Shovkovy},\ and\ \citenamefont {Sukhachov}}]{gorbar2018anomalous}%
  \BibitemOpen
  \bibfield  {author} {\bibinfo {author} {\bibfnamefont {E.}~\bibnamefont {Gorbar}}, \bibinfo {author} {\bibfnamefont {V.}~\bibnamefont {Miransky}}, \bibinfo {author} {\bibfnamefont {I.}~\bibnamefont {Shovkovy}},\ and\ \bibinfo {author} {\bibfnamefont {P.}~\bibnamefont {Sukhachov}},\ }\bibfield  {title} {\bibinfo {title} {Anomalous transport properties of dirac and weyl semimetals},\ }\href@noop {} {\bibfield  {journal} {\bibinfo  {journal} {Low Temperature Physics}\ }\textbf {\bibinfo {volume} {44}},\ \bibinfo {pages} {487} (\bibinfo {year} {2018}{\natexlab{a}})}\BibitemShut {NoStop}%
\bibitem [{\citenamefont {Gorbar}\ \emph {et~al.}(2021)\citenamefont {Gorbar}, \citenamefont {Miransky}, \citenamefont {Shovkovy},\ and\ \citenamefont {Sukhachov}}]{gorbar2021electronic}%
  \BibitemOpen
  \bibfield  {author} {\bibinfo {author} {\bibfnamefont {E.~V.}\ \bibnamefont {Gorbar}}, \bibinfo {author} {\bibfnamefont {V.~A.}\ \bibnamefont {Miransky}}, \bibinfo {author} {\bibfnamefont {I.~A.}\ \bibnamefont {Shovkovy}},\ and\ \bibinfo {author} {\bibfnamefont {P.~O.}\ \bibnamefont {Sukhachov}},\ }\href@noop {} {\emph {\bibinfo {title} {Electronic properties of Dirac and Weyl semimetals}}}\ (\bibinfo  {publisher} {World Scientific},\ \bibinfo {year} {2021})\BibitemShut {NoStop}%
\bibitem [{\citenamefont {Ahmad}\ \emph {et~al.}(2025)\citenamefont {Ahmad}, \citenamefont {K},\ and\ \citenamefont {Sharma}}]{ahmad2025geometry}%
  \BibitemOpen
  \bibfield  {author} {\bibinfo {author} {\bibfnamefont {A.}~\bibnamefont {Ahmad}}, \bibinfo {author} {\bibfnamefont {G.~V.}\ \bibnamefont {K}},\ and\ \bibinfo {author} {\bibfnamefont {G.}~\bibnamefont {Sharma}},\ }\bibfield  {title} {\bibinfo {title} {Geometry, anomaly, topology, and transport in weyl fermions},\ }\href@noop {} {\bibfield  {journal} {\bibinfo  {journal} {Journal of Physics: Condensed Matter}\ }\textbf {\bibinfo {volume} {37}},\ \bibinfo {pages} {043001} (\bibinfo {year} {2025})}\BibitemShut {NoStop}%
\bibitem [{\citenamefont {Laughlin}(1981)}]{laughlin1981quantized}%
  \BibitemOpen
  \bibfield  {author} {\bibinfo {author} {\bibfnamefont {R.~B.}\ \bibnamefont {Laughlin}},\ }\bibfield  {title} {\bibinfo {title} {Quantized hall conductivity in two dimensions},\ }\href@noop {} {\bibfield  {journal} {\bibinfo  {journal} {Physical Review B}\ }\textbf {\bibinfo {volume} {23}},\ \bibinfo {pages} {5632} (\bibinfo {year} {1981})}\BibitemShut {NoStop}%
\bibitem [{\citenamefont {Klitzing}\ \emph {et~al.}(1980)\citenamefont {Klitzing}, \citenamefont {Dorda},\ and\ \citenamefont {Pepper}}]{klitzing1980new}%
  \BibitemOpen
  \bibfield  {author} {\bibinfo {author} {\bibfnamefont {K.~v.}\ \bibnamefont {Klitzing}}, \bibinfo {author} {\bibfnamefont {G.}~\bibnamefont {Dorda}},\ and\ \bibinfo {author} {\bibfnamefont {M.}~\bibnamefont {Pepper}},\ }\bibfield  {title} {\bibinfo {title} {New method for high-accuracy determination of the fine-structure constant based on quantized hall resistance},\ }\href@noop {} {\bibfield  {journal} {\bibinfo  {journal} {Physical review letters}\ }\textbf {\bibinfo {volume} {45}},\ \bibinfo {pages} {494} (\bibinfo {year} {1980})}\BibitemShut {NoStop}%
\bibitem [{\citenamefont {Tsui}\ \emph {et~al.}(1982)\citenamefont {Tsui}, \citenamefont {Stormer},\ and\ \citenamefont {Gossard}}]{tsui1982two}%
  \BibitemOpen
  \bibfield  {author} {\bibinfo {author} {\bibfnamefont {D.~C.}\ \bibnamefont {Tsui}}, \bibinfo {author} {\bibfnamefont {H.~L.}\ \bibnamefont {Stormer}},\ and\ \bibinfo {author} {\bibfnamefont {A.~C.}\ \bibnamefont {Gossard}},\ }\bibfield  {title} {\bibinfo {title} {Two-dimensional magnetotransport in the extreme quantum limit},\ }\href@noop {} {\bibfield  {journal} {\bibinfo  {journal} {Physical Review Letters}\ }\textbf {\bibinfo {volume} {48}},\ \bibinfo {pages} {1559} (\bibinfo {year} {1982})}\BibitemShut {NoStop}%
\bibitem [{\citenamefont {Jackiw}(1984)}]{jackiw1984fractional}%
  \BibitemOpen
  \bibfield  {author} {\bibinfo {author} {\bibfnamefont {R.}~\bibnamefont {Jackiw}},\ }\bibfield  {title} {\bibinfo {title} {Fractional charge and zero modes for planar systems in a magnetic field},\ }\href@noop {} {\bibfield  {journal} {\bibinfo  {journal} {Physical Review D}\ }\textbf {\bibinfo {volume} {29}},\ \bibinfo {pages} {2375} (\bibinfo {year} {1984})}\BibitemShut {NoStop}%
\bibitem [{\citenamefont {Novoselov}\ \emph {et~al.}(2005)\citenamefont {Novoselov}, \citenamefont {Geim}, \citenamefont {Morozov}, \citenamefont {Jiang}, \citenamefont {Katsnelson}, \citenamefont {Grigorieva}, \citenamefont {Dubonos},\ and\ \citenamefont {Firsov}}]{novoselov2005two}%
  \BibitemOpen
  \bibfield  {author} {\bibinfo {author} {\bibfnamefont {K.~S.}\ \bibnamefont {Novoselov}}, \bibinfo {author} {\bibfnamefont {A.~K.}\ \bibnamefont {Geim}}, \bibinfo {author} {\bibfnamefont {S.~V.}\ \bibnamefont {Morozov}}, \bibinfo {author} {\bibfnamefont {D.}~\bibnamefont {Jiang}}, \bibinfo {author} {\bibfnamefont {M.~I.}\ \bibnamefont {Katsnelson}}, \bibinfo {author} {\bibfnamefont {I.~V.}\ \bibnamefont {Grigorieva}}, \bibinfo {author} {\bibfnamefont {S.~V.}\ \bibnamefont {Dubonos}},\ and\ \bibinfo {author} {\bibfnamefont {A.~A.}\ \bibnamefont {Firsov}},\ }\bibfield  {title} {\bibinfo {title} {Two-dimensional gas of massless dirac fermions in graphene},\ }\href@noop {} {\bibfield  {journal} {\bibinfo  {journal} {nature}\ }\textbf {\bibinfo {volume} {438}},\ \bibinfo {pages} {197} (\bibinfo {year} {2005})}\BibitemShut {NoStop}%
\bibitem [{\citenamefont {Zhang}\ \emph {et~al.}(2005)\citenamefont {Zhang}, \citenamefont {Tan}, \citenamefont {Stormer},\ and\ \citenamefont {Kim}}]{zhang2005experimental}%
  \BibitemOpen
  \bibfield  {author} {\bibinfo {author} {\bibfnamefont {Y.}~\bibnamefont {Zhang}}, \bibinfo {author} {\bibfnamefont {Y.-W.}\ \bibnamefont {Tan}}, \bibinfo {author} {\bibfnamefont {H.~L.}\ \bibnamefont {Stormer}},\ and\ \bibinfo {author} {\bibfnamefont {P.}~\bibnamefont {Kim}},\ }\bibfield  {title} {\bibinfo {title} {Experimental observation of the quantum hall effect and berry's phase in graphene},\ }\href@noop {} {\bibfield  {journal} {\bibinfo  {journal} {nature}\ }\textbf {\bibinfo {volume} {438}},\ \bibinfo {pages} {201} (\bibinfo {year} {2005})}\BibitemShut {NoStop}%
\bibitem [{\citenamefont {Yang}\ \emph {et~al.}(2011)\citenamefont {Yang}, \citenamefont {Lu},\ and\ \citenamefont {Ran}}]{Yang2011prb}%
  \BibitemOpen
  \bibfield  {author} {\bibinfo {author} {\bibfnamefont {K.-Y.}\ \bibnamefont {Yang}}, \bibinfo {author} {\bibfnamefont {Y.-M.}\ \bibnamefont {Lu}},\ and\ \bibinfo {author} {\bibfnamefont {Y.}~\bibnamefont {Ran}},\ }\bibfield  {title} {\bibinfo {title} {Quantum hall effects in a weyl semimetal: Possible application in pyrochlore iridates},\ }\href {https://doi.org/10.1103/PhysRevB.84.075129} {\bibfield  {journal} {\bibinfo  {journal} {Phys. Rev. B}\ }\textbf {\bibinfo {volume} {84}},\ \bibinfo {pages} {075129} (\bibinfo {year} {2011})}\BibitemShut {NoStop}%
\bibitem [{\citenamefont {Grushin}\ \emph {et~al.}(2016)\citenamefont {Grushin}, \citenamefont {Venderbos}, \citenamefont {Vishwanath},\ and\ \citenamefont {Ilan}}]{Grushin2016PRX}%
  \BibitemOpen
  \bibfield  {author} {\bibinfo {author} {\bibfnamefont {A.~G.}\ \bibnamefont {Grushin}}, \bibinfo {author} {\bibfnamefont {J.~W.~F.}\ \bibnamefont {Venderbos}}, \bibinfo {author} {\bibfnamefont {A.}~\bibnamefont {Vishwanath}},\ and\ \bibinfo {author} {\bibfnamefont {R.}~\bibnamefont {Ilan}},\ }\bibfield  {title} {\bibinfo {title} {Inhomogeneous weyl and dirac semimetals: Transport in axial magnetic fields and fermi arc surface states from pseudo-landau levels},\ }\href {https://doi.org/10.1103/PhysRevX.6.041046} {\bibfield  {journal} {\bibinfo  {journal} {Phys. Rev. X}\ }\textbf {\bibinfo {volume} {6}},\ \bibinfo {pages} {041046} (\bibinfo {year} {2016})}\BibitemShut {NoStop}%
\bibitem [{\citenamefont {Kurebayashi}\ and\ \citenamefont {Nomura}(2019)}]{kurebayashi2019theory}%
  \BibitemOpen
  \bibfield  {author} {\bibinfo {author} {\bibfnamefont {D.}~\bibnamefont {Kurebayashi}}\ and\ \bibinfo {author} {\bibfnamefont {K.}~\bibnamefont {Nomura}},\ }\bibfield  {title} {\bibinfo {title} {Theory for spin torque in weyl semimetal with magnetic texture},\ }\href {https://doi.org/10.1038/s41598-019-41776-z} {\bibfield  {journal} {\bibinfo  {journal} {Scientific reports}\ }\textbf {\bibinfo {volume} {9}},\ \bibinfo {pages} {5365} (\bibinfo {year} {2019})}\BibitemShut {NoStop}%
\bibitem [{\citenamefont {Kurebayashi}\ \emph {et~al.}(2021)\citenamefont {Kurebayashi}, \citenamefont {Araki},\ and\ \citenamefont {Nomura}}]{kurebayashi2021jpsj}%
  \BibitemOpen
  \bibfield  {author} {\bibinfo {author} {\bibfnamefont {D.}~\bibnamefont {Kurebayashi}}, \bibinfo {author} {\bibfnamefont {Y.}~\bibnamefont {Araki}},\ and\ \bibinfo {author} {\bibfnamefont {K.}~\bibnamefont {Nomura}},\ }\bibfield  {title} {\bibinfo {title} {Microscopic theory of electrically induced spin torques in magnetic weyl semimetals},\ }\href {https://doi.org/10.7566/JPSJ.90.084702} {\bibfield  {journal} {\bibinfo  {journal} {Journal of the Physical Society of Japan}\ }\textbf {\bibinfo {volume} {90}},\ \bibinfo {pages} {084702} (\bibinfo {year} {2021})}\BibitemShut {NoStop}%
\bibitem [{\citenamefont {Zhang}\ and\ \citenamefont {Nagaosa}(2024{\natexlab{b}})}]{Zhang2024}%
  \BibitemOpen
  \bibfield  {author} {\bibinfo {author} {\bibfnamefont {X.-X.}\ \bibnamefont {Zhang}}\ and\ \bibinfo {author} {\bibfnamefont {N.}~\bibnamefont {Nagaosa}},\ }\bibfield  {title} {\bibinfo {title} {Nonmonotonic hall effect of weyl semimetals under a magnetic field},\ }\href@noop {} {\bibfield  {journal} {\bibinfo  {journal} {Phys. Rev. Lett.}\ }\textbf {\bibinfo {volume} {133}},\ \bibinfo {pages} {166301} (\bibinfo {year} {2024}{\natexlab{b}})}\BibitemShut {NoStop}%
\bibitem [{\citenamefont {Metlitski}\ and\ \citenamefont {Zhitnitsky}(2005)}]{metlitski2005anomalous}%
  \BibitemOpen
  \bibfield  {author} {\bibinfo {author} {\bibfnamefont {M.~A.}\ \bibnamefont {Metlitski}}\ and\ \bibinfo {author} {\bibfnamefont {A.~R.}\ \bibnamefont {Zhitnitsky}},\ }\bibfield  {title} {\bibinfo {title} {Anomalous axion interactions and topological currents in dense matter},\ }\href@noop {} {\bibfield  {journal} {\bibinfo  {journal} {Physical Review D—Particles, Fields, Gravitation, and Cosmology}\ }\textbf {\bibinfo {volume} {72}},\ \bibinfo {pages} {045011} (\bibinfo {year} {2005})}\BibitemShut {NoStop}%
\bibitem [{\citenamefont {Fukushima}\ \emph {et~al.}(2008)\citenamefont {Fukushima}, \citenamefont {Kharzeev},\ and\ \citenamefont {Warringa}}]{Fukushima2008}%
  \BibitemOpen
  \bibfield  {author} {\bibinfo {author} {\bibfnamefont {K.}~\bibnamefont {Fukushima}}, \bibinfo {author} {\bibfnamefont {D.~E.}\ \bibnamefont {Kharzeev}},\ and\ \bibinfo {author} {\bibfnamefont {H.~J.}\ \bibnamefont {Warringa}},\ }\bibfield  {title} {\bibinfo {title} {Chiral magnetic effect},\ }\href {https://doi.org/10.1103/PhysRevD.78.074033} {\bibfield  {journal} {\bibinfo  {journal} {Phys. Rev. D}\ }\textbf {\bibinfo {volume} {78}},\ \bibinfo {pages} {074033} (\bibinfo {year} {2008})}\BibitemShut {NoStop}%
\bibitem [{\citenamefont {Kharzeev}\ and\ \citenamefont {Warringa}(2009)}]{kharzeev2009chiral}%
  \BibitemOpen
  \bibfield  {author} {\bibinfo {author} {\bibfnamefont {D.~E.}\ \bibnamefont {Kharzeev}}\ and\ \bibinfo {author} {\bibfnamefont {H.~J.}\ \bibnamefont {Warringa}},\ }\bibfield  {title} {\bibinfo {title} {Chiral magnetic conductivity},\ }\href@noop {} {\bibfield  {journal} {\bibinfo  {journal} {Physical Review D—Particles, Fields, Gravitation, and Cosmology}\ }\textbf {\bibinfo {volume} {80}},\ \bibinfo {pages} {034028} (\bibinfo {year} {2009})}\BibitemShut {NoStop}%
\bibitem [{\citenamefont {Son}\ and\ \citenamefont {Yamamoto}(2013)}]{son2013kinetic}%
  \BibitemOpen
  \bibfield  {author} {\bibinfo {author} {\bibfnamefont {D.~T.}\ \bibnamefont {Son}}\ and\ \bibinfo {author} {\bibfnamefont {N.}~\bibnamefont {Yamamoto}},\ }\bibfield  {title} {\bibinfo {title} {Kinetic theory with berry curvature from quantum field theories},\ }\href@noop {} {\bibfield  {journal} {\bibinfo  {journal} {Physical Review D—Particles, Fields, Gravitation, and Cosmology}\ }\textbf {\bibinfo {volume} {87}},\ \bibinfo {pages} {085016} (\bibinfo {year} {2013})}\BibitemShut {NoStop}%
\bibitem [{\citenamefont {Goswami}\ and\ \citenamefont {Tewari}(2013)}]{goswami2013axionic}%
  \BibitemOpen
  \bibfield  {author} {\bibinfo {author} {\bibfnamefont {P.}~\bibnamefont {Goswami}}\ and\ \bibinfo {author} {\bibfnamefont {S.}~\bibnamefont {Tewari}},\ }\bibfield  {title} {\bibinfo {title} {Axionic field theory of (3+ 1)-dimensional weyl semimetals},\ }\href@noop {} {\bibfield  {journal} {\bibinfo  {journal} {Physical Review B—Condensed Matter and Materials Physics}\ }\textbf {\bibinfo {volume} {88}},\ \bibinfo {pages} {245107} (\bibinfo {year} {2013})}\BibitemShut {NoStop}%
\bibitem [{\citenamefont {Chen}\ \emph {et~al.}(2013{\natexlab{a}})\citenamefont {Chen}, \citenamefont {Wu},\ and\ \citenamefont {Burkov}}]{chen2013axion}%
  \BibitemOpen
  \bibfield  {author} {\bibinfo {author} {\bibfnamefont {Y.}~\bibnamefont {Chen}}, \bibinfo {author} {\bibfnamefont {S.}~\bibnamefont {Wu}},\ and\ \bibinfo {author} {\bibfnamefont {A.}~\bibnamefont {Burkov}},\ }\bibfield  {title} {\bibinfo {title} {Axion response in weyl semimetals},\ }\href@noop {} {\bibfield  {journal} {\bibinfo  {journal} {Physical Review B—Condensed Matter and Materials Physics}\ }\textbf {\bibinfo {volume} {88}},\ \bibinfo {pages} {125105} (\bibinfo {year} {2013}{\natexlab{a}})}\BibitemShut {NoStop}%
\bibitem [{\citenamefont {Vazifeh}\ and\ \citenamefont {Franz}(2013)}]{Vazifeh2013}%
  \BibitemOpen
  \bibfield  {author} {\bibinfo {author} {\bibfnamefont {M.~M.}\ \bibnamefont {Vazifeh}}\ and\ \bibinfo {author} {\bibfnamefont {M.}~\bibnamefont {Franz}},\ }\bibfield  {title} {\bibinfo {title} {Electromagnetic response of weyl semimetals},\ }\href {https://doi.org/10.1103/PhysRevLett.111.027201} {\bibfield  {journal} {\bibinfo  {journal} {Phys. Rev. Lett.}\ }\textbf {\bibinfo {volume} {111}},\ \bibinfo {pages} {027201} (\bibinfo {year} {2013})}\BibitemShut {NoStop}%
\bibitem [{\citenamefont {Yamamoto}(2015)}]{yamamoto2015generalized}%
  \BibitemOpen
  \bibfield  {author} {\bibinfo {author} {\bibfnamefont {N.}~\bibnamefont {Yamamoto}},\ }\bibfield  {title} {\bibinfo {title} {Generalized bloch theorem and chiral transport phenomena},\ }\href@noop {} {\bibfield  {journal} {\bibinfo  {journal} {Physical Review D}\ }\textbf {\bibinfo {volume} {92}},\ \bibinfo {pages} {085011} (\bibinfo {year} {2015})}\BibitemShut {NoStop}%
\bibitem [{\citenamefont {Sekine}\ and\ \citenamefont {Nomura}(2016)}]{sekine2016chiral}%
  \BibitemOpen
  \bibfield  {author} {\bibinfo {author} {\bibfnamefont {A.}~\bibnamefont {Sekine}}\ and\ \bibinfo {author} {\bibfnamefont {K.}~\bibnamefont {Nomura}},\ }\bibfield  {title} {\bibinfo {title} {Chiral magnetic effect and anomalous hall effect in antiferromagnetic insulators with spin-orbit coupling},\ }\href@noop {} {\bibfield  {journal} {\bibinfo  {journal} {Physical review letters}\ }\textbf {\bibinfo {volume} {116}},\ \bibinfo {pages} {096401} (\bibinfo {year} {2016})}\BibitemShut {NoStop}%
\bibitem [{\citenamefont {Zhong}\ \emph {et~al.}(2016)\citenamefont {Zhong}, \citenamefont {Moore},\ and\ \citenamefont {Souza}}]{zhong2016gyrotropic}%
  \BibitemOpen
  \bibfield  {author} {\bibinfo {author} {\bibfnamefont {S.}~\bibnamefont {Zhong}}, \bibinfo {author} {\bibfnamefont {J.~E.}\ \bibnamefont {Moore}},\ and\ \bibinfo {author} {\bibfnamefont {I.}~\bibnamefont {Souza}},\ }\bibfield  {title} {\bibinfo {title} {Gyrotropic magnetic effect and the magnetic moment on the fermi surface},\ }\href@noop {} {\bibfield  {journal} {\bibinfo  {journal} {Physical review letters}\ }\textbf {\bibinfo {volume} {116}},\ \bibinfo {pages} {077201} (\bibinfo {year} {2016})}\BibitemShut {NoStop}%
\bibitem [{\citenamefont {Ma}\ and\ \citenamefont {Pesin}(2015)}]{ma2015chiral}%
  \BibitemOpen
  \bibfield  {author} {\bibinfo {author} {\bibfnamefont {J.}~\bibnamefont {Ma}}\ and\ \bibinfo {author} {\bibfnamefont {D.}~\bibnamefont {Pesin}},\ }\bibfield  {title} {\bibinfo {title} {Chiral magnetic effect and natural optical activity in metals with or without weyl points},\ }\href@noop {} {\bibfield  {journal} {\bibinfo  {journal} {Physical Review B}\ }\textbf {\bibinfo {volume} {92}},\ \bibinfo {pages} {235205} (\bibinfo {year} {2015})}\BibitemShut {NoStop}%
\bibitem [{\citenamefont {Zhou}\ \emph {et~al.}(2013)\citenamefont {Zhou}, \citenamefont {Jiang}, \citenamefont {Niu},\ and\ \citenamefont {Shi}}]{zhou2013topological}%
  \BibitemOpen
  \bibfield  {author} {\bibinfo {author} {\bibfnamefont {J.-H.}\ \bibnamefont {Zhou}}, \bibinfo {author} {\bibfnamefont {H.}~\bibnamefont {Jiang}}, \bibinfo {author} {\bibfnamefont {Q.}~\bibnamefont {Niu}},\ and\ \bibinfo {author} {\bibfnamefont {J.-R.}\ \bibnamefont {Shi}},\ }\bibfield  {title} {\bibinfo {title} {Topological invariants of metals and the related physical effects},\ }\href@noop {} {\bibfield  {journal} {\bibinfo  {journal} {Chinese Physics Letters}\ }\textbf {\bibinfo {volume} {30}},\ \bibinfo {pages} {027101} (\bibinfo {year} {2013})}\BibitemShut {NoStop}%
\bibitem [{\citenamefont {Huang}\ \emph {et~al.}(2017)\citenamefont {Huang}, \citenamefont {Zhou},\ and\ \citenamefont {Shen}}]{huang2017topological}%
  \BibitemOpen
  \bibfield  {author} {\bibinfo {author} {\bibfnamefont {Z.-M.}\ \bibnamefont {Huang}}, \bibinfo {author} {\bibfnamefont {J.}~\bibnamefont {Zhou}},\ and\ \bibinfo {author} {\bibfnamefont {S.-Q.}\ \bibnamefont {Shen}},\ }\bibfield  {title} {\bibinfo {title} {Topological responses from chiral anomaly in multi-weyl semimetals},\ }\href@noop {} {\bibfield  {journal} {\bibinfo  {journal} {Physical Review B}\ }\textbf {\bibinfo {volume} {96}},\ \bibinfo {pages} {085201} (\bibinfo {year} {2017})}\BibitemShut {NoStop}%
\bibitem [{\citenamefont {Shitade}\ and\ \citenamefont {Araki}(2021)}]{Shitade2021prb}%
  \BibitemOpen
  \bibfield  {author} {\bibinfo {author} {\bibfnamefont {A.}~\bibnamefont {Shitade}}\ and\ \bibinfo {author} {\bibfnamefont {Y.}~\bibnamefont {Araki}},\ }\bibfield  {title} {\bibinfo {title} {Magnetization energy current in the axial magnetic effect},\ }\href {https://doi.org/10.1103/PhysRevB.103.155202} {\bibfield  {journal} {\bibinfo  {journal} {Phys. Rev. B}\ }\textbf {\bibinfo {volume} {103}},\ \bibinfo {pages} {155202} (\bibinfo {year} {2021})}\BibitemShut {NoStop}%
\bibitem [{\citenamefont {Adler}(1969)}]{Adler1969PhysRev}%
  \BibitemOpen
  \bibfield  {author} {\bibinfo {author} {\bibfnamefont {S.~L.}\ \bibnamefont {Adler}},\ }\bibfield  {title} {\bibinfo {title} {Axial-vector vertex in spinor electrodynamics},\ }\href {https://doi.org/10.1103/PhysRev.177.2426} {\bibfield  {journal} {\bibinfo  {journal} {Phys. Rev.}\ }\textbf {\bibinfo {volume} {177}},\ \bibinfo {pages} {2426} (\bibinfo {year} {1969})}\BibitemShut {NoStop}%
\bibitem [{\citenamefont {Bell}\ and\ \citenamefont {Jackiw}(1969)}]{Bell1969pcac}%
  \BibitemOpen
  \bibfield  {author} {\bibinfo {author} {\bibfnamefont {J.~S.}\ \bibnamefont {Bell}}\ and\ \bibinfo {author} {\bibfnamefont {R.~W.}\ \bibnamefont {Jackiw}},\ }\bibfield  {title} {\bibinfo {title} {A pcac puzzle},\ }\href {https://doi.org/10.1007/BF02823296} {\bibfield  {journal} {\bibinfo  {journal} {Nuovo cimento}\ }\textbf {\bibinfo {volume} {60}},\ \bibinfo {pages} {47} (\bibinfo {year} {1969})}\BibitemShut {NoStop}%
\bibitem [{\citenamefont {Son}\ and\ \citenamefont {Spivak}(2013)}]{son2013chiral}%
  \BibitemOpen
  \bibfield  {author} {\bibinfo {author} {\bibfnamefont {D.}~\bibnamefont {Son}}\ and\ \bibinfo {author} {\bibfnamefont {B.}~\bibnamefont {Spivak}},\ }\bibfield  {title} {\bibinfo {title} {Chiral anomaly and classical negative magnetoresistance of weyl metals},\ }\href@noop {} {\bibfield  {journal} {\bibinfo  {journal} {Physical Review B—Condensed Matter and Materials Physics}\ }\textbf {\bibinfo {volume} {88}},\ \bibinfo {pages} {104412} (\bibinfo {year} {2013})}\BibitemShut {NoStop}%
\bibitem [{\citenamefont {Liang}\ \emph {et~al.}(2018{\natexlab{a}})\citenamefont {Liang}, \citenamefont {Lin}, \citenamefont {Kushwaha}, \citenamefont {Xing}, \citenamefont {Ni}, \citenamefont {Cava},\ and\ \citenamefont {Ong}}]{Liang2018}%
  \BibitemOpen
  \bibfield  {author} {\bibinfo {author} {\bibfnamefont {S.}~\bibnamefont {Liang}}, \bibinfo {author} {\bibfnamefont {J.}~\bibnamefont {Lin}}, \bibinfo {author} {\bibfnamefont {S.}~\bibnamefont {Kushwaha}}, \bibinfo {author} {\bibfnamefont {J.}~\bibnamefont {Xing}}, \bibinfo {author} {\bibfnamefont {N.}~\bibnamefont {Ni}}, \bibinfo {author} {\bibfnamefont {R.~J.}\ \bibnamefont {Cava}},\ and\ \bibinfo {author} {\bibfnamefont {N.~P.}\ \bibnamefont {Ong}},\ }\bibfield  {title} {\bibinfo {title} {Experimental tests of the chiral anomaly magnetoresistance in the dirac-weyl semimetals {Na3Bi} and {GdPtBi}},\ }\href@noop {} {\bibfield  {journal} {\bibinfo  {journal} {Phys. Rev. X.}\ }\textbf {\bibinfo {volume} {8}} (\bibinfo {year} {2018}{\natexlab{a}})}\BibitemShut {NoStop}%
\bibitem [{\citenamefont {Bardeen}\ and\ \citenamefont {Zumino}(1984)}]{bardeen1984consistent}%
  \BibitemOpen
  \bibfield  {author} {\bibinfo {author} {\bibfnamefont {W.~A.}\ \bibnamefont {Bardeen}}\ and\ \bibinfo {author} {\bibfnamefont {B.}~\bibnamefont {Zumino}},\ }\bibfield  {title} {\bibinfo {title} {Consistent and covariant anomalies in gauge and gravitational theories},\ }\href@noop {} {\bibfield  {journal} {\bibinfo  {journal} {Nuclear Physics B}\ }\textbf {\bibinfo {volume} {244}},\ \bibinfo {pages} {421} (\bibinfo {year} {1984})}\BibitemShut {NoStop}%
\bibitem [{\citenamefont {Gorbar}\ \emph {et~al.}(2017{\natexlab{a}})\citenamefont {Gorbar}, \citenamefont {Miransky}, \citenamefont {Shovkovy},\ and\ \citenamefont {Sukhachov}}]{gorbar2017origin}%
  \BibitemOpen
  \bibfield  {author} {\bibinfo {author} {\bibfnamefont {E.}~\bibnamefont {Gorbar}}, \bibinfo {author} {\bibfnamefont {V.}~\bibnamefont {Miransky}}, \bibinfo {author} {\bibfnamefont {I.}~\bibnamefont {Shovkovy}},\ and\ \bibinfo {author} {\bibfnamefont {P.}~\bibnamefont {Sukhachov}},\ }\bibfield  {title} {\bibinfo {title} {Origin of bardeen-zumino current in lattice models of weyl semimetals},\ }\href@noop {} {\bibfield  {journal} {\bibinfo  {journal} {Physical Review B}\ }\textbf {\bibinfo {volume} {96}},\ \bibinfo {pages} {085130} (\bibinfo {year} {2017}{\natexlab{a}})}\BibitemShut {NoStop}%
\bibitem [{\citenamefont {Behrends}\ \emph {et~al.}(2019)\citenamefont {Behrends}, \citenamefont {Roy}, \citenamefont {Kolodrubetz}, \citenamefont {Bardarson},\ and\ \citenamefont {Grushin}}]{Behrends2019}%
  \BibitemOpen
  \bibfield  {author} {\bibinfo {author} {\bibfnamefont {J.}~\bibnamefont {Behrends}}, \bibinfo {author} {\bibfnamefont {S.}~\bibnamefont {Roy}}, \bibinfo {author} {\bibfnamefont {M.~H.}\ \bibnamefont {Kolodrubetz}}, \bibinfo {author} {\bibfnamefont {J.~H.}\ \bibnamefont {Bardarson}},\ and\ \bibinfo {author} {\bibfnamefont {A.~G.}\ \bibnamefont {Grushin}},\ }\bibfield  {title} {\bibinfo {title} {Landau levels, bardeen polynomials, and fermi arcs in weyl semimetals: Lattice-based approach to the chiral anomaly},\ }\href {https://doi.org/10.1103/PhysRevB.99.140201} {\bibfield  {journal} {\bibinfo  {journal} {Phys. Rev. B}\ }\textbf {\bibinfo {volume} {99}},\ \bibinfo {pages} {140201} (\bibinfo {year} {2019})}\BibitemShut {NoStop}%
\bibitem [{\citenamefont {Gorbar}\ \emph {et~al.}(2017{\natexlab{b}})\citenamefont {Gorbar}, \citenamefont {Miransky}, \citenamefont {Shovkovy},\ and\ \citenamefont {Sukhachov}}]{gorbar2017consistent}%
  \BibitemOpen
  \bibfield  {author} {\bibinfo {author} {\bibfnamefont {E.}~\bibnamefont {Gorbar}}, \bibinfo {author} {\bibfnamefont {V.}~\bibnamefont {Miransky}}, \bibinfo {author} {\bibfnamefont {I.}~\bibnamefont {Shovkovy}},\ and\ \bibinfo {author} {\bibfnamefont {P.}~\bibnamefont {Sukhachov}},\ }\bibfield  {title} {\bibinfo {title} {Consistent chiral kinetic theory in weyl materials: chiral magnetic plasmons},\ }\href@noop {} {\bibfield  {journal} {\bibinfo  {journal} {Physical review letters}\ }\textbf {\bibinfo {volume} {118}},\ \bibinfo {pages} {127601} (\bibinfo {year} {2017}{\natexlab{b}})}\BibitemShut {NoStop}%
\bibitem [{\citenamefont {Gorbar}\ \emph {et~al.}(2017{\natexlab{c}})\citenamefont {Gorbar}, \citenamefont {Miransky}, \citenamefont {Shovkovy},\ and\ \citenamefont {Sukhachov}}]{gorbar2017chiral}%
  \BibitemOpen
  \bibfield  {author} {\bibinfo {author} {\bibfnamefont {E.}~\bibnamefont {Gorbar}}, \bibinfo {author} {\bibfnamefont {V.}~\bibnamefont {Miransky}}, \bibinfo {author} {\bibfnamefont {I.}~\bibnamefont {Shovkovy}},\ and\ \bibinfo {author} {\bibfnamefont {P.}~\bibnamefont {Sukhachov}},\ }\bibfield  {title} {\bibinfo {title} {Chiral magnetic plasmons in anomalous relativistic matter},\ }\href@noop {} {\bibfield  {journal} {\bibinfo  {journal} {Physical Review B}\ }\textbf {\bibinfo {volume} {95}},\ \bibinfo {pages} {115202} (\bibinfo {year} {2017}{\natexlab{c}})}\BibitemShut {NoStop}%
\bibitem [{\citenamefont {Gorbar}\ \emph {et~al.}(2017{\natexlab{d}})\citenamefont {Gorbar}, \citenamefont {Miransky}, \citenamefont {Shovkovy},\ and\ \citenamefont {Sukhachov}}]{gorbar2017second}%
  \BibitemOpen
  \bibfield  {author} {\bibinfo {author} {\bibfnamefont {E.}~\bibnamefont {Gorbar}}, \bibinfo {author} {\bibfnamefont {V.}~\bibnamefont {Miransky}}, \bibinfo {author} {\bibfnamefont {I.}~\bibnamefont {Shovkovy}},\ and\ \bibinfo {author} {\bibfnamefont {P.}~\bibnamefont {Sukhachov}},\ }\bibfield  {title} {\bibinfo {title} {Second-order chiral kinetic theory: Chiral magnetic and pseudomagnetic waves},\ }\href@noop {} {\bibfield  {journal} {\bibinfo  {journal} {Physical Review B}\ }\textbf {\bibinfo {volume} {95}},\ \bibinfo {pages} {205141} (\bibinfo {year} {2017}{\natexlab{d}})}\BibitemShut {NoStop}%
\bibitem [{\citenamefont {Gorbar}\ \emph {et~al.}(2018{\natexlab{b}})\citenamefont {Gorbar}, \citenamefont {Miransky}, \citenamefont {Shovkovy},\ and\ \citenamefont {Sukhachov}}]{gorbar2018consistent}%
  \BibitemOpen
  \bibfield  {author} {\bibinfo {author} {\bibfnamefont {E.}~\bibnamefont {Gorbar}}, \bibinfo {author} {\bibfnamefont {V.}~\bibnamefont {Miransky}}, \bibinfo {author} {\bibfnamefont {I.}~\bibnamefont {Shovkovy}},\ and\ \bibinfo {author} {\bibfnamefont {P.}~\bibnamefont {Sukhachov}},\ }\bibfield  {title} {\bibinfo {title} {Consistent hydrodynamic theory of chiral electrons in weyl semimetals},\ }\href@noop {} {\bibfield  {journal} {\bibinfo  {journal} {Physical Review B}\ }\textbf {\bibinfo {volume} {97}},\ \bibinfo {pages} {121105} (\bibinfo {year} {2018}{\natexlab{b}})}\BibitemShut {NoStop}%
\bibitem [{\citenamefont {Gorbar}\ \emph {et~al.}(2018{\natexlab{c}})\citenamefont {Gorbar}, \citenamefont {Miransky}, \citenamefont {Shovkovy},\ and\ \citenamefont {Sukhachov}}]{gorbar2018hydrodynamic}%
  \BibitemOpen
  \bibfield  {author} {\bibinfo {author} {\bibfnamefont {E.}~\bibnamefont {Gorbar}}, \bibinfo {author} {\bibfnamefont {V.}~\bibnamefont {Miransky}}, \bibinfo {author} {\bibfnamefont {I.}~\bibnamefont {Shovkovy}},\ and\ \bibinfo {author} {\bibfnamefont {P.}~\bibnamefont {Sukhachov}},\ }\bibfield  {title} {\bibinfo {title} {Hydrodynamic electron flow in a weyl semimetal slab: Role of chern-simons terms},\ }\href@noop {} {\bibfield  {journal} {\bibinfo  {journal} {Physical Review B}\ }\textbf {\bibinfo {volume} {97}},\ \bibinfo {pages} {205119} (\bibinfo {year} {2018}{\natexlab{c}})}\BibitemShut {NoStop}%
\bibitem [{\citenamefont {Chernodub}\ \emph {et~al.}(2022)\citenamefont {Chernodub}, \citenamefont {Ferreiros}, \citenamefont {Grushin}, \citenamefont {Landsteiner},\ and\ \citenamefont {Vozmediano}}]{chernodub2022thermal}%
  \BibitemOpen
  \bibfield  {author} {\bibinfo {author} {\bibfnamefont {M.~N.}\ \bibnamefont {Chernodub}}, \bibinfo {author} {\bibfnamefont {Y.}~\bibnamefont {Ferreiros}}, \bibinfo {author} {\bibfnamefont {A.~G.}\ \bibnamefont {Grushin}}, \bibinfo {author} {\bibfnamefont {K.}~\bibnamefont {Landsteiner}},\ and\ \bibinfo {author} {\bibfnamefont {M.~A.}\ \bibnamefont {Vozmediano}},\ }\bibfield  {title} {\bibinfo {title} {Thermal transport, geometry, and anomalies},\ }\href@noop {} {\bibfield  {journal} {\bibinfo  {journal} {Physics Reports}\ }\textbf {\bibinfo {volume} {977}},\ \bibinfo {pages} {1} (\bibinfo {year} {2022})}\BibitemShut {NoStop}%
\bibitem [{\citenamefont {Sekine}\ and\ \citenamefont {Nomura}(2020)}]{Sekine2020-km}%
  \BibitemOpen
  \bibfield  {author} {\bibinfo {author} {\bibfnamefont {A.}~\bibnamefont {Sekine}}\ and\ \bibinfo {author} {\bibfnamefont {K.}~\bibnamefont {Nomura}},\ }\bibfield  {title} {\bibinfo {title} {Axion electrodynamics in topological materials},\ }\href@noop {} {\bibfield  {journal} {\bibinfo  {journal} {J. Appl. Phys.}\ }\textbf {\bibinfo {volume} {129}},\ \bibinfo {pages} {141101} (\bibinfo {year} {2020})}\BibitemShut {NoStop}%
\bibitem [{\citenamefont {Hirschberger}\ \emph {et~al.}(2016)\citenamefont {Hirschberger}, \citenamefont {Kushwaha}, \citenamefont {Wang}, \citenamefont {Gibson}, \citenamefont {Liang}, \citenamefont {Belvin}, \citenamefont {Bernevig}, \citenamefont {Cava},\ and\ \citenamefont {Ong}}]{hirschberger2016}%
  \BibitemOpen
  \bibfield  {author} {\bibinfo {author} {\bibfnamefont {M.}~\bibnamefont {Hirschberger}}, \bibinfo {author} {\bibfnamefont {S.}~\bibnamefont {Kushwaha}}, \bibinfo {author} {\bibfnamefont {Z.}~\bibnamefont {Wang}}, \bibinfo {author} {\bibfnamefont {Q.}~\bibnamefont {Gibson}}, \bibinfo {author} {\bibfnamefont {S.}~\bibnamefont {Liang}}, \bibinfo {author} {\bibfnamefont {C.~A.}\ \bibnamefont {Belvin}}, \bibinfo {author} {\bibfnamefont {B.~A.}\ \bibnamefont {Bernevig}}, \bibinfo {author} {\bibfnamefont {R.~J.}\ \bibnamefont {Cava}},\ and\ \bibinfo {author} {\bibfnamefont {N.~P.}\ \bibnamefont {Ong}},\ }\bibfield  {title} {\bibinfo {title} {The chiral anomaly and thermopower of weyl fermions in the half-heusler gdptbi},\ }\href@noop {} {\bibfield  {journal} {\bibinfo  {journal} {Nature materials}\ }\textbf {\bibinfo {volume} {15}},\ \bibinfo {pages} {1161} (\bibinfo {year} {2016})}\BibitemShut {NoStop}%
\bibitem [{\citenamefont {Sakai}\ \emph {et~al.}(2018)\citenamefont {Sakai}, \citenamefont {Mizuta}, \citenamefont {Nugroho}, \citenamefont {Sihombing}, \citenamefont {Koretsune}, \citenamefont {Suzuki}, \citenamefont {Takemori}, \citenamefont {Ishii}, \citenamefont {Nishio-Hamane}, \citenamefont {Arita} \emph {et~al.}}]{Sakai2018}%
  \BibitemOpen
  \bibfield  {author} {\bibinfo {author} {\bibfnamefont {A.}~\bibnamefont {Sakai}}, \bibinfo {author} {\bibfnamefont {Y.~P.}\ \bibnamefont {Mizuta}}, \bibinfo {author} {\bibfnamefont {A.~A.}\ \bibnamefont {Nugroho}}, \bibinfo {author} {\bibfnamefont {R.}~\bibnamefont {Sihombing}}, \bibinfo {author} {\bibfnamefont {T.}~\bibnamefont {Koretsune}}, \bibinfo {author} {\bibfnamefont {M.-T.}\ \bibnamefont {Suzuki}}, \bibinfo {author} {\bibfnamefont {N.}~\bibnamefont {Takemori}}, \bibinfo {author} {\bibfnamefont {R.}~\bibnamefont {Ishii}}, \bibinfo {author} {\bibfnamefont {D.}~\bibnamefont {Nishio-Hamane}}, \bibinfo {author} {\bibfnamefont {R.}~\bibnamefont {Arita}}, \emph {et~al.},\ }\bibfield  {title} {\bibinfo {title} {Giant anomalous nernst effect and quantum-critical scaling in a ferromagnetic semimetal},\ }\href@noop {} {\bibfield  {journal} {\bibinfo  {journal} {Nature Physics}\ }\textbf {\bibinfo {volume} {14}},\ \bibinfo {pages} {1119} (\bibinfo {year} {2018})}\BibitemShut {NoStop}%
\bibitem [{\citenamefont {Soh}\ \emph {et~al.}(2018)\citenamefont {Soh}, \citenamefont {Donnerer}, \citenamefont {Hughes}, \citenamefont {Schierle}, \citenamefont {Weschke}, \citenamefont {Prabhakaran},\ and\ \citenamefont {Boothroyd}}]{soh2018magnetic}%
  \BibitemOpen
  \bibfield  {author} {\bibinfo {author} {\bibfnamefont {J.-R.}\ \bibnamefont {Soh}}, \bibinfo {author} {\bibfnamefont {C.}~\bibnamefont {Donnerer}}, \bibinfo {author} {\bibfnamefont {K.}~\bibnamefont {Hughes}}, \bibinfo {author} {\bibfnamefont {E.}~\bibnamefont {Schierle}}, \bibinfo {author} {\bibfnamefont {E.}~\bibnamefont {Weschke}}, \bibinfo {author} {\bibfnamefont {D.}~\bibnamefont {Prabhakaran}},\ and\ \bibinfo {author} {\bibfnamefont {A.}~\bibnamefont {Boothroyd}},\ }\bibfield  {title} {\bibinfo {title} {Magnetic and electronic structure of the layered rare-earth pnictide eucd 2 sb 2},\ }\href@noop {} {\bibfield  {journal} {\bibinfo  {journal} {Physical Review B}\ }\textbf {\bibinfo {volume} {98}},\ \bibinfo {pages} {064419} (\bibinfo {year} {2018})}\BibitemShut {NoStop}%
\bibitem [{\citenamefont {Gaudet}\ \emph {et~al.}(2021)\citenamefont {Gaudet}, \citenamefont {Yang}, \citenamefont {Baidya}, \citenamefont {Lu}, \citenamefont {Xu}, \citenamefont {Zhao}, \citenamefont {Rodriguez-Rivera}, \citenamefont {Hoffmann}, \citenamefont {Graf}, \citenamefont {Torchinsky} \emph {et~al.}}]{Gaudet2021}%
  \BibitemOpen
  \bibfield  {author} {\bibinfo {author} {\bibfnamefont {J.}~\bibnamefont {Gaudet}}, \bibinfo {author} {\bibfnamefont {H.-Y.}\ \bibnamefont {Yang}}, \bibinfo {author} {\bibfnamefont {S.}~\bibnamefont {Baidya}}, \bibinfo {author} {\bibfnamefont {B.}~\bibnamefont {Lu}}, \bibinfo {author} {\bibfnamefont {G.}~\bibnamefont {Xu}}, \bibinfo {author} {\bibfnamefont {Y.}~\bibnamefont {Zhao}}, \bibinfo {author} {\bibfnamefont {J.~A.}\ \bibnamefont {Rodriguez-Rivera}}, \bibinfo {author} {\bibfnamefont {C.~M.}\ \bibnamefont {Hoffmann}}, \bibinfo {author} {\bibfnamefont {D.~E.}\ \bibnamefont {Graf}}, \bibinfo {author} {\bibfnamefont {D.~H.}\ \bibnamefont {Torchinsky}}, \emph {et~al.},\ }\bibfield  {title} {\bibinfo {title} {Weyl-mediated helical magnetism in ndalsi},\ }\href@noop {} {\bibfield  {journal} {\bibinfo  {journal} {Nature materials}\ }\textbf {\bibinfo {volume} {20}},\ \bibinfo {pages} {1650} (\bibinfo {year} {2021})}\BibitemShut {NoStop}%
\bibitem [{\citenamefont {Belopolski}\ \emph {et~al.}(2025)\citenamefont {Belopolski}, \citenamefont {Watanabe}, \citenamefont {Sato}, \citenamefont {Yoshimi}, \citenamefont {Kawamura}, \citenamefont {Nagahama}, \citenamefont {Zhao}, \citenamefont {Shao}, \citenamefont {Jin}, \citenamefont {Kato} \emph {et~al.}}]{belopolski2024took}%
  \BibitemOpen
  \bibfield  {author} {\bibinfo {author} {\bibfnamefont {I.}~\bibnamefont {Belopolski}}, \bibinfo {author} {\bibfnamefont {R.}~\bibnamefont {Watanabe}}, \bibinfo {author} {\bibfnamefont {Y.}~\bibnamefont {Sato}}, \bibinfo {author} {\bibfnamefont {R.}~\bibnamefont {Yoshimi}}, \bibinfo {author} {\bibfnamefont {M.}~\bibnamefont {Kawamura}}, \bibinfo {author} {\bibfnamefont {S.}~\bibnamefont {Nagahama}}, \bibinfo {author} {\bibfnamefont {Y.}~\bibnamefont {Zhao}}, \bibinfo {author} {\bibfnamefont {S.}~\bibnamefont {Shao}}, \bibinfo {author} {\bibfnamefont {Y.}~\bibnamefont {Jin}}, \bibinfo {author} {\bibfnamefont {Y.}~\bibnamefont {Kato}}, \emph {et~al.},\ }\bibfield  {title} {\bibinfo {title} {Synthesis of a semimetallic weyl ferromagnet with point fermi surface},\ }\href@noop {} {\bibfield  {journal} {\bibinfo  {journal} {Nature}\ }\textbf {\bibinfo {volume} {637}},\ \bibinfo {pages} {1078} (\bibinfo {year} {2025})}\BibitemShut {NoStop}%
\bibitem [{\citenamefont {Xu}\ \emph {et~al.}(2015{\natexlab{b}})\citenamefont {Xu}, \citenamefont {Belopolski}, \citenamefont {Alidoust}, \citenamefont {Neupane}, \citenamefont {Bian}, \citenamefont {Zhang}, \citenamefont {Sankar}, \citenamefont {Chang}, \citenamefont {Yuan},\ and\ \citenamefont {C.-C.~Lee}}]{Xu2015_1}%
  \BibitemOpen
  \bibfield  {author} {\bibinfo {author} {\bibfnamefont {S.-Y.}\ \bibnamefont {Xu}}, \bibinfo {author} {\bibfnamefont {I.}~\bibnamefont {Belopolski}}, \bibinfo {author} {\bibfnamefont {N.}~\bibnamefont {Alidoust}}, \bibinfo {author} {\bibfnamefont {M.}~\bibnamefont {Neupane}}, \bibinfo {author} {\bibfnamefont {G.}~\bibnamefont {Bian}}, \bibinfo {author} {\bibfnamefont {C.}~\bibnamefont {Zhang}}, \bibinfo {author} {\bibfnamefont {R.}~\bibnamefont {Sankar}}, \bibinfo {author} {\bibfnamefont {G.}~\bibnamefont {Chang}}, \bibinfo {author} {\bibfnamefont {Z.}~\bibnamefont {Yuan}},\ and\ \bibinfo {author} {\bibfnamefont {e.~a.}\ \bibnamefont {C.-C.~Lee}},\ }\bibfield  {title} {\bibinfo {title} {Discovery of a weyl fermion semimetal and topological fermi arcs},\ }\href@noop {} {\bibfield  {journal} {\bibinfo  {journal} {Science}\ }\textbf {\bibinfo {volume} {349}} (\bibinfo {year} {2015}{\natexlab{b}})}\BibitemShut {NoStop}%
\bibitem [{\citenamefont {Xu}\ \emph {et~al.}(2015{\natexlab{c}})\citenamefont {Xu}, \citenamefont {Belopolski}, \citenamefont {Sanchez}, \citenamefont {Zhang}, \citenamefont {Chang}, \citenamefont {Guo}, \citenamefont {Bian}, \citenamefont {Yuan}, \citenamefont {Lu},\ and\ \citenamefont {T.-R.~Chang}}]{Xu2015_2}%
  \BibitemOpen
  \bibfield  {author} {\bibinfo {author} {\bibfnamefont {S.-Y.}\ \bibnamefont {Xu}}, \bibinfo {author} {\bibfnamefont {I.}~\bibnamefont {Belopolski}}, \bibinfo {author} {\bibfnamefont {D.~S.}\ \bibnamefont {Sanchez}}, \bibinfo {author} {\bibfnamefont {C.}~\bibnamefont {Zhang}}, \bibinfo {author} {\bibfnamefont {G.}~\bibnamefont {Chang}}, \bibinfo {author} {\bibfnamefont {C.}~\bibnamefont {Guo}}, \bibinfo {author} {\bibfnamefont {G.}~\bibnamefont {Bian}}, \bibinfo {author} {\bibfnamefont {Z.}~\bibnamefont {Yuan}}, \bibinfo {author} {\bibfnamefont {H.}~\bibnamefont {Lu}},\ and\ \bibinfo {author} {\bibfnamefont {e.~a.}\ \bibnamefont {T.-R.~Chang}},\ }\bibfield  {title} {\bibinfo {title} {Experimental discovery of a topological weyl semimetal state in tap},\ }\href@noop {} {\bibfield  {journal} {\bibinfo  {journal} {Science advances}\ }\textbf {\bibinfo {volume} {1}} (\bibinfo {year} {2015}{\natexlab{c}})}\BibitemShut {NoStop}%
\bibitem [{\citenamefont {Singh}(1996)}]{Singh1996}%
  \BibitemOpen
  \bibfield  {author} {\bibinfo {author} {\bibfnamefont {D.~J.}\ \bibnamefont {Singh}},\ }\bibfield  {title} {\bibinfo {title} {{Electronic and magnetic properties of the 4d itinerant ferromagnet SrRuO3}},\ }\href {https://doi.org/10.1063/1.361618} {\bibfield  {journal} {\bibinfo  {journal} {Journal of Applied Physics}\ }\textbf {\bibinfo {volume} {79}},\ \bibinfo {pages} {4818} (\bibinfo {year} {1996})}\BibitemShut {NoStop}%
\bibitem [{\citenamefont {Mazin}\ \emph {et~al.}(2000)\citenamefont {Mazin}, \citenamefont {Papaconstantopoulos},\ and\ \citenamefont {Singh}}]{Mazin2000}%
  \BibitemOpen
  \bibfield  {author} {\bibinfo {author} {\bibfnamefont {I.~I.}\ \bibnamefont {Mazin}}, \bibinfo {author} {\bibfnamefont {D.~A.}\ \bibnamefont {Papaconstantopoulos}},\ and\ \bibinfo {author} {\bibfnamefont {D.~J.}\ \bibnamefont {Singh}},\ }\bibfield  {title} {\bibinfo {title} {Tight-binding hamiltonians for sr-filled ruthenates: Application to the gap anisotropy and hall coefficient in ${\mathrm{sr}}_{2}{\mathrm{ruo}}_{4}$},\ }\href {https://doi.org/10.1103/PhysRevB.61.5223} {\bibfield  {journal} {\bibinfo  {journal} {Phys. Rev. B}\ }\textbf {\bibinfo {volume} {61}},\ \bibinfo {pages} {5223} (\bibinfo {year} {2000})}\BibitemShut {NoStop}%
\bibitem [{\citenamefont {Rondinelli}\ \emph {et~al.}(2008)\citenamefont {Rondinelli}, \citenamefont {Caffrey}, \citenamefont {Sanvito},\ and\ \citenamefont {Spaldin}}]{Rondinelli2008}%
  \BibitemOpen
  \bibfield  {author} {\bibinfo {author} {\bibfnamefont {J.~M.}\ \bibnamefont {Rondinelli}}, \bibinfo {author} {\bibfnamefont {N.~M.}\ \bibnamefont {Caffrey}}, \bibinfo {author} {\bibfnamefont {S.}~\bibnamefont {Sanvito}},\ and\ \bibinfo {author} {\bibfnamefont {N.~A.}\ \bibnamefont {Spaldin}},\ }\bibfield  {title} {\bibinfo {title} {Electronic properties of bulk and thin film ${\text{srruo}}_{3}$: Search for the metal-insulator transition},\ }\href {https://doi.org/10.1103/PhysRevB.78.155107} {\bibfield  {journal} {\bibinfo  {journal} {Phys. Rev. B}\ }\textbf {\bibinfo {volume} {78}},\ \bibinfo {pages} {155107} (\bibinfo {year} {2008})}\BibitemShut {NoStop}%
\bibitem [{\citenamefont {Toyota}\ \emph {et~al.}(2005)\citenamefont {Toyota}, \citenamefont {Ohkubo}, \citenamefont {Kumigashira}, \citenamefont {Oshima}, \citenamefont {Ohnishi}, \citenamefont {Lippmaa}, \citenamefont {Takizawa}, \citenamefont {Fujimori}, \citenamefont {Ono}, \citenamefont {Kawasaki},\ and\ \citenamefont {Koinuma}}]{Toyota2005}%
  \BibitemOpen
  \bibfield  {author} {\bibinfo {author} {\bibfnamefont {D.}~\bibnamefont {Toyota}}, \bibinfo {author} {\bibfnamefont {I.}~\bibnamefont {Ohkubo}}, \bibinfo {author} {\bibfnamefont {H.}~\bibnamefont {Kumigashira}}, \bibinfo {author} {\bibfnamefont {M.}~\bibnamefont {Oshima}}, \bibinfo {author} {\bibfnamefont {T.}~\bibnamefont {Ohnishi}}, \bibinfo {author} {\bibfnamefont {M.}~\bibnamefont {Lippmaa}}, \bibinfo {author} {\bibfnamefont {M.}~\bibnamefont {Takizawa}}, \bibinfo {author} {\bibfnamefont {A.}~\bibnamefont {Fujimori}}, \bibinfo {author} {\bibfnamefont {K.}~\bibnamefont {Ono}}, \bibinfo {author} {\bibfnamefont {M.}~\bibnamefont {Kawasaki}},\ and\ \bibinfo {author} {\bibfnamefont {H.}~\bibnamefont {Koinuma}},\ }\bibfield  {title} {\bibinfo {title} {{Thickness-dependent electronic structure of ultrathin SrRuO3 films studied by in situ photoemission spectroscopy}},\ }\href {https://doi.org/10.1063/1.2108123} {\bibfield  {journal} {\bibinfo  {journal} {Applied Physics Letters}\ }\textbf {\bibinfo {volume}
  {87}},\ \bibinfo {pages} {162508} (\bibinfo {year} {2005})}\BibitemShut {NoStop}%
\bibitem [{\citenamefont {Kim}\ \emph {et~al.}(2004)\citenamefont {Kim}, \citenamefont {Noh}, \citenamefont {Kim},\ and\ \citenamefont {Oh}}]{Kim2004}%
  \BibitemOpen
  \bibfield  {author} {\bibinfo {author} {\bibfnamefont {H.-D.}\ \bibnamefont {Kim}}, \bibinfo {author} {\bibfnamefont {H.-J.}\ \bibnamefont {Noh}}, \bibinfo {author} {\bibfnamefont {K.~H.}\ \bibnamefont {Kim}},\ and\ \bibinfo {author} {\bibfnamefont {S.-J.}\ \bibnamefont {Oh}},\ }\bibfield  {title} {\bibinfo {title} {Core-level x-ray photoemission satellites in ruthenates: A new mechanism revealing the mott transition},\ }\href {https://doi.org/10.1103/PhysRevLett.93.126404} {\bibfield  {journal} {\bibinfo  {journal} {Phys. Rev. Lett.}\ }\textbf {\bibinfo {volume} {93}},\ \bibinfo {pages} {126404} (\bibinfo {year} {2004})}\BibitemShut {NoStop}%
\bibitem [{\citenamefont {Chen}\ \emph {et~al.}(2013{\natexlab{b}})\citenamefont {Chen}, \citenamefont {Bergman},\ and\ \citenamefont {Burkov}}]{Chen2013}%
  \BibitemOpen
  \bibfield  {author} {\bibinfo {author} {\bibfnamefont {Y.}~\bibnamefont {Chen}}, \bibinfo {author} {\bibfnamefont {D.~L.}\ \bibnamefont {Bergman}},\ and\ \bibinfo {author} {\bibfnamefont {A.~A.}\ \bibnamefont {Burkov}},\ }\bibfield  {title} {\bibinfo {title} {Weyl fermions and the anomalous hall effect in metallic ferromagnets},\ }\href {https://doi.org/10.1103/PhysRevB.88.125110} {\bibfield  {journal} {\bibinfo  {journal} {Phys. Rev. B}\ }\textbf {\bibinfo {volume} {88}},\ \bibinfo {pages} {125110} (\bibinfo {year} {2013}{\natexlab{b}})}\BibitemShut {NoStop}%
\bibitem [{\citenamefont {Itoh}\ \emph {et~al.}(2016)\citenamefont {Itoh}, \citenamefont {Endoh}, \citenamefont {Yokoo}, \citenamefont {Ibuka}, \citenamefont {Park}, \citenamefont {Kaneko}, \citenamefont {Takahashi}, \citenamefont {Tokura},\ and\ \citenamefont {Nagaosa}}]{Itoh2016}%
  \BibitemOpen
  \bibfield  {author} {\bibinfo {author} {\bibfnamefont {S.}~\bibnamefont {Itoh}}, \bibinfo {author} {\bibfnamefont {Y.}~\bibnamefont {Endoh}}, \bibinfo {author} {\bibfnamefont {T.}~\bibnamefont {Yokoo}}, \bibinfo {author} {\bibfnamefont {S.}~\bibnamefont {Ibuka}}, \bibinfo {author} {\bibfnamefont {J.-G.}\ \bibnamefont {Park}}, \bibinfo {author} {\bibfnamefont {Y.}~\bibnamefont {Kaneko}}, \bibinfo {author} {\bibfnamefont {K.~S.}\ \bibnamefont {Takahashi}}, \bibinfo {author} {\bibfnamefont {Y.}~\bibnamefont {Tokura}},\ and\ \bibinfo {author} {\bibfnamefont {N.}~\bibnamefont {Nagaosa}},\ }\bibfield  {title} {\bibinfo {title} {Weyl fermions and spin dynamics of metallic ferromagnet srruo3},\ }\href {https://api.semanticscholar.org/CorpusID:18499969} {\bibfield  {journal} {\bibinfo  {journal} {Nature Communications}\ }\textbf {\bibinfo {volume} {7}} (\bibinfo {year} {2016})}\BibitemShut {NoStop}%
\bibitem [{\citenamefont {Onoda}\ \emph {et~al.}(2008)\citenamefont {Onoda}, \citenamefont {S.~Mishchenko},\ and\ \citenamefont {Nagaosa}}]{Onoda2008}%
  \BibitemOpen
  \bibfield  {author} {\bibinfo {author} {\bibfnamefont {M.}~\bibnamefont {Onoda}}, \bibinfo {author} {\bibfnamefont {A.}~\bibnamefont {S.~Mishchenko}},\ and\ \bibinfo {author} {\bibfnamefont {N.}~\bibnamefont {Nagaosa}},\ }\bibfield  {title} {\bibinfo {title} {Left-handed spin wave excitation in ferromagnet},\ }\href {https://doi.org/10.1143/JPSJ.77.013702} {\bibfield  {journal} {\bibinfo  {journal} {Journal of the Physical Society of Japan}\ }\textbf {\bibinfo {volume} {77}},\ \bibinfo {pages} {013702} (\bibinfo {year} {2008})}\BibitemShut {NoStop}%
\bibitem [{\citenamefont {Feigenson}\ \emph {et~al.}(2007)\citenamefont {Feigenson}, \citenamefont {Reiner},\ and\ \citenamefont {Klein}}]{Feigenson2007}%
  \BibitemOpen
  \bibfield  {author} {\bibinfo {author} {\bibfnamefont {M.}~\bibnamefont {Feigenson}}, \bibinfo {author} {\bibfnamefont {J.~W.}\ \bibnamefont {Reiner}},\ and\ \bibinfo {author} {\bibfnamefont {L.}~\bibnamefont {Klein}},\ }\bibfield  {title} {\bibinfo {title} {Efficient current-induced domain-wall displacement in ${\mathrm{srruo}}_{3}$},\ }\href {https://doi.org/10.1103/PhysRevLett.98.247204} {\bibfield  {journal} {\bibinfo  {journal} {Phys. Rev. Lett.}\ }\textbf {\bibinfo {volume} {98}},\ \bibinfo {pages} {247204} (\bibinfo {year} {2007})}\BibitemShut {NoStop}%
\bibitem [{\citenamefont {Yamanouchi}\ \emph {et~al.}(2019)\citenamefont {Yamanouchi}, \citenamefont {Oyamada}, \citenamefont {Sato}, \citenamefont {Ohta},\ and\ \citenamefont {Ieda}}]{Yamanouchi2019}%
  \BibitemOpen
  \bibfield  {author} {\bibinfo {author} {\bibfnamefont {M.}~\bibnamefont {Yamanouchi}}, \bibinfo {author} {\bibfnamefont {T.}~\bibnamefont {Oyamada}}, \bibinfo {author} {\bibfnamefont {K.}~\bibnamefont {Sato}}, \bibinfo {author} {\bibfnamefont {H.}~\bibnamefont {Ohta}},\ and\ \bibinfo {author} {\bibfnamefont {J.}~\bibnamefont {Ieda}},\ }\bibfield  {title} {\bibinfo {title} {Current-induced modulation of coercive field in the ferromagnetic oxide srruo3},\ }\href {https://doi.org/10.1109/TMAG.2019.2894897} {\bibfield  {journal} {\bibinfo  {journal} {IEEE Transactions on Magnetics}\ }\textbf {\bibinfo {volume} {55}},\ \bibinfo {pages} {1} (\bibinfo {year} {2019})}\BibitemShut {NoStop}%
\bibitem [{\citenamefont {Yamanouchi}\ \emph {et~al.}(2022)\citenamefont {Yamanouchi}, \citenamefont {Araki}, \citenamefont {Sakai}, \citenamefont {Uemura}, \citenamefont {Ohta},\ and\ \citenamefont {Ieda}}]{yamanouchi2022sciadv}%
  \BibitemOpen
  \bibfield  {author} {\bibinfo {author} {\bibfnamefont {M.}~\bibnamefont {Yamanouchi}}, \bibinfo {author} {\bibfnamefont {Y.}~\bibnamefont {Araki}}, \bibinfo {author} {\bibfnamefont {T.}~\bibnamefont {Sakai}}, \bibinfo {author} {\bibfnamefont {T.}~\bibnamefont {Uemura}}, \bibinfo {author} {\bibfnamefont {H.}~\bibnamefont {Ohta}},\ and\ \bibinfo {author} {\bibfnamefont {J.}~\bibnamefont {Ieda}},\ }\bibfield  {title} {\bibinfo {title} {Observation of topological hall torque exerted on a domain wall in the ferromagnetic oxide srruo3},\ }\href@noop {} {\bibfield  {journal} {\bibinfo  {journal} {Science advances}\ }\textbf {\bibinfo {volume} {8}},\ \bibinfo {pages} {eabl6192} (\bibinfo {year} {2022})}\BibitemShut {NoStop}%
\bibitem [{\citenamefont {Araki}\ and\ \citenamefont {Ieda}(2021)}]{Araki2021prl}%
  \BibitemOpen
  \bibfield  {author} {\bibinfo {author} {\bibfnamefont {Y.}~\bibnamefont {Araki}}\ and\ \bibinfo {author} {\bibfnamefont {J.}~\bibnamefont {Ieda}},\ }\bibfield  {title} {\bibinfo {title} {Intrinsic torques emerging from anomalous velocity in magnetic textures},\ }\href {https://doi.org/10.1103/PhysRevLett.127.277205} {\bibfield  {journal} {\bibinfo  {journal} {Phys. Rev. Lett.}\ }\textbf {\bibinfo {volume} {127}},\ \bibinfo {pages} {277205} (\bibinfo {year} {2021})}\BibitemShut {NoStop}%
\bibitem [{\citenamefont {Takiguchi}\ \emph {et~al.}(2020)\citenamefont {Takiguchi}, \citenamefont {Wakabayashi}, \citenamefont {Irie}, \citenamefont {Krockenberger}, \citenamefont {Otsuka}, \citenamefont {Sawada}, \citenamefont {Nikolaev}, \citenamefont {Das}, \citenamefont {Tanaka}, \citenamefont {Taniyasu} \emph {et~al.}}]{Takiguchi2020}%
  \BibitemOpen
  \bibfield  {author} {\bibinfo {author} {\bibfnamefont {K.}~\bibnamefont {Takiguchi}}, \bibinfo {author} {\bibfnamefont {Y.~K.}\ \bibnamefont {Wakabayashi}}, \bibinfo {author} {\bibfnamefont {H.}~\bibnamefont {Irie}}, \bibinfo {author} {\bibfnamefont {Y.}~\bibnamefont {Krockenberger}}, \bibinfo {author} {\bibfnamefont {T.}~\bibnamefont {Otsuka}}, \bibinfo {author} {\bibfnamefont {H.}~\bibnamefont {Sawada}}, \bibinfo {author} {\bibfnamefont {S.~A.}\ \bibnamefont {Nikolaev}}, \bibinfo {author} {\bibfnamefont {H.}~\bibnamefont {Das}}, \bibinfo {author} {\bibfnamefont {M.}~\bibnamefont {Tanaka}}, \bibinfo {author} {\bibfnamefont {Y.}~\bibnamefont {Taniyasu}}, \emph {et~al.},\ }\bibfield  {title} {\bibinfo {title} {Quantum transport evidence of weyl fermions in an epitaxial ferromagnetic oxide},\ }\href@noop {} {\bibfield  {journal} {\bibinfo  {journal} {Nature communications}\ }\textbf {\bibinfo {volume} {11}},\ \bibinfo {pages} {4969} (\bibinfo {year} {2020})}\BibitemShut {NoStop}%
\bibitem [{\citenamefont {Kaneta-Takada}\ \emph {et~al.}(2022)\citenamefont {Kaneta-Takada}, \citenamefont {Wakabayashi}, \citenamefont {Krockenberger}, \citenamefont {Nomura}, \citenamefont {Kohama}, \citenamefont {Nikolaev}, \citenamefont {Das}, \citenamefont {Irie}, \citenamefont {Takiguchi}, \citenamefont {Ohya} \emph {et~al.}}]{Kaneta-Takada2022}%
  \BibitemOpen
  \bibfield  {author} {\bibinfo {author} {\bibfnamefont {S.}~\bibnamefont {Kaneta-Takada}}, \bibinfo {author} {\bibfnamefont {Y.~K.}\ \bibnamefont {Wakabayashi}}, \bibinfo {author} {\bibfnamefont {Y.}~\bibnamefont {Krockenberger}}, \bibinfo {author} {\bibfnamefont {T.}~\bibnamefont {Nomura}}, \bibinfo {author} {\bibfnamefont {Y.}~\bibnamefont {Kohama}}, \bibinfo {author} {\bibfnamefont {S.~A.}\ \bibnamefont {Nikolaev}}, \bibinfo {author} {\bibfnamefont {H.}~\bibnamefont {Das}}, \bibinfo {author} {\bibfnamefont {H.}~\bibnamefont {Irie}}, \bibinfo {author} {\bibfnamefont {K.}~\bibnamefont {Takiguchi}}, \bibinfo {author} {\bibfnamefont {S.}~\bibnamefont {Ohya}}, \emph {et~al.},\ }\bibfield  {title} {\bibinfo {title} {High-mobility two-dimensional carriers from surface fermi arcs in magnetic weyl semimetal films},\ }\href@noop {} {\bibfield  {journal} {\bibinfo  {journal} {npj Quantum Materials}\ }\textbf {\bibinfo {volume} {7}},\ \bibinfo {pages} {102} (\bibinfo {year} {2022})}\BibitemShut {NoStop}%
\bibitem [{\citenamefont {Chang}\ \emph {et~al.}(2017)\citenamefont {Chang}, \citenamefont {Xu}, \citenamefont {Zhou}, \citenamefont {Huang}, \citenamefont {Singh}, \citenamefont {Wang}, \citenamefont {Belopolski}, \citenamefont {Yin}, \citenamefont {Zhang}, \citenamefont {Bansil} \emph {et~al.}}]{chang2017topological}%
  \BibitemOpen
  \bibfield  {author} {\bibinfo {author} {\bibfnamefont {G.}~\bibnamefont {Chang}}, \bibinfo {author} {\bibfnamefont {S.-Y.}\ \bibnamefont {Xu}}, \bibinfo {author} {\bibfnamefont {X.}~\bibnamefont {Zhou}}, \bibinfo {author} {\bibfnamefont {S.-M.}\ \bibnamefont {Huang}}, \bibinfo {author} {\bibfnamefont {B.}~\bibnamefont {Singh}}, \bibinfo {author} {\bibfnamefont {B.}~\bibnamefont {Wang}}, \bibinfo {author} {\bibfnamefont {I.}~\bibnamefont {Belopolski}}, \bibinfo {author} {\bibfnamefont {J.}~\bibnamefont {Yin}}, \bibinfo {author} {\bibfnamefont {S.}~\bibnamefont {Zhang}}, \bibinfo {author} {\bibfnamefont {A.}~\bibnamefont {Bansil}}, \emph {et~al.},\ }\bibfield  {title} {\bibinfo {title} {Topological hopf and chain link semimetal states and their application to co 2 mn g a},\ }\href@noop {} {\bibfield  {journal} {\bibinfo  {journal} {Physical review letters}\ }\textbf {\bibinfo {volume} {119}},\ \bibinfo {pages} {156401} (\bibinfo {year} {2017})}\BibitemShut {NoStop}%
\bibitem [{\citenamefont {Felser}\ \emph {et~al.}(2007)\citenamefont {Felser}, \citenamefont {Fecher},\ and\ \citenamefont {Balke}}]{felser2007spintronics}%
  \BibitemOpen
  \bibfield  {author} {\bibinfo {author} {\bibfnamefont {C.}~\bibnamefont {Felser}}, \bibinfo {author} {\bibfnamefont {G.~H.}\ \bibnamefont {Fecher}},\ and\ \bibinfo {author} {\bibfnamefont {B.}~\bibnamefont {Balke}},\ }\bibfield  {title} {\bibinfo {title} {Spintronics: a challenge for materials science and solid-state chemistry},\ }\href@noop {} {\bibfield  {journal} {\bibinfo  {journal} {Angewandte Chemie International Edition}\ }\textbf {\bibinfo {volume} {46}},\ \bibinfo {pages} {668} (\bibinfo {year} {2007})}\BibitemShut {NoStop}%
\bibitem [{\citenamefont {K{\"u}bler}\ \emph {et~al.}(2007)\citenamefont {K{\"u}bler}, \citenamefont {Fecher},\ and\ \citenamefont {Felser}}]{kubler2007understanding}%
  \BibitemOpen
  \bibfield  {author} {\bibinfo {author} {\bibfnamefont {J.}~\bibnamefont {K{\"u}bler}}, \bibinfo {author} {\bibfnamefont {G.}~\bibnamefont {Fecher}},\ and\ \bibinfo {author} {\bibfnamefont {C.}~\bibnamefont {Felser}},\ }\bibfield  {title} {\bibinfo {title} {Understanding the trend in the curie temperatures of co 2-based heusler compounds: Ab initio calculations},\ }\href@noop {} {\bibfield  {journal} {\bibinfo  {journal} {Physical Review B—Condensed Matter and Materials Physics}\ }\textbf {\bibinfo {volume} {76}},\ \bibinfo {pages} {024414} (\bibinfo {year} {2007})}\BibitemShut {NoStop}%
\bibitem [{\citenamefont {Umetsu}\ \emph {et~al.}(2008)\citenamefont {Umetsu}, \citenamefont {Kobayashi}, \citenamefont {Fujita}, \citenamefont {Kainuma},\ and\ \citenamefont {Ishida}}]{umetsu2008magnetic}%
  \BibitemOpen
  \bibfield  {author} {\bibinfo {author} {\bibfnamefont {R.}~\bibnamefont {Umetsu}}, \bibinfo {author} {\bibfnamefont {K.}~\bibnamefont {Kobayashi}}, \bibinfo {author} {\bibfnamefont {A.}~\bibnamefont {Fujita}}, \bibinfo {author} {\bibfnamefont {R.}~\bibnamefont {Kainuma}},\ and\ \bibinfo {author} {\bibfnamefont {K.}~\bibnamefont {Ishida}},\ }\bibfield  {title} {\bibinfo {title} {Magnetic properties and stability of l21 and b2 phases in the co2mnal heusler alloy},\ }\href@noop {} {\bibfield  {journal} {\bibinfo  {journal} {Journal of Applied Physics}\ }\textbf {\bibinfo {volume} {103}} (\bibinfo {year} {2008})}\BibitemShut {NoStop}%
\bibitem [{\citenamefont {Wang}\ \emph {et~al.}(2016{\natexlab{a}})\citenamefont {Wang}, \citenamefont {Vergniory}, \citenamefont {Kushwaha}, \citenamefont {Hirschberger}, \citenamefont {Chulkov}, \citenamefont {Ernst}, \citenamefont {Ong}, \citenamefont {Cava},\ and\ \citenamefont {Bernevig}}]{wang2016time}%
  \BibitemOpen
  \bibfield  {author} {\bibinfo {author} {\bibfnamefont {Z.}~\bibnamefont {Wang}}, \bibinfo {author} {\bibfnamefont {M.}~\bibnamefont {Vergniory}}, \bibinfo {author} {\bibfnamefont {S.}~\bibnamefont {Kushwaha}}, \bibinfo {author} {\bibfnamefont {M.}~\bibnamefont {Hirschberger}}, \bibinfo {author} {\bibfnamefont {E.}~\bibnamefont {Chulkov}}, \bibinfo {author} {\bibfnamefont {A.}~\bibnamefont {Ernst}}, \bibinfo {author} {\bibfnamefont {N.~P.}\ \bibnamefont {Ong}}, \bibinfo {author} {\bibfnamefont {R.~J.}\ \bibnamefont {Cava}},\ and\ \bibinfo {author} {\bibfnamefont {B.~A.}\ \bibnamefont {Bernevig}},\ }\bibfield  {title} {\bibinfo {title} {Time-reversal-breaking weyl fermions in magnetic heusler alloys},\ }\href@noop {} {\bibfield  {journal} {\bibinfo  {journal} {Physical review letters}\ }\textbf {\bibinfo {volume} {117}},\ \bibinfo {pages} {236401} (\bibinfo {year} {2016}{\natexlab{a}})}\BibitemShut {NoStop}%
\bibitem [{\citenamefont {Manna}\ \emph {et~al.}(2018)\citenamefont {Manna}, \citenamefont {Muechler}, \citenamefont {Kao}, \citenamefont {Stinshoff}, \citenamefont {Zhang}, \citenamefont {Gooth}, \citenamefont {Kumar}, \citenamefont {Kreiner}, \citenamefont {Koepernik}, \citenamefont {Car} \emph {et~al.}}]{manna2018colossal}%
  \BibitemOpen
  \bibfield  {author} {\bibinfo {author} {\bibfnamefont {K.}~\bibnamefont {Manna}}, \bibinfo {author} {\bibfnamefont {L.}~\bibnamefont {Muechler}}, \bibinfo {author} {\bibfnamefont {T.-H.}\ \bibnamefont {Kao}}, \bibinfo {author} {\bibfnamefont {R.}~\bibnamefont {Stinshoff}}, \bibinfo {author} {\bibfnamefont {Y.}~\bibnamefont {Zhang}}, \bibinfo {author} {\bibfnamefont {J.}~\bibnamefont {Gooth}}, \bibinfo {author} {\bibfnamefont {N.}~\bibnamefont {Kumar}}, \bibinfo {author} {\bibfnamefont {G.}~\bibnamefont {Kreiner}}, \bibinfo {author} {\bibfnamefont {K.}~\bibnamefont {Koepernik}}, \bibinfo {author} {\bibfnamefont {R.}~\bibnamefont {Car}}, \emph {et~al.},\ }\bibfield  {title} {\bibinfo {title} {From colossal to zero: controlling the anomalous hall effect in magnetic heusler compounds via berry curvature design},\ }\href@noop {} {\bibfield  {journal} {\bibinfo  {journal} {Physical Review X}\ }\textbf {\bibinfo {volume} {8}},\ \bibinfo {pages} {041045} (\bibinfo {year} {2018})}\BibitemShut {NoStop}%
\bibitem [{\citenamefont {Sumida}\ \emph {et~al.}(2020)\citenamefont {Sumida}, \citenamefont {Sakuraba}, \citenamefont {Masuda}, \citenamefont {Kono}, \citenamefont {Kakoki}, \citenamefont {Goto}, \citenamefont {Zhou}, \citenamefont {Miyamoto}, \citenamefont {Miura}, \citenamefont {Okuda} \emph {et~al.}}]{sumida2020spin}%
  \BibitemOpen
  \bibfield  {author} {\bibinfo {author} {\bibfnamefont {K.}~\bibnamefont {Sumida}}, \bibinfo {author} {\bibfnamefont {Y.}~\bibnamefont {Sakuraba}}, \bibinfo {author} {\bibfnamefont {K.}~\bibnamefont {Masuda}}, \bibinfo {author} {\bibfnamefont {T.}~\bibnamefont {Kono}}, \bibinfo {author} {\bibfnamefont {M.}~\bibnamefont {Kakoki}}, \bibinfo {author} {\bibfnamefont {K.}~\bibnamefont {Goto}}, \bibinfo {author} {\bibfnamefont {W.}~\bibnamefont {Zhou}}, \bibinfo {author} {\bibfnamefont {K.}~\bibnamefont {Miyamoto}}, \bibinfo {author} {\bibfnamefont {Y.}~\bibnamefont {Miura}}, \bibinfo {author} {\bibfnamefont {T.}~\bibnamefont {Okuda}}, \emph {et~al.},\ }\bibfield  {title} {\bibinfo {title} {Spin-polarized weyl cones and giant anomalous nernst effect in ferromagnetic heusler films},\ }\href@noop {} {\bibfield  {journal} {\bibinfo  {journal} {Communications Materials}\ }\textbf {\bibinfo {volume} {1}},\ \bibinfo {pages} {89} (\bibinfo {year} {2020})}\BibitemShut {NoStop}%
\bibitem [{\citenamefont {Xu}\ \emph {et~al.}(2020)\citenamefont {Xu}, \citenamefont {Li}, \citenamefont {Ding}, \citenamefont {Chen}, \citenamefont {Sakai}, \citenamefont {Fauqu{\'e}}, \citenamefont {Nakatsuji}, \citenamefont {Zhu},\ and\ \citenamefont {Behnia}}]{xu2020anomalous}%
  \BibitemOpen
  \bibfield  {author} {\bibinfo {author} {\bibfnamefont {L.}~\bibnamefont {Xu}}, \bibinfo {author} {\bibfnamefont {X.}~\bibnamefont {Li}}, \bibinfo {author} {\bibfnamefont {L.}~\bibnamefont {Ding}}, \bibinfo {author} {\bibfnamefont {T.}~\bibnamefont {Chen}}, \bibinfo {author} {\bibfnamefont {A.}~\bibnamefont {Sakai}}, \bibinfo {author} {\bibfnamefont {B.}~\bibnamefont {Fauqu{\'e}}}, \bibinfo {author} {\bibfnamefont {S.}~\bibnamefont {Nakatsuji}}, \bibinfo {author} {\bibfnamefont {Z.}~\bibnamefont {Zhu}},\ and\ \bibinfo {author} {\bibfnamefont {K.}~\bibnamefont {Behnia}},\ }\bibfield  {title} {\bibinfo {title} {Anomalous transverse response of co 2 mnga and universality of the room-temperature $\alpha$ ij a/$\sigma$ ij a ratio across topological magnets},\ }\href@noop {} {\bibfield  {journal} {\bibinfo  {journal} {Physical Review B}\ }\textbf {\bibinfo {volume} {101}},\ \bibinfo {pages} {180404} (\bibinfo {year} {2020})}\BibitemShut {NoStop}%
\bibitem [{\citenamefont {Isshiki}\ \emph {et~al.}(2022)\citenamefont {Isshiki}, \citenamefont {Zhu}, \citenamefont {Mizuno}, \citenamefont {Uesugi}, \citenamefont {Higo}, \citenamefont {Nakatsuji},\ and\ \citenamefont {Otani}}]{isshiki2022determination}%
  \BibitemOpen
  \bibfield  {author} {\bibinfo {author} {\bibfnamefont {H.}~\bibnamefont {Isshiki}}, \bibinfo {author} {\bibfnamefont {Z.}~\bibnamefont {Zhu}}, \bibinfo {author} {\bibfnamefont {H.}~\bibnamefont {Mizuno}}, \bibinfo {author} {\bibfnamefont {R.}~\bibnamefont {Uesugi}}, \bibinfo {author} {\bibfnamefont {T.}~\bibnamefont {Higo}}, \bibinfo {author} {\bibfnamefont {S.}~\bibnamefont {Nakatsuji}},\ and\ \bibinfo {author} {\bibfnamefont {Y.}~\bibnamefont {Otani}},\ }\bibfield  {title} {\bibinfo {title} {Determination of spin hall angle in the weyl ferromagnet co 2 mnga by taking into account the thermoelectric contributions},\ }\href@noop {} {\bibfield  {journal} {\bibinfo  {journal} {Physical Review Materials}\ }\textbf {\bibinfo {volume} {6}},\ \bibinfo {pages} {084411} (\bibinfo {year} {2022})}\BibitemShut {NoStop}%
\bibitem [{\citenamefont {Belopolski}\ \emph {et~al.}(2019)\citenamefont {Belopolski}, \citenamefont {Manna}, \citenamefont {Sanchez}, \citenamefont {Chang}, \citenamefont {Ernst}, \citenamefont {Yin}, \citenamefont {Zhang}, \citenamefont {Cochran}, \citenamefont {Shumiya}, \citenamefont {Zheng} \emph {et~al.}}]{belopolski2019discovery}%
  \BibitemOpen
  \bibfield  {author} {\bibinfo {author} {\bibfnamefont {I.}~\bibnamefont {Belopolski}}, \bibinfo {author} {\bibfnamefont {K.}~\bibnamefont {Manna}}, \bibinfo {author} {\bibfnamefont {D.~S.}\ \bibnamefont {Sanchez}}, \bibinfo {author} {\bibfnamefont {G.}~\bibnamefont {Chang}}, \bibinfo {author} {\bibfnamefont {B.}~\bibnamefont {Ernst}}, \bibinfo {author} {\bibfnamefont {J.}~\bibnamefont {Yin}}, \bibinfo {author} {\bibfnamefont {S.~S.}\ \bibnamefont {Zhang}}, \bibinfo {author} {\bibfnamefont {T.}~\bibnamefont {Cochran}}, \bibinfo {author} {\bibfnamefont {N.}~\bibnamefont {Shumiya}}, \bibinfo {author} {\bibfnamefont {H.}~\bibnamefont {Zheng}}, \emph {et~al.},\ }\bibfield  {title} {\bibinfo {title} {Discovery of topological weyl fermion lines and drumhead surface states in a room temperature magnet},\ }\href@noop {} {\bibfield  {journal} {\bibinfo  {journal} {Science}\ }\textbf {\bibinfo {volume} {365}},\ \bibinfo {pages} {1278} (\bibinfo {year} {2019})}\BibitemShut {NoStop}%
\bibitem [{\citenamefont {Chen}\ \emph {et~al.}(2004)\citenamefont {Chen}, \citenamefont {Basiaga}, \citenamefont {O’Brien},\ and\ \citenamefont {Heiman}}]{chen2004anomalous}%
  \BibitemOpen
  \bibfield  {author} {\bibinfo {author} {\bibfnamefont {Y.}~\bibnamefont {Chen}}, \bibinfo {author} {\bibfnamefont {D.}~\bibnamefont {Basiaga}}, \bibinfo {author} {\bibfnamefont {J.}~\bibnamefont {O’Brien}},\ and\ \bibinfo {author} {\bibfnamefont {D.}~\bibnamefont {Heiman}},\ }\bibfield  {title} {\bibinfo {title} {Anomalous magnetic properties and hall effect in ferromagnetic co 2 mnal epilayers},\ }\href@noop {} {\bibfield  {journal} {\bibinfo  {journal} {Applied physics letters}\ }\textbf {\bibinfo {volume} {84}},\ \bibinfo {pages} {4301} (\bibinfo {year} {2004})}\BibitemShut {NoStop}%
\bibitem [{\citenamefont {Vilanova~Vidal}\ \emph {et~al.}(2011)\citenamefont {Vilanova~Vidal}, \citenamefont {Stryganyuk}, \citenamefont {Schneider}, \citenamefont {Felser},\ and\ \citenamefont {Jakob}}]{vilanova2011exploring}%
  \BibitemOpen
  \bibfield  {author} {\bibinfo {author} {\bibfnamefont {E.}~\bibnamefont {Vilanova~Vidal}}, \bibinfo {author} {\bibfnamefont {G.}~\bibnamefont {Stryganyuk}}, \bibinfo {author} {\bibfnamefont {H.}~\bibnamefont {Schneider}}, \bibinfo {author} {\bibfnamefont {C.}~\bibnamefont {Felser}},\ and\ \bibinfo {author} {\bibfnamefont {G.}~\bibnamefont {Jakob}},\ }\bibfield  {title} {\bibinfo {title} {Exploring co2mnal heusler compound for anomalous hall effect sensors},\ }\href@noop {} {\bibfield  {journal} {\bibinfo  {journal} {Applied Physics Letters}\ }\textbf {\bibinfo {volume} {99}} (\bibinfo {year} {2011})}\BibitemShut {NoStop}%
\bibitem [{\citenamefont {K{\"u}bler}\ and\ \citenamefont {Felser}(2012)}]{kubler2012berry}%
  \BibitemOpen
  \bibfield  {author} {\bibinfo {author} {\bibfnamefont {J.}~\bibnamefont {K{\"u}bler}}\ and\ \bibinfo {author} {\bibfnamefont {C.}~\bibnamefont {Felser}},\ }\bibfield  {title} {\bibinfo {title} {Berry curvature and the anomalous hall effect in heusler compounds},\ }\href@noop {} {\bibfield  {journal} {\bibinfo  {journal} {Physical Review B—Condensed Matter and Materials Physics}\ }\textbf {\bibinfo {volume} {85}},\ \bibinfo {pages} {012405} (\bibinfo {year} {2012})}\BibitemShut {NoStop}%
\bibitem [{\citenamefont {K{\"u}bler}\ and\ \citenamefont {Felser}(2016)}]{kubler2016weyl}%
  \BibitemOpen
  \bibfield  {author} {\bibinfo {author} {\bibfnamefont {J.}~\bibnamefont {K{\"u}bler}}\ and\ \bibinfo {author} {\bibfnamefont {C.}~\bibnamefont {Felser}},\ }\bibfield  {title} {\bibinfo {title} {Weyl points in the ferromagnetic heusler compound co2mnal},\ }\href@noop {} {\bibfield  {journal} {\bibinfo  {journal} {Europhysics Letters}\ }\textbf {\bibinfo {volume} {114}},\ \bibinfo {pages} {47005} (\bibinfo {year} {2016})}\BibitemShut {NoStop}%
\bibitem [{\citenamefont {Li}\ \emph {et~al.}(2020)\citenamefont {Li}, \citenamefont {Koo}, \citenamefont {Ning}, \citenamefont {Li}, \citenamefont {Miao}, \citenamefont {Min}, \citenamefont {Zhu}, \citenamefont {Wang}, \citenamefont {Alem}, \citenamefont {Liu} \emph {et~al.}}]{li2020giant}%
  \BibitemOpen
  \bibfield  {author} {\bibinfo {author} {\bibfnamefont {P.}~\bibnamefont {Li}}, \bibinfo {author} {\bibfnamefont {J.}~\bibnamefont {Koo}}, \bibinfo {author} {\bibfnamefont {W.}~\bibnamefont {Ning}}, \bibinfo {author} {\bibfnamefont {J.}~\bibnamefont {Li}}, \bibinfo {author} {\bibfnamefont {L.}~\bibnamefont {Miao}}, \bibinfo {author} {\bibfnamefont {L.}~\bibnamefont {Min}}, \bibinfo {author} {\bibfnamefont {Y.}~\bibnamefont {Zhu}}, \bibinfo {author} {\bibfnamefont {Y.}~\bibnamefont {Wang}}, \bibinfo {author} {\bibfnamefont {N.}~\bibnamefont {Alem}}, \bibinfo {author} {\bibfnamefont {C.-X.}\ \bibnamefont {Liu}}, \emph {et~al.},\ }\bibfield  {title} {\bibinfo {title} {Giant room temperature anomalous hall effect and tunable topology in a ferromagnetic topological semimetal co2mnal},\ }\href@noop {} {\bibfield  {journal} {\bibinfo  {journal} {Nature communications}\ }\textbf {\bibinfo {volume} {11}},\ \bibinfo {pages} {3476} (\bibinfo {year} {2020})}\BibitemShut {NoStop}%
\bibitem [{\citenamefont {Muechler}\ \emph {et~al.}(2020)\citenamefont {Muechler}, \citenamefont {Liu}, \citenamefont {Gayles}, \citenamefont {Xu}, \citenamefont {Felser},\ and\ \citenamefont {Sun}}]{Muechler2020}%
  \BibitemOpen
  \bibfield  {author} {\bibinfo {author} {\bibfnamefont {L.}~\bibnamefont {Muechler}}, \bibinfo {author} {\bibfnamefont {E.}~\bibnamefont {Liu}}, \bibinfo {author} {\bibfnamefont {J.}~\bibnamefont {Gayles}}, \bibinfo {author} {\bibfnamefont {Q.}~\bibnamefont {Xu}}, \bibinfo {author} {\bibfnamefont {C.}~\bibnamefont {Felser}},\ and\ \bibinfo {author} {\bibfnamefont {Y.}~\bibnamefont {Sun}},\ }\bibfield  {title} {\bibinfo {title} {Emerging chiral edge states from the confinement of a magnetic weyl semimetal in ${\mathrm{co}}_{3}{\mathrm{sn}}_{2}{\mathrm{s}}_{2}$},\ }\href {https://doi.org/10.1103/PhysRevB.101.115106} {\bibfield  {journal} {\bibinfo  {journal} {Phys. Rev. B}\ }\textbf {\bibinfo {volume} {101}},\ \bibinfo {pages} {115106} (\bibinfo {year} {2020})}\BibitemShut {NoStop}%
\bibitem [{\citenamefont {Ozawa}\ and\ \citenamefont {Nomura}(2019)}]{Ozawa2019}%
  \BibitemOpen
  \bibfield  {author} {\bibinfo {author} {\bibfnamefont {A.}~\bibnamefont {Ozawa}}\ and\ \bibinfo {author} {\bibfnamefont {K.}~\bibnamefont {Nomura}},\ }\bibfield  {title} {\bibinfo {title} {Two-orbital effective model for magnetic weyl semimetal in kagome-lattice shandite},\ }\href@noop {} {\bibfield  {journal} {\bibinfo  {journal} {J. Phys. Soc. Jpn.}\ }\textbf {\bibinfo {volume} {88}},\ \bibinfo {pages} {123703} (\bibinfo {year} {2019})}\BibitemShut {NoStop}%
\bibitem [{\citenamefont {Ozawa}\ \emph {et~al.}(2024{\natexlab{a}})\citenamefont {Ozawa}, \citenamefont {Araki},\ and\ \citenamefont {Nomura}}]{Ozawa2024}%
  \BibitemOpen
  \bibfield  {author} {\bibinfo {author} {\bibfnamefont {A.}~\bibnamefont {Ozawa}}, \bibinfo {author} {\bibfnamefont {Y.}~\bibnamefont {Araki}},\ and\ \bibinfo {author} {\bibfnamefont {K.}~\bibnamefont {Nomura}},\ }\bibfield  {title} {\bibinfo {title} {Chiral gauge field in fully spin-polarized magnetic weyl semimetal with magnetic domain walls},\ }\href {https://doi.org/10.7566/JPSJ.93.094704} {\bibfield  {journal} {\bibinfo  {journal} {Journal of the Physical Society of Japan}\ }\textbf {\bibinfo {volume} {93}},\ \bibinfo {pages} {094704} (\bibinfo {year} {2024}{\natexlab{a}})}\BibitemShut {NoStop}%
\bibitem [{\citenamefont {Ozawa}\ \emph {et~al.}(2024{\natexlab{b}})\citenamefont {Ozawa}, \citenamefont {Kobayashi},\ and\ \citenamefont {Nomura}}]{ozawa2024effective}%
  \BibitemOpen
  \bibfield  {author} {\bibinfo {author} {\bibfnamefont {A.}~\bibnamefont {Ozawa}}, \bibinfo {author} {\bibfnamefont {K.}~\bibnamefont {Kobayashi}},\ and\ \bibinfo {author} {\bibfnamefont {K.}~\bibnamefont {Nomura}},\ }\bibfield  {title} {\bibinfo {title} {{Effective Model Analysis of Intrinsic Spin Hall Effect with Magnetism in the Stacked Kagome Weyl Semimetal \ce{Co3Sn2S2}}},\ }\href@noop {} {\bibfield  {journal} {\bibinfo  {journal} {Physical Review Applied}\ }\textbf {\bibinfo {volume} {21}},\ \bibinfo {pages} {014041} (\bibinfo {year} {2024}{\natexlab{b}})}\BibitemShut {NoStop}%
\bibitem [{\citenamefont {Huang}\ \emph {et~al.}(2025)\citenamefont {Huang}, \citenamefont {Xing}, \citenamefont {Zheng}, \citenamefont {Lv}, \citenamefont {Chen}, \citenamefont {Chen}, \citenamefont {Yang}, \citenamefont {Ji},\ and\ \citenamefont {Gao}}]{huang2025co3sn2s2}%
  \BibitemOpen
  \bibfield  {author} {\bibinfo {author} {\bibfnamefont {L.}~\bibnamefont {Huang}}, \bibinfo {author} {\bibfnamefont {Y.}~\bibnamefont {Xing}}, \bibinfo {author} {\bibfnamefont {Q.}~\bibnamefont {Zheng}}, \bibinfo {author} {\bibfnamefont {S.}~\bibnamefont {Lv}}, \bibinfo {author} {\bibfnamefont {L.}~\bibnamefont {Chen}}, \bibinfo {author} {\bibfnamefont {H.}~\bibnamefont {Chen}}, \bibinfo {author} {\bibfnamefont {H.}~\bibnamefont {Yang}}, \bibinfo {author} {\bibfnamefont {W.}~\bibnamefont {Ji}},\ and\ \bibinfo {author} {\bibfnamefont {H.-J.}\ \bibnamefont {Gao}},\ }\bibfield  {title} {\bibinfo {title} {On co3sn2s2 surfaces: crystal growth, surface recognition, atomic engineering and novel quantum structures},\ }\href@noop {} {\bibfield  {journal} {\bibinfo  {journal} {Journal of Physics: Condensed Matter}\ }\textbf {\bibinfo {volume} {37}},\ \bibinfo {pages} {473005} (\bibinfo {year} {2025})}\BibitemShut {NoStop}%
\bibitem [{\citenamefont {Schnelle}\ \emph {et~al.}(2013)\citenamefont {Schnelle}, \citenamefont {Leithe-Jasper}, \citenamefont {Rosner}, \citenamefont {Schappacher}, \citenamefont {P{\"o}ttgen}, \citenamefont {Pielnhofer},\ and\ \citenamefont {Weihrich}}]{schnelle2013ferromagnetic}%
  \BibitemOpen
  \bibfield  {author} {\bibinfo {author} {\bibfnamefont {W.}~\bibnamefont {Schnelle}}, \bibinfo {author} {\bibfnamefont {A.}~\bibnamefont {Leithe-Jasper}}, \bibinfo {author} {\bibfnamefont {H.}~\bibnamefont {Rosner}}, \bibinfo {author} {\bibfnamefont {F.}~\bibnamefont {Schappacher}}, \bibinfo {author} {\bibfnamefont {R.}~\bibnamefont {P{\"o}ttgen}}, \bibinfo {author} {\bibfnamefont {F.}~\bibnamefont {Pielnhofer}},\ and\ \bibinfo {author} {\bibfnamefont {R.}~\bibnamefont {Weihrich}},\ }\bibfield  {title} {\bibinfo {title} {Ferromagnetic ordering and half-metallic state of sn 2 co 3 s 2 with the shandite-type structure},\ }\href@noop {} {\bibfield  {journal} {\bibinfo  {journal} {Physical Review B—Condensed Matter and Materials Physics}\ }\textbf {\bibinfo {volume} {88}},\ \bibinfo {pages} {144404} (\bibinfo {year} {2013})}\BibitemShut {NoStop}%
\bibitem [{\citenamefont {Kassem}\ \emph {et~al.}(2016)\citenamefont {Kassem}, \citenamefont {Tabata}, \citenamefont {Waki},\ and\ \citenamefont {Nakamura}}]{kassem2016quasi}%
  \BibitemOpen
  \bibfield  {author} {\bibinfo {author} {\bibfnamefont {M.~A.}\ \bibnamefont {Kassem}}, \bibinfo {author} {\bibfnamefont {Y.}~\bibnamefont {Tabata}}, \bibinfo {author} {\bibfnamefont {T.}~\bibnamefont {Waki}},\ and\ \bibinfo {author} {\bibfnamefont {H.}~\bibnamefont {Nakamura}},\ }\bibfield  {title} {\bibinfo {title} {Quasi-two-dimensional magnetism in co-based shandites},\ }\href@noop {} {\bibfield  {journal} {\bibinfo  {journal} {Journal of the Physical Society of Japan}\ }\textbf {\bibinfo {volume} {85}},\ \bibinfo {pages} {064706} (\bibinfo {year} {2016})}\BibitemShut {NoStop}%
\bibitem [{\citenamefont {Weihrich}\ \emph {et~al.}(2004)\citenamefont {Weihrich}, \citenamefont {St{\"u}ckl}, \citenamefont {Zabel},\ and\ \citenamefont {Schnelle}}]{weihrich2004magnetischer}%
  \BibitemOpen
  \bibfield  {author} {\bibinfo {author} {\bibfnamefont {R.}~\bibnamefont {Weihrich}}, \bibinfo {author} {\bibfnamefont {A.}~\bibnamefont {St{\"u}ckl}}, \bibinfo {author} {\bibfnamefont {M.}~\bibnamefont {Zabel}},\ and\ \bibinfo {author} {\bibfnamefont {W.}~\bibnamefont {Schnelle}},\ }\bibfield  {title} {\bibinfo {title} {Magnetischer phasenuebergang des \ce{Co3Sn2S2}},\ }\href@noop {} {\bibfield  {journal} {\bibinfo  {journal} {Zeitschrift f{\"u}r anorganische und allgemeine Chemie}\ }\textbf {\bibinfo {volume} {630}},\ \bibinfo {pages} {1767} (\bibinfo {year} {2004})}\BibitemShut {NoStop}%
\bibitem [{\citenamefont {Weihrich}\ \emph {et~al.}(2005)\citenamefont {Weihrich}, \citenamefont {Anusca},\ and\ \citenamefont {Zabel}}]{weihrich2005half}%
  \BibitemOpen
  \bibfield  {author} {\bibinfo {author} {\bibfnamefont {R.}~\bibnamefont {Weihrich}}, \bibinfo {author} {\bibfnamefont {I.}~\bibnamefont {Anusca}},\ and\ \bibinfo {author} {\bibfnamefont {M.}~\bibnamefont {Zabel}},\ }\bibfield  {title} {\bibinfo {title} {Half-antiperovskites: Structure and type-antitype relations of shandites m3/2as (m: Co, ni; a: In, sn).},\ }\href@noop {} {\bibfield  {journal} {\bibinfo  {journal} {ChemInform}\ }\textbf {\bibinfo {volume} {36}},\ \bibinfo {pages} {no} (\bibinfo {year} {2005})}\BibitemShut {NoStop}%
\bibitem [{\citenamefont {Weihrich}\ and\ \citenamefont {Anusca}(2006)}]{weihrich2006half}%
  \BibitemOpen
  \bibfield  {author} {\bibinfo {author} {\bibfnamefont {R.}~\bibnamefont {Weihrich}}\ and\ \bibinfo {author} {\bibfnamefont {I.}~\bibnamefont {Anusca}},\ }\bibfield  {title} {\bibinfo {title} {Half antiperovskites. iii. crystallographic and electronic structure effects in sn2- xinxco3s2},\ }\href@noop {} {\bibfield  {journal} {\bibinfo  {journal} {Zeitschrift f{\"u}r anorganische und allgemeine Chemie}\ }\textbf {\bibinfo {volume} {632}},\ \bibinfo {pages} {1531} (\bibinfo {year} {2006})}\BibitemShut {NoStop}%
\bibitem [{\citenamefont {Weihrich}\ \emph {et~al.}(2007)\citenamefont {Weihrich}, \citenamefont {Matar}, \citenamefont {Eyert}, \citenamefont {Rau}, \citenamefont {Zabel}, \citenamefont {Andratschke}, \citenamefont {Anusca},\ and\ \citenamefont {Bernert}}]{weihrich2007structure}%
  \BibitemOpen
  \bibfield  {author} {\bibinfo {author} {\bibfnamefont {R.}~\bibnamefont {Weihrich}}, \bibinfo {author} {\bibfnamefont {S.~F.}\ \bibnamefont {Matar}}, \bibinfo {author} {\bibfnamefont {V.}~\bibnamefont {Eyert}}, \bibinfo {author} {\bibfnamefont {F.}~\bibnamefont {Rau}}, \bibinfo {author} {\bibfnamefont {M.}~\bibnamefont {Zabel}}, \bibinfo {author} {\bibfnamefont {M.}~\bibnamefont {Andratschke}}, \bibinfo {author} {\bibfnamefont {I.}~\bibnamefont {Anusca}},\ and\ \bibinfo {author} {\bibfnamefont {T.}~\bibnamefont {Bernert}},\ }\bibfield  {title} {\bibinfo {title} {Structure, ordering, and bonding of half antiperovskites: Pbni3/2s and bipd3/2s},\ }\href@noop {} {\bibfield  {journal} {\bibinfo  {journal} {Progress in solid state chemistry}\ }\textbf {\bibinfo {volume} {35}},\ \bibinfo {pages} {309} (\bibinfo {year} {2007})}\BibitemShut {NoStop}%
\bibitem [{\citenamefont {Kassem}\ \emph {et~al.}(2015)\citenamefont {Kassem}, \citenamefont {Tabata}, \citenamefont {Waki},\ and\ \citenamefont {Nakamura}}]{kassem2015single}%
  \BibitemOpen
  \bibfield  {author} {\bibinfo {author} {\bibfnamefont {M.~A.}\ \bibnamefont {Kassem}}, \bibinfo {author} {\bibfnamefont {Y.}~\bibnamefont {Tabata}}, \bibinfo {author} {\bibfnamefont {T.}~\bibnamefont {Waki}},\ and\ \bibinfo {author} {\bibfnamefont {H.}~\bibnamefont {Nakamura}},\ }\bibfield  {title} {\bibinfo {title} {Single crystal growth and characterization of kagom{\'e}-lattice shandites co3sn2- xinxs2},\ }\href@noop {} {\bibfield  {journal} {\bibinfo  {journal} {Journal of Crystal Growth}\ }\textbf {\bibinfo {volume} {426}},\ \bibinfo {pages} {208} (\bibinfo {year} {2015})}\BibitemShut {NoStop}%
\bibitem [{\citenamefont {Kassem}\ \emph {et~al.}(2017)\citenamefont {Kassem}, \citenamefont {Tabata}, \citenamefont {Waki},\ and\ \citenamefont {Nakamura}}]{kassem2017low}%
  \BibitemOpen
  \bibfield  {author} {\bibinfo {author} {\bibfnamefont {M.~A.}\ \bibnamefont {Kassem}}, \bibinfo {author} {\bibfnamefont {Y.}~\bibnamefont {Tabata}}, \bibinfo {author} {\bibfnamefont {T.}~\bibnamefont {Waki}},\ and\ \bibinfo {author} {\bibfnamefont {H.}~\bibnamefont {Nakamura}},\ }\bibfield  {title} {\bibinfo {title} {Low-field anomalous magnetic phase in the kagome-lattice shandite c o 3 s n 2 s 2},\ }\href@noop {} {\bibfield  {journal} {\bibinfo  {journal} {Physical Review B}\ }\textbf {\bibinfo {volume} {96}},\ \bibinfo {pages} {014429} (\bibinfo {year} {2017})}\BibitemShut {NoStop}%
\bibitem [{\citenamefont {Q.~Xu}\ \emph {et~al.}(2018)\citenamefont {Q.~Xu}, \citenamefont {Shi}, \citenamefont {Muechler}, \citenamefont {Gayles}, \citenamefont {Felser},\ and\ \citenamefont {Sun}}]{Xu2018}%
  \BibitemOpen
  \bibfield  {author} {\bibinfo {author} {\bibfnamefont {E.}~\bibnamefont {Q.~Xu}, \bibfnamefont {Liu}}, \bibinfo {author} {\bibfnamefont {W.}~\bibnamefont {Shi}}, \bibinfo {author} {\bibfnamefont {L.}~\bibnamefont {Muechler}}, \bibinfo {author} {\bibfnamefont {J.}~\bibnamefont {Gayles}}, \bibinfo {author} {\bibfnamefont {C.}~\bibnamefont {Felser}},\ and\ \bibinfo {author} {\bibfnamefont {Y.}~\bibnamefont {Sun}},\ }\bibfield  {title} {\bibinfo {title} {Topological surface {Fermi} arcs in the magnetic {Weyl} semimetal $\mathrm{Co}_{3}\mathrm{Sn}_{2}\mathrm{S}_{2}$},\ }\href@noop {} {\bibfield  {journal} {\bibinfo  {journal} {Phys. Rev. B}\ }\textbf {\bibinfo {volume} {97}},\ \bibinfo {pages} {235416} (\bibinfo {year} {2018})}\BibitemShut {NoStop}%
\bibitem [{\citenamefont {Wang}\ \emph {et~al.}(2018)\citenamefont {Wang}, \citenamefont {Xu}, \citenamefont {Lou}, \citenamefont {Liu}, \citenamefont {Li}, \citenamefont {Huang}, \citenamefont {Shen}, \citenamefont {Weng}, \citenamefont {Wang},\ and\ \citenamefont {Lei}}]{Wang2018}%
  \BibitemOpen
  \bibfield  {author} {\bibinfo {author} {\bibfnamefont {Q.}~\bibnamefont {Wang}}, \bibinfo {author} {\bibfnamefont {Y.}~\bibnamefont {Xu}}, \bibinfo {author} {\bibfnamefont {R.}~\bibnamefont {Lou}}, \bibinfo {author} {\bibfnamefont {Z.}~\bibnamefont {Liu}}, \bibinfo {author} {\bibfnamefont {M.}~\bibnamefont {Li}}, \bibinfo {author} {\bibfnamefont {Y.}~\bibnamefont {Huang}}, \bibinfo {author} {\bibfnamefont {D.}~\bibnamefont {Shen}}, \bibinfo {author} {\bibfnamefont {H.}~\bibnamefont {Weng}}, \bibinfo {author} {\bibfnamefont {S.}~\bibnamefont {Wang}},\ and\ \bibinfo {author} {\bibfnamefont {H.}~\bibnamefont {Lei}},\ }\bibfield  {title} {\bibinfo {title} {Large intrinsic anomalous {Hall} effect in half-metallic ferromagnet $\mathrm{Co}_{3}\mathrm{Sn}_{2}\mathrm{S}_{2}$ with magnetic {Weyl} fermions},\ }\href@noop {} {\bibfield  {journal} {\bibinfo  {journal} {Nat.Commun.}\ }\textbf {\bibinfo {volume} {9}} (\bibinfo {year} {2018})}\BibitemShut {NoStop}%
\bibitem [{\citenamefont {Liu}\ \emph {et~al.}(2019)\citenamefont {Liu}, \citenamefont {Liang}, \citenamefont {Liu}, \citenamefont {Xu}, \citenamefont {Li}, \citenamefont {Chen}, \citenamefont {Pei}, \citenamefont {Shi}, \citenamefont {Mo}, \citenamefont {Dudin}, \citenamefont {Kim}, \citenamefont {Cacho}, \citenamefont {Li}, \citenamefont {Sun}, \citenamefont {Yang}, \citenamefont {Liu}, \citenamefont {Parkin}, \citenamefont {Felser},\ and\ \citenamefont {Chen}}]{Liu2019}%
  \BibitemOpen
  \bibfield  {author} {\bibinfo {author} {\bibfnamefont {D.}~\bibnamefont {Liu}}, \bibinfo {author} {\bibfnamefont {A.}~\bibnamefont {Liang}}, \bibinfo {author} {\bibfnamefont {E.}~\bibnamefont {Liu}}, \bibinfo {author} {\bibfnamefont {Q.}~\bibnamefont {Xu}}, \bibinfo {author} {\bibfnamefont {Y.}~\bibnamefont {Li}}, \bibinfo {author} {\bibfnamefont {C.}~\bibnamefont {Chen}}, \bibinfo {author} {\bibfnamefont {D.}~\bibnamefont {Pei}}, \bibinfo {author} {\bibfnamefont {W.}~\bibnamefont {Shi}}, \bibinfo {author} {\bibfnamefont {S.}~\bibnamefont {Mo}}, \bibinfo {author} {\bibfnamefont {P.}~\bibnamefont {Dudin}}, \bibinfo {author} {\bibfnamefont {T.}~\bibnamefont {Kim}}, \bibinfo {author} {\bibfnamefont {C.}~\bibnamefont {Cacho}}, \bibinfo {author} {\bibfnamefont {G.}~\bibnamefont {Li}}, \bibinfo {author} {\bibfnamefont {Y.}~\bibnamefont {Sun}}, \bibinfo {author} {\bibfnamefont {L.}~\bibnamefont {Yang}}, \bibinfo {author} {\bibfnamefont {Z.~K.}\ \bibnamefont {Liu}}, \bibinfo {author} {\bibfnamefont
  {S.}~\bibnamefont {Parkin}}, \bibinfo {author} {\bibfnamefont {C.}~\bibnamefont {Felser}},\ and\ \bibinfo {author} {\bibfnamefont {Y.}~\bibnamefont {Chen}},\ }\bibfield  {title} {\bibinfo {title} {Magnetic {Weyl} semimetal phase in a kagom{\'e} crystal},\ }\href@noop {} {\bibfield  {journal} {\bibinfo  {journal} {Science}\ }\textbf {\bibinfo {volume} {365}},\ \bibinfo {pages} {1282} (\bibinfo {year} {2019})}\BibitemShut {NoStop}%
\bibitem [{\citenamefont {Guin}\ \emph {et~al.}(2019)\citenamefont {Guin}, \citenamefont {Vir}, \citenamefont {Zhang}, \citenamefont {Kumar}, \citenamefont {Watzman}, \citenamefont {Fu}, \citenamefont {Liu}, \citenamefont {Manna}, \citenamefont {Schnelle}, \citenamefont {Gooth} \emph {et~al.}}]{guin2019}%
  \BibitemOpen
  \bibfield  {author} {\bibinfo {author} {\bibfnamefont {S.~N.}\ \bibnamefont {Guin}}, \bibinfo {author} {\bibfnamefont {P.}~\bibnamefont {Vir}}, \bibinfo {author} {\bibfnamefont {Y.}~\bibnamefont {Zhang}}, \bibinfo {author} {\bibfnamefont {N.}~\bibnamefont {Kumar}}, \bibinfo {author} {\bibfnamefont {S.~J.}\ \bibnamefont {Watzman}}, \bibinfo {author} {\bibfnamefont {C.}~\bibnamefont {Fu}}, \bibinfo {author} {\bibfnamefont {E.}~\bibnamefont {Liu}}, \bibinfo {author} {\bibfnamefont {K.}~\bibnamefont {Manna}}, \bibinfo {author} {\bibfnamefont {W.}~\bibnamefont {Schnelle}}, \bibinfo {author} {\bibfnamefont {J.}~\bibnamefont {Gooth}}, \emph {et~al.},\ }\bibfield  {title} {\bibinfo {title} {Zero-field nernst effect in a ferromagnetic kagome-lattice weyl-semimetal co3sn2s2},\ }\href@noop {} {\bibfield  {journal} {\bibinfo  {journal} {Advanced Materials}\ }\textbf {\bibinfo {volume} {31}},\ \bibinfo {pages} {1806622} (\bibinfo {year} {2019})}\BibitemShut {NoStop}%
\bibitem [{\citenamefont {Yang}\ \emph {et~al.}(2020{\natexlab{a}})\citenamefont {Yang}, \citenamefont {You}, \citenamefont {Wang}, \citenamefont {Huang}, \citenamefont {Xi}, \citenamefont {Xu}, \citenamefont {Cao}, \citenamefont {Tian}, \citenamefont {Xu}, \citenamefont {Dai},\ and\ \citenamefont {Li}}]{Yang2020}%
  \BibitemOpen
  \bibfield  {author} {\bibinfo {author} {\bibfnamefont {H.}~\bibnamefont {Yang}}, \bibinfo {author} {\bibfnamefont {W.}~\bibnamefont {You}}, \bibinfo {author} {\bibfnamefont {J.}~\bibnamefont {Wang}}, \bibinfo {author} {\bibfnamefont {J.}~\bibnamefont {Huang}}, \bibinfo {author} {\bibfnamefont {C.}~\bibnamefont {Xi}}, \bibinfo {author} {\bibfnamefont {X.}~\bibnamefont {Xu}}, \bibinfo {author} {\bibfnamefont {C.}~\bibnamefont {Cao}}, \bibinfo {author} {\bibfnamefont {M.}~\bibnamefont {Tian}}, \bibinfo {author} {\bibfnamefont {Z.-A.}\ \bibnamefont {Xu}}, \bibinfo {author} {\bibfnamefont {J.}~\bibnamefont {Dai}},\ and\ \bibinfo {author} {\bibfnamefont {Y.}~\bibnamefont {Li}},\ }\bibfield  {title} {\bibinfo {title} {Giant anomalous nernst effect in the magnetic weyl semimetal ${\mathrm{co}}_{3}{\mathrm{sn}}_{2}{\mathrm{s}}_{2}$},\ }\href {https://doi.org/10.1103/PhysRevMaterials.4.024202} {\bibfield  {journal} {\bibinfo  {journal} {Phys. Rev. Mater.}\ }\textbf {\bibinfo {volume} {4}},\ \bibinfo {pages} {024202}
  (\bibinfo {year} {2020}{\natexlab{a}})}\BibitemShut {NoStop}%
\bibitem [{\citenamefont {Yanagi}\ \emph {et~al.}(2021)\citenamefont {Yanagi}, \citenamefont {Ikeda}, \citenamefont {Fujiwara}, \citenamefont {Nomura}, \citenamefont {Tsukazaki},\ and\ \citenamefont {Suzuki}}]{yanagi2021prb}%
  \BibitemOpen
  \bibfield  {author} {\bibinfo {author} {\bibfnamefont {Y.}~\bibnamefont {Yanagi}}, \bibinfo {author} {\bibfnamefont {J.}~\bibnamefont {Ikeda}}, \bibinfo {author} {\bibfnamefont {K.}~\bibnamefont {Fujiwara}}, \bibinfo {author} {\bibfnamefont {K.}~\bibnamefont {Nomura}}, \bibinfo {author} {\bibfnamefont {A.}~\bibnamefont {Tsukazaki}},\ and\ \bibinfo {author} {\bibfnamefont {M.-T.}\ \bibnamefont {Suzuki}},\ }\bibfield  {title} {\bibinfo {title} {First-principles investigation of magnetic and transport properties in hole-doped shandite compounds ${\mathrm{co}}_{3}{\mathrm{in}}_{x}{\mathrm{sn}}_{2\ensuremath{-}x}{\mathrm{s}}_{2}$},\ }\href {https://doi.org/10.1103/PhysRevB.103.205112} {\bibfield  {journal} {\bibinfo  {journal} {Phys. Rev. B}\ }\textbf {\bibinfo {volume} {103}},\ \bibinfo {pages} {205112} (\bibinfo {year} {2021})}\BibitemShut {NoStop}%
\bibitem [{\citenamefont {Noguchi}\ \emph {et~al.}(2024)\citenamefont {Noguchi}, \citenamefont {Fujiwara}, \citenamefont {Yanagi}, \citenamefont {Suzuki}, \citenamefont {Hirai}, \citenamefont {Seki}, \citenamefont {Uchida},\ and\ \citenamefont {Tsukazaki}}]{noguchi2024bipolarity}%
  \BibitemOpen
  \bibfield  {author} {\bibinfo {author} {\bibfnamefont {S.}~\bibnamefont {Noguchi}}, \bibinfo {author} {\bibfnamefont {K.}~\bibnamefont {Fujiwara}}, \bibinfo {author} {\bibfnamefont {Y.}~\bibnamefont {Yanagi}}, \bibinfo {author} {\bibfnamefont {M.-T.}\ \bibnamefont {Suzuki}}, \bibinfo {author} {\bibfnamefont {T.}~\bibnamefont {Hirai}}, \bibinfo {author} {\bibfnamefont {T.}~\bibnamefont {Seki}}, \bibinfo {author} {\bibfnamefont {K.-i.}\ \bibnamefont {Uchida}},\ and\ \bibinfo {author} {\bibfnamefont {A.}~\bibnamefont {Tsukazaki}},\ }\bibfield  {title} {\bibinfo {title} {Bipolarity of large anomalous nernst effect in weyl magnet-based alloy films},\ }\href@noop {} {\bibfield  {journal} {\bibinfo  {journal} {Nature Physics}\ }\textbf {\bibinfo {volume} {20}},\ \bibinfo {pages} {254} (\bibinfo {year} {2024})}\BibitemShut {NoStop}%
\bibitem [{\citenamefont {Seki}\ \emph {et~al.}(2023)\citenamefont {Seki}, \citenamefont {Lau}, \citenamefont {Ikeda}, \citenamefont {Fujiwara}, \citenamefont {Ozawa}, \citenamefont {Iihama}, \citenamefont {Nomura},\ and\ \citenamefont {Tsukazaki}}]{seki2023enhancement}%
  \BibitemOpen
  \bibfield  {author} {\bibinfo {author} {\bibfnamefont {T.}~\bibnamefont {Seki}}, \bibinfo {author} {\bibfnamefont {Y.-C.}\ \bibnamefont {Lau}}, \bibinfo {author} {\bibfnamefont {J.}~\bibnamefont {Ikeda}}, \bibinfo {author} {\bibfnamefont {K.}~\bibnamefont {Fujiwara}}, \bibinfo {author} {\bibfnamefont {A.}~\bibnamefont {Ozawa}}, \bibinfo {author} {\bibfnamefont {S.}~\bibnamefont {Iihama}}, \bibinfo {author} {\bibfnamefont {K.}~\bibnamefont {Nomura}},\ and\ \bibinfo {author} {\bibfnamefont {A.}~\bibnamefont {Tsukazaki}},\ }\bibfield  {title} {\bibinfo {title} {Enhancement of spin-charge conversion efficiency for \ce{Co3Sn2S2} across transition from paramagnetic to ferromagnetic phase},\ }\href@noop {} {\bibfield  {journal} {\bibinfo  {journal} {Physical Review Research}\ }\textbf {\bibinfo {volume} {5}},\ \bibinfo {pages} {013222} (\bibinfo {year} {2023})}\BibitemShut {NoStop}%
\bibitem [{\citenamefont {Sugawara}\ \emph {et~al.}(2019)\citenamefont {Sugawara}, \citenamefont {Akashi}, \citenamefont {Kassem}, \citenamefont {Tabata}, \citenamefont {Waki},\ and\ \citenamefont {Nakamura}}]{Sugawara2019}%
  \BibitemOpen
  \bibfield  {author} {\bibinfo {author} {\bibfnamefont {A.}~\bibnamefont {Sugawara}}, \bibinfo {author} {\bibfnamefont {T.}~\bibnamefont {Akashi}}, \bibinfo {author} {\bibfnamefont {M.~A.}\ \bibnamefont {Kassem}}, \bibinfo {author} {\bibfnamefont {Y.}~\bibnamefont {Tabata}}, \bibinfo {author} {\bibfnamefont {T.}~\bibnamefont {Waki}},\ and\ \bibinfo {author} {\bibfnamefont {H.}~\bibnamefont {Nakamura}},\ }\bibfield  {title} {\bibinfo {title} {Magnetic domain structure within half-metallic ferromagnetic kagome compound $\mathrm{C}{\mathrm{o}}_{3}\mathrm{S}{\mathrm{n}}_{2}{\mathrm{s}}_{2}$},\ }\href {https://doi.org/10.1103/PhysRevMaterials.3.104421} {\bibfield  {journal} {\bibinfo  {journal} {Phys. Rev. Mater.}\ }\textbf {\bibinfo {volume} {3}},\ \bibinfo {pages} {104421} (\bibinfo {year} {2019})}\BibitemShut {NoStop}%
\bibitem [{\citenamefont {Lee}\ \emph {et~al.}(2022)\citenamefont {Lee}, \citenamefont {Vir}, \citenamefont {Manna}, \citenamefont {Shekhar}, \citenamefont {Moore}, \citenamefont {Kastner}, \citenamefont {Felser},\ and\ \citenamefont {Orenstein}}]{Lee2022}%
  \BibitemOpen
  \bibfield  {author} {\bibinfo {author} {\bibfnamefont {C.}~\bibnamefont {Lee}}, \bibinfo {author} {\bibfnamefont {P.}~\bibnamefont {Vir}}, \bibinfo {author} {\bibfnamefont {K.}~\bibnamefont {Manna}}, \bibinfo {author} {\bibfnamefont {C.}~\bibnamefont {Shekhar}}, \bibinfo {author} {\bibfnamefont {J.~E.}\ \bibnamefont {Moore}}, \bibinfo {author} {\bibfnamefont {M.}~\bibnamefont {Kastner}}, \bibinfo {author} {\bibfnamefont {C.}~\bibnamefont {Felser}},\ and\ \bibinfo {author} {\bibfnamefont {J.}~\bibnamefont {Orenstein}},\ }\bibfield  {title} {\bibinfo {title} {Observation of a phase transition within the domain walls of ferromagnetic co3sn2s2},\ }\href@noop {} {\bibfield  {journal} {\bibinfo  {journal} {Nature Communications}\ }\textbf {\bibinfo {volume} {13}},\ \bibinfo {pages} {3000} (\bibinfo {year} {2022})}\BibitemShut {NoStop}%
\bibitem [{\citenamefont {Shiogai}\ \emph {et~al.}(2022)\citenamefont {Shiogai}, \citenamefont {Ikeda}, \citenamefont {Fujiwara}, \citenamefont {Seki}, \citenamefont {Takanashi},\ and\ \citenamefont {Tsukazaki}}]{Shiogai2022}%
  \BibitemOpen
  \bibfield  {author} {\bibinfo {author} {\bibfnamefont {J.}~\bibnamefont {Shiogai}}, \bibinfo {author} {\bibfnamefont {J.}~\bibnamefont {Ikeda}}, \bibinfo {author} {\bibfnamefont {K.}~\bibnamefont {Fujiwara}}, \bibinfo {author} {\bibfnamefont {T.}~\bibnamefont {Seki}}, \bibinfo {author} {\bibfnamefont {K.}~\bibnamefont {Takanashi}},\ and\ \bibinfo {author} {\bibfnamefont {A.}~\bibnamefont {Tsukazaki}},\ }\bibfield  {title} {\bibinfo {title} {Electrical detection of domain evolution in magnetic weyl semimetal ${\mathrm{co}}_{3}{\mathrm{sn}}_{2}{\mathrm{s}}_{2}$ submicrometer-wide wire devices},\ }\href {https://doi.org/10.1103/PhysRevMaterials.6.114203} {\bibfield  {journal} {\bibinfo  {journal} {Phys. Rev. Mater.}\ }\textbf {\bibinfo {volume} {6}},\ \bibinfo {pages} {114203} (\bibinfo {year} {2022})}\BibitemShut {NoStop}%
\bibitem [{\citenamefont {Yoshikawa}\ \emph {et~al.}(2022{\natexlab{a}})\citenamefont {Yoshikawa}, \citenamefont {Ogawa}, \citenamefont {Hirai}, \citenamefont {Fujiwara}, \citenamefont {Ikeda}, \citenamefont {Tsukazaki},\ and\ \citenamefont {Shimano}}]{yoshikawa2022non}%
  \BibitemOpen
  \bibfield  {author} {\bibinfo {author} {\bibfnamefont {N.}~\bibnamefont {Yoshikawa}}, \bibinfo {author} {\bibfnamefont {K.}~\bibnamefont {Ogawa}}, \bibinfo {author} {\bibfnamefont {Y.}~\bibnamefont {Hirai}}, \bibinfo {author} {\bibfnamefont {K.}~\bibnamefont {Fujiwara}}, \bibinfo {author} {\bibfnamefont {J.}~\bibnamefont {Ikeda}}, \bibinfo {author} {\bibfnamefont {A.}~\bibnamefont {Tsukazaki}},\ and\ \bibinfo {author} {\bibfnamefont {R.}~\bibnamefont {Shimano}},\ }\bibfield  {title} {\bibinfo {title} {Non-volatile chirality switching by all-optical magnetization reversal in ferromagnetic weyl semimetal co3sn2s2},\ }\href@noop {} {\bibfield  {journal} {\bibinfo  {journal} {Communications Physics}\ }\textbf {\bibinfo {volume} {5}},\ \bibinfo {pages} {328} (\bibinfo {year} {2022}{\natexlab{a}})}\BibitemShut {NoStop}%
\bibitem [{\citenamefont {Yoshikawa}\ \emph {et~al.}(2025{\natexlab{a}})\citenamefont {Yoshikawa}, \citenamefont {Ogawa},\ and\ \citenamefont {Shimano}}]{yoshikawa2025all}%
  \BibitemOpen
  \bibfield  {author} {\bibinfo {author} {\bibfnamefont {N.}~\bibnamefont {Yoshikawa}}, \bibinfo {author} {\bibfnamefont {K.}~\bibnamefont {Ogawa}},\ and\ \bibinfo {author} {\bibfnamefont {R.}~\bibnamefont {Shimano}},\ }\bibfield  {title} {\bibinfo {title} {All-optical switching in ferromagnets and its application to magnetic weyl semimetals},\ }\href@noop {} {\bibfield  {journal} {\bibinfo  {journal} {Journal of the Physical Society of Japan}\ }\textbf {\bibinfo {volume} {94}},\ \bibinfo {pages} {111005} (\bibinfo {year} {2025}{\natexlab{a}})}\BibitemShut {NoStop}%
\bibitem [{\citenamefont {Wang}\ \emph {et~al.}(2023{\natexlab{a}})\citenamefont {Wang}, \citenamefont {Zeng}, \citenamefont {Yuan}, \citenamefont {Zeng}, \citenamefont {Gu}, \citenamefont {Xu}, \citenamefont {Wang}, \citenamefont {Han}, \citenamefont {Nomura}, \citenamefont {Wang} \emph {et~al.}}]{wang2023magnetism}%
  \BibitemOpen
  \bibfield  {author} {\bibinfo {author} {\bibfnamefont {Q.}~\bibnamefont {Wang}}, \bibinfo {author} {\bibfnamefont {Y.}~\bibnamefont {Zeng}}, \bibinfo {author} {\bibfnamefont {K.}~\bibnamefont {Yuan}}, \bibinfo {author} {\bibfnamefont {Q.}~\bibnamefont {Zeng}}, \bibinfo {author} {\bibfnamefont {P.}~\bibnamefont {Gu}}, \bibinfo {author} {\bibfnamefont {X.}~\bibnamefont {Xu}}, \bibinfo {author} {\bibfnamefont {H.}~\bibnamefont {Wang}}, \bibinfo {author} {\bibfnamefont {Z.}~\bibnamefont {Han}}, \bibinfo {author} {\bibfnamefont {K.}~\bibnamefont {Nomura}}, \bibinfo {author} {\bibfnamefont {W.}~\bibnamefont {Wang}}, \emph {et~al.},\ }\bibfield  {title} {\bibinfo {title} {Magnetism modulation in co3sn2s2 by current-assisted domain wall motion},\ }\href@noop {} {\bibfield  {journal} {\bibinfo  {journal} {Nature Electronics}\ }\textbf {\bibinfo {volume} {6}},\ \bibinfo {pages} {119} (\bibinfo {year} {2023}{\natexlab{a}})}\BibitemShut {NoStop}%
\bibitem [{\citenamefont {Ikeda}\ \emph {et~al.}(2021{\natexlab{a}})\citenamefont {Ikeda}, \citenamefont {Fujiwara}, \citenamefont {Shiogai}, \citenamefont {Seki}, \citenamefont {Nomura}, \citenamefont {Takanashi},\ and\ \citenamefont {Tsukazaki}}]{Ikeda2021}%
  \BibitemOpen
  \bibfield  {author} {\bibinfo {author} {\bibfnamefont {J.}~\bibnamefont {Ikeda}}, \bibinfo {author} {\bibfnamefont {K.}~\bibnamefont {Fujiwara}}, \bibinfo {author} {\bibfnamefont {J.}~\bibnamefont {Shiogai}}, \bibinfo {author} {\bibfnamefont {T.}~\bibnamefont {Seki}}, \bibinfo {author} {\bibfnamefont {K.}~\bibnamefont {Nomura}}, \bibinfo {author} {\bibfnamefont {K.}~\bibnamefont {Takanashi}},\ and\ \bibinfo {author} {\bibfnamefont {A.}~\bibnamefont {Tsukazaki}},\ }\bibfield  {title} {\bibinfo {title} {Critical thickness for the emergence of {Weyl} features in $\mathrm{Co}_{3}\mathrm{Sn}_{2}\mathrm{S}_{2}$ thin films},\ }\href@noop {} {\bibfield  {journal} {\bibinfo  {journal} {Commun. Mater.}\ }\textbf {\bibinfo {volume} {2}},\ \bibinfo {pages} {1} (\bibinfo {year} {2021}{\natexlab{a}})}\BibitemShut {NoStop}%
\bibitem [{\citenamefont {Ikeda}\ \emph {et~al.}(2021{\natexlab{b}})\citenamefont {Ikeda}, \citenamefont {Fujiwara}, \citenamefont {Shiogai}, \citenamefont {Seki}, \citenamefont {Nomura}, \citenamefont {Takanashi},\ and\ \citenamefont {Tsukazaki}}]{ikeda2021two}%
  \BibitemOpen
  \bibfield  {author} {\bibinfo {author} {\bibfnamefont {J.}~\bibnamefont {Ikeda}}, \bibinfo {author} {\bibfnamefont {K.}~\bibnamefont {Fujiwara}}, \bibinfo {author} {\bibfnamefont {J.}~\bibnamefont {Shiogai}}, \bibinfo {author} {\bibfnamefont {T.}~\bibnamefont {Seki}}, \bibinfo {author} {\bibfnamefont {K.}~\bibnamefont {Nomura}}, \bibinfo {author} {\bibfnamefont {K.}~\bibnamefont {Takanashi}},\ and\ \bibinfo {author} {\bibfnamefont {A.}~\bibnamefont {Tsukazaki}},\ }\bibfield  {title} {\bibinfo {title} {Two-dimensionality of metallic surface conduction in \ce{Co3Sn2S2} thin films},\ }\href@noop {} {\bibfield  {journal} {\bibinfo  {journal} {Communications Physics}\ }\textbf {\bibinfo {volume} {4}},\ \bibinfo {pages} {117} (\bibinfo {year} {2021}{\natexlab{b}})}\BibitemShut {NoStop}%
\bibitem [{\citenamefont {Tanaka}\ \emph {et~al.}(2020)\citenamefont {Tanaka}, \citenamefont {Fujishiro}, \citenamefont {Mogi}, \citenamefont {Kaneko}, \citenamefont {Yokosawa}, \citenamefont {Kanazawa}, \citenamefont {Minami}, \citenamefont {Koretsune}, \citenamefont {Arita}, \citenamefont {Tarucha} \emph {et~al.}}]{tanaka2020topological}%
  \BibitemOpen
  \bibfield  {author} {\bibinfo {author} {\bibfnamefont {M.}~\bibnamefont {Tanaka}}, \bibinfo {author} {\bibfnamefont {Y.}~\bibnamefont {Fujishiro}}, \bibinfo {author} {\bibfnamefont {M.}~\bibnamefont {Mogi}}, \bibinfo {author} {\bibfnamefont {Y.}~\bibnamefont {Kaneko}}, \bibinfo {author} {\bibfnamefont {T.}~\bibnamefont {Yokosawa}}, \bibinfo {author} {\bibfnamefont {N.}~\bibnamefont {Kanazawa}}, \bibinfo {author} {\bibfnamefont {S.}~\bibnamefont {Minami}}, \bibinfo {author} {\bibfnamefont {T.}~\bibnamefont {Koretsune}}, \bibinfo {author} {\bibfnamefont {R.}~\bibnamefont {Arita}}, \bibinfo {author} {\bibfnamefont {S.}~\bibnamefont {Tarucha}}, \emph {et~al.},\ }\bibfield  {title} {\bibinfo {title} {Topological kagome magnet co3sn2s2 thin flakes with high electron mobility and large anomalous hall effect},\ }\href@noop {} {\bibfield  {journal} {\bibinfo  {journal} {Nano Letters}\ }\textbf {\bibinfo {volume} {20}},\ \bibinfo {pages} {7476} (\bibinfo {year} {2020})}\BibitemShut {NoStop}%
\bibitem [{\citenamefont {Zhang}\ \emph {et~al.}(2025)\citenamefont {Zhang}, \citenamefont {Si}, \citenamefont {Wang}, \citenamefont {Zhao}, \citenamefont {Li}, \citenamefont {Wei}, \citenamefont {Yang}, \citenamefont {Tang}, \citenamefont {Liu}, \citenamefont {Wu},\ and\ \citenamefont {Gong}}]{Zhang2025-xk}%
  \BibitemOpen
  \bibfield  {author} {\bibinfo {author} {\bibfnamefont {P.}~\bibnamefont {Zhang}}, \bibinfo {author} {\bibfnamefont {K.}~\bibnamefont {Si}}, \bibinfo {author} {\bibfnamefont {X.}~\bibnamefont {Wang}}, \bibinfo {author} {\bibfnamefont {F.}~\bibnamefont {Zhao}}, \bibinfo {author} {\bibfnamefont {B.}~\bibnamefont {Li}}, \bibinfo {author} {\bibfnamefont {J.}~\bibnamefont {Wei}}, \bibinfo {author} {\bibfnamefont {Y.}~\bibnamefont {Yang}}, \bibinfo {author} {\bibfnamefont {P.}~\bibnamefont {Tang}}, \bibinfo {author} {\bibfnamefont {Z.}~\bibnamefont {Liu}}, \bibinfo {author} {\bibfnamefont {K.}~\bibnamefont {Wu}},\ and\ \bibinfo {author} {\bibfnamefont {Y.}~\bibnamefont {Gong}},\ }\bibfield  {title} {\bibinfo {title} {Synthesis engineering of {2D} {Co3Sn2S2} with tunable anomalous hall effect},\ }\href@noop {} {\bibfield  {journal} {\bibinfo  {journal} {Adv. Mater.}\ ,\ \bibinfo {pages} {e09261}} (\bibinfo {year} {2025})}\BibitemShut {NoStop}%
\bibitem [{\citenamefont {Nakazawa}\ \emph {et~al.}(2024{\natexlab{a}})\citenamefont {Nakazawa}, \citenamefont {Kato},\ and\ \citenamefont {Motome}}]{Nakazawa2024magnetic}%
  \BibitemOpen
  \bibfield  {author} {\bibinfo {author} {\bibfnamefont {K.}~\bibnamefont {Nakazawa}}, \bibinfo {author} {\bibfnamefont {Y.}~\bibnamefont {Kato}},\ and\ \bibinfo {author} {\bibfnamefont {Y.}~\bibnamefont {Motome}},\ }\bibfield  {title} {\bibinfo {title} {Magnetic, transport and topological properties of co-based shandite thin films},\ }\href {https://doi.org/10.1038/s42005-024-01534-8} {\bibfield  {journal} {\bibinfo  {journal} {Communications Physics}\ }\textbf {\bibinfo {volume} {7}},\ \bibinfo {pages} {48} (\bibinfo {year} {2024}{\natexlab{a}})}\BibitemShut {NoStop}%
\bibitem [{\citenamefont {Nakazawa}\ \emph {et~al.}(2024{\natexlab{b}})\citenamefont {Nakazawa}, \citenamefont {Kato},\ and\ \citenamefont {Motome}}]{Nakazawa2024topo}%
  \BibitemOpen
  \bibfield  {author} {\bibinfo {author} {\bibfnamefont {K.}~\bibnamefont {Nakazawa}}, \bibinfo {author} {\bibfnamefont {Y.}~\bibnamefont {Kato}},\ and\ \bibinfo {author} {\bibfnamefont {Y.}~\bibnamefont {Motome}},\ }\bibfield  {title} {\bibinfo {title} {Topological transitions by magnetization rotation in kagome monolayers of the ferromagnetic weyl semimetal co-based shandite},\ }\href {https://doi.org/10.1103/PhysRevB.110.085112} {\bibfield  {journal} {\bibinfo  {journal} {Phys. Rev. B}\ }\textbf {\bibinfo {volume} {110}},\ \bibinfo {pages} {085112} (\bibinfo {year} {2024}{\natexlab{b}})}\BibitemShut {NoStop}%
\bibitem [{\citenamefont {Kurebayashi}\ and\ \citenamefont {Nomura}(2014)}]{kurebayashi2014weyl}%
  \BibitemOpen
  \bibfield  {author} {\bibinfo {author} {\bibfnamefont {D.}~\bibnamefont {Kurebayashi}}\ and\ \bibinfo {author} {\bibfnamefont {K.}~\bibnamefont {Nomura}},\ }\bibfield  {title} {\bibinfo {title} {Weyl semimetal phase in solid-solution narrow-gap semiconductors},\ }\href@noop {} {\bibfield  {journal} {\bibinfo  {journal} {journal of the physical society of japan}\ }\textbf {\bibinfo {volume} {83}},\ \bibinfo {pages} {063709} (\bibinfo {year} {2014})}\BibitemShut {NoStop}%
\bibitem [{\citenamefont {Tokura}\ \emph {et~al.}(2019)\citenamefont {Tokura}, \citenamefont {Yasuda},\ and\ \citenamefont {Tsukazaki}}]{tokura2019magnetic}%
  \BibitemOpen
  \bibfield  {author} {\bibinfo {author} {\bibfnamefont {Y.}~\bibnamefont {Tokura}}, \bibinfo {author} {\bibfnamefont {K.}~\bibnamefont {Yasuda}},\ and\ \bibinfo {author} {\bibfnamefont {A.}~\bibnamefont {Tsukazaki}},\ }\bibfield  {title} {\bibinfo {title} {Magnetic topological insulators},\ }\href@noop {} {\bibfield  {journal} {\bibinfo  {journal} {Nature Reviews Physics}\ }\textbf {\bibinfo {volume} {1}},\ \bibinfo {pages} {126} (\bibinfo {year} {2019})}\BibitemShut {NoStop}%
\bibitem [{\citenamefont {Bernevig}\ \emph {et~al.}(2022)\citenamefont {Bernevig}, \citenamefont {Felser},\ and\ \citenamefont {Beidenkopf}}]{Bernevig2022}%
  \BibitemOpen
  \bibfield  {author} {\bibinfo {author} {\bibfnamefont {B.~A.}\ \bibnamefont {Bernevig}}, \bibinfo {author} {\bibfnamefont {C.}~\bibnamefont {Felser}},\ and\ \bibinfo {author} {\bibfnamefont {H.}~\bibnamefont {Beidenkopf}},\ }\bibfield  {title} {\bibinfo {title} {Progress and prospects in magnetic topological materials},\ }\href@noop {} {\bibfield  {journal} {\bibinfo  {journal} {Nature}\ }\textbf {\bibinfo {volume} {603}},\ \bibinfo {pages} {41} (\bibinfo {year} {2022})}\BibitemShut {NoStop}%
\bibitem [{\citenamefont {Li}\ \emph {et~al.}(2015{\natexlab{a}})\citenamefont {Li}, \citenamefont {Su}, \citenamefont {Yang},\ and\ \citenamefont {Zhang}}]{li2015electronic}%
  \BibitemOpen
  \bibfield  {author} {\bibinfo {author} {\bibfnamefont {Z.}~\bibnamefont {Li}}, \bibinfo {author} {\bibfnamefont {H.}~\bibnamefont {Su}}, \bibinfo {author} {\bibfnamefont {X.}~\bibnamefont {Yang}},\ and\ \bibinfo {author} {\bibfnamefont {J.}~\bibnamefont {Zhang}},\ }\bibfield  {title} {\bibinfo {title} {Electronic structure of the antiferromagnetic topological insulator candidate gdbipt},\ }\href@noop {} {\bibfield  {journal} {\bibinfo  {journal} {Physical Review B}\ }\textbf {\bibinfo {volume} {91}},\ \bibinfo {pages} {235128} (\bibinfo {year} {2015}{\natexlab{a}})}\BibitemShut {NoStop}%
\bibitem [{\citenamefont {Liang}\ \emph {et~al.}(2018{\natexlab{b}})\citenamefont {Liang}, \citenamefont {Lin}, \citenamefont {Kushwaha}, \citenamefont {Xing}, \citenamefont {Ni}, \citenamefont {Cava},\ and\ \citenamefont {Ong}}]{liang2018experimental}%
  \BibitemOpen
  \bibfield  {author} {\bibinfo {author} {\bibfnamefont {S.}~\bibnamefont {Liang}}, \bibinfo {author} {\bibfnamefont {J.}~\bibnamefont {Lin}}, \bibinfo {author} {\bibfnamefont {S.}~\bibnamefont {Kushwaha}}, \bibinfo {author} {\bibfnamefont {J.}~\bibnamefont {Xing}}, \bibinfo {author} {\bibfnamefont {N.}~\bibnamefont {Ni}}, \bibinfo {author} {\bibfnamefont {R.~J.}\ \bibnamefont {Cava}},\ and\ \bibinfo {author} {\bibfnamefont {N.~P.}\ \bibnamefont {Ong}},\ }\bibfield  {title} {\bibinfo {title} {Experimental tests of the chiral anomaly magnetoresistance in the dirac-weyl semimetals na 3 bi and gdptbi},\ }\href@noop {} {\bibfield  {journal} {\bibinfo  {journal} {Physical Review X}\ }\textbf {\bibinfo {volume} {8}},\ \bibinfo {pages} {031002} (\bibinfo {year} {2018}{\natexlab{b}})}\BibitemShut {NoStop}%
\bibitem [{\citenamefont {Kumar}\ \emph {et~al.}(2018)\citenamefont {Kumar}, \citenamefont {Guin}, \citenamefont {Felser},\ and\ \citenamefont {Shekhar}}]{kumar2018planar}%
  \BibitemOpen
  \bibfield  {author} {\bibinfo {author} {\bibfnamefont {N.}~\bibnamefont {Kumar}}, \bibinfo {author} {\bibfnamefont {S.~N.}\ \bibnamefont {Guin}}, \bibinfo {author} {\bibfnamefont {C.}~\bibnamefont {Felser}},\ and\ \bibinfo {author} {\bibfnamefont {C.}~\bibnamefont {Shekhar}},\ }\bibfield  {title} {\bibinfo {title} {Planar hall effect in the weyl semimetal gdptbi},\ }\href@noop {} {\bibfield  {journal} {\bibinfo  {journal} {Physical Review B}\ }\textbf {\bibinfo {volume} {98}},\ \bibinfo {pages} {041103} (\bibinfo {year} {2018})}\BibitemShut {NoStop}%
\bibitem [{\citenamefont {Suzuki}\ \emph {et~al.}(2016)\citenamefont {Suzuki}, \citenamefont {Chisnell}, \citenamefont {Devarakonda}, \citenamefont {Liu}, \citenamefont {Feng}, \citenamefont {Xiao}, \citenamefont {Lynn},\ and\ \citenamefont {Checkelsky}}]{suzuki2016large}%
  \BibitemOpen
  \bibfield  {author} {\bibinfo {author} {\bibfnamefont {T.}~\bibnamefont {Suzuki}}, \bibinfo {author} {\bibfnamefont {R.}~\bibnamefont {Chisnell}}, \bibinfo {author} {\bibfnamefont {A.}~\bibnamefont {Devarakonda}}, \bibinfo {author} {\bibfnamefont {Y.-T.}\ \bibnamefont {Liu}}, \bibinfo {author} {\bibfnamefont {W.}~\bibnamefont {Feng}}, \bibinfo {author} {\bibfnamefont {D.}~\bibnamefont {Xiao}}, \bibinfo {author} {\bibfnamefont {J.~W.}\ \bibnamefont {Lynn}},\ and\ \bibinfo {author} {\bibfnamefont {J.}~\bibnamefont {Checkelsky}},\ }\bibfield  {title} {\bibinfo {title} {Large anomalous hall effect in a half-heusler antiferromagnet},\ }\href@noop {} {\bibfield  {journal} {\bibinfo  {journal} {Nature Physics}\ }\textbf {\bibinfo {volume} {12}},\ \bibinfo {pages} {1119} (\bibinfo {year} {2016})}\BibitemShut {NoStop}%
\bibitem [{\citenamefont {Sun}\ \emph {et~al.}(2021{\natexlab{a}})\citenamefont {Sun}, \citenamefont {Peng}, \citenamefont {Cui}, \citenamefont {Zhu}, \citenamefont {Zhuo}, \citenamefont {Wang},\ and\ \citenamefont {Chen}}]{sun2021pressure}%
  \BibitemOpen
  \bibfield  {author} {\bibinfo {author} {\bibfnamefont {Z.}~\bibnamefont {Sun}}, \bibinfo {author} {\bibfnamefont {K.}~\bibnamefont {Peng}}, \bibinfo {author} {\bibfnamefont {J.}~\bibnamefont {Cui}}, \bibinfo {author} {\bibfnamefont {C.}~\bibnamefont {Zhu}}, \bibinfo {author} {\bibfnamefont {W.}~\bibnamefont {Zhuo}}, \bibinfo {author} {\bibfnamefont {Z.}~\bibnamefont {Wang}},\ and\ \bibinfo {author} {\bibfnamefont {X.}~\bibnamefont {Chen}},\ }\bibfield  {title} {\bibinfo {title} {Pressure-controlled anomalous hall conductivity in the half-heusler antiferromagnet gdptbi},\ }\href@noop {} {\bibfield  {journal} {\bibinfo  {journal} {Physical Review B}\ }\textbf {\bibinfo {volume} {103}},\ \bibinfo {pages} {085116} (\bibinfo {year} {2021}{\natexlab{a}})}\BibitemShut {NoStop}%
\bibitem [{\citenamefont {Mun}\ and\ \citenamefont {Bud’ko}(2022)}]{mun2022r}%
  \BibitemOpen
  \bibfield  {author} {\bibinfo {author} {\bibfnamefont {E.}~\bibnamefont {Mun}}\ and\ \bibinfo {author} {\bibfnamefont {S.~L.}\ \bibnamefont {Bud’ko}},\ }\bibfield  {title} {\bibinfo {title} {R ptbi: Magnetism and topology},\ }\href@noop {} {\bibfield  {journal} {\bibinfo  {journal} {MRS Bulletin}\ }\textbf {\bibinfo {volume} {47}},\ \bibinfo {pages} {609} (\bibinfo {year} {2022})}\BibitemShut {NoStop}%
\bibitem [{\citenamefont {Pavlosiuk}\ \emph {et~al.}(2020)\citenamefont {Pavlosiuk}, \citenamefont {Fa{\l}at}, \citenamefont {Kaczorowski},\ and\ \citenamefont {Wi{\'s}niewski}}]{pavlosiuk2020anomalous}%
  \BibitemOpen
  \bibfield  {author} {\bibinfo {author} {\bibfnamefont {O.}~\bibnamefont {Pavlosiuk}}, \bibinfo {author} {\bibfnamefont {P.}~\bibnamefont {Fa{\l}at}}, \bibinfo {author} {\bibfnamefont {D.}~\bibnamefont {Kaczorowski}},\ and\ \bibinfo {author} {\bibfnamefont {P.}~\bibnamefont {Wi{\'s}niewski}},\ }\bibfield  {title} {\bibinfo {title} {Anomalous hall effect and negative longitudinal magnetoresistance in half-heusler topological semimetal candidates tbptbi and hoptbi},\ }\href@noop {} {\bibfield  {journal} {\bibinfo  {journal} {APL Materials}\ }\textbf {\bibinfo {volume} {8}} (\bibinfo {year} {2020})}\BibitemShut {NoStop}%
\bibitem [{\citenamefont {Bhattacharya}\ \emph {et~al.}(2023)\citenamefont {Bhattacharya}, \citenamefont {Bhardwaj}, \citenamefont {Bhogra}, \citenamefont {Mani}, \citenamefont {Waghmare},\ and\ \citenamefont {Chatterjee}}]{bhattacharya2023first}%
  \BibitemOpen
  \bibfield  {author} {\bibinfo {author} {\bibfnamefont {A.}~\bibnamefont {Bhattacharya}}, \bibinfo {author} {\bibfnamefont {V.}~\bibnamefont {Bhardwaj}}, \bibinfo {author} {\bibfnamefont {M.}~\bibnamefont {Bhogra}}, \bibinfo {author} {\bibfnamefont {B.}~\bibnamefont {Mani}}, \bibinfo {author} {\bibfnamefont {U.~V.}\ \bibnamefont {Waghmare}},\ and\ \bibinfo {author} {\bibfnamefont {R.}~\bibnamefont {Chatterjee}},\ }\bibfield  {title} {\bibinfo {title} {First-principles theoretical analysis of magnetically tunable topological semimetallic states in antiferromagnetic dypdbi},\ }\href@noop {} {\bibfield  {journal} {\bibinfo  {journal} {Physical Review B}\ }\textbf {\bibinfo {volume} {107}},\ \bibinfo {pages} {075144} (\bibinfo {year} {2023})}\BibitemShut {NoStop}%
\bibitem [{\citenamefont {Wang}\ \emph {et~al.}(2016{\natexlab{b}})\citenamefont {Wang}, \citenamefont {Wu}, \citenamefont {Shi},\ and\ \citenamefont {Wang}}]{wang2016anisotropic}%
  \BibitemOpen
  \bibfield  {author} {\bibinfo {author} {\bibfnamefont {H.}~\bibnamefont {Wang}}, \bibinfo {author} {\bibfnamefont {D.}~\bibnamefont {Wu}}, \bibinfo {author} {\bibfnamefont {Y.}~\bibnamefont {Shi}},\ and\ \bibinfo {author} {\bibfnamefont {N.}~\bibnamefont {Wang}},\ }\bibfield  {title} {\bibinfo {title} {Anisotropic transport and optical spectroscopy study on antiferromagnetic triangular lattice eucd 2 as 2: An interplay between magnetism and charge transport properties},\ }\href@noop {} {\bibfield  {journal} {\bibinfo  {journal} {Physical Review B}\ }\textbf {\bibinfo {volume} {94}},\ \bibinfo {pages} {045112} (\bibinfo {year} {2016}{\natexlab{b}})}\BibitemShut {NoStop}%
\bibitem [{\citenamefont {Rahn}\ \emph {et~al.}(2018)\citenamefont {Rahn}, \citenamefont {Soh}, \citenamefont {Francoual}, \citenamefont {Veiga}, \citenamefont {Strempfer}, \citenamefont {Mardegan}, \citenamefont {Yan}, \citenamefont {Guo}, \citenamefont {Shi},\ and\ \citenamefont {Boothroyd}}]{rahn2018coupling}%
  \BibitemOpen
  \bibfield  {author} {\bibinfo {author} {\bibfnamefont {M.}~\bibnamefont {Rahn}}, \bibinfo {author} {\bibfnamefont {J.-R.}\ \bibnamefont {Soh}}, \bibinfo {author} {\bibfnamefont {S.}~\bibnamefont {Francoual}}, \bibinfo {author} {\bibfnamefont {L.}~\bibnamefont {Veiga}}, \bibinfo {author} {\bibfnamefont {J.}~\bibnamefont {Strempfer}}, \bibinfo {author} {\bibfnamefont {J.}~\bibnamefont {Mardegan}}, \bibinfo {author} {\bibfnamefont {D.}~\bibnamefont {Yan}}, \bibinfo {author} {\bibfnamefont {Y.}~\bibnamefont {Guo}}, \bibinfo {author} {\bibfnamefont {Y.}~\bibnamefont {Shi}},\ and\ \bibinfo {author} {\bibfnamefont {A.}~\bibnamefont {Boothroyd}},\ }\bibfield  {title} {\bibinfo {title} {Coupling of magnetic order and charge transport in the candidate dirac semimetal eucd 2 as 2},\ }\href@noop {} {\bibfield  {journal} {\bibinfo  {journal} {Physical Review B}\ }\textbf {\bibinfo {volume} {97}},\ \bibinfo {pages} {214422} (\bibinfo {year} {2018})}\BibitemShut {NoStop}%
\bibitem [{\citenamefont {Li}\ \emph {et~al.}(2021)\citenamefont {Li}, \citenamefont {Wang}, \citenamefont {Mao}, \citenamefont {Ma}, \citenamefont {Huang}, \citenamefont {Dai},\ and\ \citenamefont {Niu}}]{li2021engineering}%
  \BibitemOpen
  \bibfield  {author} {\bibinfo {author} {\bibfnamefont {R.}~\bibnamefont {Li}}, \bibinfo {author} {\bibfnamefont {H.}~\bibnamefont {Wang}}, \bibinfo {author} {\bibfnamefont {N.}~\bibnamefont {Mao}}, \bibinfo {author} {\bibfnamefont {H.}~\bibnamefont {Ma}}, \bibinfo {author} {\bibfnamefont {B.}~\bibnamefont {Huang}}, \bibinfo {author} {\bibfnamefont {Y.}~\bibnamefont {Dai}},\ and\ \bibinfo {author} {\bibfnamefont {C.}~\bibnamefont {Niu}},\ }\bibfield  {title} {\bibinfo {title} {Engineering antiferromagnetic topological insulator by strain in two-dimensional rare-earth pnictide eucd2sb2},\ }\href@noop {} {\bibfield  {journal} {\bibinfo  {journal} {Applied Physics Letters}\ }\textbf {\bibinfo {volume} {119}} (\bibinfo {year} {2021})}\BibitemShut {NoStop}%
\bibitem [{\citenamefont {Wang}\ \emph {et~al.}(2019{\natexlab{a}})\citenamefont {Wang}, \citenamefont {Jo}, \citenamefont {Kuthanazhi}, \citenamefont {Wu}, \citenamefont {McQueeney}, \citenamefont {Kaminski},\ and\ \citenamefont {Canfield}}]{wang2019single}%
  \BibitemOpen
  \bibfield  {author} {\bibinfo {author} {\bibfnamefont {L.-L.}\ \bibnamefont {Wang}}, \bibinfo {author} {\bibfnamefont {N.~H.}\ \bibnamefont {Jo}}, \bibinfo {author} {\bibfnamefont {B.}~\bibnamefont {Kuthanazhi}}, \bibinfo {author} {\bibfnamefont {Y.}~\bibnamefont {Wu}}, \bibinfo {author} {\bibfnamefont {R.~J.}\ \bibnamefont {McQueeney}}, \bibinfo {author} {\bibfnamefont {A.}~\bibnamefont {Kaminski}},\ and\ \bibinfo {author} {\bibfnamefont {P.~C.}\ \bibnamefont {Canfield}},\ }\bibfield  {title} {\bibinfo {title} {Single pair of weyl fermions in the half-metallic semimetal euc d 2 a s 2},\ }\href@noop {} {\bibfield  {journal} {\bibinfo  {journal} {Physical Review B}\ }\textbf {\bibinfo {volume} {99}},\ \bibinfo {pages} {245147} (\bibinfo {year} {2019}{\natexlab{a}})}\BibitemShut {NoStop}%
\bibitem [{\citenamefont {Su}\ \emph {et~al.}(2020)\citenamefont {Su}, \citenamefont {Gong}, \citenamefont {Shi}, \citenamefont {Yang}, \citenamefont {Wang}, \citenamefont {Xia}, \citenamefont {Yu}, \citenamefont {Guo}, \citenamefont {Wang}, \citenamefont {Ding} \emph {et~al.}}]{su2020magnetic}%
  \BibitemOpen
  \bibfield  {author} {\bibinfo {author} {\bibfnamefont {H.}~\bibnamefont {Su}}, \bibinfo {author} {\bibfnamefont {B.}~\bibnamefont {Gong}}, \bibinfo {author} {\bibfnamefont {W.}~\bibnamefont {Shi}}, \bibinfo {author} {\bibfnamefont {H.}~\bibnamefont {Yang}}, \bibinfo {author} {\bibfnamefont {H.}~\bibnamefont {Wang}}, \bibinfo {author} {\bibfnamefont {W.}~\bibnamefont {Xia}}, \bibinfo {author} {\bibfnamefont {Z.}~\bibnamefont {Yu}}, \bibinfo {author} {\bibfnamefont {P.-J.}\ \bibnamefont {Guo}}, \bibinfo {author} {\bibfnamefont {J.}~\bibnamefont {Wang}}, \bibinfo {author} {\bibfnamefont {L.}~\bibnamefont {Ding}}, \emph {et~al.},\ }\bibfield  {title} {\bibinfo {title} {Magnetic exchange induced weyl state in a semimetal eucd2sb2},\ }\href@noop {} {\bibfield  {journal} {\bibinfo  {journal} {APL Materials}\ }\textbf {\bibinfo {volume} {8}} (\bibinfo {year} {2020})}\BibitemShut {NoStop}%
\bibitem [{\citenamefont {Soh}\ \emph {et~al.}(2019)\citenamefont {Soh}, \citenamefont {de~Juan}, \citenamefont {Vergniory}, \citenamefont {Schr\"oter}, \citenamefont {Rahn}, \citenamefont {Yan}, \citenamefont {Jiang}, \citenamefont {Bristow}, \citenamefont {Reiss}, \citenamefont {Blandy}, \citenamefont {Guo}, \citenamefont {Shi}, \citenamefont {Kim}, \citenamefont {McCollam}, \citenamefont {Simon}, \citenamefont {Chen}, \citenamefont {Coldea},\ and\ \citenamefont {Boothroyd}}]{Soh2019}%
  \BibitemOpen
  \bibfield  {author} {\bibinfo {author} {\bibfnamefont {J.-R.}\ \bibnamefont {Soh}}, \bibinfo {author} {\bibfnamefont {F.}~\bibnamefont {de~Juan}}, \bibinfo {author} {\bibfnamefont {M.~G.}\ \bibnamefont {Vergniory}}, \bibinfo {author} {\bibfnamefont {N.~B.~M.}\ \bibnamefont {Schr\"oter}}, \bibinfo {author} {\bibfnamefont {M.~C.}\ \bibnamefont {Rahn}}, \bibinfo {author} {\bibfnamefont {D.~Y.}\ \bibnamefont {Yan}}, \bibinfo {author} {\bibfnamefont {J.}~\bibnamefont {Jiang}}, \bibinfo {author} {\bibfnamefont {M.}~\bibnamefont {Bristow}}, \bibinfo {author} {\bibfnamefont {P.}~\bibnamefont {Reiss}}, \bibinfo {author} {\bibfnamefont {J.~N.}\ \bibnamefont {Blandy}}, \bibinfo {author} {\bibfnamefont {Y.~F.}\ \bibnamefont {Guo}}, \bibinfo {author} {\bibfnamefont {Y.~G.}\ \bibnamefont {Shi}}, \bibinfo {author} {\bibfnamefont {T.~K.}\ \bibnamefont {Kim}}, \bibinfo {author} {\bibfnamefont {A.}~\bibnamefont {McCollam}}, \bibinfo {author} {\bibfnamefont {S.~H.}\ \bibnamefont {Simon}}, \bibinfo {author} {\bibfnamefont
  {Y.}~\bibnamefont {Chen}}, \bibinfo {author} {\bibfnamefont {A.~I.}\ \bibnamefont {Coldea}},\ and\ \bibinfo {author} {\bibfnamefont {A.~T.}\ \bibnamefont {Boothroyd}},\ }\bibfield  {title} {\bibinfo {title} {Ideal weyl semimetal induced by magnetic exchange},\ }\href {https://doi.org/10.1103/PhysRevB.100.201102} {\bibfield  {journal} {\bibinfo  {journal} {Phys. Rev. B}\ }\textbf {\bibinfo {volume} {100}},\ \bibinfo {pages} {201102} (\bibinfo {year} {2019})}\BibitemShut {NoStop}%
\bibitem [{\citenamefont {Roychowdhury}\ \emph {et~al.}(2023)\citenamefont {Roychowdhury}, \citenamefont {Yao}, \citenamefont {Samanta}, \citenamefont {Bae}, \citenamefont {Chen}, \citenamefont {Ju}, \citenamefont {Raghavan}, \citenamefont {Kumar}, \citenamefont {Constantinou}, \citenamefont {Guin} \emph {et~al.}}]{roychowdhury2023anomalous}%
  \BibitemOpen
  \bibfield  {author} {\bibinfo {author} {\bibfnamefont {S.}~\bibnamefont {Roychowdhury}}, \bibinfo {author} {\bibfnamefont {M.}~\bibnamefont {Yao}}, \bibinfo {author} {\bibfnamefont {K.}~\bibnamefont {Samanta}}, \bibinfo {author} {\bibfnamefont {S.}~\bibnamefont {Bae}}, \bibinfo {author} {\bibfnamefont {D.}~\bibnamefont {Chen}}, \bibinfo {author} {\bibfnamefont {S.}~\bibnamefont {Ju}}, \bibinfo {author} {\bibfnamefont {A.}~\bibnamefont {Raghavan}}, \bibinfo {author} {\bibfnamefont {N.}~\bibnamefont {Kumar}}, \bibinfo {author} {\bibfnamefont {P.}~\bibnamefont {Constantinou}}, \bibinfo {author} {\bibfnamefont {S.~N.}\ \bibnamefont {Guin}}, \emph {et~al.},\ }\bibfield  {title} {\bibinfo {title} {Anomalous hall conductivity and nernst effect of the ideal weyl semimetallic ferromagnet eucd2as2},\ }\href@noop {} {\bibfield  {journal} {\bibinfo  {journal} {Advanced Science}\ }\textbf {\bibinfo {volume} {10}},\ \bibinfo {pages} {2207121} (\bibinfo {year} {2023})}\BibitemShut {NoStop}%
\bibitem [{\citenamefont {Ohno}\ \emph {et~al.}(2022)\citenamefont {Ohno}, \citenamefont {Minami}, \citenamefont {Nakazawa}, \citenamefont {Sato}, \citenamefont {Kriener}, \citenamefont {Arita}, \citenamefont {Kawasaki},\ and\ \citenamefont {Uchida}}]{ohno2022maximizing}%
  \BibitemOpen
  \bibfield  {author} {\bibinfo {author} {\bibfnamefont {M.}~\bibnamefont {Ohno}}, \bibinfo {author} {\bibfnamefont {S.}~\bibnamefont {Minami}}, \bibinfo {author} {\bibfnamefont {Y.}~\bibnamefont {Nakazawa}}, \bibinfo {author} {\bibfnamefont {S.}~\bibnamefont {Sato}}, \bibinfo {author} {\bibfnamefont {M.}~\bibnamefont {Kriener}}, \bibinfo {author} {\bibfnamefont {R.}~\bibnamefont {Arita}}, \bibinfo {author} {\bibfnamefont {M.}~\bibnamefont {Kawasaki}},\ and\ \bibinfo {author} {\bibfnamefont {M.}~\bibnamefont {Uchida}},\ }\bibfield  {title} {\bibinfo {title} {Maximizing intrinsic anomalous hall effect by controlling the fermi level in simple weyl semimetal films},\ }\href@noop {} {\bibfield  {journal} {\bibinfo  {journal} {Physical Review B}\ }\textbf {\bibinfo {volume} {105}},\ \bibinfo {pages} {L201101} (\bibinfo {year} {2022})}\BibitemShut {NoStop}%
\bibitem [{\citenamefont {Nakamura}\ \emph {et~al.}(2024{\natexlab{a}})\citenamefont {Nakamura}, \citenamefont {Nishihaya}, \citenamefont {Ishizuka}, \citenamefont {Kriener}, \citenamefont {Ohno}, \citenamefont {Watanabe}, \citenamefont {Kawasaki},\ and\ \citenamefont {Uchida}}]{nakamura2024berry}%
  \BibitemOpen
  \bibfield  {author} {\bibinfo {author} {\bibfnamefont {A.}~\bibnamefont {Nakamura}}, \bibinfo {author} {\bibfnamefont {S.}~\bibnamefont {Nishihaya}}, \bibinfo {author} {\bibfnamefont {H.}~\bibnamefont {Ishizuka}}, \bibinfo {author} {\bibfnamefont {M.}~\bibnamefont {Kriener}}, \bibinfo {author} {\bibfnamefont {M.}~\bibnamefont {Ohno}}, \bibinfo {author} {\bibfnamefont {Y.}~\bibnamefont {Watanabe}}, \bibinfo {author} {\bibfnamefont {M.}~\bibnamefont {Kawasaki}},\ and\ \bibinfo {author} {\bibfnamefont {M.}~\bibnamefont {Uchida}},\ }\bibfield  {title} {\bibinfo {title} {Berry curvature derived negative magnetoconductivity observed in type-ii magnetic weyl semimetal films},\ }\href@noop {} {\bibfield  {journal} {\bibinfo  {journal} {Physical Review B}\ }\textbf {\bibinfo {volume} {109}},\ \bibinfo {pages} {L121108} (\bibinfo {year} {2024}{\natexlab{a}})}\BibitemShut {NoStop}%
\bibitem [{\citenamefont {Nakamura}\ \emph {et~al.}(2024{\natexlab{b}})\citenamefont {Nakamura}, \citenamefont {Nishihaya}, \citenamefont {Ishizuka}, \citenamefont {Kriener}, \citenamefont {Watanabe},\ and\ \citenamefont {Uchida}}]{nakamura2024plane}%
  \BibitemOpen
  \bibfield  {author} {\bibinfo {author} {\bibfnamefont {A.}~\bibnamefont {Nakamura}}, \bibinfo {author} {\bibfnamefont {S.}~\bibnamefont {Nishihaya}}, \bibinfo {author} {\bibfnamefont {H.}~\bibnamefont {Ishizuka}}, \bibinfo {author} {\bibfnamefont {M.}~\bibnamefont {Kriener}}, \bibinfo {author} {\bibfnamefont {Y.}~\bibnamefont {Watanabe}},\ and\ \bibinfo {author} {\bibfnamefont {M.}~\bibnamefont {Uchida}},\ }\bibfield  {title} {\bibinfo {title} {In-plane anomalous hall effect associated with orbital magnetization: Measurements of low-carrier density films of a magnetic weyl semimetal},\ }\href@noop {} {\bibfield  {journal} {\bibinfo  {journal} {Physical Review Letters}\ }\textbf {\bibinfo {volume} {133}},\ \bibinfo {pages} {236602} (\bibinfo {year} {2024}{\natexlab{b}})}\BibitemShut {NoStop}%
\bibitem [{\citenamefont {Ma}\ \emph {et~al.}(2019)\citenamefont {Ma}, \citenamefont {Nie}, \citenamefont {Yi}, \citenamefont {Jandke}, \citenamefont {Shang}, \citenamefont {Yao}, \citenamefont {Naamneh}, \citenamefont {Yan}, \citenamefont {Sun}, \citenamefont {Chikina} \emph {et~al.}}]{Ma2019}%
  \BibitemOpen
  \bibfield  {author} {\bibinfo {author} {\bibfnamefont {J.-Z.}\ \bibnamefont {Ma}}, \bibinfo {author} {\bibfnamefont {S.}~\bibnamefont {Nie}}, \bibinfo {author} {\bibfnamefont {C.}~\bibnamefont {Yi}}, \bibinfo {author} {\bibfnamefont {J.}~\bibnamefont {Jandke}}, \bibinfo {author} {\bibfnamefont {T.}~\bibnamefont {Shang}}, \bibinfo {author} {\bibfnamefont {M.-Y.}\ \bibnamefont {Yao}}, \bibinfo {author} {\bibfnamefont {M.}~\bibnamefont {Naamneh}}, \bibinfo {author} {\bibfnamefont {L.}~\bibnamefont {Yan}}, \bibinfo {author} {\bibfnamefont {Y.}~\bibnamefont {Sun}}, \bibinfo {author} {\bibfnamefont {A.}~\bibnamefont {Chikina}}, \emph {et~al.},\ }\bibfield  {title} {\bibinfo {title} {Spin fluctuation induced weyl semimetal state in the paramagnetic phase of eucd2as2},\ }\href@noop {} {\bibfield  {journal} {\bibinfo  {journal} {Science advances}\ }\textbf {\bibinfo {volume} {5}},\ \bibinfo {pages} {eaaw4718} (\bibinfo {year} {2019})}\BibitemShut {NoStop}%
\bibitem [{\citenamefont {Ito}\ and\ \citenamefont {Nomura}(2017)}]{ito2017anomalous}%
  \BibitemOpen
  \bibfield  {author} {\bibinfo {author} {\bibfnamefont {N.}~\bibnamefont {Ito}}\ and\ \bibinfo {author} {\bibfnamefont {K.}~\bibnamefont {Nomura}},\ }\bibfield  {title} {\bibinfo {title} {Anomalous hall effect and spontaneous orbital magnetization in antiferromagnetic weyl metal},\ }\href@noop {} {\bibfield  {journal} {\bibinfo  {journal} {journal of the physical society of japan}\ }\textbf {\bibinfo {volume} {86}},\ \bibinfo {pages} {063703} (\bibinfo {year} {2017})}\BibitemShut {NoStop}%
\bibitem [{\citenamefont {Kr{\'e}n}\ \emph {et~al.}(1975)\citenamefont {Kr{\'e}n}, \citenamefont {Paitz}, \citenamefont {Zimmer},\ and\ \citenamefont {Zsoldos}}]{kren1975study}%
  \BibitemOpen
  \bibfield  {author} {\bibinfo {author} {\bibfnamefont {E.}~\bibnamefont {Kr{\'e}n}}, \bibinfo {author} {\bibfnamefont {J.}~\bibnamefont {Paitz}}, \bibinfo {author} {\bibfnamefont {G.}~\bibnamefont {Zimmer}},\ and\ \bibinfo {author} {\bibfnamefont {{\'E}.}~\bibnamefont {Zsoldos}},\ }\bibfield  {title} {\bibinfo {title} {Study of the magnetic phase transformation in the mn3sn phase},\ }\href@noop {} {\bibfield  {journal} {\bibinfo  {journal} {Physica B+ C}\ }\textbf {\bibinfo {volume} {80}},\ \bibinfo {pages} {226} (\bibinfo {year} {1975})}\BibitemShut {NoStop}%
\bibitem [{\citenamefont {Nagamiya}\ \emph {et~al.}(1982)\citenamefont {Nagamiya}, \citenamefont {Tomiyoshi},\ and\ \citenamefont {Yamaguchi}}]{nagamiya1982triangular}%
  \BibitemOpen
  \bibfield  {author} {\bibinfo {author} {\bibfnamefont {T.}~\bibnamefont {Nagamiya}}, \bibinfo {author} {\bibfnamefont {S.}~\bibnamefont {Tomiyoshi}},\ and\ \bibinfo {author} {\bibfnamefont {Y.}~\bibnamefont {Yamaguchi}},\ }\bibfield  {title} {\bibinfo {title} {Triangular spin configuration and weak ferromagnetism of mn3sn and mn3ge},\ }\href@noop {} {\bibfield  {journal} {\bibinfo  {journal} {Solid State Communications}\ }\textbf {\bibinfo {volume} {42}},\ \bibinfo {pages} {385} (\bibinfo {year} {1982})}\BibitemShut {NoStop}%
\bibitem [{\citenamefont {Tomiyoshi}\ and\ \citenamefont {Yamaguchi}(1982)}]{tomiyoshi1982magnetic}%
  \BibitemOpen
  \bibfield  {author} {\bibinfo {author} {\bibfnamefont {S.}~\bibnamefont {Tomiyoshi}}\ and\ \bibinfo {author} {\bibfnamefont {Y.}~\bibnamefont {Yamaguchi}},\ }\bibfield  {title} {\bibinfo {title} {Magnetic structure and weak ferromagnetism of mn3sn studied by polarized neutron diffraction},\ }\href@noop {} {\bibfield  {journal} {\bibinfo  {journal} {Journal of the Physical Society of Japan}\ }\textbf {\bibinfo {volume} {51}},\ \bibinfo {pages} {2478} (\bibinfo {year} {1982})}\BibitemShut {NoStop}%
\bibitem [{\citenamefont {Brown}\ \emph {et~al.}(1990)\citenamefont {Brown}, \citenamefont {Nunez}, \citenamefont {Tasset}, \citenamefont {Forsyth},\ and\ \citenamefont {Radhakrishna}}]{brown1990determination}%
  \BibitemOpen
  \bibfield  {author} {\bibinfo {author} {\bibfnamefont {P.}~\bibnamefont {Brown}}, \bibinfo {author} {\bibfnamefont {V.}~\bibnamefont {Nunez}}, \bibinfo {author} {\bibfnamefont {F.}~\bibnamefont {Tasset}}, \bibinfo {author} {\bibfnamefont {J.}~\bibnamefont {Forsyth}},\ and\ \bibinfo {author} {\bibfnamefont {P.}~\bibnamefont {Radhakrishna}},\ }\bibfield  {title} {\bibinfo {title} {Determination of the magnetic structure of mn3sn using generalized neutron polarization analysis},\ }\href@noop {} {\bibfield  {journal} {\bibinfo  {journal} {Journal of Physics: Condensed Matter}\ }\textbf {\bibinfo {volume} {2}},\ \bibinfo {pages} {9409} (\bibinfo {year} {1990})}\BibitemShut {NoStop}%
\bibitem [{\citenamefont {Chen}\ \emph {et~al.}(2014)\citenamefont {Chen}, \citenamefont {Niu},\ and\ \citenamefont {MacDonald}}]{Chen2014}%
  \BibitemOpen
  \bibfield  {author} {\bibinfo {author} {\bibfnamefont {H.}~\bibnamefont {Chen}}, \bibinfo {author} {\bibfnamefont {Q.}~\bibnamefont {Niu}},\ and\ \bibinfo {author} {\bibfnamefont {A.~H.}\ \bibnamefont {MacDonald}},\ }\bibfield  {title} {\bibinfo {title} {Anomalous hall effect arising from noncollinear antiferromagnetism},\ }\href {https://doi.org/10.1103/PhysRevLett.112.017205} {\bibfield  {journal} {\bibinfo  {journal} {Phys. Rev. Lett.}\ }\textbf {\bibinfo {volume} {112}},\ \bibinfo {pages} {017205} (\bibinfo {year} {2014})}\BibitemShut {NoStop}%
\bibitem [{\citenamefont {K{\"u}bler}\ and\ \citenamefont {Felser}(2014)}]{kubler2014non}%
  \BibitemOpen
  \bibfield  {author} {\bibinfo {author} {\bibfnamefont {J.}~\bibnamefont {K{\"u}bler}}\ and\ \bibinfo {author} {\bibfnamefont {C.}~\bibnamefont {Felser}},\ }\bibfield  {title} {\bibinfo {title} {Non-collinear antiferromagnets and the anomalous hall effect},\ }\href@noop {} {\bibfield  {journal} {\bibinfo  {journal} {Europhysics Letters}\ }\textbf {\bibinfo {volume} {108}},\ \bibinfo {pages} {67001} (\bibinfo {year} {2014})}\BibitemShut {NoStop}%
\bibitem [{\citenamefont {Nakatsuji}\ \emph {et~al.}(2015)\citenamefont {Nakatsuji}, \citenamefont {Kiyohara},\ and\ \citenamefont {Higo}}]{nakatsuji2015large}%
  \BibitemOpen
  \bibfield  {author} {\bibinfo {author} {\bibfnamefont {S.}~\bibnamefont {Nakatsuji}}, \bibinfo {author} {\bibfnamefont {N.}~\bibnamefont {Kiyohara}},\ and\ \bibinfo {author} {\bibfnamefont {T.}~\bibnamefont {Higo}},\ }\bibfield  {title} {\bibinfo {title} {Large anomalous hall effect in a non-collinear antiferromagnet at room temperature},\ }\href@noop {} {\bibfield  {journal} {\bibinfo  {journal} {Nature}\ }\textbf {\bibinfo {volume} {527}},\ \bibinfo {pages} {212} (\bibinfo {year} {2015})}\BibitemShut {NoStop}%
\bibitem [{\citenamefont {Yang}\ \emph {et~al.}(2017)\citenamefont {Yang}, \citenamefont {Sun}, \citenamefont {Zhang}, \citenamefont {Shi}, \citenamefont {Parkin},\ and\ \citenamefont {Yan}}]{Yang2017}%
  \BibitemOpen
  \bibfield  {author} {\bibinfo {author} {\bibfnamefont {H.}~\bibnamefont {Yang}}, \bibinfo {author} {\bibfnamefont {Y.}~\bibnamefont {Sun}}, \bibinfo {author} {\bibfnamefont {Y.}~\bibnamefont {Zhang}}, \bibinfo {author} {\bibfnamefont {W.-J.}\ \bibnamefont {Shi}}, \bibinfo {author} {\bibfnamefont {S.~S.}\ \bibnamefont {Parkin}},\ and\ \bibinfo {author} {\bibfnamefont {B.}~\bibnamefont {Yan}},\ }\bibfield  {title} {\bibinfo {title} {Topological weyl semimetals in the chiral antiferromagnetic materials mn3ge and mn3sn},\ }\href@noop {} {\bibfield  {journal} {\bibinfo  {journal} {New Journal of Physics}\ }\textbf {\bibinfo {volume} {19}},\ \bibinfo {pages} {015008} (\bibinfo {year} {2017})}\BibitemShut {NoStop}%
\bibitem [{\citenamefont {Liu}\ and\ \citenamefont {Balents}(2017)}]{Liu2017prl}%
  \BibitemOpen
  \bibfield  {author} {\bibinfo {author} {\bibfnamefont {J.}~\bibnamefont {Liu}}\ and\ \bibinfo {author} {\bibfnamefont {L.}~\bibnamefont {Balents}},\ }\bibfield  {title} {\bibinfo {title} {Anomalous hall effect and topological defects in antiferromagnetic weyl semimetals: ${\mathrm{mn}}_{3}\mathrm{Sn}/\mathrm{Ge}$},\ }\href {https://doi.org/10.1103/PhysRevLett.119.087202} {\bibfield  {journal} {\bibinfo  {journal} {Phys. Rev. Lett.}\ }\textbf {\bibinfo {volume} {119}},\ \bibinfo {pages} {087202} (\bibinfo {year} {2017})}\BibitemShut {NoStop}%
\bibitem [{\citenamefont {Suzuki}\ \emph {et~al.}(2017)\citenamefont {Suzuki}, \citenamefont {Koretsune}, \citenamefont {Ochi},\ and\ \citenamefont {Arita}}]{Suzuki2017}%
  \BibitemOpen
  \bibfield  {author} {\bibinfo {author} {\bibfnamefont {M.-T.}\ \bibnamefont {Suzuki}}, \bibinfo {author} {\bibfnamefont {T.}~\bibnamefont {Koretsune}}, \bibinfo {author} {\bibfnamefont {M.}~\bibnamefont {Ochi}},\ and\ \bibinfo {author} {\bibfnamefont {R.}~\bibnamefont {Arita}},\ }\bibfield  {title} {\bibinfo {title} {Cluster multipole theory for anomalous hall effect in antiferromagnets},\ }\href {https://doi.org/10.1103/PhysRevB.95.094406} {\bibfield  {journal} {\bibinfo  {journal} {Phys. Rev. B}\ }\textbf {\bibinfo {volume} {95}},\ \bibinfo {pages} {094406} (\bibinfo {year} {2017})}\BibitemShut {NoStop}%
\bibitem [{\citenamefont {Higo}\ \emph {et~al.}(2018)\citenamefont {Higo}, \citenamefont {Man}, \citenamefont {Gopman}, \citenamefont {Wu}, \citenamefont {Koretsune}, \citenamefont {van’t Erve}, \citenamefont {Kabanov}, \citenamefont {Rees}, \citenamefont {Li}, \citenamefont {Suzuki} \emph {et~al.}}]{Higo2018}%
  \BibitemOpen
  \bibfield  {author} {\bibinfo {author} {\bibfnamefont {T.}~\bibnamefont {Higo}}, \bibinfo {author} {\bibfnamefont {H.}~\bibnamefont {Man}}, \bibinfo {author} {\bibfnamefont {D.~B.}\ \bibnamefont {Gopman}}, \bibinfo {author} {\bibfnamefont {L.}~\bibnamefont {Wu}}, \bibinfo {author} {\bibfnamefont {T.}~\bibnamefont {Koretsune}}, \bibinfo {author} {\bibfnamefont {O.~M.}\ \bibnamefont {van’t Erve}}, \bibinfo {author} {\bibfnamefont {Y.~P.}\ \bibnamefont {Kabanov}}, \bibinfo {author} {\bibfnamefont {D.}~\bibnamefont {Rees}}, \bibinfo {author} {\bibfnamefont {Y.}~\bibnamefont {Li}}, \bibinfo {author} {\bibfnamefont {M.-T.}\ \bibnamefont {Suzuki}}, \emph {et~al.},\ }\bibfield  {title} {\bibinfo {title} {Large magneto-optical kerr effect and imaging of magnetic octupole domains in an antiferromagnetic metal},\ }\href@noop {} {\bibfield  {journal} {\bibinfo  {journal} {Nature photonics}\ }\textbf {\bibinfo {volume} {12}},\ \bibinfo {pages} {73} (\bibinfo {year} {2018})}\BibitemShut {NoStop}%
\bibitem [{\citenamefont {Uchimura}\ \emph {et~al.}(2022)\citenamefont {Uchimura}, \citenamefont {Yoon}, \citenamefont {Sato}, \citenamefont {Takeuchi}, \citenamefont {Kanai}, \citenamefont {Takechi}, \citenamefont {Kishi}, \citenamefont {Yamane}, \citenamefont {DuttaGupta}, \citenamefont {Ieda} \emph {et~al.}}]{uchimura2022observation}%
  \BibitemOpen
  \bibfield  {author} {\bibinfo {author} {\bibfnamefont {T.}~\bibnamefont {Uchimura}}, \bibinfo {author} {\bibfnamefont {J.-Y.}\ \bibnamefont {Yoon}}, \bibinfo {author} {\bibfnamefont {Y.}~\bibnamefont {Sato}}, \bibinfo {author} {\bibfnamefont {Y.}~\bibnamefont {Takeuchi}}, \bibinfo {author} {\bibfnamefont {S.}~\bibnamefont {Kanai}}, \bibinfo {author} {\bibfnamefont {R.}~\bibnamefont {Takechi}}, \bibinfo {author} {\bibfnamefont {K.}~\bibnamefont {Kishi}}, \bibinfo {author} {\bibfnamefont {Y.}~\bibnamefont {Yamane}}, \bibinfo {author} {\bibfnamefont {S.}~\bibnamefont {DuttaGupta}}, \bibinfo {author} {\bibfnamefont {J.}~\bibnamefont {Ieda}}, \emph {et~al.},\ }\bibfield  {title} {\bibinfo {title} {Observation of domain structure in non-collinear antiferromagnetic mn3sn thin films by magneto-optical kerr effect},\ }\href@noop {} {\bibfield  {journal} {\bibinfo  {journal} {Applied Physics Letters}\ }\textbf {\bibinfo {volume} {120}} (\bibinfo {year} {2022})}\BibitemShut {NoStop}%
\bibitem [{\citenamefont {Ikhlas}\ \emph {et~al.}(2017)\citenamefont {Ikhlas}, \citenamefont {Tomita}, \citenamefont {Koretsune}, \citenamefont {Suzuki}, \citenamefont {Nishio-Hamane}, \citenamefont {Arita}, \citenamefont {Otani},\ and\ \citenamefont {Nakatsuji}}]{ikhlas2017large}%
  \BibitemOpen
  \bibfield  {author} {\bibinfo {author} {\bibfnamefont {M.}~\bibnamefont {Ikhlas}}, \bibinfo {author} {\bibfnamefont {T.}~\bibnamefont {Tomita}}, \bibinfo {author} {\bibfnamefont {T.}~\bibnamefont {Koretsune}}, \bibinfo {author} {\bibfnamefont {M.-T.}\ \bibnamefont {Suzuki}}, \bibinfo {author} {\bibfnamefont {D.}~\bibnamefont {Nishio-Hamane}}, \bibinfo {author} {\bibfnamefont {R.}~\bibnamefont {Arita}}, \bibinfo {author} {\bibfnamefont {Y.}~\bibnamefont {Otani}},\ and\ \bibinfo {author} {\bibfnamefont {S.}~\bibnamefont {Nakatsuji}},\ }\bibfield  {title} {\bibinfo {title} {Large anomalous nernst effect at room temperature in a chiral antiferromagnet},\ }\href@noop {} {\bibfield  {journal} {\bibinfo  {journal} {Nature Physics}\ }\textbf {\bibinfo {volume} {13}},\ \bibinfo {pages} {1085} (\bibinfo {year} {2017})}\BibitemShut {NoStop}%
\bibitem [{\citenamefont {Isshiki}\ \emph {et~al.}(2024)\citenamefont {Isshiki}, \citenamefont {Budai}, \citenamefont {Kobayashi}, \citenamefont {Uesugi}, \citenamefont {Higo}, \citenamefont {Nakatsuji},\ and\ \citenamefont {Otani}}]{isshiki2024observation}%
  \BibitemOpen
  \bibfield  {author} {\bibinfo {author} {\bibfnamefont {H.}~\bibnamefont {Isshiki}}, \bibinfo {author} {\bibfnamefont {N.}~\bibnamefont {Budai}}, \bibinfo {author} {\bibfnamefont {A.}~\bibnamefont {Kobayashi}}, \bibinfo {author} {\bibfnamefont {R.}~\bibnamefont {Uesugi}}, \bibinfo {author} {\bibfnamefont {T.}~\bibnamefont {Higo}}, \bibinfo {author} {\bibfnamefont {S.}~\bibnamefont {Nakatsuji}},\ and\ \bibinfo {author} {\bibfnamefont {Y.}~\bibnamefont {Otani}},\ }\bibfield  {title} {\bibinfo {title} {Observation of cluster magnetic octupole domains in the antiferromagnetic weyl semimetal mn 3 sn nanowire},\ }\href@noop {} {\bibfield  {journal} {\bibinfo  {journal} {Physical Review Letters}\ }\textbf {\bibinfo {volume} {132}},\ \bibinfo {pages} {216702} (\bibinfo {year} {2024})}\BibitemShut {NoStop}%
\bibitem [{\citenamefont {Reichlova}\ \emph {et~al.}(2019)\citenamefont {Reichlova}, \citenamefont {Janda}, \citenamefont {Godinho}, \citenamefont {Markou}, \citenamefont {Kriegner}, \citenamefont {Schlitz}, \citenamefont {Zelezny}, \citenamefont {Soban}, \citenamefont {Bejarano}, \citenamefont {Schultheiss} \emph {et~al.}}]{reichlova2019imaging}%
  \BibitemOpen
  \bibfield  {author} {\bibinfo {author} {\bibfnamefont {H.}~\bibnamefont {Reichlova}}, \bibinfo {author} {\bibfnamefont {T.}~\bibnamefont {Janda}}, \bibinfo {author} {\bibfnamefont {J.}~\bibnamefont {Godinho}}, \bibinfo {author} {\bibfnamefont {A.}~\bibnamefont {Markou}}, \bibinfo {author} {\bibfnamefont {D.}~\bibnamefont {Kriegner}}, \bibinfo {author} {\bibfnamefont {R.}~\bibnamefont {Schlitz}}, \bibinfo {author} {\bibfnamefont {J.}~\bibnamefont {Zelezny}}, \bibinfo {author} {\bibfnamefont {Z.}~\bibnamefont {Soban}}, \bibinfo {author} {\bibfnamefont {M.}~\bibnamefont {Bejarano}}, \bibinfo {author} {\bibfnamefont {H.}~\bibnamefont {Schultheiss}}, \emph {et~al.},\ }\bibfield  {title} {\bibinfo {title} {Imaging and writing magnetic domains in the non-collinear antiferromagnet mn3sn},\ }\href@noop {} {\bibfield  {journal} {\bibinfo  {journal} {Nature communications}\ }\textbf {\bibinfo {volume} {10}},\ \bibinfo {pages} {5459} (\bibinfo {year} {2019})}\BibitemShut {NoStop}%
\bibitem [{\citenamefont {Kimata}\ \emph {et~al.}(2019)\citenamefont {Kimata}, \citenamefont {Chen}, \citenamefont {Kondou}, \citenamefont {Sugimoto}, \citenamefont {Muduli}, \citenamefont {Ikhlas}, \citenamefont {Omori}, \citenamefont {Tomita}, \citenamefont {MacDonald}, \citenamefont {Nakatsuji} \emph {et~al.}}]{kimata2019magnetic}%
  \BibitemOpen
  \bibfield  {author} {\bibinfo {author} {\bibfnamefont {M.}~\bibnamefont {Kimata}}, \bibinfo {author} {\bibfnamefont {H.}~\bibnamefont {Chen}}, \bibinfo {author} {\bibfnamefont {K.}~\bibnamefont {Kondou}}, \bibinfo {author} {\bibfnamefont {S.}~\bibnamefont {Sugimoto}}, \bibinfo {author} {\bibfnamefont {P.~K.}\ \bibnamefont {Muduli}}, \bibinfo {author} {\bibfnamefont {M.}~\bibnamefont {Ikhlas}}, \bibinfo {author} {\bibfnamefont {Y.}~\bibnamefont {Omori}}, \bibinfo {author} {\bibfnamefont {T.}~\bibnamefont {Tomita}}, \bibinfo {author} {\bibfnamefont {A.~H.}\ \bibnamefont {MacDonald}}, \bibinfo {author} {\bibfnamefont {S.}~\bibnamefont {Nakatsuji}}, \emph {et~al.},\ }\bibfield  {title} {\bibinfo {title} {Magnetic and magnetic inverse spin hall effects in a non-collinear antiferromagnet},\ }\href@noop {} {\bibfield  {journal} {\bibinfo  {journal} {Nature}\ }\textbf {\bibinfo {volume} {565}},\ \bibinfo {pages} {627} (\bibinfo {year} {2019})}\BibitemShut {NoStop}%
\bibitem [{\citenamefont {Resta}(2011)}]{resta2011insulating}%
  \BibitemOpen
  \bibfield  {author} {\bibinfo {author} {\bibfnamefont {R.}~\bibnamefont {Resta}},\ }\bibfield  {title} {\bibinfo {title} {The insulating state of matter: a geometrical theory},\ }\href@noop {} {\bibfield  {journal} {\bibinfo  {journal} {The European Physical Journal B}\ }\textbf {\bibinfo {volume} {79}},\ \bibinfo {pages} {121} (\bibinfo {year} {2011})}\BibitemShut {NoStop}%
\bibitem [{\citenamefont {Mera}\ and\ \citenamefont {Ozawa}(2021)}]{mera2021kahler}%
  \BibitemOpen
  \bibfield  {author} {\bibinfo {author} {\bibfnamefont {B.}~\bibnamefont {Mera}}\ and\ \bibinfo {author} {\bibfnamefont {T.}~\bibnamefont {Ozawa}},\ }\bibfield  {title} {\bibinfo {title} {K{\"a}hler geometry and chern insulators: Relations between topology and the quantum metric},\ }\href@noop {} {\bibfield  {journal} {\bibinfo  {journal} {Physical Review B}\ }\textbf {\bibinfo {volume} {104}},\ \bibinfo {pages} {045104} (\bibinfo {year} {2021})}\BibitemShut {NoStop}%
\bibitem [{\citenamefont {Ozawa}\ and\ \citenamefont {Mera}(2021)}]{ozawa2021relations}%
  \BibitemOpen
  \bibfield  {author} {\bibinfo {author} {\bibfnamefont {T.}~\bibnamefont {Ozawa}}\ and\ \bibinfo {author} {\bibfnamefont {B.}~\bibnamefont {Mera}},\ }\bibfield  {title} {\bibinfo {title} {Relations between topology and the quantum metric for chern insulators},\ }\href@noop {} {\bibfield  {journal} {\bibinfo  {journal} {Physical Review B}\ }\textbf {\bibinfo {volume} {104}},\ \bibinfo {pages} {045103} (\bibinfo {year} {2021})}\BibitemShut {NoStop}%
\bibitem [{\citenamefont {Gao}\ \emph {et~al.}(2014)\citenamefont {Gao}, \citenamefont {Yang},\ and\ \citenamefont {Niu}}]{gao2014field}%
  \BibitemOpen
  \bibfield  {author} {\bibinfo {author} {\bibfnamefont {Y.}~\bibnamefont {Gao}}, \bibinfo {author} {\bibfnamefont {S.~A.}\ \bibnamefont {Yang}},\ and\ \bibinfo {author} {\bibfnamefont {Q.}~\bibnamefont {Niu}},\ }\bibfield  {title} {\bibinfo {title} {Field induced positional shift of bloch electrons and its dynamical implications},\ }\href@noop {} {\bibfield  {journal} {\bibinfo  {journal} {Physical review letters}\ }\textbf {\bibinfo {volume} {112}},\ \bibinfo {pages} {166601} (\bibinfo {year} {2014})}\BibitemShut {NoStop}%
\bibitem [{\citenamefont {Han}\ \emph {et~al.}(2024)\citenamefont {Han}, \citenamefont {Uchimura}, \citenamefont {Araki}, \citenamefont {Yoon}, \citenamefont {Takeuchi}, \citenamefont {Yamane}, \citenamefont {Kanai}, \citenamefont {Ieda}, \citenamefont {Ohno},\ and\ \citenamefont {Fukami}}]{han2024room}%
  \BibitemOpen
  \bibfield  {author} {\bibinfo {author} {\bibfnamefont {J.}~\bibnamefont {Han}}, \bibinfo {author} {\bibfnamefont {T.}~\bibnamefont {Uchimura}}, \bibinfo {author} {\bibfnamefont {Y.}~\bibnamefont {Araki}}, \bibinfo {author} {\bibfnamefont {J.-Y.}\ \bibnamefont {Yoon}}, \bibinfo {author} {\bibfnamefont {Y.}~\bibnamefont {Takeuchi}}, \bibinfo {author} {\bibfnamefont {Y.}~\bibnamefont {Yamane}}, \bibinfo {author} {\bibfnamefont {S.}~\bibnamefont {Kanai}}, \bibinfo {author} {\bibfnamefont {J.}~\bibnamefont {Ieda}}, \bibinfo {author} {\bibfnamefont {H.}~\bibnamefont {Ohno}},\ and\ \bibinfo {author} {\bibfnamefont {S.}~\bibnamefont {Fukami}},\ }\bibfield  {title} {\bibinfo {title} {Room-temperature flexible manipulation of the quantum-metric structure in a topological chiral antiferromagnet},\ }\href@noop {} {\bibfield  {journal} {\bibinfo  {journal} {Nature Physics}\ ,\ \bibinfo {pages} {1}} (\bibinfo {year} {2024})}\BibitemShut {NoStop}%
\bibitem [{\citenamefont {Wang}\ \emph {et~al.}(2019{\natexlab{b}})\citenamefont {Wang}, \citenamefont {Lian},\ and\ \citenamefont {Zhang}}]{wang2019generation}%
  \BibitemOpen
  \bibfield  {author} {\bibinfo {author} {\bibfnamefont {J.}~\bibnamefont {Wang}}, \bibinfo {author} {\bibfnamefont {B.}~\bibnamefont {Lian}},\ and\ \bibinfo {author} {\bibfnamefont {S.-C.}\ \bibnamefont {Zhang}},\ }\bibfield  {title} {\bibinfo {title} {Generation of spin currents by magnetic field in t-and p-broken materials},\ }in\ \href@noop {} {\emph {\bibinfo {booktitle} {Spin}}},\ Vol.~\bibinfo {volume} {9}\ (\bibinfo {organization} {World Scientific},\ \bibinfo {year} {2019})\ p.\ \bibinfo {pages} {1940013}\BibitemShut {NoStop}%
\bibitem [{\citenamefont {Meguro}\ \emph {et~al.}(2025)\citenamefont {Meguro}, \citenamefont {Ozawa}, \citenamefont {Kobayashi}, \citenamefont {Araki},\ and\ \citenamefont {Nomura}}]{meguro2025topological}%
  \BibitemOpen
  \bibfield  {author} {\bibinfo {author} {\bibfnamefont {T.}~\bibnamefont {Meguro}}, \bibinfo {author} {\bibfnamefont {A.}~\bibnamefont {Ozawa}}, \bibinfo {author} {\bibfnamefont {K.}~\bibnamefont {Kobayashi}}, \bibinfo {author} {\bibfnamefont {Y.}~\bibnamefont {Araki}},\ and\ \bibinfo {author} {\bibfnamefont {K.}~\bibnamefont {Nomura}},\ }\bibfield  {title} {\bibinfo {title} {Topological spin-orbit torque in ferrimagnetic weyl semimetal},\ }\href@noop {} {\bibfield  {journal} {\bibinfo  {journal} {Physical Review Research}\ }\textbf {\bibinfo {volume} {7}},\ \bibinfo {pages} {L022065} (\bibinfo {year} {2025})}\BibitemShut {NoStop}%
\bibitem [{\citenamefont {Xu}\ \emph {et~al.}(2017)\citenamefont {Xu}, \citenamefont {Alidoust}, \citenamefont {Chang}, \citenamefont {Lu}, \citenamefont {Singh}, \citenamefont {Belopolski}, \citenamefont {Sanchez}, \citenamefont {Zhang}, \citenamefont {Bian}, \citenamefont {Zheng} \emph {et~al.}}]{xu2017discovery}%
  \BibitemOpen
  \bibfield  {author} {\bibinfo {author} {\bibfnamefont {S.-Y.}\ \bibnamefont {Xu}}, \bibinfo {author} {\bibfnamefont {N.}~\bibnamefont {Alidoust}}, \bibinfo {author} {\bibfnamefont {G.}~\bibnamefont {Chang}}, \bibinfo {author} {\bibfnamefont {H.}~\bibnamefont {Lu}}, \bibinfo {author} {\bibfnamefont {B.}~\bibnamefont {Singh}}, \bibinfo {author} {\bibfnamefont {I.}~\bibnamefont {Belopolski}}, \bibinfo {author} {\bibfnamefont {D.~S.}\ \bibnamefont {Sanchez}}, \bibinfo {author} {\bibfnamefont {X.}~\bibnamefont {Zhang}}, \bibinfo {author} {\bibfnamefont {G.}~\bibnamefont {Bian}}, \bibinfo {author} {\bibfnamefont {H.}~\bibnamefont {Zheng}}, \emph {et~al.},\ }\bibfield  {title} {\bibinfo {title} {Discovery of lorentz-violating type ii weyl fermions in laalge},\ }\href@noop {} {\bibfield  {journal} {\bibinfo  {journal} {Science advances}\ }\textbf {\bibinfo {volume} {3}},\ \bibinfo {pages} {e1603266} (\bibinfo {year} {2017})}\BibitemShut {NoStop}%
\bibitem [{\citenamefont {Wang}\ \emph {et~al.}(2022)\citenamefont {Wang}, \citenamefont {Dong}, \citenamefont {Guo}, \citenamefont {Lv}, \citenamefont {Huang}, \citenamefont {Xiang}, \citenamefont {Ren}, \citenamefont {Wang}, \citenamefont {Sun}, \citenamefont {Li} \emph {et~al.}}]{wang2022ndalsi}%
  \BibitemOpen
  \bibfield  {author} {\bibinfo {author} {\bibfnamefont {J.-F.}\ \bibnamefont {Wang}}, \bibinfo {author} {\bibfnamefont {Q.-X.}\ \bibnamefont {Dong}}, \bibinfo {author} {\bibfnamefont {Z.-P.}\ \bibnamefont {Guo}}, \bibinfo {author} {\bibfnamefont {M.}~\bibnamefont {Lv}}, \bibinfo {author} {\bibfnamefont {Y.-F.}\ \bibnamefont {Huang}}, \bibinfo {author} {\bibfnamefont {J.-S.}\ \bibnamefont {Xiang}}, \bibinfo {author} {\bibfnamefont {Z.-A.}\ \bibnamefont {Ren}}, \bibinfo {author} {\bibfnamefont {Z.-J.}\ \bibnamefont {Wang}}, \bibinfo {author} {\bibfnamefont {P.-J.}\ \bibnamefont {Sun}}, \bibinfo {author} {\bibfnamefont {G.}~\bibnamefont {Li}}, \emph {et~al.},\ }\bibfield  {title} {\bibinfo {title} {Ndalsi: A magnetic weyl semimetal candidate with rich magnetic phases and atypical transport properties},\ }\href@noop {} {\bibfield  {journal} {\bibinfo  {journal} {Physical Review B}\ }\textbf {\bibinfo {volume} {105}},\ \bibinfo {pages} {144435} (\bibinfo {year} {2022})}\BibitemShut {NoStop}%
\bibitem [{\citenamefont {Nikoli{\'c}}(2021{\natexlab{a}})}]{nikolic2021dynamics}%
  \BibitemOpen
  \bibfield  {author} {\bibinfo {author} {\bibfnamefont {P.}~\bibnamefont {Nikoli{\'c}}},\ }\bibfield  {title} {\bibinfo {title} {Dynamics of local magnetic moments induced by itinerant weyl electrons},\ }\href@noop {} {\bibfield  {journal} {\bibinfo  {journal} {Physical Review B}\ }\textbf {\bibinfo {volume} {103}},\ \bibinfo {pages} {155151} (\bibinfo {year} {2021}{\natexlab{a}})}\BibitemShut {NoStop}%
\bibitem [{\citenamefont {Li}\ \emph {et~al.}(2023)\citenamefont {Li}, \citenamefont {Zhang}, \citenamefont {Wang}, \citenamefont {Liu}, \citenamefont {Guo}, \citenamefont {Rienks}, \citenamefont {Chen}, \citenamefont {Bertran}, \citenamefont {Yang}, \citenamefont {Phuyal} \emph {et~al.}}]{li2023emergence}%
  \BibitemOpen
  \bibfield  {author} {\bibinfo {author} {\bibfnamefont {C.}~\bibnamefont {Li}}, \bibinfo {author} {\bibfnamefont {J.}~\bibnamefont {Zhang}}, \bibinfo {author} {\bibfnamefont {Y.}~\bibnamefont {Wang}}, \bibinfo {author} {\bibfnamefont {H.}~\bibnamefont {Liu}}, \bibinfo {author} {\bibfnamefont {Q.}~\bibnamefont {Guo}}, \bibinfo {author} {\bibfnamefont {E.}~\bibnamefont {Rienks}}, \bibinfo {author} {\bibfnamefont {W.}~\bibnamefont {Chen}}, \bibinfo {author} {\bibfnamefont {F.}~\bibnamefont {Bertran}}, \bibinfo {author} {\bibfnamefont {H.}~\bibnamefont {Yang}}, \bibinfo {author} {\bibfnamefont {D.}~\bibnamefont {Phuyal}}, \emph {et~al.},\ }\bibfield  {title} {\bibinfo {title} {Emergence of weyl fermions by ferrimagnetism in a noncentrosymmetric magnetic weyl semimetal},\ }\href@noop {} {\bibfield  {journal} {\bibinfo  {journal} {Nature Communications}\ }\textbf {\bibinfo {volume} {14}},\ \bibinfo {pages} {7185} (\bibinfo {year} {2023})}\BibitemShut {NoStop}%
\bibitem [{\citenamefont {Yamada}\ \emph {et~al.}(2024)\citenamefont {Yamada}, \citenamefont {Nomoto}, \citenamefont {Miyake}, \citenamefont {Terakawa}, \citenamefont {Kikkawa}, \citenamefont {Arita}, \citenamefont {Tokunaga}, \citenamefont {Taguchi}, \citenamefont {Tokura},\ and\ \citenamefont {Hirschberger}}]{yamada2024nernst}%
  \BibitemOpen
  \bibfield  {author} {\bibinfo {author} {\bibfnamefont {R.}~\bibnamefont {Yamada}}, \bibinfo {author} {\bibfnamefont {T.}~\bibnamefont {Nomoto}}, \bibinfo {author} {\bibfnamefont {A.}~\bibnamefont {Miyake}}, \bibinfo {author} {\bibfnamefont {T.}~\bibnamefont {Terakawa}}, \bibinfo {author} {\bibfnamefont {A.}~\bibnamefont {Kikkawa}}, \bibinfo {author} {\bibfnamefont {R.}~\bibnamefont {Arita}}, \bibinfo {author} {\bibfnamefont {M.}~\bibnamefont {Tokunaga}}, \bibinfo {author} {\bibfnamefont {Y.}~\bibnamefont {Taguchi}}, \bibinfo {author} {\bibfnamefont {Y.}~\bibnamefont {Tokura}},\ and\ \bibinfo {author} {\bibfnamefont {M.}~\bibnamefont {Hirschberger}},\ }\bibfield  {title} {\bibinfo {title} {Nernst effect of high-mobility weyl electrons in ndalsi enhanced by a fermi surface nesting instability},\ }\href@noop {} {\bibfield  {journal} {\bibinfo  {journal} {Physical Review X}\ }\textbf {\bibinfo {volume} {14}},\ \bibinfo {pages} {021012} (\bibinfo {year} {2024})}\BibitemShut {NoStop}%
\bibitem [{\citenamefont {Zhang}\ \emph {et~al.}(2024)\citenamefont {Zhang}, \citenamefont {Tu}, \citenamefont {Li}, \citenamefont {Tang}, \citenamefont {Nie}, \citenamefont {Li}, \citenamefont {Li}, \citenamefont {Qi}, \citenamefont {Wu}, \citenamefont {Zhou} \emph {et~al.}}]{zhang2024abnormally}%
  \BibitemOpen
  \bibfield  {author} {\bibinfo {author} {\bibfnamefont {N.}~\bibnamefont {Zhang}}, \bibinfo {author} {\bibfnamefont {D.}~\bibnamefont {Tu}}, \bibinfo {author} {\bibfnamefont {D.}~\bibnamefont {Li}}, \bibinfo {author} {\bibfnamefont {K.}~\bibnamefont {Tang}}, \bibinfo {author} {\bibfnamefont {L.}~\bibnamefont {Nie}}, \bibinfo {author} {\bibfnamefont {H.}~\bibnamefont {Li}}, \bibinfo {author} {\bibfnamefont {H.}~\bibnamefont {Li}}, \bibinfo {author} {\bibfnamefont {T.}~\bibnamefont {Qi}}, \bibinfo {author} {\bibfnamefont {T.}~\bibnamefont {Wu}}, \bibinfo {author} {\bibfnamefont {J.}~\bibnamefont {Zhou}}, \emph {et~al.},\ }\bibfield  {title} {\bibinfo {title} {Abnormally enhanced hall lorenz number in the magnetic weyl semimetal ndalsi},\ }\href@noop {} {\bibfield  {journal} {\bibinfo  {journal} {Nature Communications}\ }\textbf {\bibinfo {volume} {15}},\ \bibinfo {pages} {10255} (\bibinfo {year} {2024})}\BibitemShut {NoStop}%
\bibitem [{\citenamefont {Barnes}\ and\ \citenamefont {Maekawa}(2007)}]{Barnes2007prl}%
  \BibitemOpen
  \bibfield  {author} {\bibinfo {author} {\bibfnamefont {S.~E.}\ \bibnamefont {Barnes}}\ and\ \bibinfo {author} {\bibfnamefont {S.}~\bibnamefont {Maekawa}},\ }\bibfield  {title} {\bibinfo {title} {Generalization of faraday's law to include nonconservative spin forces},\ }\href {https://doi.org/10.1103/PhysRevLett.98.246601} {\bibfield  {journal} {\bibinfo  {journal} {Phys. Rev. Lett.}\ }\textbf {\bibinfo {volume} {98}},\ \bibinfo {pages} {246601} (\bibinfo {year} {2007})}\BibitemShut {NoStop}%
\bibitem [{\citenamefont {Yamada}\ \emph {et~al.}(2026)\citenamefont {Yamada}, \citenamefont {Kurebayashi}, \citenamefont {Fujishiro}, \citenamefont {Okumura}, \citenamefont {Nakamura}, \citenamefont {Yasin}, \citenamefont {Nakajima}, \citenamefont {Yokouchi}, \citenamefont {Kikkawa}, \citenamefont {Taguchi} \emph {et~al.}}]{yamada2026emergent}%
  \BibitemOpen
  \bibfield  {author} {\bibinfo {author} {\bibfnamefont {R.}~\bibnamefont {Yamada}}, \bibinfo {author} {\bibfnamefont {D.}~\bibnamefont {Kurebayashi}}, \bibinfo {author} {\bibfnamefont {Y.}~\bibnamefont {Fujishiro}}, \bibinfo {author} {\bibfnamefont {S.}~\bibnamefont {Okumura}}, \bibinfo {author} {\bibfnamefont {D.}~\bibnamefont {Nakamura}}, \bibinfo {author} {\bibfnamefont {F.~S.}\ \bibnamefont {Yasin}}, \bibinfo {author} {\bibfnamefont {T.}~\bibnamefont {Nakajima}}, \bibinfo {author} {\bibfnamefont {T.}~\bibnamefont {Yokouchi}}, \bibinfo {author} {\bibfnamefont {A.}~\bibnamefont {Kikkawa}}, \bibinfo {author} {\bibfnamefont {Y.}~\bibnamefont {Taguchi}}, \emph {et~al.},\ }\bibfield  {title} {\bibinfo {title} {Emergent electric field induced by dissipative sliding dynamics of domain walls in a weyl magnet},\ }\href@noop {} {\bibfield  {journal} {\bibinfo  {journal} {Nature Physics}\ ,\ \bibinfo {pages} {1}} (\bibinfo {year} {2026})}\BibitemShut {NoStop}%
\bibitem [{\citenamefont {Wang}\ \emph {et~al.}(2020)\citenamefont {Wang}, \citenamefont {Guo}, \citenamefont {Wang},\ and\ \citenamefont {Yang}}]{wang2020correlation}%
  \BibitemOpen
  \bibfield  {author} {\bibinfo {author} {\bibfnamefont {T.}~\bibnamefont {Wang}}, \bibinfo {author} {\bibfnamefont {Y.}~\bibnamefont {Guo}}, \bibinfo {author} {\bibfnamefont {C.}~\bibnamefont {Wang}},\ and\ \bibinfo {author} {\bibfnamefont {S.}~\bibnamefont {Yang}},\ }\bibfield  {title} {\bibinfo {title} {Correlation between non-centrosymmetic structure and magnetic properties in weyl semimetal ndalge},\ }\href@noop {} {\bibfield  {journal} {\bibinfo  {journal} {Solid State Communications}\ }\textbf {\bibinfo {volume} {321}},\ \bibinfo {pages} {114041} (\bibinfo {year} {2020})}\BibitemShut {NoStop}%
\bibitem [{\citenamefont {Zhao}\ \emph {et~al.}(2022)\citenamefont {Zhao}, \citenamefont {Liu}, \citenamefont {Rahman}, \citenamefont {Meng}, \citenamefont {Ling}, \citenamefont {Xi}, \citenamefont {Tong}, \citenamefont {Bai}, \citenamefont {Tian}, \citenamefont {Zhong} \emph {et~al.}}]{zhao2022field}%
  \BibitemOpen
  \bibfield  {author} {\bibinfo {author} {\bibfnamefont {J.}~\bibnamefont {Zhao}}, \bibinfo {author} {\bibfnamefont {W.}~\bibnamefont {Liu}}, \bibinfo {author} {\bibfnamefont {A.}~\bibnamefont {Rahman}}, \bibinfo {author} {\bibfnamefont {F.}~\bibnamefont {Meng}}, \bibinfo {author} {\bibfnamefont {L.}~\bibnamefont {Ling}}, \bibinfo {author} {\bibfnamefont {C.}~\bibnamefont {Xi}}, \bibinfo {author} {\bibfnamefont {W.}~\bibnamefont {Tong}}, \bibinfo {author} {\bibfnamefont {Y.}~\bibnamefont {Bai}}, \bibinfo {author} {\bibfnamefont {Z.}~\bibnamefont {Tian}}, \bibinfo {author} {\bibfnamefont {Y.}~\bibnamefont {Zhong}}, \emph {et~al.},\ }\bibfield  {title} {\bibinfo {title} {Field-induced tricritical phenomenon and magnetic structures in magnetic weyl semimetal candidate ndalge},\ }\href@noop {} {\bibfield  {journal} {\bibinfo  {journal} {New Journal of Physics}\ }\textbf {\bibinfo {volume} {24}},\ \bibinfo {pages} {013010} (\bibinfo {year} {2022})}\BibitemShut {NoStop}%
\bibitem [{\citenamefont {Yang}\ \emph {et~al.}(2023)\citenamefont {Yang}, \citenamefont {Gaudet}, \citenamefont {Verma}, \citenamefont {Baidya}, \citenamefont {Bahrami}, \citenamefont {Yao}, \citenamefont {Huang}, \citenamefont {DeBeer-Schmitt}, \citenamefont {Aczel}, \citenamefont {Xu} \emph {et~al.}}]{yang2023stripe}%
  \BibitemOpen
  \bibfield  {author} {\bibinfo {author} {\bibfnamefont {H.-Y.}\ \bibnamefont {Yang}}, \bibinfo {author} {\bibfnamefont {J.}~\bibnamefont {Gaudet}}, \bibinfo {author} {\bibfnamefont {R.}~\bibnamefont {Verma}}, \bibinfo {author} {\bibfnamefont {S.}~\bibnamefont {Baidya}}, \bibinfo {author} {\bibfnamefont {F.}~\bibnamefont {Bahrami}}, \bibinfo {author} {\bibfnamefont {X.}~\bibnamefont {Yao}}, \bibinfo {author} {\bibfnamefont {C.-Y.}\ \bibnamefont {Huang}}, \bibinfo {author} {\bibfnamefont {L.}~\bibnamefont {DeBeer-Schmitt}}, \bibinfo {author} {\bibfnamefont {A.~A.}\ \bibnamefont {Aczel}}, \bibinfo {author} {\bibfnamefont {G.}~\bibnamefont {Xu}}, \emph {et~al.},\ }\bibfield  {title} {\bibinfo {title} {Stripe helical magnetism and two regimes of anomalous hall effect in ndalge},\ }\href@noop {} {\bibfield  {journal} {\bibinfo  {journal} {Physical Review Materials}\ }\textbf {\bibinfo {volume} {7}},\ \bibinfo {pages} {034202} (\bibinfo {year} {2023})}\BibitemShut {NoStop}%
\bibitem [{\citenamefont {Dhital}\ \emph {et~al.}(2023)\citenamefont {Dhital}, \citenamefont {Dally}, \citenamefont {Ruvalcaba}, \citenamefont {Gonzalez-Hernandez}, \citenamefont {Guerrero-Sanchez}, \citenamefont {Cao}, \citenamefont {Zhang}, \citenamefont {Tian}, \citenamefont {Wu}, \citenamefont {Frontzek} \emph {et~al.}}]{dhital2023multi}%
  \BibitemOpen
  \bibfield  {author} {\bibinfo {author} {\bibfnamefont {C.}~\bibnamefont {Dhital}}, \bibinfo {author} {\bibfnamefont {R.~L.}\ \bibnamefont {Dally}}, \bibinfo {author} {\bibfnamefont {R.}~\bibnamefont {Ruvalcaba}}, \bibinfo {author} {\bibfnamefont {R.}~\bibnamefont {Gonzalez-Hernandez}}, \bibinfo {author} {\bibfnamefont {J.}~\bibnamefont {Guerrero-Sanchez}}, \bibinfo {author} {\bibfnamefont {H.~B.}\ \bibnamefont {Cao}}, \bibinfo {author} {\bibfnamefont {Q.}~\bibnamefont {Zhang}}, \bibinfo {author} {\bibfnamefont {W.}~\bibnamefont {Tian}}, \bibinfo {author} {\bibfnamefont {Y.}~\bibnamefont {Wu}}, \bibinfo {author} {\bibfnamefont {M.~D.}\ \bibnamefont {Frontzek}}, \emph {et~al.},\ }\bibfield  {title} {\bibinfo {title} {Multi-k magnetic structure and large anomalous hall effect in candidate magnetic weyl semimetal ndalge},\ }\href@noop {} {\bibfield  {journal} {\bibinfo  {journal} {Physical Review B}\ }\textbf {\bibinfo {volume} {107}},\ \bibinfo {pages} {224414} (\bibinfo {year} {2023})}\BibitemShut {NoStop}%
\bibitem [{\citenamefont {Kikugawa}\ \emph {et~al.}(2024)\citenamefont {Kikugawa}, \citenamefont {Uji},\ and\ \citenamefont {Terashima}}]{kikugawa2024anomalous}%
  \BibitemOpen
  \bibfield  {author} {\bibinfo {author} {\bibfnamefont {N.}~\bibnamefont {Kikugawa}}, \bibinfo {author} {\bibfnamefont {S.}~\bibnamefont {Uji}},\ and\ \bibinfo {author} {\bibfnamefont {T.}~\bibnamefont {Terashima}},\ }\bibfield  {title} {\bibinfo {title} {Anomalous hall effect in the magnetic weyl semimetal ndalge with plateaus observed at low temperatures},\ }\href@noop {} {\bibfield  {journal} {\bibinfo  {journal} {Physical Review B}\ }\textbf {\bibinfo {volume} {109}},\ \bibinfo {pages} {035143} (\bibinfo {year} {2024})}\BibitemShut {NoStop}%
\bibitem [{\citenamefont {Meng}\ \emph {et~al.}(2019)\citenamefont {Meng}, \citenamefont {Wu}, \citenamefont {Qiu}, \citenamefont {Wang}, \citenamefont {Liu}, \citenamefont {Xia}, \citenamefont {Yuan}, \citenamefont {Chang},\ and\ \citenamefont {Tian}}]{meng2019large}%
  \BibitemOpen
  \bibfield  {author} {\bibinfo {author} {\bibfnamefont {B.}~\bibnamefont {Meng}}, \bibinfo {author} {\bibfnamefont {H.}~\bibnamefont {Wu}}, \bibinfo {author} {\bibfnamefont {Y.}~\bibnamefont {Qiu}}, \bibinfo {author} {\bibfnamefont {C.}~\bibnamefont {Wang}}, \bibinfo {author} {\bibfnamefont {Y.}~\bibnamefont {Liu}}, \bibinfo {author} {\bibfnamefont {Z.}~\bibnamefont {Xia}}, \bibinfo {author} {\bibfnamefont {S.}~\bibnamefont {Yuan}}, \bibinfo {author} {\bibfnamefont {H.}~\bibnamefont {Chang}},\ and\ \bibinfo {author} {\bibfnamefont {Z.}~\bibnamefont {Tian}},\ }\bibfield  {title} {\bibinfo {title} {Large anomalous hall effect in ferromagnetic weyl semimetal candidate pralge},\ }\href@noop {} {\bibfield  {journal} {\bibinfo  {journal} {APL Materials}\ }\textbf {\bibinfo {volume} {7}} (\bibinfo {year} {2019})}\BibitemShut {NoStop}%
\bibitem [{\citenamefont {Yang}\ \emph {et~al.}(2020{\natexlab{b}})\citenamefont {Yang}, \citenamefont {Singh}, \citenamefont {Lu}, \citenamefont {Huang}, \citenamefont {Bahrami}, \citenamefont {Chiu}, \citenamefont {Graf}, \citenamefont {Huang}, \citenamefont {Wang}, \citenamefont {Lin} \emph {et~al.}}]{yang2020transition}%
  \BibitemOpen
  \bibfield  {author} {\bibinfo {author} {\bibfnamefont {H.-Y.}\ \bibnamefont {Yang}}, \bibinfo {author} {\bibfnamefont {B.}~\bibnamefont {Singh}}, \bibinfo {author} {\bibfnamefont {B.}~\bibnamefont {Lu}}, \bibinfo {author} {\bibfnamefont {C.-Y.}\ \bibnamefont {Huang}}, \bibinfo {author} {\bibfnamefont {F.}~\bibnamefont {Bahrami}}, \bibinfo {author} {\bibfnamefont {W.-C.}\ \bibnamefont {Chiu}}, \bibinfo {author} {\bibfnamefont {D.}~\bibnamefont {Graf}}, \bibinfo {author} {\bibfnamefont {S.-M.}\ \bibnamefont {Huang}}, \bibinfo {author} {\bibfnamefont {B.}~\bibnamefont {Wang}}, \bibinfo {author} {\bibfnamefont {H.}~\bibnamefont {Lin}}, \emph {et~al.},\ }\bibfield  {title} {\bibinfo {title} {Transition from intrinsic to extrinsic anomalous hall effect in the ferromagnetic weyl semimetal pralge1- xsix},\ }\href@noop {} {\bibfield  {journal} {\bibinfo  {journal} {Apl Materials}\ }\textbf {\bibinfo {volume} {8}} (\bibinfo {year} {2020}{\natexlab{b}})}\BibitemShut {NoStop}%
\bibitem [{\citenamefont {Destraz}\ \emph {et~al.}(2020)\citenamefont {Destraz}, \citenamefont {Das}, \citenamefont {Tsirkin}, \citenamefont {Xu}, \citenamefont {Neupert}, \citenamefont {Chang}, \citenamefont {Schilling}, \citenamefont {Grushin}, \citenamefont {Kohlbrecher}, \citenamefont {Keller} \emph {et~al.}}]{destraz2020magnetism}%
  \BibitemOpen
  \bibfield  {author} {\bibinfo {author} {\bibfnamefont {D.}~\bibnamefont {Destraz}}, \bibinfo {author} {\bibfnamefont {L.}~\bibnamefont {Das}}, \bibinfo {author} {\bibfnamefont {S.~S.}\ \bibnamefont {Tsirkin}}, \bibinfo {author} {\bibfnamefont {Y.}~\bibnamefont {Xu}}, \bibinfo {author} {\bibfnamefont {T.}~\bibnamefont {Neupert}}, \bibinfo {author} {\bibfnamefont {J.}~\bibnamefont {Chang}}, \bibinfo {author} {\bibfnamefont {A.}~\bibnamefont {Schilling}}, \bibinfo {author} {\bibfnamefont {A.~G.}\ \bibnamefont {Grushin}}, \bibinfo {author} {\bibfnamefont {J.}~\bibnamefont {Kohlbrecher}}, \bibinfo {author} {\bibfnamefont {L.}~\bibnamefont {Keller}}, \emph {et~al.},\ }\bibfield  {title} {\bibinfo {title} {Magnetism and anomalous transport in the weyl semimetal pralge: possible route to axial gauge fields},\ }\href@noop {} {\bibfield  {journal} {\bibinfo  {journal} {npj Quantum Materials}\ }\textbf {\bibinfo {volume} {5}},\ \bibinfo {pages} {5} (\bibinfo {year} {2020})}\BibitemShut {NoStop}%
\bibitem [{\citenamefont {Sanchez}\ \emph {et~al.}(2020)\citenamefont {Sanchez}, \citenamefont {Chang}, \citenamefont {Belopolski}, \citenamefont {Lu}, \citenamefont {Yin}, \citenamefont {Alidoust}, \citenamefont {Xu}, \citenamefont {Cochran}, \citenamefont {Zhang}, \citenamefont {Bian} \emph {et~al.}}]{sanchez2020observation}%
  \BibitemOpen
  \bibfield  {author} {\bibinfo {author} {\bibfnamefont {D.~S.}\ \bibnamefont {Sanchez}}, \bibinfo {author} {\bibfnamefont {G.}~\bibnamefont {Chang}}, \bibinfo {author} {\bibfnamefont {I.}~\bibnamefont {Belopolski}}, \bibinfo {author} {\bibfnamefont {H.}~\bibnamefont {Lu}}, \bibinfo {author} {\bibfnamefont {J.-X.}\ \bibnamefont {Yin}}, \bibinfo {author} {\bibfnamefont {N.}~\bibnamefont {Alidoust}}, \bibinfo {author} {\bibfnamefont {X.}~\bibnamefont {Xu}}, \bibinfo {author} {\bibfnamefont {T.~A.}\ \bibnamefont {Cochran}}, \bibinfo {author} {\bibfnamefont {X.}~\bibnamefont {Zhang}}, \bibinfo {author} {\bibfnamefont {Y.}~\bibnamefont {Bian}}, \emph {et~al.},\ }\bibfield  {title} {\bibinfo {title} {Observation of weyl fermions in a magnetic non-centrosymmetric crystal},\ }\href@noop {} {\bibfield  {journal} {\bibinfo  {journal} {Nature communications}\ }\textbf {\bibinfo {volume} {11}},\ \bibinfo {pages} {3356} (\bibinfo {year} {2020})}\BibitemShut {NoStop}%
\bibitem [{\citenamefont {Shoriki}\ \emph {et~al.}(2024)\citenamefont {Shoriki}, \citenamefont {Moriishi}, \citenamefont {Okamura}, \citenamefont {Yokoi}, \citenamefont {Usui}, \citenamefont {Murakawa}, \citenamefont {Sakai}, \citenamefont {Hanasaki}, \citenamefont {Tokura},\ and\ \citenamefont {Takahashi}}]{shoriki2024large}%
  \BibitemOpen
  \bibfield  {author} {\bibinfo {author} {\bibfnamefont {K.}~\bibnamefont {Shoriki}}, \bibinfo {author} {\bibfnamefont {K.}~\bibnamefont {Moriishi}}, \bibinfo {author} {\bibfnamefont {Y.}~\bibnamefont {Okamura}}, \bibinfo {author} {\bibfnamefont {K.}~\bibnamefont {Yokoi}}, \bibinfo {author} {\bibfnamefont {H.}~\bibnamefont {Usui}}, \bibinfo {author} {\bibfnamefont {H.}~\bibnamefont {Murakawa}}, \bibinfo {author} {\bibfnamefont {H.}~\bibnamefont {Sakai}}, \bibinfo {author} {\bibfnamefont {N.}~\bibnamefont {Hanasaki}}, \bibinfo {author} {\bibfnamefont {Y.}~\bibnamefont {Tokura}},\ and\ \bibinfo {author} {\bibfnamefont {Y.}~\bibnamefont {Takahashi}},\ }\bibfield  {title} {\bibinfo {title} {Large nonlinear optical magnetoelectric response in a noncentrosymmetric magnetic weyl semimetal},\ }\href@noop {} {\bibfield  {journal} {\bibinfo  {journal} {Proceedings of the National Academy of Sciences}\ }\textbf {\bibinfo {volume} {121}},\ \bibinfo {pages} {e2316910121} (\bibinfo {year} {2024})}\BibitemShut {NoStop}%
\bibitem [{\citenamefont {Yang}\ \emph {et~al.}(2021)\citenamefont {Yang}, \citenamefont {Singh}, \citenamefont {Gaudet}, \citenamefont {Lu}, \citenamefont {Huang}, \citenamefont {Chiu}, \citenamefont {Huang}, \citenamefont {Wang}, \citenamefont {Bahrami}, \citenamefont {Xu} \emph {et~al.}}]{yang2021noncollinear}%
  \BibitemOpen
  \bibfield  {author} {\bibinfo {author} {\bibfnamefont {H.-Y.}\ \bibnamefont {Yang}}, \bibinfo {author} {\bibfnamefont {B.}~\bibnamefont {Singh}}, \bibinfo {author} {\bibfnamefont {J.}~\bibnamefont {Gaudet}}, \bibinfo {author} {\bibfnamefont {B.}~\bibnamefont {Lu}}, \bibinfo {author} {\bibfnamefont {C.-Y.}\ \bibnamefont {Huang}}, \bibinfo {author} {\bibfnamefont {W.-C.}\ \bibnamefont {Chiu}}, \bibinfo {author} {\bibfnamefont {S.-M.}\ \bibnamefont {Huang}}, \bibinfo {author} {\bibfnamefont {B.}~\bibnamefont {Wang}}, \bibinfo {author} {\bibfnamefont {F.}~\bibnamefont {Bahrami}}, \bibinfo {author} {\bibfnamefont {B.}~\bibnamefont {Xu}}, \emph {et~al.},\ }\bibfield  {title} {\bibinfo {title} {Noncollinear ferromagnetic weyl semimetal with anisotropic anomalous hall effect},\ }\href@noop {} {\bibfield  {journal} {\bibinfo  {journal} {Physical Review B}\ }\textbf {\bibinfo {volume} {103}},\ \bibinfo {pages} {115143} (\bibinfo {year} {2021})}\BibitemShut {NoStop}%
\bibitem [{\citenamefont {Sakhya}\ \emph {et~al.}(2023)\citenamefont {Sakhya}, \citenamefont {Huang}, \citenamefont {Dhakal}, \citenamefont {Gao}, \citenamefont {Regmi}, \citenamefont {Wang}, \citenamefont {Wen}, \citenamefont {He}, \citenamefont {Yao}, \citenamefont {Smith} \emph {et~al.}}]{sakhya2023observation}%
  \BibitemOpen
  \bibfield  {author} {\bibinfo {author} {\bibfnamefont {A.~P.}\ \bibnamefont {Sakhya}}, \bibinfo {author} {\bibfnamefont {C.-Y.}\ \bibnamefont {Huang}}, \bibinfo {author} {\bibfnamefont {G.}~\bibnamefont {Dhakal}}, \bibinfo {author} {\bibfnamefont {X.-J.}\ \bibnamefont {Gao}}, \bibinfo {author} {\bibfnamefont {S.}~\bibnamefont {Regmi}}, \bibinfo {author} {\bibfnamefont {B.}~\bibnamefont {Wang}}, \bibinfo {author} {\bibfnamefont {W.}~\bibnamefont {Wen}}, \bibinfo {author} {\bibfnamefont {R.-H.}\ \bibnamefont {He}}, \bibinfo {author} {\bibfnamefont {X.}~\bibnamefont {Yao}}, \bibinfo {author} {\bibfnamefont {R.}~\bibnamefont {Smith}}, \emph {et~al.},\ }\bibfield  {title} {\bibinfo {title} {Observation of fermi arcs and weyl nodes in a noncentrosymmetric magnetic weyl semimetal},\ }\href@noop {} {\bibfield  {journal} {\bibinfo  {journal} {Physical Review Materials}\ }\textbf {\bibinfo {volume} {7}},\ \bibinfo {pages} {L051202} (\bibinfo {year} {2023})}\BibitemShut {NoStop}%
\bibitem [{\citenamefont {Cheng}\ \emph {et~al.}(2024)\citenamefont {Cheng}, \citenamefont {Yan}, \citenamefont {Shi}, \citenamefont {Lou}, \citenamefont {Fedorov}, \citenamefont {Behnami}, \citenamefont {Yuan}, \citenamefont {Yang}, \citenamefont {Wang}, \citenamefont {Cheng} \emph {et~al.}}]{cheng2024tunable}%
  \BibitemOpen
  \bibfield  {author} {\bibinfo {author} {\bibfnamefont {E.}~\bibnamefont {Cheng}}, \bibinfo {author} {\bibfnamefont {L.}~\bibnamefont {Yan}}, \bibinfo {author} {\bibfnamefont {X.}~\bibnamefont {Shi}}, \bibinfo {author} {\bibfnamefont {R.}~\bibnamefont {Lou}}, \bibinfo {author} {\bibfnamefont {A.}~\bibnamefont {Fedorov}}, \bibinfo {author} {\bibfnamefont {M.}~\bibnamefont {Behnami}}, \bibinfo {author} {\bibfnamefont {J.}~\bibnamefont {Yuan}}, \bibinfo {author} {\bibfnamefont {P.}~\bibnamefont {Yang}}, \bibinfo {author} {\bibfnamefont {B.}~\bibnamefont {Wang}}, \bibinfo {author} {\bibfnamefont {J.-G.}\ \bibnamefont {Cheng}}, \emph {et~al.},\ }\bibfield  {title} {\bibinfo {title} {Tunable positions of weyl nodes via magnetism and pressure in the ferromagnetic weyl semimetal cealsi},\ }\href@noop {} {\bibfield  {journal} {\bibinfo  {journal} {Nature Communications}\ }\textbf {\bibinfo {volume} {15}},\ \bibinfo {pages} {1467} (\bibinfo {year} {2024})}\BibitemShut {NoStop}%
\bibitem [{\citenamefont {Sun}\ \emph {et~al.}(2021{\natexlab{b}})\citenamefont {Sun}, \citenamefont {Lee}, \citenamefont {Yang}, \citenamefont {Torchinsky}, \citenamefont {Tafti},\ and\ \citenamefont {Orenstein}}]{Sun2021}%
  \BibitemOpen
  \bibfield  {author} {\bibinfo {author} {\bibfnamefont {Y.}~\bibnamefont {Sun}}, \bibinfo {author} {\bibfnamefont {C.}~\bibnamefont {Lee}}, \bibinfo {author} {\bibfnamefont {H.-Y.}\ \bibnamefont {Yang}}, \bibinfo {author} {\bibfnamefont {D.~H.}\ \bibnamefont {Torchinsky}}, \bibinfo {author} {\bibfnamefont {F.}~\bibnamefont {Tafti}},\ and\ \bibinfo {author} {\bibfnamefont {J.}~\bibnamefont {Orenstein}},\ }\bibfield  {title} {\bibinfo {title} {Mapping domain-wall topology in the magnetic weyl semimetal cealsi},\ }\href {https://doi.org/10.1103/PhysRevB.104.235119} {\bibfield  {journal} {\bibinfo  {journal} {Phys. Rev. B}\ }\textbf {\bibinfo {volume} {104}},\ \bibinfo {pages} {235119} (\bibinfo {year} {2021}{\natexlab{b}})}\BibitemShut {NoStop}%
\bibitem [{\citenamefont {Piva}\ \emph {et~al.}(2023{\natexlab{a}})\citenamefont {Piva}, \citenamefont {Souza}, \citenamefont {Brousseau-Couture}, \citenamefont {Sorn}, \citenamefont {Pakuszewski}, \citenamefont {John}, \citenamefont {Adriano}, \citenamefont {C{\^o}t{\'e}}, \citenamefont {Pagliuso}, \citenamefont {Paramekanti} \emph {et~al.}}]{Piva2023}%
  \BibitemOpen
  \bibfield  {author} {\bibinfo {author} {\bibfnamefont {M.~M.}\ \bibnamefont {Piva}}, \bibinfo {author} {\bibfnamefont {J.}~\bibnamefont {Souza}}, \bibinfo {author} {\bibfnamefont {V.}~\bibnamefont {Brousseau-Couture}}, \bibinfo {author} {\bibfnamefont {S.}~\bibnamefont {Sorn}}, \bibinfo {author} {\bibfnamefont {K.}~\bibnamefont {Pakuszewski}}, \bibinfo {author} {\bibfnamefont {J.~K.}\ \bibnamefont {John}}, \bibinfo {author} {\bibfnamefont {C.}~\bibnamefont {Adriano}}, \bibinfo {author} {\bibfnamefont {M.}~\bibnamefont {C{\^o}t{\'e}}}, \bibinfo {author} {\bibfnamefont {P.}~\bibnamefont {Pagliuso}}, \bibinfo {author} {\bibfnamefont {A.}~\bibnamefont {Paramekanti}}, \emph {et~al.},\ }\bibfield  {title} {\bibinfo {title} {Topological features in the ferromagnetic weyl semimetal cealsi: Role of domain walls},\ }\href@noop {} {\bibfield  {journal} {\bibinfo  {journal} {Physical Review Research}\ }\textbf {\bibinfo {volume} {5}},\ \bibinfo {pages} {013068} (\bibinfo {year} {2023}{\natexlab{a}})}\BibitemShut
  {NoStop}%
\bibitem [{\citenamefont {Yao}\ \emph {et~al.}(2023)\citenamefont {Yao}, \citenamefont {Gaudet}, \citenamefont {Verma}, \citenamefont {Graf}, \citenamefont {Yang}, \citenamefont {Bahrami}, \citenamefont {Zhang}, \citenamefont {Aczel}, \citenamefont {Subedi}, \citenamefont {Torchinsky} \emph {et~al.}}]{yao2023large}%
  \BibitemOpen
  \bibfield  {author} {\bibinfo {author} {\bibfnamefont {X.}~\bibnamefont {Yao}}, \bibinfo {author} {\bibfnamefont {J.}~\bibnamefont {Gaudet}}, \bibinfo {author} {\bibfnamefont {R.}~\bibnamefont {Verma}}, \bibinfo {author} {\bibfnamefont {D.~E.}\ \bibnamefont {Graf}}, \bibinfo {author} {\bibfnamefont {H.-Y.}\ \bibnamefont {Yang}}, \bibinfo {author} {\bibfnamefont {F.}~\bibnamefont {Bahrami}}, \bibinfo {author} {\bibfnamefont {R.}~\bibnamefont {Zhang}}, \bibinfo {author} {\bibfnamefont {A.~A.}\ \bibnamefont {Aczel}}, \bibinfo {author} {\bibfnamefont {S.}~\bibnamefont {Subedi}}, \bibinfo {author} {\bibfnamefont {D.~H.}\ \bibnamefont {Torchinsky}}, \emph {et~al.},\ }\bibfield  {title} {\bibinfo {title} {Large topological hall effect and spiral magnetic order in the weyl semimetal smalsi},\ }\href@noop {} {\bibfield  {journal} {\bibinfo  {journal} {Physical Review X}\ }\textbf {\bibinfo {volume} {13}},\ \bibinfo {pages} {011035} (\bibinfo {year} {2023})}\BibitemShut {NoStop}%
\bibitem [{\citenamefont {Puphal}\ \emph {et~al.}(2020)\citenamefont {Puphal}, \citenamefont {Pomjakushin}, \citenamefont {Kanazawa}, \citenamefont {Ukleev}, \citenamefont {Gawryluk}, \citenamefont {Ma}, \citenamefont {Naamneh}, \citenamefont {Plumb}, \citenamefont {Keller}, \citenamefont {Cubitt}, \citenamefont {Pomjakushina},\ and\ \citenamefont {White}}]{Puphal2020}%
  \BibitemOpen
  \bibfield  {author} {\bibinfo {author} {\bibfnamefont {P.}~\bibnamefont {Puphal}}, \bibinfo {author} {\bibfnamefont {V.}~\bibnamefont {Pomjakushin}}, \bibinfo {author} {\bibfnamefont {N.}~\bibnamefont {Kanazawa}}, \bibinfo {author} {\bibfnamefont {V.}~\bibnamefont {Ukleev}}, \bibinfo {author} {\bibfnamefont {D.~J.}\ \bibnamefont {Gawryluk}}, \bibinfo {author} {\bibfnamefont {J.}~\bibnamefont {Ma}}, \bibinfo {author} {\bibfnamefont {M.}~\bibnamefont {Naamneh}}, \bibinfo {author} {\bibfnamefont {N.~C.}\ \bibnamefont {Plumb}}, \bibinfo {author} {\bibfnamefont {L.}~\bibnamefont {Keller}}, \bibinfo {author} {\bibfnamefont {R.}~\bibnamefont {Cubitt}}, \bibinfo {author} {\bibfnamefont {E.}~\bibnamefont {Pomjakushina}},\ and\ \bibinfo {author} {\bibfnamefont {J.~S.}\ \bibnamefont {White}},\ }\bibfield  {title} {\bibinfo {title} {Topological magnetic phase in the candidate weyl semimetal cealge},\ }\href {https://doi.org/10.1103/PhysRevLett.124.017202} {\bibfield  {journal} {\bibinfo  {journal} {Phys. Rev. Lett.}\
  }\textbf {\bibinfo {volume} {124}},\ \bibinfo {pages} {017202} (\bibinfo {year} {2020})}\BibitemShut {NoStop}%
\bibitem [{\citenamefont {Drucker}\ \emph {et~al.}(2023)\citenamefont {Drucker}, \citenamefont {Nguyen}, \citenamefont {Han}, \citenamefont {Siriviboon}, \citenamefont {Luo}, \citenamefont {Andrejevic}, \citenamefont {Zhu}, \citenamefont {Bednik}, \citenamefont {Nguyen}, \citenamefont {Chen} \emph {et~al.}}]{drucker2023topology}%
  \BibitemOpen
  \bibfield  {author} {\bibinfo {author} {\bibfnamefont {N.~C.}\ \bibnamefont {Drucker}}, \bibinfo {author} {\bibfnamefont {T.}~\bibnamefont {Nguyen}}, \bibinfo {author} {\bibfnamefont {F.}~\bibnamefont {Han}}, \bibinfo {author} {\bibfnamefont {P.}~\bibnamefont {Siriviboon}}, \bibinfo {author} {\bibfnamefont {X.}~\bibnamefont {Luo}}, \bibinfo {author} {\bibfnamefont {N.}~\bibnamefont {Andrejevic}}, \bibinfo {author} {\bibfnamefont {Z.}~\bibnamefont {Zhu}}, \bibinfo {author} {\bibfnamefont {G.}~\bibnamefont {Bednik}}, \bibinfo {author} {\bibfnamefont {Q.~T.}\ \bibnamefont {Nguyen}}, \bibinfo {author} {\bibfnamefont {Z.}~\bibnamefont {Chen}}, \emph {et~al.},\ }\bibfield  {title} {\bibinfo {title} {Topology stabilized fluctuations in a magnetic nodal semimetal},\ }\href@noop {} {\bibfield  {journal} {\bibinfo  {journal} {Nature communications}\ }\textbf {\bibinfo {volume} {14}},\ \bibinfo {pages} {5182} (\bibinfo {year} {2023})}\BibitemShut {NoStop}%
\bibitem [{\citenamefont {Piva}\ \emph {et~al.}(2023{\natexlab{b}})\citenamefont {Piva}, \citenamefont {Souza}, \citenamefont {Lombardi}, \citenamefont {Pakuszewski}, \citenamefont {Adriano}, \citenamefont {Pagliuso},\ and\ \citenamefont {Nicklas}}]{piva2023topological}%
  \BibitemOpen
  \bibfield  {author} {\bibinfo {author} {\bibfnamefont {M.}~\bibnamefont {Piva}}, \bibinfo {author} {\bibfnamefont {J.}~\bibnamefont {Souza}}, \bibinfo {author} {\bibfnamefont {G.}~\bibnamefont {Lombardi}}, \bibinfo {author} {\bibfnamefont {K.}~\bibnamefont {Pakuszewski}}, \bibinfo {author} {\bibfnamefont {C.}~\bibnamefont {Adriano}}, \bibinfo {author} {\bibfnamefont {P.}~\bibnamefont {Pagliuso}},\ and\ \bibinfo {author} {\bibfnamefont {M.}~\bibnamefont {Nicklas}},\ }\bibfield  {title} {\bibinfo {title} {Topological hall effect in cealge},\ }\href@noop {} {\bibfield  {journal} {\bibinfo  {journal} {Physical Review Materials}\ }\textbf {\bibinfo {volume} {7}},\ \bibinfo {pages} {074204} (\bibinfo {year} {2023}{\natexlab{b}})}\BibitemShut {NoStop}%
\bibitem [{\citenamefont {Meguro}\ \emph {et~al.}(2024)\citenamefont {Meguro}, \citenamefont {Ozawa}, \citenamefont {Kobayashi},\ and\ \citenamefont {Nomura}}]{meguro2024effective}%
  \BibitemOpen
  \bibfield  {author} {\bibinfo {author} {\bibfnamefont {T.}~\bibnamefont {Meguro}}, \bibinfo {author} {\bibfnamefont {A.}~\bibnamefont {Ozawa}}, \bibinfo {author} {\bibfnamefont {K.}~\bibnamefont {Kobayashi}},\ and\ \bibinfo {author} {\bibfnamefont {K.}~\bibnamefont {Nomura}},\ }\bibfield  {title} {\bibinfo {title} {Effective tight-binding model of compensated ferrimagnetic weyl semimetal with spontaneous orbital magnetization},\ }\href@noop {} {\bibfield  {journal} {\bibinfo  {journal} {Journal of the Physical Society of Japan}\ }\textbf {\bibinfo {volume} {93}},\ \bibinfo {pages} {034703} (\bibinfo {year} {2024})}\BibitemShut {NoStop}%
\bibitem [{\citenamefont {Skaftouros}\ \emph {et~al.}(2013{\natexlab{a}})\citenamefont {Skaftouros}, \citenamefont {{\"O}zdo{\u{g}}an}, \citenamefont {{\c{S}}a{\c{s}}{\i}o{\u{g}}lu},\ and\ \citenamefont {Galanakis}}]{skaftouros2013search}%
  \BibitemOpen
  \bibfield  {author} {\bibinfo {author} {\bibfnamefont {S.}~\bibnamefont {Skaftouros}}, \bibinfo {author} {\bibfnamefont {K.}~\bibnamefont {{\"O}zdo{\u{g}}an}}, \bibinfo {author} {\bibfnamefont {E.}~\bibnamefont {{\c{S}}a{\c{s}}{\i}o{\u{g}}lu}},\ and\ \bibinfo {author} {\bibfnamefont {I.}~\bibnamefont {Galanakis}},\ }\bibfield  {title} {\bibinfo {title} {Search for spin gapless semiconductors: The case of inverse heusler compounds},\ }\href@noop {} {\bibfield  {journal} {\bibinfo  {journal} {Applied physics letters}\ }\textbf {\bibinfo {volume} {102}} (\bibinfo {year} {2013}{\natexlab{a}})}\BibitemShut {NoStop}%
\bibitem [{\citenamefont {Skaftouros}\ \emph {et~al.}(2013{\natexlab{b}})\citenamefont {Skaftouros}, \citenamefont {{\"O}zdo{\u{g}}an}, \citenamefont {{\c{S}}a{\c{s}}{\i}o{\u{g}}lu},\ and\ \citenamefont {Galanakis}}]{skaftouros2013generalized}%
  \BibitemOpen
  \bibfield  {author} {\bibinfo {author} {\bibfnamefont {S.}~\bibnamefont {Skaftouros}}, \bibinfo {author} {\bibfnamefont {K.}~\bibnamefont {{\"O}zdo{\u{g}}an}}, \bibinfo {author} {\bibfnamefont {E.}~\bibnamefont {{\c{S}}a{\c{s}}{\i}o{\u{g}}lu}},\ and\ \bibinfo {author} {\bibfnamefont {I.}~\bibnamefont {Galanakis}},\ }\bibfield  {title} {\bibinfo {title} {Generalized slater-pauling rule for the inverse heusler compounds},\ }\href@noop {} {\bibfield  {journal} {\bibinfo  {journal} {Physical Review B—Condensed Matter and Materials Physics}\ }\textbf {\bibinfo {volume} {87}},\ \bibinfo {pages} {024420} (\bibinfo {year} {2013}{\natexlab{b}})}\BibitemShut {NoStop}%
\bibitem [{\citenamefont {Fang}\ \emph {et~al.}(2014)\citenamefont {Fang}, \citenamefont {Zhang},\ and\ \citenamefont {Xu}}]{fang2014magnetic}%
  \BibitemOpen
  \bibfield  {author} {\bibinfo {author} {\bibfnamefont {Q.-L.}\ \bibnamefont {Fang}}, \bibinfo {author} {\bibfnamefont {J.-M.}\ \bibnamefont {Zhang}},\ and\ \bibinfo {author} {\bibfnamefont {K.-W.}\ \bibnamefont {Xu}},\ }\bibfield  {title} {\bibinfo {title} {Magnetic properties and origin of the half-metallicity of ti2mnz (z= al, ga, in, si, ge, sn) heusler alloys with the hg2cuti-type structure},\ }\href@noop {} {\bibfield  {journal} {\bibinfo  {journal} {Journal of magnetism and magnetic materials}\ }\textbf {\bibinfo {volume} {349}},\ \bibinfo {pages} {104} (\bibinfo {year} {2014})}\BibitemShut {NoStop}%
\bibitem [{\citenamefont {Feng}\ \emph {et~al.}(2015)\citenamefont {Feng}, \citenamefont {Fu},\ and\ \citenamefont {Wan}}]{feng2015z}%
  \BibitemOpen
  \bibfield  {author} {\bibinfo {author} {\bibfnamefont {W.}~\bibnamefont {Feng}}, \bibinfo {author} {\bibfnamefont {X.}~\bibnamefont {Fu}},\ and\ \bibinfo {author} {\bibfnamefont {C.}~\bibnamefont {Wan}},\ }\bibfield  {title} {\bibinfo {title} {Z. h. yuan, x. han, nv quang, s. cho},\ }\href@noop {} {\bibfield  {journal} {\bibinfo  {journal} {Phys. Status Solidi RRL}\ }\textbf {\bibinfo {volume} {11}},\ \bibinfo {pages} {641} (\bibinfo {year} {2015})}\BibitemShut {NoStop}%
\bibitem [{\citenamefont {Yosida}(1996)}]{yosida1996theory}%
  \BibitemOpen
  \bibfield  {author} {\bibinfo {author} {\bibfnamefont {K.}~\bibnamefont {Yosida}},\ }\href@noop {} {\emph {\bibinfo {title} {Theory of magnetism.: Edition en anglais}}},\ Vol.\ \bibinfo {volume} {122}\ (\bibinfo  {publisher} {Springer Science \& Business Media},\ \bibinfo {year} {1996})\BibitemShut {NoStop}%
\bibitem [{\citenamefont {Stoner}(1938)}]{stoner1938collective}%
  \BibitemOpen
  \bibfield  {author} {\bibinfo {author} {\bibfnamefont {E.~C.}\ \bibnamefont {Stoner}},\ }\bibfield  {title} {\bibinfo {title} {Collective electron ferromagnetism},\ }\href@noop {} {\bibfield  {journal} {\bibinfo  {journal} {Proceedings of the Royal Society of London. Series A. Mathematical and Physical Sciences}\ }\textbf {\bibinfo {volume} {165}},\ \bibinfo {pages} {372} (\bibinfo {year} {1938})}\BibitemShut {NoStop}%
\bibitem [{\citenamefont {Altland}\ and\ \citenamefont {Simons}(2010)}]{altland2010condensed}%
  \BibitemOpen
  \bibfield  {author} {\bibinfo {author} {\bibfnamefont {A.}~\bibnamefont {Altland}}\ and\ \bibinfo {author} {\bibfnamefont {B.~D.}\ \bibnamefont {Simons}},\ }\href@noop {} {\emph {\bibinfo {title} {Condensed matter field theory}}}\ (\bibinfo  {publisher} {Cambridge university press},\ \bibinfo {year} {2010})\BibitemShut {NoStop}%
\bibitem [{\citenamefont {Ozawa}\ and\ \citenamefont {Nomura}(2022)}]{Ozawa2022}%
  \BibitemOpen
  \bibfield  {author} {\bibinfo {author} {\bibfnamefont {A.}~\bibnamefont {Ozawa}}\ and\ \bibinfo {author} {\bibfnamefont {K.}~\bibnamefont {Nomura}},\ }\bibfield  {title} {\bibinfo {title} {Self-consistent analysis of doping effect for magnetic ordering in stacked-kagome weyl system},\ }\href@noop {} {\bibfield  {journal} {\bibinfo  {journal} {Phys.~Rev.~Materials}\ }\textbf {\bibinfo {volume} {6}},\ \bibinfo {pages} {024202} (\bibinfo {year} {2022})}\BibitemShut {NoStop}%
\bibitem [{\citenamefont {Kubodera}\ \emph {et~al.}(2006)\citenamefont {Kubodera}, \citenamefont {Okabe}, \citenamefont {Kamihara},\ and\ \citenamefont {Matoba}}]{kubodera2006ni}%
  \BibitemOpen
  \bibfield  {author} {\bibinfo {author} {\bibfnamefont {T.}~\bibnamefont {Kubodera}}, \bibinfo {author} {\bibfnamefont {H.}~\bibnamefont {Okabe}}, \bibinfo {author} {\bibfnamefont {Y.}~\bibnamefont {Kamihara}},\ and\ \bibinfo {author} {\bibfnamefont {M.}~\bibnamefont {Matoba}},\ }\bibfield  {title} {\bibinfo {title} {Ni substitution effect on magnetic and transport properties in metallic ferromagnet co3sn2s2},\ }\href@noop {} {\bibfield  {journal} {\bibinfo  {journal} {Physica B: Condensed Matter}\ }\textbf {\bibinfo {volume} {378}},\ \bibinfo {pages} {1142} (\bibinfo {year} {2006})}\BibitemShut {NoStop}%
\bibitem [{\citenamefont {Yu}\ \emph {et~al.}(2010)\citenamefont {Yu}, \citenamefont {Zhang}, \citenamefont {Zhang}, \citenamefont {Zhang}, \citenamefont {Dai},\ and\ \citenamefont {Fang}}]{yu2010quantized}%
  \BibitemOpen
  \bibfield  {author} {\bibinfo {author} {\bibfnamefont {R.}~\bibnamefont {Yu}}, \bibinfo {author} {\bibfnamefont {W.}~\bibnamefont {Zhang}}, \bibinfo {author} {\bibfnamefont {H.-J.}\ \bibnamefont {Zhang}}, \bibinfo {author} {\bibfnamefont {S.-C.}\ \bibnamefont {Zhang}}, \bibinfo {author} {\bibfnamefont {X.}~\bibnamefont {Dai}},\ and\ \bibinfo {author} {\bibfnamefont {Z.}~\bibnamefont {Fang}},\ }\bibfield  {title} {\bibinfo {title} {Quantized anomalous hall effect in magnetic topological insulators},\ }\href@noop {} {\bibfield  {journal} {\bibinfo  {journal} {science}\ }\textbf {\bibinfo {volume} {329}},\ \bibinfo {pages} {61} (\bibinfo {year} {2010})}\BibitemShut {NoStop}%
\bibitem [{\citenamefont {Munekata}\ \emph {et~al.}(1989)\citenamefont {Munekata}, \citenamefont {Ohno}, \citenamefont {Von~Molnar}, \citenamefont {Segm{\"u}ller}, \citenamefont {Chang},\ and\ \citenamefont {Esaki}}]{munekata1989diluted}%
  \BibitemOpen
  \bibfield  {author} {\bibinfo {author} {\bibfnamefont {H.}~\bibnamefont {Munekata}}, \bibinfo {author} {\bibfnamefont {H.}~\bibnamefont {Ohno}}, \bibinfo {author} {\bibfnamefont {S.}~\bibnamefont {Von~Molnar}}, \bibinfo {author} {\bibfnamefont {A.}~\bibnamefont {Segm{\"u}ller}}, \bibinfo {author} {\bibfnamefont {L.}~\bibnamefont {Chang}},\ and\ \bibinfo {author} {\bibfnamefont {L.}~\bibnamefont {Esaki}},\ }\bibfield  {title} {\bibinfo {title} {Diluted magnetic iii-v semiconductors},\ }\href@noop {} {\bibfield  {journal} {\bibinfo  {journal} {Physical Review Letters}\ }\textbf {\bibinfo {volume} {63}},\ \bibinfo {pages} {1849} (\bibinfo {year} {1989})}\BibitemShut {NoStop}%
\bibitem [{\citenamefont {Ohno}\ \emph {et~al.}(1991)\citenamefont {Ohno}, \citenamefont {Munekata}, \citenamefont {Von~Moln{\'a}r},\ and\ \citenamefont {Chang}}]{ohno1991new}%
  \BibitemOpen
  \bibfield  {author} {\bibinfo {author} {\bibfnamefont {H.}~\bibnamefont {Ohno}}, \bibinfo {author} {\bibfnamefont {H.}~\bibnamefont {Munekata}}, \bibinfo {author} {\bibfnamefont {S.}~\bibnamefont {Von~Moln{\'a}r}},\ and\ \bibinfo {author} {\bibfnamefont {L.}~\bibnamefont {Chang}},\ }\bibfield  {title} {\bibinfo {title} {New iii-v diluted magnetic semiconductors},\ }\href@noop {} {\bibfield  {journal} {\bibinfo  {journal} {Journal of applied physics}\ }\textbf {\bibinfo {volume} {69}},\ \bibinfo {pages} {6103} (\bibinfo {year} {1991})}\BibitemShut {NoStop}%
\bibitem [{\citenamefont {Ohno}\ \emph {et~al.}(1992)\citenamefont {Ohno}, \citenamefont {Munekata}, \citenamefont {Penney}, \citenamefont {Von~Molnar},\ and\ \citenamefont {Chang}}]{ohno1992magnetotransport}%
  \BibitemOpen
  \bibfield  {author} {\bibinfo {author} {\bibfnamefont {H.}~\bibnamefont {Ohno}}, \bibinfo {author} {\bibfnamefont {H.}~\bibnamefont {Munekata}}, \bibinfo {author} {\bibfnamefont {T.}~\bibnamefont {Penney}}, \bibinfo {author} {\bibfnamefont {S.}~\bibnamefont {Von~Molnar}},\ and\ \bibinfo {author} {\bibfnamefont {L.}~\bibnamefont {Chang}},\ }\bibfield  {title} {\bibinfo {title} {Magnetotransport properties of p-type (in, mn) as diluted magnetic iii-v semiconductors},\ }\href@noop {} {\bibfield  {journal} {\bibinfo  {journal} {Physical Review Letters}\ }\textbf {\bibinfo {volume} {68}},\ \bibinfo {pages} {2664} (\bibinfo {year} {1992})}\BibitemShut {NoStop}%
\bibitem [{\citenamefont {Ohno}\ \emph {et~al.}(1996)\citenamefont {Ohno}, \citenamefont {Shen}, \citenamefont {Matsukura}, \citenamefont {Oiwa}, \citenamefont {Endo}, \citenamefont {Katsumoto},\ and\ \citenamefont {Iye}}]{ohno1996}%
  \BibitemOpen
  \bibfield  {author} {\bibinfo {author} {\bibfnamefont {H.}~\bibnamefont {Ohno}}, \bibinfo {author} {\bibfnamefont {A.}~\bibnamefont {Shen}}, \bibinfo {author} {\bibfnamefont {F.}~\bibnamefont {Matsukura}}, \bibinfo {author} {\bibfnamefont {A.}~\bibnamefont {Oiwa}}, \bibinfo {author} {\bibfnamefont {A.}~\bibnamefont {Endo}}, \bibinfo {author} {\bibfnamefont {S.}~\bibnamefont {Katsumoto}},\ and\ \bibinfo {author} {\bibfnamefont {Y.}~\bibnamefont {Iye}},\ }\bibfield  {title} {\bibinfo {title} {(ga, mn) as: a new diluted magnetic semiconductor based on gaas},\ }\href@noop {} {\bibfield  {journal} {\bibinfo  {journal} {Applied Physics Letters}\ }\textbf {\bibinfo {volume} {69}},\ \bibinfo {pages} {363} (\bibinfo {year} {1996})}\BibitemShut {NoStop}%
\bibitem [{\citenamefont {van Vleck}(1932)}]{vanvleck1932}%
  \BibitemOpen
  \bibfield  {author} {\bibinfo {author} {\bibfnamefont {J.~H.}\ \bibnamefont {van Vleck}},\ }\href@noop {} {\emph {\bibinfo {title} {The Theory of Electric and Magnetic Susceptibilities}}}\ (\bibinfo  {publisher} {Oxford University Press, London},\ \bibinfo {year} {1932})\BibitemShut {NoStop}%
\bibitem [{\citenamefont {Chen}\ \emph {et~al.}(2010)\citenamefont {Chen}, \citenamefont {Chu}, \citenamefont {Analytis}, \citenamefont {Liu}, \citenamefont {Igarashi}, \citenamefont {Kuo}, \citenamefont {Qi}, \citenamefont {Mo}, \citenamefont {Moore}, \citenamefont {Lu} \emph {et~al.}}]{chen2010massive}%
  \BibitemOpen
  \bibfield  {author} {\bibinfo {author} {\bibfnamefont {Y.}~\bibnamefont {Chen}}, \bibinfo {author} {\bibfnamefont {J.-H.}\ \bibnamefont {Chu}}, \bibinfo {author} {\bibfnamefont {J.}~\bibnamefont {Analytis}}, \bibinfo {author} {\bibfnamefont {Z.}~\bibnamefont {Liu}}, \bibinfo {author} {\bibfnamefont {K.}~\bibnamefont {Igarashi}}, \bibinfo {author} {\bibfnamefont {H.-H.}\ \bibnamefont {Kuo}}, \bibinfo {author} {\bibfnamefont {X.}~\bibnamefont {Qi}}, \bibinfo {author} {\bibfnamefont {S.-K.}\ \bibnamefont {Mo}}, \bibinfo {author} {\bibfnamefont {R.}~\bibnamefont {Moore}}, \bibinfo {author} {\bibfnamefont {D.}~\bibnamefont {Lu}}, \emph {et~al.},\ }\bibfield  {title} {\bibinfo {title} {Massive dirac fermion on the surface of a magnetically doped topological insulator},\ }\href@noop {} {\bibfield  {journal} {\bibinfo  {journal} {Science}\ }\textbf {\bibinfo {volume} {329}},\ \bibinfo {pages} {659} (\bibinfo {year} {2010})}\BibitemShut {NoStop}%
\bibitem [{\citenamefont {Chang}\ \emph {et~al.}(2013)\citenamefont {Chang}, \citenamefont {Zhang}, \citenamefont {Feng}, \citenamefont {Shen}, \citenamefont {Zhang}, \citenamefont {Guo}, \citenamefont {Li}, \citenamefont {Ou}, \citenamefont {Wei}, \citenamefont {Wang} \emph {et~al.}}]{chang2013experimental}%
  \BibitemOpen
  \bibfield  {author} {\bibinfo {author} {\bibfnamefont {C.-Z.}\ \bibnamefont {Chang}}, \bibinfo {author} {\bibfnamefont {J.}~\bibnamefont {Zhang}}, \bibinfo {author} {\bibfnamefont {X.}~\bibnamefont {Feng}}, \bibinfo {author} {\bibfnamefont {J.}~\bibnamefont {Shen}}, \bibinfo {author} {\bibfnamefont {Z.}~\bibnamefont {Zhang}}, \bibinfo {author} {\bibfnamefont {M.}~\bibnamefont {Guo}}, \bibinfo {author} {\bibfnamefont {K.}~\bibnamefont {Li}}, \bibinfo {author} {\bibfnamefont {Y.}~\bibnamefont {Ou}}, \bibinfo {author} {\bibfnamefont {P.}~\bibnamefont {Wei}}, \bibinfo {author} {\bibfnamefont {L.-L.}\ \bibnamefont {Wang}}, \emph {et~al.},\ }\bibfield  {title} {\bibinfo {title} {Experimental observation of the quantum anomalous hall effect in a magnetic topological insulator},\ }\href@noop {} {\bibfield  {journal} {\bibinfo  {journal} {Science}\ }\textbf {\bibinfo {volume} {340}},\ \bibinfo {pages} {167} (\bibinfo {year} {2013})}\BibitemShut {NoStop}%
\bibitem [{\citenamefont {Zhang}\ \emph {et~al.}(2013)\citenamefont {Zhang}, \citenamefont {Chang}, \citenamefont {Tang}, \citenamefont {Zhang}, \citenamefont {Feng}, \citenamefont {Li}, \citenamefont {Wang}, \citenamefont {Chen}, \citenamefont {Liu}, \citenamefont {Duan} \emph {et~al.}}]{zhang2013topology}%
  \BibitemOpen
  \bibfield  {author} {\bibinfo {author} {\bibfnamefont {J.}~\bibnamefont {Zhang}}, \bibinfo {author} {\bibfnamefont {C.-Z.}\ \bibnamefont {Chang}}, \bibinfo {author} {\bibfnamefont {P.}~\bibnamefont {Tang}}, \bibinfo {author} {\bibfnamefont {Z.}~\bibnamefont {Zhang}}, \bibinfo {author} {\bibfnamefont {X.}~\bibnamefont {Feng}}, \bibinfo {author} {\bibfnamefont {K.}~\bibnamefont {Li}}, \bibinfo {author} {\bibfnamefont {L.-l.}\ \bibnamefont {Wang}}, \bibinfo {author} {\bibfnamefont {X.}~\bibnamefont {Chen}}, \bibinfo {author} {\bibfnamefont {C.}~\bibnamefont {Liu}}, \bibinfo {author} {\bibfnamefont {W.}~\bibnamefont {Duan}}, \emph {et~al.},\ }\bibfield  {title} {\bibinfo {title} {Topology-driven magnetic quantum phase transition in topological insulators},\ }\href@noop {} {\bibfield  {journal} {\bibinfo  {journal} {Science}\ }\textbf {\bibinfo {volume} {339}},\ \bibinfo {pages} {1582} (\bibinfo {year} {2013})}\BibitemShut {NoStop}%
\bibitem [{\citenamefont {Chang}\ \emph {et~al.}(2015{\natexlab{a}})\citenamefont {Chang}, \citenamefont {Zhao}, \citenamefont {Kim}, \citenamefont {Zhang}, \citenamefont {Assaf}, \citenamefont {Heiman}, \citenamefont {Zhang}, \citenamefont {Liu}, \citenamefont {Chan},\ and\ \citenamefont {Moodera}}]{chang2015high}%
  \BibitemOpen
  \bibfield  {author} {\bibinfo {author} {\bibfnamefont {C.-Z.}\ \bibnamefont {Chang}}, \bibinfo {author} {\bibfnamefont {W.}~\bibnamefont {Zhao}}, \bibinfo {author} {\bibfnamefont {D.~Y.}\ \bibnamefont {Kim}}, \bibinfo {author} {\bibfnamefont {H.}~\bibnamefont {Zhang}}, \bibinfo {author} {\bibfnamefont {B.~A.}\ \bibnamefont {Assaf}}, \bibinfo {author} {\bibfnamefont {D.}~\bibnamefont {Heiman}}, \bibinfo {author} {\bibfnamefont {S.-C.}\ \bibnamefont {Zhang}}, \bibinfo {author} {\bibfnamefont {C.}~\bibnamefont {Liu}}, \bibinfo {author} {\bibfnamefont {M.~H.}\ \bibnamefont {Chan}},\ and\ \bibinfo {author} {\bibfnamefont {J.~S.}\ \bibnamefont {Moodera}},\ }\bibfield  {title} {\bibinfo {title} {High-precision realization of robust quantum anomalous hall state in a hard ferromagnetic topological insulator},\ }\href@noop {} {\bibfield  {journal} {\bibinfo  {journal} {Nature materials}\ }\textbf {\bibinfo {volume} {14}},\ \bibinfo {pages} {473} (\bibinfo {year} {2015}{\natexlab{a}})}\BibitemShut {NoStop}%
\bibitem [{\citenamefont {Li}\ \emph {et~al.}(2015{\natexlab{b}})\citenamefont {Li}, \citenamefont {Chang}, \citenamefont {Wu}, \citenamefont {Tao}, \citenamefont {Zhao}, \citenamefont {Chan}, \citenamefont {Moodera}, \citenamefont {Li},\ and\ \citenamefont {Zhu}}]{Li2015-mj}%
  \BibitemOpen
  \bibfield  {author} {\bibinfo {author} {\bibfnamefont {M.}~\bibnamefont {Li}}, \bibinfo {author} {\bibfnamefont {C.-Z.}\ \bibnamefont {Chang}}, \bibinfo {author} {\bibfnamefont {L.}~\bibnamefont {Wu}}, \bibinfo {author} {\bibfnamefont {J.}~\bibnamefont {Tao}}, \bibinfo {author} {\bibfnamefont {W.}~\bibnamefont {Zhao}}, \bibinfo {author} {\bibfnamefont {M.~H.~W.}\ \bibnamefont {Chan}}, \bibinfo {author} {\bibfnamefont {J.~S.}\ \bibnamefont {Moodera}}, \bibinfo {author} {\bibfnamefont {J.}~\bibnamefont {Li}},\ and\ \bibinfo {author} {\bibfnamefont {Y.}~\bibnamefont {Zhu}},\ }\bibfield  {title} {\bibinfo {title} {Experimental verification of the van vleck nature of long-range ferromagnetic order in the vanadium-doped three-dimensional topological insulator sb(2)te(3)},\ }\href@noop {} {\bibfield  {journal} {\bibinfo  {journal} {Phys. Rev. Lett.}\ }\textbf {\bibinfo {volume} {114}},\ \bibinfo {pages} {146802} (\bibinfo {year} {2015}{\natexlab{b}})}\BibitemShut {NoStop}%
\bibitem [{\citenamefont {Brooks}(1940)}]{brooks1940ferromagnetic}%
  \BibitemOpen
  \bibfield  {author} {\bibinfo {author} {\bibfnamefont {H.}~\bibnamefont {Brooks}},\ }\bibfield  {title} {\bibinfo {title} {Ferromagnetic anisotropy and the itinerant electron model},\ }\href@noop {} {\bibfield  {journal} {\bibinfo  {journal} {Physical Review}\ }\textbf {\bibinfo {volume} {58}},\ \bibinfo {pages} {909} (\bibinfo {year} {1940})}\BibitemShut {NoStop}%
\bibitem [{\citenamefont {Holstein}\ and\ \citenamefont {Primakoff}(1940)}]{holstein1940field}%
  \BibitemOpen
  \bibfield  {author} {\bibinfo {author} {\bibfnamefont {T.}~\bibnamefont {Holstein}}\ and\ \bibinfo {author} {\bibfnamefont {H.}~\bibnamefont {Primakoff}},\ }\bibfield  {title} {\bibinfo {title} {Field dependence of the intrinsic domain magnetization of a ferromagnet},\ }\href@noop {} {\bibfield  {journal} {\bibinfo  {journal} {Physical Review}\ }\textbf {\bibinfo {volume} {58}},\ \bibinfo {pages} {1098} (\bibinfo {year} {1940})}\BibitemShut {NoStop}%
\bibitem [{\citenamefont {Watanabe}\ \emph {et~al.}(2022)\citenamefont {Watanabe}, \citenamefont {Araki}, \citenamefont {Kobayashi}, \citenamefont {Ozawa},\ and\ \citenamefont {Nomura}}]{watanabe2022magnetic}%
  \BibitemOpen
  \bibfield  {author} {\bibinfo {author} {\bibfnamefont {J.}~\bibnamefont {Watanabe}}, \bibinfo {author} {\bibfnamefont {Y.}~\bibnamefont {Araki}}, \bibinfo {author} {\bibfnamefont {K.}~\bibnamefont {Kobayashi}}, \bibinfo {author} {\bibfnamefont {A.}~\bibnamefont {Ozawa}},\ and\ \bibinfo {author} {\bibfnamefont {K.}~\bibnamefont {Nomura}},\ }\bibfield  {title} {\bibinfo {title} {Magnetic orderings from spin--orbit coupled electrons on kagome lattice},\ }\href@noop {} {\bibfield  {journal} {\bibinfo  {journal} {journal of the physical society of japan}\ }\textbf {\bibinfo {volume} {91}},\ \bibinfo {pages} {083702} (\bibinfo {year} {2022})}\BibitemShut {NoStop}%
\bibitem [{\citenamefont {Kanbayasi}(1976)}]{kanbayasi1976magnetocrystalline}%
  \BibitemOpen
  \bibfield  {author} {\bibinfo {author} {\bibfnamefont {A.}~\bibnamefont {Kanbayasi}},\ }\bibfield  {title} {\bibinfo {title} {Magnetocrystalline anisotropy of srruo3},\ }\href@noop {} {\bibfield  {journal} {\bibinfo  {journal} {Journal of the Physical Society of Japan}\ }\textbf {\bibinfo {volume} {41}},\ \bibinfo {pages} {1879} (\bibinfo {year} {1976})}\BibitemShut {NoStop}%
\bibitem [{\citenamefont {Ikhlas}\ \emph {et~al.}(2022)\citenamefont {Ikhlas}, \citenamefont {Dasgupta}, \citenamefont {Theuss}, \citenamefont {Higo}, \citenamefont {Kittaka}, \citenamefont {Ramshaw}, \citenamefont {Tchernyshyov}, \citenamefont {Hicks},\ and\ \citenamefont {Nakatsuji}}]{ikhlas2022piezomagnetic}%
  \BibitemOpen
  \bibfield  {author} {\bibinfo {author} {\bibfnamefont {M.}~\bibnamefont {Ikhlas}}, \bibinfo {author} {\bibfnamefont {S.}~\bibnamefont {Dasgupta}}, \bibinfo {author} {\bibfnamefont {F.}~\bibnamefont {Theuss}}, \bibinfo {author} {\bibfnamefont {T.}~\bibnamefont {Higo}}, \bibinfo {author} {\bibfnamefont {S.}~\bibnamefont {Kittaka}}, \bibinfo {author} {\bibfnamefont {B.}~\bibnamefont {Ramshaw}}, \bibinfo {author} {\bibfnamefont {O.}~\bibnamefont {Tchernyshyov}}, \bibinfo {author} {\bibfnamefont {C.}~\bibnamefont {Hicks}},\ and\ \bibinfo {author} {\bibfnamefont {S.}~\bibnamefont {Nakatsuji}},\ }\bibfield  {title} {\bibinfo {title} {Piezomagnetic switching of the anomalous hall effect in an antiferromagnet at room temperature},\ }\href@noop {} {\bibfield  {journal} {\bibinfo  {journal} {Nature Physics}\ }\textbf {\bibinfo {volume} {18}},\ \bibinfo {pages} {1086} (\bibinfo {year} {2022})}\BibitemShut {NoStop}%
\bibitem [{\citenamefont {Yoon}\ \emph {et~al.}(2023)\citenamefont {Yoon}, \citenamefont {Zhang}, \citenamefont {Chou}, \citenamefont {Takeuchi}, \citenamefont {Uchimura}, \citenamefont {Hou}, \citenamefont {Han}, \citenamefont {Kanai}, \citenamefont {Ohno}, \citenamefont {Fukami} \emph {et~al.}}]{yoon2023handedness}%
  \BibitemOpen
  \bibfield  {author} {\bibinfo {author} {\bibfnamefont {J.-Y.}\ \bibnamefont {Yoon}}, \bibinfo {author} {\bibfnamefont {P.}~\bibnamefont {Zhang}}, \bibinfo {author} {\bibfnamefont {C.-T.}\ \bibnamefont {Chou}}, \bibinfo {author} {\bibfnamefont {Y.}~\bibnamefont {Takeuchi}}, \bibinfo {author} {\bibfnamefont {T.}~\bibnamefont {Uchimura}}, \bibinfo {author} {\bibfnamefont {J.~T.}\ \bibnamefont {Hou}}, \bibinfo {author} {\bibfnamefont {J.}~\bibnamefont {Han}}, \bibinfo {author} {\bibfnamefont {S.}~\bibnamefont {Kanai}}, \bibinfo {author} {\bibfnamefont {H.}~\bibnamefont {Ohno}}, \bibinfo {author} {\bibfnamefont {S.}~\bibnamefont {Fukami}}, \emph {et~al.},\ }\bibfield  {title} {\bibinfo {title} {Handedness anomaly in a non-collinear antiferromagnet under spin--orbit torque},\ }\href@noop {} {\bibfield  {journal} {\bibinfo  {journal} {Nature Materials}\ }\textbf {\bibinfo {volume} {22}},\ \bibinfo {pages} {1106} (\bibinfo {year} {2023})}\BibitemShut {NoStop}%
\bibitem [{\citenamefont {Ruderman}\ and\ \citenamefont {Kittel}(1954)}]{ruderman1954indirect}%
  \BibitemOpen
  \bibfield  {author} {\bibinfo {author} {\bibfnamefont {M.~A.}\ \bibnamefont {Ruderman}}\ and\ \bibinfo {author} {\bibfnamefont {C.}~\bibnamefont {Kittel}},\ }\bibfield  {title} {\bibinfo {title} {Indirect exchange coupling of nuclear magnetic moments by conduction electrons},\ }\href@noop {} {\bibfield  {journal} {\bibinfo  {journal} {Physical Review}\ }\textbf {\bibinfo {volume} {96}},\ \bibinfo {pages} {99} (\bibinfo {year} {1954})}\BibitemShut {NoStop}%
\bibitem [{\citenamefont {Kasuya}(1956)}]{kasuya1956theory}%
  \BibitemOpen
  \bibfield  {author} {\bibinfo {author} {\bibfnamefont {T.}~\bibnamefont {Kasuya}},\ }\bibfield  {title} {\bibinfo {title} {A theory of metallic ferro-and antiferromagnetism on zener's model},\ }\href@noop {} {\bibfield  {journal} {\bibinfo  {journal} {Progress of theoretical physics}\ }\textbf {\bibinfo {volume} {16}},\ \bibinfo {pages} {45} (\bibinfo {year} {1956})}\BibitemShut {NoStop}%
\bibitem [{\citenamefont {Yosida}(1957)}]{yosida1957magnetic}%
  \BibitemOpen
  \bibfield  {author} {\bibinfo {author} {\bibfnamefont {K.}~\bibnamefont {Yosida}},\ }\bibfield  {title} {\bibinfo {title} {Magnetic properties of cu-mn alloys},\ }\href@noop {} {\bibfield  {journal} {\bibinfo  {journal} {Physical Review}\ }\textbf {\bibinfo {volume} {106}},\ \bibinfo {pages} {893} (\bibinfo {year} {1957})}\BibitemShut {NoStop}%
\bibitem [{\citenamefont {Chang}\ \emph {et~al.}(2015{\natexlab{b}})\citenamefont {Chang}, \citenamefont {Zhou}, \citenamefont {Wang}, \citenamefont {Shan},\ and\ \citenamefont {Xiao}}]{Chang2015rkky}%
  \BibitemOpen
  \bibfield  {author} {\bibinfo {author} {\bibfnamefont {H.-R.}\ \bibnamefont {Chang}}, \bibinfo {author} {\bibfnamefont {J.}~\bibnamefont {Zhou}}, \bibinfo {author} {\bibfnamefont {S.-X.}\ \bibnamefont {Wang}}, \bibinfo {author} {\bibfnamefont {W.-Y.}\ \bibnamefont {Shan}},\ and\ \bibinfo {author} {\bibfnamefont {D.}~\bibnamefont {Xiao}},\ }\bibfield  {title} {\bibinfo {title} {Rkky interaction of magnetic impurities in dirac and weyl semimetals},\ }\href {https://doi.org/10.1103/PhysRevB.92.241103} {\bibfield  {journal} {\bibinfo  {journal} {Phys. Rev. B}\ }\textbf {\bibinfo {volume} {92}},\ \bibinfo {pages} {241103} (\bibinfo {year} {2015}{\natexlab{b}})}\BibitemShut {NoStop}%
\bibitem [{\citenamefont {Hosseini}\ and\ \citenamefont {Askari}(2015)}]{hosseini2015ruderman}%
  \BibitemOpen
  \bibfield  {author} {\bibinfo {author} {\bibfnamefont {M.~V.}\ \bibnamefont {Hosseini}}\ and\ \bibinfo {author} {\bibfnamefont {M.}~\bibnamefont {Askari}},\ }\bibfield  {title} {\bibinfo {title} {Ruderman-kittel-kasuya-yosida interaction in weyl semimetals},\ }\href@noop {} {\bibfield  {journal} {\bibinfo  {journal} {Physical Review B}\ }\textbf {\bibinfo {volume} {92}},\ \bibinfo {pages} {224435} (\bibinfo {year} {2015})}\BibitemShut {NoStop}%
\bibitem [{\citenamefont {Araki}\ and\ \citenamefont {Nomura}(2016)}]{araki2016spin}%
  \BibitemOpen
  \bibfield  {author} {\bibinfo {author} {\bibfnamefont {Y.}~\bibnamefont {Araki}}\ and\ \bibinfo {author} {\bibfnamefont {K.}~\bibnamefont {Nomura}},\ }\bibfield  {title} {\bibinfo {title} {Spin textures and spin-wave excitations in doped dirac-weyl semimetals},\ }\href {https://doi.org/10.1103/PhysRevB.93.094438} {\bibfield  {journal} {\bibinfo  {journal} {Phys. Rev. B}\ }\textbf {\bibinfo {volume} {93}},\ \bibinfo {pages} {094438} (\bibinfo {year} {2016})}\BibitemShut {NoStop}%
\bibitem [{\citenamefont {Duan}\ \emph {et~al.}(2019)\citenamefont {Duan}, \citenamefont {Zheng}, \citenamefont {Wang}, \citenamefont {Deng},\ and\ \citenamefont {Yang}}]{duan2019signature}%
  \BibitemOpen
  \bibfield  {author} {\bibinfo {author} {\bibfnamefont {H.-J.}\ \bibnamefont {Duan}}, \bibinfo {author} {\bibfnamefont {S.-H.}\ \bibnamefont {Zheng}}, \bibinfo {author} {\bibfnamefont {R.-Q.}\ \bibnamefont {Wang}}, \bibinfo {author} {\bibfnamefont {M.-X.}\ \bibnamefont {Deng}},\ and\ \bibinfo {author} {\bibfnamefont {M.}~\bibnamefont {Yang}},\ }\bibfield  {title} {\bibinfo {title} {{Signature of indirect magnetic interaction in the crossover from type-I to type-II Weyl semimetals}},\ }\href@noop {} {\bibfield  {journal} {\bibinfo  {journal} {Physical Review B}\ }\textbf {\bibinfo {volume} {99}},\ \bibinfo {pages} {165111} (\bibinfo {year} {2019})}\BibitemShut {NoStop}%
\bibitem [{\citenamefont {Bloch}(1930)}]{bloch1930theorie}%
  \BibitemOpen
  \bibfield  {author} {\bibinfo {author} {\bibfnamefont {F.}~\bibnamefont {Bloch}},\ }\bibfield  {title} {\bibinfo {title} {Zur theorie des ferromagnetismus},\ }\href@noop {} {\bibfield  {journal} {\bibinfo  {journal} {Zeitschrift f{\"u}r Physik}\ }\textbf {\bibinfo {volume} {61}},\ \bibinfo {pages} {206} (\bibinfo {year} {1930})}\BibitemShut {NoStop}%
\bibitem [{\citenamefont {Chumak}\ \emph {et~al.}(2015)\citenamefont {Chumak}, \citenamefont {Vasyuchka}, \citenamefont {Serga},\ and\ \citenamefont {Hillebrands}}]{chumak2015magnon}%
  \BibitemOpen
  \bibfield  {author} {\bibinfo {author} {\bibfnamefont {A.~V.}\ \bibnamefont {Chumak}}, \bibinfo {author} {\bibfnamefont {V.~I.}\ \bibnamefont {Vasyuchka}}, \bibinfo {author} {\bibfnamefont {A.~A.}\ \bibnamefont {Serga}},\ and\ \bibinfo {author} {\bibfnamefont {B.}~\bibnamefont {Hillebrands}},\ }\bibfield  {title} {\bibinfo {title} {Magnon spintronics},\ }\href@noop {} {\bibfield  {journal} {\bibinfo  {journal} {Nature physics}\ }\textbf {\bibinfo {volume} {11}},\ \bibinfo {pages} {453} (\bibinfo {year} {2015})}\BibitemShut {NoStop}%
\bibitem [{\citenamefont {Nikoli{\'c}}(2021{\natexlab{b}})}]{nikolic2021universal}%
  \BibitemOpen
  \bibfield  {author} {\bibinfo {author} {\bibfnamefont {P.}~\bibnamefont {Nikoli{\'c}}},\ }\bibfield  {title} {\bibinfo {title} {Universal spin wave damping in magnetic weyl semimetals},\ }\href@noop {} {\bibfield  {journal} {\bibinfo  {journal} {Physical Review B}\ }\textbf {\bibinfo {volume} {104}},\ \bibinfo {pages} {024414} (\bibinfo {year} {2021}{\natexlab{b}})}\BibitemShut {NoStop}%
\bibitem [{\citenamefont {Jenni}\ \emph {et~al.}(2019)\citenamefont {Jenni}, \citenamefont {Kunkem{\"o}ller}, \citenamefont {Br{\"u}ning}, \citenamefont {Lorenz}, \citenamefont {Sidis}, \citenamefont {Schneidewind}, \citenamefont {Nugroho}, \citenamefont {Rosch}, \citenamefont {Khomskii},\ and\ \citenamefont {Braden}}]{jenni2019interplay}%
  \BibitemOpen
  \bibfield  {author} {\bibinfo {author} {\bibfnamefont {K.}~\bibnamefont {Jenni}}, \bibinfo {author} {\bibfnamefont {S.}~\bibnamefont {Kunkem{\"o}ller}}, \bibinfo {author} {\bibfnamefont {D.}~\bibnamefont {Br{\"u}ning}}, \bibinfo {author} {\bibfnamefont {T.}~\bibnamefont {Lorenz}}, \bibinfo {author} {\bibfnamefont {Y.}~\bibnamefont {Sidis}}, \bibinfo {author} {\bibfnamefont {A.}~\bibnamefont {Schneidewind}}, \bibinfo {author} {\bibfnamefont {A.~A.}\ \bibnamefont {Nugroho}}, \bibinfo {author} {\bibfnamefont {A.}~\bibnamefont {Rosch}}, \bibinfo {author} {\bibfnamefont {D.~I.}\ \bibnamefont {Khomskii}},\ and\ \bibinfo {author} {\bibfnamefont {M.}~\bibnamefont {Braden}},\ }\bibfield  {title} {\bibinfo {title} {Interplay of electronic and spin degrees in ferromagnetic \ce{SrRuO3}: Anomalous softening of the magnon gap and stiffness},\ }\href@noop {} {\bibfield  {journal} {\bibinfo  {journal} {Physical review letters}\ }\textbf {\bibinfo {volume} {123}},\ \bibinfo {pages} {017202} (\bibinfo {year}
  {2019})}\BibitemShut {NoStop}%
\bibitem [{\citenamefont {Liu}\ \emph {et~al.}(2021)\citenamefont {Liu}, \citenamefont {Shen}, \citenamefont {Gao}, \citenamefont {Yi}, \citenamefont {Liu}, \citenamefont {Xie}, \citenamefont {Yang}, \citenamefont {Danilkin}, \citenamefont {Deng}, \citenamefont {Wang} \emph {et~al.}}]{liu2021spin}%
  \BibitemOpen
  \bibfield  {author} {\bibinfo {author} {\bibfnamefont {C.}~\bibnamefont {Liu}}, \bibinfo {author} {\bibfnamefont {J.}~\bibnamefont {Shen}}, \bibinfo {author} {\bibfnamefont {J.}~\bibnamefont {Gao}}, \bibinfo {author} {\bibfnamefont {C.}~\bibnamefont {Yi}}, \bibinfo {author} {\bibfnamefont {D.}~\bibnamefont {Liu}}, \bibinfo {author} {\bibfnamefont {T.}~\bibnamefont {Xie}}, \bibinfo {author} {\bibfnamefont {L.}~\bibnamefont {Yang}}, \bibinfo {author} {\bibfnamefont {S.}~\bibnamefont {Danilkin}}, \bibinfo {author} {\bibfnamefont {G.}~\bibnamefont {Deng}}, \bibinfo {author} {\bibfnamefont {W.}~\bibnamefont {Wang}}, \emph {et~al.},\ }\bibfield  {title} {\bibinfo {title} {Spin excitations and spin wave gap in the ferromagnetic weyl semimetal co3sn2s2},\ }\href@noop {} {\bibfield  {journal} {\bibinfo  {journal} {Science China Physics, Mechanics \& Astronomy}\ }\textbf {\bibinfo {volume} {64}},\ \bibinfo {pages} {217062} (\bibinfo {year} {2021})}\BibitemShut {NoStop}%
\bibitem [{\citenamefont {Neubauer}\ \emph {et~al.}(2022)\citenamefont {Neubauer}, \citenamefont {Ye}, \citenamefont {Shi}, \citenamefont {Malinowski}, \citenamefont {Gao}, \citenamefont {Taddei}, \citenamefont {Bourges}, \citenamefont {Ivanov}, \citenamefont {Chu},\ and\ \citenamefont {Dai}}]{neubauer2022spin}%
  \BibitemOpen
  \bibfield  {author} {\bibinfo {author} {\bibfnamefont {K.~J.}\ \bibnamefont {Neubauer}}, \bibinfo {author} {\bibfnamefont {F.}~\bibnamefont {Ye}}, \bibinfo {author} {\bibfnamefont {Y.}~\bibnamefont {Shi}}, \bibinfo {author} {\bibfnamefont {P.}~\bibnamefont {Malinowski}}, \bibinfo {author} {\bibfnamefont {B.}~\bibnamefont {Gao}}, \bibinfo {author} {\bibfnamefont {K.~M.}\ \bibnamefont {Taddei}}, \bibinfo {author} {\bibfnamefont {P.}~\bibnamefont {Bourges}}, \bibinfo {author} {\bibfnamefont {A.}~\bibnamefont {Ivanov}}, \bibinfo {author} {\bibfnamefont {J.-H.}\ \bibnamefont {Chu}},\ and\ \bibinfo {author} {\bibfnamefont {P.}~\bibnamefont {Dai}},\ }\bibfield  {title} {\bibinfo {title} {Spin structure and dynamics of the topological semimetal co3sn2-x in x s2},\ }\href@noop {} {\bibfield  {journal} {\bibinfo  {journal} {npj Quantum Materials}\ }\textbf {\bibinfo {volume} {7}},\ \bibinfo {pages} {112} (\bibinfo {year} {2022})}\BibitemShut {NoStop}%
\bibitem [{\citenamefont {Toyoda}\ \emph {et~al.}(2022)\citenamefont {Toyoda}, \citenamefont {Yamada}, \citenamefont {Kaneko}, \citenamefont {Tokura},\ and\ \citenamefont {Ogawa}}]{toyoda2022weyl}%
  \BibitemOpen
  \bibfield  {author} {\bibinfo {author} {\bibfnamefont {S.}~\bibnamefont {Toyoda}}, \bibinfo {author} {\bibfnamefont {R.}~\bibnamefont {Yamada}}, \bibinfo {author} {\bibfnamefont {Y.}~\bibnamefont {Kaneko}}, \bibinfo {author} {\bibfnamefont {Y.}~\bibnamefont {Tokura}},\ and\ \bibinfo {author} {\bibfnamefont {N.}~\bibnamefont {Ogawa}},\ }\bibfield  {title} {\bibinfo {title} {Weyl fermions in srruo3 detected by brillouin light scattering},\ }\href@noop {} {\bibfield  {journal} {\bibinfo  {journal} {Applied Physics Letters}\ }\textbf {\bibinfo {volume} {120}} (\bibinfo {year} {2022})}\BibitemShut {NoStop}%
\bibitem [{\citenamefont {Parkin}\ \emph {et~al.}(2008)\citenamefont {Parkin}, \citenamefont {Hayashi},\ and\ \citenamefont {Thomas}}]{Parkin2008}%
  \BibitemOpen
  \bibfield  {author} {\bibinfo {author} {\bibfnamefont {S.~S.~P.}\ \bibnamefont {Parkin}}, \bibinfo {author} {\bibfnamefont {M.}~\bibnamefont {Hayashi}},\ and\ \bibinfo {author} {\bibfnamefont {L.}~\bibnamefont {Thomas}},\ }\bibfield  {title} {\bibinfo {title} {Magnetic domain-wall racetrack memory},\ }\href@noop {} {\bibfield  {journal} {\bibinfo  {journal} {Science}\ }\textbf {\bibinfo {volume} {320}},\ \bibinfo {pages} {190} (\bibinfo {year} {2008})}\BibitemShut {NoStop}%
\bibitem [{\citenamefont {Nagaosa}\ and\ \citenamefont {Tokura}(2013)}]{nagaosa2013topological}%
  \BibitemOpen
  \bibfield  {author} {\bibinfo {author} {\bibfnamefont {N.}~\bibnamefont {Nagaosa}}\ and\ \bibinfo {author} {\bibfnamefont {Y.}~\bibnamefont {Tokura}},\ }\bibfield  {title} {\bibinfo {title} {Topological properties and dynamics of magnetic skyrmions},\ }\href@noop {} {\bibfield  {journal} {\bibinfo  {journal} {Nature nanotechnology}\ }\textbf {\bibinfo {volume} {8}},\ \bibinfo {pages} {899} (\bibinfo {year} {2013})}\BibitemShut {NoStop}%
\bibitem [{\citenamefont {Maekawa}\ \emph {et~al.}(2017)\citenamefont {Maekawa}, \citenamefont {Valenzuela},\ and\ \citenamefont {Saitoh}}]{Maekawa2017}%
  \BibitemOpen
  \bibfield  {author} {\bibinfo {author} {\bibfnamefont {S.}~\bibnamefont {Maekawa}}, \bibinfo {author} {\bibfnamefont {S.~O.}\ \bibnamefont {Valenzuela}},\ and\ \bibinfo {author} {\bibfnamefont {E.}~\bibnamefont {Saitoh}},\ }\href@noop {} {\emph {\bibinfo {title} {Spin current}}}\ (\bibinfo  {publisher} {Oxford University Press},\ \bibinfo {year} {2017})\BibitemShut {NoStop}%
\bibitem [{\citenamefont {Lee}\ \emph {et~al.}(2009)\citenamefont {Lee}, \citenamefont {Kang}, \citenamefont {Onose}, \citenamefont {Tokura},\ and\ \citenamefont {Ong}}]{lee2009unusual}%
  \BibitemOpen
  \bibfield  {author} {\bibinfo {author} {\bibfnamefont {M.}~\bibnamefont {Lee}}, \bibinfo {author} {\bibfnamefont {W.}~\bibnamefont {Kang}}, \bibinfo {author} {\bibfnamefont {Y.}~\bibnamefont {Onose}}, \bibinfo {author} {\bibfnamefont {Y.}~\bibnamefont {Tokura}},\ and\ \bibinfo {author} {\bibfnamefont {N.~P.}\ \bibnamefont {Ong}},\ }\bibfield  {title} {\bibinfo {title} {Unusual hall effect anomaly in mnsi under pressure},\ }\href@noop {} {\bibfield  {journal} {\bibinfo  {journal} {Physical review letters}\ }\textbf {\bibinfo {volume} {102}},\ \bibinfo {pages} {186601} (\bibinfo {year} {2009})}\BibitemShut {NoStop}%
\bibitem [{\citenamefont {Neubauer}\ \emph {et~al.}(2009)\citenamefont {Neubauer}, \citenamefont {Pfleiderer}, \citenamefont {Binz}, \citenamefont {Rosch}, \citenamefont {Ritz}, \citenamefont {Niklowitz},\ and\ \citenamefont {B{\"o}ni}}]{neubauer2009topological}%
  \BibitemOpen
  \bibfield  {author} {\bibinfo {author} {\bibfnamefont {A.}~\bibnamefont {Neubauer}}, \bibinfo {author} {\bibfnamefont {C.}~\bibnamefont {Pfleiderer}}, \bibinfo {author} {\bibfnamefont {B.}~\bibnamefont {Binz}}, \bibinfo {author} {\bibfnamefont {A.}~\bibnamefont {Rosch}}, \bibinfo {author} {\bibfnamefont {R.}~\bibnamefont {Ritz}}, \bibinfo {author} {\bibfnamefont {P.}~\bibnamefont {Niklowitz}},\ and\ \bibinfo {author} {\bibfnamefont {P.}~\bibnamefont {B{\"o}ni}},\ }\bibfield  {title} {\bibinfo {title} {Topological hall effect in the a phase of mnsi},\ }\href@noop {} {\bibfield  {journal} {\bibinfo  {journal} {Physical review letters}\ }\textbf {\bibinfo {volume} {102}},\ \bibinfo {pages} {186602} (\bibinfo {year} {2009})}\BibitemShut {NoStop}%
\bibitem [{\citenamefont {Kanazawa}\ \emph {et~al.}(2011)\citenamefont {Kanazawa}, \citenamefont {Onose}, \citenamefont {Arima}, \citenamefont {Okuyama}, \citenamefont {Ohoyama}, \citenamefont {Wakimoto}, \citenamefont {Kakurai}, \citenamefont {Ishiwata},\ and\ \citenamefont {Tokura}}]{kanazawa2011large}%
  \BibitemOpen
  \bibfield  {author} {\bibinfo {author} {\bibfnamefont {N.}~\bibnamefont {Kanazawa}}, \bibinfo {author} {\bibfnamefont {Y.}~\bibnamefont {Onose}}, \bibinfo {author} {\bibfnamefont {T.}~\bibnamefont {Arima}}, \bibinfo {author} {\bibfnamefont {D.}~\bibnamefont {Okuyama}}, \bibinfo {author} {\bibfnamefont {K.}~\bibnamefont {Ohoyama}}, \bibinfo {author} {\bibfnamefont {S.}~\bibnamefont {Wakimoto}}, \bibinfo {author} {\bibfnamefont {K.}~\bibnamefont {Kakurai}}, \bibinfo {author} {\bibfnamefont {S.}~\bibnamefont {Ishiwata}},\ and\ \bibinfo {author} {\bibfnamefont {Y.}~\bibnamefont {Tokura}},\ }\bibfield  {title} {\bibinfo {title} {Large topological hall effect in a short-period helimagnet mnge},\ }\href@noop {} {\bibfield  {journal} {\bibinfo  {journal} {Physical review letters}\ }\textbf {\bibinfo {volume} {106}},\ \bibinfo {pages} {156603} (\bibinfo {year} {2011})}\BibitemShut {NoStop}%
\bibitem [{\citenamefont {Slonczewski}(1989)}]{Slonczewski1989prb}%
  \BibitemOpen
  \bibfield  {author} {\bibinfo {author} {\bibfnamefont {J.~C.}\ \bibnamefont {Slonczewski}},\ }\bibfield  {title} {\bibinfo {title} {Conductance and exchange coupling of two ferromagnets separated by a tunneling barrier},\ }\href {https://doi.org/10.1103/PhysRevB.39.6995} {\bibfield  {journal} {\bibinfo  {journal} {Phys. Rev. B}\ }\textbf {\bibinfo {volume} {39}},\ \bibinfo {pages} {6995} (\bibinfo {year} {1989})}\BibitemShut {NoStop}%
\bibitem [{\citenamefont {Berger}(1996)}]{Berger1996prb}%
  \BibitemOpen
  \bibfield  {author} {\bibinfo {author} {\bibfnamefont {L.}~\bibnamefont {Berger}},\ }\bibfield  {title} {\bibinfo {title} {Emission of spin waves by a magnetic multilayer traversed by a current},\ }\href {https://doi.org/10.1103/PhysRevB.54.9353} {\bibfield  {journal} {\bibinfo  {journal} {Phys. Rev. B}\ }\textbf {\bibinfo {volume} {54}},\ \bibinfo {pages} {9353} (\bibinfo {year} {1996})}\BibitemShut {NoStop}%
\bibitem [{\citenamefont {Tserkovnyak}\ \emph {et~al.}(2005)\citenamefont {Tserkovnyak}, \citenamefont {Brataas}, \citenamefont {Bauer},\ and\ \citenamefont {Halperin}}]{Tserkovnyak2005rmp}%
  \BibitemOpen
  \bibfield  {author} {\bibinfo {author} {\bibfnamefont {Y.}~\bibnamefont {Tserkovnyak}}, \bibinfo {author} {\bibfnamefont {A.}~\bibnamefont {Brataas}}, \bibinfo {author} {\bibfnamefont {G.~E.~W.}\ \bibnamefont {Bauer}},\ and\ \bibinfo {author} {\bibfnamefont {B.~I.}\ \bibnamefont {Halperin}},\ }\bibfield  {title} {\bibinfo {title} {Nonlocal magnetization dynamics in ferromagnetic heterostructures},\ }\href {https://doi.org/10.1103/RevModPhys.77.1375} {\bibfield  {journal} {\bibinfo  {journal} {Rev. Mod. Phys.}\ }\textbf {\bibinfo {volume} {77}},\ \bibinfo {pages} {1375} (\bibinfo {year} {2005})}\BibitemShut {NoStop}%
\bibitem [{\citenamefont {{Volovik}}(1987)}]{Volovik1987jopc}%
  \BibitemOpen
  \bibfield  {author} {\bibinfo {author} {\bibfnamefont {G.~E.}\ \bibnamefont {{Volovik}}},\ }\bibfield  {title} {\bibinfo {title} {{Linear momentum in ferromagnets}},\ }\href {https://doi.org/10.1088/0022-3719/20/7/003} {\bibfield  {journal} {\bibinfo  {journal} {Journal of Physics C Solid State Physics}\ }\textbf {\bibinfo {volume} {20}},\ \bibinfo {pages} {L83} (\bibinfo {year} {1987})}\BibitemShut {NoStop}%
\bibitem [{\citenamefont {Stern}(1992)}]{Stern1992prl}%
  \BibitemOpen
  \bibfield  {author} {\bibinfo {author} {\bibfnamefont {A.}~\bibnamefont {Stern}},\ }\bibfield  {title} {\bibinfo {title} {Berry's phase, motive forces, and mesoscopic conductivity},\ }\href {https://doi.org/10.1103/PhysRevLett.68.1022} {\bibfield  {journal} {\bibinfo  {journal} {Phys. Rev. Lett.}\ }\textbf {\bibinfo {volume} {68}},\ \bibinfo {pages} {1022} (\bibinfo {year} {1992})}\BibitemShut {NoStop}%
\bibitem [{\citenamefont {Nagaosa}\ and\ \citenamefont {Tokura}(2012)}]{nagaosa2012physscr}%
  \BibitemOpen
  \bibfield  {author} {\bibinfo {author} {\bibfnamefont {N.}~\bibnamefont {Nagaosa}}\ and\ \bibinfo {author} {\bibfnamefont {Y.}~\bibnamefont {Tokura}},\ }\bibfield  {title} {\bibinfo {title} {Emergent electromagnetism in solids},\ }\href@noop {} {\bibfield  {journal} {\bibinfo  {journal} {Physica Scripta}\ }\textbf {\bibinfo {volume} {2012}},\ \bibinfo {pages} {014020} (\bibinfo {year} {2012})}\BibitemShut {NoStop}%
\bibitem [{\citenamefont {Tatara}\ \emph {et~al.}(2008)\citenamefont {Tatara}, \citenamefont {Kohno},\ and\ \citenamefont {Shibata}}]{Tatara2008}%
  \BibitemOpen
  \bibfield  {author} {\bibinfo {author} {\bibfnamefont {G.}~\bibnamefont {Tatara}}, \bibinfo {author} {\bibfnamefont {H.}~\bibnamefont {Kohno}},\ and\ \bibinfo {author} {\bibfnamefont {J.}~\bibnamefont {Shibata}},\ }\bibfield  {title} {\bibinfo {title} {Microscopic approach to current-driven domain wall dynamics},\ }\href@noop {} {\bibfield  {journal} {\bibinfo  {journal} {Physics Reports}\ }\textbf {\bibinfo {volume} {468}},\ \bibinfo {pages} {213} (\bibinfo {year} {2008})}\BibitemShut {NoStop}%
\bibitem [{\citenamefont {Howlader}\ \emph {et~al.}(2020)\citenamefont {Howlader}, \citenamefont {Ramachandran}, \citenamefont {Singh}, \citenamefont {Sheet} \emph {et~al.}}]{Howlader2020}%
  \BibitemOpen
  \bibfield  {author} {\bibinfo {author} {\bibfnamefont {S.}~\bibnamefont {Howlader}}, \bibinfo {author} {\bibfnamefont {R.}~\bibnamefont {Ramachandran}}, \bibinfo {author} {\bibfnamefont {Y.}~\bibnamefont {Singh}}, \bibinfo {author} {\bibfnamefont {G.}~\bibnamefont {Sheet}}, \emph {et~al.},\ }\bibfield  {title} {\bibinfo {title} {Domain structure evolution in the ferromagnetic kagome-lattice weyl semimetal co3sn2s2},\ }\href@noop {} {\bibfield  {journal} {\bibinfo  {journal} {Journal of Physics: Condensed Matter}\ }\textbf {\bibinfo {volume} {33}},\ \bibinfo {pages} {075801} (\bibinfo {year} {2020})}\BibitemShut {NoStop}%
\bibitem [{\citenamefont {Li}\ \emph {et~al.}(2022)\citenamefont {Li}, \citenamefont {Ma}, \citenamefont {Zhang}, \citenamefont {Guo}, \citenamefont {Liu}, \citenamefont {Wang}, \citenamefont {Zhou},\ and\ \citenamefont {Shen}}]{Li2022}%
  \BibitemOpen
  \bibfield  {author} {\bibinfo {author} {\bibfnamefont {Q.}~\bibnamefont {Li}}, \bibinfo {author} {\bibfnamefont {H.}~\bibnamefont {Ma}}, \bibinfo {author} {\bibfnamefont {X.}~\bibnamefont {Zhang}}, \bibinfo {author} {\bibfnamefont {Y.}~\bibnamefont {Guo}}, \bibinfo {author} {\bibfnamefont {J.}~\bibnamefont {Liu}}, \bibinfo {author} {\bibfnamefont {W.}~\bibnamefont {Wang}}, \bibinfo {author} {\bibfnamefont {X.}~\bibnamefont {Zhou}},\ and\ \bibinfo {author} {\bibfnamefont {J.}~\bibnamefont {Shen}},\ }\bibfield  {title} {\bibinfo {title} {Magnetic domain structure and domain-wall bound states of the topological semimetal ${\mathrm{eub}}_{6}$},\ }\href {https://doi.org/10.1103/PhysRevB.106.235108} {\bibfield  {journal} {\bibinfo  {journal} {Phys. Rev. B}\ }\textbf {\bibinfo {volume} {106}},\ \bibinfo {pages} {235108} (\bibinfo {year} {2022})}\BibitemShut {NoStop}%
\bibitem [{\citenamefont {Xu}\ \emph {et~al.}(2021)\citenamefont {Xu}, \citenamefont {Franklin}, \citenamefont {Jayakody}, \citenamefont {Yang}, \citenamefont {Tafti},\ and\ \citenamefont {Sochnikov}}]{Xu2021}%
  \BibitemOpen
  \bibfield  {author} {\bibinfo {author} {\bibfnamefont {B.}~\bibnamefont {Xu}}, \bibinfo {author} {\bibfnamefont {J.}~\bibnamefont {Franklin}}, \bibinfo {author} {\bibfnamefont {A.}~\bibnamefont {Jayakody}}, \bibinfo {author} {\bibfnamefont {H.-Y.}\ \bibnamefont {Yang}}, \bibinfo {author} {\bibfnamefont {F.}~\bibnamefont {Tafti}},\ and\ \bibinfo {author} {\bibfnamefont {I.}~\bibnamefont {Sochnikov}},\ }\bibfield  {title} {\bibinfo {title} {Picoscale magnetoelasticity governs heterogeneous magnetic domains in a noncentrosymmetric ferromagnetic weyl semimetal},\ }\href@noop {} {\bibfield  {journal} {\bibinfo  {journal} {Advanced Quantum Technologies}\ }\textbf {\bibinfo {volume} {4}},\ \bibinfo {pages} {2000101} (\bibinfo {year} {2021})}\BibitemShut {NoStop}%
\bibitem [{\citenamefont {Tsukamoto}\ \emph {et~al.}(2025)\citenamefont {Tsukamoto}, \citenamefont {Xu}, \citenamefont {Higo}, \citenamefont {Kondou}, \citenamefont {Sasaki}, \citenamefont {Asakura}, \citenamefont {Sakamoto}, \citenamefont {Gambardella}, \citenamefont {Miwa}, \citenamefont {Otani}, \citenamefont {Nakatsuji}, \citenamefont {Degen},\ and\ \citenamefont {Kobayashi}}]{tsukamoto2025}%
  \BibitemOpen
  \bibfield  {author} {\bibinfo {author} {\bibfnamefont {M.}~\bibnamefont {Tsukamoto}}, \bibinfo {author} {\bibfnamefont {Z.}~\bibnamefont {Xu}}, \bibinfo {author} {\bibfnamefont {T.}~\bibnamefont {Higo}}, \bibinfo {author} {\bibfnamefont {K.}~\bibnamefont {Kondou}}, \bibinfo {author} {\bibfnamefont {K.}~\bibnamefont {Sasaki}}, \bibinfo {author} {\bibfnamefont {M.}~\bibnamefont {Asakura}}, \bibinfo {author} {\bibfnamefont {S.}~\bibnamefont {Sakamoto}}, \bibinfo {author} {\bibfnamefont {P.}~\bibnamefont {Gambardella}}, \bibinfo {author} {\bibfnamefont {S.}~\bibnamefont {Miwa}}, \bibinfo {author} {\bibfnamefont {Y.}~\bibnamefont {Otani}}, \bibinfo {author} {\bibfnamefont {S.}~\bibnamefont {Nakatsuji}}, \bibinfo {author} {\bibfnamefont {C.~L.}\ \bibnamefont {Degen}},\ and\ \bibinfo {author} {\bibfnamefont {K.}~\bibnamefont {Kobayashi}},\ }\bibfield  {title} {\bibinfo {title} {Observation of chiral domain walls in an octupole-ordered antiferromagnet},\ }\href {https://doi.org/10.1103/njnm-nl6n} {\bibfield
  {journal} {\bibinfo  {journal} {Phys. Rev. B}\ }\textbf {\bibinfo {volume} {112}},\ \bibinfo {pages} {L020404} (\bibinfo {year} {2025})}\BibitemShut {NoStop}%
\bibitem [{\citenamefont {Kargarian}\ \emph {et~al.}(2015)\citenamefont {Kargarian}, \citenamefont {Randeria},\ and\ \citenamefont {Trivedi}}]{Kargarian2015-uq}%
  \BibitemOpen
  \bibfield  {author} {\bibinfo {author} {\bibfnamefont {M.}~\bibnamefont {Kargarian}}, \bibinfo {author} {\bibfnamefont {M.}~\bibnamefont {Randeria}},\ and\ \bibinfo {author} {\bibfnamefont {N.}~\bibnamefont {Trivedi}},\ }\bibfield  {title} {\bibinfo {title} {Theory of kerr and faraday rotations and linear dichroism in topological weyl semimetals},\ }\href@noop {} {\bibfield  {journal} {\bibinfo  {journal} {Sci. Rep.}\ }\textbf {\bibinfo {volume} {5}},\ \bibinfo {pages} {12683} (\bibinfo {year} {2015})}\BibitemShut {NoStop}%
\bibitem [{\citenamefont {Cheskis}(2020)}]{Cheskis2020-oi}%
  \BibitemOpen
  \bibfield  {author} {\bibinfo {author} {\bibfnamefont {D.}~\bibnamefont {Cheskis}},\ }\bibfield  {title} {\bibinfo {title} {Magneto-optical tools to study effects in dirac and weyl semimetals},\ }\href@noop {} {\bibfield  {journal} {\bibinfo  {journal} {Symmetry (Basel)}\ }\textbf {\bibinfo {volume} {12}},\ \bibinfo {pages} {1412} (\bibinfo {year} {2020})}\BibitemShut {NoStop}%
\bibitem [{\citenamefont {Yoshikawa}\ \emph {et~al.}(2022{\natexlab{b}})\citenamefont {Yoshikawa}, \citenamefont {Ogawa}, \citenamefont {Hirai}, \citenamefont {Fujiwara}, \citenamefont {Ikeda}, \citenamefont {Tsukazaki},\ and\ \citenamefont {Shimano}}]{yoshikawa2022commphys}%
  \BibitemOpen
  \bibfield  {author} {\bibinfo {author} {\bibfnamefont {N.}~\bibnamefont {Yoshikawa}}, \bibinfo {author} {\bibfnamefont {K.}~\bibnamefont {Ogawa}}, \bibinfo {author} {\bibfnamefont {Y.}~\bibnamefont {Hirai}}, \bibinfo {author} {\bibfnamefont {K.}~\bibnamefont {Fujiwara}}, \bibinfo {author} {\bibfnamefont {J.}~\bibnamefont {Ikeda}}, \bibinfo {author} {\bibfnamefont {A.}~\bibnamefont {Tsukazaki}},\ and\ \bibinfo {author} {\bibfnamefont {R.}~\bibnamefont {Shimano}},\ }\bibfield  {title} {\bibinfo {title} {Non-volatile chirality switching by all-optical magnetization reversal in ferromagnetic weyl semimetal co3sn2s2},\ }\href@noop {} {\bibfield  {journal} {\bibinfo  {journal} {Communications Physics}\ }\textbf {\bibinfo {volume} {5}},\ \bibinfo {pages} {328} (\bibinfo {year} {2022}{\natexlab{b}})}\BibitemShut {NoStop}%
\bibitem [{\citenamefont {Budai}\ \emph {et~al.}(2023)\citenamefont {Budai}, \citenamefont {Isshiki}, \citenamefont {Uesugi}, \citenamefont {Zhu}, \citenamefont {Higo}, \citenamefont {Nakatsuji},\ and\ \citenamefont {Otani}}]{Budai2023}%
  \BibitemOpen
  \bibfield  {author} {\bibinfo {author} {\bibfnamefont {N.}~\bibnamefont {Budai}}, \bibinfo {author} {\bibfnamefont {H.}~\bibnamefont {Isshiki}}, \bibinfo {author} {\bibfnamefont {R.}~\bibnamefont {Uesugi}}, \bibinfo {author} {\bibfnamefont {Z.}~\bibnamefont {Zhu}}, \bibinfo {author} {\bibfnamefont {T.}~\bibnamefont {Higo}}, \bibinfo {author} {\bibfnamefont {S.}~\bibnamefont {Nakatsuji}},\ and\ \bibinfo {author} {\bibfnamefont {Y.}~\bibnamefont {Otani}},\ }\bibfield  {title} {\bibinfo {title} {{High-resolution magnetic imaging by mapping the locally induced anomalous Nernst effect using atomic force microscopy}},\ }\href {https://doi.org/10.1063/5.0136613} {\bibfield  {journal} {\bibinfo  {journal} {Applied Physics Letters}\ }\textbf {\bibinfo {volume} {122}},\ \bibinfo {pages} {102401} (\bibinfo {year} {2023})}\BibitemShut {NoStop}%
\bibitem [{\citenamefont {Fujiwara}\ \emph {et~al.}(2024)\citenamefont {Fujiwara}, \citenamefont {Ogawa}, \citenamefont {Yoshikawa}, \citenamefont {Kobayashi}, \citenamefont {Nomura}, \citenamefont {Shimano},\ and\ \citenamefont {Tsukazaki}}]{Fujiwara2024-my}%
  \BibitemOpen
  \bibfield  {author} {\bibinfo {author} {\bibfnamefont {K.}~\bibnamefont {Fujiwara}}, \bibinfo {author} {\bibfnamefont {K.}~\bibnamefont {Ogawa}}, \bibinfo {author} {\bibfnamefont {N.}~\bibnamefont {Yoshikawa}}, \bibinfo {author} {\bibfnamefont {K.}~\bibnamefont {Kobayashi}}, \bibinfo {author} {\bibfnamefont {K.}~\bibnamefont {Nomura}}, \bibinfo {author} {\bibfnamefont {R.}~\bibnamefont {Shimano}},\ and\ \bibinfo {author} {\bibfnamefont {A.}~\bibnamefont {Tsukazaki}},\ }\bibfield  {title} {\bibinfo {title} {{Giant antisymmetric magnetoresistance arising across optically controlled domain walls in the magnetic Weyl semimetal \ce{Co3Sn2S2}}},\ }\href@noop {} {\bibfield  {journal} {\bibinfo  {journal} {Commun. Mater.}\ }\textbf {\bibinfo {volume} {5}},\ \bibinfo {pages} {239} (\bibinfo {year} {2024})}\BibitemShut {NoStop}%
\bibitem [{\citenamefont {Wang}\ \emph {et~al.}(2023{\natexlab{b}})\citenamefont {Wang}, \citenamefont {Wang}, \citenamefont {Li}, \citenamefont {Zhu}, \citenamefont {Wang},\ and\ \citenamefont {Shi}}]{Wang2023jmmm}%
  \BibitemOpen
  \bibfield  {author} {\bibinfo {author} {\bibfnamefont {S.-Q.}\ \bibnamefont {Wang}}, \bibinfo {author} {\bibfnamefont {M.-Z.}\ \bibnamefont {Wang}}, \bibinfo {author} {\bibfnamefont {Y.-F.}\ \bibnamefont {Li}}, \bibinfo {author} {\bibfnamefont {W.}~\bibnamefont {Zhu}}, \bibinfo {author} {\bibfnamefont {Z.-G.}\ \bibnamefont {Wang}},\ and\ \bibinfo {author} {\bibfnamefont {Z.}~\bibnamefont {Shi}},\ }\bibfield  {title} {\bibinfo {title} {Sign change of intrinsic domain wall resistance in epitaxial heusler alloy co2mnal thin films},\ }\href@noop {} {\bibfield  {journal} {\bibinfo  {journal} {Journal of Magnetism and Magnetic Materials}\ }\textbf {\bibinfo {volume} {579}},\ \bibinfo {pages} {170771} (\bibinfo {year} {2023}{\natexlab{b}})}\BibitemShut {NoStop}%
\bibitem [{\citenamefont {Wu}\ \emph {et~al.}(2020)\citenamefont {Wu}, \citenamefont {Sun}, \citenamefont {Hsieh}, \citenamefont {Chen}, \citenamefont {Kakarla}, \citenamefont {Deng}, \citenamefont {Chu},\ and\ \citenamefont {Yang}}]{Wu2020}%
  \BibitemOpen
  \bibfield  {author} {\bibinfo {author} {\bibfnamefont {H.}~\bibnamefont {Wu}}, \bibinfo {author} {\bibfnamefont {P.}~\bibnamefont {Sun}}, \bibinfo {author} {\bibfnamefont {D.}~\bibnamefont {Hsieh}}, \bibinfo {author} {\bibfnamefont {H.}~\bibnamefont {Chen}}, \bibinfo {author} {\bibfnamefont {D.~C.}\ \bibnamefont {Kakarla}}, \bibinfo {author} {\bibfnamefont {L.}~\bibnamefont {Deng}}, \bibinfo {author} {\bibfnamefont {C.}~\bibnamefont {Chu}},\ and\ \bibinfo {author} {\bibfnamefont {H.}~\bibnamefont {Yang}},\ }\bibfield  {title} {\bibinfo {title} {Observation of skyrmion-like magnetism in magnetic weyl semimetal co3sn2s2},\ }\href {https://doi.org/https://doi.org/10.1016/j.mtphys.2020.100189} {\bibfield  {journal} {\bibinfo  {journal} {Materials Today Physics}\ }\textbf {\bibinfo {volume} {12}},\ \bibinfo {pages} {100189} (\bibinfo {year} {2020})}\BibitemShut {NoStop}%
\bibitem [{\citenamefont {Araki}\ and\ \citenamefont {Nomura}(2018)}]{Araki2018prapplied}%
  \BibitemOpen
  \bibfield  {author} {\bibinfo {author} {\bibfnamefont {Y.}~\bibnamefont {Araki}}\ and\ \bibinfo {author} {\bibfnamefont {K.}~\bibnamefont {Nomura}},\ }\bibfield  {title} {\bibinfo {title} {Charge pumping induced by magnetic texture dynamics in weyl semimetals},\ }\href {https://doi.org/10.1103/PhysRevApplied.10.014007} {\bibfield  {journal} {\bibinfo  {journal} {Phys. Rev. Appl.}\ }\textbf {\bibinfo {volume} {10}},\ \bibinfo {pages} {014007} (\bibinfo {year} {2018})}\BibitemShut {NoStop}%
\bibitem [{\citenamefont {Araki}(2020)}]{Araki2020adp}%
  \BibitemOpen
  \bibfield  {author} {\bibinfo {author} {\bibfnamefont {Y.}~\bibnamefont {Araki}},\ }\bibfield  {title} {\bibinfo {title} {Magnetic textures and dynamics in magnetic weyl semimetals},\ }\href {https://doi.org/https://doi.org/10.1002/andp.201900287} {\bibfield  {journal} {\bibinfo  {journal} {Annalen der Physik}\ }\textbf {\bibinfo {volume} {532}},\ \bibinfo {pages} {1900287} (\bibinfo {year} {2020})}\BibitemShut {NoStop}%
\bibitem [{\citenamefont {Ghimire}\ \emph {et~al.}(2019)\citenamefont {Ghimire}, \citenamefont {Facio}, \citenamefont {You}, \citenamefont {Ye}, \citenamefont {Checkelsky}, \citenamefont {Fang}, \citenamefont {Kaxiras}, \citenamefont {Richter},\ and\ \citenamefont {van~den Brink}}]{Ghimire2019}%
  \BibitemOpen
  \bibfield  {author} {\bibinfo {author} {\bibfnamefont {M.~P.}\ \bibnamefont {Ghimire}}, \bibinfo {author} {\bibfnamefont {J.~I.}\ \bibnamefont {Facio}}, \bibinfo {author} {\bibfnamefont {J.-S.}\ \bibnamefont {You}}, \bibinfo {author} {\bibfnamefont {L.}~\bibnamefont {Ye}}, \bibinfo {author} {\bibfnamefont {J.~G.}\ \bibnamefont {Checkelsky}}, \bibinfo {author} {\bibfnamefont {S.}~\bibnamefont {Fang}}, \bibinfo {author} {\bibfnamefont {E.}~\bibnamefont {Kaxiras}}, \bibinfo {author} {\bibfnamefont {M.}~\bibnamefont {Richter}},\ and\ \bibinfo {author} {\bibfnamefont {J.}~\bibnamefont {van~den Brink}},\ }\bibfield  {title} {\bibinfo {title} {Creating weyl nodes and controlling their energy by magnetization rotation},\ }\href {https://doi.org/10.1103/PhysRevResearch.1.032044} {\bibfield  {journal} {\bibinfo  {journal} {Phys. Rev. Res.}\ }\textbf {\bibinfo {volume} {1}},\ \bibinfo {pages} {032044} (\bibinfo {year} {2019})}\BibitemShut {NoStop}%
\bibitem [{\citenamefont {Araki}\ \emph {et~al.}(2016)\citenamefont {Araki}, \citenamefont {Yoshida},\ and\ \citenamefont {Nomura}}]{Araki2016prb}%
  \BibitemOpen
  \bibfield  {author} {\bibinfo {author} {\bibfnamefont {Y.}~\bibnamefont {Araki}}, \bibinfo {author} {\bibfnamefont {A.}~\bibnamefont {Yoshida}},\ and\ \bibinfo {author} {\bibfnamefont {K.}~\bibnamefont {Nomura}},\ }\bibfield  {title} {\bibinfo {title} {Universal charge and current on magnetic domain walls in weyl semimetals},\ }\href {https://doi.org/10.1103/PhysRevB.94.115312} {\bibfield  {journal} {\bibinfo  {journal} {Phys. Rev. B}\ }\textbf {\bibinfo {volume} {94}},\ \bibinfo {pages} {115312} (\bibinfo {year} {2016})}\BibitemShut {NoStop}%
\bibitem [{\citenamefont {Araki}\ \emph {et~al.}(2018)\citenamefont {Araki}, \citenamefont {Yoshida},\ and\ \citenamefont {Nomura}}]{Araki2018prb}%
  \BibitemOpen
  \bibfield  {author} {\bibinfo {author} {\bibfnamefont {Y.}~\bibnamefont {Araki}}, \bibinfo {author} {\bibfnamefont {A.}~\bibnamefont {Yoshida}},\ and\ \bibinfo {author} {\bibfnamefont {K.}~\bibnamefont {Nomura}},\ }\bibfield  {title} {\bibinfo {title} {Localized charge in various configurations of magnetic domain wall in a weyl semimetal},\ }\href {https://doi.org/10.1103/PhysRevB.98.045302} {\bibfield  {journal} {\bibinfo  {journal} {Phys. Rev. B}\ }\textbf {\bibinfo {volume} {98}},\ \bibinfo {pages} {045302} (\bibinfo {year} {2018})}\BibitemShut {NoStop}%
\bibitem [{\citenamefont {Araki}\ \emph {et~al.}(2021)\citenamefont {Araki}, \citenamefont {Watanabe},\ and\ \citenamefont {Nomura}}]{araki2021jpsj}%
  \BibitemOpen
  \bibfield  {author} {\bibinfo {author} {\bibfnamefont {Y.}~\bibnamefont {Araki}}, \bibinfo {author} {\bibfnamefont {J.}~\bibnamefont {Watanabe}},\ and\ \bibinfo {author} {\bibfnamefont {K.}~\bibnamefont {Nomura}},\ }\bibfield  {title} {\bibinfo {title} {Nodal lines and boundary modes in topological dirac semimetals with magnetism},\ }\href@noop {} {\bibfield  {journal} {\bibinfo  {journal} {journal of the physical society of japan}\ }\textbf {\bibinfo {volume} {90}},\ \bibinfo {pages} {094702} (\bibinfo {year} {2021})}\BibitemShut {NoStop}%
\bibitem [{\citenamefont {Kim}\ \emph {et~al.}(2015)\citenamefont {Kim}, \citenamefont {Wieder}, \citenamefont {Kane},\ and\ \citenamefont {Rappe}}]{Kim2015}%
  \BibitemOpen
  \bibfield  {author} {\bibinfo {author} {\bibfnamefont {Y.}~\bibnamefont {Kim}}, \bibinfo {author} {\bibfnamefont {B.~J.}\ \bibnamefont {Wieder}}, \bibinfo {author} {\bibfnamefont {C.~L.}\ \bibnamefont {Kane}},\ and\ \bibinfo {author} {\bibfnamefont {A.~M.}\ \bibnamefont {Rappe}},\ }\bibfield  {title} {\bibinfo {title} {Dirac line nodes in inversion-symmetric crystals},\ }\href {https://doi.org/10.1103/PhysRevLett.115.036806} {\bibfield  {journal} {\bibinfo  {journal} {Phys. Rev. Lett.}\ }\textbf {\bibinfo {volume} {115}},\ \bibinfo {pages} {036806} (\bibinfo {year} {2015})}\BibitemShut {NoStop}%
\bibitem [{\citenamefont {Chan}\ \emph {et~al.}(2016)\citenamefont {Chan}, \citenamefont {Chiu}, \citenamefont {Chou},\ and\ \citenamefont {Schnyder}}]{Chan2016prb}%
  \BibitemOpen
  \bibfield  {author} {\bibinfo {author} {\bibfnamefont {Y.-H.}\ \bibnamefont {Chan}}, \bibinfo {author} {\bibfnamefont {C.-K.}\ \bibnamefont {Chiu}}, \bibinfo {author} {\bibfnamefont {M.~Y.}\ \bibnamefont {Chou}},\ and\ \bibinfo {author} {\bibfnamefont {A.~P.}\ \bibnamefont {Schnyder}},\ }\bibfield  {title} {\bibinfo {title} {${\mathrm{ca}}_{3}{\mathrm{p}}_{2}$ and other topological semimetals with line nodes and drumhead surface states},\ }\href {https://doi.org/10.1103/PhysRevB.93.205132} {\bibfield  {journal} {\bibinfo  {journal} {Phys. Rev. B}\ }\textbf {\bibinfo {volume} {93}},\ \bibinfo {pages} {205132} (\bibinfo {year} {2016})}\BibitemShut {NoStop}%
\bibitem [{\citenamefont {Yamakage}\ \emph {et~al.}(2016)\citenamefont {Yamakage}, \citenamefont {Yamakawa}, \citenamefont {Tanaka},\ and\ \citenamefont {Okamoto}}]{Yamakage2016jpsj}%
  \BibitemOpen
  \bibfield  {author} {\bibinfo {author} {\bibfnamefont {A.}~\bibnamefont {Yamakage}}, \bibinfo {author} {\bibfnamefont {Y.}~\bibnamefont {Yamakawa}}, \bibinfo {author} {\bibfnamefont {Y.}~\bibnamefont {Tanaka}},\ and\ \bibinfo {author} {\bibfnamefont {Y.}~\bibnamefont {Okamoto}},\ }\bibfield  {title} {\bibinfo {title} {Line-node dirac semimetal and topological insulating phase in noncentrosymmetric pnictides caagx (x = p, as)},\ }\href {https://doi.org/10.7566/JPSJ.85.013708} {\bibfield  {journal} {\bibinfo  {journal} {Journal of the Physical Society of Japan}\ }\textbf {\bibinfo {volume} {85}},\ \bibinfo {pages} {013708} (\bibinfo {year} {2016})}\BibitemShut {NoStop}%
\bibitem [{\citenamefont {Saslow}(2007)}]{Saslow2007prb}%
  \BibitemOpen
  \bibfield  {author} {\bibinfo {author} {\bibfnamefont {W.~M.}\ \bibnamefont {Saslow}},\ }\bibfield  {title} {\bibinfo {title} {Spin pumping of current in non-uniform conducting magnets},\ }\href {https://doi.org/10.1103/PhysRevB.76.184434} {\bibfield  {journal} {\bibinfo  {journal} {Phys. Rev. B}\ }\textbf {\bibinfo {volume} {76}},\ \bibinfo {pages} {184434} (\bibinfo {year} {2007})}\BibitemShut {NoStop}%
\bibitem [{\citenamefont {Duine}(2008)}]{Duine2008prb}%
  \BibitemOpen
  \bibfield  {author} {\bibinfo {author} {\bibfnamefont {R.~A.}\ \bibnamefont {Duine}},\ }\bibfield  {title} {\bibinfo {title} {Spin pumping by a field-driven domain wall},\ }\href {https://doi.org/10.1103/PhysRevB.77.014409} {\bibfield  {journal} {\bibinfo  {journal} {Phys. Rev. B}\ }\textbf {\bibinfo {volume} {77}},\ \bibinfo {pages} {014409} (\bibinfo {year} {2008})}\BibitemShut {NoStop}%
\bibitem [{\citenamefont {Duine}(2009)}]{Duine2009prb}%
  \BibitemOpen
  \bibfield  {author} {\bibinfo {author} {\bibfnamefont {R.~A.}\ \bibnamefont {Duine}},\ }\bibfield  {title} {\bibinfo {title} {Effects of nonadiabaticity on the voltage generated by a moving domain wall},\ }\href {https://doi.org/10.1103/PhysRevB.79.014407} {\bibfield  {journal} {\bibinfo  {journal} {Phys. Rev. B}\ }\textbf {\bibinfo {volume} {79}},\ \bibinfo {pages} {014407} (\bibinfo {year} {2009})}\BibitemShut {NoStop}%
\bibitem [{\citenamefont {Tserkovnyak}\ and\ \citenamefont {Mecklenburg}(2008)}]{Tserkovnyak2008prb}%
  \BibitemOpen
  \bibfield  {author} {\bibinfo {author} {\bibfnamefont {Y.}~\bibnamefont {Tserkovnyak}}\ and\ \bibinfo {author} {\bibfnamefont {M.}~\bibnamefont {Mecklenburg}},\ }\bibfield  {title} {\bibinfo {title} {Electron transport driven by nonequilibrium magnetic textures},\ }\href {https://doi.org/10.1103/PhysRevB.77.134407} {\bibfield  {journal} {\bibinfo  {journal} {Phys. Rev. B}\ }\textbf {\bibinfo {volume} {77}},\ \bibinfo {pages} {134407} (\bibinfo {year} {2008})}\BibitemShut {NoStop}%
\bibitem [{\citenamefont {Ryu}(1996)}]{Ryu1996prl}%
  \BibitemOpen
  \bibfield  {author} {\bibinfo {author} {\bibfnamefont {C.-M.}\ \bibnamefont {Ryu}},\ }\bibfield  {title} {\bibinfo {title} {Spin motive force and faraday law for electrons in mesoscopic rings},\ }\href {https://doi.org/10.1103/PhysRevLett.76.968} {\bibfield  {journal} {\bibinfo  {journal} {Phys. Rev. Lett.}\ }\textbf {\bibinfo {volume} {76}},\ \bibinfo {pages} {968} (\bibinfo {year} {1996})}\BibitemShut {NoStop}%
\bibitem [{\citenamefont {Kim}\ \emph {et~al.}(2012)\citenamefont {Kim}, \citenamefont {Moon}, \citenamefont {Lee},\ and\ \citenamefont {Lee}}]{Kim2012prl}%
  \BibitemOpen
  \bibfield  {author} {\bibinfo {author} {\bibfnamefont {K.-W.}\ \bibnamefont {Kim}}, \bibinfo {author} {\bibfnamefont {J.-H.}\ \bibnamefont {Moon}}, \bibinfo {author} {\bibfnamefont {K.-J.}\ \bibnamefont {Lee}},\ and\ \bibinfo {author} {\bibfnamefont {H.-W.}\ \bibnamefont {Lee}},\ }\bibfield  {title} {\bibinfo {title} {Prediction of giant spin motive force due to rashba spin-orbit coupling},\ }\href {https://doi.org/10.1103/PhysRevLett.108.217202} {\bibfield  {journal} {\bibinfo  {journal} {Phys. Rev. Lett.}\ }\textbf {\bibinfo {volume} {108}},\ \bibinfo {pages} {217202} (\bibinfo {year} {2012})}\BibitemShut {NoStop}%
\bibitem [{\citenamefont {Tatara}\ \emph {et~al.}(2013)\citenamefont {Tatara}, \citenamefont {Nakabayashi},\ and\ \citenamefont {Lee}}]{Tatara2013prb}%
  \BibitemOpen
  \bibfield  {author} {\bibinfo {author} {\bibfnamefont {G.}~\bibnamefont {Tatara}}, \bibinfo {author} {\bibfnamefont {N.}~\bibnamefont {Nakabayashi}},\ and\ \bibinfo {author} {\bibfnamefont {K.-J.}\ \bibnamefont {Lee}},\ }\bibfield  {title} {\bibinfo {title} {Spin motive force induced by rashba interaction in the strong $sd$ coupling regime},\ }\href {https://doi.org/10.1103/PhysRevB.87.054403} {\bibfield  {journal} {\bibinfo  {journal} {Phys. Rev. B}\ }\textbf {\bibinfo {volume} {87}},\ \bibinfo {pages} {054403} (\bibinfo {year} {2013})}\BibitemShut {NoStop}%
\bibitem [{\citenamefont {Yamane}\ \emph {et~al.}(2013)\citenamefont {Yamane}, \citenamefont {Ieda},\ and\ \citenamefont {Maekawa}}]{Yamane2013prb}%
  \BibitemOpen
  \bibfield  {author} {\bibinfo {author} {\bibfnamefont {Y.}~\bibnamefont {Yamane}}, \bibinfo {author} {\bibfnamefont {J.}~\bibnamefont {Ieda}},\ and\ \bibinfo {author} {\bibfnamefont {S.}~\bibnamefont {Maekawa}},\ }\bibfield  {title} {\bibinfo {title} {Spinmotive force with static and uniform magnetization induced by a time-varying electric field},\ }\href {https://doi.org/10.1103/PhysRevB.88.014430} {\bibfield  {journal} {\bibinfo  {journal} {Phys. Rev. B}\ }\textbf {\bibinfo {volume} {88}},\ \bibinfo {pages} {014430} (\bibinfo {year} {2013})}\BibitemShut {NoStop}%
\bibitem [{\citenamefont {Ciccarelli}\ \emph {et~al.}(2015)\citenamefont {Ciccarelli}, \citenamefont {Hals}, \citenamefont {Irvine}, \citenamefont {Novak}, \citenamefont {Tserkovnyak}, \citenamefont {Kurebayashi}, \citenamefont {Brataas},\ and\ \citenamefont {Ferguson}}]{ciccarelli2015magnonic}%
  \BibitemOpen
  \bibfield  {author} {\bibinfo {author} {\bibfnamefont {C.}~\bibnamefont {Ciccarelli}}, \bibinfo {author} {\bibfnamefont {K.~M.}\ \bibnamefont {Hals}}, \bibinfo {author} {\bibfnamefont {A.}~\bibnamefont {Irvine}}, \bibinfo {author} {\bibfnamefont {V.}~\bibnamefont {Novak}}, \bibinfo {author} {\bibfnamefont {Y.}~\bibnamefont {Tserkovnyak}}, \bibinfo {author} {\bibfnamefont {H.}~\bibnamefont {Kurebayashi}}, \bibinfo {author} {\bibfnamefont {A.}~\bibnamefont {Brataas}},\ and\ \bibinfo {author} {\bibfnamefont {A.}~\bibnamefont {Ferguson}},\ }\bibfield  {title} {\bibinfo {title} {Magnonic charge pumping via spin--orbit coupling},\ }\href@noop {} {\bibfield  {journal} {\bibinfo  {journal} {Nature nanotechnology}\ }\textbf {\bibinfo {volume} {10}},\ \bibinfo {pages} {50} (\bibinfo {year} {2015})}\BibitemShut {NoStop}%
\bibitem [{\citenamefont {Nagaosa}(2019)}]{Nagaosa2019jjap}%
  \BibitemOpen
  \bibfield  {author} {\bibinfo {author} {\bibfnamefont {N.}~\bibnamefont {Nagaosa}},\ }\bibfield  {title} {\bibinfo {title} {Emergent inductor by spiral magnets},\ }\href {https://doi.org/10.7567/1347-4065/ab5294} {\bibfield  {journal} {\bibinfo  {journal} {Japanese Journal of Applied Physics}\ }\textbf {\bibinfo {volume} {58}},\ \bibinfo {pages} {120909} (\bibinfo {year} {2019})}\BibitemShut {NoStop}%
\bibitem [{\citenamefont {Kurebayashi}\ and\ \citenamefont {Nagaosa}(2021)}]{kurebayashi2021electromagnetic}%
  \BibitemOpen
  \bibfield  {author} {\bibinfo {author} {\bibfnamefont {D.}~\bibnamefont {Kurebayashi}}\ and\ \bibinfo {author} {\bibfnamefont {N.}~\bibnamefont {Nagaosa}},\ }\bibfield  {title} {\bibinfo {title} {Electromagnetic response in spiral magnets and emergent inductance},\ }\href {https://doi.org/10.1038/s42005-021-00765-3} {\bibfield  {journal} {\bibinfo  {journal} {Communications Physics}\ }\textbf {\bibinfo {volume} {4}},\ \bibinfo {pages} {260} (\bibinfo {year} {2021})}\BibitemShut {NoStop}%
\bibitem [{\citenamefont {Ieda}\ and\ \citenamefont {Yamane}(2021)}]{Ieda2021prb}%
  \BibitemOpen
  \bibfield  {author} {\bibinfo {author} {\bibfnamefont {J.}~\bibnamefont {Ieda}}\ and\ \bibinfo {author} {\bibfnamefont {Y.}~\bibnamefont {Yamane}},\ }\bibfield  {title} {\bibinfo {title} {Intrinsic and extrinsic tunability of rashba spin-orbit coupled emergent inductors},\ }\href {https://doi.org/10.1103/PhysRevB.103.L100402} {\bibfield  {journal} {\bibinfo  {journal} {Phys. Rev. B}\ }\textbf {\bibinfo {volume} {103}},\ \bibinfo {pages} {L100402} (\bibinfo {year} {2021})}\BibitemShut {NoStop}%
\bibitem [{\citenamefont {Yamane}\ \emph {et~al.}(2022)\citenamefont {Yamane}, \citenamefont {Fukami},\ and\ \citenamefont {Ieda}}]{Yamane2022prl}%
  \BibitemOpen
  \bibfield  {author} {\bibinfo {author} {\bibfnamefont {Y.}~\bibnamefont {Yamane}}, \bibinfo {author} {\bibfnamefont {S.}~\bibnamefont {Fukami}},\ and\ \bibinfo {author} {\bibfnamefont {J.}~\bibnamefont {Ieda}},\ }\bibfield  {title} {\bibinfo {title} {Theory of emergent inductance with spin-orbit coupling effects},\ }\href {https://doi.org/10.1103/PhysRevLett.128.147201} {\bibfield  {journal} {\bibinfo  {journal} {Phys. Rev. Lett.}\ }\textbf {\bibinfo {volume} {128}},\ \bibinfo {pages} {147201} (\bibinfo {year} {2022})}\BibitemShut {NoStop}%
\bibitem [{\citenamefont {Yokouchi}\ \emph {et~al.}(2020)\citenamefont {Yokouchi}, \citenamefont {Kagawa}, \citenamefont {Hirschberger}, \citenamefont {Otani}, \citenamefont {Nagaosa},\ and\ \citenamefont {Tokura}}]{yokouchi2020emergent}%
  \BibitemOpen
  \bibfield  {author} {\bibinfo {author} {\bibfnamefont {T.}~\bibnamefont {Yokouchi}}, \bibinfo {author} {\bibfnamefont {F.}~\bibnamefont {Kagawa}}, \bibinfo {author} {\bibfnamefont {M.}~\bibnamefont {Hirschberger}}, \bibinfo {author} {\bibfnamefont {Y.}~\bibnamefont {Otani}}, \bibinfo {author} {\bibfnamefont {N.}~\bibnamefont {Nagaosa}},\ and\ \bibinfo {author} {\bibfnamefont {Y.}~\bibnamefont {Tokura}},\ }\bibfield  {title} {\bibinfo {title} {Emergent electromagnetic induction in a helical-spin magnet},\ }\href@noop {} {\bibfield  {journal} {\bibinfo  {journal} {Nature}\ }\textbf {\bibinfo {volume} {586}},\ \bibinfo {pages} {232} (\bibinfo {year} {2020})}\BibitemShut {NoStop}%
\bibitem [{\citenamefont {Anan}\ and\ \citenamefont {Morimoto}(2025)}]{anan2025emergent}%
  \BibitemOpen
  \bibfield  {author} {\bibinfo {author} {\bibfnamefont {T.}~\bibnamefont {Anan}}\ and\ \bibinfo {author} {\bibfnamefont {T.}~\bibnamefont {Morimoto}},\ }\bibfield  {title} {\bibinfo {title} {Emergent inductors in nonhelical magnets},\ }\href@noop {} {\bibfield  {journal} {\bibinfo  {journal} {Physical Review B}\ }\textbf {\bibinfo {volume} {111}},\ \bibinfo {pages} {184424} (\bibinfo {year} {2025})}\BibitemShut {NoStop}%
\bibitem [{\citenamefont {Oh}\ and\ \citenamefont {Nagaosa}(2024)}]{oh2024emergent}%
  \BibitemOpen
  \bibfield  {author} {\bibinfo {author} {\bibfnamefont {T.}~\bibnamefont {Oh}}\ and\ \bibinfo {author} {\bibfnamefont {N.}~\bibnamefont {Nagaosa}},\ }\bibfield  {title} {\bibinfo {title} {Emergent inductance from spin fluctuations in strongly correlated magnets},\ }\href@noop {} {\bibfield  {journal} {\bibinfo  {journal} {Physical Review Letters}\ }\textbf {\bibinfo {volume} {132}},\ \bibinfo {pages} {116501} (\bibinfo {year} {2024})}\BibitemShut {NoStop}%
\bibitem [{\citenamefont {Ishizuka}\ \emph {et~al.}(2016)\citenamefont {Ishizuka}, \citenamefont {Hayata}, \citenamefont {Ueda},\ and\ \citenamefont {Nagaosa}}]{Ishizuka2016prl}%
  \BibitemOpen
  \bibfield  {author} {\bibinfo {author} {\bibfnamefont {H.}~\bibnamefont {Ishizuka}}, \bibinfo {author} {\bibfnamefont {T.}~\bibnamefont {Hayata}}, \bibinfo {author} {\bibfnamefont {M.}~\bibnamefont {Ueda}},\ and\ \bibinfo {author} {\bibfnamefont {N.}~\bibnamefont {Nagaosa}},\ }\bibfield  {title} {\bibinfo {title} {Emergent electromagnetic induction and adiabatic charge pumping in noncentrosymmetric weyl semimetals},\ }\href {https://doi.org/10.1103/PhysRevLett.117.216601} {\bibfield  {journal} {\bibinfo  {journal} {Phys. Rev. Lett.}\ }\textbf {\bibinfo {volume} {117}},\ \bibinfo {pages} {216601} (\bibinfo {year} {2016})}\BibitemShut {NoStop}%
\bibitem [{\citenamefont {Ishizuka}\ \emph {et~al.}(2017)\citenamefont {Ishizuka}, \citenamefont {Hayata}, \citenamefont {Ueda},\ and\ \citenamefont {Nagaosa}}]{Ishizuka2017prb}%
  \BibitemOpen
  \bibfield  {author} {\bibinfo {author} {\bibfnamefont {H.}~\bibnamefont {Ishizuka}}, \bibinfo {author} {\bibfnamefont {T.}~\bibnamefont {Hayata}}, \bibinfo {author} {\bibfnamefont {M.}~\bibnamefont {Ueda}},\ and\ \bibinfo {author} {\bibfnamefont {N.}~\bibnamefont {Nagaosa}},\ }\bibfield  {title} {\bibinfo {title} {Momentum-space electromagnetic induction in weyl semimetals},\ }\href {https://doi.org/10.1103/PhysRevB.95.245211} {\bibfield  {journal} {\bibinfo  {journal} {Phys. Rev. B}\ }\textbf {\bibinfo {volume} {95}},\ \bibinfo {pages} {245211} (\bibinfo {year} {2017})}\BibitemShut {NoStop}%
\bibitem [{\citenamefont {Harada}\ and\ \citenamefont {Ishizuka}(2023)}]{Harada2023prb}%
  \BibitemOpen
  \bibfield  {author} {\bibinfo {author} {\bibfnamefont {A.}~\bibnamefont {Harada}}\ and\ \bibinfo {author} {\bibfnamefont {H.}~\bibnamefont {Ishizuka}},\ }\bibfield  {title} {\bibinfo {title} {Spin motive force by the momentum-space berry phase in magnetic weyl semimetals},\ }\href {https://doi.org/10.1103/PhysRevB.107.195202} {\bibfield  {journal} {\bibinfo  {journal} {Phys. Rev. B}\ }\textbf {\bibinfo {volume} {107}},\ \bibinfo {pages} {195202} (\bibinfo {year} {2023})}\BibitemShut {NoStop}%
\bibitem [{\citenamefont {Hannukainen}\ \emph {et~al.}(2020)\citenamefont {Hannukainen}, \citenamefont {Ferreiros}, \citenamefont {Cortijo},\ and\ \citenamefont {Bardarson}}]{Hannukainen2020}%
  \BibitemOpen
  \bibfield  {author} {\bibinfo {author} {\bibfnamefont {J.~D.}\ \bibnamefont {Hannukainen}}, \bibinfo {author} {\bibfnamefont {Y.}~\bibnamefont {Ferreiros}}, \bibinfo {author} {\bibfnamefont {A.}~\bibnamefont {Cortijo}},\ and\ \bibinfo {author} {\bibfnamefont {J.~H.}\ \bibnamefont {Bardarson}},\ }\bibfield  {title} {\bibinfo {title} {Axial anomaly generation by domain wall motion in weyl semimetals},\ }\href {https://doi.org/10.1103/PhysRevB.102.241401} {\bibfield  {journal} {\bibinfo  {journal} {Phys. Rev. B}\ }\textbf {\bibinfo {volume} {102}},\ \bibinfo {pages} {241401} (\bibinfo {year} {2020})}\BibitemShut {NoStop}%
\bibitem [{\citenamefont {Heidari}\ \emph {et~al.}(2023)\citenamefont {Heidari}, \citenamefont {Asgari},\ and\ \citenamefont {Culcer}}]{Heidari2023}%
  \BibitemOpen
  \bibfield  {author} {\bibinfo {author} {\bibfnamefont {S.}~\bibnamefont {Heidari}}, \bibinfo {author} {\bibfnamefont {R.}~\bibnamefont {Asgari}},\ and\ \bibinfo {author} {\bibfnamefont {D.}~\bibnamefont {Culcer}},\ }\bibfield  {title} {\bibinfo {title} {Probing domain wall dynamics in magnetic weyl semimetals via the nonlinear anomalous hall effect},\ }\href {https://doi.org/10.1103/PhysRevB.107.214450} {\bibfield  {journal} {\bibinfo  {journal} {Phys. Rev. B}\ }\textbf {\bibinfo {volume} {107}},\ \bibinfo {pages} {214450} (\bibinfo {year} {2023})}\BibitemShut {NoStop}%
\bibitem [{\citenamefont {Zhang}\ and\ \citenamefont {Li}(2004)}]{Zhang2005prl}%
  \BibitemOpen
  \bibfield  {author} {\bibinfo {author} {\bibfnamefont {S.}~\bibnamefont {Zhang}}\ and\ \bibinfo {author} {\bibfnamefont {Z.}~\bibnamefont {Li}},\ }\bibfield  {title} {\bibinfo {title} {Roles of nonequilibrium conduction electrons on the magnetization dynamics of ferromagnets},\ }\href {https://doi.org/10.1103/PhysRevLett.93.127204} {\bibfield  {journal} {\bibinfo  {journal} {Phys. Rev. Lett.}\ }\textbf {\bibinfo {volume} {93}},\ \bibinfo {pages} {127204} (\bibinfo {year} {2004})}\BibitemShut {NoStop}%
\bibitem [{\citenamefont {Barnes}\ and\ \citenamefont {Maekawa}(2005)}]{Barnes2005prl}%
  \BibitemOpen
  \bibfield  {author} {\bibinfo {author} {\bibfnamefont {S.~E.}\ \bibnamefont {Barnes}}\ and\ \bibinfo {author} {\bibfnamefont {S.}~\bibnamefont {Maekawa}},\ }\bibfield  {title} {\bibinfo {title} {Current-spin coupling for ferromagnetic domain walls in fine wires},\ }\href {https://doi.org/10.1103/PhysRevLett.95.107204} {\bibfield  {journal} {\bibinfo  {journal} {Phys. Rev. Lett.}\ }\textbf {\bibinfo {volume} {95}},\ \bibinfo {pages} {107204} (\bibinfo {year} {2005})}\BibitemShut {NoStop}%
\bibitem [{\citenamefont {Thiaville}\ \emph {et~al.}(2005)\citenamefont {Thiaville}, \citenamefont {Nakatani}, \citenamefont {Miltat},\ and\ \citenamefont {Suzuki}}]{thiaville2005micromagnetic}%
  \BibitemOpen
  \bibfield  {author} {\bibinfo {author} {\bibfnamefont {A.}~\bibnamefont {Thiaville}}, \bibinfo {author} {\bibfnamefont {Y.}~\bibnamefont {Nakatani}}, \bibinfo {author} {\bibfnamefont {J.}~\bibnamefont {Miltat}},\ and\ \bibinfo {author} {\bibfnamefont {Y.}~\bibnamefont {Suzuki}},\ }\bibfield  {title} {\bibinfo {title} {Micromagnetic understanding of current-driven domain wall motion in patterned nanowires},\ }\href {https://doi.org/10.1209/epl/i2004-10452-6} {\bibfield  {journal} {\bibinfo  {journal} {Europhysics Letters}\ }\textbf {\bibinfo {volume} {69}},\ \bibinfo {pages} {990} (\bibinfo {year} {2005})}\BibitemShut {NoStop}%
\bibitem [{\citenamefont {Tserkovnyak}\ \emph {et~al.}(2006)\citenamefont {Tserkovnyak}, \citenamefont {Skadsem}, \citenamefont {Brataas},\ and\ \citenamefont {Bauer}}]{Tserkovnyak2006prb}%
  \BibitemOpen
  \bibfield  {author} {\bibinfo {author} {\bibfnamefont {Y.}~\bibnamefont {Tserkovnyak}}, \bibinfo {author} {\bibfnamefont {H.~J.}\ \bibnamefont {Skadsem}}, \bibinfo {author} {\bibfnamefont {A.}~\bibnamefont {Brataas}},\ and\ \bibinfo {author} {\bibfnamefont {G.~E.~W.}\ \bibnamefont {Bauer}},\ }\bibfield  {title} {\bibinfo {title} {Current-induced magnetization dynamics in disordered itinerant ferromagnets},\ }\href {https://doi.org/10.1103/PhysRevB.74.144405} {\bibfield  {journal} {\bibinfo  {journal} {Phys. Rev. B}\ }\textbf {\bibinfo {volume} {74}},\ \bibinfo {pages} {144405} (\bibinfo {year} {2006})}\BibitemShut {NoStop}%
\bibitem [{\citenamefont {Kohno}\ \emph {et~al.}(2006)\citenamefont {Kohno}, \citenamefont {Tatara},\ and\ \citenamefont {Shibata}}]{Kohno2006jpsj}%
  \BibitemOpen
  \bibfield  {author} {\bibinfo {author} {\bibfnamefont {H.}~\bibnamefont {Kohno}}, \bibinfo {author} {\bibfnamefont {G.}~\bibnamefont {Tatara}},\ and\ \bibinfo {author} {\bibfnamefont {J.}~\bibnamefont {Shibata}},\ }\bibfield  {title} {\bibinfo {title} {Microscopic calculation of spin torques in disordered ferromagnets},\ }\href {https://doi.org/10.1143/JPSJ.75.113706} {\bibfield  {journal} {\bibinfo  {journal} {Journal of the Physical Society of Japan}\ }\textbf {\bibinfo {volume} {75}},\ \bibinfo {pages} {113706} (\bibinfo {year} {2006})}\BibitemShut {NoStop}%
\bibitem [{\citenamefont {Obata}\ and\ \citenamefont {Tatara}(2008)}]{obata2008prb}%
  \BibitemOpen
  \bibfield  {author} {\bibinfo {author} {\bibfnamefont {K.}~\bibnamefont {Obata}}\ and\ \bibinfo {author} {\bibfnamefont {G.}~\bibnamefont {Tatara}},\ }\bibfield  {title} {\bibinfo {title} {Current-induced domain wall motion in rashba spin-orbit system},\ }\href {https://doi.org/10.1103/PhysRevB.77.214429} {\bibfield  {journal} {\bibinfo  {journal} {Phys. Rev. B}\ }\textbf {\bibinfo {volume} {77}},\ \bibinfo {pages} {214429} (\bibinfo {year} {2008})}\BibitemShut {NoStop}%
\bibitem [{\citenamefont {Manchon}\ and\ \citenamefont {Zhang}(2008)}]{manchon2008prb}%
  \BibitemOpen
  \bibfield  {author} {\bibinfo {author} {\bibfnamefont {A.}~\bibnamefont {Manchon}}\ and\ \bibinfo {author} {\bibfnamefont {S.}~\bibnamefont {Zhang}},\ }\bibfield  {title} {\bibinfo {title} {Theory of nonequilibrium intrinsic spin torque in a single nanomagnet},\ }\href {https://doi.org/10.1103/PhysRevB.78.212405} {\bibfield  {journal} {\bibinfo  {journal} {Phys. Rev. B}\ }\textbf {\bibinfo {volume} {78}},\ \bibinfo {pages} {212405} (\bibinfo {year} {2008})}\BibitemShut {NoStop}%
\bibitem [{\citenamefont {Mihai~Miron}\ \emph {et~al.}(2010)\citenamefont {Mihai~Miron}, \citenamefont {Gaudin}, \citenamefont {Auffret}, \citenamefont {Rodmacq}, \citenamefont {Schuhl}, \citenamefont {Pizzini}, \citenamefont {Vogel},\ and\ \citenamefont {Gambardella}}]{miron2010current}%
  \BibitemOpen
  \bibfield  {author} {\bibinfo {author} {\bibfnamefont {I.}~\bibnamefont {Mihai~Miron}}, \bibinfo {author} {\bibfnamefont {G.}~\bibnamefont {Gaudin}}, \bibinfo {author} {\bibfnamefont {S.}~\bibnamefont {Auffret}}, \bibinfo {author} {\bibfnamefont {B.}~\bibnamefont {Rodmacq}}, \bibinfo {author} {\bibfnamefont {A.}~\bibnamefont {Schuhl}}, \bibinfo {author} {\bibfnamefont {S.}~\bibnamefont {Pizzini}}, \bibinfo {author} {\bibfnamefont {J.}~\bibnamefont {Vogel}},\ and\ \bibinfo {author} {\bibfnamefont {P.}~\bibnamefont {Gambardella}},\ }\bibfield  {title} {\bibinfo {title} {Current-driven spin torque induced by the rashba effect in a ferromagnetic metal layer},\ }\href@noop {} {\bibfield  {journal} {\bibinfo  {journal} {Nature materials}\ }\textbf {\bibinfo {volume} {9}},\ \bibinfo {pages} {230} (\bibinfo {year} {2010})}\BibitemShut {NoStop}%
\bibitem [{\citenamefont {Miron}\ \emph {et~al.}(2011)\citenamefont {Miron}, \citenamefont {Garello}, \citenamefont {Gaudin}, \citenamefont {Zermatten}, \citenamefont {Costache}, \citenamefont {Auffret}, \citenamefont {Bandiera}, \citenamefont {Rodmacq}, \citenamefont {Schuhl},\ and\ \citenamefont {Gambardella}}]{miron2011perpendicular}%
  \BibitemOpen
  \bibfield  {author} {\bibinfo {author} {\bibfnamefont {I.~M.}\ \bibnamefont {Miron}}, \bibinfo {author} {\bibfnamefont {K.}~\bibnamefont {Garello}}, \bibinfo {author} {\bibfnamefont {G.}~\bibnamefont {Gaudin}}, \bibinfo {author} {\bibfnamefont {P.-J.}\ \bibnamefont {Zermatten}}, \bibinfo {author} {\bibfnamefont {M.~V.}\ \bibnamefont {Costache}}, \bibinfo {author} {\bibfnamefont {S.}~\bibnamefont {Auffret}}, \bibinfo {author} {\bibfnamefont {S.}~\bibnamefont {Bandiera}}, \bibinfo {author} {\bibfnamefont {B.}~\bibnamefont {Rodmacq}}, \bibinfo {author} {\bibfnamefont {A.}~\bibnamefont {Schuhl}},\ and\ \bibinfo {author} {\bibfnamefont {P.}~\bibnamefont {Gambardella}},\ }\bibfield  {title} {\bibinfo {title} {Perpendicular switching of a single ferromagnetic layer induced by in-plane current injection},\ }\href@noop {} {\bibfield  {journal} {\bibinfo  {journal} {Nature}\ }\textbf {\bibinfo {volume} {476}},\ \bibinfo {pages} {189} (\bibinfo {year} {2011})}\BibitemShut {NoStop}%
\bibitem [{\citenamefont {Liu}\ \emph {et~al.}(2012{\natexlab{a}})\citenamefont {Liu}, \citenamefont {Lee}, \citenamefont {Gudmundsen}, \citenamefont {Ralph},\ and\ \citenamefont {Buhrman}}]{liu2012prb}%
  \BibitemOpen
  \bibfield  {author} {\bibinfo {author} {\bibfnamefont {L.}~\bibnamefont {Liu}}, \bibinfo {author} {\bibfnamefont {O.~J.}\ \bibnamefont {Lee}}, \bibinfo {author} {\bibfnamefont {T.~J.}\ \bibnamefont {Gudmundsen}}, \bibinfo {author} {\bibfnamefont {D.~C.}\ \bibnamefont {Ralph}},\ and\ \bibinfo {author} {\bibfnamefont {R.~A.}\ \bibnamefont {Buhrman}},\ }\bibfield  {title} {\bibinfo {title} {Current-induced switching of perpendicularly magnetized magnetic layers using spin torque from the spin hall effect},\ }\href {https://doi.org/10.1103/PhysRevLett.109.096602} {\bibfield  {journal} {\bibinfo  {journal} {Phys. Rev. Lett.}\ }\textbf {\bibinfo {volume} {109}},\ \bibinfo {pages} {096602} (\bibinfo {year} {2012}{\natexlab{a}})}\BibitemShut {NoStop}%
\bibitem [{\citenamefont {Liu}\ \emph {et~al.}(2012{\natexlab{b}})\citenamefont {Liu}, \citenamefont {Pai}, \citenamefont {Li}, \citenamefont {Tseng}, \citenamefont {Ralph},\ and\ \citenamefont {Buhrman}}]{liu2012spin}%
  \BibitemOpen
  \bibfield  {author} {\bibinfo {author} {\bibfnamefont {L.}~\bibnamefont {Liu}}, \bibinfo {author} {\bibfnamefont {C.-F.}\ \bibnamefont {Pai}}, \bibinfo {author} {\bibfnamefont {Y.}~\bibnamefont {Li}}, \bibinfo {author} {\bibfnamefont {H.}~\bibnamefont {Tseng}}, \bibinfo {author} {\bibfnamefont {D.}~\bibnamefont {Ralph}},\ and\ \bibinfo {author} {\bibfnamefont {R.}~\bibnamefont {Buhrman}},\ }\bibfield  {title} {\bibinfo {title} {Spin-torque switching with the giant spin hall effect of tantalum},\ }\href@noop {} {\bibfield  {journal} {\bibinfo  {journal} {Science}\ }\textbf {\bibinfo {volume} {336}},\ \bibinfo {pages} {555} (\bibinfo {year} {2012}{\natexlab{b}})}\BibitemShut {NoStop}%
\bibitem [{\citenamefont {Hirsch}(1999)}]{hirsch1999spin}%
  \BibitemOpen
  \bibfield  {author} {\bibinfo {author} {\bibfnamefont {J.}~\bibnamefont {Hirsch}},\ }\bibfield  {title} {\bibinfo {title} {Spin hall effect},\ }\href@noop {} {\bibfield  {journal} {\bibinfo  {journal} {Physical review letters}\ }\textbf {\bibinfo {volume} {83}},\ \bibinfo {pages} {1834} (\bibinfo {year} {1999})}\BibitemShut {NoStop}%
\bibitem [{\citenamefont {Edelstein}(1990)}]{edelstein1990spin}%
  \BibitemOpen
  \bibfield  {author} {\bibinfo {author} {\bibfnamefont {V.~M.}\ \bibnamefont {Edelstein}},\ }\bibfield  {title} {\bibinfo {title} {Spin polarization of conduction electrons induced by electric current in two-dimensional asymmetric electron systems},\ }\href@noop {} {\bibfield  {journal} {\bibinfo  {journal} {Solid State Communications}\ }\textbf {\bibinfo {volume} {73}},\ \bibinfo {pages} {233} (\bibinfo {year} {1990})}\BibitemShut {NoStop}%
\bibitem [{\citenamefont {{\v{Z}}elezn{\`y}}\ \emph {et~al.}(2017)\citenamefont {{\v{Z}}elezn{\`y}}, \citenamefont {Gao}, \citenamefont {Manchon}, \citenamefont {Freimuth}, \citenamefont {Mokrousov}, \citenamefont {Zemen}, \citenamefont {Ma{\v{s}}ek}, \citenamefont {Sinova},\ and\ \citenamefont {Jungwirth}}]{vzelezny2017spin}%
  \BibitemOpen
  \bibfield  {author} {\bibinfo {author} {\bibfnamefont {J.}~\bibnamefont {{\v{Z}}elezn{\`y}}}, \bibinfo {author} {\bibfnamefont {H.}~\bibnamefont {Gao}}, \bibinfo {author} {\bibfnamefont {A.}~\bibnamefont {Manchon}}, \bibinfo {author} {\bibfnamefont {F.}~\bibnamefont {Freimuth}}, \bibinfo {author} {\bibfnamefont {Y.}~\bibnamefont {Mokrousov}}, \bibinfo {author} {\bibfnamefont {J.}~\bibnamefont {Zemen}}, \bibinfo {author} {\bibfnamefont {J.}~\bibnamefont {Ma{\v{s}}ek}}, \bibinfo {author} {\bibfnamefont {J.}~\bibnamefont {Sinova}},\ and\ \bibinfo {author} {\bibfnamefont {T.}~\bibnamefont {Jungwirth}},\ }\bibfield  {title} {\bibinfo {title} {Spin-orbit torques in locally and globally noncentrosymmetric crystals: Antiferromagnets and ferromagnets},\ }\href@noop {} {\bibfield  {journal} {\bibinfo  {journal} {Physical Review B}\ }\textbf {\bibinfo {volume} {95}},\ \bibinfo {pages} {014403} (\bibinfo {year} {2017})}\BibitemShut {NoStop}%
\bibitem [{\citenamefont {Freimuth}\ \emph {et~al.}(2014)\citenamefont {Freimuth}, \citenamefont {Bl{\"u}gel},\ and\ \citenamefont {Mokrousov}}]{freimuth2014spin}%
  \BibitemOpen
  \bibfield  {author} {\bibinfo {author} {\bibfnamefont {F.}~\bibnamefont {Freimuth}}, \bibinfo {author} {\bibfnamefont {S.}~\bibnamefont {Bl{\"u}gel}},\ and\ \bibinfo {author} {\bibfnamefont {Y.}~\bibnamefont {Mokrousov}},\ }\bibfield  {title} {\bibinfo {title} {Spin-orbit torques in co/pt (111) and mn/w (001) magnetic bilayers from first principles},\ }\href@noop {} {\bibfield  {journal} {\bibinfo  {journal} {Physical Review B}\ }\textbf {\bibinfo {volume} {90}},\ \bibinfo {pages} {174423} (\bibinfo {year} {2014})}\BibitemShut {NoStop}%
\bibitem [{\citenamefont {Hanke}\ \emph {et~al.}(2017)\citenamefont {Hanke}, \citenamefont {Freimuth}, \citenamefont {Niu}, \citenamefont {Bl{\"u}gel},\ and\ \citenamefont {Mokrousov}}]{hanke2017mixed}%
  \BibitemOpen
  \bibfield  {author} {\bibinfo {author} {\bibfnamefont {J.-P.}\ \bibnamefont {Hanke}}, \bibinfo {author} {\bibfnamefont {F.}~\bibnamefont {Freimuth}}, \bibinfo {author} {\bibfnamefont {C.}~\bibnamefont {Niu}}, \bibinfo {author} {\bibfnamefont {S.}~\bibnamefont {Bl{\"u}gel}},\ and\ \bibinfo {author} {\bibfnamefont {Y.}~\bibnamefont {Mokrousov}},\ }\bibfield  {title} {\bibinfo {title} {Mixed weyl semimetals and low-dissipation magnetization control in insulators by spin--orbit torques},\ }\href@noop {} {\bibfield  {journal} {\bibinfo  {journal} {Nature Communications}\ }\textbf {\bibinfo {volume} {8}},\ \bibinfo {pages} {1479} (\bibinfo {year} {2017})}\BibitemShut {NoStop}%
\bibitem [{\citenamefont {Li}\ \emph {et~al.}(2015{\natexlab{c}})\citenamefont {Li}, \citenamefont {Gao}, \citenamefont {Z{\^a}rbo}, \citenamefont {V{\`y}born{\`y}}, \citenamefont {Wang}, \citenamefont {Garate}, \citenamefont {Doǧan}, \citenamefont {{\v{C}}ejchan}, \citenamefont {Sinova}, \citenamefont {Jungwirth} \emph {et~al.}}]{li2015intraband}%
  \BibitemOpen
  \bibfield  {author} {\bibinfo {author} {\bibfnamefont {H.}~\bibnamefont {Li}}, \bibinfo {author} {\bibfnamefont {H.}~\bibnamefont {Gao}}, \bibinfo {author} {\bibfnamefont {L.~P.}\ \bibnamefont {Z{\^a}rbo}}, \bibinfo {author} {\bibfnamefont {K.}~\bibnamefont {V{\`y}born{\`y}}}, \bibinfo {author} {\bibfnamefont {X.}~\bibnamefont {Wang}}, \bibinfo {author} {\bibfnamefont {I.}~\bibnamefont {Garate}}, \bibinfo {author} {\bibfnamefont {F.}~\bibnamefont {Doǧan}}, \bibinfo {author} {\bibfnamefont {A.}~\bibnamefont {{\v{C}}ejchan}}, \bibinfo {author} {\bibfnamefont {J.}~\bibnamefont {Sinova}}, \bibinfo {author} {\bibfnamefont {T.}~\bibnamefont {Jungwirth}}, \emph {et~al.},\ }\bibfield  {title} {\bibinfo {title} {Intraband and interband spin-orbit torques in noncentrosymmetric ferromagnets},\ }\href@noop {} {\bibfield  {journal} {\bibinfo  {journal} {Physical Review B}\ }\textbf {\bibinfo {volume} {91}},\ \bibinfo {pages} {134402} (\bibinfo {year} {2015}{\natexlab{c}})}\BibitemShut {NoStop}%
\bibitem [{\citenamefont {Kurebayashi}\ \emph {et~al.}(2014)\citenamefont {Kurebayashi}, \citenamefont {Sinova}, \citenamefont {Fang}, \citenamefont {Irvine}, \citenamefont {Skinner}, \citenamefont {Wunderlich}, \citenamefont {Nov{\'a}k}, \citenamefont {Campion}, \citenamefont {Gallagher}, \citenamefont {Vehstedt} \emph {et~al.}}]{kurebayashi2014antidamping}%
  \BibitemOpen
  \bibfield  {author} {\bibinfo {author} {\bibfnamefont {H.}~\bibnamefont {Kurebayashi}}, \bibinfo {author} {\bibfnamefont {J.}~\bibnamefont {Sinova}}, \bibinfo {author} {\bibfnamefont {D.}~\bibnamefont {Fang}}, \bibinfo {author} {\bibfnamefont {A.}~\bibnamefont {Irvine}}, \bibinfo {author} {\bibfnamefont {T.}~\bibnamefont {Skinner}}, \bibinfo {author} {\bibfnamefont {J.}~\bibnamefont {Wunderlich}}, \bibinfo {author} {\bibfnamefont {V.}~\bibnamefont {Nov{\'a}k}}, \bibinfo {author} {\bibfnamefont {R.}~\bibnamefont {Campion}}, \bibinfo {author} {\bibfnamefont {B.}~\bibnamefont {Gallagher}}, \bibinfo {author} {\bibfnamefont {E.}~\bibnamefont {Vehstedt}}, \emph {et~al.},\ }\bibfield  {title} {\bibinfo {title} {An antidamping spin--orbit torque originating from the berry curvature},\ }\href@noop {} {\bibfield  {journal} {\bibinfo  {journal} {Nature nanotechnology}\ }\textbf {\bibinfo {volume} {9}},\ \bibinfo {pages} {211} (\bibinfo {year} {2014})}\BibitemShut {NoStop}%
\bibitem [{\citenamefont {{\v{Z}}elezn{\`y}}\ \emph {et~al.}(2014)\citenamefont {{\v{Z}}elezn{\`y}}, \citenamefont {Gao}, \citenamefont {V{\`y}born{\`y}}, \citenamefont {Zemen}, \citenamefont {Ma{\v{s}}ek}, \citenamefont {Manchon}, \citenamefont {Wunderlich}, \citenamefont {Sinova},\ and\ \citenamefont {Jungwirth}}]{vzelezny2014relativistic}%
  \BibitemOpen
  \bibfield  {author} {\bibinfo {author} {\bibfnamefont {J.}~\bibnamefont {{\v{Z}}elezn{\`y}}}, \bibinfo {author} {\bibfnamefont {H.}~\bibnamefont {Gao}}, \bibinfo {author} {\bibfnamefont {K.}~\bibnamefont {V{\`y}born{\`y}}}, \bibinfo {author} {\bibfnamefont {J.}~\bibnamefont {Zemen}}, \bibinfo {author} {\bibfnamefont {J.}~\bibnamefont {Ma{\v{s}}ek}}, \bibinfo {author} {\bibfnamefont {A.}~\bibnamefont {Manchon}}, \bibinfo {author} {\bibfnamefont {J.}~\bibnamefont {Wunderlich}}, \bibinfo {author} {\bibfnamefont {J.}~\bibnamefont {Sinova}},\ and\ \bibinfo {author} {\bibfnamefont {T.}~\bibnamefont {Jungwirth}},\ }\bibfield  {title} {\bibinfo {title} {Relativistic n{\'e}el-order fields induced by electrical current in antiferromagnets},\ }\href@noop {} {\bibfield  {journal} {\bibinfo  {journal} {Physical review letters}\ }\textbf {\bibinfo {volume} {113}},\ \bibinfo {pages} {157201} (\bibinfo {year} {2014})}\BibitemShut {NoStop}%
\bibitem [{\citenamefont {Wadley}\ \emph {et~al.}(2016)\citenamefont {Wadley}, \citenamefont {Howells}, \citenamefont {{\v{Z}}elezn{\`y}}, \citenamefont {Andrews}, \citenamefont {Hills}, \citenamefont {Campion}, \citenamefont {Nov{\'a}k}, \citenamefont {Olejn{\'\i}k}, \citenamefont {Maccherozzi}, \citenamefont {Dhesi} \emph {et~al.}}]{wadley2016electrical}%
  \BibitemOpen
  \bibfield  {author} {\bibinfo {author} {\bibfnamefont {P.}~\bibnamefont {Wadley}}, \bibinfo {author} {\bibfnamefont {B.}~\bibnamefont {Howells}}, \bibinfo {author} {\bibfnamefont {J.}~\bibnamefont {{\v{Z}}elezn{\`y}}}, \bibinfo {author} {\bibfnamefont {C.}~\bibnamefont {Andrews}}, \bibinfo {author} {\bibfnamefont {V.}~\bibnamefont {Hills}}, \bibinfo {author} {\bibfnamefont {R.~P.}\ \bibnamefont {Campion}}, \bibinfo {author} {\bibfnamefont {V.}~\bibnamefont {Nov{\'a}k}}, \bibinfo {author} {\bibfnamefont {K.}~\bibnamefont {Olejn{\'\i}k}}, \bibinfo {author} {\bibfnamefont {F.}~\bibnamefont {Maccherozzi}}, \bibinfo {author} {\bibfnamefont {S.}~\bibnamefont {Dhesi}}, \emph {et~al.},\ }\bibfield  {title} {\bibinfo {title} {Electrical switching of an antiferromagnet},\ }\href@noop {} {\bibfield  {journal} {\bibinfo  {journal} {Science}\ }\textbf {\bibinfo {volume} {351}},\ \bibinfo {pages} {587} (\bibinfo {year} {2016})}\BibitemShut {NoStop}%
\bibitem [{\citenamefont {Sugimoto}\ \emph {et~al.}(2020)\citenamefont {Sugimoto}, \citenamefont {Nakatani}, \citenamefont {Yamane}, \citenamefont {Ikhlas}, \citenamefont {Kondou}, \citenamefont {Kimata}, \citenamefont {Tomita}, \citenamefont {Nakatsuji},\ and\ \citenamefont {Otani}}]{Sugimoto2020}%
  \BibitemOpen
  \bibfield  {author} {\bibinfo {author} {\bibfnamefont {S.}~\bibnamefont {Sugimoto}}, \bibinfo {author} {\bibfnamefont {Y.}~\bibnamefont {Nakatani}}, \bibinfo {author} {\bibfnamefont {Y.}~\bibnamefont {Yamane}}, \bibinfo {author} {\bibfnamefont {M.}~\bibnamefont {Ikhlas}}, \bibinfo {author} {\bibfnamefont {K.}~\bibnamefont {Kondou}}, \bibinfo {author} {\bibfnamefont {M.}~\bibnamefont {Kimata}}, \bibinfo {author} {\bibfnamefont {T.}~\bibnamefont {Tomita}}, \bibinfo {author} {\bibfnamefont {S.}~\bibnamefont {Nakatsuji}},\ and\ \bibinfo {author} {\bibfnamefont {Y.}~\bibnamefont {Otani}},\ }\bibfield  {title} {\bibinfo {title} {Electrical nucleation, displacement, and detection of antiferromagnetic domain walls in the chiral antiferromagnet mn3sn},\ }\href@noop {} {\bibfield  {journal} {\bibinfo  {journal} {Communications Physics}\ }\textbf {\bibinfo {volume} {3}},\ \bibinfo {pages} {111} (\bibinfo {year} {2020})}\BibitemShut {NoStop}%
\bibitem [{\citenamefont {Datta}(1997)}]{datta1997electronic}%
  \BibitemOpen
  \bibfield  {author} {\bibinfo {author} {\bibfnamefont {S.}~\bibnamefont {Datta}},\ }\href@noop {} {\emph {\bibinfo {title} {Electronic transport in mesoscopic systems}}}\ (\bibinfo  {publisher} {Cambridge university press},\ \bibinfo {year} {1997})\BibitemShut {NoStop}%
\bibitem [{\citenamefont {Fradkin}(1986{\natexlab{a}})}]{fradkin1986critical1}%
  \BibitemOpen
  \bibfield  {author} {\bibinfo {author} {\bibfnamefont {E.}~\bibnamefont {Fradkin}},\ }\bibfield  {title} {\bibinfo {title} {Critical behavior of disordered degenerate semiconductors. i. models, symmetries, and formalism},\ }\href@noop {} {\bibfield  {journal} {\bibinfo  {journal} {Physical Review B}\ }\textbf {\bibinfo {volume} {33}},\ \bibinfo {pages} {3257} (\bibinfo {year} {1986}{\natexlab{a}})}\BibitemShut {NoStop}%
\bibitem [{\citenamefont {Fradkin}(1986{\natexlab{b}})}]{fradkin1986critical2}%
  \BibitemOpen
  \bibfield  {author} {\bibinfo {author} {\bibfnamefont {E.}~\bibnamefont {Fradkin}},\ }\bibfield  {title} {\bibinfo {title} {Critical behavior of disordered degenerate semiconductors. ii. spectrum and transport properties in mean-field theory},\ }\href@noop {} {\bibfield  {journal} {\bibinfo  {journal} {Physical review B}\ }\textbf {\bibinfo {volume} {33}},\ \bibinfo {pages} {3263} (\bibinfo {year} {1986}{\natexlab{b}})}\BibitemShut {NoStop}%
\bibitem [{\citenamefont {Ludwig}\ \emph {et~al.}(1994)\citenamefont {Ludwig}, \citenamefont {Fisher}, \citenamefont {Shankar},\ and\ \citenamefont {Grinstein}}]{Ludwig1994-af}%
  \BibitemOpen
  \bibfield  {author} {\bibinfo {author} {\bibfnamefont {A.~W.}\ \bibnamefont {Ludwig}}, \bibinfo {author} {\bibfnamefont {M.~P.}\ \bibnamefont {Fisher}}, \bibinfo {author} {\bibfnamefont {R.}~\bibnamefont {Shankar}},\ and\ \bibinfo {author} {\bibfnamefont {G.}~\bibnamefont {Grinstein}},\ }\bibfield  {title} {\bibinfo {title} {Integer quantum hall transition: An alternative approach and exact results},\ }\href@noop {} {\bibfield  {journal} {\bibinfo  {journal} {Phys. Rev. B Condens. Matter}\ }\textbf {\bibinfo {volume} {50}},\ \bibinfo {pages} {7526} (\bibinfo {year} {1994})}\BibitemShut {NoStop}%
\bibitem [{\citenamefont {Shon}\ and\ \citenamefont {Ando}(1998)}]{Shon1998-ab}%
  \BibitemOpen
  \bibfield  {author} {\bibinfo {author} {\bibfnamefont {N.~H.}\ \bibnamefont {Shon}}\ and\ \bibinfo {author} {\bibfnamefont {T.}~\bibnamefont {Ando}},\ }\bibfield  {title} {\bibinfo {title} {Quantum transport in two-dimensional graphite system},\ }\href@noop {} {\bibfield  {journal} {\bibinfo  {journal} {Journal of the Physical Society of Japan}\ } (\bibinfo {year} {1998})}\BibitemShut {NoStop}%
\bibitem [{\citenamefont {Zheng}\ and\ \citenamefont {Ando}(2002)}]{zheng2002hall}%
  \BibitemOpen
  \bibfield  {author} {\bibinfo {author} {\bibfnamefont {Y.}~\bibnamefont {Zheng}}\ and\ \bibinfo {author} {\bibfnamefont {T.}~\bibnamefont {Ando}},\ }\bibfield  {title} {\bibinfo {title} {Hall conductivity of a two-dimensional graphite system},\ }\href@noop {} {\bibfield  {journal} {\bibinfo  {journal} {Physical Review B}\ }\textbf {\bibinfo {volume} {65}},\ \bibinfo {pages} {245420} (\bibinfo {year} {2002})}\BibitemShut {NoStop}%
\bibitem [{\citenamefont {Gusynin}\ and\ \citenamefont {Sharapov}(2005)}]{Gusynin2005}%
  \BibitemOpen
  \bibfield  {author} {\bibinfo {author} {\bibfnamefont {V.~P.}\ \bibnamefont {Gusynin}}\ and\ \bibinfo {author} {\bibfnamefont {S.~G.}\ \bibnamefont {Sharapov}},\ }\bibfield  {title} {\bibinfo {title} {Unconventional integer quantum hall effect in graphene},\ }\href@noop {} {\bibfield  {journal} {\bibinfo  {journal} {Phys. Rev. Lett.}\ }\textbf {\bibinfo {volume} {95}},\ \bibinfo {pages} {146801} (\bibinfo {year} {2005})}\BibitemShut {NoStop}%
\bibitem [{\citenamefont {Hasan}\ and\ \citenamefont {Kane}(2010)}]{Hasan10topological}%
  \BibitemOpen
  \bibfield  {author} {\bibinfo {author} {\bibfnamefont {M.~Z.}\ \bibnamefont {Hasan}}\ and\ \bibinfo {author} {\bibfnamefont {C.~L.}\ \bibnamefont {Kane}},\ }\bibfield  {title} {\bibinfo {title} {{Colloquium: Topological Insulators}},\ }\href {https://doi.org/10.1103/RevModPhys.82.3045} {\bibfield  {journal} {\bibinfo  {journal} {Rev. Mod. Phys.}\ }\textbf {\bibinfo {volume} {82}},\ \bibinfo {pages} {3045} (\bibinfo {year} {2010})}\BibitemShut {NoStop}%
\bibitem [{\citenamefont {Kajita}\ \emph {et~al.}(2014)\citenamefont {Kajita}, \citenamefont {Nishio}, \citenamefont {Tajima}, \citenamefont {Suzumura},\ and\ \citenamefont {Kobayashi}}]{Kajita14molecular}%
  \BibitemOpen
  \bibfield  {author} {\bibinfo {author} {\bibfnamefont {K.}~\bibnamefont {Kajita}}, \bibinfo {author} {\bibfnamefont {Y.}~\bibnamefont {Nishio}}, \bibinfo {author} {\bibfnamefont {N.}~\bibnamefont {Tajima}}, \bibinfo {author} {\bibfnamefont {Y.}~\bibnamefont {Suzumura}},\ and\ \bibinfo {author} {\bibfnamefont {A.}~\bibnamefont {Kobayashi}},\ }\bibfield  {title} {\bibinfo {title} {{Molecular Dirac Fermion Systems — Theoretical and Experimental Approaches —}},\ }\href {https://doi.org/10.7566/JPSJ.83.072002} {\bibfield  {journal} {\bibinfo  {journal} {J. Phys. Soc. Jpn.}\ }\textbf {\bibinfo {volume} {83}},\ \bibinfo {pages} {072002} (\bibinfo {year} {2014})}\BibitemShut {NoStop}%
\bibitem [{\citenamefont {Nandkishore}\ \emph {et~al.}(2014)\citenamefont {Nandkishore}, \citenamefont {Huse},\ and\ \citenamefont {Sondhi}}]{Nandkishore14rare}%
  \BibitemOpen
  \bibfield  {author} {\bibinfo {author} {\bibfnamefont {R.}~\bibnamefont {Nandkishore}}, \bibinfo {author} {\bibfnamefont {D.~A.}\ \bibnamefont {Huse}},\ and\ \bibinfo {author} {\bibfnamefont {S.~L.}\ \bibnamefont {Sondhi}},\ }\bibfield  {title} {\bibinfo {title} {{Rare region effects dominate weakly disordered three-dimensional Dirac points}},\ }\href@noop {} {\bibfield  {journal} {\bibinfo  {journal} {Phys. Rev. B}\ }\textbf {\bibinfo {volume} {89}},\ \bibinfo {pages} {245110} (\bibinfo {year} {2014})}\BibitemShut {NoStop}%
\bibitem [{\citenamefont {Skinner}(2014)}]{Skinner14coulomb}%
  \BibitemOpen
  \bibfield  {author} {\bibinfo {author} {\bibfnamefont {B.}~\bibnamefont {Skinner}},\ }\bibfield  {title} {\bibinfo {title} {{Coulomb disorder in three-dimensional Dirac systems}},\ }\href@noop {} {\bibfield  {journal} {\bibinfo  {journal} {Phys. Rev. B}\ }\textbf {\bibinfo {volume} {90}},\ \bibinfo {pages} {060202(R)} (\bibinfo {year} {2014})}\BibitemShut {NoStop}%
\bibitem [{\citenamefont {Ominato}\ and\ \citenamefont {Koshino}(2015)}]{Ominato15quantum}%
  \BibitemOpen
  \bibfield  {author} {\bibinfo {author} {\bibfnamefont {Y.}~\bibnamefont {Ominato}}\ and\ \bibinfo {author} {\bibfnamefont {M.}~\bibnamefont {Koshino}},\ }\bibfield  {title} {\bibinfo {title} {{Quantum transport in three-dimensional Weyl electron system in the presence of charged impurity scattering}},\ }\href@noop {} {\bibfield  {journal} {\bibinfo  {journal} {Phys. Rev. B}\ }\textbf {\bibinfo {volume} {91}},\ \bibinfo {pages} {035202} (\bibinfo {year} {2015})}\BibitemShut {NoStop}%
\bibitem [{\citenamefont {Syzranov}\ \emph {et~al.}(2015{\natexlab{a}})\citenamefont {Syzranov}, \citenamefont {Gurarie},\ and\ \citenamefont {Radzihovsky}}]{Syzranov15unconventional}%
  \BibitemOpen
  \bibfield  {author} {\bibinfo {author} {\bibfnamefont {S.~V.}\ \bibnamefont {Syzranov}}, \bibinfo {author} {\bibfnamefont {V.}~\bibnamefont {Gurarie}},\ and\ \bibinfo {author} {\bibfnamefont {L.}~\bibnamefont {Radzihovsky}},\ }\bibfield  {title} {\bibinfo {title} {{Unconventional localisation transition in high dimensions}},\ }\href@noop {} {\bibfield  {journal} {\bibinfo  {journal} {Phys. Rev. B}\ }\textbf {\bibinfo {volume} {91}},\ \bibinfo {pages} {035133} (\bibinfo {year} {2015}{\natexlab{a}})}\BibitemShut {NoStop}%
\bibitem [{\citenamefont {Pixley}\ \emph {et~al.}(2016{\natexlab{a}})\citenamefont {Pixley}, \citenamefont {Huse},\ and\ \citenamefont {Sarma}}]{Pixley16rare}%
  \BibitemOpen
  \bibfield  {author} {\bibinfo {author} {\bibfnamefont {J.~H.}\ \bibnamefont {Pixley}}, \bibinfo {author} {\bibfnamefont {D.~A.}\ \bibnamefont {Huse}},\ and\ \bibinfo {author} {\bibfnamefont {S.~D.}\ \bibnamefont {Sarma}},\ }\bibfield  {title} {\bibinfo {title} {{Rare-Region-Induced Avoided Quantum Criticality in Disordered Three-Dimensional Dirac and Weyl Semimetals}},\ }\href@noop {} {\bibfield  {journal} {\bibinfo  {journal} {Phys. Rev. X}\ }\textbf {\bibinfo {volume} {6}},\ \bibinfo {pages} {021042} (\bibinfo {year} {2016}{\natexlab{a}})}\BibitemShut {NoStop}%
\bibitem [{\citenamefont {Louvet}\ \emph {et~al.}(2017)\citenamefont {Louvet}, \citenamefont {Carpentier},\ and\ \citenamefont {Fedorenko}}]{Louvet17new}%
  \BibitemOpen
  \bibfield  {author} {\bibinfo {author} {\bibfnamefont {T.}~\bibnamefont {Louvet}}, \bibinfo {author} {\bibfnamefont {D.}~\bibnamefont {Carpentier}},\ and\ \bibinfo {author} {\bibfnamefont {A.~A.}\ \bibnamefont {Fedorenko}},\ }\bibfield  {title} {\bibinfo {title} {{New quantum transition in Weyl semimetals with correlated disorder}},\ }\href@noop {} {\bibfield  {journal} {\bibinfo  {journal} {Phys. Rev. B}\ }\textbf {\bibinfo {volume} {95}},\ \bibinfo {pages} {014204} (\bibinfo {year} {2017})}\BibitemShut {NoStop}%
\bibitem [{\citenamefont {Goswami}\ and\ \citenamefont {Chakravarty}(2011)}]{Goswami11quantum}%
  \BibitemOpen
  \bibfield  {author} {\bibinfo {author} {\bibfnamefont {P.}~\bibnamefont {Goswami}}\ and\ \bibinfo {author} {\bibfnamefont {S.}~\bibnamefont {Chakravarty}},\ }\bibfield  {title} {\bibinfo {title} {{Quantum Criticality between Topological and Band Insulators in 3+1 Dimensions}},\ }\href {https://doi.org/10.1103/PhysRevLett.107.196803} {\bibfield  {journal} {\bibinfo  {journal} {Phys. Rev. Lett.}\ }\textbf {\bibinfo {volume} {107}},\ \bibinfo {pages} {196803} (\bibinfo {year} {2011})}\BibitemShut {NoStop}%
\bibitem [{\citenamefont {Kobayashi}\ \emph {et~al.}(2014)\citenamefont {Kobayashi}, \citenamefont {Ohtsuki}, \citenamefont {Imura},\ and\ \citenamefont {Herbut}}]{Kobayashi14density}%
  \BibitemOpen
  \bibfield  {author} {\bibinfo {author} {\bibfnamefont {K.}~\bibnamefont {Kobayashi}}, \bibinfo {author} {\bibfnamefont {T.}~\bibnamefont {Ohtsuki}}, \bibinfo {author} {\bibfnamefont {K.-I.}\ \bibnamefont {Imura}},\ and\ \bibinfo {author} {\bibfnamefont {I.~F.}\ \bibnamefont {Herbut}},\ }\bibfield  {title} {\bibinfo {title} {{Density of States Scaling at the Semimetal to Metal Transition in Three Dimensional Topological Insulators}},\ }\href {https://doi.org/10.1103/PhysRevLett.112.016402} {\bibfield  {journal} {\bibinfo  {journal} {Phys. Rev. Lett.}\ }\textbf {\bibinfo {volume} {112}},\ \bibinfo {pages} {016402} (\bibinfo {year} {2014})}\BibitemShut {NoStop}%
\bibitem [{\citenamefont {Ominato}\ and\ \citenamefont {Koshino}(2014)}]{Ominato14quantum}%
  \BibitemOpen
  \bibfield  {author} {\bibinfo {author} {\bibfnamefont {Y.}~\bibnamefont {Ominato}}\ and\ \bibinfo {author} {\bibfnamefont {M.}~\bibnamefont {Koshino}},\ }\bibfield  {title} {\bibinfo {title} {{Quantum transport in a three-dimensional Weyl electron system}},\ }\href@noop {} {\bibfield  {journal} {\bibinfo  {journal} {Phys. Rev. B}\ }\textbf {\bibinfo {volume} {89}},\ \bibinfo {pages} {054202} (\bibinfo {year} {2014})}\BibitemShut {NoStop}%
\bibitem [{\citenamefont {Sbierski}\ and\ \citenamefont {Brouwer}(2014)}]{Sbierski14z2}%
  \BibitemOpen
  \bibfield  {author} {\bibinfo {author} {\bibfnamefont {B.}~\bibnamefont {Sbierski}}\ and\ \bibinfo {author} {\bibfnamefont {P.~W.}\ \bibnamefont {Brouwer}},\ }\bibfield  {title} {\bibinfo {title} {{$\mathbb{Z}2$ phase diagram of three-dimensional disordered topological insulators via a scattering matrix approach}},\ }\href@noop {} {\bibfield  {journal} {\bibinfo  {journal} {Phys. Rev. B}\ }\textbf {\bibinfo {volume} {89}},\ \bibinfo {pages} {155311} (\bibinfo {year} {2014})}\BibitemShut {NoStop}%
\bibitem [{\citenamefont {Sbierski}\ \emph {et~al.}(2014)\citenamefont {Sbierski}, \citenamefont {Pohl}, \citenamefont {Bergholtz},\ and\ \citenamefont {Brouwer}}]{Sbierski14quantum}%
  \BibitemOpen
  \bibfield  {author} {\bibinfo {author} {\bibfnamefont {B.}~\bibnamefont {Sbierski}}, \bibinfo {author} {\bibfnamefont {G.}~\bibnamefont {Pohl}}, \bibinfo {author} {\bibfnamefont {E.~J.}\ \bibnamefont {Bergholtz}},\ and\ \bibinfo {author} {\bibfnamefont {P.~W.}\ \bibnamefont {Brouwer}},\ }\bibfield  {title} {\bibinfo {title} {{Quantum Transport of Disordered Weyl Semimetals at the Nodal Point}},\ }\href@noop {} {\bibfield  {journal} {\bibinfo  {journal} {Phys. Rev. Lett.}\ }\textbf {\bibinfo {volume} {113}},\ \bibinfo {pages} {026602} (\bibinfo {year} {2014})}\BibitemShut {NoStop}%
\bibitem [{\citenamefont {Syzranov}\ \emph {et~al.}(2015{\natexlab{b}})\citenamefont {Syzranov}, \citenamefont {Radzihovsky},\ and\ \citenamefont {Gurarie}}]{Syzranov15criticalTransport}%
  \BibitemOpen
  \bibfield  {author} {\bibinfo {author} {\bibfnamefont {S.~V.}\ \bibnamefont {Syzranov}}, \bibinfo {author} {\bibfnamefont {L.}~\bibnamefont {Radzihovsky}},\ and\ \bibinfo {author} {\bibfnamefont {V.}~\bibnamefont {Gurarie}},\ }\bibfield  {title} {\bibinfo {title} {{Critical transport in weakly disordered semiconductors and semimetals}},\ }\href@noop {} {\bibfield  {journal} {\bibinfo  {journal} {Phys. Rev. Lett.}\ }\textbf {\bibinfo {volume} {114}},\ \bibinfo {pages} {166601} (\bibinfo {year} {2015}{\natexlab{b}})}\BibitemShut {NoStop}%
\bibitem [{\citenamefont {Roy}\ and\ \citenamefont {Sarma}(2014)}]{Roy14diffusive}%
  \BibitemOpen
  \bibfield  {author} {\bibinfo {author} {\bibfnamefont {B.}~\bibnamefont {Roy}}\ and\ \bibinfo {author} {\bibfnamefont {S.~D.}\ \bibnamefont {Sarma}},\ }\bibfield  {title} {\bibinfo {title} {{Diffusive quantum criticality in three-dimensional disordered Dirac semimetals}},\ }\href@noop {} {\bibfield  {journal} {\bibinfo  {journal} {Phys. Rev. B}\ }\textbf {\bibinfo {volume} {90}},\ \bibinfo {pages} {241112(R)} (\bibinfo {year} {2014})}\BibitemShut {NoStop}%
\bibitem [{\citenamefont {Roy}\ and\ \citenamefont {Sarma}(2016)}]{Roy16erratum}%
  \BibitemOpen
  \bibfield  {author} {\bibinfo {author} {\bibfnamefont {B.}~\bibnamefont {Roy}}\ and\ \bibinfo {author} {\bibfnamefont {S.~D.}\ \bibnamefont {Sarma}},\ }\bibfield  {title} {\bibinfo {title} {{Erratum: Diffusive quantum criticality in three-dimensional disordered Dirac semimetals [Phys. Rev. B 90, 241112(R) (2014)]}},\ }\href@noop {} {\bibfield  {journal} {\bibinfo  {journal} {Phys. Rev. B}\ }\textbf {\bibinfo {volume} {93}},\ \bibinfo {pages} {119911(E)} (\bibinfo {year} {2016})}\BibitemShut {NoStop}%
\bibitem [{\citenamefont {Pixley}\ \emph {et~al.}(2015)\citenamefont {Pixley}, \citenamefont {Goswami},\ and\ \citenamefont {Sarma}}]{Pixley15anderson}%
  \BibitemOpen
  \bibfield  {author} {\bibinfo {author} {\bibfnamefont {J.~H.}\ \bibnamefont {Pixley}}, \bibinfo {author} {\bibfnamefont {P.}~\bibnamefont {Goswami}},\ and\ \bibinfo {author} {\bibfnamefont {S.~D.}\ \bibnamefont {Sarma}},\ }\bibfield  {title} {\bibinfo {title} {{Anderson Localization and the Quantum Phase Diagram of Three Dimensional Disordered Dirac Semimetals}},\ }\href@noop {} {\bibfield  {journal} {\bibinfo  {journal} {Phys. Rev. Lett.}\ }\textbf {\bibinfo {volume} {115}},\ \bibinfo {pages} {076601} (\bibinfo {year} {2015})}\BibitemShut {NoStop}%
\bibitem [{\citenamefont {Sbierski}\ \emph {et~al.}(2015)\citenamefont {Sbierski}, \citenamefont {Bergholtz},\ and\ \citenamefont {Brouwer}}]{Sbierski15quantum}%
  \BibitemOpen
  \bibfield  {author} {\bibinfo {author} {\bibfnamefont {B.}~\bibnamefont {Sbierski}}, \bibinfo {author} {\bibfnamefont {E.~J.}\ \bibnamefont {Bergholtz}},\ and\ \bibinfo {author} {\bibfnamefont {P.~W.}\ \bibnamefont {Brouwer}},\ }\bibfield  {title} {\bibinfo {title} {{Quantum critical exponents for a disordered three-dimensional Weyl node}},\ }\href@noop {} {\bibfield  {journal} {\bibinfo  {journal} {Phys. Rev. B}\ }\textbf {\bibinfo {volume} {92}},\ \bibinfo {pages} {115145} (\bibinfo {year} {2015})}\BibitemShut {NoStop}%
\bibitem [{\citenamefont {Chen}\ \emph {et~al.}(2015)\citenamefont {Chen}, \citenamefont {Song}, \citenamefont {Jiang}, \citenamefont {Sun}, \citenamefont {Wang},\ and\ \citenamefont {Xie}}]{Chen15disorder}%
  \BibitemOpen
  \bibfield  {author} {\bibinfo {author} {\bibfnamefont {C.-Z.}\ \bibnamefont {Chen}}, \bibinfo {author} {\bibfnamefont {J.}~\bibnamefont {Song}}, \bibinfo {author} {\bibfnamefont {H.}~\bibnamefont {Jiang}}, \bibinfo {author} {\bibfnamefont {Q.-F.}\ \bibnamefont {Sun}}, \bibinfo {author} {\bibfnamefont {Z.}~\bibnamefont {Wang}},\ and\ \bibinfo {author} {\bibfnamefont {X.-C.}\ \bibnamefont {Xie}},\ }\bibfield  {title} {\bibinfo {title} {{Disorder and Metal-Insulator Transitions in Weyl Semimetals}},\ }\href {https://doi.org/10.1103/PhysRevLett.115.246603} {\bibfield  {journal} {\bibinfo  {journal} {Phys. Rev. Lett.}\ }\textbf {\bibinfo {volume} {115}},\ \bibinfo {pages} {246603} (\bibinfo {year} {2015})}\BibitemShut {NoStop}%
\bibitem [{\citenamefont {Liu}\ \emph {et~al.}(2016)\citenamefont {Liu}, \citenamefont {Ohtsuki},\ and\ \citenamefont {Shindou}}]{Liu16effect}%
  \BibitemOpen
  \bibfield  {author} {\bibinfo {author} {\bibfnamefont {S.}~\bibnamefont {Liu}}, \bibinfo {author} {\bibfnamefont {T.}~\bibnamefont {Ohtsuki}},\ and\ \bibinfo {author} {\bibfnamefont {R.}~\bibnamefont {Shindou}},\ }\bibfield  {title} {\bibinfo {title} {{Effect of Disorder in a Three-Dimensional Layered Chern Insulator}},\ }\href {https://doi.org/10.1103/PhysRevLett.116.066401} {\bibfield  {journal} {\bibinfo  {journal} {Phys. Rev. Lett.}\ }\textbf {\bibinfo {volume} {116}},\ \bibinfo {pages} {066401} (\bibinfo {year} {2016})}\BibitemShut {NoStop}%
\bibitem [{\citenamefont {Bera}\ \emph {et~al.}(2016)\citenamefont {Bera}, \citenamefont {Sau},\ and\ \citenamefont {Roy}}]{Bera16dirty}%
  \BibitemOpen
  \bibfield  {author} {\bibinfo {author} {\bibfnamefont {S.}~\bibnamefont {Bera}}, \bibinfo {author} {\bibfnamefont {J.~D.}\ \bibnamefont {Sau}},\ and\ \bibinfo {author} {\bibfnamefont {B.}~\bibnamefont {Roy}},\ }\bibfield  {title} {\bibinfo {title} {{Dirty Weyl semimetals: Stability, phase transition, and quantum criticality}},\ }\href@noop {} {\bibfield  {journal} {\bibinfo  {journal} {Phys. Rev. B}\ }\textbf {\bibinfo {volume} {93}},\ \bibinfo {pages} {201302(R)} (\bibinfo {year} {2016})}\BibitemShut {NoStop}%
\bibitem [{\citenamefont {Pixley}\ \emph {et~al.}(2016{\natexlab{b}})\citenamefont {Pixley}, \citenamefont {Goswami},\ and\ \citenamefont {Sarma}}]{Pixley16disorder}%
  \BibitemOpen
  \bibfield  {author} {\bibinfo {author} {\bibfnamefont {J.~H.}\ \bibnamefont {Pixley}}, \bibinfo {author} {\bibfnamefont {P.}~\bibnamefont {Goswami}},\ and\ \bibinfo {author} {\bibfnamefont {S.~D.}\ \bibnamefont {Sarma}},\ }\bibfield  {title} {\bibinfo {title} {{Disorder-driven itinerant quantum criticality of three-dimensional massless Dirac fermions}},\ }\href@noop {} {\bibfield  {journal} {\bibinfo  {journal} {Phys. Rev. B}\ }\textbf {\bibinfo {volume} {93}},\ \bibinfo {pages} {085103} (\bibinfo {year} {2016}{\natexlab{b}})}\BibitemShut {NoStop}%
\bibitem [{\citenamefont {Shapourian}\ and\ \citenamefont {Hughes}(2016)}]{Shapourian16phase}%
  \BibitemOpen
  \bibfield  {author} {\bibinfo {author} {\bibfnamefont {H.}~\bibnamefont {Shapourian}}\ and\ \bibinfo {author} {\bibfnamefont {T.~L.}\ \bibnamefont {Hughes}},\ }\bibfield  {title} {\bibinfo {title} {{Phase diagrams of disordered Weyl semimetals}},\ }\href {https://doi.org/10.1103/PhysRevB.93.075108} {\bibfield  {journal} {\bibinfo  {journal} {Phys. Rev. B}\ }\textbf {\bibinfo {volume} {93}},\ \bibinfo {pages} {075108} (\bibinfo {year} {2016})}\BibitemShut {NoStop}%
\bibitem [{\citenamefont {Syzranov}\ \emph {et~al.}(2016{\natexlab{a}})\citenamefont {Syzranov}, \citenamefont {Ostrovsky}, \citenamefont {Gurarie},\ and\ \citenamefont {Radzihovsky}}]{Syzranov16criticalExponents}%
  \BibitemOpen
  \bibfield  {author} {\bibinfo {author} {\bibfnamefont {S.~V.}\ \bibnamefont {Syzranov}}, \bibinfo {author} {\bibfnamefont {P.~M.}\ \bibnamefont {Ostrovsky}}, \bibinfo {author} {\bibfnamefont {V.}~\bibnamefont {Gurarie}},\ and\ \bibinfo {author} {\bibfnamefont {L.}~\bibnamefont {Radzihovsky}},\ }\bibfield  {title} {\bibinfo {title} {{Critical exponents at the unconventional disorder-driven transition in a Weyl semimetal}},\ }\href@noop {} {\bibfield  {journal} {\bibinfo  {journal} {Phys. Rev. B}\ }\textbf {\bibinfo {volume} {93}},\ \bibinfo {pages} {155113} (\bibinfo {year} {2016}{\natexlab{a}})}\BibitemShut {NoStop}%
\bibitem [{\citenamefont {Louvet}\ \emph {et~al.}(2016)\citenamefont {Louvet}, \citenamefont {Carpentier},\ and\ \citenamefont {Fedorenko}}]{Louvet16on}%
  \BibitemOpen
  \bibfield  {author} {\bibinfo {author} {\bibfnamefont {T.}~\bibnamefont {Louvet}}, \bibinfo {author} {\bibfnamefont {D.}~\bibnamefont {Carpentier}},\ and\ \bibinfo {author} {\bibfnamefont {A.~A.}\ \bibnamefont {Fedorenko}},\ }\bibfield  {title} {\bibinfo {title} {{On the disorder-driven quantum transition in three-dimensional relativistic metals}},\ }\href@noop {} {\bibfield  {journal} {\bibinfo  {journal} {Phys. Rev. B}\ }\textbf {\bibinfo {volume} {94}},\ \bibinfo {pages} {220201(R)} (\bibinfo {year} {2016})}\BibitemShut {NoStop}%
\bibitem [{\citenamefont {Pixley}\ \emph {et~al.}(2016{\natexlab{c}})\citenamefont {Pixley}, \citenamefont {Huse},\ and\ \citenamefont {Sarma}}]{Pixley16uncovering}%
  \BibitemOpen
  \bibfield  {author} {\bibinfo {author} {\bibfnamefont {J.~H.}\ \bibnamefont {Pixley}}, \bibinfo {author} {\bibfnamefont {D.~A.}\ \bibnamefont {Huse}},\ and\ \bibinfo {author} {\bibfnamefont {S.~D.}\ \bibnamefont {Sarma}},\ }\bibfield  {title} {\bibinfo {title} {{Uncovering the hidden quantum critical point in disordered massless Dirac and Weyl semimetals}},\ }\href@noop {} {\bibfield  {journal} {\bibinfo  {journal} {Phys. Rev. B}\ }\textbf {\bibinfo {volume} {94}},\ \bibinfo {pages} {121107(R)} (\bibinfo {year} {2016}{\natexlab{c}})}\BibitemShut {NoStop}%
\bibitem [{\citenamefont {Syzranov}\ \emph {et~al.}(2016{\natexlab{b}})\citenamefont {Syzranov}, \citenamefont {Gurarie},\ and\ \citenamefont {Radzihovsky}}]{Syzranov16multifractality}%
  \BibitemOpen
  \bibfield  {author} {\bibinfo {author} {\bibfnamefont {S.~V.}\ \bibnamefont {Syzranov}}, \bibinfo {author} {\bibfnamefont {V.}~\bibnamefont {Gurarie}},\ and\ \bibinfo {author} {\bibfnamefont {L.}~\bibnamefont {Radzihovsky}},\ }\bibfield  {title} {\bibinfo {title} {{Multifractality at non-Anderson disorder-driven transitions in Weyl semimetals and other systems}},\ }\href@noop {} {\bibfield  {journal} {\bibinfo  {journal} {Ann. Phys.}\ }\textbf {\bibinfo {volume} {373}},\ \bibinfo {pages} {694} (\bibinfo {year} {2016}{\natexlab{b}})}\BibitemShut {NoStop}%
\bibitem [{\citenamefont {Syzranov}\ and\ \citenamefont {Radzihovsky}(2018)}]{Syzranov18high}%
  \BibitemOpen
  \bibfield  {author} {\bibinfo {author} {\bibfnamefont {S.~V.}\ \bibnamefont {Syzranov}}\ and\ \bibinfo {author} {\bibfnamefont {L.}~\bibnamefont {Radzihovsky}},\ }\bibfield  {title} {\bibinfo {title} {{High-Dimensional Disorder-Driven Phenomena in Weyl Semimetals, Semiconductors and Related Systems}},\ }\href@noop {} {\bibfield  {journal} {\bibinfo  {journal} {Ann. Rev. Cond. Mat. Phys.}\ }\textbf {\bibinfo {volume} {9}},\ \bibinfo {pages} {35} (\bibinfo {year} {2018})}\BibitemShut {NoStop}%
\bibitem [{\citenamefont {Fu}\ \emph {et~al.}(2017)\citenamefont {Fu}, \citenamefont {Zhu}, \citenamefont {Shi}, \citenamefont {Li}, \citenamefont {Yang},\ and\ \citenamefont {Zhang}}]{Fu17accurate}%
  \BibitemOpen
  \bibfield  {author} {\bibinfo {author} {\bibfnamefont {B.}~\bibnamefont {Fu}}, \bibinfo {author} {\bibfnamefont {W.}~\bibnamefont {Zhu}}, \bibinfo {author} {\bibfnamefont {Q.}~\bibnamefont {Shi}}, \bibinfo {author} {\bibfnamefont {Q.}~\bibnamefont {Li}}, \bibinfo {author} {\bibfnamefont {J.}~\bibnamefont {Yang}},\ and\ \bibinfo {author} {\bibfnamefont {Z.}~\bibnamefont {Zhang}},\ }\bibfield  {title} {\bibinfo {title} {{Accurate Determination of the Quasiparticle and Scaling Properties Surrounding the Quantum Critical Point of Disordered Three-Dimensional Dirac Semimetals}},\ }\href@noop {} {\bibfield  {journal} {\bibinfo  {journal} {Phys. Rev. Lett.}\ }\textbf {\bibinfo {volume} {118}},\ \bibinfo {pages} {146401} (\bibinfo {year} {2017})}\BibitemShut {NoStop}%
\bibitem [{\citenamefont {Roy}\ \emph {et~al.}(2018)\citenamefont {Roy}, \citenamefont {Slager},\ and\ \citenamefont {Juri\v{c}i\'{c}}}]{Roy18global}%
  \BibitemOpen
  \bibfield  {author} {\bibinfo {author} {\bibfnamefont {B.}~\bibnamefont {Roy}}, \bibinfo {author} {\bibfnamefont {R.-J.}\ \bibnamefont {Slager}},\ and\ \bibinfo {author} {\bibfnamefont {V.}~\bibnamefont {Juri\v{c}i\'{c}}},\ }\bibfield  {title} {\bibinfo {title} {{Global Phase Diagram of a Dirty Weyl Liquid and Emergent Superuniversality}},\ }\href@noop {} {\bibfield  {journal} {\bibinfo  {journal} {Phys. Rev. X}\ }\textbf {\bibinfo {volume} {8}},\ \bibinfo {pages} {031076} (\bibinfo {year} {2018})}\BibitemShut {NoStop}%
\bibitem [{\citenamefont {Kobayashi}\ \emph {et~al.}(2020)\citenamefont {Kobayashi}, \citenamefont {Wada},\ and\ \citenamefont {Ohtsuki}}]{Kobayashi20ballistic}%
  \BibitemOpen
  \bibfield  {author} {\bibinfo {author} {\bibfnamefont {K.}~\bibnamefont {Kobayashi}}, \bibinfo {author} {\bibfnamefont {M.}~\bibnamefont {Wada}},\ and\ \bibinfo {author} {\bibfnamefont {T.}~\bibnamefont {Ohtsuki}},\ }\bibfield  {title} {\bibinfo {title} {{Ballistic transport in disordered Dirac and Weyl semimetals}},\ }\href {https://doi.org/10.1103/PhysRevResearch.2.022061} {\bibfield  {journal} {\bibinfo  {journal} {Phys. Rev. Res.}\ }\textbf {\bibinfo {volume} {2}},\ \bibinfo {pages} {022061(R)} (\bibinfo {year} {2020})}\BibitemShut {NoStop}%
\bibitem [{\citenamefont {Hosur}\ \emph {et~al.}(2012)\citenamefont {Hosur}, \citenamefont {Parameswaran},\ and\ \citenamefont {Vishwanath}}]{Hosur12charge}%
  \BibitemOpen
  \bibfield  {author} {\bibinfo {author} {\bibfnamefont {P.}~\bibnamefont {Hosur}}, \bibinfo {author} {\bibfnamefont {S.~A.}\ \bibnamefont {Parameswaran}},\ and\ \bibinfo {author} {\bibfnamefont {A.}~\bibnamefont {Vishwanath}},\ }\bibfield  {title} {\bibinfo {title} {{Charge Transport in Weyl Semimetals}},\ }\href@noop {} {\bibfield  {journal} {\bibinfo  {journal} {Phys. Rev. Lett.}\ }\textbf {\bibinfo {volume} {108}},\ \bibinfo {pages} {046602} (\bibinfo {year} {2012})}\BibitemShut {NoStop}%
\bibitem [{\citenamefont {Ziegler}(2016)}]{Ziegler16quantum}%
  \BibitemOpen
  \bibfield  {author} {\bibinfo {author} {\bibfnamefont {K.}~\bibnamefont {Ziegler}},\ }\bibfield  {title} {\bibinfo {title} {{Quantum transport in 3D Weyl semimetals: Is there a metal-insulator transition?}},\ }\href@noop {} {\bibfield  {journal} {\bibinfo  {journal} {Eur. Phys. J. B}\ }\textbf {\bibinfo {volume} {89}},\ \bibinfo {pages} {268} (\bibinfo {year} {2016})}\BibitemShut {NoStop}%
\bibitem [{\citenamefont {Klier}\ \emph {et~al.}(2019)\citenamefont {Klier}, \citenamefont {Gornyi},\ and\ \citenamefont {Mirlin}}]{Klier19from}%
  \BibitemOpen
  \bibfield  {author} {\bibinfo {author} {\bibfnamefont {J.}~\bibnamefont {Klier}}, \bibinfo {author} {\bibfnamefont {I.~V.}\ \bibnamefont {Gornyi}},\ and\ \bibinfo {author} {\bibfnamefont {A.~D.}\ \bibnamefont {Mirlin}},\ }\bibfield  {title} {\bibinfo {title} {{From weak to strong disorder in Weyl semimetals: Self-consistent Born approximation}},\ }\href@noop {} {\bibfield  {journal} {\bibinfo  {journal} {Phys. Rev. B}\ }\textbf {\bibinfo {volume} {100}},\ \bibinfo {pages} {125160} (\bibinfo {year} {2019})}\BibitemShut {NoStop}%
\bibitem [{\citenamefont {\v{Z}uti\'{c}}\ \emph {et~al.}(2004)\citenamefont {\v{Z}uti\'{c}}, \citenamefont {Fabian},\ and\ \citenamefont {Sarma}}]{Zutic2004spintronics}%
  \BibitemOpen
  \bibfield  {author} {\bibinfo {author} {\bibfnamefont {I.}~\bibnamefont {\v{Z}uti\'{c}}}, \bibinfo {author} {\bibfnamefont {J.}~\bibnamefont {Fabian}},\ and\ \bibinfo {author} {\bibfnamefont {S.~D.}\ \bibnamefont {Sarma}},\ }\bibfield  {title} {\bibinfo {title} {{Spintronics: Fundamentals and applications}},\ }\href {https://doi.org/10.1103/RevModPhys.76.323} {\bibfield  {journal} {\bibinfo  {journal} {Rev. Mod. Phys.}\ }\textbf {\bibinfo {volume} {76}},\ \bibinfo {pages} {323} (\bibinfo {year} {2004})}\BibitemShut {NoStop}%
\bibitem [{\citenamefont {Ominato}\ \emph {et~al.}(2017)\citenamefont {Ominato}, \citenamefont {Kobayashi},\ and\ \citenamefont {Nomura}}]{Ominato17anisotropic}%
  \BibitemOpen
  \bibfield  {author} {\bibinfo {author} {\bibfnamefont {Y.}~\bibnamefont {Ominato}}, \bibinfo {author} {\bibfnamefont {K.}~\bibnamefont {Kobayashi}},\ and\ \bibinfo {author} {\bibfnamefont {K.}~\bibnamefont {Nomura}},\ }\bibfield  {title} {\bibinfo {title} {{Anisotropic magnetotransport in Dirac-Weyl magnetic junctions}},\ }\href@noop {} {\bibfield  {journal} {\bibinfo  {journal} {Phys. Rev. B}\ }\textbf {\bibinfo {volume} {95}},\ \bibinfo {pages} {085308} (\bibinfo {year} {2017})}\BibitemShut {NoStop}%
\bibitem [{\citenamefont {Kobayashi}\ \emph {et~al.}(2018)\citenamefont {Kobayashi}, \citenamefont {Ominato},\ and\ \citenamefont {Nomura}}]{Kobayashi2018helicity}%
  \BibitemOpen
  \bibfield  {author} {\bibinfo {author} {\bibfnamefont {K.}~\bibnamefont {Kobayashi}}, \bibinfo {author} {\bibfnamefont {Y.}~\bibnamefont {Ominato}},\ and\ \bibinfo {author} {\bibfnamefont {K.}~\bibnamefont {Nomura}},\ }\bibfield  {title} {\bibinfo {title} {{Helicity-protected domain-wall magnetoresistance in ferromagnetic Weyl semimetal}},\ }\href@noop {} {\bibfield  {journal} {\bibinfo  {journal} {J. Phys. Soc. Jpn.}\ }\textbf {\bibinfo {volume} {87}},\ \bibinfo {pages} {073707} (\bibinfo {year} {2018})}\BibitemShut {NoStop}%
\bibitem [{\citenamefont {Kent}\ \emph {et~al.}(2001)\citenamefont {Kent}, \citenamefont {Yu}, \citenamefont {Rüdiger},\ and\ \citenamefont {Parkin}}]{Kent2001}%
  \BibitemOpen
  \bibfield  {author} {\bibinfo {author} {\bibfnamefont {A.~D.}\ \bibnamefont {Kent}}, \bibinfo {author} {\bibfnamefont {J.}~\bibnamefont {Yu}}, \bibinfo {author} {\bibfnamefont {U.}~\bibnamefont {Rüdiger}},\ and\ \bibinfo {author} {\bibfnamefont {S.~S.~P.}\ \bibnamefont {Parkin}},\ }\bibfield  {title} {\bibinfo {title} {Domain wall resistivity in epitaxial thin film microstructures},\ }\href@noop {} {\bibfield  {journal} {\bibinfo  {journal} {J. Phys. Condens. Matter}\ }\textbf {\bibinfo {volume} {13}},\ \bibinfo {pages} {R461} (\bibinfo {year} {2001})}\BibitemShut {NoStop}%
\bibitem [{\citenamefont {Pendry}\ \emph {et~al.}(1992)\citenamefont {Pendry}, \citenamefont {MacKinnon},\ and\ \citenamefont {Roberts}}]{Pendry1992universality}%
  \BibitemOpen
  \bibfield  {author} {\bibinfo {author} {\bibfnamefont {J.~B.}\ \bibnamefont {Pendry}}, \bibinfo {author} {\bibfnamefont {A.}~\bibnamefont {MacKinnon}},\ and\ \bibinfo {author} {\bibfnamefont {P.~J.}\ \bibnamefont {Roberts}},\ }\bibfield  {title} {\bibinfo {title} {{Universality classes and fluctuations in disordered systems}},\ }\href {https://doi.org/10.1098/rspa.1992.0047} {\bibfield  {journal} {\bibinfo  {journal} {Proc. R. Soc. Lond. A}\ }\textbf {\bibinfo {volume} {437}},\ \bibinfo {pages} {67} (\bibinfo {year} {1992})}\BibitemShut {NoStop}%
\bibitem [{\citenamefont {Slevin}\ and\ \citenamefont {Ohtsuki}(2001)}]{Slevin2001numerical}%
  \BibitemOpen
  \bibfield  {author} {\bibinfo {author} {\bibfnamefont {K.}~\bibnamefont {Slevin}}\ and\ \bibinfo {author} {\bibfnamefont {T.}~\bibnamefont {Ohtsuki}},\ }\bibfield  {title} {\bibinfo {title} {{Numerical verification of universality for the Anderson transition}},\ }\href@noop {} {\bibfield  {journal} {\bibinfo  {journal} {Phys. Rev. B}\ }\textbf {\bibinfo {volume} {63}},\ \bibinfo {pages} {045108} (\bibinfo {year} {2001})}\BibitemShut {NoStop}%
\bibitem [{\citenamefont {Kobayashi}\ \emph {et~al.}(2013)\citenamefont {Kobayashi}, \citenamefont {Ohtsuki},\ and\ \citenamefont {Imura}}]{Kobayashi2013disordered}%
  \BibitemOpen
  \bibfield  {author} {\bibinfo {author} {\bibfnamefont {K.}~\bibnamefont {Kobayashi}}, \bibinfo {author} {\bibfnamefont {T.}~\bibnamefont {Ohtsuki}},\ and\ \bibinfo {author} {\bibfnamefont {K.-I.}\ \bibnamefont {Imura}},\ }\bibfield  {title} {\bibinfo {title} {{Disordered weak and strong topological insulators}},\ }\href {https://doi.org/10.1103/PhysRevLett.110.236803} {\bibfield  {journal} {\bibinfo  {journal} {Phys. Rev. Lett.}\ }\textbf {\bibinfo {volume} {110}},\ \bibinfo {pages} {236803} (\bibinfo {year} {2013})}\BibitemShut {NoStop}%
\bibitem [{\citenamefont {Kobayashi}\ and\ \citenamefont {Ohtsuki}(2025)}]{Kobayashi2025backsolution}%
  \BibitemOpen
  \bibfield  {author} {\bibinfo {author} {\bibfnamefont {K.}~\bibnamefont {Kobayashi}}\ and\ \bibinfo {author} {\bibfnamefont {T.}~\bibnamefont {Ohtsuki}},\ }\bibfield  {title} {\bibinfo {title} {{Backsolution: A Framework for Solving Inverse Problems via Automatic Differentiation}},\ }\href {https://doi.org/10.48550/arXiv.2506.13210} {\bibfield  {journal} {\bibinfo  {journal} {arXiv:2506.13210}\ } (\bibinfo {year} {2025})}\BibitemShut {NoStop}%
\bibitem [{\citenamefont {Kobayashi}\ \emph {et~al.}(2019)\citenamefont {Kobayashi}, \citenamefont {Takagaki},\ and\ \citenamefont {Nomura}}]{Kobayashi19robust}%
  \BibitemOpen
  \bibfield  {author} {\bibinfo {author} {\bibfnamefont {K.}~\bibnamefont {Kobayashi}}, \bibinfo {author} {\bibfnamefont {M.}~\bibnamefont {Takagaki}},\ and\ \bibinfo {author} {\bibfnamefont {K.}~\bibnamefont {Nomura}},\ }\bibfield  {title} {\bibinfo {title} {{Robust magnetotransport in disordered ferromagnetic kagome layers with quantum anomalous Hall effect}},\ }\href@noop {} {\bibfield  {journal} {\bibinfo  {journal} {Phys. Rev. B}\ }\textbf {\bibinfo {volume} {100}},\ \bibinfo {pages} {161301(R)} (\bibinfo {year} {2019})}\BibitemShut {NoStop}%
\bibitem [{\citenamefont {Kobayashi}\ and\ \citenamefont {Nomura}(2021{\natexlab{a}})}]{Kobayashi2021intrinsic}%
  \BibitemOpen
  \bibfield  {author} {\bibinfo {author} {\bibfnamefont {K.}~\bibnamefont {Kobayashi}}\ and\ \bibinfo {author} {\bibfnamefont {K.}~\bibnamefont {Nomura}},\ }\bibfield  {title} {\bibinfo {title} {{Intrinsic and extrinsic anomalous Hall effects in disordered magnetic Weyl semimetal}},\ }\href@noop {} {\bibfield  {journal} {\bibinfo  {journal} {J. Phys. Soc. Jpn.}\ }\textbf {\bibinfo {volume} {91}},\ \bibinfo {pages} {013703} (\bibinfo {year} {2021}{\natexlab{a}})}\BibitemShut {NoStop}%
\bibitem [{\citenamefont {Aoki}(1987)}]{Aoki87quantized}%
  \BibitemOpen
  \bibfield  {author} {\bibinfo {author} {\bibfnamefont {H.}~\bibnamefont {Aoki}},\ }\bibfield  {title} {\bibinfo {title} {{Quantised Hall effect}},\ }\href {https://doi.org/10.1088/0034-4885/50/6/002} {\bibfield  {journal} {\bibinfo  {journal} {Rep. Prog. Phys}\ }\textbf {\bibinfo {volume} {50}},\ \bibinfo {pages} {655} (\bibinfo {year} {1987})}\BibitemShut {NoStop}%
\bibitem [{\citenamefont {Huckestein}(1995)}]{Huckestein95scaling}%
  \BibitemOpen
  \bibfield  {author} {\bibinfo {author} {\bibfnamefont {B.}~\bibnamefont {Huckestein}},\ }\bibfield  {title} {\bibinfo {title} {{Scaling theory of the integer quantum Hall effect}},\ }\href {https://doi.org/10.1103/RevModPhys.67.357} {\bibfield  {journal} {\bibinfo  {journal} {Rev. Mod. Phys.}\ }\textbf {\bibinfo {volume} {67}},\ \bibinfo {pages} {357} (\bibinfo {year} {1995})}\BibitemShut {NoStop}%
\bibitem [{\citenamefont {Ando}(2015)}]{Ando15theory}%
  \BibitemOpen
  \bibfield  {author} {\bibinfo {author} {\bibfnamefont {T.}~\bibnamefont {Ando}},\ }\bibfield  {title} {\bibinfo {title} {{Theory of Valley Hall Conductivity in Graphene with Gap}},\ }\href {https://doi.org/10.7566/JPSJ.84.114705} {\bibfield  {journal} {\bibinfo  {journal} {J. Phys. Soc. Jpn.}\ }\textbf {\bibinfo {volume} {84}},\ \bibinfo {pages} {114705} (\bibinfo {year} {2015})}\BibitemShut {NoStop}%
\bibitem [{\citenamefont {Nikoli{\'c}}\ and\ \citenamefont {Z{\^a}rbo}(2007)}]{Nikolic07extrinsically}%
  \BibitemOpen
  \bibfield  {author} {\bibinfo {author} {\bibfnamefont {B.}~\bibnamefont {Nikoli{\'c}}}\ and\ \bibinfo {author} {\bibfnamefont {L.}~\bibnamefont {Z{\^a}rbo}},\ }\bibfield  {title} {\bibinfo {title} {Extrinsically vs. intrinsically driven spin hall effect in disordered mesoscopic multiterminal bars},\ }\href@noop {} {\bibfield  {journal} {\bibinfo  {journal} {EPL (Europhysics Letters)}\ }\textbf {\bibinfo {volume} {77}},\ \bibinfo {pages} {47004} (\bibinfo {year} {2007})}\BibitemShut {NoStop}%
\bibitem [{\citenamefont {Datta}(2005)}]{Datta2005quantum}%
  \BibitemOpen
  \bibfield  {author} {\bibinfo {author} {\bibfnamefont {S.}~\bibnamefont {Datta}},\ }\href@noop {} {\emph {\bibinfo {title} {{Quantum Transport: Atom to Transistor}}}}\ (\bibinfo  {publisher} {Cambridge University Press},\ \bibinfo {year} {2005})\BibitemShut {NoStop}%
\bibitem [{\citenamefont {Takane}(2016)}]{Takane16disorder}%
  \BibitemOpen
  \bibfield  {author} {\bibinfo {author} {\bibfnamefont {Y.}~\bibnamefont {Takane}},\ }\bibfield  {title} {\bibinfo {title} {{Disorder Effect on Chiral Edge Modes and Anomalous Hall Conductance in Weyl Semimetals}},\ }\href {http://dx.doi.org/10.7566/JPSJ.85.124711} {\bibfield  {journal} {\bibinfo  {journal} {J. Phys. Soc. Jpn.}\ }\textbf {\bibinfo {volume} {85}},\ \bibinfo {pages} {124711} (\bibinfo {year} {2016})}\BibitemShut {NoStop}%
\bibitem [{\citenamefont {Sinova}\ \emph {et~al.}(2015)\citenamefont {Sinova}, \citenamefont {Valenzuela}, \citenamefont {Wunderlich}, \citenamefont {Back},\ and\ \citenamefont {Jungwirth}}]{sinova2015spin}%
  \BibitemOpen
  \bibfield  {author} {\bibinfo {author} {\bibfnamefont {J.}~\bibnamefont {Sinova}}, \bibinfo {author} {\bibfnamefont {S.~O.}\ \bibnamefont {Valenzuela}}, \bibinfo {author} {\bibfnamefont {J.}~\bibnamefont {Wunderlich}}, \bibinfo {author} {\bibfnamefont {C.}~\bibnamefont {Back}},\ and\ \bibinfo {author} {\bibfnamefont {T.}~\bibnamefont {Jungwirth}},\ }\bibfield  {title} {\bibinfo {title} {Spin hall effects},\ }\href@noop {} {\bibfield  {journal} {\bibinfo  {journal} {Reviews of modern physics}\ }\textbf {\bibinfo {volume} {87}},\ \bibinfo {pages} {1213} (\bibinfo {year} {2015})}\BibitemShut {NoStop}%
\bibitem [{\citenamefont {Saitoh}\ \emph {et~al.}(2006)\citenamefont {Saitoh}, \citenamefont {Ueda}, \citenamefont {Miyajima},\ and\ \citenamefont {Tatara}}]{Saitoh2006}%
  \BibitemOpen
  \bibfield  {author} {\bibinfo {author} {\bibfnamefont {E.}~\bibnamefont {Saitoh}}, \bibinfo {author} {\bibfnamefont {M.}~\bibnamefont {Ueda}}, \bibinfo {author} {\bibfnamefont {H.}~\bibnamefont {Miyajima}},\ and\ \bibinfo {author} {\bibfnamefont {G.}~\bibnamefont {Tatara}},\ }\bibfield  {title} {\bibinfo {title} {Conversion of spin current into charge current at room temperature: Inverse spin-hall effect},\ }\href@noop {} {\bibfield  {journal} {\bibinfo  {journal} {Appl. Phys. Lett.}\ }\textbf {\bibinfo {volume} {88}},\ \bibinfo {pages} {182509} (\bibinfo {year} {2006})}\BibitemShut {NoStop}%
\bibitem [{\citenamefont {Guo}\ \emph {et~al.}(2008)\citenamefont {Guo}, \citenamefont {Murakami}, \citenamefont {Chen},\ and\ \citenamefont {Nagaosa}}]{Guo2008}%
  \BibitemOpen
  \bibfield  {author} {\bibinfo {author} {\bibfnamefont {G.~Y.}\ \bibnamefont {Guo}}, \bibinfo {author} {\bibfnamefont {S.}~\bibnamefont {Murakami}}, \bibinfo {author} {\bibfnamefont {T.-W.}\ \bibnamefont {Chen}},\ and\ \bibinfo {author} {\bibfnamefont {N.}~\bibnamefont {Nagaosa}},\ }\bibfield  {title} {\bibinfo {title} {Intrinsic spin hall effect in platinum: First-principles calculations},\ }\href@noop {} {\bibfield  {journal} {\bibinfo  {journal} {Phys. Rev. Lett.}\ }\textbf {\bibinfo {volume} {100}} (\bibinfo {year} {2008})}\BibitemShut {NoStop}%
\bibitem [{\citenamefont {Liu}\ \emph {et~al.}(2012{\natexlab{c}})\citenamefont {Liu}, \citenamefont {Pai}, \citenamefont {Li}, \citenamefont {Tseng}, \citenamefont {Ralph},\ and\ \citenamefont {Buhrman}}]{Liu2012-pw}%
  \BibitemOpen
  \bibfield  {author} {\bibinfo {author} {\bibfnamefont {L.}~\bibnamefont {Liu}}, \bibinfo {author} {\bibfnamefont {C.-F.}\ \bibnamefont {Pai}}, \bibinfo {author} {\bibfnamefont {Y.}~\bibnamefont {Li}}, \bibinfo {author} {\bibfnamefont {H.~W.}\ \bibnamefont {Tseng}}, \bibinfo {author} {\bibfnamefont {D.~C.}\ \bibnamefont {Ralph}},\ and\ \bibinfo {author} {\bibfnamefont {R.~A.}\ \bibnamefont {Buhrman}},\ }\bibfield  {title} {\bibinfo {title} {Spin-torque switching with the giant spin hall effect of tantalum},\ }\href@noop {} {\bibfield  {journal} {\bibinfo  {journal} {Science}\ }\textbf {\bibinfo {volume} {336}},\ \bibinfo {pages} {555} (\bibinfo {year} {2012}{\natexlab{c}})}\BibitemShut {NoStop}%
\bibitem [{\citenamefont {Pai}\ \emph {et~al.}(2012)\citenamefont {Pai}, \citenamefont {Liu}, \citenamefont {Li}, \citenamefont {Tseng}, \citenamefont {Ralph},\ and\ \citenamefont {Buhrman}}]{pai2012spin}%
  \BibitemOpen
  \bibfield  {author} {\bibinfo {author} {\bibfnamefont {C.-F.}\ \bibnamefont {Pai}}, \bibinfo {author} {\bibfnamefont {L.}~\bibnamefont {Liu}}, \bibinfo {author} {\bibfnamefont {Y.}~\bibnamefont {Li}}, \bibinfo {author} {\bibfnamefont {H.}~\bibnamefont {Tseng}}, \bibinfo {author} {\bibfnamefont {D.}~\bibnamefont {Ralph}},\ and\ \bibinfo {author} {\bibfnamefont {R.}~\bibnamefont {Buhrman}},\ }\bibfield  {title} {\bibinfo {title} {Spin transfer torque devices utilizing the giant spin hall effect of tungsten},\ }\href@noop {} {\bibfield  {journal} {\bibinfo  {journal} {Applied Physics Letters}\ }\textbf {\bibinfo {volume} {101}} (\bibinfo {year} {2012})}\BibitemShut {NoStop}%
\bibitem [{\citenamefont {Dyakonov}(1971)}]{dyakonov1971possibility}%
  \BibitemOpen
  \bibfield  {author} {\bibinfo {author} {\bibfnamefont {M.~I.}\ \bibnamefont {Dyakonov}},\ }\bibfield  {title} {\bibinfo {title} {Possibility of orienting electron spins with current},\ }\href@noop {} {\bibfield  {journal} {\bibinfo  {journal} {JETP Lett. USSR}\ }\textbf {\bibinfo {volume} {13}},\ \bibinfo {pages} {467} (\bibinfo {year} {1971})}\BibitemShut {NoStop}%
\bibitem [{\citenamefont {Murakami}\ \emph {et~al.}(2003)\citenamefont {Murakami}, \citenamefont {Nagaosa},\ and\ \citenamefont {Zhang}}]{murakami2003dissipationless}%
  \BibitemOpen
  \bibfield  {author} {\bibinfo {author} {\bibfnamefont {S.}~\bibnamefont {Murakami}}, \bibinfo {author} {\bibfnamefont {N.}~\bibnamefont {Nagaosa}},\ and\ \bibinfo {author} {\bibfnamefont {S.-C.}\ \bibnamefont {Zhang}},\ }\bibfield  {title} {\bibinfo {title} {Dissipationless quantum spin current at room temperature},\ }\href@noop {} {\bibfield  {journal} {\bibinfo  {journal} {Science}\ }\textbf {\bibinfo {volume} {301}},\ \bibinfo {pages} {1348} (\bibinfo {year} {2003})}\BibitemShut {NoStop}%
\bibitem [{\citenamefont {Murakami}\ \emph {et~al.}(2004)\citenamefont {Murakami}, \citenamefont {Nagaosa},\ and\ \citenamefont {Zhang}}]{murakami2004spin}%
  \BibitemOpen
  \bibfield  {author} {\bibinfo {author} {\bibfnamefont {S.}~\bibnamefont {Murakami}}, \bibinfo {author} {\bibfnamefont {N.}~\bibnamefont {Nagaosa}},\ and\ \bibinfo {author} {\bibfnamefont {S.-C.}\ \bibnamefont {Zhang}},\ }\bibfield  {title} {\bibinfo {title} {Spin-hall insulator},\ }\href@noop {} {\bibfield  {journal} {\bibinfo  {journal} {Physical review letters}\ }\textbf {\bibinfo {volume} {93}},\ \bibinfo {pages} {156804} (\bibinfo {year} {2004})}\BibitemShut {NoStop}%
\bibitem [{\citenamefont {Shi}\ \emph {et~al.}(2006)\citenamefont {Shi}, \citenamefont {Zhang}, \citenamefont {Xiao},\ and\ \citenamefont {Niu}}]{Shi2006}%
  \BibitemOpen
  \bibfield  {author} {\bibinfo {author} {\bibfnamefont {J.}~\bibnamefont {Shi}}, \bibinfo {author} {\bibfnamefont {P.}~\bibnamefont {Zhang}}, \bibinfo {author} {\bibfnamefont {D.}~\bibnamefont {Xiao}},\ and\ \bibinfo {author} {\bibfnamefont {Q.}~\bibnamefont {Niu}},\ }\bibfield  {title} {\bibinfo {title} {Proper definition of spin current in spin-orbit coupled systems},\ }\href@noop {} {\bibfield  {journal} {\bibinfo  {journal} {Phys. Rev. Lett.}\ }\textbf {\bibinfo {volume} {96}},\ \bibinfo {pages} {076604} (\bibinfo {year} {2006})}\BibitemShut {NoStop}%
\bibitem [{\citenamefont {Kane}\ and\ \citenamefont {Mele}(2005)}]{Kane2005}%
  \BibitemOpen
  \bibfield  {author} {\bibinfo {author} {\bibfnamefont {C.~L.}\ \bibnamefont {Kane}}\ and\ \bibinfo {author} {\bibfnamefont {E.~J.}\ \bibnamefont {Mele}},\ }\bibfield  {title} {\bibinfo {title} {Quantum spin hall effect in graphene},\ }\href {https://doi.org/10.1103/PhysRevLett.95.226801} {\bibfield  {journal} {\bibinfo  {journal} {Phys. Rev. Lett.}\ }\textbf {\bibinfo {volume} {95}},\ \bibinfo {pages} {226801} (\bibinfo {year} {2005})}\BibitemShut {NoStop}%
\bibitem [{\citenamefont {Bernevig}\ \emph {et~al.}(2006)\citenamefont {Bernevig}, \citenamefont {Hughes},\ and\ \citenamefont {Zhang}}]{bernevig2006quantum}%
  \BibitemOpen
  \bibfield  {author} {\bibinfo {author} {\bibfnamefont {B.~A.}\ \bibnamefont {Bernevig}}, \bibinfo {author} {\bibfnamefont {T.~L.}\ \bibnamefont {Hughes}},\ and\ \bibinfo {author} {\bibfnamefont {S.-C.}\ \bibnamefont {Zhang}},\ }\bibfield  {title} {\bibinfo {title} {Quantum spin hall effect and topological phase transition in hgte quantum wells},\ }\href@noop {} {\bibfield  {journal} {\bibinfo  {journal} {science}\ }\textbf {\bibinfo {volume} {314}},\ \bibinfo {pages} {1757} (\bibinfo {year} {2006})}\BibitemShut {NoStop}%
\bibitem [{\citenamefont {Wu}\ \emph {et~al.}(2006)\citenamefont {Wu}, \citenamefont {Bernevig},\ and\ \citenamefont {Zhang}}]{wu2006helical}%
  \BibitemOpen
  \bibfield  {author} {\bibinfo {author} {\bibfnamefont {C.}~\bibnamefont {Wu}}, \bibinfo {author} {\bibfnamefont {B.~A.}\ \bibnamefont {Bernevig}},\ and\ \bibinfo {author} {\bibfnamefont {S.-C.}\ \bibnamefont {Zhang}},\ }\bibfield  {title} {\bibinfo {title} {Helical liquid and the edge of quantum spin hall systems},\ }\href@noop {} {\bibfield  {journal} {\bibinfo  {journal} {Physical review letters}\ }\textbf {\bibinfo {volume} {96}},\ \bibinfo {pages} {106401} (\bibinfo {year} {2006})}\BibitemShut {NoStop}%
\bibitem [{\citenamefont {Burkov}\ and\ \citenamefont {Kim}(2016)}]{burkov2016z}%
  \BibitemOpen
  \bibfield  {author} {\bibinfo {author} {\bibfnamefont {A.~A.}\ \bibnamefont {Burkov}}\ and\ \bibinfo {author} {\bibfnamefont {Y.~B.}\ \bibnamefont {Kim}},\ }\bibfield  {title} {\bibinfo {title} {Z 2 and chiral anomalies in topological dirac semimetals},\ }\href@noop {} {\bibfield  {journal} {\bibinfo  {journal} {Physical Review Letters}\ }\textbf {\bibinfo {volume} {117}},\ \bibinfo {pages} {136602} (\bibinfo {year} {2016})}\BibitemShut {NoStop}%
\bibitem [{\citenamefont {Taguchi}\ \emph {et~al.}(2020)\citenamefont {Taguchi}, \citenamefont {Oshima}, \citenamefont {Yamaguchi}, \citenamefont {Hashimoto}, \citenamefont {Tanaka},\ and\ \citenamefont {Sato}}]{taguchi2020spin}%
  \BibitemOpen
  \bibfield  {author} {\bibinfo {author} {\bibfnamefont {K.}~\bibnamefont {Taguchi}}, \bibinfo {author} {\bibfnamefont {D.}~\bibnamefont {Oshima}}, \bibinfo {author} {\bibfnamefont {Y.}~\bibnamefont {Yamaguchi}}, \bibinfo {author} {\bibfnamefont {T.}~\bibnamefont {Hashimoto}}, \bibinfo {author} {\bibfnamefont {Y.}~\bibnamefont {Tanaka}},\ and\ \bibinfo {author} {\bibfnamefont {M.}~\bibnamefont {Sato}},\ }\bibfield  {title} {\bibinfo {title} {Spin hall conductivity in topological dirac semimetals},\ }\href@noop {} {\bibfield  {journal} {\bibinfo  {journal} {Physical Review B}\ }\textbf {\bibinfo {volume} {101}},\ \bibinfo {pages} {235201} (\bibinfo {year} {2020})}\BibitemShut {NoStop}%
\bibitem [{\citenamefont {Lau}\ \emph {et~al.}(2023)\citenamefont {Lau}, \citenamefont {Ikeda}, \citenamefont {Fujiwara}, \citenamefont {Ozawa}, \citenamefont {Zheng}, \citenamefont {Seki}, \citenamefont {Nomura}, \citenamefont {Du}, \citenamefont {Wu}, \citenamefont {Tsukazaki},\ and\ \citenamefont {Takanashi}}]{Lau2023}%
  \BibitemOpen
  \bibfield  {author} {\bibinfo {author} {\bibfnamefont {Y.-C.}\ \bibnamefont {Lau}}, \bibinfo {author} {\bibfnamefont {J.}~\bibnamefont {Ikeda}}, \bibinfo {author} {\bibfnamefont {K.}~\bibnamefont {Fujiwara}}, \bibinfo {author} {\bibfnamefont {A.}~\bibnamefont {Ozawa}}, \bibinfo {author} {\bibfnamefont {J.}~\bibnamefont {Zheng}}, \bibinfo {author} {\bibfnamefont {T.}~\bibnamefont {Seki}}, \bibinfo {author} {\bibfnamefont {K.}~\bibnamefont {Nomura}}, \bibinfo {author} {\bibfnamefont {L.}~\bibnamefont {Du}}, \bibinfo {author} {\bibfnamefont {Q.}~\bibnamefont {Wu}}, \bibinfo {author} {\bibfnamefont {A.}~\bibnamefont {Tsukazaki}},\ and\ \bibinfo {author} {\bibfnamefont {K.}~\bibnamefont {Takanashi}},\ }\bibfield  {title} {\bibinfo {title} {Intercorrelated anomalous hall and spin hall effect in kagome-lattice ${\mathrm{co}}_{3}{\mathrm{sn}}_{2}{\mathrm{s}}_{2}$-based shandite films},\ }\href {https://doi.org/10.1103/PhysRevB.108.064429} {\bibfield  {journal} {\bibinfo  {journal} {Phys. Rev. B}\ }\textbf {\bibinfo
  {volume} {108}},\ \bibinfo {pages} {064429} (\bibinfo {year} {2023})}\BibitemShut {NoStop}%
\bibitem [{\citenamefont {Horiuchi}\ \emph {et~al.}(2025)\citenamefont {Horiuchi}, \citenamefont {Araki}, \citenamefont {Wakabayashi}, \citenamefont {Ieda}, \citenamefont {Yamanouchi}, \citenamefont {Sato}, \citenamefont {Kaneta-Takada}, \citenamefont {Taniyasu}, \citenamefont {Yamamoto}, \citenamefont {Krockenberger} \emph {et~al.}}]{horiuchi2025single}%
  \BibitemOpen
  \bibfield  {author} {\bibinfo {author} {\bibfnamefont {H.}~\bibnamefont {Horiuchi}}, \bibinfo {author} {\bibfnamefont {Y.}~\bibnamefont {Araki}}, \bibinfo {author} {\bibfnamefont {Y.~K.}\ \bibnamefont {Wakabayashi}}, \bibinfo {author} {\bibfnamefont {J.}~\bibnamefont {Ieda}}, \bibinfo {author} {\bibfnamefont {M.}~\bibnamefont {Yamanouchi}}, \bibinfo {author} {\bibfnamefont {Y.}~\bibnamefont {Sato}}, \bibinfo {author} {\bibfnamefont {S.}~\bibnamefont {Kaneta-Takada}}, \bibinfo {author} {\bibfnamefont {Y.}~\bibnamefont {Taniyasu}}, \bibinfo {author} {\bibfnamefont {H.}~\bibnamefont {Yamamoto}}, \bibinfo {author} {\bibfnamefont {Y.}~\bibnamefont {Krockenberger}}, \emph {et~al.},\ }\bibfield  {title} {\bibinfo {title} {Single-layer spin-orbit-torque magnetization switching due to spin berry curvature generated by minute spontaneous atomic displacement in a weyl oxide},\ }\href@noop {} {\bibfield  {journal} {\bibinfo  {journal} {Advanced Materials}\ ,\ \bibinfo {pages} {2416091}} (\bibinfo {year}
  {2025})}\BibitemShut {NoStop}%
\bibitem [{\citenamefont {Fujimoto}\ \emph {et~al.}(2024)\citenamefont {Fujimoto}, \citenamefont {Matsushita},\ and\ \citenamefont {Ogata}}]{fujimoto2024microscopic}%
  \BibitemOpen
  \bibfield  {author} {\bibinfo {author} {\bibfnamefont {J.}~\bibnamefont {Fujimoto}}, \bibinfo {author} {\bibfnamefont {T.}~\bibnamefont {Matsushita}},\ and\ \bibinfo {author} {\bibfnamefont {M.}~\bibnamefont {Ogata}},\ }\bibfield  {title} {\bibinfo {title} {Microscopic theory of spin nernst effect},\ }\href@noop {} {\bibfield  {journal} {\bibinfo  {journal} {Physical Review B}\ }\textbf {\bibinfo {volume} {110}},\ \bibinfo {pages} {174411} (\bibinfo {year} {2024})}\BibitemShut {NoStop}%
\bibitem [{\citenamefont {Bauer}\ \emph {et~al.}(2012)\citenamefont {Bauer}, \citenamefont {Saitoh},\ and\ \citenamefont {Van~Wees}}]{bauer2012spin}%
  \BibitemOpen
  \bibfield  {author} {\bibinfo {author} {\bibfnamefont {G.~E.}\ \bibnamefont {Bauer}}, \bibinfo {author} {\bibfnamefont {E.}~\bibnamefont {Saitoh}},\ and\ \bibinfo {author} {\bibfnamefont {B.~J.}\ \bibnamefont {Van~Wees}},\ }\bibfield  {title} {\bibinfo {title} {Spin caloritronics},\ }\href@noop {} {\bibfield  {journal} {\bibinfo  {journal} {Nature materials}\ }\textbf {\bibinfo {volume} {11}},\ \bibinfo {pages} {391} (\bibinfo {year} {2012})}\BibitemShut {NoStop}%
\bibitem [{\citenamefont {Rammer}\ and\ \citenamefont {Smith}(1986)}]{rammer1986quantum}%
  \BibitemOpen
  \bibfield  {author} {\bibinfo {author} {\bibfnamefont {J.}~\bibnamefont {Rammer}}\ and\ \bibinfo {author} {\bibfnamefont {H.}~\bibnamefont {Smith}},\ }\bibfield  {title} {\bibinfo {title} {Quantum field-theoretical methods in transport theory of metals},\ }\href@noop {} {\bibfield  {journal} {\bibinfo  {journal} {Reviews of modern physics}\ }\textbf {\bibinfo {volume} {58}},\ \bibinfo {pages} {323} (\bibinfo {year} {1986})}\BibitemShut {NoStop}%
\bibitem [{\citenamefont {Oka}\ and\ \citenamefont {Kitamura}(2019)}]{Oka2019}%
  \BibitemOpen
  \bibfield  {author} {\bibinfo {author} {\bibfnamefont {T.}~\bibnamefont {Oka}}\ and\ \bibinfo {author} {\bibfnamefont {S.}~\bibnamefont {Kitamura}},\ }\bibfield  {title} {\bibinfo {title} {Floquet engineering of quantum materials},\ }\href@noop {} {\bibfield  {journal} {\bibinfo  {journal} {Annu. Rev. Condens. Matter Phys.}\ }\textbf {\bibinfo {volume} {10}},\ \bibinfo {pages} {387} (\bibinfo {year} {2019})}\BibitemShut {NoStop}%
\bibitem [{\citenamefont {Yoshimura}\ \emph {et~al.}(2016)\citenamefont {Yoshimura}, \citenamefont {Onishi}, \citenamefont {Kobayashi}, \citenamefont {Ohtsuki},\ and\ \citenamefont {Imura}}]{Yoshimura2016comparative}%
  \BibitemOpen
  \bibfield  {author} {\bibinfo {author} {\bibfnamefont {Y.}~\bibnamefont {Yoshimura}}, \bibinfo {author} {\bibfnamefont {W.}~\bibnamefont {Onishi}}, \bibinfo {author} {\bibfnamefont {K.}~\bibnamefont {Kobayashi}}, \bibinfo {author} {\bibfnamefont {T.}~\bibnamefont {Ohtsuki}},\ and\ \bibinfo {author} {\bibfnamefont {K.-I.}\ \bibnamefont {Imura}},\ }\bibfield  {title} {\bibinfo {title} {{Comparative study of Weyl semimetal and topological/Chern insulators: Thin-film point of view}},\ }\href {https://doi.org/10.1103/PhysRevB.94.235414} {\bibfield  {journal} {\bibinfo  {journal} {Phys. Rev. B}\ }\textbf {\bibinfo {volume} {94}},\ \bibinfo {pages} {235414} (\bibinfo {year} {2016})}\BibitemShut {NoStop}%
\bibitem [{\citenamefont {Qi}\ \emph {et~al.}(2008)\citenamefont {Qi}, \citenamefont {Hughes},\ and\ \citenamefont {Zhang}}]{Qi08topological}%
  \BibitemOpen
  \bibfield  {author} {\bibinfo {author} {\bibfnamefont {X.-L.}\ \bibnamefont {Qi}}, \bibinfo {author} {\bibfnamefont {T.~L.}\ \bibnamefont {Hughes}},\ and\ \bibinfo {author} {\bibfnamefont {S.-C.}\ \bibnamefont {Zhang}},\ }\bibfield  {title} {\bibinfo {title} {Topological field theory of time-reversal invariant insulators},\ }\href@noop {} {\bibfield  {journal} {\bibinfo  {journal} {Physical Review B—Condensed Matter and Materials Physics}\ }\textbf {\bibinfo {volume} {78}},\ \bibinfo {pages} {195424} (\bibinfo {year} {2008})}\BibitemShut {NoStop}%
\bibitem [{\citenamefont {Liu}\ \emph {et~al.}(2010)\citenamefont {Liu}, \citenamefont {Qi}, \citenamefont {Zhang}, \citenamefont {Dai}, \citenamefont {Fang},\ and\ \citenamefont {Zhang}}]{liu2010model}%
  \BibitemOpen
  \bibfield  {author} {\bibinfo {author} {\bibfnamefont {C.-X.}\ \bibnamefont {Liu}}, \bibinfo {author} {\bibfnamefont {X.-L.}\ \bibnamefont {Qi}}, \bibinfo {author} {\bibfnamefont {H.}~\bibnamefont {Zhang}}, \bibinfo {author} {\bibfnamefont {X.}~\bibnamefont {Dai}}, \bibinfo {author} {\bibfnamefont {Z.}~\bibnamefont {Fang}},\ and\ \bibinfo {author} {\bibfnamefont {S.-C.}\ \bibnamefont {Zhang}},\ }\bibfield  {title} {\bibinfo {title} {Model hamiltonian for topological insulators},\ }\href@noop {} {\bibfield  {journal} {\bibinfo  {journal} {Physical Review B—Condensed Matter and Materials Physics}\ }\textbf {\bibinfo {volume} {82}},\ \bibinfo {pages} {045122} (\bibinfo {year} {2010})}\BibitemShut {NoStop}%
\bibitem [{\citenamefont {Ryu}\ and\ \citenamefont {Nomura}(2012)}]{Ryu12disorder}%
  \BibitemOpen
  \bibfield  {author} {\bibinfo {author} {\bibfnamefont {S.}~\bibnamefont {Ryu}}\ and\ \bibinfo {author} {\bibfnamefont {K.}~\bibnamefont {Nomura}},\ }\bibfield  {title} {\bibinfo {title} {Disorder-induced quantum phase transitions in three-dimensional topological insulators and superconductors},\ }\href@noop {} {\bibfield  {journal} {\bibinfo  {journal} {Physical Review B—Condensed Matter and Materials Physics}\ }\textbf {\bibinfo {volume} {85}},\ \bibinfo {pages} {155138} (\bibinfo {year} {2012})}\BibitemShut {NoStop}%
\bibitem [{\citenamefont {Yan}\ \emph {et~al.}(2010)\citenamefont {Yan}, \citenamefont {Liu}, \citenamefont {Zhang}, \citenamefont {Yam}, \citenamefont {Qi}, \citenamefont {Frauenheim},\ and\ \citenamefont {Zhang}}]{yan2010theoretical}%
  \BibitemOpen
  \bibfield  {author} {\bibinfo {author} {\bibfnamefont {B.}~\bibnamefont {Yan}}, \bibinfo {author} {\bibfnamefont {C.-X.}\ \bibnamefont {Liu}}, \bibinfo {author} {\bibfnamefont {H.-J.}\ \bibnamefont {Zhang}}, \bibinfo {author} {\bibfnamefont {C.-Y.}\ \bibnamefont {Yam}}, \bibinfo {author} {\bibfnamefont {X.-L.}\ \bibnamefont {Qi}}, \bibinfo {author} {\bibfnamefont {T.}~\bibnamefont {Frauenheim}},\ and\ \bibinfo {author} {\bibfnamefont {S.-C.}\ \bibnamefont {Zhang}},\ }\bibfield  {title} {\bibinfo {title} {Theoretical prediction of topological insulators in thallium-based iii-v-vi2 ternary chalcogenides},\ }\href@noop {} {\bibfield  {journal} {\bibinfo  {journal} {Europhysics Letters}\ }\textbf {\bibinfo {volume} {90}},\ \bibinfo {pages} {37002} (\bibinfo {year} {2010})}\BibitemShut {NoStop}%
\bibitem [{\citenamefont {Ominato}\ \emph {et~al.}(2019)\citenamefont {Ominato}, \citenamefont {Yamakage},\ and\ \citenamefont {Nomura}}]{ominato2019phase}%
  \BibitemOpen
  \bibfield  {author} {\bibinfo {author} {\bibfnamefont {Y.}~\bibnamefont {Ominato}}, \bibinfo {author} {\bibfnamefont {A.}~\bibnamefont {Yamakage}},\ and\ \bibinfo {author} {\bibfnamefont {K.}~\bibnamefont {Nomura}},\ }\bibfield  {title} {\bibinfo {title} {Phase diagram of a magnetic topological nodal semimetal: stable nodal line in an easy-plane ferromagnet},\ }\href@noop {} {\bibfield  {journal} {\bibinfo  {journal} {journal of the physical society of japan}\ }\textbf {\bibinfo {volume} {88}},\ \bibinfo {pages} {114701} (\bibinfo {year} {2019})}\BibitemShut {NoStop}%
\bibitem [{\citenamefont {Kobayashi}\ and\ \citenamefont {Nomura}(2021{\natexlab{b}})}]{Kobayashi2021ferromagnetic}%
  \BibitemOpen
  \bibfield  {author} {\bibinfo {author} {\bibfnamefont {K.}~\bibnamefont {Kobayashi}}\ and\ \bibinfo {author} {\bibfnamefont {K.}~\bibnamefont {Nomura}},\ }\bibfield  {title} {\bibinfo {title} {Ferromagnetic-electrodes-induced hall effect in topological dirac semimetals},\ }\href@noop {} {\bibfield  {journal} {\bibinfo  {journal} {Physical Review Research}\ }\textbf {\bibinfo {volume} {3}},\ \bibinfo {pages} {033023} (\bibinfo {year} {2021}{\natexlab{b}})}\BibitemShut {NoStop}%
\bibitem [{\citenamefont {Yoshikawa}\ \emph {et~al.}(2025{\natexlab{b}})\citenamefont {Yoshikawa}, \citenamefont {Okumura}, \citenamefont {Hirai}, \citenamefont {Ogawa}, \citenamefont {Fujiwara}, \citenamefont {Ikeda}, \citenamefont {Ozawa}, \citenamefont {Koretsune}, \citenamefont {Arita}, \citenamefont {Mitra}, \citenamefont {Tsukazaki}, \citenamefont {Oka},\ and\ \citenamefont {Shimano}}]{yoshikawa2025}%
  \BibitemOpen
  \bibfield  {author} {\bibinfo {author} {\bibfnamefont {N.}~\bibnamefont {Yoshikawa}}, \bibinfo {author} {\bibfnamefont {S.}~\bibnamefont {Okumura}}, \bibinfo {author} {\bibfnamefont {Y.}~\bibnamefont {Hirai}}, \bibinfo {author} {\bibfnamefont {K.}~\bibnamefont {Ogawa}}, \bibinfo {author} {\bibfnamefont {K.}~\bibnamefont {Fujiwara}}, \bibinfo {author} {\bibfnamefont {J.}~\bibnamefont {Ikeda}}, \bibinfo {author} {\bibfnamefont {A.}~\bibnamefont {Ozawa}}, \bibinfo {author} {\bibfnamefont {T.}~\bibnamefont {Koretsune}}, \bibinfo {author} {\bibfnamefont {R.}~\bibnamefont {Arita}}, \bibinfo {author} {\bibfnamefont {A.}~\bibnamefont {Mitra}}, \bibinfo {author} {\bibfnamefont {A.}~\bibnamefont {Tsukazaki}}, \bibinfo {author} {\bibfnamefont {T.}~\bibnamefont {Oka}},\ and\ \bibinfo {author} {\bibfnamefont {R.}~\bibnamefont {Shimano}},\ }\bibfield  {title} {\bibinfo {title} {Light-induced anomalous hall conductivity in the massive three-dimensional dirac semimetal {Co3Sn2S2}},\ }\href@noop {} {\bibfield  {journal}
  {\bibinfo  {journal} {Phys. Rev. B.}\ }\textbf {\bibinfo {volume} {111}},\ \bibinfo {pages} {245104} (\bibinfo {year} {2025}{\natexlab{b}})}\BibitemShut {NoStop}%
\bibitem [{\citenamefont {Jia}\ \emph {et~al.}(2026)\citenamefont {Jia}, \citenamefont {Li}, \citenamefont {Zou}, \citenamefont {Geng},\ and\ \citenamefont {Jiang}}]{jia2026nonlinear}%
  \BibitemOpen
  \bibfield  {author} {\bibinfo {author} {\bibfnamefont {K.}~\bibnamefont {Jia}}, \bibinfo {author} {\bibfnamefont {H.}~\bibnamefont {Li}}, \bibinfo {author} {\bibfnamefont {M.}~\bibnamefont {Zou}}, \bibinfo {author} {\bibfnamefont {H.}~\bibnamefont {Geng}},\ and\ \bibinfo {author} {\bibfnamefont {H.}~\bibnamefont {Jiang}},\ }\bibfield  {title} {\bibinfo {title} {Nonlinear transport fingerprints of tunable fermi-arc connectivity in magnetic weyl semimetal co3sn2s2},\ }\href@noop {} {\bibfield  {journal} {\bibinfo  {journal} {Reports on Progress in Physics}\ }\textbf {\bibinfo {volume} {89}},\ \bibinfo {pages} {020503} (\bibinfo {year} {2026})}\BibitemShut {NoStop}%
\bibitem [{\citenamefont {Guo}\ and\ \citenamefont {Franz}(2009)}]{Guo2009}%
  \BibitemOpen
  \bibfield  {author} {\bibinfo {author} {\bibfnamefont {H.-M.}\ \bibnamefont {Guo}}\ and\ \bibinfo {author} {\bibfnamefont {M.}~\bibnamefont {Franz}},\ }\bibfield  {title} {\bibinfo {title} {Topological insulator on the kagome lattice},\ }\href {https://doi.org/10.1103/PhysRevB.80.113102} {\bibfield  {journal} {\bibinfo  {journal} {Phys. Rev. B}\ }\textbf {\bibinfo {volume} {80}},\ \bibinfo {pages} {113102} (\bibinfo {year} {2009})}\BibitemShut {NoStop}%
\bibitem [{\citenamefont {Setyawan}\ and\ \citenamefont {Curtarolo}(2010)}]{setyawan2010high}%
  \BibitemOpen
  \bibfield  {author} {\bibinfo {author} {\bibfnamefont {W.}~\bibnamefont {Setyawan}}\ and\ \bibinfo {author} {\bibfnamefont {S.}~\bibnamefont {Curtarolo}},\ }\bibfield  {title} {\bibinfo {title} {High-throughput electronic band structure calculations: Challenges and tools},\ }\href@noop {} {\bibfield  {journal} {\bibinfo  {journal} {Computational materials science}\ }\textbf {\bibinfo {volume} {49}},\ \bibinfo {pages} {299} (\bibinfo {year} {2010})}\BibitemShut {NoStop}%
\end{thebibliography}%
\end{document}